\documentclass[sigplan,10pt,screen]{acmart}
\usepackage[utf8]{inputenc}
\usepackage{ textcomp }

\usepackage{newunicodechar}
\newunicodechar{λ}{\ensuremath{\mathnormal\lambda}}
\newunicodechar{∉}{\ensuremath{\mathnormal\not\in}}
\newunicodechar{∘}{\ensuremath{\textdegree}}
\newunicodechar{Σ}{\ensuremath{\mathnormal\Sigma}}

\usepackage{balance}
\usepackage{csquotes}
\usepackage{agda}
\usepackage{amssymb}
\usepackage{amsmath}
\usepackage{amsthm}
\usepackage{natbib}
\usepackage{xspace}
\usepackage{setspace}
\usepackage{newunicodechar}
\newunicodechar{∈}{\ensuremath{\mathnormal\in}}
\newunicodechar{∀}{\ensuremath{\mathnormal\forall}}
\newunicodechar{₁}{$_1$}
\newunicodechar{₂}{$_2$}
\newunicodechar{₃}{$_3$}
\newunicodechar{≡}{\ensuremath{\mathnormal\equiv}}
\newunicodechar{∷}{\ensuremath{\mathnormal\::}}
\newunicodechar{∃}{\ensuremath{\mathnormal\exists}}
\newunicodechar{⊎}{\ensuremath{\mathnormal\uplus}}
\newunicodechar{λ}{\ensuremath{\mathnormal\lambda}}
\newunicodechar{⊥}{\ensuremath{\mathnormal\bot}}
\newunicodechar{↭}{\ensuremath{\mathnormal\leftrightsquigarrow}}

\begin{code}[hide]%
\>[0]\AgdaKeyword{module}\AgdaSpace{}%
\AgdaModule{Functional.State}\AgdaSpace{}%
\AgdaKeyword{where}\<%
\\
\\[\AgdaEmptyExtraSkip]%
\>[0]\AgdaKeyword{open}\AgdaSpace{}%
\AgdaKeyword{import}\AgdaSpace{}%
\AgdaModule{Agda.Builtin.Equality}\<%
\\
\>[0]\AgdaKeyword{open}\AgdaSpace{}%
\AgdaKeyword{import}\AgdaSpace{}%
\AgdaModule{Data.Bool}\AgdaSpace{}%
\AgdaKeyword{using}\AgdaSpace{}%
\AgdaSymbol{(}\AgdaDatatype{Bool}\AgdaSpace{}%
\AgdaSymbol{;}\AgdaSpace{}%
\AgdaOperator{\AgdaFunction{if\AgdaUnderscore{}then\AgdaUnderscore{}else\AgdaUnderscore{}}}\AgdaSymbol{)}\<%
\\
\>[0]\AgdaKeyword{open}\AgdaSpace{}%
\AgdaKeyword{import}\AgdaSpace{}%
\AgdaModule{Data.List}\AgdaSpace{}%
\AgdaKeyword{using}\AgdaSpace{}%
\AgdaSymbol{(}\AgdaDatatype{List}\AgdaSpace{}%
\AgdaSymbol{;}\AgdaSpace{}%
\AgdaFunction{map}\AgdaSpace{}%
\AgdaSymbol{;}\AgdaSpace{}%
\AgdaFunction{foldr}\AgdaSpace{}%
\AgdaSymbol{;}\AgdaSpace{}%
\AgdaInductiveConstructor{[]}\AgdaSpace{}%
\AgdaSymbol{;}\AgdaSpace{}%
\AgdaOperator{\AgdaInductiveConstructor{\AgdaUnderscore{}∷\AgdaUnderscore{}}}\AgdaSymbol{)}\<%
\\
\>[0]\AgdaKeyword{open}\AgdaSpace{}%
\AgdaKeyword{import}\AgdaSpace{}%
\AgdaModule{Data.Maybe}\AgdaSpace{}%
\AgdaKeyword{using}\AgdaSpace{}%
\AgdaSymbol{(}\AgdaDatatype{Maybe}\AgdaSpace{}%
\AgdaSymbol{;}\AgdaSpace{}%
\AgdaInductiveConstructor{nothing}\AgdaSpace{}%
\AgdaSymbol{;}\AgdaSpace{}%
\AgdaInductiveConstructor{just}\AgdaSymbol{)}\<%
\\
\>[0]\AgdaKeyword{open}\AgdaSpace{}%
\AgdaKeyword{import}\AgdaSpace{}%
\AgdaModule{Data.Product}\AgdaSpace{}%
\AgdaKeyword{using}\AgdaSpace{}%
\AgdaSymbol{(}\AgdaOperator{\AgdaFunction{\AgdaUnderscore{}×\AgdaUnderscore{}}}\AgdaSpace{}%
\AgdaSymbol{;}\AgdaSpace{}%
\AgdaOperator{\AgdaInductiveConstructor{\AgdaUnderscore{},\AgdaUnderscore{}}}\AgdaSpace{}%
\AgdaSymbol{;}\AgdaSpace{}%
\AgdaField{proj₁}\AgdaSpace{}%
\AgdaSymbol{;}\AgdaSpace{}%
\AgdaField{proj₂}\AgdaSpace{}%
\AgdaSymbol{;}\AgdaSpace{}%
\AgdaFunction{Σ-syntax}\AgdaSymbol{)}\<%
\\
\>[0]\AgdaKeyword{open}\AgdaSpace{}%
\AgdaKeyword{import}\AgdaSpace{}%
\AgdaModule{Data.String}\AgdaSpace{}%
\AgdaKeyword{using}\AgdaSpace{}%
\AgdaSymbol{(}\AgdaPostulate{String}\AgdaSpace{}%
\AgdaSymbol{;}\AgdaSpace{}%
\AgdaOperator{\AgdaFunction{\AgdaUnderscore{}==\AgdaUnderscore{}}}\AgdaSymbol{)}\<%
\\
\>[0]\AgdaKeyword{open}\AgdaSpace{}%
\AgdaKeyword{import}\AgdaSpace{}%
\AgdaModule{Data.String.Properties}\AgdaSpace{}%
\AgdaKeyword{using}\AgdaSpace{}%
\AgdaSymbol{(}\AgdaFunction{≡-setoid}\AgdaSpace{}%
\AgdaSymbol{;}\AgdaSpace{}%
\AgdaOperator{\AgdaFunction{\AgdaUnderscore{}≟\AgdaUnderscore{}}}\AgdaSymbol{)}\<%
\\
\>[0]\AgdaKeyword{open}\AgdaSpace{}%
\AgdaKeyword{import}\AgdaSpace{}%
\AgdaModule{Functional.File}\AgdaSpace{}%
\AgdaKeyword{using}\AgdaSpace{}%
\AgdaSymbol{(}\AgdaFunction{File}\AgdaSpace{}%
\AgdaSymbol{;}\AgdaSpace{}%
\AgdaFunction{FileName}\AgdaSpace{}%
\AgdaSymbol{;}\AgdaSpace{}%
\AgdaFunction{FileContent}\AgdaSpace{}%
\AgdaSymbol{;}\AgdaSpace{}%
\AgdaFunction{Files}\AgdaSpace{}%
\AgdaSymbol{;}\AgdaSpace{}%
\AgdaFunction{FileNames}\AgdaSymbol{)}\<%
\\
\>[0]\AgdaKeyword{open}\AgdaSpace{}%
\AgdaKeyword{import}\AgdaSpace{}%
\AgdaModule{Data.List.Membership.Setoid}\AgdaSpace{}%
\AgdaSymbol{(}\AgdaFunction{≡-setoid}\AgdaSymbol{)}\AgdaSpace{}%
\AgdaKeyword{using}\AgdaSpace{}%
\AgdaSymbol{(}\AgdaUnderscore{}∈\AgdaUnderscore{}\AgdaSymbol{)}\<%
\\
\>[0]\AgdaKeyword{open}\AgdaSpace{}%
\AgdaKeyword{import}\AgdaSpace{}%
\AgdaModule{Data.List.Relation.Unary.All}\AgdaSpace{}%
\AgdaKeyword{using}\AgdaSpace{}%
\AgdaSymbol{(}\AgdaDatatype{All}\AgdaSpace{}%
\AgdaSymbol{;}\AgdaSpace{}%
\AgdaFunction{lookup}\AgdaSymbol{)}\<%
\\
\>[0]\AgdaKeyword{open}\AgdaSpace{}%
\AgdaKeyword{import}\AgdaSpace{}%
\AgdaModule{Relation.Binary.PropositionalEquality}\AgdaSpace{}%
\AgdaKeyword{using}\AgdaSpace{}%
\AgdaSymbol{(}\AgdaFunction{decSetoid}\AgdaSpace{}%
\AgdaSymbol{;}\AgdaSpace{}%
\AgdaFunction{trans}\AgdaSpace{}%
\AgdaSymbol{;}\AgdaSpace{}%
\AgdaFunction{sym}\AgdaSpace{}%
\AgdaSymbol{;}\AgdaSpace{}%
\AgdaFunction{subst}\AgdaSymbol{)}\<%
\\
\>[0]\AgdaKeyword{open}\AgdaSpace{}%
\AgdaKeyword{import}\AgdaSpace{}%
\AgdaModule{Data.List.Membership.DecSetoid}\AgdaSpace{}%
\AgdaSymbol{(}\AgdaFunction{decSetoid}\AgdaSpace{}%
\AgdaOperator{\AgdaFunction{\AgdaUnderscore{}≟\AgdaUnderscore{}}}\AgdaSymbol{)}\AgdaSpace{}%
\AgdaKeyword{using}\AgdaSpace{}%
\AgdaSymbol{(}\AgdaUnderscore{}∈?\AgdaUnderscore{} \AgdaSymbol{;} \AgdaUnderscore{}∉\AgdaUnderscore{}\AgdaSymbol{)}\<%
\\
\>[0]\AgdaKeyword{open}\AgdaSpace{}%
\AgdaKeyword{import}\AgdaSpace{}%
\AgdaModule{Relation.Nullary}\AgdaSpace{}%
\AgdaKeyword{using}\AgdaSpace{}%
\AgdaSymbol{(}\AgdaInductiveConstructor{yes}\AgdaSpace{}%
\AgdaSymbol{;}\AgdaSpace{}%
\AgdaInductiveConstructor{no}\AgdaSymbol{)}\<%
\\
\>[0]\AgdaKeyword{open}\AgdaSpace{}%
\AgdaKeyword{import}\AgdaSpace{}%
\AgdaModule{Relation.Nullary.Negation}\AgdaSpace{}%
\AgdaKeyword{using}\AgdaSpace{}%
\AgdaSymbol{(}\AgdaFunction{contradiction}\AgdaSymbol{)}\<%
\\
\>[0]\AgdaKeyword{open}\AgdaSpace{}%
\AgdaKeyword{import}\AgdaSpace{}%
\AgdaModule{Data.List.Relation.Unary.Any}\AgdaSpace{}%
\AgdaKeyword{using}\AgdaSpace{}%
\AgdaSymbol{(}\AgdaInductiveConstructor{here}\AgdaSpace{}%
\AgdaSymbol{;}\AgdaSpace{}%
\AgdaInductiveConstructor{there}\AgdaSpace{}%
\AgdaSymbol{;}\AgdaSpace{}%
\AgdaFunction{tail}\AgdaSymbol{)}\<%
\\
\>[0]\AgdaKeyword{open}\AgdaSpace{}%
\AgdaKeyword{import}\AgdaSpace{}%
\AgdaModule{Data.List.Relation.Binary.Disjoint.Propositional}\AgdaSpace{}%
\AgdaKeyword{using}\AgdaSpace{}%
\AgdaSymbol{(}\AgdaFunction{Disjoint}\AgdaSymbol{)}\<%
\end{code}

\newcommand{\cmdF}{%
\begin{code}%
\>[0]\AgdaFunction{CmdFunction}\AgdaSpace{}%
\AgdaSymbol{:}\AgdaSpace{}%
\AgdaPrimitiveType{Set}\<%
\\
\>[0]\AgdaFunction{CmdFunction}\AgdaSpace{}%
\AgdaSymbol{=}\AgdaSpace{}%
\AgdaFunction{FileSystem}\AgdaSpace{}%
\AgdaSymbol{→}\AgdaSpace{}%
\AgdaDatatype{List}\AgdaSpace{}%
\AgdaFunction{File}\AgdaSpace{}%
\AgdaOperator{\AgdaFunction{×}}\AgdaSpace{}%
\AgdaDatatype{List}\AgdaSpace{}%
\AgdaFunction{File}\<%
\end{code}}

\newcommand{\cmdP}{%
\begin{code}%
\>[0]\AgdaComment{-- names of files read according to cmdFunction}\<%
\\
\>[0]\AgdaFunction{reads}\AgdaSpace{}%
\AgdaSymbol{:}\AgdaSpace{}%
\AgdaFunction{CmdFunction}\AgdaSpace{}%
\AgdaSymbol{→}\AgdaSpace{}%
\AgdaFunction{FileSystem}\AgdaSpace{}%
\AgdaSymbol{→}\AgdaSpace{}%
\AgdaDatatype{List}\AgdaSpace{}%
\AgdaFunction{FileName}\<%
\\
\>[0]\AgdaFunction{reads}\AgdaSpace{}%
\AgdaBound{f}\AgdaSpace{}%
\AgdaBound{s}\AgdaSpace{}%
\AgdaSymbol{=}\AgdaSpace{}%
\AgdaFunction{map}\AgdaSpace{}%
\AgdaField{proj₁}\AgdaSpace{}%
\AgdaSymbol{(}\AgdaField{proj₁}\AgdaSpace{}%
\AgdaSymbol{(}\AgdaBound{f}\AgdaSpace{}%
\AgdaBound{s}\AgdaSymbol{))}\<%
\\
\\[\AgdaEmptyExtraSkip]%
\>[0]\AgdaFunction{CmdProof}\AgdaSpace{}%
\AgdaSymbol{:}\AgdaSpace{}%
\AgdaFunction{CmdFunction}\AgdaSpace{}%
\AgdaSymbol{→}\AgdaSpace{}%
\AgdaPrimitiveType{Set}\<%
\\
\>[0]\AgdaFunction{CmdProof}\AgdaSpace{}%
\AgdaBound{f}\AgdaSpace{}%
\AgdaSymbol{=}\AgdaSpace{}%
\AgdaSymbol{∀}\AgdaSpace{}%
\AgdaBound{s₁}\AgdaSpace{}%
\AgdaBound{s₂}\<%
\\
\>[0][@{}l@{\AgdaIndent{0}}]%
\>[2]\AgdaSymbol{→}\AgdaSpace{}%
\AgdaSymbol{(∀}\AgdaSpace{}%
\AgdaBound{g₁}\AgdaSpace{}%
\AgdaSymbol{→}\AgdaSpace{}%
\AgdaBound{g₁}\AgdaSpace{}%
\AgdaOperator{\AgdaFunction{∈}}\AgdaSpace{}%
\AgdaFunction{reads}\AgdaSpace{}%
\AgdaBound{f}\AgdaSpace{}%
\AgdaBound{s₁}\AgdaSpace{}%
\AgdaSymbol{→}\AgdaSpace{}%
\AgdaBound{s₁}\AgdaSpace{}%
\AgdaBound{g₁}\AgdaSpace{}%
\AgdaOperator{\AgdaDatatype{≡}}\AgdaSpace{}%
\AgdaBound{s₂}\AgdaSpace{}%
\AgdaBound{g₁}\AgdaSymbol{)}\<%
\\
\>[2]\AgdaSymbol{→}\AgdaSpace{}%
\AgdaBound{f}\AgdaSpace{}%
\AgdaBound{s₁}\AgdaSpace{}%
\AgdaOperator{\AgdaDatatype{≡}}\AgdaSpace{}%
\AgdaBound{f}\AgdaSpace{}%
\AgdaBound{s₂}\<%
\end{code}}

\newcommand{\oracle}{%
\begin{code}%
\>[0]\AgdaFunction{Oracle}\AgdaSpace{}%
\AgdaSymbol{:}\AgdaSpace{}%
\AgdaPrimitiveType{Set}\<%
\\
\>[0]\AgdaFunction{Oracle}\AgdaSpace{}%
\AgdaSymbol{=}\AgdaSpace{}%
\AgdaFunction{Cmd}\AgdaSpace{}%
\AgdaSymbol{->}\AgdaSpace{}%
\AgdaFunction{Σ[}\AgdaSpace{}%
\AgdaBound{f}\AgdaSpace{}%
\AgdaFunction{∈}\AgdaSpace{}%
\AgdaFunction{CmdFunction}\AgdaSpace{}%
\AgdaFunction{]}\AgdaSymbol{(}\AgdaFunction{CmdProof}\AgdaSpace{}%
\AgdaBound{f}\AgdaSymbol{)}\<%
\end{code}}

\begin{code}[hide]%
\>[0]\AgdaFunction{extend}\AgdaSpace{}%
\AgdaSymbol{:}\AgdaSpace{}%
\AgdaFunction{File}\AgdaSpace{}%
\AgdaSymbol{->}\AgdaSpace{}%
\AgdaFunction{FileSystem}\AgdaSpace{}%
\AgdaSymbol{->}\AgdaSpace{}%
\AgdaFunction{FileSystem}\<%
\\
\>[0]\AgdaFunction{extend}\AgdaSpace{}%
\AgdaSymbol{(}\AgdaBound{k}\AgdaSpace{}%
\AgdaOperator{\AgdaInductiveConstructor{,}}\AgdaSpace{}%
\AgdaBound{v}\AgdaSymbol{)}\AgdaSpace{}%
\AgdaBound{st}\AgdaSpace{}%
\AgdaSymbol{=}\AgdaSpace{}%
\AgdaSymbol{\textbackslash{}}\AgdaSpace{}%
\AgdaBound{k₁}\AgdaSpace{}%
\AgdaSymbol{->}\AgdaSpace{}%
\AgdaOperator{\AgdaFunction{if}}\AgdaSpace{}%
\AgdaSymbol{(}\AgdaBound{k}\AgdaSpace{}%
\AgdaOperator{\AgdaFunction{==}}\AgdaSpace{}%
\AgdaBound{k₁}\AgdaSymbol{)}\AgdaSpace{}%
\AgdaOperator{\AgdaFunction{then}}\AgdaSpace{}%
\AgdaInductiveConstructor{just}\AgdaSpace{}%
\AgdaBound{v}\AgdaSpace{}%
\AgdaOperator{\AgdaFunction{else}}\AgdaSpace{}%
\AgdaBound{st}\AgdaSpace{}%
\AgdaBound{k₁}\<%
\\
\\[\AgdaEmptyExtraSkip]%
\>[0]\AgdaFunction{read}\AgdaSpace{}%
\AgdaSymbol{:}\AgdaSpace{}%
\AgdaDatatype{List}\AgdaSpace{}%
\AgdaFunction{FileName}\AgdaSpace{}%
\AgdaSymbol{->}\AgdaSpace{}%
\AgdaFunction{FileSystem}\AgdaSpace{}%
\AgdaSymbol{->}\AgdaSpace{}%
\AgdaDatatype{List}\AgdaSpace{}%
\AgdaSymbol{(}\AgdaFunction{FileName}\AgdaSpace{}%
\AgdaOperator{\AgdaFunction{×}}\AgdaSpace{}%
\AgdaDatatype{Maybe}\AgdaSpace{}%
\AgdaFunction{FileContent}\AgdaSymbol{)}\<%
\\
\>[0]\AgdaFunction{read}\AgdaSpace{}%
\AgdaBound{fs}\AgdaSpace{}%
\AgdaBound{sys}\AgdaSpace{}%
\AgdaSymbol{=}\AgdaSpace{}%
\AgdaFunction{map}\AgdaSpace{}%
\AgdaSymbol{(\textbackslash{}}\AgdaSpace{}%
\AgdaBound{x}\AgdaSpace{}%
\AgdaSymbol{->}\AgdaSpace{}%
\AgdaSymbol{(}\AgdaBound{x}\AgdaSpace{}%
\AgdaOperator{\AgdaInductiveConstructor{,}}\AgdaSpace{}%
\AgdaBound{sys}\AgdaSpace{}%
\AgdaBound{x}\AgdaSymbol{))}\AgdaSpace{}%
\AgdaBound{fs}\<%
\end{code}

\begin{code}[hide]%
\>[0]\AgdaFunction{save}\AgdaSpace{}%
\AgdaSymbol{:}\AgdaSpace{}%
\AgdaFunction{Cmd}\AgdaSpace{}%
\AgdaSymbol{->}\AgdaSpace{}%
\AgdaDatatype{List}\AgdaSpace{}%
\AgdaFunction{FileName}\AgdaSpace{}%
\AgdaSymbol{->}\AgdaSpace{}%
\AgdaFunction{FileSystem}\AgdaSpace{}%
\AgdaSymbol{->}\AgdaSpace{}%
\AgdaFunction{Memory}\AgdaSpace{}%
\AgdaSymbol{->}\AgdaSpace{}%
\AgdaFunction{Memory}\<%
\\
\>[0]\AgdaFunction{save}\AgdaSpace{}%
\AgdaBound{cmd}\AgdaSpace{}%
\AgdaBound{files}\AgdaSpace{}%
\AgdaBound{sys}\AgdaSpace{}%
\AgdaBound{mm}\AgdaSpace{}%
\AgdaSymbol{=}\AgdaSpace{}%
\AgdaSymbol{(}\AgdaBound{cmd}\AgdaSpace{}%
\AgdaOperator{\AgdaInductiveConstructor{,}}\AgdaSpace{}%
\AgdaFunction{map}\AgdaSpace{}%
\AgdaSymbol{(\textbackslash{}}\AgdaBound{f}\AgdaSpace{}%
\AgdaSymbol{->}\AgdaSpace{}%
\AgdaBound{f}\AgdaSpace{}%
\AgdaOperator{\AgdaInductiveConstructor{,}}\AgdaSpace{}%
\AgdaBound{sys}\AgdaSpace{}%
\AgdaBound{f}\AgdaSymbol{)}\AgdaSpace{}%
\AgdaBound{files}\AgdaSymbol{)}\AgdaSpace{}%
\AgdaOperator{\AgdaInductiveConstructor{∷}}\AgdaSpace{}%
\AgdaBound{mm}\<%
\end{code}

\begin{code}[hide]%
\>[0]\AgdaKeyword{open}\AgdaSpace{}%
\AgdaKeyword{import}\AgdaSpace{}%
\AgdaModule{Functional.State}\AgdaSpace{}%
\AgdaKeyword{using}\AgdaSpace{}%
\AgdaSymbol{(}\AgdaFunction{Oracle}\AgdaSpace{}%
\AgdaSymbol{;}\AgdaSpace{}%
\AgdaFunction{Cmd}\AgdaSpace{}%
\AgdaSymbol{;}\AgdaSpace{}%
\AgdaFunction{FileSystem}\AgdaSpace{}%
\AgdaSymbol{;}\AgdaSpace{}%
\AgdaFunction{CmdProof}\AgdaSpace{}%
\AgdaSymbol{;}\AgdaSpace{}%
\AgdaFunction{CmdFunction}\AgdaSpace{}%
\AgdaSymbol{;}\AgdaSpace{}%
\AgdaFunction{extend}\AgdaSymbol{)}\<%
\\
\\[\AgdaEmptyExtraSkip]%
\>[0]\AgdaKeyword{module}\AgdaSpace{}%
\AgdaModule{Functional.State.Helpers}\AgdaSpace{}%
\AgdaSymbol{(}\AgdaBound{oracle}\AgdaSpace{}%
\AgdaSymbol{:}\AgdaSpace{}%
\AgdaFunction{Oracle}\AgdaSymbol{)}\AgdaSpace{}%
\AgdaKeyword{where}\<%
\\
\\[\AgdaEmptyExtraSkip]%
\>[0]\AgdaKeyword{open}\AgdaSpace{}%
\AgdaKeyword{import}\AgdaSpace{}%
\AgdaModule{Functional.File}\AgdaSpace{}%
\AgdaKeyword{using}\AgdaSpace{}%
\AgdaSymbol{(}\AgdaFunction{Files}\AgdaSpace{}%
\AgdaSymbol{;}\AgdaSpace{}%
\AgdaFunction{File}\AgdaSpace{}%
\AgdaSymbol{;}\AgdaSpace{}%
\AgdaFunction{FileNames}\AgdaSymbol{)}\<%
\\
\>[0]\AgdaKeyword{open}\AgdaSpace{}%
\AgdaKeyword{import}\AgdaSpace{}%
\AgdaModule{Data.Product}\AgdaSpace{}%
\AgdaKeyword{using}\AgdaSpace{}%
\AgdaSymbol{(}\AgdaFunction{map}\AgdaSpace{}%
\AgdaSymbol{;}\AgdaSpace{}%
\AgdaField{proj₁}\AgdaSpace{}%
\AgdaSymbol{;}\AgdaSpace{}%
\AgdaField{proj₂}\AgdaSpace{}%
\AgdaSymbol{;}\AgdaSpace{}%
\AgdaOperator{\AgdaFunction{\AgdaUnderscore{}×\AgdaUnderscore{}}}\AgdaSymbol{)}\<%
\\
\>[0]\AgdaKeyword{open}\AgdaSpace{}%
\AgdaKeyword{import}\AgdaSpace{}%
\AgdaModule{Function}\AgdaSpace{}%
\AgdaKeyword{using}\AgdaSpace{}%
\AgdaSymbol{(}\AgdaOperator{\AgdaFunction{\AgdaUnderscore{}∘\AgdaUnderscore{}}}\AgdaSpace{}%
\AgdaSymbol{;}\AgdaSpace{}%
\AgdaOperator{\AgdaFunction{\AgdaUnderscore{}∘₂\AgdaUnderscore{}}}\AgdaSymbol{)}\<%
\\
\>[0]\AgdaKeyword{open}\AgdaSpace{}%
\AgdaKeyword{import}\AgdaSpace{}%
\AgdaModule{Data.List}\AgdaSpace{}%
\AgdaSymbol{as}\AgdaSpace{}%
\AgdaModule{L}\AgdaSpace{}%
\AgdaKeyword{hiding}\AgdaSpace{}%
\AgdaSymbol{(}\AgdaFunction{map}\AgdaSymbol{)}\<%
\\
\\[\AgdaEmptyExtraSkip]%
\>[0]\AgdaComment{-- commands for accessing State related data structures. In it's own file so we can parameterize}\<%
\\
\>[0]\AgdaComment{-- over F so each function doesn't need to take the oracle as an argument.}\<%
\\
\\[\AgdaEmptyExtraSkip]%
\>[0]\AgdaFunction{getNames}\AgdaSpace{}%
\AgdaSymbol{:}\AgdaSpace{}%
\AgdaFunction{Files}\AgdaSpace{}%
\AgdaSymbol{→}\AgdaSpace{}%
\AgdaFunction{FileNames}\<%
\\
\>[0]\AgdaFunction{getNames}\AgdaSpace{}%
\AgdaSymbol{=}\AgdaSpace{}%
\AgdaFunction{L.map}\AgdaSpace{}%
\AgdaField{proj₁}\<%
\\
\\[\AgdaEmptyExtraSkip]%
\>[0]\AgdaFunction{getCmdFunction}\AgdaSpace{}%
\AgdaSymbol{:}\AgdaSpace{}%
\AgdaFunction{Cmd}\AgdaSpace{}%
\AgdaSymbol{→}\AgdaSpace{}%
\AgdaFunction{CmdFunction}\<%
\\
\>[0]\AgdaFunction{getCmdFunction}\AgdaSpace{}%
\AgdaSymbol{=}\AgdaSpace{}%
\AgdaField{proj₁}\AgdaSpace{}%
\AgdaOperator{\AgdaFunction{∘}}\AgdaSpace{}%
\AgdaBound{oracle}\<%
\\
\\[\AgdaEmptyExtraSkip]%
\>[0]\AgdaFunction{cmdReads}\AgdaSpace{}%
\AgdaSymbol{:}\AgdaSpace{}%
\AgdaFunction{Cmd}\AgdaSpace{}%
\AgdaSymbol{→}\AgdaSpace{}%
\AgdaFunction{FileSystem}\AgdaSpace{}%
\AgdaSymbol{→}\AgdaSpace{}%
\AgdaFunction{Files}\<%
\\
\>[0]\AgdaFunction{cmdReads}\AgdaSpace{}%
\AgdaSymbol{=}\AgdaSpace{}%
\AgdaField{proj₁}\AgdaSpace{}%
\AgdaOperator{\AgdaFunction{∘₂}}\AgdaSpace{}%
\AgdaFunction{getCmdFunction}\<%
\\
\\[\AgdaEmptyExtraSkip]%
\>[0]\AgdaFunction{cmdWrites}\AgdaSpace{}%
\AgdaSymbol{:}\AgdaSpace{}%
\AgdaFunction{Cmd}\AgdaSpace{}%
\AgdaSymbol{→}\AgdaSpace{}%
\AgdaFunction{FileSystem}\AgdaSpace{}%
\AgdaSymbol{→}\AgdaSpace{}%
\AgdaFunction{Files}\<%
\\
\>[0]\AgdaFunction{cmdWrites}\AgdaSpace{}%
\AgdaSymbol{=}\AgdaSpace{}%
\AgdaField{proj₂}\AgdaSpace{}%
\AgdaOperator{\AgdaFunction{∘₂}}\AgdaSpace{}%
\AgdaFunction{getCmdFunction}\<%
\\
\\[\AgdaEmptyExtraSkip]%
\>[0]\AgdaFunction{cmdWriteNames}\AgdaSpace{}%
\AgdaSymbol{:}\AgdaSpace{}%
\AgdaFunction{Cmd}\AgdaSpace{}%
\AgdaSymbol{→}\AgdaSpace{}%
\AgdaFunction{FileSystem}\AgdaSpace{}%
\AgdaSymbol{→}\AgdaSpace{}%
\AgdaFunction{FileNames}\<%
\\
\>[0]\AgdaFunction{cmdWriteNames}\AgdaSpace{}%
\AgdaSymbol{=}\AgdaSpace{}%
\AgdaFunction{getNames}\AgdaSpace{}%
\AgdaOperator{\AgdaFunction{∘₂}}\AgdaSpace{}%
\AgdaFunction{cmdWrites}\<%
\\
\\[\AgdaEmptyExtraSkip]%
\>[0]\AgdaFunction{cmdFiles}\AgdaSpace{}%
\AgdaSymbol{:}\AgdaSpace{}%
\AgdaFunction{Cmd}\AgdaSpace{}%
\AgdaSymbol{→}\AgdaSpace{}%
\AgdaFunction{FileSystem}\AgdaSpace{}%
\AgdaSymbol{→}\AgdaSpace{}%
\AgdaFunction{Files}\AgdaSpace{}%
\AgdaOperator{\AgdaFunction{×}}\AgdaSpace{}%
\AgdaFunction{Files}\<%
\\
\>[0]\AgdaFunction{cmdFiles}\AgdaSpace{}%
\AgdaSymbol{=}\AgdaSpace{}%
\AgdaFunction{getCmdFunction}\<%
\\
\\[\AgdaEmptyExtraSkip]%
\>[0]\AgdaFunction{cmdReadNames}\AgdaSpace{}%
\AgdaSymbol{:}\AgdaSpace{}%
\AgdaFunction{Cmd}\AgdaSpace{}%
\AgdaSymbol{→}\AgdaSpace{}%
\AgdaFunction{FileSystem}\AgdaSpace{}%
\AgdaSymbol{→}\AgdaSpace{}%
\AgdaFunction{FileNames}\<%
\\
\>[0]\AgdaFunction{cmdReadNames}\AgdaSpace{}%
\AgdaSymbol{=}\AgdaSpace{}%
\AgdaFunction{getNames}\AgdaSpace{}%
\AgdaOperator{\AgdaFunction{∘₂}}\AgdaSpace{}%
\AgdaFunction{cmdReads}\<%
\\
\\[\AgdaEmptyExtraSkip]%
\>[0]\AgdaFunction{cmdReadWriteNames}\AgdaSpace{}%
\AgdaSymbol{:}\AgdaSpace{}%
\AgdaFunction{Cmd}\AgdaSpace{}%
\AgdaSymbol{→}\AgdaSpace{}%
\AgdaFunction{FileSystem}\AgdaSpace{}%
\AgdaSymbol{→}\AgdaSpace{}%
\AgdaFunction{FileNames}\<%
\\
\>[0]\AgdaFunction{cmdReadWriteNames}\AgdaSpace{}%
\AgdaBound{x}\AgdaSpace{}%
\AgdaBound{s}\AgdaSpace{}%
\AgdaSymbol{=}\AgdaSpace{}%
\AgdaSymbol{(}\AgdaFunction{cmdReadNames}\AgdaSpace{}%
\AgdaBound{x}\AgdaSpace{}%
\AgdaBound{s}\AgdaSymbol{)}\AgdaSpace{}%
\AgdaOperator{\AgdaFunction{++}}\AgdaSpace{}%
\AgdaSymbol{(}\AgdaFunction{cmdWriteNames}\AgdaSpace{}%
\AgdaBound{x}\AgdaSpace{}%
\AgdaBound{s}\AgdaSymbol{)}\<%
\\
\\[\AgdaEmptyExtraSkip]%
\>[0]\AgdaFunction{trace}\AgdaSpace{}%
\AgdaSymbol{:}\AgdaSpace{}%
\AgdaFunction{Cmd}\AgdaSpace{}%
\AgdaSymbol{→}\AgdaSpace{}%
\AgdaFunction{FileSystem}\AgdaSpace{}%
\AgdaSymbol{→}\AgdaSpace{}%
\AgdaFunction{FileNames}\AgdaSpace{}%
\AgdaOperator{\AgdaFunction{×}}\AgdaSpace{}%
\AgdaFunction{FileNames}\<%
\\
\>[0]\AgdaFunction{trace}\AgdaSpace{}%
\AgdaSymbol{=}\AgdaSpace{}%
\AgdaSymbol{(}\AgdaFunction{map}\AgdaSpace{}%
\AgdaFunction{getNames}\AgdaSpace{}%
\AgdaFunction{getNames}\AgdaSymbol{)}\AgdaSpace{}%
\AgdaOperator{\AgdaFunction{∘₂}}\AgdaSpace{}%
\AgdaFunction{cmdFiles}\<%
\end{code}

\newcommand{\run}{%
\begin{code}%
\>[0]\AgdaComment{-- writes according to Cmd's CmdFunction}\<%
\\
\>[0]\AgdaFunction{writes}\AgdaSpace{}%
\AgdaSymbol{:}\AgdaSpace{}%
\AgdaFunction{Cmd}\AgdaSpace{}%
\AgdaSymbol{→}\AgdaSpace{}%
\AgdaFunction{FileSystem}\AgdaSpace{}%
\AgdaSymbol{→}\AgdaSpace{}%
\AgdaDatatype{List}\AgdaSpace{}%
\AgdaFunction{File}\<%
\\
\>[0]\AgdaFunction{writes}\AgdaSpace{}%
\AgdaSymbol{=}\AgdaSpace{}%
\AgdaField{proj₂}\AgdaSpace{}%
\AgdaOperator{\AgdaFunction{∘₂}}\AgdaSpace{}%
\AgdaSymbol{(}\AgdaField{proj₁}\AgdaSpace{}%
\AgdaOperator{\AgdaFunction{∘}}\AgdaSpace{}%
\AgdaBound{oracle}\AgdaSymbol{)}\<%
\\
\\[\AgdaEmptyExtraSkip]%
\>[0]\AgdaFunction{run}\AgdaSpace{}%
\AgdaSymbol{:}\AgdaSpace{}%
\AgdaFunction{Cmd}\AgdaSpace{}%
\AgdaSymbol{→}\AgdaSpace{}%
\AgdaFunction{FileSystem}\AgdaSpace{}%
\AgdaSymbol{→}\AgdaSpace{}%
\AgdaFunction{FileSystem}\<%
\\
\>[0]\AgdaFunction{run}\AgdaSpace{}%
\AgdaBound{x}\AgdaSpace{}%
\AgdaBound{s}\AgdaSpace{}%
\AgdaSymbol{=}\AgdaSpace{}%
\AgdaFunction{foldr}\AgdaSpace{}%
\AgdaFunction{extend}\AgdaSpace{}%
\AgdaBound{s}\AgdaSpace{}%
\AgdaSymbol{(}\AgdaFunction{writes}\AgdaSpace{}%
\AgdaBound{x}\AgdaSpace{}%
\AgdaBound{s}\AgdaSymbol{)}\<%
\end{code}}

\begin{code}[hide]%
\>[0]\AgdaKeyword{open}\AgdaSpace{}%
\AgdaKeyword{import}\AgdaSpace{}%
\AgdaModule{Functional.State}\AgdaSpace{}%
\AgdaKeyword{using}\AgdaSpace{}%
\AgdaSymbol{(}\AgdaFunction{Oracle}\AgdaSpace{}%
\AgdaSymbol{;}\AgdaSpace{}%
\AgdaFunction{Cmd}\AgdaSpace{}%
\AgdaSymbol{;}\AgdaSpace{}%
\AgdaFunction{FileSystem}\AgdaSymbol{)}\<%
\\
\\[\AgdaEmptyExtraSkip]%
\>[0]\AgdaKeyword{module}\AgdaSpace{}%
\AgdaModule{Functional.Build}\AgdaSpace{}%
\AgdaSymbol{(}\AgdaBound{oracle}\AgdaSpace{}%
\AgdaSymbol{:}\AgdaSpace{}%
\AgdaFunction{Oracle}\AgdaSymbol{)}\AgdaSpace{}%
\AgdaKeyword{where}\<%
\\
\\[\AgdaEmptyExtraSkip]%
\>[0]\AgdaKeyword{open}\AgdaSpace{}%
\AgdaKeyword{import}\AgdaSpace{}%
\AgdaModule{Data.List}\AgdaSpace{}%
\AgdaKeyword{using}\AgdaSpace{}%
\AgdaSymbol{(}\AgdaDatatype{List}\AgdaSpace{}%
\AgdaSymbol{;}\AgdaSpace{}%
\AgdaInductiveConstructor{[]}\AgdaSpace{}%
\AgdaSymbol{;}\AgdaSpace{}%
\AgdaOperator{\AgdaInductiveConstructor{\AgdaUnderscore{}∷\AgdaUnderscore{}}}\AgdaSymbol{)}\<%
\\
\>[0]\AgdaKeyword{open}\AgdaSpace{}%
\AgdaKeyword{import}\AgdaSpace{}%
\AgdaModule{Functional.State.Helpers}\AgdaSpace{}%
\AgdaSymbol{(}\AgdaBound{oracle}\AgdaSymbol{)}\AgdaSpace{}%
\AgdaKeyword{using}\AgdaSpace{}%
\AgdaSymbol{(}cmdReadNames \AgdaSymbol{;} cmdWriteNames \AgdaSymbol{;} run\AgdaSymbol{)}\<%
\\
\>[0]\AgdaKeyword{open}\AgdaSpace{}%
\AgdaKeyword{import}\AgdaSpace{}%
\AgdaModule{Data.List.Relation.Unary.Unique.Propositional}\AgdaSpace{}%
\AgdaKeyword{using}\AgdaSpace{}%
\AgdaSymbol{(}\AgdaDatatype{Unique}\AgdaSymbol{)}\<%
\\
\>[0]\AgdaKeyword{open}\AgdaSpace{}%
\AgdaKeyword{import}\AgdaSpace{}%
\AgdaModule{Data.List.Relation.Binary.Disjoint.Propositional}\AgdaSpace{}%
\AgdaKeyword{using}\AgdaSpace{}%
\AgdaSymbol{(}\AgdaFunction{Disjoint}\AgdaSymbol{)}\<%
\\
\>[0]\AgdaKeyword{open}\AgdaSpace{}%
\AgdaKeyword{import}\AgdaSpace{}%
\AgdaModule{Data.Product}\AgdaSpace{}%
\AgdaKeyword{using}\AgdaSpace{}%
\AgdaSymbol{(}\AgdaOperator{\AgdaFunction{\AgdaUnderscore{}×\AgdaUnderscore{}}}\AgdaSymbol{)}\<%
\\
\>[0]\AgdaKeyword{open}\AgdaSpace{}%
\AgdaKeyword{import}\AgdaSpace{}%
\AgdaModule{Data.List.Relation.Binary.Permutation.Propositional}\AgdaSpace{}%
\AgdaKeyword{using}\AgdaSpace{}%
\AgdaSymbol{(}\AgdaOperator{\AgdaDatatype{\AgdaUnderscore{}↭\AgdaUnderscore{}}}\AgdaSymbol{)}\<%
\end{code}

\begin{code}[hide]%
\>[0]\AgdaFunction{UniqueEvidence}\AgdaSpace{}%
\AgdaSymbol{:}\AgdaSpace{}%
\AgdaFunction{Build}\AgdaSpace{}%
\AgdaSymbol{→}\AgdaSpace{}%
\AgdaFunction{Build}\AgdaSpace{}%
\AgdaSymbol{→}\AgdaSpace{}%
\AgdaDatatype{List}\AgdaSpace{}%
\AgdaFunction{Cmd}\AgdaSpace{}%
\AgdaSymbol{→}\AgdaSpace{}%
\AgdaPrimitiveType{Set}\<%
\\
\>[0]\AgdaFunction{UniqueEvidence}\AgdaSpace{}%
\AgdaBound{b₁}\AgdaSpace{}%
\AgdaBound{b₂}\AgdaSpace{}%
\AgdaBound{ls}\AgdaSpace{}%
\AgdaSymbol{=}\AgdaSpace{}%
\AgdaDatatype{Unique}\AgdaSpace{}%
\AgdaBound{b₁}\AgdaSpace{}%
\AgdaOperator{\AgdaFunction{×}}\AgdaSpace{}%
\AgdaDatatype{Unique}\AgdaSpace{}%
\AgdaBound{b₂}\AgdaSpace{}%
\AgdaOperator{\AgdaFunction{×}}\AgdaSpace{}%
\AgdaDatatype{Unique}\AgdaSpace{}%
\AgdaBound{ls}\AgdaSpace{}%
\AgdaOperator{\AgdaFunction{×}}\AgdaSpace{}%
\AgdaFunction{Disjoint}\AgdaSpace{}%
\AgdaBound{b₁}\AgdaSpace{}%
\AgdaBound{ls}\<%
\\
\\[\AgdaEmptyExtraSkip]%
\\[\AgdaEmptyExtraSkip]%
\>[0]\AgdaKeyword{data}\AgdaSpace{}%
\AgdaDatatype{DisjointBuild}\AgdaSpace{}%
\AgdaSymbol{:}\AgdaSpace{}%
\AgdaFunction{FileSystem}\AgdaSpace{}%
\AgdaSymbol{->}\AgdaSpace{}%
\AgdaFunction{Build}\AgdaSpace{}%
\AgdaSymbol{->}\AgdaSpace{}%
\AgdaPrimitiveType{Set}\AgdaSpace{}%
\AgdaKeyword{where}\<%
\\
\>[0][@{}l@{\AgdaIndent{0}}]%
\>[2]\AgdaInductiveConstructor{Null}\AgdaSpace{}%
\AgdaSymbol{:}\AgdaSpace{}%
\AgdaSymbol{∀}\AgdaSpace{}%
\AgdaSymbol{\{}\AgdaBound{s}\AgdaSymbol{\}}\AgdaSpace{}%
\AgdaSymbol{→}\AgdaSpace{}%
\AgdaDatatype{DisjointBuild}\AgdaSpace{}%
\AgdaBound{s}\AgdaSpace{}%
\AgdaInductiveConstructor{[]}\<%
\\
\>[2]\AgdaInductiveConstructor{Cons}\AgdaSpace{}%
\AgdaSymbol{:}\AgdaSpace{}%
\AgdaSymbol{∀}\AgdaSpace{}%
\AgdaSymbol{\{}\AgdaBound{s}\AgdaSymbol{\}}\AgdaSpace{}%
\AgdaBound{x}\AgdaSpace{}%
\AgdaSymbol{->}\AgdaSpace{}%
\AgdaFunction{Disjoint}\AgdaSpace{}%
\AgdaSymbol{(}\AgdaFunction{cmdReadNames}\AgdaSpace{}%
\AgdaBound{x}\AgdaSpace{}%
\AgdaBound{s}\AgdaSymbol{)}\AgdaSpace{}%
\AgdaSymbol{(}\AgdaFunction{cmdWriteNames}\AgdaSpace{}%
\AgdaBound{x}\AgdaSpace{}%
\AgdaBound{s}\AgdaSymbol{)}\AgdaSpace{}%
\AgdaSymbol{->}\AgdaSpace{}%
\AgdaSymbol{(}\AgdaBound{b}\AgdaSpace{}%
\AgdaSymbol{:}\AgdaSpace{}%
\AgdaFunction{Build}\AgdaSymbol{)}\AgdaSpace{}%
\AgdaSymbol{->}\AgdaSpace{}%
\AgdaDatatype{DisjointBuild}\AgdaSpace{}%
\AgdaSymbol{(}\AgdaFunction{run}\AgdaSpace{}%
\AgdaBound{x}\AgdaSpace{}%
\AgdaBound{s}\AgdaSymbol{)}\AgdaSpace{}%
\AgdaBound{b}\AgdaSpace{}%
\AgdaSymbol{->}\AgdaSpace{}%
\AgdaDatatype{DisjointBuild}\AgdaSpace{}%
\AgdaBound{s}\AgdaSpace{}%
\AgdaSymbol{(}\AgdaBound{x}\AgdaSpace{}%
\AgdaOperator{\AgdaInductiveConstructor{∷}}\AgdaSpace{}%
\AgdaBound{b}\AgdaSymbol{)}\<%
\\
\\[\AgdaEmptyExtraSkip]%
\\[\AgdaEmptyExtraSkip]%
\>[0]\AgdaFunction{PreCond}\AgdaSpace{}%
\AgdaSymbol{:}\AgdaSpace{}%
\AgdaFunction{FileSystem}\AgdaSpace{}%
\AgdaSymbol{→}\AgdaSpace{}%
\AgdaFunction{Build}\AgdaSpace{}%
\AgdaSymbol{→}\AgdaSpace{}%
\AgdaFunction{Build}\AgdaSpace{}%
\AgdaSymbol{→}\AgdaSpace{}%
\AgdaPrimitiveType{Set}\<%
\\
\>[0]\AgdaFunction{PreCond}\AgdaSpace{}%
\AgdaBound{s}\AgdaSpace{}%
\AgdaBound{br}\AgdaSpace{}%
\AgdaBound{bc}\AgdaSpace{}%
\AgdaSymbol{=}\AgdaSpace{}%
\AgdaDatatype{DisjointBuild}\AgdaSpace{}%
\AgdaBound{s}\AgdaSpace{}%
\AgdaBound{br}\AgdaSpace{}%
\AgdaOperator{\AgdaFunction{×}}\AgdaSpace{}%
\AgdaDatatype{Unique}\AgdaSpace{}%
\AgdaBound{br}\AgdaSpace{}%
\AgdaOperator{\AgdaFunction{×}}\AgdaSpace{}%
\AgdaDatatype{Unique}\AgdaSpace{}%
\AgdaBound{bc}\AgdaSpace{}%
\AgdaOperator{\AgdaFunction{×}}\AgdaSpace{}%
\AgdaBound{bc}\AgdaSpace{}%
\AgdaOperator{\AgdaDatatype{↭}}\AgdaSpace{}%
\AgdaBound{br}\<%
\end{code}

\begin{code}[hide]%
\>[0]\AgdaKeyword{open}\AgdaSpace{}%
\AgdaKeyword{import}\AgdaSpace{}%
\AgdaModule{Functional.State}\AgdaSpace{}%
\AgdaKeyword{using}\AgdaSpace{}%
\AgdaSymbol{(}\AgdaFunction{Oracle}\AgdaSpace{}%
\AgdaSymbol{;}\AgdaSpace{}%
\AgdaFunction{Cmd}\AgdaSpace{}%
\AgdaSymbol{;}\AgdaSpace{}%
\AgdaFunction{FileSystem}\AgdaSymbol{)}\<%
\\
\\[\AgdaEmptyExtraSkip]%
\>[0]\AgdaKeyword{module}\AgdaSpace{}%
\AgdaModule{Functional.Script.Hazard.Base}\AgdaSpace{}%
\AgdaSymbol{(}\AgdaBound{oracle}\AgdaSpace{}%
\AgdaSymbol{:}\AgdaSpace{}%
\AgdaFunction{Oracle}\AgdaSymbol{)}\AgdaSpace{}%
\AgdaKeyword{where}\<%
\\
\\[\AgdaEmptyExtraSkip]%
\>[0]\AgdaKeyword{open}\AgdaSpace{}%
\AgdaKeyword{import}\AgdaSpace{}%
\AgdaModule{Agda.Builtin.Equality}\<%
\\
\>[0]\AgdaKeyword{open}\AgdaSpace{}%
\AgdaKeyword{import}\AgdaSpace{}%
\AgdaModule{Functional.State.Helpers}\AgdaSpace{}%
\AgdaSymbol{(}\AgdaBound{oracle}\AgdaSymbol{)}\AgdaSpace{}%
\AgdaKeyword{using}\AgdaSpace{}%
\AgdaSymbol{(}cmdWriteNames \AgdaSymbol{;} cmdReadNames \AgdaSymbol{;} run\AgdaSymbol{)}\<%
\\
\>[0]\AgdaKeyword{open}\AgdaSpace{}%
\AgdaKeyword{import}\AgdaSpace{}%
\AgdaModule{Agda.Builtin.Nat}\AgdaSpace{}%
\AgdaKeyword{renaming}\AgdaSpace{}%
\AgdaSymbol{(}\AgdaDatatype{Nat}\AgdaSpace{}%
\AgdaSymbol{to}\AgdaSpace{}%
\AgdaDatatype{ℕ}\AgdaSymbol{)}\<%
\\
\>[0]\AgdaKeyword{open}\AgdaSpace{}%
\AgdaKeyword{import}\AgdaSpace{}%
\AgdaModule{Data.Nat.Properties}\AgdaSpace{}%
\AgdaSymbol{as}\AgdaSpace{}%
\AgdaModule{N}\AgdaSpace{}%
\AgdaKeyword{using}\AgdaSpace{}%
\AgdaSymbol{(}\AgdaFunction{1+n≢n}\AgdaSpace{}%
\AgdaSymbol{;}\AgdaSpace{}%
\AgdaFunction{≤-refl}\AgdaSpace{}%
\AgdaSymbol{;}\AgdaSpace{}%
\AgdaFunction{≤-step}\AgdaSymbol{)}\<%
\\
\>[0]\AgdaKeyword{open}\AgdaSpace{}%
\AgdaKeyword{import}\AgdaSpace{}%
\AgdaModule{Data.Bool}\AgdaSpace{}%
\AgdaKeyword{using}\AgdaSpace{}%
\AgdaSymbol{(}\AgdaInductiveConstructor{true}\AgdaSpace{}%
\AgdaSymbol{;}\AgdaSpace{}%
\AgdaInductiveConstructor{false}\AgdaSymbol{)}\<%
\\
\>[0]\AgdaKeyword{open}\AgdaSpace{}%
\AgdaKeyword{import}\AgdaSpace{}%
\AgdaModule{Data.Nat}\AgdaSpace{}%
\AgdaKeyword{using}\AgdaSpace{}%
\AgdaSymbol{(}\AgdaOperator{\AgdaFunction{\AgdaUnderscore{}>\AgdaUnderscore{}}}\AgdaSymbol{)}\<%
\\
\>[0]\AgdaKeyword{open}\AgdaSpace{}%
\AgdaKeyword{import}\AgdaSpace{}%
\AgdaModule{Data.List}\AgdaSpace{}%
\AgdaKeyword{using}\AgdaSpace{}%
\AgdaSymbol{(}\AgdaDatatype{List}\AgdaSpace{}%
\AgdaSymbol{;}\AgdaSpace{}%
\AgdaInductiveConstructor{[]}\AgdaSpace{}%
\AgdaSymbol{;}\AgdaSpace{}%
\AgdaOperator{\AgdaInductiveConstructor{\AgdaUnderscore{}∷\AgdaUnderscore{}}}\AgdaSpace{}%
\AgdaSymbol{;}\AgdaSpace{}%
\AgdaOperator{\AgdaFunction{\AgdaUnderscore{}∷ʳ\AgdaUnderscore{}}}%
\>[53]\AgdaSymbol{;}\AgdaSpace{}%
\AgdaOperator{\AgdaFunction{\AgdaUnderscore{}++\AgdaUnderscore{}}}\AgdaSpace{}%
\AgdaSymbol{;}\AgdaSpace{}%
\AgdaFunction{map}\AgdaSpace{}%
\AgdaSymbol{;}\AgdaSpace{}%
\AgdaFunction{foldl}\AgdaSpace{}%
\AgdaSymbol{;}\AgdaSpace{}%
\AgdaFunction{filter}\AgdaSpace{}%
\AgdaSymbol{;}\AgdaSpace{}%
\AgdaFunction{concatMap}\AgdaSpace{}%
\AgdaSymbol{;}\AgdaSpace{}%
\AgdaFunction{length}\AgdaSpace{}%
\AgdaSymbol{;}\AgdaSpace{}%
\AgdaFunction{concat}\AgdaSymbol{)}\<%
\\
\>[0]\AgdaKeyword{open}\AgdaSpace{}%
\AgdaKeyword{import}\AgdaSpace{}%
\AgdaModule{Data.List.Properties}\AgdaSpace{}%
\AgdaKeyword{using}\AgdaSpace{}%
\AgdaSymbol{(}\AgdaFunction{concat-++}\AgdaSpace{}%
\AgdaSymbol{;}\AgdaSpace{}%
\AgdaFunction{map-++-commute}\AgdaSpace{}%
\AgdaSymbol{;}\AgdaSpace{}%
\AgdaFunction{∷-injectiveˡ}\AgdaSpace{}%
\AgdaSymbol{;}\AgdaSpace{}%
\AgdaFunction{∷-injectiveʳ}\AgdaSymbol{)}\<%
\\
\>[0]\AgdaKeyword{open}\AgdaSpace{}%
\AgdaKeyword{import}\AgdaSpace{}%
\AgdaModule{Data.Product}\AgdaSpace{}%
\AgdaKeyword{using}\AgdaSpace{}%
\AgdaSymbol{(}\AgdaOperator{\AgdaFunction{\AgdaUnderscore{}×\AgdaUnderscore{}}}\AgdaSpace{}%
\AgdaSymbol{;}\AgdaSpace{}%
\AgdaField{proj₁}\AgdaSpace{}%
\AgdaSymbol{;}\AgdaSpace{}%
\AgdaField{proj₂}\AgdaSpace{}%
\AgdaSymbol{;}\AgdaSpace{}%
\AgdaOperator{\AgdaInductiveConstructor{\AgdaUnderscore{},\AgdaUnderscore{}}}\AgdaSpace{}%
\AgdaSymbol{;}\AgdaSpace{}%
\AgdaFunction{Σ-syntax}\AgdaSpace{}%
\AgdaSymbol{;}\AgdaSpace{}%
\AgdaFunction{∃-syntax}\AgdaSymbol{)}\<%
\\
\>[0]\AgdaKeyword{open}\AgdaSpace{}%
\AgdaKeyword{import}\AgdaSpace{}%
\AgdaModule{Data.Product.Properties}\AgdaSpace{}%
\AgdaKeyword{using}\AgdaSpace{}%
\AgdaSymbol{(}\AgdaFunction{,-injectiveˡ}\AgdaSpace{}%
\AgdaSymbol{;}\AgdaSpace{}%
\AgdaFunction{,-injectiveʳ}\AgdaSymbol{)}\<%
\\
\>[0]\AgdaKeyword{open}\AgdaSpace{}%
\AgdaKeyword{import}\AgdaSpace{}%
\AgdaModule{Functional.File}\AgdaSpace{}%
\AgdaKeyword{using}\AgdaSpace{}%
\AgdaSymbol{(}\AgdaFunction{FileName}\AgdaSymbol{)}\<%
\\
\>[0]\AgdaKeyword{open}\AgdaSpace{}%
\AgdaKeyword{import}\AgdaSpace{}%
\AgdaModule{Functional.Build}\AgdaSpace{}%
\AgdaSymbol{(}\AgdaBound{oracle}\AgdaSymbol{)}\AgdaSpace{}%
\AgdaKeyword{using}\AgdaSpace{}%
\AgdaSymbol{(}Build\AgdaSymbol{)}\<%
\\
\>[0]\AgdaKeyword{open}\AgdaSpace{}%
\AgdaKeyword{import}\AgdaSpace{}%
\AgdaModule{Data.List.Membership.Propositional}\AgdaSpace{}%
\AgdaKeyword{using}\AgdaSpace{}%
\AgdaSymbol{(}\AgdaOperator{\AgdaFunction{\AgdaUnderscore{}∈\AgdaUnderscore{}}}\AgdaSpace{}%
\AgdaSymbol{;}\AgdaSpace{}%
\AgdaOperator{\AgdaFunction{\AgdaUnderscore{}∉\AgdaUnderscore{}}}\AgdaSymbol{)}\<%
\\
\\[\AgdaEmptyExtraSkip]%
\>[0]\AgdaKeyword{open}\AgdaSpace{}%
\AgdaKeyword{import}\AgdaSpace{}%
\AgdaModule{Data.List.Membership.Propositional.Properties}\AgdaSpace{}%
\AgdaKeyword{using}\AgdaSpace{}%
\AgdaSymbol{(}\AgdaFunction{∈-++⁺ˡ}\AgdaSpace{}%
\AgdaSymbol{;}\AgdaSpace{}%
\AgdaFunction{∈-++⁺ʳ}\AgdaSymbol{)}\<%
\\
\>[0]\AgdaKeyword{open}\AgdaSpace{}%
\AgdaKeyword{import}\AgdaSpace{}%
\AgdaModule{Relation.Nullary}\AgdaSpace{}%
\AgdaKeyword{using}\AgdaSpace{}%
\AgdaSymbol{(}\AgdaInductiveConstructor{yes}\AgdaSpace{}%
\AgdaSymbol{;}\AgdaSpace{}%
\AgdaInductiveConstructor{no}\AgdaSpace{}%
\AgdaSymbol{;}\AgdaSpace{}%
\AgdaOperator{\AgdaFunction{¬\AgdaUnderscore{}}}\AgdaSpace{}%
\AgdaSymbol{;}\AgdaSpace{}%
\AgdaRecord{Dec}\AgdaSymbol{)}\<%
\\
\>[0]\AgdaKeyword{open}\AgdaSpace{}%
\AgdaKeyword{import}\AgdaSpace{}%
\AgdaModule{Relation.Nullary.Negation}\AgdaSpace{}%
\AgdaKeyword{using}\AgdaSpace{}%
\AgdaSymbol{(}\AgdaFunction{contradiction}\AgdaSymbol{)}\<%
\\
\>[0]\AgdaKeyword{open}\AgdaSpace{}%
\AgdaKeyword{import}\AgdaSpace{}%
\AgdaModule{Data.List.Relation.Binary.Disjoint.Propositional}\AgdaSpace{}%
\AgdaKeyword{using}\AgdaSpace{}%
\AgdaSymbol{(}\AgdaFunction{Disjoint}\AgdaSymbol{)}\<%
\\
\>[0]\AgdaKeyword{open}\AgdaSpace{}%
\AgdaKeyword{import}\AgdaSpace{}%
\AgdaModule{Common.List.Properties}\AgdaSpace{}%
\AgdaKeyword{using}\AgdaSpace{}%
\AgdaSymbol{(}\AgdaOperator{\AgdaFunction{\AgdaUnderscore{}before\AgdaUnderscore{}∈\AgdaUnderscore{}}}\AgdaSymbol{)}\<%
\\
\>[0]\AgdaKeyword{open}\AgdaSpace{}%
\AgdaKeyword{import}\AgdaSpace{}%
\AgdaModule{Function}\AgdaSpace{}%
\AgdaKeyword{using}\AgdaSpace{}%
\AgdaSymbol{(}\AgdaOperator{\AgdaFunction{\AgdaUnderscore{}∘\AgdaUnderscore{}}}\AgdaSymbol{)}\<%
\\
\>[0]\AgdaKeyword{open}\AgdaSpace{}%
\AgdaKeyword{import}\AgdaSpace{}%
\AgdaModule{Data.String.Properties}\AgdaSpace{}%
\AgdaKeyword{using}\AgdaSpace{}%
\AgdaSymbol{(}\AgdaOperator{\AgdaFunction{\AgdaUnderscore{}≟\AgdaUnderscore{}}}\AgdaSpace{}%
\AgdaSymbol{;}\AgdaSpace{}%
\AgdaFunction{≡-decSetoid}\AgdaSymbol{)}\<%
\\
\>[0]\AgdaKeyword{open}\AgdaSpace{}%
\AgdaKeyword{import}\AgdaSpace{}%
\AgdaModule{Data.List.Membership.DecSetoid}\AgdaSpace{}%
\AgdaSymbol{(}\AgdaFunction{≡-decSetoid}\AgdaSymbol{)}\AgdaSpace{}%
\AgdaKeyword{using}\AgdaSpace{}%
\AgdaSymbol{(}\AgdaUnderscore{}∈?\AgdaUnderscore{}\AgdaSymbol{)}\<%
\\
\>[0]\AgdaKeyword{open}\AgdaSpace{}%
\AgdaKeyword{import}\AgdaSpace{}%
\AgdaModule{Data.List.Relation.Unary.Any}\AgdaSpace{}%
\AgdaKeyword{using}\AgdaSpace{}%
\AgdaSymbol{(}\AgdaFunction{tail}\AgdaSymbol{)}\<%
\\
\>[0]\AgdaKeyword{open}\AgdaSpace{}%
\AgdaKeyword{import}\AgdaSpace{}%
\AgdaModule{Data.List.Relation.Unary.All}\AgdaSpace{}%
\AgdaKeyword{using}\AgdaSpace{}%
\AgdaSymbol{(}\AgdaDatatype{All}\AgdaSpace{}%
\AgdaSymbol{;}\AgdaSpace{}%
\AgdaFunction{lookup}\AgdaSpace{}%
\AgdaSymbol{;}\AgdaSpace{}%
\AgdaOperator{\AgdaInductiveConstructor{\AgdaUnderscore{}∷\AgdaUnderscore{}}}\AgdaSymbol{)}\<%
\\
\>[0]\AgdaKeyword{open}\AgdaSpace{}%
\AgdaKeyword{import}\AgdaSpace{}%
\AgdaModule{Data.List.Relation.Binary.Subset.Propositional}\AgdaSpace{}%
\AgdaKeyword{using}\AgdaSpace{}%
\AgdaSymbol{(}\AgdaOperator{\AgdaFunction{\AgdaUnderscore{}⊆\AgdaUnderscore{}}}\AgdaSymbol{)}\<%
\\
\>[0]\AgdaKeyword{open}\AgdaSpace{}%
\AgdaKeyword{import}\AgdaSpace{}%
\AgdaModule{Data.List.Relation.Binary.Subset.Propositional.Properties}\AgdaSpace{}%
\AgdaKeyword{using}\AgdaSpace{}%
\AgdaSymbol{(}\AgdaFunction{concat⁺}\AgdaSpace{}%
\AgdaSymbol{;}\AgdaSpace{}%
\AgdaFunction{map⁺}\AgdaSpace{}%
\AgdaSymbol{;}\AgdaSpace{}%
\AgdaFunction{filter-⊆}\AgdaSpace{}%
\AgdaSymbol{;}\AgdaSpace{}%
\AgdaFunction{filter⁺′}\AgdaSpace{}%
\AgdaSymbol{;}\AgdaSpace{}%
\AgdaFunction{++⁺}\AgdaSymbol{)}\<%
\\
\>[0]\AgdaKeyword{open}\AgdaSpace{}%
\AgdaKeyword{import}\AgdaSpace{}%
\AgdaModule{Data.List.Relation.Unary.Any}\AgdaSpace{}%
\AgdaKeyword{using}\AgdaSpace{}%
\AgdaSymbol{(}\AgdaInductiveConstructor{there}\AgdaSpace{}%
\AgdaSymbol{;}\AgdaSpace{}%
\AgdaInductiveConstructor{here}\AgdaSymbol{)}\<%
\\
\>[0]\AgdaKeyword{open}\AgdaSpace{}%
\AgdaKeyword{import}\AgdaSpace{}%
\AgdaModule{Data.List.Membership.Propositional.Properties}\AgdaSpace{}%
\AgdaKeyword{using}\AgdaSpace{}%
\AgdaSymbol{(}\AgdaFunction{∈-++⁻}\AgdaSymbol{)}\<%
\\
\>[0]\AgdaKeyword{open}\AgdaSpace{}%
\AgdaKeyword{import}\AgdaSpace{}%
\AgdaModule{Data.Sum}\AgdaSpace{}%
\AgdaKeyword{using}\AgdaSpace{}%
\AgdaSymbol{(}\AgdaOperator{\AgdaDatatype{\AgdaUnderscore{}⊎\AgdaUnderscore{}}}\AgdaSpace{}%
\AgdaSymbol{;}\AgdaSpace{}%
\AgdaInductiveConstructor{inj₁}\AgdaSpace{}%
\AgdaSymbol{;}\AgdaSpace{}%
\AgdaInductiveConstructor{inj₂}\AgdaSymbol{)}\<%
\\
\>[0]\AgdaKeyword{open}\AgdaSpace{}%
\AgdaKeyword{import}\AgdaSpace{}%
\AgdaModule{Data.Product.Relation.Binary.Pointwise.NonDependent}\AgdaSpace{}%
\AgdaKeyword{using}\AgdaSpace{}%
\AgdaSymbol{(}\AgdaFunction{×-decidable}\AgdaSpace{}%
\AgdaSymbol{;}\AgdaSpace{}%
\AgdaFunction{≡×≡⇒≡}\AgdaSymbol{)}\<%
\\
\>[0]\AgdaKeyword{open}\AgdaSpace{}%
\AgdaKeyword{import}\AgdaSpace{}%
\AgdaModule{Data.List.Relation.Binary.Equality.DecPropositional}\AgdaSpace{}%
\AgdaSymbol{(}\AgdaOperator{\AgdaFunction{\AgdaUnderscore{}≟\AgdaUnderscore{}}}\AgdaSymbol{)}\AgdaSpace{}%
\AgdaKeyword{using}\AgdaSpace{}%
\AgdaSymbol{(}\AgdaUnderscore{}≡?\AgdaUnderscore{}\AgdaSymbol{)}\<%
\\
\>[0]\AgdaKeyword{open}\AgdaSpace{}%
\AgdaKeyword{import}\AgdaSpace{}%
\AgdaModule{Relation.Binary.PropositionalEquality}\AgdaSpace{}%
\AgdaKeyword{using}\AgdaSpace{}%
\AgdaSymbol{(}\AgdaFunction{subst}\AgdaSpace{}%
\AgdaSymbol{;}\AgdaSpace{}%
\AgdaFunction{subst₂}\AgdaSpace{}%
\AgdaSymbol{;}\AgdaSpace{}%
\AgdaFunction{cong}\AgdaSpace{}%
\AgdaSymbol{;}\AgdaSpace{}%
\AgdaFunction{sym}\AgdaSpace{}%
\AgdaSymbol{;}\AgdaSpace{}%
\AgdaFunction{trans}\AgdaSymbol{)}\<%
\\
\>[0]\AgdaKeyword{open}\AgdaSpace{}%
\AgdaKeyword{import}\AgdaSpace{}%
\AgdaModule{Data.Empty}\AgdaSpace{}%
\AgdaKeyword{using}\AgdaSpace{}%
\AgdaSymbol{(}\AgdaDatatype{⊥}\AgdaSymbol{)}\<%
\\
\>[0]\AgdaKeyword{open}\AgdaSpace{}%
\AgdaKeyword{import}\AgdaSpace{}%
\AgdaModule{Data.List.Relation.Unary.AllPairs}\AgdaSpace{}%
\AgdaKeyword{using}\AgdaSpace{}%
\AgdaSymbol{(}\AgdaDatatype{AllPairs}\AgdaSpace{}%
\AgdaSymbol{;}\AgdaSpace{}%
\AgdaOperator{\AgdaInductiveConstructor{\AgdaUnderscore{}∷\AgdaUnderscore{}}}\AgdaSymbol{)}\<%
\\
\>[0]\AgdaKeyword{open}\AgdaSpace{}%
\AgdaKeyword{import}\AgdaSpace{}%
\AgdaModule{Data.List.Relation.Unary.Unique.Propositional}\AgdaSpace{}%
\AgdaKeyword{using}\AgdaSpace{}%
\AgdaSymbol{(}\AgdaDatatype{Unique}\AgdaSymbol{)}\<%
\\
\\[\AgdaEmptyExtraSkip]%
\\[\AgdaEmptyExtraSkip]%
\>[0]\AgdaComment{\{- Hazards:

Do not depend on a reference order:

   Read before Write
   Write before Write

Does depend on a reference order:
  
  Speculative Write before Read

---------------------------------------

Goals (what we need):

execWError says there is either a hazard or a state

completeness takes evidence that the build has no hazards and proves that execWerror will return 
a state and not a hazard.  fundamentally this works by producing a contradiction when execWError
returns a hazard with the hazardfree evidence. 

What would be ideal is if hazard and hazardfree are set up in such a way that we can just say
contradiction 'hazard free evidnece' 'hazard evidence' without anymore manipulation

Existing hazardfree defs:

HazardFree is recursive and says the writes of the command at the head of the list are disjoint from the writes/reads that happened previously (exist in some list)

Would be nice to stick with this so i dont have to completely redo a lot of my proofs.  

HazardFreeReordering says 2 builds are permutations of one another. which could jsut be passed as its own argument to a function

Says both of the builds are hazardFree (by above def). 
And says there is not speculativewrhazard in the first of the 2 builds.  

speculativewrhazard says there are two commands, such that the first command is first in the original build, and the 2nd command writes to something before the first command in the build that is running.

this definition is just so long and ugly.  

-\}}\<%
\\
\>[0]\AgdaFunction{FileNames}\AgdaSpace{}%
\AgdaSymbol{:}\AgdaSpace{}%
\AgdaPrimitiveType{Set}\<%
\\
\>[0]\AgdaFunction{FileNames}\AgdaSpace{}%
\AgdaSymbol{=}\AgdaSpace{}%
\AgdaDatatype{List}\AgdaSpace{}%
\AgdaFunction{FileName}\<%
\end{code}

\newcommand{\fileinfo}{%
\begin{code}%
\>[0]\AgdaComment{-- FileNames == List FileName}\<%
\\
\>[0]\AgdaFunction{FileInfo}\AgdaSpace{}%
\AgdaSymbol{:}\AgdaSpace{}%
\AgdaPrimitiveType{Set}\<%
\\
\>[0]\AgdaFunction{FileInfo}\AgdaSpace{}%
\AgdaSymbol{=}\AgdaSpace{}%
\AgdaDatatype{List}\AgdaSpace{}%
\AgdaSymbol{(}\AgdaFunction{Cmd}\AgdaSpace{}%
\AgdaOperator{\AgdaFunction{×}}\AgdaSpace{}%
\AgdaFunction{FileNames}\AgdaSpace{}%
\AgdaOperator{\AgdaFunction{×}}\AgdaSpace{}%
\AgdaFunction{FileNames}\AgdaSymbol{)}\<%
\end{code}}

\newcommand{\save}{%
\begin{code}%
\>[0]\AgdaComment{-- extends FileInfo with a new entry for the Cmd}\<%
\\
\>[0]\AgdaFunction{save}\AgdaSpace{}%
\AgdaSymbol{:}\AgdaSpace{}%
\AgdaFunction{FileSystem}\AgdaSpace{}%
\AgdaSymbol{→}\AgdaSpace{}%
\AgdaFunction{Cmd}\AgdaSpace{}%
\AgdaSymbol{→}\AgdaSpace{}%
\AgdaFunction{FileInfo}\AgdaSpace{}%
\AgdaSymbol{→}\AgdaSpace{}%
\AgdaFunction{FileInfo}\<%
\\
\>[0]\AgdaFunction{save}\AgdaSpace{}%
\AgdaBound{s}\AgdaSpace{}%
\AgdaBound{x}\AgdaSpace{}%
\AgdaBound{fi}\AgdaSpace{}%
\AgdaSymbol{=}\AgdaSpace{}%
\AgdaSymbol{(}\AgdaBound{x}\AgdaSpace{}%
\AgdaOperator{\AgdaInductiveConstructor{,}}\AgdaSpace{}%
\AgdaSymbol{(}\AgdaPostulate{cmdReadNames}\AgdaSpace{}%
\AgdaBound{x}\AgdaSpace{}%
\AgdaBound{s}\AgdaSymbol{)}\AgdaSpace{}%
\AgdaOperator{\AgdaInductiveConstructor{,}}\AgdaSpace{}%
\AgdaSymbol{(}\AgdaPostulate{cmdWriteNames}\AgdaSpace{}%
\AgdaBound{x}\AgdaSpace{}%
\AgdaBound{s}\AgdaSymbol{))}\AgdaSpace{}%
\AgdaOperator{\AgdaInductiveConstructor{∷}}\AgdaSpace{}%
\AgdaBound{fi}\<%
\end{code}}

\begin{code}[hide]%
\>[0]\AgdaFunction{cmdsRun}\AgdaSpace{}%
\AgdaSymbol{:}\AgdaSpace{}%
\AgdaFunction{FileInfo}\AgdaSpace{}%
\AgdaSymbol{→}\AgdaSpace{}%
\AgdaDatatype{List}\AgdaSpace{}%
\AgdaFunction{Cmd}\<%
\\
\>[0]\AgdaFunction{cmdsRun}\AgdaSpace{}%
\AgdaSymbol{=}\AgdaSpace{}%
\AgdaFunction{map}\AgdaSpace{}%
\AgdaField{proj₁}\<%
\\
\\[\AgdaEmptyExtraSkip]%
\>[0]\AgdaFunction{filesRead}\AgdaSpace{}%
\AgdaSymbol{:}\AgdaSpace{}%
\AgdaFunction{FileInfo}\AgdaSpace{}%
\AgdaSymbol{→}\AgdaSpace{}%
\AgdaDatatype{List}\AgdaSpace{}%
\AgdaFunction{FileName}\<%
\\
\>[0]\AgdaFunction{filesRead}\AgdaSpace{}%
\AgdaSymbol{=}\AgdaSpace{}%
\AgdaFunction{concatMap}\AgdaSpace{}%
\AgdaSymbol{(}\AgdaField{proj₁}\AgdaSpace{}%
\AgdaOperator{\AgdaFunction{∘}}\AgdaSpace{}%
\AgdaField{proj₂}\AgdaSymbol{)}\<%
\\
\\[\AgdaEmptyExtraSkip]%
\>[0]\AgdaFunction{∈-concatMap-++}\AgdaSpace{}%
\AgdaSymbol{:}\AgdaSpace{}%
\AgdaSymbol{∀}\AgdaSpace{}%
\AgdaSymbol{\{}\AgdaBound{v}\AgdaSpace{}%
\AgdaSymbol{:}\AgdaSpace{}%
\AgdaFunction{FileName}\AgdaSymbol{\}}\AgdaSpace{}%
\AgdaBound{f}\AgdaSpace{}%
\AgdaSymbol{(}\AgdaBound{xs}\AgdaSpace{}%
\AgdaSymbol{:}\AgdaSpace{}%
\AgdaFunction{FileInfo}\AgdaSymbol{)}\AgdaSpace{}%
\AgdaBound{ys}\AgdaSpace{}%
\AgdaBound{zs}\AgdaSpace{}%
\AgdaSymbol{→}\AgdaSpace{}%
\AgdaBound{v}\AgdaSpace{}%
\AgdaOperator{\AgdaFunction{∈}}\AgdaSpace{}%
\AgdaFunction{concatMap}\AgdaSpace{}%
\AgdaBound{f}\AgdaSpace{}%
\AgdaSymbol{(}\AgdaBound{xs}\AgdaSpace{}%
\AgdaOperator{\AgdaFunction{++}}\AgdaSpace{}%
\AgdaBound{zs}\AgdaSymbol{)}\AgdaSpace{}%
\AgdaSymbol{→}\AgdaSpace{}%
\AgdaBound{v}\AgdaSpace{}%
\AgdaOperator{\AgdaFunction{∈}}\AgdaSpace{}%
\AgdaFunction{concatMap}\AgdaSpace{}%
\AgdaBound{f}\AgdaSpace{}%
\AgdaSymbol{(}\AgdaBound{xs}\AgdaSpace{}%
\AgdaOperator{\AgdaFunction{++}}\AgdaSpace{}%
\AgdaBound{ys}\AgdaSpace{}%
\AgdaOperator{\AgdaFunction{++}}\AgdaSpace{}%
\AgdaBound{zs}\AgdaSymbol{)}\<%
\\
\>[0]\AgdaFunction{∈-concatMap-++}\AgdaSpace{}%
\AgdaBound{f}\AgdaSpace{}%
\AgdaBound{xs}\AgdaSpace{}%
\AgdaBound{ys}\AgdaSpace{}%
\AgdaBound{zs}\AgdaSpace{}%
\AgdaBound{v∈}\AgdaSpace{}%
\AgdaKeyword{with}\AgdaSpace{}%
\AgdaFunction{∈-++⁻}\AgdaSpace{}%
\AgdaSymbol{(}\AgdaFunction{concat}\AgdaSpace{}%
\AgdaSymbol{(}\AgdaFunction{map}\AgdaSpace{}%
\AgdaBound{f}\AgdaSpace{}%
\AgdaBound{xs}\AgdaSymbol{))}\AgdaSpace{}%
\AgdaSymbol{(}\AgdaFunction{subst}\AgdaSpace{}%
\AgdaSymbol{(λ}\AgdaSpace{}%
\AgdaBound{x}\AgdaSpace{}%
\AgdaSymbol{→}\AgdaSpace{}%
\AgdaSymbol{\AgdaUnderscore{}}\AgdaSpace{}%
\AgdaOperator{\AgdaFunction{∈}}\AgdaSpace{}%
\AgdaBound{x}\AgdaSymbol{)}\AgdaSpace{}%
\AgdaFunction{≡₁}\AgdaSpace{}%
\AgdaBound{v∈}\AgdaSymbol{)}\<%
\\
\>[0][@{}l@{\AgdaIndent{0}}]%
\>[2]\AgdaKeyword{where}%
\>[344I]\AgdaFunction{≡₁}\AgdaSpace{}%
\AgdaSymbol{:}\AgdaSpace{}%
\AgdaFunction{concatMap}\AgdaSpace{}%
\AgdaBound{f}\AgdaSpace{}%
\AgdaSymbol{(}\AgdaBound{xs}\AgdaSpace{}%
\AgdaOperator{\AgdaFunction{++}}\AgdaSpace{}%
\AgdaBound{zs}\AgdaSymbol{)}\AgdaSpace{}%
\AgdaOperator{\AgdaDatatype{≡}}\AgdaSpace{}%
\AgdaFunction{concat}\AgdaSpace{}%
\AgdaSymbol{(}\AgdaFunction{map}\AgdaSpace{}%
\AgdaBound{f}\AgdaSpace{}%
\AgdaBound{xs}\AgdaSymbol{)}\AgdaSpace{}%
\AgdaOperator{\AgdaFunction{++}}\AgdaSpace{}%
\AgdaFunction{concat}\AgdaSpace{}%
\AgdaSymbol{(}\AgdaFunction{map}\AgdaSpace{}%
\AgdaBound{f}\AgdaSpace{}%
\AgdaBound{zs}\AgdaSymbol{)}\<%
\\
\>[.][@{}l@{}]\<[344I]%
\>[8]\AgdaFunction{≡₁}\AgdaSpace{}%
\AgdaSymbol{=}\AgdaSpace{}%
\AgdaFunction{trans}\AgdaSpace{}%
\AgdaSymbol{(}\AgdaFunction{cong}\AgdaSpace{}%
\AgdaFunction{concat}\AgdaSpace{}%
\AgdaSymbol{(}\AgdaFunction{map-++-commute}\AgdaSpace{}%
\AgdaBound{f}\AgdaSpace{}%
\AgdaBound{xs}\AgdaSpace{}%
\AgdaBound{zs}\AgdaSymbol{))}\AgdaSpace{}%
\AgdaSymbol{(}\AgdaFunction{sym}\AgdaSpace{}%
\AgdaSymbol{(}\AgdaFunction{concat-++}\AgdaSpace{}%
\AgdaSymbol{(}\AgdaFunction{map}\AgdaSpace{}%
\AgdaBound{f}\AgdaSpace{}%
\AgdaBound{xs}\AgdaSymbol{)}\AgdaSpace{}%
\AgdaSymbol{(}\AgdaFunction{map}\AgdaSpace{}%
\AgdaBound{f}\AgdaSpace{}%
\AgdaBound{zs}\AgdaSymbol{)))}\<%
\\
\>[0]\AgdaSymbol{...}\AgdaSpace{}%
\AgdaSymbol{|}\AgdaSpace{}%
\AgdaInductiveConstructor{inj₁}\AgdaSpace{}%
\AgdaBound{v∈xs}\AgdaSpace{}%
\AgdaSymbol{=}\AgdaSpace{}%
\AgdaFunction{subst}\AgdaSpace{}%
\AgdaSymbol{(λ}\AgdaSpace{}%
\AgdaBound{x}\AgdaSpace{}%
\AgdaSymbol{→}\AgdaSpace{}%
\AgdaSymbol{\AgdaUnderscore{}}\AgdaSpace{}%
\AgdaOperator{\AgdaFunction{∈}}\AgdaSpace{}%
\AgdaBound{x}\AgdaSymbol{)}\AgdaSpace{}%
\AgdaFunction{≡₁}\AgdaSpace{}%
\AgdaSymbol{(}\AgdaFunction{∈-++⁺ˡ}\AgdaSpace{}%
\AgdaBound{v∈xs}\AgdaSymbol{)}\<%
\\
\>[0][@{}l@{\AgdaIndent{0}}]%
\>[2]\AgdaKeyword{where}%
\>[391I]\AgdaFunction{≡₁}\AgdaSpace{}%
\AgdaSymbol{:}\AgdaSpace{}%
\AgdaFunction{concat}\AgdaSpace{}%
\AgdaSymbol{(}\AgdaFunction{map}\AgdaSpace{}%
\AgdaBound{f}\AgdaSpace{}%
\AgdaBound{xs}\AgdaSymbol{)}\AgdaSpace{}%
\AgdaOperator{\AgdaFunction{++}}\AgdaSpace{}%
\AgdaFunction{concat}\AgdaSpace{}%
\AgdaSymbol{(}\AgdaFunction{map}\AgdaSpace{}%
\AgdaBound{f}\AgdaSpace{}%
\AgdaSymbol{(}\AgdaBound{ys}\AgdaSpace{}%
\AgdaOperator{\AgdaFunction{++}}\AgdaSpace{}%
\AgdaBound{zs}\AgdaSymbol{))}\AgdaSpace{}%
\AgdaOperator{\AgdaDatatype{≡}}\AgdaSpace{}%
\AgdaFunction{concat}\AgdaSpace{}%
\AgdaSymbol{(}\AgdaFunction{map}\AgdaSpace{}%
\AgdaBound{f}\AgdaSpace{}%
\AgdaSymbol{(}\AgdaBound{xs}\AgdaSpace{}%
\AgdaOperator{\AgdaFunction{++}}\AgdaSpace{}%
\AgdaBound{ys}\AgdaSpace{}%
\AgdaOperator{\AgdaFunction{++}}\AgdaSpace{}%
\AgdaBound{zs}\AgdaSymbol{))}\<%
\\
\>[.][@{}l@{}]\<[391I]%
\>[8]\AgdaFunction{≡₁}\AgdaSpace{}%
\AgdaSymbol{=}\AgdaSpace{}%
\AgdaFunction{trans}\AgdaSpace{}%
\AgdaSymbol{(}\AgdaFunction{concat-++}\AgdaSpace{}%
\AgdaSymbol{(}\AgdaFunction{map}\AgdaSpace{}%
\AgdaBound{f}\AgdaSpace{}%
\AgdaBound{xs}\AgdaSymbol{)}\AgdaSpace{}%
\AgdaSymbol{(}\AgdaFunction{map}\AgdaSpace{}%
\AgdaBound{f}\AgdaSpace{}%
\AgdaSymbol{(}\AgdaBound{ys}\AgdaSpace{}%
\AgdaOperator{\AgdaFunction{++}}\AgdaSpace{}%
\AgdaBound{zs}\AgdaSymbol{)))}\AgdaSpace{}%
\AgdaSymbol{(}\AgdaFunction{cong}\AgdaSpace{}%
\AgdaFunction{concat}\AgdaSpace{}%
\AgdaSymbol{(}\AgdaFunction{sym}\AgdaSpace{}%
\AgdaSymbol{(}\AgdaFunction{map-++-commute}\AgdaSpace{}%
\AgdaBound{f}\AgdaSpace{}%
\AgdaBound{xs}\AgdaSpace{}%
\AgdaSymbol{(}\AgdaBound{ys}\AgdaSpace{}%
\AgdaOperator{\AgdaFunction{++}}\AgdaSpace{}%
\AgdaBound{zs}\AgdaSymbol{))))}\<%
\\
\>[0]\AgdaSymbol{...}\AgdaSpace{}%
\AgdaSymbol{|}\AgdaSpace{}%
\AgdaInductiveConstructor{inj₂}\AgdaSpace{}%
\AgdaBound{v∈ys}\AgdaSpace{}%
\AgdaSymbol{=}\AgdaSpace{}%
\AgdaFunction{subst}\AgdaSpace{}%
\AgdaSymbol{(λ}\AgdaSpace{}%
\AgdaBound{x}\AgdaSpace{}%
\AgdaSymbol{→}\AgdaSpace{}%
\AgdaSymbol{\AgdaUnderscore{}}\AgdaSpace{}%
\AgdaOperator{\AgdaFunction{∈}}\AgdaSpace{}%
\AgdaBound{x}\AgdaSymbol{)}\AgdaSpace{}%
\AgdaFunction{≡₁}\AgdaSpace{}%
\AgdaSymbol{(}\AgdaFunction{∈-++⁺ʳ}\AgdaSpace{}%
\AgdaSymbol{(}\AgdaFunction{concat}\AgdaSpace{}%
\AgdaSymbol{(}\AgdaFunction{map}\AgdaSpace{}%
\AgdaBound{f}\AgdaSpace{}%
\AgdaBound{xs}\AgdaSymbol{))}\AgdaSpace{}%
\AgdaSymbol{(}\AgdaFunction{∈-++⁺ʳ}\AgdaSpace{}%
\AgdaSymbol{(}\AgdaFunction{concat}\AgdaSpace{}%
\AgdaSymbol{(}\AgdaFunction{map}\AgdaSpace{}%
\AgdaBound{f}\AgdaSpace{}%
\AgdaBound{ys}\AgdaSymbol{))}\AgdaSpace{}%
\AgdaBound{v∈ys}\AgdaSymbol{))}\<%
\\
\>[0][@{}l@{\AgdaIndent{0}}]%
\>[2]\AgdaKeyword{where}%
\>[456I]\AgdaFunction{≡₁}\AgdaSpace{}%
\AgdaSymbol{:}\AgdaSpace{}%
\AgdaFunction{concat}\AgdaSpace{}%
\AgdaSymbol{(}\AgdaFunction{map}\AgdaSpace{}%
\AgdaBound{f}\AgdaSpace{}%
\AgdaBound{xs}\AgdaSymbol{)}\AgdaSpace{}%
\AgdaOperator{\AgdaFunction{++}}\AgdaSpace{}%
\AgdaFunction{concat}\AgdaSpace{}%
\AgdaSymbol{(}\AgdaFunction{map}\AgdaSpace{}%
\AgdaBound{f}\AgdaSpace{}%
\AgdaBound{ys}\AgdaSymbol{)}\AgdaSpace{}%
\AgdaOperator{\AgdaFunction{++}}\AgdaSpace{}%
\AgdaFunction{concat}\AgdaSpace{}%
\AgdaSymbol{(}\AgdaFunction{map}\AgdaSpace{}%
\AgdaBound{f}\AgdaSpace{}%
\AgdaBound{zs}\AgdaSymbol{)}\AgdaSpace{}%
\AgdaOperator{\AgdaDatatype{≡}}\AgdaSpace{}%
\AgdaFunction{concat}\AgdaSpace{}%
\AgdaSymbol{(}\AgdaFunction{map}\AgdaSpace{}%
\AgdaBound{f}\AgdaSpace{}%
\AgdaSymbol{(}\AgdaBound{xs}\AgdaSpace{}%
\AgdaOperator{\AgdaFunction{++}}\AgdaSpace{}%
\AgdaBound{ys}\AgdaSpace{}%
\AgdaOperator{\AgdaFunction{++}}\AgdaSpace{}%
\AgdaBound{zs}\AgdaSymbol{))}\<%
\\
\>[.][@{}l@{}]\<[456I]%
\>[8]\AgdaFunction{≡₁}\AgdaSpace{}%
\AgdaSymbol{=}\AgdaSpace{}%
\AgdaFunction{trans}%
\>[483I]\AgdaSymbol{(}\AgdaFunction{cong}\AgdaSpace{}%
\AgdaSymbol{(}\AgdaFunction{concat}\AgdaSpace{}%
\AgdaSymbol{(}\AgdaFunction{map}\AgdaSpace{}%
\AgdaBound{f}\AgdaSpace{}%
\AgdaBound{xs}\AgdaSymbol{)}\AgdaSpace{}%
\AgdaOperator{\AgdaFunction{++\AgdaUnderscore{}}}\AgdaSymbol{)}\AgdaSpace{}%
\AgdaSymbol{(}\AgdaFunction{concat-++}\AgdaSpace{}%
\AgdaSymbol{(}\AgdaFunction{map}\AgdaSpace{}%
\AgdaBound{f}\AgdaSpace{}%
\AgdaBound{ys}\AgdaSymbol{)}\AgdaSpace{}%
\AgdaSymbol{(}\AgdaFunction{map}\AgdaSpace{}%
\AgdaBound{f}\AgdaSpace{}%
\AgdaBound{zs}\AgdaSymbol{)))}\<%
\\
\>[.][@{}l@{}]\<[483I]%
\>[19]\AgdaSymbol{(}\AgdaFunction{trans}%
\>[496I]\AgdaSymbol{(}\AgdaFunction{concat-++}\AgdaSpace{}%
\AgdaSymbol{(}\AgdaFunction{map}\AgdaSpace{}%
\AgdaBound{f}\AgdaSpace{}%
\AgdaBound{xs}\AgdaSymbol{)}\AgdaSpace{}%
\AgdaSymbol{(}\AgdaFunction{map}\AgdaSpace{}%
\AgdaBound{f}\AgdaSpace{}%
\AgdaBound{ys}\AgdaSpace{}%
\AgdaOperator{\AgdaFunction{++}}\AgdaSpace{}%
\AgdaFunction{map}\AgdaSpace{}%
\AgdaBound{f}\AgdaSpace{}%
\AgdaBound{zs}\AgdaSymbol{))}\<%
\\
\>[.][@{}l@{}]\<[496I]%
\>[26]\AgdaSymbol{(}\AgdaFunction{cong}\AgdaSpace{}%
\AgdaFunction{concat}%
\>[508I]\AgdaSymbol{(}\AgdaFunction{trans}\AgdaSpace{}%
\AgdaSymbol{(}\AgdaFunction{cong}\AgdaSpace{}%
\AgdaSymbol{(}\AgdaFunction{map}\AgdaSpace{}%
\AgdaBound{f}\AgdaSpace{}%
\AgdaBound{xs}\AgdaSpace{}%
\AgdaOperator{\AgdaFunction{++\AgdaUnderscore{}}}\AgdaSymbol{)}\AgdaSpace{}%
\AgdaSymbol{(}\AgdaFunction{sym}\AgdaSpace{}%
\AgdaSymbol{(}\AgdaFunction{map-++-commute}\AgdaSpace{}%
\AgdaBound{f}\AgdaSpace{}%
\AgdaBound{ys}\AgdaSpace{}%
\AgdaBound{zs}\AgdaSymbol{)))}\<%
\\
\>[.][@{}l@{}]\<[508I]%
\>[39]\AgdaSymbol{(}\AgdaFunction{sym}\AgdaSpace{}%
\AgdaSymbol{(}\AgdaFunction{map-++-commute}\AgdaSpace{}%
\AgdaBound{f}\AgdaSpace{}%
\AgdaBound{xs}\AgdaSpace{}%
\AgdaSymbol{(}\AgdaBound{ys}\AgdaSpace{}%
\AgdaOperator{\AgdaFunction{++}}\AgdaSpace{}%
\AgdaBound{zs}\AgdaSymbol{))))))}\<%
\\
\\[\AgdaEmptyExtraSkip]%
\\[\AgdaEmptyExtraSkip]%
\>[0]\AgdaFunction{∈-filesRead-++}\AgdaSpace{}%
\AgdaSymbol{:}\AgdaSpace{}%
\AgdaSymbol{∀}\AgdaSpace{}%
\AgdaSymbol{\{}\AgdaBound{v}\AgdaSymbol{\}}\AgdaSpace{}%
\AgdaBound{xs}\AgdaSpace{}%
\AgdaBound{ys}\AgdaSpace{}%
\AgdaBound{zs}\AgdaSpace{}%
\AgdaSymbol{→}\AgdaSpace{}%
\AgdaBound{v}\AgdaSpace{}%
\AgdaOperator{\AgdaFunction{∈}}\AgdaSpace{}%
\AgdaFunction{filesRead}\AgdaSpace{}%
\AgdaSymbol{(}\AgdaBound{xs}\AgdaSpace{}%
\AgdaOperator{\AgdaFunction{++}}\AgdaSpace{}%
\AgdaBound{zs}\AgdaSymbol{)}\AgdaSpace{}%
\AgdaSymbol{→}\AgdaSpace{}%
\AgdaBound{v}\AgdaSpace{}%
\AgdaOperator{\AgdaFunction{∈}}\AgdaSpace{}%
\AgdaFunction{filesRead}\AgdaSpace{}%
\AgdaSymbol{(}\AgdaBound{xs}\AgdaSpace{}%
\AgdaOperator{\AgdaFunction{++}}\AgdaSpace{}%
\AgdaBound{ys}\AgdaSpace{}%
\AgdaOperator{\AgdaFunction{++}}\AgdaSpace{}%
\AgdaBound{zs}\AgdaSymbol{)}\<%
\\
\>[0]\AgdaFunction{∈-filesRead-++}\AgdaSpace{}%
\AgdaSymbol{=}\AgdaSpace{}%
\AgdaFunction{∈-concatMap-++}\AgdaSpace{}%
\AgdaSymbol{(}\AgdaField{proj₁}\AgdaSpace{}%
\AgdaOperator{\AgdaFunction{∘}}\AgdaSpace{}%
\AgdaField{proj₂}\AgdaSymbol{)}\<%
\\
\\[\AgdaEmptyExtraSkip]%
\>[0]\AgdaFunction{filesWrote}\AgdaSpace{}%
\AgdaSymbol{:}\AgdaSpace{}%
\AgdaFunction{FileInfo}\AgdaSpace{}%
\AgdaSymbol{→}\AgdaSpace{}%
\AgdaDatatype{List}\AgdaSpace{}%
\AgdaFunction{FileName}\<%
\\
\>[0]\AgdaFunction{filesWrote}\AgdaSpace{}%
\AgdaSymbol{=}\AgdaSpace{}%
\AgdaFunction{concatMap}\AgdaSpace{}%
\AgdaSymbol{(}\AgdaField{proj₂}\AgdaSpace{}%
\AgdaOperator{\AgdaFunction{∘}}\AgdaSpace{}%
\AgdaField{proj₂}\AgdaSymbol{)}\<%
\\
\\[\AgdaEmptyExtraSkip]%
\>[0]\AgdaFunction{∈-filesWrote-++}\AgdaSpace{}%
\AgdaSymbol{:}\AgdaSpace{}%
\AgdaSymbol{∀}\AgdaSpace{}%
\AgdaSymbol{\{}\AgdaBound{v}\AgdaSymbol{\}}\AgdaSpace{}%
\AgdaBound{xs}\AgdaSpace{}%
\AgdaBound{ys}\AgdaSpace{}%
\AgdaBound{zs}\AgdaSpace{}%
\AgdaSymbol{→}\AgdaSpace{}%
\AgdaBound{v}\AgdaSpace{}%
\AgdaOperator{\AgdaFunction{∈}}\AgdaSpace{}%
\AgdaFunction{filesWrote}\AgdaSpace{}%
\AgdaSymbol{(}\AgdaBound{xs}\AgdaSpace{}%
\AgdaOperator{\AgdaFunction{++}}\AgdaSpace{}%
\AgdaBound{zs}\AgdaSymbol{)}\AgdaSpace{}%
\AgdaSymbol{→}\AgdaSpace{}%
\AgdaBound{v}\AgdaSpace{}%
\AgdaOperator{\AgdaFunction{∈}}\AgdaSpace{}%
\AgdaFunction{filesWrote}\AgdaSpace{}%
\AgdaSymbol{(}\AgdaBound{xs}\AgdaSpace{}%
\AgdaOperator{\AgdaFunction{++}}\AgdaSpace{}%
\AgdaBound{ys}\AgdaSpace{}%
\AgdaOperator{\AgdaFunction{++}}\AgdaSpace{}%
\AgdaBound{zs}\AgdaSymbol{)}\<%
\\
\>[0]\AgdaFunction{∈-filesWrote-++}\AgdaSpace{}%
\AgdaSymbol{=}\AgdaSpace{}%
\AgdaFunction{∈-concatMap-++}\AgdaSpace{}%
\AgdaSymbol{(}\AgdaField{proj₂}\AgdaSpace{}%
\AgdaOperator{\AgdaFunction{∘}}\AgdaSpace{}%
\AgdaField{proj₂}\AgdaSymbol{)}\<%
\\
\\[\AgdaEmptyExtraSkip]%
\>[0]\AgdaFunction{files}\AgdaSpace{}%
\AgdaSymbol{:}\AgdaSpace{}%
\AgdaFunction{FileInfo}\AgdaSpace{}%
\AgdaSymbol{→}\AgdaSpace{}%
\AgdaDatatype{List}\AgdaSpace{}%
\AgdaFunction{FileName}\<%
\\
\>[0]\AgdaFunction{files}\AgdaSpace{}%
\AgdaBound{ls}\AgdaSpace{}%
\AgdaSymbol{=}\AgdaSpace{}%
\AgdaSymbol{(}\AgdaFunction{filesRead}\AgdaSpace{}%
\AgdaBound{ls}\AgdaSymbol{)}\AgdaSpace{}%
\AgdaOperator{\AgdaFunction{++}}\AgdaSpace{}%
\AgdaSymbol{(}\AgdaFunction{filesWrote}\AgdaSpace{}%
\AgdaBound{ls}\AgdaSymbol{)}\<%
\\
\\[\AgdaEmptyExtraSkip]%
\>[0]\AgdaFunction{∈-files-++}\AgdaSpace{}%
\AgdaSymbol{:}\AgdaSpace{}%
\AgdaSymbol{∀}\AgdaSpace{}%
\AgdaSymbol{\{}\AgdaBound{v}\AgdaSymbol{\}}\AgdaSpace{}%
\AgdaBound{xs}\AgdaSpace{}%
\AgdaBound{ys}\AgdaSpace{}%
\AgdaBound{zs}\AgdaSpace{}%
\AgdaSymbol{→}\AgdaSpace{}%
\AgdaBound{v}\AgdaSpace{}%
\AgdaOperator{\AgdaFunction{∈}}\AgdaSpace{}%
\AgdaFunction{files}\AgdaSpace{}%
\AgdaSymbol{(}\AgdaBound{xs}\AgdaSpace{}%
\AgdaOperator{\AgdaFunction{++}}\AgdaSpace{}%
\AgdaBound{zs}\AgdaSymbol{)}\AgdaSpace{}%
\AgdaSymbol{→}\AgdaSpace{}%
\AgdaBound{v}\AgdaSpace{}%
\AgdaOperator{\AgdaFunction{∈}}\AgdaSpace{}%
\AgdaFunction{files}\AgdaSpace{}%
\AgdaSymbol{(}\AgdaBound{xs}\AgdaSpace{}%
\AgdaOperator{\AgdaFunction{++}}\AgdaSpace{}%
\AgdaBound{ys}\AgdaSpace{}%
\AgdaOperator{\AgdaFunction{++}}\AgdaSpace{}%
\AgdaBound{zs}\AgdaSymbol{)}\<%
\\
\>[0]\AgdaFunction{∈-files-++}\AgdaSpace{}%
\AgdaBound{xs}\AgdaSpace{}%
\AgdaBound{ys}\AgdaSpace{}%
\AgdaBound{zs}\AgdaSpace{}%
\AgdaBound{v∈files}\AgdaSpace{}%
\AgdaKeyword{with}\AgdaSpace{}%
\AgdaFunction{∈-++⁻}\AgdaSpace{}%
\AgdaSymbol{(}\AgdaFunction{filesRead}\AgdaSpace{}%
\AgdaSymbol{(}\AgdaBound{xs}\AgdaSpace{}%
\AgdaOperator{\AgdaFunction{++}}\AgdaSpace{}%
\AgdaBound{zs}\AgdaSymbol{))}\AgdaSpace{}%
\AgdaBound{v∈files}\<%
\\
\>[0]\AgdaSymbol{...}\AgdaSpace{}%
\AgdaSymbol{|}\AgdaSpace{}%
\AgdaInductiveConstructor{inj₁}\AgdaSpace{}%
\AgdaBound{v∈reads}\AgdaSpace{}%
\AgdaSymbol{=}\AgdaSpace{}%
\AgdaFunction{∈-++⁺ˡ}\AgdaSpace{}%
\AgdaSymbol{(}\AgdaFunction{∈-filesRead-++}\AgdaSpace{}%
\AgdaBound{xs}\AgdaSpace{}%
\AgdaBound{ys}\AgdaSpace{}%
\AgdaBound{zs}\AgdaSpace{}%
\AgdaBound{v∈reads}\AgdaSymbol{)}\<%
\\
\>[0]\AgdaSymbol{...}\AgdaSpace{}%
\AgdaSymbol{|}\AgdaSpace{}%
\AgdaInductiveConstructor{inj₂}\AgdaSpace{}%
\AgdaBound{v∈writes}\AgdaSpace{}%
\AgdaSymbol{=}\AgdaSpace{}%
\AgdaFunction{∈-++⁺ʳ}\AgdaSpace{}%
\AgdaSymbol{(}\AgdaFunction{filesRead}\AgdaSpace{}%
\AgdaSymbol{(}\AgdaBound{xs}\AgdaSpace{}%
\AgdaOperator{\AgdaFunction{++}}\AgdaSpace{}%
\AgdaBound{ys}\AgdaSpace{}%
\AgdaOperator{\AgdaFunction{++}}\AgdaSpace{}%
\AgdaBound{zs}\AgdaSymbol{))}\AgdaSpace{}%
\AgdaSymbol{(}\AgdaFunction{∈-filesWrote-++}\AgdaSpace{}%
\AgdaBound{xs}\AgdaSpace{}%
\AgdaBound{ys}\AgdaSpace{}%
\AgdaBound{zs}\AgdaSpace{}%
\AgdaBound{v∈writes}\AgdaSymbol{)}\<%
\end{code}

\newcommand{\cmdWrote}{%
\begin{code}%
\>[0]\AgdaComment{-- The FileNames the Cmd wrote according to the FileInfo}\<%
\\
\>[0]\AgdaFunction{cmdWrote}\AgdaSpace{}%
\AgdaSymbol{:}\AgdaSpace{}%
\AgdaFunction{FileInfo}\AgdaSpace{}%
\AgdaSymbol{→}\AgdaSpace{}%
\AgdaFunction{Cmd}\AgdaSpace{}%
\AgdaSymbol{→}\AgdaSpace{}%
\AgdaDatatype{List}\AgdaSpace{}%
\AgdaFunction{FileName}\<%
\end{code}}
\begin{code}[hide]%
\>[0]\AgdaFunction{cmdWrote}\AgdaSpace{}%
\AgdaInductiveConstructor{[]}\AgdaSpace{}%
\AgdaBound{p}\AgdaSpace{}%
\AgdaSymbol{=}\AgdaSpace{}%
\AgdaInductiveConstructor{[]}\<%
\\
\>[0]\AgdaFunction{cmdWrote}\AgdaSpace{}%
\AgdaSymbol{(}\AgdaBound{x}\AgdaSpace{}%
\AgdaOperator{\AgdaInductiveConstructor{∷}}\AgdaSpace{}%
\AgdaBound{ls}\AgdaSymbol{)}\AgdaSpace{}%
\AgdaBound{p}\AgdaSpace{}%
\AgdaKeyword{with}\AgdaSpace{}%
\AgdaSymbol{(}\AgdaField{proj₁}\AgdaSpace{}%
\AgdaBound{x}\AgdaSymbol{)}\AgdaSpace{}%
\AgdaOperator{\AgdaFunction{≟}}\AgdaSpace{}%
\AgdaBound{p}\<%
\\
\>[0]\AgdaSymbol{...}\AgdaSpace{}%
\AgdaSymbol{|}\AgdaSpace{}%
\AgdaInductiveConstructor{yes}\AgdaSpace{}%
\AgdaBound{x≡p}\AgdaSpace{}%
\AgdaSymbol{=}\AgdaSpace{}%
\AgdaField{proj₂}\AgdaSpace{}%
\AgdaSymbol{(}\AgdaField{proj₂}\AgdaSpace{}%
\AgdaBound{x}\AgdaSymbol{)}\<%
\\
\>[0]\AgdaSymbol{...}\AgdaSpace{}%
\AgdaSymbol{|}\AgdaSpace{}%
\AgdaInductiveConstructor{no}\AgdaSpace{}%
\AgdaBound{¬x≡p}\AgdaSpace{}%
\AgdaSymbol{=}\AgdaSpace{}%
\AgdaFunction{cmdWrote}\AgdaSpace{}%
\AgdaBound{ls}\AgdaSpace{}%
\AgdaBound{p}\<%
\\
\\[\AgdaEmptyExtraSkip]%
\>[0]\AgdaFunction{cmdWrote∷-≡}\AgdaSpace{}%
\AgdaSymbol{:}\AgdaSpace{}%
\AgdaSymbol{∀}\AgdaSpace{}%
\AgdaBound{x}\AgdaSpace{}%
\AgdaBound{ls}\AgdaSpace{}%
\AgdaSymbol{→}\AgdaSpace{}%
\AgdaFunction{cmdWrote}\AgdaSpace{}%
\AgdaSymbol{(}\AgdaBound{x}\AgdaSpace{}%
\AgdaOperator{\AgdaInductiveConstructor{∷}}\AgdaSpace{}%
\AgdaBound{ls}\AgdaSymbol{)}\AgdaSpace{}%
\AgdaSymbol{(}\AgdaField{proj₁}\AgdaSpace{}%
\AgdaBound{x}\AgdaSymbol{)}\AgdaSpace{}%
\AgdaOperator{\AgdaDatatype{≡}}\AgdaSpace{}%
\AgdaField{proj₂}\AgdaSpace{}%
\AgdaSymbol{(}\AgdaField{proj₂}\AgdaSpace{}%
\AgdaBound{x}\AgdaSymbol{)}\<%
\\
\>[0]\AgdaFunction{cmdWrote∷-≡}\AgdaSpace{}%
\AgdaBound{x}\AgdaSpace{}%
\AgdaBound{ls}\AgdaSpace{}%
\AgdaKeyword{with}\AgdaSpace{}%
\AgdaSymbol{(}\AgdaField{proj₁}\AgdaSpace{}%
\AgdaBound{x}\AgdaSymbol{)}\AgdaSpace{}%
\AgdaOperator{\AgdaFunction{≟}}\AgdaSpace{}%
\AgdaSymbol{(}\AgdaField{proj₁}\AgdaSpace{}%
\AgdaBound{x}\AgdaSymbol{)}\<%
\\
\>[0]\AgdaSymbol{...}\AgdaSpace{}%
\AgdaSymbol{|}\AgdaSpace{}%
\AgdaInductiveConstructor{yes}\AgdaSpace{}%
\AgdaBound{x≡x}\AgdaSpace{}%
\AgdaSymbol{=}\AgdaSpace{}%
\AgdaInductiveConstructor{refl}\<%
\\
\>[0]\AgdaSymbol{...}\AgdaSpace{}%
\AgdaSymbol{|}\AgdaSpace{}%
\AgdaInductiveConstructor{no}\AgdaSpace{}%
\AgdaBound{¬x≡x}\AgdaSpace{}%
\AgdaSymbol{=}\AgdaSpace{}%
\AgdaFunction{contradiction}\AgdaSpace{}%
\AgdaInductiveConstructor{refl}\AgdaSpace{}%
\AgdaBound{¬x≡x}\<%
\\
\\[\AgdaEmptyExtraSkip]%
\>[0]\AgdaFunction{∈-filesWrote}\AgdaSpace{}%
\AgdaSymbol{:}\AgdaSpace{}%
\AgdaSymbol{∀}\AgdaSpace{}%
\AgdaSymbol{\{}\AgdaBound{v}\AgdaSymbol{\}}\AgdaSpace{}%
\AgdaSymbol{\{}\AgdaBound{x}\AgdaSymbol{\}}\AgdaSpace{}%
\AgdaBound{ls}\AgdaSpace{}%
\AgdaSymbol{→}\AgdaSpace{}%
\AgdaBound{v}\AgdaSpace{}%
\AgdaOperator{\AgdaFunction{∈}}\AgdaSpace{}%
\AgdaFunction{cmdWrote}\AgdaSpace{}%
\AgdaBound{ls}\AgdaSpace{}%
\AgdaBound{x}\AgdaSpace{}%
\AgdaSymbol{→}\AgdaSpace{}%
\AgdaBound{v}\AgdaSpace{}%
\AgdaOperator{\AgdaFunction{∈}}\AgdaSpace{}%
\AgdaFunction{filesWrote}\AgdaSpace{}%
\AgdaBound{ls}\<%
\\
\>[0]\AgdaFunction{∈-filesWrote}\AgdaSpace{}%
\AgdaSymbol{\{}\AgdaBound{v}\AgdaSymbol{\}}\AgdaSpace{}%
\AgdaSymbol{\{}\AgdaBound{x}\AgdaSymbol{\}}\AgdaSpace{}%
\AgdaSymbol{((}\AgdaBound{x₁}\AgdaSpace{}%
\AgdaOperator{\AgdaInductiveConstructor{,}}\AgdaSpace{}%
\AgdaSymbol{\AgdaUnderscore{}}\AgdaSpace{}%
\AgdaOperator{\AgdaInductiveConstructor{,}}\AgdaSpace{}%
\AgdaBound{ws}\AgdaSymbol{)}\AgdaSpace{}%
\AgdaOperator{\AgdaInductiveConstructor{∷}}\AgdaSpace{}%
\AgdaBound{ls}\AgdaSymbol{)}\AgdaSpace{}%
\AgdaBound{v∈}\AgdaSpace{}%
\AgdaKeyword{with}\AgdaSpace{}%
\AgdaBound{x₁}\AgdaSpace{}%
\AgdaOperator{\AgdaFunction{≟}}\AgdaSpace{}%
\AgdaBound{x}\<%
\\
\>[0]\AgdaSymbol{...}\AgdaSpace{}%
\AgdaSymbol{|}\AgdaSpace{}%
\AgdaInductiveConstructor{yes}\AgdaSpace{}%
\AgdaBound{x₁≡x}\AgdaSpace{}%
\AgdaSymbol{=}\AgdaSpace{}%
\AgdaFunction{∈-++⁺ˡ}\AgdaSpace{}%
\AgdaBound{v∈}\<%
\\
\>[0]\AgdaSymbol{...}\AgdaSpace{}%
\AgdaSymbol{|}\AgdaSpace{}%
\AgdaInductiveConstructor{no}\AgdaSpace{}%
\AgdaBound{¬x₁≡x}\AgdaSpace{}%
\AgdaSymbol{=}\AgdaSpace{}%
\AgdaFunction{∈-++⁺ʳ}\AgdaSpace{}%
\AgdaBound{ws}\AgdaSpace{}%
\AgdaSymbol{(}\AgdaFunction{∈-filesWrote}\AgdaSpace{}%
\AgdaBound{ls}\AgdaSpace{}%
\AgdaBound{v∈}\AgdaSymbol{)}\<%
\\
\\[\AgdaEmptyExtraSkip]%
\>[0]\AgdaFunction{∈-cmdWrote∷l}\AgdaSpace{}%
\AgdaSymbol{:}\AgdaSpace{}%
\AgdaSymbol{∀}\AgdaSpace{}%
\AgdaSymbol{\{}\AgdaBound{v}\AgdaSymbol{\}}\AgdaSpace{}%
\AgdaBound{x}\AgdaSpace{}%
\AgdaBound{ls}\AgdaSpace{}%
\AgdaSymbol{→}\AgdaSpace{}%
\AgdaBound{v}\AgdaSpace{}%
\AgdaOperator{\AgdaFunction{∈}}\AgdaSpace{}%
\AgdaField{proj₂}\AgdaSpace{}%
\AgdaSymbol{(}\AgdaField{proj₂}\AgdaSpace{}%
\AgdaBound{x}\AgdaSymbol{)}\AgdaSpace{}%
\AgdaSymbol{→}\AgdaSpace{}%
\AgdaBound{v}\AgdaSpace{}%
\AgdaOperator{\AgdaFunction{∈}}\AgdaSpace{}%
\AgdaFunction{cmdWrote}\AgdaSpace{}%
\AgdaSymbol{(}\AgdaBound{x}\AgdaSpace{}%
\AgdaOperator{\AgdaInductiveConstructor{∷}}\AgdaSpace{}%
\AgdaBound{ls}\AgdaSymbol{)}\AgdaSpace{}%
\AgdaSymbol{(}\AgdaField{proj₁}\AgdaSpace{}%
\AgdaBound{x}\AgdaSymbol{)}\<%
\\
\>[0]\AgdaFunction{∈-cmdWrote∷l}\AgdaSpace{}%
\AgdaBound{x}\AgdaSpace{}%
\AgdaBound{ls}\AgdaSpace{}%
\AgdaBound{v∈}\AgdaSpace{}%
\AgdaKeyword{with}\AgdaSpace{}%
\AgdaSymbol{(}\AgdaField{proj₁}\AgdaSpace{}%
\AgdaBound{x}\AgdaSymbol{)}\AgdaSpace{}%
\AgdaOperator{\AgdaFunction{≟}}\AgdaSpace{}%
\AgdaSymbol{(}\AgdaField{proj₁}\AgdaSpace{}%
\AgdaBound{x}\AgdaSymbol{)}\<%
\\
\>[0]\AgdaSymbol{...}\AgdaSpace{}%
\AgdaSymbol{|}\AgdaSpace{}%
\AgdaInductiveConstructor{no}\AgdaSpace{}%
\AgdaBound{¬x≡x}\AgdaSpace{}%
\AgdaSymbol{=}\AgdaSpace{}%
\AgdaFunction{contradiction}\AgdaSpace{}%
\AgdaInductiveConstructor{refl}\AgdaSpace{}%
\AgdaBound{¬x≡x}\<%
\\
\>[0]\AgdaSymbol{...}\AgdaSpace{}%
\AgdaSymbol{|}\AgdaSpace{}%
\AgdaInductiveConstructor{yes}\AgdaSpace{}%
\AgdaBound{x≡x}\AgdaSpace{}%
\AgdaSymbol{=}\AgdaSpace{}%
\AgdaBound{v∈}\<%
\\
\\[\AgdaEmptyExtraSkip]%
\>[0]\AgdaFunction{∈-cmdWrote∷}\AgdaSpace{}%
\AgdaSymbol{:}\AgdaSpace{}%
\AgdaSymbol{∀}\AgdaSpace{}%
\AgdaSymbol{\{}\AgdaBound{v}\AgdaSymbol{\}}\AgdaSpace{}%
\AgdaBound{x}\AgdaSpace{}%
\AgdaBound{x₁}\AgdaSpace{}%
\AgdaBound{ls}\AgdaSpace{}%
\AgdaSymbol{→}\AgdaSpace{}%
\AgdaBound{v}\AgdaSpace{}%
\AgdaOperator{\AgdaFunction{∈}}\AgdaSpace{}%
\AgdaFunction{cmdWrote}\AgdaSpace{}%
\AgdaBound{ls}\AgdaSpace{}%
\AgdaBound{x₁}\AgdaSpace{}%
\AgdaSymbol{→}\AgdaSpace{}%
\AgdaOperator{\AgdaFunction{¬}}\AgdaSpace{}%
\AgdaSymbol{(}\AgdaField{proj₁}\AgdaSpace{}%
\AgdaBound{x}\AgdaSymbol{)}\AgdaSpace{}%
\AgdaOperator{\AgdaDatatype{≡}}\AgdaSpace{}%
\AgdaBound{x₁}\AgdaSpace{}%
\AgdaSymbol{→}\AgdaSpace{}%
\AgdaBound{v}\AgdaSpace{}%
\AgdaOperator{\AgdaFunction{∈}}\AgdaSpace{}%
\AgdaFunction{cmdWrote}\AgdaSpace{}%
\AgdaSymbol{(}\AgdaBound{x}\AgdaSpace{}%
\AgdaOperator{\AgdaInductiveConstructor{∷}}\AgdaSpace{}%
\AgdaBound{ls}\AgdaSymbol{)}\AgdaSpace{}%
\AgdaBound{x₁}\<%
\\
\>[0]\AgdaFunction{∈-cmdWrote∷}\AgdaSpace{}%
\AgdaBound{x}\AgdaSpace{}%
\AgdaBound{x₁}\AgdaSpace{}%
\AgdaBound{ls}\AgdaSpace{}%
\AgdaBound{v∈}\AgdaSpace{}%
\AgdaBound{¬≡}\AgdaSpace{}%
\AgdaKeyword{with}\AgdaSpace{}%
\AgdaSymbol{(}\AgdaField{proj₁}\AgdaSpace{}%
\AgdaBound{x}\AgdaSymbol{)}\AgdaSpace{}%
\AgdaOperator{\AgdaFunction{≟}}\AgdaSpace{}%
\AgdaBound{x₁}\<%
\\
\>[0]\AgdaSymbol{...}\AgdaSpace{}%
\AgdaSymbol{|}\AgdaSpace{}%
\AgdaInductiveConstructor{yes}\AgdaSpace{}%
\AgdaBound{x≡x₁}\AgdaSpace{}%
\AgdaSymbol{=}\AgdaSpace{}%
\AgdaFunction{contradiction}\AgdaSpace{}%
\AgdaBound{x≡x₁}\AgdaSpace{}%
\AgdaBound{¬≡}\<%
\\
\>[0]\AgdaSymbol{...}\AgdaSpace{}%
\AgdaSymbol{|}\AgdaSpace{}%
\AgdaInductiveConstructor{no}\AgdaSpace{}%
\AgdaBound{¬x≡x₁}\AgdaSpace{}%
\AgdaSymbol{=}\AgdaSpace{}%
\AgdaBound{v∈}\<%
\\
\\[\AgdaEmptyExtraSkip]%
\>[0]\AgdaComment{-- if there is a file in the cmdRead for x , then x ∈ xs}\<%
\\
\>[0]\AgdaFunction{lemma2}\AgdaSpace{}%
\AgdaSymbol{:}\AgdaSpace{}%
\AgdaSymbol{∀}\AgdaSpace{}%
\AgdaSymbol{\{}\AgdaBound{v}\AgdaSymbol{\}}\AgdaSpace{}%
\AgdaBound{x}\AgdaSpace{}%
\AgdaBound{xs}\AgdaSpace{}%
\AgdaSymbol{→}\AgdaSpace{}%
\AgdaBound{v}\AgdaSpace{}%
\AgdaOperator{\AgdaFunction{∈}}\AgdaSpace{}%
\AgdaFunction{cmdWrote}\AgdaSpace{}%
\AgdaBound{xs}\AgdaSpace{}%
\AgdaBound{x}\AgdaSpace{}%
\AgdaSymbol{→}\AgdaSpace{}%
\AgdaBound{x}\AgdaSpace{}%
\AgdaOperator{\AgdaFunction{∈}}\AgdaSpace{}%
\AgdaFunction{map}\AgdaSpace{}%
\AgdaField{proj₁}\AgdaSpace{}%
\AgdaBound{xs}\<%
\\
\>[0]\AgdaFunction{lemma2}\AgdaSpace{}%
\AgdaBound{x}\AgdaSpace{}%
\AgdaSymbol{(}\AgdaBound{x₁}\AgdaSpace{}%
\AgdaOperator{\AgdaInductiveConstructor{∷}}\AgdaSpace{}%
\AgdaBound{xs}\AgdaSymbol{)}\AgdaSpace{}%
\AgdaBound{v∈}\AgdaSpace{}%
\AgdaKeyword{with}\AgdaSpace{}%
\AgdaSymbol{(}\AgdaField{proj₁}\AgdaSpace{}%
\AgdaBound{x₁}\AgdaSymbol{)}\AgdaSpace{}%
\AgdaOperator{\AgdaFunction{≟}}\AgdaSpace{}%
\AgdaBound{x}\<%
\\
\>[0]\AgdaSymbol{...}\AgdaSpace{}%
\AgdaSymbol{|}\AgdaSpace{}%
\AgdaInductiveConstructor{yes}\AgdaSpace{}%
\AgdaBound{x₁≡x}\AgdaSpace{}%
\AgdaSymbol{=}\AgdaSpace{}%
\AgdaInductiveConstructor{here}\AgdaSpace{}%
\AgdaSymbol{(}\AgdaFunction{sym}\AgdaSpace{}%
\AgdaBound{x₁≡x}\AgdaSymbol{)}\<%
\\
\>[0]\AgdaSymbol{...}\AgdaSpace{}%
\AgdaSymbol{|}\AgdaSpace{}%
\AgdaInductiveConstructor{no}\AgdaSpace{}%
\AgdaBound{¬x₁≡x}\AgdaSpace{}%
\AgdaSymbol{=}\AgdaSpace{}%
\AgdaInductiveConstructor{there}\AgdaSpace{}%
\AgdaSymbol{(}\AgdaFunction{lemma2}\AgdaSpace{}%
\AgdaBound{x}\AgdaSpace{}%
\AgdaBound{xs}\AgdaSpace{}%
\AgdaBound{v∈}\AgdaSymbol{)}\<%
\\
\\[\AgdaEmptyExtraSkip]%
\>[0]\AgdaFunction{∈-cmdWrote++⁺ʳ}\AgdaSpace{}%
\AgdaSymbol{:}\AgdaSpace{}%
\AgdaSymbol{∀}\AgdaSpace{}%
\AgdaSymbol{\{}\AgdaBound{v}\AgdaSymbol{\}}\AgdaSpace{}%
\AgdaBound{x}\AgdaSpace{}%
\AgdaBound{xs}\AgdaSpace{}%
\AgdaBound{ys}\AgdaSpace{}%
\AgdaSymbol{→}\AgdaSpace{}%
\AgdaDatatype{Unique}\AgdaSpace{}%
\AgdaSymbol{(}\AgdaFunction{map}\AgdaSpace{}%
\AgdaField{proj₁}\AgdaSpace{}%
\AgdaSymbol{(}\AgdaBound{xs}\AgdaSpace{}%
\AgdaOperator{\AgdaFunction{++}}\AgdaSpace{}%
\AgdaBound{ys}\AgdaSymbol{))}\AgdaSpace{}%
\AgdaSymbol{→}\AgdaSpace{}%
\AgdaBound{v}\AgdaSpace{}%
\AgdaOperator{\AgdaFunction{∈}}\AgdaSpace{}%
\AgdaFunction{cmdWrote}\AgdaSpace{}%
\AgdaBound{ys}\AgdaSpace{}%
\AgdaBound{x}\AgdaSpace{}%
\AgdaSymbol{→}\AgdaSpace{}%
\AgdaBound{v}\AgdaSpace{}%
\AgdaOperator{\AgdaFunction{∈}}\AgdaSpace{}%
\AgdaFunction{cmdWrote}\AgdaSpace{}%
\AgdaSymbol{(}\AgdaBound{xs}\AgdaSpace{}%
\AgdaOperator{\AgdaFunction{++}}\AgdaSpace{}%
\AgdaBound{ys}\AgdaSymbol{)}\AgdaSpace{}%
\AgdaBound{x}\<%
\\
\>[0]\AgdaFunction{∈-cmdWrote++⁺ʳ}\AgdaSpace{}%
\AgdaBound{x}\AgdaSpace{}%
\AgdaInductiveConstructor{[]}\AgdaSpace{}%
\AgdaBound{ys}\AgdaSpace{}%
\AgdaBound{u}\AgdaSpace{}%
\AgdaBound{v∈}\AgdaSpace{}%
\AgdaSymbol{=}\AgdaSpace{}%
\AgdaBound{v∈}\<%
\\
\>[0]\AgdaFunction{∈-cmdWrote++⁺ʳ}\AgdaSpace{}%
\AgdaBound{x}\AgdaSpace{}%
\AgdaSymbol{(}\AgdaBound{x₁}\AgdaSpace{}%
\AgdaOperator{\AgdaInductiveConstructor{∷}}\AgdaSpace{}%
\AgdaBound{xs}\AgdaSymbol{)}\AgdaSpace{}%
\AgdaBound{ys}\AgdaSpace{}%
\AgdaSymbol{(}\AgdaBound{px₁}\AgdaSpace{}%
\AgdaOperator{\AgdaInductiveConstructor{∷}}\AgdaSpace{}%
\AgdaBound{u}\AgdaSymbol{)}\AgdaSpace{}%
\AgdaBound{v∈}\AgdaSpace{}%
\AgdaKeyword{with}\AgdaSpace{}%
\AgdaSymbol{(}\AgdaField{proj₁}\AgdaSpace{}%
\AgdaBound{x₁}\AgdaSymbol{)}\AgdaSpace{}%
\AgdaOperator{\AgdaFunction{≟}}\AgdaSpace{}%
\AgdaBound{x}\<%
\\
\>[0]\AgdaSymbol{...}\AgdaSpace{}%
\AgdaSymbol{|}\AgdaSpace{}%
\AgdaInductiveConstructor{yes}\AgdaSpace{}%
\AgdaBound{x₁≡x}\AgdaSpace{}%
\AgdaSymbol{=}\AgdaSpace{}%
\AgdaFunction{contradiction}\AgdaSpace{}%
\AgdaBound{x₁≡x}\AgdaSpace{}%
\AgdaSymbol{(}\AgdaFunction{lookup}\AgdaSpace{}%
\AgdaBound{px₁}\AgdaSpace{}%
\AgdaFunction{x∈xs++ys}\AgdaSymbol{)}\<%
\\
\>[0][@{}l@{\AgdaIndent{0}}]%
\>[2]\AgdaKeyword{where}%
\>[963I]\AgdaFunction{x∈xs++ys}\AgdaSpace{}%
\AgdaSymbol{:}\AgdaSpace{}%
\AgdaBound{x}\AgdaSpace{}%
\AgdaOperator{\AgdaFunction{∈}}\AgdaSpace{}%
\AgdaFunction{map}\AgdaSpace{}%
\AgdaField{proj₁}\AgdaSpace{}%
\AgdaSymbol{(}\AgdaBound{xs}\AgdaSpace{}%
\AgdaOperator{\AgdaFunction{++}}\AgdaSpace{}%
\AgdaBound{ys}\AgdaSymbol{)}\<%
\\
\>[.][@{}l@{}]\<[963I]%
\>[8]\AgdaFunction{x∈xs++ys}\AgdaSpace{}%
\AgdaSymbol{=}\AgdaSpace{}%
\AgdaFunction{subst}\AgdaSpace{}%
\AgdaSymbol{(λ}\AgdaSpace{}%
\AgdaBound{ls}\AgdaSpace{}%
\AgdaSymbol{→}\AgdaSpace{}%
\AgdaSymbol{\AgdaUnderscore{}}\AgdaSpace{}%
\AgdaOperator{\AgdaFunction{∈}}\AgdaSpace{}%
\AgdaBound{ls}\AgdaSymbol{)}\AgdaSpace{}%
\AgdaSymbol{(}\AgdaFunction{sym}\AgdaSpace{}%
\AgdaSymbol{(}\AgdaFunction{map-++-commute}\AgdaSpace{}%
\AgdaField{proj₁}\AgdaSpace{}%
\AgdaBound{xs}\AgdaSpace{}%
\AgdaBound{ys}\AgdaSymbol{))}\AgdaSpace{}%
\AgdaSymbol{(}\AgdaFunction{∈-++⁺ʳ}\AgdaSpace{}%
\AgdaSymbol{(}\AgdaFunction{map}\AgdaSpace{}%
\AgdaField{proj₁}\AgdaSpace{}%
\AgdaBound{xs}\AgdaSymbol{)}\AgdaSpace{}%
\AgdaSymbol{(}\AgdaFunction{lemma2}\AgdaSpace{}%
\AgdaBound{x}\AgdaSpace{}%
\AgdaBound{ys}\AgdaSpace{}%
\AgdaBound{v∈}\AgdaSymbol{))}\<%
\\
\>[0]\AgdaSymbol{...}\AgdaSpace{}%
\AgdaSymbol{|}\AgdaSpace{}%
\AgdaInductiveConstructor{no}\AgdaSpace{}%
\AgdaBound{¬x₁≡x}\AgdaSpace{}%
\AgdaSymbol{=}\AgdaSpace{}%
\AgdaFunction{∈-cmdWrote++⁺ʳ}\AgdaSpace{}%
\AgdaBound{x}\AgdaSpace{}%
\AgdaBound{xs}\AgdaSpace{}%
\AgdaBound{ys}\AgdaSpace{}%
\AgdaBound{u}\AgdaSpace{}%
\AgdaBound{v∈}\<%
\\
\\[\AgdaEmptyExtraSkip]%
\>[0]\AgdaFunction{∈-cmdWrote++mid}\AgdaSpace{}%
\AgdaSymbol{:}\AgdaSpace{}%
\AgdaSymbol{∀}\AgdaSpace{}%
\AgdaSymbol{\{}\AgdaBound{v}\AgdaSymbol{\}}\AgdaSpace{}%
\AgdaBound{x}\AgdaSpace{}%
\AgdaBound{xs}\AgdaSpace{}%
\AgdaBound{ys}\AgdaSpace{}%
\AgdaBound{zs}\AgdaSpace{}%
\AgdaSymbol{→}\AgdaSpace{}%
\AgdaDatatype{Unique}\AgdaSpace{}%
\AgdaSymbol{(}\AgdaFunction{map}\AgdaSpace{}%
\AgdaField{proj₁}\AgdaSpace{}%
\AgdaSymbol{(}\AgdaBound{xs}\AgdaSpace{}%
\AgdaOperator{\AgdaFunction{++}}\AgdaSpace{}%
\AgdaBound{ys}\AgdaSpace{}%
\AgdaOperator{\AgdaFunction{++}}\AgdaSpace{}%
\AgdaBound{zs}\AgdaSymbol{))}\AgdaSpace{}%
\AgdaSymbol{→}\AgdaSpace{}%
\AgdaBound{v}\AgdaSpace{}%
\AgdaOperator{\AgdaFunction{∈}}\AgdaSpace{}%
\AgdaFunction{cmdWrote}\AgdaSpace{}%
\AgdaSymbol{(}\AgdaBound{xs}\AgdaSpace{}%
\AgdaOperator{\AgdaFunction{++}}\AgdaSpace{}%
\AgdaBound{zs}\AgdaSymbol{)}\AgdaSpace{}%
\AgdaBound{x}\AgdaSpace{}%
\AgdaSymbol{→}\AgdaSpace{}%
\AgdaBound{v}\AgdaSpace{}%
\AgdaOperator{\AgdaFunction{∈}}\AgdaSpace{}%
\AgdaFunction{cmdWrote}\AgdaSpace{}%
\AgdaSymbol{(}\AgdaBound{xs}\AgdaSpace{}%
\AgdaOperator{\AgdaFunction{++}}\AgdaSpace{}%
\AgdaBound{ys}\AgdaSpace{}%
\AgdaOperator{\AgdaFunction{++}}\AgdaSpace{}%
\AgdaBound{zs}\AgdaSymbol{)}\AgdaSpace{}%
\AgdaBound{x}\<%
\\
\>[0]\AgdaFunction{∈-cmdWrote++mid}\AgdaSpace{}%
\AgdaBound{x}\AgdaSpace{}%
\AgdaInductiveConstructor{[]}\AgdaSpace{}%
\AgdaBound{ys}\AgdaSpace{}%
\AgdaBound{zs}\AgdaSpace{}%
\AgdaBound{u}\AgdaSpace{}%
\AgdaBound{v∈}\AgdaSpace{}%
\AgdaSymbol{=}\AgdaSpace{}%
\AgdaFunction{∈-cmdWrote++⁺ʳ}\AgdaSpace{}%
\AgdaBound{x}\AgdaSpace{}%
\AgdaBound{ys}\AgdaSpace{}%
\AgdaBound{zs}\AgdaSpace{}%
\AgdaBound{u}\AgdaSpace{}%
\AgdaBound{v∈}\<%
\\
\>[0]\AgdaFunction{∈-cmdWrote++mid}\AgdaSpace{}%
\AgdaBound{x}\AgdaSpace{}%
\AgdaSymbol{(}\AgdaBound{x₁}\AgdaSpace{}%
\AgdaOperator{\AgdaInductiveConstructor{∷}}\AgdaSpace{}%
\AgdaBound{xs}\AgdaSymbol{)}\AgdaSpace{}%
\AgdaBound{ys}\AgdaSpace{}%
\AgdaBound{zs}\AgdaSpace{}%
\AgdaSymbol{(}\AgdaBound{px₁}\AgdaSpace{}%
\AgdaOperator{\AgdaInductiveConstructor{∷}}\AgdaSpace{}%
\AgdaBound{u}\AgdaSymbol{)}\AgdaSpace{}%
\AgdaBound{v∈}\AgdaSpace{}%
\AgdaKeyword{with}\AgdaSpace{}%
\AgdaSymbol{(}\AgdaField{proj₁}\AgdaSpace{}%
\AgdaBound{x₁}\AgdaSymbol{)}\AgdaSpace{}%
\AgdaOperator{\AgdaFunction{≟}}\AgdaSpace{}%
\AgdaBound{x}\<%
\\
\>[0]\AgdaSymbol{...}\AgdaSpace{}%
\AgdaSymbol{|}\AgdaSpace{}%
\AgdaInductiveConstructor{yes}\AgdaSpace{}%
\AgdaBound{x₁≡x}\AgdaSpace{}%
\AgdaSymbol{=}\AgdaSpace{}%
\AgdaBound{v∈}\<%
\\
\>[0]\AgdaSymbol{...}\AgdaSpace{}%
\AgdaSymbol{|}\AgdaSpace{}%
\AgdaInductiveConstructor{no}\AgdaSpace{}%
\AgdaBound{¬x₁≡x}\AgdaSpace{}%
\AgdaSymbol{=}\AgdaSpace{}%
\AgdaFunction{∈-cmdWrote++mid}\AgdaSpace{}%
\AgdaBound{x}\AgdaSpace{}%
\AgdaBound{xs}\AgdaSpace{}%
\AgdaBound{ys}\AgdaSpace{}%
\AgdaBound{zs}\AgdaSpace{}%
\AgdaBound{u}\AgdaSpace{}%
\AgdaBound{v∈}\<%
\end{code}

\newcommand{\cmdRead}{%
\begin{code}%
\>[0]\AgdaComment{-- The FileNames the Cmd read according to the FileInfo}\<%
\\
\>[0]\AgdaFunction{cmdRead}\AgdaSpace{}%
\AgdaSymbol{:}\AgdaSpace{}%
\AgdaFunction{FileInfo}\AgdaSpace{}%
\AgdaSymbol{→}\AgdaSpace{}%
\AgdaFunction{Cmd}\AgdaSpace{}%
\AgdaSymbol{→}\AgdaSpace{}%
\AgdaDatatype{List}\AgdaSpace{}%
\AgdaFunction{FileName}\<%
\end{code}}
\begin{code}[hide]%
\>[0]\AgdaFunction{cmdRead}\AgdaSpace{}%
\AgdaInductiveConstructor{[]}\AgdaSpace{}%
\AgdaBound{p}\AgdaSpace{}%
\AgdaSymbol{=}\AgdaSpace{}%
\AgdaInductiveConstructor{[]}\<%
\\
\>[0]\AgdaFunction{cmdRead}\AgdaSpace{}%
\AgdaSymbol{(}\AgdaBound{x}\AgdaSpace{}%
\AgdaOperator{\AgdaInductiveConstructor{∷}}\AgdaSpace{}%
\AgdaBound{ls}\AgdaSymbol{)}\AgdaSpace{}%
\AgdaBound{p}\AgdaSpace{}%
\AgdaKeyword{with}\AgdaSpace{}%
\AgdaSymbol{(}\AgdaField{proj₁}\AgdaSpace{}%
\AgdaBound{x}\AgdaSymbol{)}\AgdaSpace{}%
\AgdaOperator{\AgdaFunction{≟}}\AgdaSpace{}%
\AgdaBound{p}\<%
\\
\>[0]\AgdaSymbol{...}\AgdaSpace{}%
\AgdaSymbol{|}\AgdaSpace{}%
\AgdaInductiveConstructor{yes}\AgdaSpace{}%
\AgdaBound{x≡p}\AgdaSpace{}%
\AgdaSymbol{=}\AgdaSpace{}%
\AgdaField{proj₁}\AgdaSpace{}%
\AgdaSymbol{(}\AgdaField{proj₂}\AgdaSpace{}%
\AgdaBound{x}\AgdaSymbol{)}\<%
\\
\>[0]\AgdaSymbol{...}\AgdaSpace{}%
\AgdaSymbol{|}\AgdaSpace{}%
\AgdaInductiveConstructor{no}\AgdaSpace{}%
\AgdaBound{¬x≡p}\AgdaSpace{}%
\AgdaSymbol{=}\AgdaSpace{}%
\AgdaFunction{cmdRead}\AgdaSpace{}%
\AgdaBound{ls}\AgdaSpace{}%
\AgdaBound{p}\<%
\\
\\[\AgdaEmptyExtraSkip]%
\>[0]\AgdaFunction{∈-cmdRead∷l}\AgdaSpace{}%
\AgdaSymbol{:}\AgdaSpace{}%
\AgdaSymbol{∀}\AgdaSpace{}%
\AgdaSymbol{\{}\AgdaBound{v}\AgdaSymbol{\}}\AgdaSpace{}%
\AgdaBound{x}\AgdaSpace{}%
\AgdaBound{ls}\AgdaSpace{}%
\AgdaSymbol{→}\AgdaSpace{}%
\AgdaBound{v}\AgdaSpace{}%
\AgdaOperator{\AgdaFunction{∈}}\AgdaSpace{}%
\AgdaField{proj₁}\AgdaSpace{}%
\AgdaSymbol{(}\AgdaField{proj₂}\AgdaSpace{}%
\AgdaBound{x}\AgdaSymbol{)}\AgdaSpace{}%
\AgdaSymbol{→}\AgdaSpace{}%
\AgdaBound{v}\AgdaSpace{}%
\AgdaOperator{\AgdaFunction{∈}}\AgdaSpace{}%
\AgdaFunction{cmdRead}\AgdaSpace{}%
\AgdaSymbol{(}\AgdaBound{x}\AgdaSpace{}%
\AgdaOperator{\AgdaInductiveConstructor{∷}}\AgdaSpace{}%
\AgdaBound{ls}\AgdaSymbol{)}\AgdaSpace{}%
\AgdaSymbol{(}\AgdaField{proj₁}\AgdaSpace{}%
\AgdaBound{x}\AgdaSymbol{)}\<%
\\
\>[0]\AgdaFunction{∈-cmdRead∷l}\AgdaSpace{}%
\AgdaBound{x}\AgdaSpace{}%
\AgdaBound{ls}\AgdaSpace{}%
\AgdaBound{v∈}\AgdaSpace{}%
\AgdaKeyword{with}\AgdaSpace{}%
\AgdaSymbol{(}\AgdaField{proj₁}\AgdaSpace{}%
\AgdaBound{x}\AgdaSymbol{)}\AgdaSpace{}%
\AgdaOperator{\AgdaFunction{≟}}\AgdaSpace{}%
\AgdaSymbol{(}\AgdaField{proj₁}\AgdaSpace{}%
\AgdaBound{x}\AgdaSymbol{)}\<%
\\
\>[0]\AgdaSymbol{...}\AgdaSpace{}%
\AgdaSymbol{|}\AgdaSpace{}%
\AgdaInductiveConstructor{no}\AgdaSpace{}%
\AgdaBound{¬x≡x}\AgdaSpace{}%
\AgdaSymbol{=}\AgdaSpace{}%
\AgdaFunction{contradiction}\AgdaSpace{}%
\AgdaInductiveConstructor{refl}\AgdaSpace{}%
\AgdaBound{¬x≡x}\<%
\\
\>[0]\AgdaSymbol{...}\AgdaSpace{}%
\AgdaSymbol{|}\AgdaSpace{}%
\AgdaInductiveConstructor{yes}\AgdaSpace{}%
\AgdaBound{x≡x}\AgdaSpace{}%
\AgdaSymbol{=}\AgdaSpace{}%
\AgdaBound{v∈}\<%
\\
\\[\AgdaEmptyExtraSkip]%
\>[0]\AgdaFunction{∈-cmdRead∷}\AgdaSpace{}%
\AgdaSymbol{:}\AgdaSpace{}%
\AgdaSymbol{∀}\AgdaSpace{}%
\AgdaSymbol{\{}\AgdaBound{v}\AgdaSymbol{\}}\AgdaSpace{}%
\AgdaBound{x}\AgdaSpace{}%
\AgdaBound{x₁}\AgdaSpace{}%
\AgdaBound{ls}\AgdaSpace{}%
\AgdaSymbol{→}\AgdaSpace{}%
\AgdaBound{v}\AgdaSpace{}%
\AgdaOperator{\AgdaFunction{∈}}\AgdaSpace{}%
\AgdaFunction{cmdRead}\AgdaSpace{}%
\AgdaBound{ls}\AgdaSpace{}%
\AgdaBound{x₁}\AgdaSpace{}%
\AgdaSymbol{→}\AgdaSpace{}%
\AgdaOperator{\AgdaFunction{¬}}\AgdaSpace{}%
\AgdaSymbol{(}\AgdaField{proj₁}\AgdaSpace{}%
\AgdaBound{x}\AgdaSymbol{)}\AgdaSpace{}%
\AgdaOperator{\AgdaDatatype{≡}}\AgdaSpace{}%
\AgdaBound{x₁}\AgdaSpace{}%
\AgdaSymbol{→}\AgdaSpace{}%
\AgdaBound{v}\AgdaSpace{}%
\AgdaOperator{\AgdaFunction{∈}}\AgdaSpace{}%
\AgdaFunction{cmdRead}\AgdaSpace{}%
\AgdaSymbol{(}\AgdaBound{x}\AgdaSpace{}%
\AgdaOperator{\AgdaInductiveConstructor{∷}}\AgdaSpace{}%
\AgdaBound{ls}\AgdaSymbol{)}\AgdaSpace{}%
\AgdaBound{x₁}\<%
\\
\>[0]\AgdaFunction{∈-cmdRead∷}\AgdaSpace{}%
\AgdaBound{x}\AgdaSpace{}%
\AgdaBound{x₁}\AgdaSpace{}%
\AgdaBound{ls}\AgdaSpace{}%
\AgdaBound{v∈}\AgdaSpace{}%
\AgdaBound{¬≡}\AgdaSpace{}%
\AgdaKeyword{with}\AgdaSpace{}%
\AgdaSymbol{(}\AgdaField{proj₁}\AgdaSpace{}%
\AgdaBound{x}\AgdaSymbol{)}\AgdaSpace{}%
\AgdaOperator{\AgdaFunction{≟}}\AgdaSpace{}%
\AgdaBound{x₁}\<%
\\
\>[0]\AgdaSymbol{...}\AgdaSpace{}%
\AgdaSymbol{|}\AgdaSpace{}%
\AgdaInductiveConstructor{yes}\AgdaSpace{}%
\AgdaBound{x≡x₁}\AgdaSpace{}%
\AgdaSymbol{=}\AgdaSpace{}%
\AgdaFunction{contradiction}\AgdaSpace{}%
\AgdaBound{x≡x₁}\AgdaSpace{}%
\AgdaBound{¬≡}\<%
\\
\>[0]\AgdaSymbol{...}\AgdaSpace{}%
\AgdaSymbol{|}\AgdaSpace{}%
\AgdaInductiveConstructor{no}\AgdaSpace{}%
\AgdaBound{¬x≡x₁}\AgdaSpace{}%
\AgdaSymbol{=}\AgdaSpace{}%
\AgdaBound{v∈}\<%
\\
\\[\AgdaEmptyExtraSkip]%
\>[0]\AgdaComment{-- if there is a file in the cmdRead for x , then x ∈ xs}\<%
\\
\>[0]\AgdaFunction{lemma1}\AgdaSpace{}%
\AgdaSymbol{:}\AgdaSpace{}%
\AgdaSymbol{∀}\AgdaSpace{}%
\AgdaSymbol{\{}\AgdaBound{v}\AgdaSymbol{\}}\AgdaSpace{}%
\AgdaBound{x}\AgdaSpace{}%
\AgdaBound{xs}\AgdaSpace{}%
\AgdaSymbol{→}\AgdaSpace{}%
\AgdaBound{v}\AgdaSpace{}%
\AgdaOperator{\AgdaFunction{∈}}\AgdaSpace{}%
\AgdaFunction{cmdRead}\AgdaSpace{}%
\AgdaBound{xs}\AgdaSpace{}%
\AgdaBound{x}\AgdaSpace{}%
\AgdaSymbol{→}\AgdaSpace{}%
\AgdaBound{x}\AgdaSpace{}%
\AgdaOperator{\AgdaFunction{∈}}\AgdaSpace{}%
\AgdaFunction{map}\AgdaSpace{}%
\AgdaField{proj₁}\AgdaSpace{}%
\AgdaBound{xs}\<%
\\
\>[0]\AgdaFunction{lemma1}\AgdaSpace{}%
\AgdaBound{x}\AgdaSpace{}%
\AgdaSymbol{(}\AgdaBound{x₁}\AgdaSpace{}%
\AgdaOperator{\AgdaInductiveConstructor{∷}}\AgdaSpace{}%
\AgdaBound{xs}\AgdaSymbol{)}\AgdaSpace{}%
\AgdaBound{v∈}\AgdaSpace{}%
\AgdaKeyword{with}\AgdaSpace{}%
\AgdaSymbol{(}\AgdaField{proj₁}\AgdaSpace{}%
\AgdaBound{x₁}\AgdaSymbol{)}\AgdaSpace{}%
\AgdaOperator{\AgdaFunction{≟}}\AgdaSpace{}%
\AgdaBound{x}\<%
\\
\>[0]\AgdaSymbol{...}\AgdaSpace{}%
\AgdaSymbol{|}\AgdaSpace{}%
\AgdaInductiveConstructor{yes}\AgdaSpace{}%
\AgdaBound{x₁≡x}\AgdaSpace{}%
\AgdaSymbol{=}\AgdaSpace{}%
\AgdaInductiveConstructor{here}\AgdaSpace{}%
\AgdaSymbol{(}\AgdaFunction{sym}\AgdaSpace{}%
\AgdaBound{x₁≡x}\AgdaSymbol{)}\<%
\\
\>[0]\AgdaSymbol{...}\AgdaSpace{}%
\AgdaSymbol{|}\AgdaSpace{}%
\AgdaInductiveConstructor{no}\AgdaSpace{}%
\AgdaBound{¬x₁≡x}\AgdaSpace{}%
\AgdaSymbol{=}\AgdaSpace{}%
\AgdaInductiveConstructor{there}\AgdaSpace{}%
\AgdaSymbol{(}\AgdaFunction{lemma1}\AgdaSpace{}%
\AgdaBound{x}\AgdaSpace{}%
\AgdaBound{xs}\AgdaSpace{}%
\AgdaBound{v∈}\AgdaSymbol{)}\<%
\\
\\[\AgdaEmptyExtraSkip]%
\>[0]\AgdaFunction{∈-cmdRead++⁺ʳ}\AgdaSpace{}%
\AgdaSymbol{:}\AgdaSpace{}%
\AgdaSymbol{∀}\AgdaSpace{}%
\AgdaSymbol{\{}\AgdaBound{v}\AgdaSymbol{\}}\AgdaSpace{}%
\AgdaBound{x}\AgdaSpace{}%
\AgdaBound{xs}\AgdaSpace{}%
\AgdaBound{ys}\AgdaSpace{}%
\AgdaSymbol{→}\AgdaSpace{}%
\AgdaDatatype{Unique}\AgdaSpace{}%
\AgdaSymbol{(}\AgdaFunction{map}\AgdaSpace{}%
\AgdaField{proj₁}\AgdaSpace{}%
\AgdaSymbol{(}\AgdaBound{xs}\AgdaSpace{}%
\AgdaOperator{\AgdaFunction{++}}\AgdaSpace{}%
\AgdaBound{ys}\AgdaSymbol{))}\AgdaSpace{}%
\AgdaSymbol{→}\AgdaSpace{}%
\AgdaBound{v}\AgdaSpace{}%
\AgdaOperator{\AgdaFunction{∈}}\AgdaSpace{}%
\AgdaFunction{cmdRead}\AgdaSpace{}%
\AgdaBound{ys}\AgdaSpace{}%
\AgdaBound{x}\AgdaSpace{}%
\AgdaSymbol{→}\AgdaSpace{}%
\AgdaBound{v}\AgdaSpace{}%
\AgdaOperator{\AgdaFunction{∈}}\AgdaSpace{}%
\AgdaFunction{cmdRead}\AgdaSpace{}%
\AgdaSymbol{(}\AgdaBound{xs}\AgdaSpace{}%
\AgdaOperator{\AgdaFunction{++}}\AgdaSpace{}%
\AgdaBound{ys}\AgdaSymbol{)}\AgdaSpace{}%
\AgdaBound{x}\<%
\\
\>[0]\AgdaFunction{∈-cmdRead++⁺ʳ}\AgdaSpace{}%
\AgdaBound{x}\AgdaSpace{}%
\AgdaInductiveConstructor{[]}\AgdaSpace{}%
\AgdaBound{ys}\AgdaSpace{}%
\AgdaBound{u}\AgdaSpace{}%
\AgdaBound{v∈}\AgdaSpace{}%
\AgdaSymbol{=}\AgdaSpace{}%
\AgdaBound{v∈}\<%
\\
\>[0]\AgdaFunction{∈-cmdRead++⁺ʳ}\AgdaSpace{}%
\AgdaBound{x}\AgdaSpace{}%
\AgdaSymbol{(}\AgdaBound{x₁}\AgdaSpace{}%
\AgdaOperator{\AgdaInductiveConstructor{∷}}\AgdaSpace{}%
\AgdaBound{xs}\AgdaSymbol{)}\AgdaSpace{}%
\AgdaBound{ys}\AgdaSpace{}%
\AgdaSymbol{(}\AgdaBound{px₁}\AgdaSpace{}%
\AgdaOperator{\AgdaInductiveConstructor{∷}}\AgdaSpace{}%
\AgdaBound{u}\AgdaSymbol{)}\AgdaSpace{}%
\AgdaBound{v∈}\AgdaSpace{}%
\AgdaKeyword{with}\AgdaSpace{}%
\AgdaSymbol{(}\AgdaField{proj₁}\AgdaSpace{}%
\AgdaBound{x₁}\AgdaSymbol{)}\AgdaSpace{}%
\AgdaOperator{\AgdaFunction{≟}}\AgdaSpace{}%
\AgdaBound{x}\<%
\\
\>[0]\AgdaSymbol{...}\AgdaSpace{}%
\AgdaSymbol{|}\AgdaSpace{}%
\AgdaInductiveConstructor{yes}\AgdaSpace{}%
\AgdaBound{x₁≡x}\AgdaSpace{}%
\AgdaSymbol{=}\AgdaSpace{}%
\AgdaFunction{contradiction}\AgdaSpace{}%
\AgdaBound{x₁≡x}\AgdaSpace{}%
\AgdaSymbol{(}\AgdaFunction{lookup}\AgdaSpace{}%
\AgdaBound{px₁}\AgdaSpace{}%
\AgdaFunction{x∈xs++ys}\AgdaSymbol{)}\<%
\\
\>[0][@{}l@{\AgdaIndent{0}}]%
\>[2]\AgdaKeyword{where}%
\>[1304I]\AgdaFunction{x∈xs++ys}\AgdaSpace{}%
\AgdaSymbol{:}\AgdaSpace{}%
\AgdaBound{x}\AgdaSpace{}%
\AgdaOperator{\AgdaFunction{∈}}\AgdaSpace{}%
\AgdaFunction{map}\AgdaSpace{}%
\AgdaField{proj₁}\AgdaSpace{}%
\AgdaSymbol{(}\AgdaBound{xs}\AgdaSpace{}%
\AgdaOperator{\AgdaFunction{++}}\AgdaSpace{}%
\AgdaBound{ys}\AgdaSymbol{)}\<%
\\
\>[.][@{}l@{}]\<[1304I]%
\>[8]\AgdaFunction{x∈xs++ys}\AgdaSpace{}%
\AgdaSymbol{=}\AgdaSpace{}%
\AgdaFunction{subst}\AgdaSpace{}%
\AgdaSymbol{(λ}\AgdaSpace{}%
\AgdaBound{ls}\AgdaSpace{}%
\AgdaSymbol{→}\AgdaSpace{}%
\AgdaSymbol{\AgdaUnderscore{}}\AgdaSpace{}%
\AgdaOperator{\AgdaFunction{∈}}\AgdaSpace{}%
\AgdaBound{ls}\AgdaSymbol{)}\AgdaSpace{}%
\AgdaSymbol{(}\AgdaFunction{sym}\AgdaSpace{}%
\AgdaSymbol{(}\AgdaFunction{map-++-commute}\AgdaSpace{}%
\AgdaField{proj₁}\AgdaSpace{}%
\AgdaBound{xs}\AgdaSpace{}%
\AgdaBound{ys}\AgdaSymbol{))}\AgdaSpace{}%
\AgdaSymbol{(}\AgdaFunction{∈-++⁺ʳ}\AgdaSpace{}%
\AgdaSymbol{(}\AgdaFunction{map}\AgdaSpace{}%
\AgdaField{proj₁}\AgdaSpace{}%
\AgdaBound{xs}\AgdaSymbol{)}\AgdaSpace{}%
\AgdaSymbol{(}\AgdaFunction{lemma1}\AgdaSpace{}%
\AgdaBound{x}\AgdaSpace{}%
\AgdaBound{ys}\AgdaSpace{}%
\AgdaBound{v∈}\AgdaSymbol{))}\<%
\\
\>[0]\AgdaSymbol{...}\AgdaSpace{}%
\AgdaSymbol{|}\AgdaSpace{}%
\AgdaInductiveConstructor{no}\AgdaSpace{}%
\AgdaBound{¬x₁≡x}\AgdaSpace{}%
\AgdaSymbol{=}\AgdaSpace{}%
\AgdaFunction{∈-cmdRead++⁺ʳ}\AgdaSpace{}%
\AgdaBound{x}\AgdaSpace{}%
\AgdaBound{xs}\AgdaSpace{}%
\AgdaBound{ys}\AgdaSpace{}%
\AgdaBound{u}\AgdaSpace{}%
\AgdaBound{v∈}\<%
\\
\\[\AgdaEmptyExtraSkip]%
\>[0]\AgdaFunction{∈-cmdRead++mid}\AgdaSpace{}%
\AgdaSymbol{:}\AgdaSpace{}%
\AgdaSymbol{∀}\AgdaSpace{}%
\AgdaSymbol{\{}\AgdaBound{v}\AgdaSymbol{\}}\AgdaSpace{}%
\AgdaBound{x}\AgdaSpace{}%
\AgdaBound{xs}\AgdaSpace{}%
\AgdaBound{ys}\AgdaSpace{}%
\AgdaBound{zs}\AgdaSpace{}%
\AgdaSymbol{→}\AgdaSpace{}%
\AgdaDatatype{Unique}\AgdaSpace{}%
\AgdaSymbol{(}\AgdaFunction{map}\AgdaSpace{}%
\AgdaField{proj₁}\AgdaSpace{}%
\AgdaSymbol{(}\AgdaBound{xs}\AgdaSpace{}%
\AgdaOperator{\AgdaFunction{++}}\AgdaSpace{}%
\AgdaBound{ys}\AgdaSpace{}%
\AgdaOperator{\AgdaFunction{++}}\AgdaSpace{}%
\AgdaBound{zs}\AgdaSymbol{))}\AgdaSpace{}%
\AgdaSymbol{→}\AgdaSpace{}%
\AgdaBound{v}\AgdaSpace{}%
\AgdaOperator{\AgdaFunction{∈}}\AgdaSpace{}%
\AgdaFunction{cmdRead}\AgdaSpace{}%
\AgdaSymbol{(}\AgdaBound{xs}\AgdaSpace{}%
\AgdaOperator{\AgdaFunction{++}}\AgdaSpace{}%
\AgdaBound{zs}\AgdaSymbol{)}\AgdaSpace{}%
\AgdaBound{x}\AgdaSpace{}%
\AgdaSymbol{→}\AgdaSpace{}%
\AgdaBound{v}\AgdaSpace{}%
\AgdaOperator{\AgdaFunction{∈}}\AgdaSpace{}%
\AgdaFunction{cmdRead}\AgdaSpace{}%
\AgdaSymbol{(}\AgdaBound{xs}\AgdaSpace{}%
\AgdaOperator{\AgdaFunction{++}}\AgdaSpace{}%
\AgdaBound{ys}\AgdaSpace{}%
\AgdaOperator{\AgdaFunction{++}}\AgdaSpace{}%
\AgdaBound{zs}\AgdaSymbol{)}\AgdaSpace{}%
\AgdaBound{x}\<%
\\
\>[0]\AgdaFunction{∈-cmdRead++mid}\AgdaSpace{}%
\AgdaBound{x}\AgdaSpace{}%
\AgdaInductiveConstructor{[]}\AgdaSpace{}%
\AgdaBound{ys}\AgdaSpace{}%
\AgdaBound{zs}\AgdaSpace{}%
\AgdaBound{u}\AgdaSpace{}%
\AgdaBound{v∈}\AgdaSpace{}%
\AgdaSymbol{=}\AgdaSpace{}%
\AgdaFunction{∈-cmdRead++⁺ʳ}\AgdaSpace{}%
\AgdaBound{x}\AgdaSpace{}%
\AgdaBound{ys}\AgdaSpace{}%
\AgdaBound{zs}\AgdaSpace{}%
\AgdaBound{u}\AgdaSpace{}%
\AgdaBound{v∈}\<%
\\
\>[0]\AgdaFunction{∈-cmdRead++mid}\AgdaSpace{}%
\AgdaBound{x}\AgdaSpace{}%
\AgdaSymbol{(}\AgdaBound{x₁}\AgdaSpace{}%
\AgdaOperator{\AgdaInductiveConstructor{∷}}\AgdaSpace{}%
\AgdaBound{xs}\AgdaSymbol{)}\AgdaSpace{}%
\AgdaBound{ys}\AgdaSpace{}%
\AgdaBound{zs}\AgdaSpace{}%
\AgdaSymbol{(}\AgdaBound{px₁}\AgdaSpace{}%
\AgdaOperator{\AgdaInductiveConstructor{∷}}\AgdaSpace{}%
\AgdaBound{u}\AgdaSymbol{)}\AgdaSpace{}%
\AgdaBound{v∈}\AgdaSpace{}%
\AgdaKeyword{with}\AgdaSpace{}%
\AgdaSymbol{(}\AgdaField{proj₁}\AgdaSpace{}%
\AgdaBound{x₁}\AgdaSymbol{)}\AgdaSpace{}%
\AgdaOperator{\AgdaFunction{≟}}\AgdaSpace{}%
\AgdaBound{x}\<%
\\
\>[0]\AgdaSymbol{...}\AgdaSpace{}%
\AgdaSymbol{|}\AgdaSpace{}%
\AgdaInductiveConstructor{yes}\AgdaSpace{}%
\AgdaBound{x₁≡x}\AgdaSpace{}%
\AgdaSymbol{=}\AgdaSpace{}%
\AgdaBound{v∈}\<%
\\
\>[0]\AgdaSymbol{...}\AgdaSpace{}%
\AgdaSymbol{|}\AgdaSpace{}%
\AgdaInductiveConstructor{no}\AgdaSpace{}%
\AgdaBound{¬x₁≡x}\AgdaSpace{}%
\AgdaSymbol{=}\AgdaSpace{}%
\AgdaFunction{∈-cmdRead++mid}\AgdaSpace{}%
\AgdaBound{x}\AgdaSpace{}%
\AgdaBound{xs}\AgdaSpace{}%
\AgdaBound{ys}\AgdaSpace{}%
\AgdaBound{zs}\AgdaSpace{}%
\AgdaBound{u}\AgdaSpace{}%
\AgdaBound{v∈}\<%
\\
\\[\AgdaEmptyExtraSkip]%
\>[0]\AgdaComment{--  ¬speculativehazard def only makes sense with unique FileINfo and build}\<%
\end{code}

\begin{code}[hide]%
\>[0]\AgdaComment{\{- we have a Speculative hazard if a required command read something a speculative command wrote to. 
 So we need to be able to determine: 
1. when a command is required : a command is required if its the first command in the list?

2. when a command is speculated
-\}}\<%
\\
\>[0]\AgdaComment{-- (cmdsRun (save x (cmdReadNames x s) (cmdWriteNames x s) ls))}\<%
\\
\>[0]\AgdaComment{-- (save x (cmdReadNames x s) (cmdWriteNames x s) ls)}\<%
\\
\>[0]\AgdaComment{-- (save x (cmdReadNames x s) (cmdWriteNames x s) ls)}\<%
\\
\\[\AgdaEmptyExtraSkip]%
\>[0]\<%
\end{code}

\newcommand{\hazard}{%
\begin{code}%
\>[0]\AgdaKeyword{data}\AgdaSpace{}%
\AgdaDatatype{Hazard}\AgdaSpace{}%
\AgdaSymbol{:}\AgdaSpace{}%
\AgdaFunction{FileSystem}\AgdaSpace{}%
\AgdaSymbol{→}\AgdaSpace{}%
\AgdaFunction{Cmd}\AgdaSpace{}%
\AgdaSymbol{→}\AgdaSpace{}%
\AgdaPostulate{Build}\AgdaSpace{}%
\AgdaSymbol{→}\AgdaSpace{}%
\AgdaFunction{FileInfo}\AgdaSpace{}%
\AgdaSymbol{→}\AgdaSpace{}%
\AgdaPrimitiveType{Set}\AgdaSpace{}%
\AgdaKeyword{where}\<%
\\
\>[0][@{}l@{\AgdaIndent{0}}]%
\>[2]\AgdaInductiveConstructor{ReadWrite}%
\>[14]\AgdaSymbol{:}\AgdaSpace{}%
\AgdaSymbol{∀}\AgdaSpace{}%
\AgdaSymbol{\{}\AgdaBound{s}\AgdaSymbol{\}}\AgdaSpace{}%
\AgdaSymbol{\{}\AgdaBound{x}\AgdaSymbol{\}}\AgdaSpace{}%
\AgdaSymbol{\{}\AgdaBound{b}\AgdaSymbol{\}}\AgdaSpace{}%
\AgdaSymbol{\{}\AgdaBound{ls}\AgdaSymbol{\}}\AgdaSpace{}%
\AgdaSymbol{\{}\AgdaBound{v}\AgdaSymbol{\}}\AgdaSpace{}%
\AgdaSymbol{→}\AgdaSpace{}%
\AgdaBound{v}\AgdaSpace{}%
\AgdaOperator{\AgdaFunction{∈}}\AgdaSpace{}%
\AgdaSymbol{(}\AgdaPostulate{cmdWriteNames}\AgdaSpace{}%
\AgdaBound{x}\AgdaSpace{}%
\AgdaBound{s}\AgdaSymbol{)}\AgdaSpace{}%
\AgdaSymbol{→}\AgdaSpace{}%
\AgdaBound{v}\AgdaSpace{}%
\AgdaOperator{\AgdaFunction{∈}}\AgdaSpace{}%
\AgdaSymbol{(}\AgdaFunction{filesRead}\AgdaSpace{}%
\AgdaBound{ls}\AgdaSymbol{)}\AgdaSpace{}%
\AgdaSymbol{→}\AgdaSpace{}%
\AgdaDatatype{Hazard}\AgdaSpace{}%
\AgdaBound{s}\AgdaSpace{}%
\AgdaBound{x}\AgdaSpace{}%
\AgdaBound{b}\AgdaSpace{}%
\AgdaBound{ls}\<%
\\
\>[2]\AgdaInductiveConstructor{WriteWrite}%
\>[14]\AgdaSymbol{:}\AgdaSpace{}%
\AgdaSymbol{∀}\AgdaSpace{}%
\AgdaSymbol{\{}\AgdaBound{s}\AgdaSymbol{\}}\AgdaSpace{}%
\AgdaSymbol{\{}\AgdaBound{x}\AgdaSymbol{\}}\AgdaSpace{}%
\AgdaSymbol{\{}\AgdaBound{b}\AgdaSymbol{\}}\AgdaSpace{}%
\AgdaSymbol{\{}\AgdaBound{ls}\AgdaSymbol{\}}\AgdaSpace{}%
\AgdaSymbol{\{}\AgdaBound{v}\AgdaSymbol{\}}\AgdaSpace{}%
\AgdaSymbol{→}\AgdaSpace{}%
\AgdaBound{v}\AgdaSpace{}%
\AgdaOperator{\AgdaFunction{∈}}\AgdaSpace{}%
\AgdaSymbol{(}\AgdaPostulate{cmdWriteNames}\AgdaSpace{}%
\AgdaBound{x}\AgdaSpace{}%
\AgdaBound{s}\AgdaSymbol{)}\AgdaSpace{}%
\AgdaSymbol{→}\AgdaSpace{}%
\AgdaBound{v}\AgdaSpace{}%
\AgdaOperator{\AgdaFunction{∈}}\AgdaSpace{}%
\AgdaSymbol{(}\AgdaFunction{filesWrote}\AgdaSpace{}%
\AgdaBound{ls}\AgdaSymbol{)}\AgdaSpace{}%
\AgdaSymbol{→}\AgdaSpace{}%
\AgdaDatatype{Hazard}\AgdaSpace{}%
\AgdaBound{s}\AgdaSpace{}%
\AgdaBound{x}\AgdaSpace{}%
\AgdaBound{b}\AgdaSpace{}%
\AgdaBound{ls}\<%
\\
\>[2]\AgdaInductiveConstructor{Speculative}\AgdaSpace{}%
\AgdaSymbol{:}%
\>[1481I]\AgdaSymbol{∀}\AgdaSpace{}%
\AgdaSymbol{\{}\AgdaBound{s}\AgdaSymbol{\}}\AgdaSpace{}%
\AgdaSymbol{\{}\AgdaBound{x}\AgdaSymbol{\}}\AgdaSpace{}%
\AgdaSymbol{\{}\AgdaBound{b}\AgdaSymbol{\}}\AgdaSpace{}%
\AgdaSymbol{\{}\AgdaBound{ls}\AgdaSymbol{\}}\AgdaSpace{}%
\AgdaSymbol{\{}\AgdaBound{v}\AgdaSymbol{\}}\AgdaSpace{}%
\AgdaBound{x₁}\AgdaSpace{}%
\AgdaBound{x₂}\AgdaSpace{}%
\AgdaSymbol{→}\AgdaSpace{}%
\AgdaBound{x₂}\AgdaSpace{}%
\AgdaOperator{\AgdaFunction{before}}\AgdaSpace{}%
\AgdaBound{x₁}\AgdaSpace{}%
\AgdaOperator{\AgdaFunction{∈}}\AgdaSpace{}%
\AgdaSymbol{(}\AgdaBound{x}\AgdaSpace{}%
\AgdaOperator{\AgdaInductiveConstructor{∷}}\AgdaSpace{}%
\AgdaSymbol{(}\AgdaFunction{cmdsRun}\AgdaSpace{}%
\AgdaBound{ls}\AgdaSymbol{))}\AgdaSpace{}%
\AgdaSymbol{→}\AgdaSpace{}%
\AgdaBound{x₂}\AgdaSpace{}%
\AgdaOperator{\AgdaFunction{∈}}\AgdaSpace{}%
\AgdaBound{b}\AgdaSpace{}%
\AgdaSymbol{→}\AgdaSpace{}%
\AgdaOperator{\AgdaFunction{¬}}\AgdaSpace{}%
\AgdaBound{x₁}\AgdaSpace{}%
\AgdaOperator{\AgdaFunction{before}}\AgdaSpace{}%
\AgdaBound{x₂}\AgdaSpace{}%
\AgdaOperator{\AgdaFunction{∈}}\AgdaSpace{}%
\AgdaBound{b}\<%
\\
\>[.][@{}l@{}]\<[1481I]%
\>[16]\AgdaSymbol{→}\AgdaSpace{}%
\AgdaBound{v}\AgdaSpace{}%
\AgdaOperator{\AgdaFunction{∈}}\AgdaSpace{}%
\AgdaFunction{cmdRead}\AgdaSpace{}%
\AgdaSymbol{(}\AgdaFunction{save}\AgdaSpace{}%
\AgdaBound{s}\AgdaSpace{}%
\AgdaBound{x}\AgdaSpace{}%
\AgdaBound{ls}\AgdaSymbol{)}\AgdaSpace{}%
\AgdaBound{x₂}\AgdaSpace{}%
\AgdaSymbol{→}\AgdaSpace{}%
\AgdaBound{v}\AgdaSpace{}%
\AgdaOperator{\AgdaFunction{∈}}\AgdaSpace{}%
\AgdaFunction{cmdWrote}\AgdaSpace{}%
\AgdaSymbol{(}\AgdaFunction{save}\AgdaSpace{}%
\AgdaBound{s}\AgdaSpace{}%
\AgdaBound{x}\AgdaSpace{}%
\AgdaBound{ls}\AgdaSymbol{)}\AgdaSpace{}%
\AgdaBound{x₁}\AgdaSpace{}%
\AgdaSymbol{→}\AgdaSpace{}%
\AgdaDatatype{Hazard}\AgdaSpace{}%
\AgdaBound{s}\AgdaSpace{}%
\AgdaBound{x}\AgdaSpace{}%
\AgdaBound{b}\AgdaSpace{}%
\AgdaBound{ls}\<%
\end{code}}

\newcommand{\exhaz}{%
\begin{code}%
\>[0]\AgdaFunction{∃Hazard}\AgdaSpace{}%
\AgdaSymbol{:}\AgdaSpace{}%
\AgdaPostulate{Build}\AgdaSpace{}%
\AgdaSymbol{→}\AgdaSpace{}%
\AgdaPrimitiveType{Set}\<%
\\
\>[0]\AgdaFunction{∃Hazard}\AgdaSpace{}%
\AgdaBound{b}\AgdaSpace{}%
\AgdaSymbol{=}\AgdaSpace{}%
\AgdaFunction{∃[}\AgdaSpace{}%
\AgdaBound{sys}\AgdaSpace{}%
\AgdaFunction{]}\AgdaSymbol{(}\AgdaFunction{∃[}\AgdaSpace{}%
\AgdaBound{x}\AgdaSpace{}%
\AgdaFunction{]}\AgdaSymbol{(}\AgdaFunction{∃[}\AgdaSpace{}%
\AgdaBound{ls}\AgdaSpace{}%
\AgdaFunction{]}\AgdaSymbol{(}\AgdaDatatype{Hazard}\AgdaSpace{}%
\AgdaBound{sys}\AgdaSpace{}%
\AgdaBound{x}\AgdaSpace{}%
\AgdaBound{b}\AgdaSpace{}%
\AgdaBound{ls}\AgdaSymbol{)))}\<%
\end{code}}

\newcommand{\hazardfree}{%
\begin{code}%
\>[0]\AgdaKeyword{data}\AgdaSpace{}%
\AgdaDatatype{HazardFree}\AgdaSpace{}%
\AgdaSymbol{:}\AgdaSpace{}%
\AgdaFunction{FileSystem}\AgdaSpace{}%
\AgdaSymbol{→}\AgdaSpace{}%
\AgdaPostulate{Build}\AgdaSpace{}%
\AgdaSymbol{→}\AgdaSpace{}%
\AgdaPostulate{Build}\AgdaSpace{}%
\AgdaSymbol{→}\AgdaSpace{}%
\AgdaFunction{FileInfo}\AgdaSpace{}%
\AgdaSymbol{→}\AgdaSpace{}%
\AgdaPrimitiveType{Set}\AgdaSpace{}%
\AgdaKeyword{where}\<%
\\
\>[0][@{}l@{\AgdaIndent{0}}]%
\>[2]\AgdaInductiveConstructor{[]}\AgdaSpace{}%
\AgdaSymbol{:}\AgdaSpace{}%
\AgdaSymbol{∀}\AgdaSpace{}%
\AgdaSymbol{\{}\AgdaBound{s}\AgdaSymbol{\}}\AgdaSpace{}%
\AgdaSymbol{\{}\AgdaBound{b}\AgdaSymbol{\}}\AgdaSpace{}%
\AgdaSymbol{\{}\AgdaBound{ls}\AgdaSymbol{\}}\AgdaSpace{}%
\AgdaSymbol{→}\AgdaSpace{}%
\AgdaDatatype{HazardFree}\AgdaSpace{}%
\AgdaBound{s}\AgdaSpace{}%
\AgdaInductiveConstructor{[]}\AgdaSpace{}%
\AgdaBound{b}\AgdaSpace{}%
\AgdaBound{ls}\<%
\\
\>[2]\AgdaOperator{\AgdaInductiveConstructor{\AgdaUnderscore{}∷\AgdaUnderscore{}}}\AgdaSpace{}%
\AgdaSymbol{:}\AgdaSpace{}%
\AgdaSymbol{∀}\AgdaSpace{}%
\AgdaSymbol{\{}\AgdaBound{s}\AgdaSymbol{\}}\AgdaSpace{}%
\AgdaSymbol{\{}\AgdaBound{x}\AgdaSymbol{\}}\AgdaSpace{}%
\AgdaSymbol{\{}\AgdaBound{b₁}\AgdaSymbol{\}}\AgdaSpace{}%
\AgdaSymbol{\{}\AgdaBound{b₂}\AgdaSymbol{\}}\AgdaSpace{}%
\AgdaSymbol{\{}\AgdaBound{ls}\AgdaSymbol{\}}\AgdaSpace{}%
\AgdaSymbol{→}\AgdaSpace{}%
\AgdaOperator{\AgdaFunction{¬}}\AgdaSpace{}%
\AgdaDatatype{Hazard}\AgdaSpace{}%
\AgdaBound{s}\AgdaSpace{}%
\AgdaBound{x}\AgdaSpace{}%
\AgdaBound{b₂}\AgdaSpace{}%
\AgdaBound{ls}\AgdaSpace{}%
\AgdaSymbol{→}\AgdaSpace{}%
\AgdaDatatype{HazardFree}\AgdaSpace{}%
\AgdaSymbol{(}\AgdaPostulate{run}\AgdaSpace{}%
\AgdaBound{x}\AgdaSpace{}%
\AgdaBound{s}\AgdaSymbol{)}\AgdaSpace{}%
\AgdaBound{b₁}\AgdaSpace{}%
\AgdaBound{b₂}\AgdaSpace{}%
\AgdaSymbol{(}\AgdaFunction{save}\AgdaSpace{}%
\AgdaBound{s}\AgdaSpace{}%
\AgdaBound{x}\AgdaSpace{}%
\AgdaBound{ls}\AgdaSymbol{)}\AgdaSpace{}%
\AgdaSymbol{→}\AgdaSpace{}%
\AgdaDatatype{HazardFree}\AgdaSpace{}%
\AgdaBound{s}\AgdaSpace{}%
\AgdaSymbol{(}\AgdaBound{x}\AgdaSpace{}%
\AgdaOperator{\AgdaInductiveConstructor{∷}}\AgdaSpace{}%
\AgdaBound{b₁}\AgdaSymbol{)}\AgdaSpace{}%
\AgdaBound{b₂}\AgdaSpace{}%
\AgdaBound{ls}\<%
\end{code}}

\begin{code}[hide]%
\>[0]\<%
\\
\>[0]\AgdaFunction{intersection?2}\AgdaSpace{}%
\AgdaSymbol{:}\AgdaSpace{}%
\AgdaSymbol{(}\AgdaBound{xs}\AgdaSpace{}%
\AgdaBound{ys}\AgdaSpace{}%
\AgdaSymbol{:}\AgdaSpace{}%
\AgdaFunction{FileNames}\AgdaSymbol{)}\AgdaSpace{}%
\AgdaSymbol{→}\AgdaSpace{}%
\AgdaRecord{Dec}\AgdaSpace{}%
\AgdaSymbol{(}\AgdaFunction{∃[}\AgdaSpace{}%
\AgdaBound{v}\AgdaSpace{}%
\AgdaFunction{]}\AgdaSymbol{(}\AgdaBound{v}\AgdaSpace{}%
\AgdaOperator{\AgdaFunction{∈}}\AgdaSpace{}%
\AgdaBound{xs}\AgdaSpace{}%
\AgdaOperator{\AgdaFunction{×}}\AgdaSpace{}%
\AgdaBound{v}\AgdaSpace{}%
\AgdaOperator{\AgdaFunction{∈}}\AgdaSpace{}%
\AgdaBound{ys}\AgdaSymbol{))}\<%
\\
\>[0]\AgdaFunction{intersection?2}\AgdaSpace{}%
\AgdaInductiveConstructor{[]}\AgdaSpace{}%
\AgdaBound{ys}\AgdaSpace{}%
\AgdaSymbol{=}\AgdaSpace{}%
\AgdaInductiveConstructor{false}\AgdaSpace{}%
\AgdaOperator{\AgdaInductiveConstructor{Relation.Nullary.because}}\AgdaSpace{}%
\AgdaInductiveConstructor{Relation.Nullary.ofⁿ}\AgdaSpace{}%
\AgdaFunction{g₁}\<%
\\
\>[0][@{}l@{\AgdaIndent{0}}]%
\>[2]\AgdaKeyword{where}%
\>[1628I]\AgdaFunction{g₁}\AgdaSpace{}%
\AgdaSymbol{:}\AgdaSpace{}%
\AgdaOperator{\AgdaFunction{¬}}\AgdaSpace{}%
\AgdaSymbol{(}\AgdaFunction{∃[}\AgdaSpace{}%
\AgdaBound{v}\AgdaSpace{}%
\AgdaFunction{]}\AgdaSymbol{(}\AgdaBound{v}\AgdaSpace{}%
\AgdaOperator{\AgdaFunction{∈}}\AgdaSpace{}%
\AgdaInductiveConstructor{[]}\AgdaSpace{}%
\AgdaOperator{\AgdaFunction{×}}\AgdaSpace{}%
\AgdaBound{v}\AgdaSpace{}%
\AgdaOperator{\AgdaFunction{∈}}\AgdaSpace{}%
\AgdaBound{ys}\AgdaSymbol{))}\<%
\\
\>[.][@{}l@{}]\<[1628I]%
\>[8]\AgdaFunction{g₁}\AgdaSpace{}%
\AgdaSymbol{()}\<%
\\
\>[0]\AgdaFunction{intersection?2}\AgdaSpace{}%
\AgdaSymbol{(}\AgdaBound{x}\AgdaSpace{}%
\AgdaOperator{\AgdaInductiveConstructor{∷}}\AgdaSpace{}%
\AgdaBound{xs}\AgdaSymbol{)}\AgdaSpace{}%
\AgdaBound{ys}\AgdaSpace{}%
\AgdaKeyword{with}\AgdaSpace{}%
\AgdaBound{x}\AgdaSpace{}%
\AgdaOperator{\AgdaPostulate{∈?}}\AgdaSpace{}%
\AgdaBound{ys}\<%
\\
\>[0]\AgdaSymbol{...}\AgdaSpace{}%
\AgdaSymbol{|}\AgdaSpace{}%
\AgdaInductiveConstructor{yes}\AgdaSpace{}%
\AgdaBound{x∈ys}\AgdaSpace{}%
\AgdaSymbol{=}\AgdaSpace{}%
\AgdaInductiveConstructor{true}\AgdaSpace{}%
\AgdaOperator{\AgdaInductiveConstructor{Relation.Nullary.because}}\AgdaSpace{}%
\AgdaInductiveConstructor{Relation.Nullary.ofʸ}\AgdaSpace{}%
\AgdaFunction{g₁}\<%
\\
\>[0][@{}l@{\AgdaIndent{0}}]%
\>[2]\AgdaKeyword{where}%
\>[1657I]\AgdaFunction{g₁}\AgdaSpace{}%
\AgdaSymbol{:}\AgdaSpace{}%
\AgdaFunction{∃[}\AgdaSpace{}%
\AgdaBound{v}\AgdaSpace{}%
\AgdaFunction{]}\AgdaSymbol{(}\AgdaBound{v}\AgdaSpace{}%
\AgdaOperator{\AgdaFunction{∈}}\AgdaSpace{}%
\AgdaBound{x}\AgdaSpace{}%
\AgdaOperator{\AgdaInductiveConstructor{∷}}\AgdaSpace{}%
\AgdaBound{xs}\AgdaSpace{}%
\AgdaOperator{\AgdaFunction{×}}\AgdaSpace{}%
\AgdaBound{v}\AgdaSpace{}%
\AgdaOperator{\AgdaFunction{∈}}\AgdaSpace{}%
\AgdaBound{ys}\AgdaSymbol{)}\<%
\\
\>[.][@{}l@{}]\<[1657I]%
\>[8]\AgdaFunction{g₁}\AgdaSpace{}%
\AgdaSymbol{=}\AgdaSpace{}%
\AgdaBound{x}\AgdaSpace{}%
\AgdaOperator{\AgdaInductiveConstructor{,}}\AgdaSpace{}%
\AgdaInductiveConstructor{here}\AgdaSpace{}%
\AgdaInductiveConstructor{refl}\AgdaSpace{}%
\AgdaOperator{\AgdaInductiveConstructor{,}}\AgdaSpace{}%
\AgdaBound{x∈ys}\<%
\\
\>[0]\AgdaSymbol{...}\AgdaSpace{}%
\AgdaSymbol{|}\AgdaSpace{}%
\AgdaInductiveConstructor{no}\AgdaSpace{}%
\AgdaBound{x∉ys}\AgdaSpace{}%
\AgdaKeyword{with}\AgdaSpace{}%
\AgdaFunction{intersection?2}\AgdaSpace{}%
\AgdaBound{xs}\AgdaSpace{}%
\AgdaBound{ys}\<%
\\
\>[0]\AgdaSymbol{...}\AgdaSpace{}%
\AgdaSymbol{|}\AgdaSpace{}%
\AgdaInductiveConstructor{yes}\AgdaSpace{}%
\AgdaSymbol{(}\AgdaBound{v}\AgdaSpace{}%
\AgdaOperator{\AgdaInductiveConstructor{,}}\AgdaSpace{}%
\AgdaBound{v∈xs}\AgdaSpace{}%
\AgdaOperator{\AgdaInductiveConstructor{,}}\AgdaSpace{}%
\AgdaBound{v∈ys}\AgdaSymbol{)}\AgdaSpace{}%
\AgdaSymbol{=}\AgdaSpace{}%
\AgdaInductiveConstructor{true}\AgdaSpace{}%
\AgdaOperator{\AgdaInductiveConstructor{Relation.Nullary.because}}\AgdaSpace{}%
\AgdaInductiveConstructor{Relation.Nullary.ofʸ}\AgdaSpace{}%
\AgdaSymbol{(}\AgdaBound{v}\AgdaSpace{}%
\AgdaOperator{\AgdaInductiveConstructor{,}}\AgdaSpace{}%
\AgdaInductiveConstructor{there}\AgdaSpace{}%
\AgdaBound{v∈xs}\AgdaSpace{}%
\AgdaOperator{\AgdaInductiveConstructor{,}}\AgdaSpace{}%
\AgdaBound{v∈ys}\AgdaSymbol{)}\<%
\\
\>[0]\AgdaSymbol{...}\AgdaSpace{}%
\AgdaSymbol{|}\AgdaSpace{}%
\AgdaInductiveConstructor{no}\AgdaSpace{}%
\AgdaBound{¬p}\AgdaSpace{}%
\AgdaSymbol{=}\AgdaSpace{}%
\AgdaInductiveConstructor{false}\AgdaSpace{}%
\AgdaOperator{\AgdaInductiveConstructor{Relation.Nullary.because}}\AgdaSpace{}%
\AgdaInductiveConstructor{Relation.Nullary.ofⁿ}\AgdaSpace{}%
\AgdaFunction{g₁}\<%
\\
\>[0][@{}l@{\AgdaIndent{0}}]%
\>[2]\AgdaKeyword{where}%
\>[1709I]\AgdaFunction{g₁}\AgdaSpace{}%
\AgdaSymbol{:}\AgdaSpace{}%
\AgdaOperator{\AgdaFunction{¬}}\AgdaSpace{}%
\AgdaSymbol{(}\AgdaFunction{∃[}\AgdaSpace{}%
\AgdaBound{v}\AgdaSpace{}%
\AgdaFunction{]}\AgdaSymbol{(}\AgdaBound{v}\AgdaSpace{}%
\AgdaOperator{\AgdaFunction{∈}}\AgdaSpace{}%
\AgdaBound{x}\AgdaSpace{}%
\AgdaOperator{\AgdaInductiveConstructor{∷}}\AgdaSpace{}%
\AgdaBound{xs}\AgdaSpace{}%
\AgdaOperator{\AgdaFunction{×}}\AgdaSpace{}%
\AgdaBound{v}\AgdaSpace{}%
\AgdaOperator{\AgdaFunction{∈}}\AgdaSpace{}%
\AgdaBound{ys}\AgdaSymbol{))}\<%
\\
\>[.][@{}l@{}]\<[1709I]%
\>[8]\AgdaFunction{g₁}\AgdaSpace{}%
\AgdaSymbol{(}\AgdaBound{v}\AgdaSpace{}%
\AgdaOperator{\AgdaInductiveConstructor{,}}\AgdaSpace{}%
\AgdaBound{v∈x∷xs}\AgdaSpace{}%
\AgdaOperator{\AgdaInductiveConstructor{,}}\AgdaSpace{}%
\AgdaBound{v∈ys}\AgdaSymbol{)}\AgdaSpace{}%
\AgdaKeyword{with}\AgdaSpace{}%
\AgdaBound{v}\AgdaSpace{}%
\AgdaOperator{\AgdaFunction{≟}}\AgdaSpace{}%
\AgdaBound{x}\<%
\\
\>[8]\AgdaSymbol{...}\AgdaSpace{}%
\AgdaSymbol{|}\AgdaSpace{}%
\AgdaInductiveConstructor{yes}\AgdaSpace{}%
\AgdaBound{v≡x}\AgdaSpace{}%
\AgdaSymbol{=}\AgdaSpace{}%
\AgdaFunction{contradiction}\AgdaSpace{}%
\AgdaSymbol{(}\AgdaFunction{subst}\AgdaSpace{}%
\AgdaSymbol{(λ}\AgdaSpace{}%
\AgdaBound{x₁}\AgdaSpace{}%
\AgdaSymbol{→}\AgdaSpace{}%
\AgdaBound{x₁}\AgdaSpace{}%
\AgdaOperator{\AgdaFunction{∈}}\AgdaSpace{}%
\AgdaBound{ys}\AgdaSymbol{)}\AgdaSpace{}%
\AgdaBound{v≡x}\AgdaSpace{}%
\AgdaBound{v∈ys}\AgdaSymbol{)}\AgdaSpace{}%
\AgdaBound{x∉ys}\<%
\\
\>[8]\AgdaSymbol{...}\AgdaSpace{}%
\AgdaSymbol{|}\AgdaSpace{}%
\AgdaInductiveConstructor{no}\AgdaSpace{}%
\AgdaBound{¬v≡x}\AgdaSpace{}%
\AgdaSymbol{=}\AgdaSpace{}%
\AgdaBound{¬p}\AgdaSpace{}%
\AgdaSymbol{(}\AgdaBound{v}\AgdaSpace{}%
\AgdaOperator{\AgdaInductiveConstructor{,}}\AgdaSpace{}%
\AgdaSymbol{(}\AgdaFunction{tail}\AgdaSpace{}%
\AgdaBound{¬v≡x}\AgdaSpace{}%
\AgdaBound{v∈x∷xs}\AgdaSymbol{)}\AgdaSpace{}%
\AgdaOperator{\AgdaInductiveConstructor{,}}\AgdaSpace{}%
\AgdaBound{v∈ys}\AgdaSymbol{)}\<%
\\
\\[\AgdaEmptyExtraSkip]%
\>[0]\AgdaFunction{intersection?}\AgdaSpace{}%
\AgdaSymbol{:}\AgdaSpace{}%
\AgdaSymbol{(}\AgdaBound{xs}\AgdaSpace{}%
\AgdaBound{ys}\AgdaSpace{}%
\AgdaSymbol{:}\AgdaSpace{}%
\AgdaFunction{FileNames}\AgdaSymbol{)}\AgdaSpace{}%
\AgdaSymbol{→}\AgdaSpace{}%
\AgdaRecord{Dec}\AgdaSpace{}%
\AgdaSymbol{(}\AgdaFunction{Disjoint}\AgdaSpace{}%
\AgdaBound{xs}\AgdaSpace{}%
\AgdaBound{ys}\AgdaSymbol{)}\<%
\\
\>[0]\AgdaFunction{intersection?}\AgdaSpace{}%
\AgdaInductiveConstructor{[]}\AgdaSpace{}%
\AgdaBound{ys}\AgdaSpace{}%
\AgdaSymbol{=}\AgdaSpace{}%
\AgdaInductiveConstructor{true}\AgdaSpace{}%
\AgdaOperator{\AgdaInductiveConstructor{Relation.Nullary.because}}\AgdaSpace{}%
\AgdaInductiveConstructor{Relation.Nullary.ofʸ}\AgdaSpace{}%
\AgdaSymbol{(λ}\AgdaSpace{}%
\AgdaSymbol{())}\<%
\\
\>[0]\AgdaFunction{intersection?}\AgdaSpace{}%
\AgdaSymbol{(}\AgdaBound{x}\AgdaSpace{}%
\AgdaOperator{\AgdaInductiveConstructor{∷}}\AgdaSpace{}%
\AgdaBound{xs}\AgdaSymbol{)}\AgdaSpace{}%
\AgdaBound{ys}\AgdaSpace{}%
\AgdaKeyword{with}\AgdaSpace{}%
\AgdaBound{x}\AgdaSpace{}%
\AgdaOperator{\AgdaPostulate{∈?}}\AgdaSpace{}%
\AgdaBound{ys}\<%
\\
\>[0]\AgdaSymbol{...}\AgdaSpace{}%
\AgdaSymbol{|}\AgdaSpace{}%
\AgdaInductiveConstructor{yes}\AgdaSpace{}%
\AgdaBound{x∈ys}\AgdaSpace{}%
\AgdaSymbol{=}\AgdaSpace{}%
\AgdaInductiveConstructor{false}\AgdaSpace{}%
\AgdaOperator{\AgdaInductiveConstructor{Relation.Nullary.because}}\AgdaSpace{}%
\AgdaInductiveConstructor{Relation.Nullary.ofⁿ}\AgdaSpace{}%
\AgdaSymbol{λ}\AgdaSpace{}%
\AgdaBound{x₁}\AgdaSpace{}%
\AgdaSymbol{→}\AgdaSpace{}%
\AgdaBound{x₁}\AgdaSpace{}%
\AgdaSymbol{((}\AgdaInductiveConstructor{here}\AgdaSpace{}%
\AgdaInductiveConstructor{refl}\AgdaSymbol{)}\AgdaSpace{}%
\AgdaOperator{\AgdaInductiveConstructor{,}}\AgdaSpace{}%
\AgdaBound{x∈ys}\AgdaSymbol{)}\<%
\\
\>[0]\AgdaSymbol{...}\AgdaSpace{}%
\AgdaSymbol{|}\AgdaSpace{}%
\AgdaInductiveConstructor{no}\AgdaSpace{}%
\AgdaBound{x∉ys}\AgdaSpace{}%
\AgdaKeyword{with}\AgdaSpace{}%
\AgdaFunction{intersection?}\AgdaSpace{}%
\AgdaBound{xs}\AgdaSpace{}%
\AgdaBound{ys}\<%
\\
\>[0]\AgdaSymbol{...}\AgdaSpace{}%
\AgdaSymbol{|}\AgdaSpace{}%
\AgdaInductiveConstructor{yes}\AgdaSpace{}%
\AgdaBound{dsj}\AgdaSpace{}%
\AgdaSymbol{=}\AgdaSpace{}%
\AgdaInductiveConstructor{true}\AgdaSpace{}%
\AgdaOperator{\AgdaInductiveConstructor{Relation.Nullary.because}}\AgdaSpace{}%
\AgdaInductiveConstructor{Relation.Nullary.ofʸ}\AgdaSpace{}%
\AgdaSymbol{λ}\AgdaSpace{}%
\AgdaBound{x₁}\AgdaSpace{}%
\AgdaSymbol{→}\AgdaSpace{}%
\AgdaFunction{g₁}\AgdaSpace{}%
\AgdaBound{x₁}\<%
\\
\>[0][@{}l@{\AgdaIndent{0}}]%
\>[2]\AgdaKeyword{where}%
\>[1819I]\AgdaFunction{g₁}\AgdaSpace{}%
\AgdaSymbol{:}\AgdaSpace{}%
\AgdaSymbol{∀}\AgdaSpace{}%
\AgdaSymbol{\{}\AgdaBound{v}\AgdaSymbol{\}}\AgdaSpace{}%
\AgdaSymbol{→}\AgdaSpace{}%
\AgdaSymbol{(}\AgdaBound{v}\AgdaSpace{}%
\AgdaOperator{\AgdaFunction{∈}}\AgdaSpace{}%
\AgdaBound{x}\AgdaSpace{}%
\AgdaOperator{\AgdaInductiveConstructor{∷}}\AgdaSpace{}%
\AgdaBound{xs}\AgdaSpace{}%
\AgdaOperator{\AgdaFunction{×}}\AgdaSpace{}%
\AgdaBound{v}\AgdaSpace{}%
\AgdaOperator{\AgdaFunction{∈}}\AgdaSpace{}%
\AgdaBound{ys}\AgdaSymbol{)}\AgdaSpace{}%
\AgdaSymbol{→}\AgdaSpace{}%
\AgdaDatatype{⊥}\<%
\\
\>[.][@{}l@{}]\<[1819I]%
\>[8]\AgdaFunction{g₁}\AgdaSpace{}%
\AgdaSymbol{\{}\AgdaBound{v}\AgdaSymbol{\}}\AgdaSpace{}%
\AgdaSymbol{(}\AgdaBound{v∈x∷xs}\AgdaSpace{}%
\AgdaOperator{\AgdaInductiveConstructor{,}}\AgdaSpace{}%
\AgdaBound{v∈ys}\AgdaSymbol{)}\AgdaSpace{}%
\AgdaKeyword{with}\AgdaSpace{}%
\AgdaBound{v}\AgdaSpace{}%
\AgdaOperator{\AgdaFunction{≟}}\AgdaSpace{}%
\AgdaBound{x}\<%
\\
\>[8]\AgdaSymbol{...}\AgdaSpace{}%
\AgdaSymbol{|}\AgdaSpace{}%
\AgdaInductiveConstructor{yes}\AgdaSpace{}%
\AgdaBound{v≡x}\AgdaSpace{}%
\AgdaSymbol{=}\AgdaSpace{}%
\AgdaBound{x∉ys}\AgdaSpace{}%
\AgdaSymbol{(}\AgdaFunction{subst}\AgdaSpace{}%
\AgdaSymbol{(λ}\AgdaSpace{}%
\AgdaBound{x₁}\AgdaSpace{}%
\AgdaSymbol{→}\AgdaSpace{}%
\AgdaBound{x₁}\AgdaSpace{}%
\AgdaOperator{\AgdaFunction{∈}}\AgdaSpace{}%
\AgdaBound{ys}\AgdaSymbol{)}\AgdaSpace{}%
\AgdaBound{v≡x}\AgdaSpace{}%
\AgdaBound{v∈ys}\AgdaSymbol{)}\<%
\\
\>[8]\AgdaSymbol{...}\AgdaSpace{}%
\AgdaSymbol{|}\AgdaSpace{}%
\AgdaInductiveConstructor{no}\AgdaSpace{}%
\AgdaBound{¬v≡x}\AgdaSpace{}%
\AgdaSymbol{=}\AgdaSpace{}%
\AgdaBound{dsj}\AgdaSpace{}%
\AgdaSymbol{(}\AgdaFunction{tail}\AgdaSpace{}%
\AgdaBound{¬v≡x}\AgdaSpace{}%
\AgdaBound{v∈x∷xs}\AgdaSpace{}%
\AgdaOperator{\AgdaInductiveConstructor{,}}\AgdaSpace{}%
\AgdaBound{v∈ys}\AgdaSymbol{)}\<%
\\
\>[0]\AgdaSymbol{...}\AgdaSpace{}%
\AgdaSymbol{|}\AgdaSpace{}%
\AgdaInductiveConstructor{no}\AgdaSpace{}%
\AgdaBound{¬dsj}\AgdaSpace{}%
\AgdaSymbol{=}\AgdaSpace{}%
\AgdaInductiveConstructor{false}\AgdaSpace{}%
\AgdaOperator{\AgdaInductiveConstructor{Relation.Nullary.because}}\AgdaSpace{}%
\AgdaInductiveConstructor{Relation.Nullary.ofⁿ}\AgdaSpace{}%
\AgdaSymbol{λ}\AgdaSpace{}%
\AgdaBound{x₁}\AgdaSpace{}%
\AgdaSymbol{→}\AgdaSpace{}%
\AgdaBound{¬dsj}\AgdaSpace{}%
\AgdaSymbol{λ}\AgdaSpace{}%
\AgdaBound{x₂}\AgdaSpace{}%
\AgdaSymbol{→}\AgdaSpace{}%
\AgdaBound{x₁}\AgdaSpace{}%
\AgdaSymbol{(}\AgdaInductiveConstructor{there}\AgdaSpace{}%
\AgdaSymbol{(}\AgdaField{proj₁}\AgdaSpace{}%
\AgdaBound{x₂}\AgdaSymbol{)}\AgdaSpace{}%
\AgdaOperator{\AgdaInductiveConstructor{,}}\AgdaSpace{}%
\AgdaField{proj₂}\AgdaSpace{}%
\AgdaBound{x₂}\AgdaSymbol{)}\<%
\\
\\[\AgdaEmptyExtraSkip]%
\>[0]\AgdaFunction{before?}\AgdaSpace{}%
\AgdaSymbol{:}\AgdaSpace{}%
\AgdaSymbol{∀}\AgdaSpace{}%
\AgdaSymbol{(}\AgdaBound{x₁}\AgdaSpace{}%
\AgdaSymbol{:}\AgdaSpace{}%
\AgdaFunction{Cmd}\AgdaSymbol{)}\AgdaSpace{}%
\AgdaBound{x}\AgdaSpace{}%
\AgdaBound{b}\AgdaSpace{}%
\AgdaSymbol{→}\AgdaSpace{}%
\AgdaRecord{Dec}\AgdaSpace{}%
\AgdaSymbol{(}\AgdaBound{x₁}\AgdaSpace{}%
\AgdaOperator{\AgdaFunction{before}}\AgdaSpace{}%
\AgdaBound{x}\AgdaSpace{}%
\AgdaOperator{\AgdaFunction{∈}}\AgdaSpace{}%
\AgdaBound{b}\AgdaSymbol{)}\<%
\\
\>[0]\AgdaFunction{before?}\AgdaSpace{}%
\AgdaBound{x₁}\AgdaSpace{}%
\AgdaBound{x}\AgdaSpace{}%
\AgdaInductiveConstructor{[]}\AgdaSpace{}%
\AgdaSymbol{=}\AgdaSpace{}%
\AgdaInductiveConstructor{false}\AgdaSpace{}%
\AgdaOperator{\AgdaInductiveConstructor{Relation.Nullary.because}}\AgdaSpace{}%
\AgdaInductiveConstructor{Relation.Nullary.ofⁿ}\AgdaSpace{}%
\AgdaFunction{g₁}\<%
\\
\>[0][@{}l@{\AgdaIndent{0}}]%
\>[2]\AgdaKeyword{where}%
\>[1910I]\AgdaFunction{g₁}\AgdaSpace{}%
\AgdaSymbol{:}\AgdaSpace{}%
\AgdaOperator{\AgdaFunction{¬}}\AgdaSpace{}%
\AgdaSymbol{(}\AgdaBound{x₁}\AgdaSpace{}%
\AgdaOperator{\AgdaFunction{before}}\AgdaSpace{}%
\AgdaBound{x}\AgdaSpace{}%
\AgdaOperator{\AgdaFunction{∈}}\AgdaSpace{}%
\AgdaInductiveConstructor{[]}\AgdaSymbol{)}\<%
\\
\>[.][@{}l@{}]\<[1910I]%
\>[8]\AgdaFunction{g₁}\AgdaSpace{}%
\AgdaSymbol{(}\AgdaBound{xs}\AgdaSpace{}%
\AgdaOperator{\AgdaInductiveConstructor{,}}\AgdaSpace{}%
\AgdaBound{ys}\AgdaSpace{}%
\AgdaOperator{\AgdaInductiveConstructor{,}}\AgdaSpace{}%
\AgdaBound{≡₁}\AgdaSpace{}%
\AgdaOperator{\AgdaInductiveConstructor{,}}\AgdaSpace{}%
\AgdaBound{x∈ys}\AgdaSymbol{)}\AgdaSpace{}%
\AgdaSymbol{=}\AgdaSpace{}%
\AgdaFunction{contradiction}\AgdaSpace{}%
\AgdaSymbol{(}\AgdaFunction{subst}\AgdaSpace{}%
\AgdaSymbol{(λ}\AgdaSpace{}%
\AgdaBound{x₂}\AgdaSpace{}%
\AgdaSymbol{→}\AgdaSpace{}%
\AgdaBound{x}\AgdaSpace{}%
\AgdaOperator{\AgdaFunction{∈}}\AgdaSpace{}%
\AgdaBound{x₂}\AgdaSymbol{)}\AgdaSpace{}%
\AgdaSymbol{(}\AgdaFunction{sym}\AgdaSpace{}%
\AgdaBound{≡₁}\AgdaSymbol{)}\AgdaSpace{}%
\AgdaSymbol{(}\AgdaFunction{∈-++⁺ʳ}\AgdaSpace{}%
\AgdaBound{xs}\AgdaSpace{}%
\AgdaSymbol{(}\AgdaInductiveConstructor{there}\AgdaSpace{}%
\AgdaBound{x∈ys}\AgdaSymbol{)))}\AgdaSpace{}%
\AgdaSymbol{λ}\AgdaSpace{}%
\AgdaSymbol{()}\<%
\\
\>[0]\AgdaFunction{before?}\AgdaSpace{}%
\AgdaBound{x₁}\AgdaSpace{}%
\AgdaBound{x}\AgdaSpace{}%
\AgdaSymbol{(}\AgdaBound{x₂}\AgdaSpace{}%
\AgdaOperator{\AgdaInductiveConstructor{∷}}\AgdaSpace{}%
\AgdaBound{b}\AgdaSymbol{)}\AgdaSpace{}%
\AgdaKeyword{with}\AgdaSpace{}%
\AgdaBound{x₁}\AgdaSpace{}%
\AgdaOperator{\AgdaFunction{≟}}\AgdaSpace{}%
\AgdaBound{x₂}\<%
\\
\>[0]\AgdaSymbol{...}\AgdaSpace{}%
\AgdaSymbol{|}\AgdaSpace{}%
\AgdaInductiveConstructor{yes}\AgdaSpace{}%
\AgdaBound{x₁≡x₂}\AgdaSpace{}%
\AgdaKeyword{with}\AgdaSpace{}%
\AgdaBound{x}\AgdaSpace{}%
\AgdaOperator{\AgdaPostulate{∈?}}\AgdaSpace{}%
\AgdaBound{b}\<%
\\
\>[0]\AgdaSymbol{...}\AgdaSpace{}%
\AgdaSymbol{|}\AgdaSpace{}%
\AgdaInductiveConstructor{yes}\AgdaSpace{}%
\AgdaBound{x∈b}\AgdaSpace{}%
\AgdaSymbol{=}\AgdaSpace{}%
\AgdaInductiveConstructor{true}\AgdaSpace{}%
\AgdaOperator{\AgdaInductiveConstructor{Relation.Nullary.because}}\AgdaSpace{}%
\AgdaInductiveConstructor{Relation.Nullary.ofʸ}\AgdaSpace{}%
\AgdaFunction{g₁}\<%
\\
\>[0][@{}l@{\AgdaIndent{0}}]%
\>[2]\AgdaKeyword{where}%
\>[1966I]\AgdaFunction{g₁}\AgdaSpace{}%
\AgdaSymbol{:}\AgdaSpace{}%
\AgdaBound{x₁}\AgdaSpace{}%
\AgdaOperator{\AgdaFunction{before}}\AgdaSpace{}%
\AgdaBound{x}\AgdaSpace{}%
\AgdaOperator{\AgdaFunction{∈}}\AgdaSpace{}%
\AgdaSymbol{(}\AgdaBound{x₂}\AgdaSpace{}%
\AgdaOperator{\AgdaInductiveConstructor{∷}}\AgdaSpace{}%
\AgdaBound{b}\AgdaSymbol{)}\<%
\\
\>[.][@{}l@{}]\<[1966I]%
\>[8]\AgdaFunction{g₁}\AgdaSpace{}%
\AgdaSymbol{=}\AgdaSpace{}%
\AgdaInductiveConstructor{[]}\AgdaSpace{}%
\AgdaOperator{\AgdaInductiveConstructor{,}}\AgdaSpace{}%
\AgdaBound{b}\AgdaSpace{}%
\AgdaOperator{\AgdaInductiveConstructor{,}}\AgdaSpace{}%
\AgdaFunction{cong}\AgdaSpace{}%
\AgdaSymbol{(}\AgdaOperator{\AgdaInductiveConstructor{\AgdaUnderscore{}∷}}\AgdaSpace{}%
\AgdaBound{b}\AgdaSymbol{)}\AgdaSpace{}%
\AgdaSymbol{(}\AgdaFunction{sym}\AgdaSpace{}%
\AgdaBound{x₁≡x₂}\AgdaSymbol{)}\AgdaSpace{}%
\AgdaOperator{\AgdaInductiveConstructor{,}}\AgdaSpace{}%
\AgdaBound{x∈b}\<%
\\
\>[0]\AgdaSymbol{...}\AgdaSpace{}%
\AgdaSymbol{|}\AgdaSpace{}%
\AgdaInductiveConstructor{no}\AgdaSpace{}%
\AgdaBound{x∉b}\AgdaSpace{}%
\AgdaSymbol{=}\AgdaSpace{}%
\AgdaInductiveConstructor{false}\AgdaSpace{}%
\AgdaOperator{\AgdaInductiveConstructor{Relation.Nullary.because}}\AgdaSpace{}%
\AgdaInductiveConstructor{Relation.Nullary.ofⁿ}\AgdaSpace{}%
\AgdaFunction{g₁}\<%
\\
\>[0][@{}l@{\AgdaIndent{0}}]%
\>[2]\AgdaKeyword{where}%
\>[1995I]\AgdaFunction{g₁}\AgdaSpace{}%
\AgdaSymbol{:}\AgdaSpace{}%
\AgdaOperator{\AgdaFunction{¬}}\AgdaSpace{}%
\AgdaSymbol{(}\AgdaBound{x₁}\AgdaSpace{}%
\AgdaOperator{\AgdaFunction{before}}\AgdaSpace{}%
\AgdaBound{x}\AgdaSpace{}%
\AgdaOperator{\AgdaFunction{∈}}\AgdaSpace{}%
\AgdaSymbol{(}\AgdaBound{x₂}\AgdaSpace{}%
\AgdaOperator{\AgdaInductiveConstructor{∷}}\AgdaSpace{}%
\AgdaBound{b}\AgdaSymbol{))}\<%
\\
\>[.][@{}l@{}]\<[1995I]%
\>[8]\AgdaFunction{g₁}\AgdaSpace{}%
\AgdaSymbol{(}\AgdaInductiveConstructor{[]}\AgdaSpace{}%
\AgdaOperator{\AgdaInductiveConstructor{,}}\AgdaSpace{}%
\AgdaBound{ys}\AgdaSpace{}%
\AgdaOperator{\AgdaInductiveConstructor{,}}\AgdaSpace{}%
\AgdaBound{x₂∷b≡x₁∷ys}\AgdaSpace{}%
\AgdaOperator{\AgdaInductiveConstructor{,}}\AgdaSpace{}%
\AgdaBound{x∈ys}\AgdaSymbol{)}\AgdaSpace{}%
\AgdaSymbol{=}\AgdaSpace{}%
\AgdaFunction{contradiction}\AgdaSpace{}%
\AgdaSymbol{(}\AgdaFunction{subst}\AgdaSpace{}%
\AgdaSymbol{(λ}\AgdaSpace{}%
\AgdaBound{x₃}\AgdaSpace{}%
\AgdaSymbol{→}\AgdaSpace{}%
\AgdaBound{x}\AgdaSpace{}%
\AgdaOperator{\AgdaFunction{∈}}\AgdaSpace{}%
\AgdaBound{x₃}\AgdaSymbol{)}\AgdaSpace{}%
\AgdaSymbol{(}\AgdaFunction{sym}\AgdaSpace{}%
\AgdaSymbol{(}\AgdaFunction{∷-injectiveʳ}\AgdaSpace{}%
\AgdaBound{x₂∷b≡x₁∷ys}\AgdaSymbol{))}\AgdaSpace{}%
\AgdaBound{x∈ys}\AgdaSymbol{)}\AgdaSpace{}%
\AgdaBound{x∉b}\<%
\\
\>[8]\AgdaFunction{g₁}\AgdaSpace{}%
\AgdaSymbol{(}\AgdaBound{x₃}\AgdaSpace{}%
\AgdaOperator{\AgdaInductiveConstructor{∷}}\AgdaSpace{}%
\AgdaBound{xs}\AgdaSpace{}%
\AgdaOperator{\AgdaInductiveConstructor{,}}\AgdaSpace{}%
\AgdaBound{ys}\AgdaSpace{}%
\AgdaOperator{\AgdaInductiveConstructor{,}}\AgdaSpace{}%
\AgdaBound{x₂∷b≡xs++x₁∷ys}\AgdaSpace{}%
\AgdaOperator{\AgdaInductiveConstructor{,}}\AgdaSpace{}%
\AgdaBound{x∈ys}\AgdaSymbol{)}\AgdaSpace{}%
\AgdaSymbol{=}\AgdaSpace{}%
\AgdaFunction{contradiction}\AgdaSpace{}%
\AgdaSymbol{(}\AgdaFunction{subst}\AgdaSpace{}%
\AgdaSymbol{(λ}\AgdaSpace{}%
\AgdaBound{x₄}\AgdaSpace{}%
\AgdaSymbol{→}\AgdaSpace{}%
\AgdaBound{x}\AgdaSpace{}%
\AgdaOperator{\AgdaFunction{∈}}\AgdaSpace{}%
\AgdaBound{x₄}\AgdaSymbol{)}\AgdaSpace{}%
\AgdaSymbol{(}\AgdaFunction{sym}\AgdaSpace{}%
\AgdaSymbol{(}\AgdaFunction{∷-injectiveʳ}\AgdaSpace{}%
\AgdaBound{x₂∷b≡xs++x₁∷ys}\AgdaSymbol{))}\AgdaSpace{}%
\AgdaSymbol{(}\AgdaFunction{∈-++⁺ʳ}\AgdaSpace{}%
\AgdaBound{xs}\AgdaSpace{}%
\AgdaSymbol{(}\AgdaInductiveConstructor{there}\AgdaSpace{}%
\AgdaBound{x∈ys}\AgdaSymbol{)))}\AgdaSpace{}%
\AgdaBound{x∉b}\<%
\\
\>[0]\AgdaFunction{before?}\AgdaSpace{}%
\AgdaBound{x₁}\AgdaSpace{}%
\AgdaBound{x}\AgdaSpace{}%
\AgdaSymbol{(}\AgdaBound{x₂}\AgdaSpace{}%
\AgdaOperator{\AgdaInductiveConstructor{∷}}\AgdaSpace{}%
\AgdaBound{b}\AgdaSymbol{)}\AgdaSpace{}%
\AgdaSymbol{|}\AgdaSpace{}%
\AgdaInductiveConstructor{no}\AgdaSpace{}%
\AgdaBound{¬x₁≡x₂}\AgdaSpace{}%
\AgdaKeyword{with}\AgdaSpace{}%
\AgdaFunction{before?}\AgdaSpace{}%
\AgdaBound{x₁}\AgdaSpace{}%
\AgdaBound{x}\AgdaSpace{}%
\AgdaBound{b}\<%
\\
\>[0]\AgdaSymbol{...}\AgdaSpace{}%
\AgdaSymbol{|}\AgdaSpace{}%
\AgdaInductiveConstructor{yes}\AgdaSpace{}%
\AgdaSymbol{(}\AgdaBound{xs}\AgdaSpace{}%
\AgdaOperator{\AgdaInductiveConstructor{,}}\AgdaSpace{}%
\AgdaBound{ys}\AgdaSpace{}%
\AgdaOperator{\AgdaInductiveConstructor{,}}\AgdaSpace{}%
\AgdaBound{≡₁}\AgdaSpace{}%
\AgdaOperator{\AgdaInductiveConstructor{,}}\AgdaSpace{}%
\AgdaBound{x∈ys}\AgdaSymbol{)}\<%
\\
\>[0][@{}l@{\AgdaIndent{0}}]%
\>[2]\AgdaSymbol{=}\AgdaSpace{}%
\AgdaInductiveConstructor{true}\AgdaSpace{}%
\AgdaOperator{\AgdaInductiveConstructor{Relation.Nullary.because}}\AgdaSpace{}%
\AgdaInductiveConstructor{Relation.Nullary.ofʸ}\AgdaSpace{}%
\AgdaSymbol{(}\AgdaBound{x₂}\AgdaSpace{}%
\AgdaOperator{\AgdaInductiveConstructor{∷}}\AgdaSpace{}%
\AgdaBound{xs}\AgdaSpace{}%
\AgdaOperator{\AgdaInductiveConstructor{,}}\AgdaSpace{}%
\AgdaBound{ys}\AgdaSpace{}%
\AgdaOperator{\AgdaInductiveConstructor{,}}\AgdaSpace{}%
\AgdaFunction{cong}\AgdaSpace{}%
\AgdaSymbol{(}\AgdaBound{x₂}\AgdaSpace{}%
\AgdaOperator{\AgdaInductiveConstructor{∷\AgdaUnderscore{}}}\AgdaSymbol{)}\AgdaSpace{}%
\AgdaBound{≡₁}\AgdaSpace{}%
\AgdaOperator{\AgdaInductiveConstructor{,}}\AgdaSpace{}%
\AgdaBound{x∈ys}\AgdaSymbol{)}\<%
\\
\>[0]\AgdaSymbol{...}\AgdaSpace{}%
\AgdaSymbol{|}\AgdaSpace{}%
\AgdaInductiveConstructor{no}\AgdaSpace{}%
\AgdaBound{¬bf}\AgdaSpace{}%
\AgdaSymbol{=}\AgdaSpace{}%
\AgdaInductiveConstructor{false}\AgdaSpace{}%
\AgdaOperator{\AgdaInductiveConstructor{Relation.Nullary.because}}\AgdaSpace{}%
\AgdaInductiveConstructor{Relation.Nullary.ofⁿ}\AgdaSpace{}%
\AgdaFunction{g₁}\<%
\\
\>[0][@{}l@{\AgdaIndent{0}}]%
\>[2]\AgdaKeyword{where}%
\>[2097I]\AgdaFunction{g₁}\AgdaSpace{}%
\AgdaSymbol{:}\AgdaSpace{}%
\AgdaOperator{\AgdaFunction{¬}}\AgdaSpace{}%
\AgdaSymbol{(}\AgdaBound{x₁}\AgdaSpace{}%
\AgdaOperator{\AgdaFunction{before}}\AgdaSpace{}%
\AgdaBound{x}\AgdaSpace{}%
\AgdaOperator{\AgdaFunction{∈}}\AgdaSpace{}%
\AgdaSymbol{(}\AgdaBound{x₂}\AgdaSpace{}%
\AgdaOperator{\AgdaInductiveConstructor{∷}}\AgdaSpace{}%
\AgdaBound{b}\AgdaSymbol{))}\<%
\\
\>[.][@{}l@{}]\<[2097I]%
\>[8]\AgdaFunction{g₁}\AgdaSpace{}%
\AgdaSymbol{(}\AgdaInductiveConstructor{[]}\AgdaSpace{}%
\AgdaOperator{\AgdaInductiveConstructor{,}}\AgdaSpace{}%
\AgdaBound{ys}\AgdaSpace{}%
\AgdaOperator{\AgdaInductiveConstructor{,}}\AgdaSpace{}%
\AgdaBound{≡₁}\AgdaSpace{}%
\AgdaOperator{\AgdaInductiveConstructor{,}}\AgdaSpace{}%
\AgdaBound{x∈ys}\AgdaSymbol{)}\AgdaSpace{}%
\AgdaSymbol{=}\AgdaSpace{}%
\AgdaFunction{contradiction}\AgdaSpace{}%
\AgdaSymbol{(}\AgdaFunction{sym}\AgdaSpace{}%
\AgdaSymbol{(}\AgdaFunction{∷-injectiveˡ}\AgdaSpace{}%
\AgdaBound{≡₁}\AgdaSymbol{))}\AgdaSpace{}%
\AgdaBound{¬x₁≡x₂}\<%
\\
\>[8]\AgdaFunction{g₁}\AgdaSpace{}%
\AgdaSymbol{(}\AgdaBound{x₃}\AgdaSpace{}%
\AgdaOperator{\AgdaInductiveConstructor{∷}}\AgdaSpace{}%
\AgdaBound{xs}\AgdaSpace{}%
\AgdaOperator{\AgdaInductiveConstructor{,}}\AgdaSpace{}%
\AgdaBound{ys}\AgdaSpace{}%
\AgdaOperator{\AgdaInductiveConstructor{,}}\AgdaSpace{}%
\AgdaBound{≡₁}\AgdaSpace{}%
\AgdaOperator{\AgdaInductiveConstructor{,}}\AgdaSpace{}%
\AgdaBound{x∈ys}\AgdaSymbol{)}\AgdaSpace{}%
\AgdaSymbol{=}\AgdaSpace{}%
\AgdaBound{¬bf}\AgdaSpace{}%
\AgdaSymbol{(}\AgdaBound{xs}\AgdaSpace{}%
\AgdaOperator{\AgdaInductiveConstructor{,}}\AgdaSpace{}%
\AgdaBound{ys}\AgdaSpace{}%
\AgdaOperator{\AgdaInductiveConstructor{,}}\AgdaSpace{}%
\AgdaFunction{∷-injectiveʳ}\AgdaSpace{}%
\AgdaBound{≡₁}\AgdaSpace{}%
\AgdaOperator{\AgdaInductiveConstructor{,}}\AgdaSpace{}%
\AgdaBound{x∈ys}\AgdaSymbol{)}\<%
\\
\\[\AgdaEmptyExtraSkip]%
\>[0]\AgdaComment{\{- does there exist a command in ls that writes to these files and is not before x in b? -\}}\<%
\\
\\[\AgdaEmptyExtraSkip]%
\>[0]\AgdaFunction{speculativeHazard-x?}\AgdaSpace{}%
\AgdaSymbol{:}\AgdaSpace{}%
\AgdaSymbol{∀}\AgdaSpace{}%
\AgdaBound{x}\AgdaSpace{}%
\AgdaBound{b₂}\AgdaSpace{}%
\AgdaBound{ls}\AgdaSpace{}%
\AgdaBound{ls₁}\AgdaSpace{}%
\AgdaBound{rs}\AgdaSpace{}%
\AgdaSymbol{→}\AgdaSpace{}%
\AgdaRecord{Dec}\AgdaSpace{}%
\AgdaSymbol{(}\AgdaFunction{∃[}\AgdaSpace{}%
\AgdaBound{x₁}\AgdaSpace{}%
\AgdaFunction{]}\AgdaSymbol{(}\AgdaFunction{∃[}\AgdaSpace{}%
\AgdaBound{v}\AgdaSpace{}%
\AgdaFunction{]}\AgdaSymbol{(}\AgdaBound{x₁}\AgdaSpace{}%
\AgdaOperator{\AgdaFunction{∈}}\AgdaSpace{}%
\AgdaBound{ls}\AgdaSpace{}%
\AgdaOperator{\AgdaFunction{×}}\AgdaSpace{}%
\AgdaOperator{\AgdaFunction{¬}}\AgdaSpace{}%
\AgdaSymbol{(}\AgdaBound{x₁}\AgdaSpace{}%
\AgdaOperator{\AgdaFunction{before}}\AgdaSpace{}%
\AgdaBound{x}\AgdaSpace{}%
\AgdaOperator{\AgdaFunction{∈}}\AgdaSpace{}%
\AgdaBound{b₂}\AgdaSymbol{)}\AgdaSpace{}%
\AgdaOperator{\AgdaFunction{×}}\AgdaSpace{}%
\AgdaBound{v}\AgdaSpace{}%
\AgdaOperator{\AgdaFunction{∈}}\AgdaSpace{}%
\AgdaBound{rs}\AgdaSpace{}%
\AgdaOperator{\AgdaFunction{×}}\AgdaSpace{}%
\AgdaBound{v}\AgdaSpace{}%
\AgdaOperator{\AgdaFunction{∈}}\AgdaSpace{}%
\AgdaFunction{cmdWrote}\AgdaSpace{}%
\AgdaBound{ls₁}\AgdaSpace{}%
\AgdaBound{x₁}\AgdaSymbol{)))}\<%
\\
\>[0]\AgdaFunction{speculativeHazard-x?}\AgdaSpace{}%
\AgdaBound{x}\AgdaSpace{}%
\AgdaBound{b₂}\AgdaSpace{}%
\AgdaInductiveConstructor{[]}\AgdaSpace{}%
\AgdaBound{ls₁}\AgdaSpace{}%
\AgdaBound{rs}\AgdaSpace{}%
\AgdaSymbol{=}\AgdaSpace{}%
\AgdaInductiveConstructor{false}\AgdaSpace{}%
\AgdaOperator{\AgdaInductiveConstructor{Relation.Nullary.because}}\AgdaSpace{}%
\AgdaInductiveConstructor{Relation.Nullary.ofⁿ}\AgdaSpace{}%
\AgdaFunction{g₁}\<%
\\
\>[0][@{}l@{\AgdaIndent{0}}]%
\>[2]\AgdaKeyword{where}%
\>[2182I]\AgdaFunction{g₁}\AgdaSpace{}%
\AgdaSymbol{:}\AgdaSpace{}%
\AgdaFunction{∃[}\AgdaSpace{}%
\AgdaBound{x₁}\AgdaSpace{}%
\AgdaFunction{]}\AgdaSymbol{(}\AgdaFunction{∃[}\AgdaSpace{}%
\AgdaBound{v}\AgdaSpace{}%
\AgdaFunction{]}\AgdaSymbol{(}\AgdaBound{x₁}\AgdaSpace{}%
\AgdaOperator{\AgdaFunction{∈}}\AgdaSpace{}%
\AgdaInductiveConstructor{[]}\AgdaSpace{}%
\AgdaOperator{\AgdaFunction{×}}\AgdaSpace{}%
\AgdaOperator{\AgdaFunction{¬}}\AgdaSpace{}%
\AgdaSymbol{(}\AgdaBound{x₁}\AgdaSpace{}%
\AgdaOperator{\AgdaFunction{before}}\AgdaSpace{}%
\AgdaBound{x}\AgdaSpace{}%
\AgdaOperator{\AgdaFunction{∈}}\AgdaSpace{}%
\AgdaBound{b₂}\AgdaSymbol{)}\AgdaSpace{}%
\AgdaOperator{\AgdaFunction{×}}\AgdaSpace{}%
\AgdaBound{v}\AgdaSpace{}%
\AgdaOperator{\AgdaFunction{∈}}\AgdaSpace{}%
\AgdaBound{rs}\AgdaSpace{}%
\AgdaOperator{\AgdaFunction{×}}\AgdaSpace{}%
\AgdaBound{v}\AgdaSpace{}%
\AgdaOperator{\AgdaFunction{∈}}\AgdaSpace{}%
\AgdaFunction{cmdWrote}\AgdaSpace{}%
\AgdaBound{ls₁}\AgdaSpace{}%
\AgdaBound{x₁}\AgdaSymbol{))}\AgdaSpace{}%
\AgdaSymbol{→}\AgdaSpace{}%
\AgdaDatatype{⊥}\<%
\\
\>[.][@{}l@{}]\<[2182I]%
\>[8]\AgdaFunction{g₁}\AgdaSpace{}%
\AgdaSymbol{()}\<%
\\
\>[0]\AgdaFunction{speculativeHazard-x?}\AgdaSpace{}%
\AgdaBound{x}\AgdaSpace{}%
\AgdaBound{b₂}\AgdaSpace{}%
\AgdaSymbol{(}\AgdaBound{x₁}\AgdaSpace{}%
\AgdaOperator{\AgdaInductiveConstructor{∷}}\AgdaSpace{}%
\AgdaBound{ls}\AgdaSymbol{)}\AgdaSpace{}%
\AgdaBound{ls₁}\AgdaSpace{}%
\AgdaBound{rs}\AgdaSpace{}%
\AgdaKeyword{with}\AgdaSpace{}%
\AgdaFunction{intersection?2}\AgdaSpace{}%
\AgdaBound{rs}\AgdaSpace{}%
\AgdaSymbol{(}\AgdaFunction{cmdWrote}\AgdaSpace{}%
\AgdaBound{ls₁}\AgdaSpace{}%
\AgdaBound{x₁}\AgdaSymbol{)}\<%
\\
\>[0]\AgdaSymbol{...}\AgdaSpace{}%
\AgdaSymbol{|}\AgdaSpace{}%
\AgdaInductiveConstructor{yes}\AgdaSpace{}%
\AgdaSymbol{(}\AgdaBound{v}\AgdaSpace{}%
\AgdaOperator{\AgdaInductiveConstructor{,}}\AgdaSpace{}%
\AgdaBound{v∈rs}\AgdaSpace{}%
\AgdaOperator{\AgdaInductiveConstructor{,}}\AgdaSpace{}%
\AgdaBound{v∈ws}\AgdaSymbol{)}\AgdaSpace{}%
\AgdaKeyword{with}\AgdaSpace{}%
\AgdaFunction{before?}\AgdaSpace{}%
\AgdaBound{x₁}\AgdaSpace{}%
\AgdaBound{x}\AgdaSpace{}%
\AgdaBound{b₂}\<%
\\
\>[0]\AgdaSymbol{...}\AgdaSpace{}%
\AgdaSymbol{|}\AgdaSpace{}%
\AgdaInductiveConstructor{no}\AgdaSpace{}%
\AgdaBound{¬bf}\AgdaSpace{}%
\AgdaSymbol{=}\AgdaSpace{}%
\AgdaInductiveConstructor{true}\AgdaSpace{}%
\AgdaOperator{\AgdaInductiveConstructor{Relation.Nullary.because}}\AgdaSpace{}%
\AgdaInductiveConstructor{Relation.Nullary.ofʸ}\AgdaSpace{}%
\AgdaFunction{g₁}\<%
\\
\>[0][@{}l@{\AgdaIndent{0}}]%
\>[2]\AgdaKeyword{where}%
\>[2244I]\AgdaFunction{g₁}\AgdaSpace{}%
\AgdaSymbol{:}\AgdaSpace{}%
\AgdaFunction{∃[}\AgdaSpace{}%
\AgdaBound{x₂}\AgdaSpace{}%
\AgdaFunction{]}\AgdaSymbol{(}\AgdaFunction{∃[}\AgdaSpace{}%
\AgdaBound{v}\AgdaSpace{}%
\AgdaFunction{]}\AgdaSymbol{(}\AgdaBound{x₂}\AgdaSpace{}%
\AgdaOperator{\AgdaFunction{∈}}\AgdaSpace{}%
\AgdaSymbol{(}\AgdaBound{x₁}\AgdaSpace{}%
\AgdaOperator{\AgdaInductiveConstructor{∷}}\AgdaSpace{}%
\AgdaBound{ls}\AgdaSymbol{)}\AgdaSpace{}%
\AgdaOperator{\AgdaFunction{×}}\AgdaSpace{}%
\AgdaOperator{\AgdaFunction{¬}}\AgdaSpace{}%
\AgdaSymbol{(}\AgdaBound{x₂}\AgdaSpace{}%
\AgdaOperator{\AgdaFunction{before}}\AgdaSpace{}%
\AgdaBound{x}\AgdaSpace{}%
\AgdaOperator{\AgdaFunction{∈}}\AgdaSpace{}%
\AgdaBound{b₂}\AgdaSymbol{)}\AgdaSpace{}%
\AgdaOperator{\AgdaFunction{×}}\AgdaSpace{}%
\AgdaBound{v}\AgdaSpace{}%
\AgdaOperator{\AgdaFunction{∈}}\AgdaSpace{}%
\AgdaBound{rs}\AgdaSpace{}%
\AgdaOperator{\AgdaFunction{×}}\AgdaSpace{}%
\AgdaBound{v}\AgdaSpace{}%
\AgdaOperator{\AgdaFunction{∈}}\AgdaSpace{}%
\AgdaFunction{cmdWrote}\AgdaSpace{}%
\AgdaBound{ls₁}\AgdaSpace{}%
\AgdaBound{x₂}\AgdaSymbol{))}\<%
\\
\>[.][@{}l@{}]\<[2244I]%
\>[8]\AgdaFunction{g₁}\AgdaSpace{}%
\AgdaSymbol{=}\AgdaSpace{}%
\AgdaBound{x₁}\AgdaSpace{}%
\AgdaOperator{\AgdaInductiveConstructor{,}}\AgdaSpace{}%
\AgdaBound{v}\AgdaSpace{}%
\AgdaOperator{\AgdaInductiveConstructor{,}}\AgdaSpace{}%
\AgdaInductiveConstructor{here}\AgdaSpace{}%
\AgdaInductiveConstructor{refl}\AgdaSpace{}%
\AgdaOperator{\AgdaInductiveConstructor{,}}\AgdaSpace{}%
\AgdaBound{¬bf}\AgdaSpace{}%
\AgdaOperator{\AgdaInductiveConstructor{,}}\AgdaSpace{}%
\AgdaBound{v∈rs}\AgdaSpace{}%
\AgdaOperator{\AgdaInductiveConstructor{,}}\AgdaSpace{}%
\AgdaBound{v∈ws}\<%
\\
\>[0]\AgdaSymbol{...}\AgdaSpace{}%
\AgdaSymbol{|}\AgdaSpace{}%
\AgdaInductiveConstructor{yes}\AgdaSpace{}%
\AgdaBound{bf}\AgdaSpace{}%
\AgdaKeyword{with}\AgdaSpace{}%
\AgdaFunction{speculativeHazard-x?}\AgdaSpace{}%
\AgdaBound{x}\AgdaSpace{}%
\AgdaBound{b₂}\AgdaSpace{}%
\AgdaBound{ls}\AgdaSpace{}%
\AgdaBound{ls₁}\AgdaSpace{}%
\AgdaBound{rs}\<%
\\
\>[0]\AgdaSymbol{...}\AgdaSpace{}%
\AgdaSymbol{|}\AgdaSpace{}%
\AgdaInductiveConstructor{yes}\AgdaSpace{}%
\AgdaSymbol{(}\AgdaBound{x₂}\AgdaSpace{}%
\AgdaOperator{\AgdaInductiveConstructor{,}}\AgdaSpace{}%
\AgdaBound{v₂}\AgdaSpace{}%
\AgdaOperator{\AgdaInductiveConstructor{,}}\AgdaSpace{}%
\AgdaBound{x₂∈}\AgdaSpace{}%
\AgdaOperator{\AgdaInductiveConstructor{,}}\AgdaSpace{}%
\AgdaBound{¬bf}\AgdaSpace{}%
\AgdaOperator{\AgdaInductiveConstructor{,}}\AgdaSpace{}%
\AgdaBound{a}\AgdaSpace{}%
\AgdaOperator{\AgdaInductiveConstructor{,}}\AgdaSpace{}%
\AgdaBound{a₁}\AgdaSymbol{)}\AgdaSpace{}%
\AgdaSymbol{=}\AgdaSpace{}%
\AgdaInductiveConstructor{true}\AgdaSpace{}%
\AgdaOperator{\AgdaInductiveConstructor{Relation.Nullary.because}}\AgdaSpace{}%
\AgdaInductiveConstructor{Relation.Nullary.ofʸ}\AgdaSpace{}%
\AgdaFunction{g₁}\<%
\\
\>[0][@{}l@{\AgdaIndent{0}}]%
\>[2]\AgdaKeyword{where}%
\>[2313I]\AgdaFunction{g₁}\AgdaSpace{}%
\AgdaSymbol{:}\AgdaSpace{}%
\AgdaFunction{∃[}\AgdaSpace{}%
\AgdaBound{x₂}\AgdaSpace{}%
\AgdaFunction{]}\AgdaSymbol{(}\AgdaFunction{∃[}\AgdaSpace{}%
\AgdaBound{v}\AgdaSpace{}%
\AgdaFunction{]}\AgdaSymbol{(}\AgdaBound{x₂}\AgdaSpace{}%
\AgdaOperator{\AgdaFunction{∈}}\AgdaSpace{}%
\AgdaSymbol{(}\AgdaBound{x₁}\AgdaSpace{}%
\AgdaOperator{\AgdaInductiveConstructor{∷}}\AgdaSpace{}%
\AgdaBound{ls}\AgdaSymbol{)}\AgdaSpace{}%
\AgdaOperator{\AgdaFunction{×}}\AgdaSpace{}%
\AgdaOperator{\AgdaFunction{¬}}\AgdaSpace{}%
\AgdaSymbol{(}\AgdaBound{x₂}\AgdaSpace{}%
\AgdaOperator{\AgdaFunction{before}}\AgdaSpace{}%
\AgdaBound{x}\AgdaSpace{}%
\AgdaOperator{\AgdaFunction{∈}}\AgdaSpace{}%
\AgdaBound{b₂}\AgdaSymbol{)}\AgdaSpace{}%
\AgdaOperator{\AgdaFunction{×}}\AgdaSpace{}%
\AgdaBound{v}\AgdaSpace{}%
\AgdaOperator{\AgdaFunction{∈}}\AgdaSpace{}%
\AgdaBound{rs}\AgdaSpace{}%
\AgdaOperator{\AgdaFunction{×}}\AgdaSpace{}%
\AgdaBound{v}\AgdaSpace{}%
\AgdaOperator{\AgdaFunction{∈}}\AgdaSpace{}%
\AgdaFunction{cmdWrote}\AgdaSpace{}%
\AgdaBound{ls₁}\AgdaSpace{}%
\AgdaBound{x₂}\AgdaSymbol{))}\<%
\\
\>[.][@{}l@{}]\<[2313I]%
\>[8]\AgdaFunction{g₁}\AgdaSpace{}%
\AgdaSymbol{=}\AgdaSpace{}%
\AgdaBound{x₂}\AgdaSpace{}%
\AgdaOperator{\AgdaInductiveConstructor{,}}\AgdaSpace{}%
\AgdaBound{v₂}\AgdaSpace{}%
\AgdaOperator{\AgdaInductiveConstructor{,}}\AgdaSpace{}%
\AgdaInductiveConstructor{there}\AgdaSpace{}%
\AgdaBound{x₂∈}\AgdaSpace{}%
\AgdaOperator{\AgdaInductiveConstructor{,}}\AgdaSpace{}%
\AgdaBound{¬bf}\AgdaSpace{}%
\AgdaOperator{\AgdaInductiveConstructor{,}}\AgdaSpace{}%
\AgdaBound{a}\AgdaSpace{}%
\AgdaOperator{\AgdaInductiveConstructor{,}}\AgdaSpace{}%
\AgdaBound{a₁}\<%
\\
\>[0]\AgdaSymbol{...}\AgdaSpace{}%
\AgdaSymbol{|}\AgdaSpace{}%
\AgdaInductiveConstructor{no}\AgdaSpace{}%
\AgdaBound{¬p}\AgdaSpace{}%
\AgdaSymbol{=}\AgdaSpace{}%
\AgdaInductiveConstructor{false}\AgdaSpace{}%
\AgdaOperator{\AgdaInductiveConstructor{Relation.Nullary.because}}\AgdaSpace{}%
\AgdaInductiveConstructor{Relation.Nullary.ofⁿ}\AgdaSpace{}%
\AgdaFunction{g₁}\<%
\\
\>[0][@{}l@{\AgdaIndent{0}}]%
\>[2]\AgdaKeyword{where}%
\>[2362I]\AgdaFunction{g₁}\AgdaSpace{}%
\AgdaSymbol{:}\AgdaSpace{}%
\AgdaOperator{\AgdaFunction{¬}}\AgdaSpace{}%
\AgdaSymbol{(}\AgdaFunction{∃[}\AgdaSpace{}%
\AgdaBound{x₂}\AgdaSpace{}%
\AgdaFunction{]}\AgdaSymbol{(}\AgdaFunction{∃[}\AgdaSpace{}%
\AgdaBound{v}\AgdaSpace{}%
\AgdaFunction{]}\AgdaSymbol{(}\AgdaBound{x₂}\AgdaSpace{}%
\AgdaOperator{\AgdaFunction{∈}}\AgdaSpace{}%
\AgdaSymbol{(}\AgdaBound{x₁}\AgdaSpace{}%
\AgdaOperator{\AgdaInductiveConstructor{∷}}\AgdaSpace{}%
\AgdaBound{ls}\AgdaSymbol{)}\AgdaSpace{}%
\AgdaOperator{\AgdaFunction{×}}\AgdaSpace{}%
\AgdaOperator{\AgdaFunction{¬}}\AgdaSpace{}%
\AgdaSymbol{(}\AgdaBound{x₂}\AgdaSpace{}%
\AgdaOperator{\AgdaFunction{before}}\AgdaSpace{}%
\AgdaBound{x}\AgdaSpace{}%
\AgdaOperator{\AgdaFunction{∈}}\AgdaSpace{}%
\AgdaBound{b₂}\AgdaSymbol{)}\AgdaSpace{}%
\AgdaOperator{\AgdaFunction{×}}\AgdaSpace{}%
\AgdaBound{v}\AgdaSpace{}%
\AgdaOperator{\AgdaFunction{∈}}\AgdaSpace{}%
\AgdaBound{rs}\AgdaSpace{}%
\AgdaOperator{\AgdaFunction{×}}\AgdaSpace{}%
\AgdaBound{v}\AgdaSpace{}%
\AgdaOperator{\AgdaFunction{∈}}\AgdaSpace{}%
\AgdaFunction{cmdWrote}\AgdaSpace{}%
\AgdaBound{ls₁}\AgdaSpace{}%
\AgdaBound{x₂}\AgdaSymbol{)))}\<%
\\
\>[.][@{}l@{}]\<[2362I]%
\>[8]\AgdaFunction{g₁}\AgdaSpace{}%
\AgdaSymbol{(}\AgdaBound{x₂}\AgdaSpace{}%
\AgdaOperator{\AgdaInductiveConstructor{,}}\AgdaSpace{}%
\AgdaBound{v}\AgdaSpace{}%
\AgdaOperator{\AgdaInductiveConstructor{,}}\AgdaSpace{}%
\AgdaBound{x₂∈}\AgdaSpace{}%
\AgdaOperator{\AgdaInductiveConstructor{,}}\AgdaSpace{}%
\AgdaBound{¬bf}\AgdaSpace{}%
\AgdaOperator{\AgdaInductiveConstructor{,}}\AgdaSpace{}%
\AgdaBound{v∈rs}\AgdaSpace{}%
\AgdaOperator{\AgdaInductiveConstructor{,}}\AgdaSpace{}%
\AgdaBound{v∈ws}\AgdaSymbol{)}\AgdaSpace{}%
\AgdaKeyword{with}\AgdaSpace{}%
\AgdaBound{x₂}\AgdaSpace{}%
\AgdaOperator{\AgdaFunction{≟}}\AgdaSpace{}%
\AgdaBound{x₁}\<%
\\
\>[8]\AgdaSymbol{...}\AgdaSpace{}%
\AgdaSymbol{|}\AgdaSpace{}%
\AgdaInductiveConstructor{yes}\AgdaSpace{}%
\AgdaBound{x₂≡x₁}\AgdaSpace{}%
\AgdaSymbol{=}\AgdaSpace{}%
\AgdaFunction{contradiction}\AgdaSpace{}%
\AgdaSymbol{(}\AgdaFunction{subst}\AgdaSpace{}%
\AgdaSymbol{(λ}\AgdaSpace{}%
\AgdaBound{x₃}\AgdaSpace{}%
\AgdaSymbol{→}\AgdaSpace{}%
\AgdaBound{x₃}\AgdaSpace{}%
\AgdaOperator{\AgdaFunction{before}}\AgdaSpace{}%
\AgdaBound{x}\AgdaSpace{}%
\AgdaOperator{\AgdaFunction{∈}}\AgdaSpace{}%
\AgdaBound{b₂}\AgdaSymbol{)}\AgdaSpace{}%
\AgdaSymbol{(}\AgdaFunction{sym}\AgdaSpace{}%
\AgdaBound{x₂≡x₁}\AgdaSymbol{)}\AgdaSpace{}%
\AgdaBound{bf}\AgdaSymbol{)}\AgdaSpace{}%
\AgdaBound{¬bf}\<%
\\
\>[8]\AgdaSymbol{...}\AgdaSpace{}%
\AgdaSymbol{|}\AgdaSpace{}%
\AgdaInductiveConstructor{no}\AgdaSpace{}%
\AgdaBound{¬x₂≡x₁}\AgdaSpace{}%
\AgdaSymbol{=}\AgdaSpace{}%
\AgdaBound{¬p}\AgdaSpace{}%
\AgdaSymbol{(}\AgdaBound{x₂}\AgdaSpace{}%
\AgdaOperator{\AgdaInductiveConstructor{,}}\AgdaSpace{}%
\AgdaBound{v}\AgdaSpace{}%
\AgdaOperator{\AgdaInductiveConstructor{,}}\AgdaSpace{}%
\AgdaFunction{tail}\AgdaSpace{}%
\AgdaBound{¬x₂≡x₁}\AgdaSpace{}%
\AgdaBound{x₂∈}\AgdaSpace{}%
\AgdaOperator{\AgdaInductiveConstructor{,}}\AgdaSpace{}%
\AgdaBound{¬bf}\AgdaSpace{}%
\AgdaOperator{\AgdaInductiveConstructor{,}}\AgdaSpace{}%
\AgdaBound{v∈rs}\AgdaSpace{}%
\AgdaOperator{\AgdaInductiveConstructor{,}}\AgdaSpace{}%
\AgdaBound{v∈ws}\AgdaSymbol{)}\<%
\\
\>[0]\AgdaFunction{speculativeHazard-x?}\AgdaSpace{}%
\AgdaBound{x}\AgdaSpace{}%
\AgdaBound{b₂}\AgdaSpace{}%
\AgdaSymbol{(}\AgdaBound{x₁}\AgdaSpace{}%
\AgdaOperator{\AgdaInductiveConstructor{∷}}\AgdaSpace{}%
\AgdaBound{ls}\AgdaSymbol{)}\AgdaSpace{}%
\AgdaBound{ls₁}\AgdaSpace{}%
\AgdaBound{rs}\AgdaSpace{}%
\AgdaSymbol{|}\AgdaSpace{}%
\AgdaInductiveConstructor{no}\AgdaSpace{}%
\AgdaBound{p₁}\AgdaSpace{}%
\AgdaKeyword{with}\AgdaSpace{}%
\AgdaFunction{speculativeHazard-x?}\AgdaSpace{}%
\AgdaBound{x}\AgdaSpace{}%
\AgdaBound{b₂}\AgdaSpace{}%
\AgdaBound{ls}\AgdaSpace{}%
\AgdaBound{ls₁}\AgdaSpace{}%
\AgdaBound{rs}\<%
\\
\>[0]\AgdaSymbol{...}\AgdaSpace{}%
\AgdaSymbol{|}\AgdaSpace{}%
\AgdaInductiveConstructor{yes}\AgdaSpace{}%
\AgdaSymbol{(}\AgdaBound{x₂}\AgdaSpace{}%
\AgdaOperator{\AgdaInductiveConstructor{,}}\AgdaSpace{}%
\AgdaBound{v₂}\AgdaSpace{}%
\AgdaOperator{\AgdaInductiveConstructor{,}}\AgdaSpace{}%
\AgdaBound{x₂∈}\AgdaSpace{}%
\AgdaOperator{\AgdaInductiveConstructor{,}}\AgdaSpace{}%
\AgdaBound{¬bf}\AgdaSpace{}%
\AgdaOperator{\AgdaInductiveConstructor{,}}\AgdaSpace{}%
\AgdaBound{a}\AgdaSpace{}%
\AgdaOperator{\AgdaInductiveConstructor{,}}\AgdaSpace{}%
\AgdaBound{a₁}\AgdaSymbol{)}\AgdaSpace{}%
\AgdaSymbol{=}\AgdaSpace{}%
\AgdaInductiveConstructor{true}\AgdaSpace{}%
\AgdaOperator{\AgdaInductiveConstructor{Relation.Nullary.because}}\AgdaSpace{}%
\AgdaInductiveConstructor{Relation.Nullary.ofʸ}\AgdaSpace{}%
\AgdaFunction{g₁}\<%
\\
\>[0][@{}l@{\AgdaIndent{0}}]%
\>[2]\AgdaKeyword{where}%
\>[2477I]\AgdaFunction{g₁}\AgdaSpace{}%
\AgdaSymbol{:}\AgdaSpace{}%
\AgdaFunction{∃[}\AgdaSpace{}%
\AgdaBound{x₂}\AgdaSpace{}%
\AgdaFunction{]}\AgdaSymbol{(}\AgdaFunction{∃[}\AgdaSpace{}%
\AgdaBound{v}\AgdaSpace{}%
\AgdaFunction{]}\AgdaSymbol{(}\AgdaBound{x₂}\AgdaSpace{}%
\AgdaOperator{\AgdaFunction{∈}}\AgdaSpace{}%
\AgdaSymbol{(}\AgdaBound{x₁}\AgdaSpace{}%
\AgdaOperator{\AgdaInductiveConstructor{∷}}\AgdaSpace{}%
\AgdaBound{ls}\AgdaSymbol{)}\AgdaSpace{}%
\AgdaOperator{\AgdaFunction{×}}\AgdaSpace{}%
\AgdaOperator{\AgdaFunction{¬}}\AgdaSpace{}%
\AgdaSymbol{(}\AgdaBound{x₂}\AgdaSpace{}%
\AgdaOperator{\AgdaFunction{before}}\AgdaSpace{}%
\AgdaBound{x}\AgdaSpace{}%
\AgdaOperator{\AgdaFunction{∈}}\AgdaSpace{}%
\AgdaBound{b₂}\AgdaSymbol{)}\AgdaSpace{}%
\AgdaOperator{\AgdaFunction{×}}\AgdaSpace{}%
\AgdaBound{v}\AgdaSpace{}%
\AgdaOperator{\AgdaFunction{∈}}\AgdaSpace{}%
\AgdaBound{rs}\AgdaSpace{}%
\AgdaOperator{\AgdaFunction{×}}\AgdaSpace{}%
\AgdaBound{v}\AgdaSpace{}%
\AgdaOperator{\AgdaFunction{∈}}\AgdaSpace{}%
\AgdaFunction{cmdWrote}\AgdaSpace{}%
\AgdaBound{ls₁}\AgdaSpace{}%
\AgdaBound{x₂}\AgdaSymbol{))}\<%
\\
\>[.][@{}l@{}]\<[2477I]%
\>[8]\AgdaFunction{g₁}\AgdaSpace{}%
\AgdaSymbol{=}\AgdaSpace{}%
\AgdaBound{x₂}\AgdaSpace{}%
\AgdaOperator{\AgdaInductiveConstructor{,}}\AgdaSpace{}%
\AgdaBound{v₂}\AgdaSpace{}%
\AgdaOperator{\AgdaInductiveConstructor{,}}\AgdaSpace{}%
\AgdaInductiveConstructor{there}\AgdaSpace{}%
\AgdaBound{x₂∈}\AgdaSpace{}%
\AgdaOperator{\AgdaInductiveConstructor{,}}\AgdaSpace{}%
\AgdaBound{¬bf}\AgdaSpace{}%
\AgdaOperator{\AgdaInductiveConstructor{,}}\AgdaSpace{}%
\AgdaBound{a}\AgdaSpace{}%
\AgdaOperator{\AgdaInductiveConstructor{,}}\AgdaSpace{}%
\AgdaBound{a₁}\<%
\\
\>[0]\AgdaSymbol{...}\AgdaSpace{}%
\AgdaSymbol{|}\AgdaSpace{}%
\AgdaInductiveConstructor{no}\AgdaSpace{}%
\AgdaBound{¬p}\AgdaSpace{}%
\AgdaSymbol{=}\AgdaSpace{}%
\AgdaInductiveConstructor{false}\AgdaSpace{}%
\AgdaOperator{\AgdaInductiveConstructor{Relation.Nullary.because}}\AgdaSpace{}%
\AgdaInductiveConstructor{Relation.Nullary.ofⁿ}\AgdaSpace{}%
\AgdaFunction{g₁}\<%
\\
\>[0][@{}l@{\AgdaIndent{0}}]%
\>[2]\AgdaKeyword{where}%
\>[2526I]\AgdaFunction{g₁}\AgdaSpace{}%
\AgdaSymbol{:}\AgdaSpace{}%
\AgdaOperator{\AgdaFunction{¬}}\AgdaSpace{}%
\AgdaSymbol{(}\AgdaFunction{∃[}\AgdaSpace{}%
\AgdaBound{x₂}\AgdaSpace{}%
\AgdaFunction{]}\AgdaSymbol{(}\AgdaFunction{∃[}\AgdaSpace{}%
\AgdaBound{v}\AgdaSpace{}%
\AgdaFunction{]}\AgdaSymbol{(}\AgdaBound{x₂}\AgdaSpace{}%
\AgdaOperator{\AgdaFunction{∈}}\AgdaSpace{}%
\AgdaSymbol{(}\AgdaBound{x₁}\AgdaSpace{}%
\AgdaOperator{\AgdaInductiveConstructor{∷}}\AgdaSpace{}%
\AgdaBound{ls}\AgdaSymbol{)}\AgdaSpace{}%
\AgdaOperator{\AgdaFunction{×}}\AgdaSpace{}%
\AgdaOperator{\AgdaFunction{¬}}\AgdaSpace{}%
\AgdaSymbol{(}\AgdaBound{x₂}\AgdaSpace{}%
\AgdaOperator{\AgdaFunction{before}}\AgdaSpace{}%
\AgdaBound{x}\AgdaSpace{}%
\AgdaOperator{\AgdaFunction{∈}}\AgdaSpace{}%
\AgdaBound{b₂}\AgdaSymbol{)}\AgdaSpace{}%
\AgdaOperator{\AgdaFunction{×}}\AgdaSpace{}%
\AgdaBound{v}\AgdaSpace{}%
\AgdaOperator{\AgdaFunction{∈}}\AgdaSpace{}%
\AgdaBound{rs}\AgdaSpace{}%
\AgdaOperator{\AgdaFunction{×}}\AgdaSpace{}%
\AgdaBound{v}\AgdaSpace{}%
\AgdaOperator{\AgdaFunction{∈}}\AgdaSpace{}%
\AgdaFunction{cmdWrote}\AgdaSpace{}%
\AgdaBound{ls₁}\AgdaSpace{}%
\AgdaBound{x₂}\AgdaSymbol{)))}\<%
\\
\>[.][@{}l@{}]\<[2526I]%
\>[8]\AgdaFunction{g₁}\AgdaSpace{}%
\AgdaSymbol{(}\AgdaBound{x₂}\AgdaSpace{}%
\AgdaOperator{\AgdaInductiveConstructor{,}}\AgdaSpace{}%
\AgdaBound{v}\AgdaSpace{}%
\AgdaOperator{\AgdaInductiveConstructor{,}}\AgdaSpace{}%
\AgdaBound{x₂∈}\AgdaSpace{}%
\AgdaOperator{\AgdaInductiveConstructor{,}}\AgdaSpace{}%
\AgdaBound{¬bf}\AgdaSpace{}%
\AgdaOperator{\AgdaInductiveConstructor{,}}\AgdaSpace{}%
\AgdaBound{v∈rs}\AgdaSpace{}%
\AgdaOperator{\AgdaInductiveConstructor{,}}\AgdaSpace{}%
\AgdaBound{v∈ws}\AgdaSymbol{)}\AgdaSpace{}%
\AgdaKeyword{with}\AgdaSpace{}%
\AgdaBound{x₂}\AgdaSpace{}%
\AgdaOperator{\AgdaFunction{≟}}\AgdaSpace{}%
\AgdaBound{x₁}\<%
\\
\>[8]\AgdaSymbol{...}\AgdaSpace{}%
\AgdaSymbol{|}\AgdaSpace{}%
\AgdaInductiveConstructor{yes}\AgdaSpace{}%
\AgdaBound{x₂≡x₁}\AgdaSpace{}%
\AgdaSymbol{=}\AgdaSpace{}%
\AgdaBound{p₁}\AgdaSpace{}%
\AgdaSymbol{(}\AgdaBound{v}\AgdaSpace{}%
\AgdaOperator{\AgdaInductiveConstructor{,}}\AgdaSpace{}%
\AgdaSymbol{(}\AgdaBound{v∈rs}\AgdaSpace{}%
\AgdaOperator{\AgdaInductiveConstructor{,}}\AgdaSpace{}%
\AgdaFunction{subst}\AgdaSpace{}%
\AgdaSymbol{(λ}\AgdaSpace{}%
\AgdaBound{x₃}\AgdaSpace{}%
\AgdaSymbol{→}\AgdaSpace{}%
\AgdaBound{v}\AgdaSpace{}%
\AgdaOperator{\AgdaFunction{∈}}\AgdaSpace{}%
\AgdaFunction{cmdWrote}\AgdaSpace{}%
\AgdaBound{ls₁}\AgdaSpace{}%
\AgdaBound{x₃}\AgdaSymbol{)}\AgdaSpace{}%
\AgdaBound{x₂≡x₁}\AgdaSpace{}%
\AgdaBound{v∈ws}\AgdaSymbol{))}\<%
\\
\>[8]\AgdaSymbol{...}\AgdaSpace{}%
\AgdaSymbol{|}\AgdaSpace{}%
\AgdaInductiveConstructor{no}\AgdaSpace{}%
\AgdaBound{¬x₂≡x₁}\AgdaSpace{}%
\AgdaSymbol{=}\AgdaSpace{}%
\AgdaBound{¬p}\AgdaSpace{}%
\AgdaSymbol{(}\AgdaBound{x₂}\AgdaSpace{}%
\AgdaOperator{\AgdaInductiveConstructor{,}}\AgdaSpace{}%
\AgdaBound{v}\AgdaSpace{}%
\AgdaOperator{\AgdaInductiveConstructor{,}}\AgdaSpace{}%
\AgdaFunction{tail}\AgdaSpace{}%
\AgdaBound{¬x₂≡x₁}\AgdaSpace{}%
\AgdaBound{x₂∈}\AgdaSpace{}%
\AgdaOperator{\AgdaInductiveConstructor{,}}\AgdaSpace{}%
\AgdaBound{¬bf}\AgdaSpace{}%
\AgdaOperator{\AgdaInductiveConstructor{,}}\AgdaSpace{}%
\AgdaBound{v∈rs}\AgdaSpace{}%
\AgdaOperator{\AgdaInductiveConstructor{,}}\AgdaSpace{}%
\AgdaBound{v∈ws}\AgdaSymbol{)}\<%
\\
\\[\AgdaEmptyExtraSkip]%
\>[0]\AgdaFunction{¬speculative?}\AgdaSpace{}%
\AgdaSymbol{:}\AgdaSpace{}%
\AgdaSymbol{∀}\AgdaSpace{}%
\AgdaBound{b₂}\AgdaSpace{}%
\AgdaBound{ls}\AgdaSpace{}%
\AgdaSymbol{→}\AgdaSpace{}%
\AgdaRecord{Dec}\AgdaSpace{}%
\AgdaSymbol{(}\AgdaFunction{∃[}\AgdaSpace{}%
\AgdaBound{x₁}\AgdaSpace{}%
\AgdaFunction{]}\AgdaSymbol{(}\AgdaFunction{∃[}\AgdaSpace{}%
\AgdaBound{x₂}\AgdaSpace{}%
\AgdaFunction{]}\AgdaSymbol{(}\AgdaFunction{∃[}\AgdaSpace{}%
\AgdaBound{v}\AgdaSpace{}%
\AgdaFunction{]}\AgdaSymbol{((}\AgdaBound{x₂}\AgdaSpace{}%
\AgdaOperator{\AgdaFunction{before}}\AgdaSpace{}%
\AgdaBound{x₁}\AgdaSpace{}%
\AgdaOperator{\AgdaFunction{∈}}\AgdaSpace{}%
\AgdaSymbol{(}\AgdaFunction{map}\AgdaSpace{}%
\AgdaField{proj₁}\AgdaSpace{}%
\AgdaBound{ls}\AgdaSymbol{))}\AgdaSpace{}%
\AgdaOperator{\AgdaFunction{×}}\AgdaSpace{}%
\AgdaBound{x₂}\AgdaSpace{}%
\AgdaOperator{\AgdaFunction{∈}}\AgdaSpace{}%
\AgdaBound{b₂}\AgdaSpace{}%
\AgdaOperator{\AgdaFunction{×}}\AgdaSpace{}%
\AgdaOperator{\AgdaFunction{¬}}\AgdaSpace{}%
\AgdaSymbol{(}\AgdaBound{x₁}\AgdaSpace{}%
\AgdaOperator{\AgdaFunction{before}}\AgdaSpace{}%
\AgdaBound{x₂}\AgdaSpace{}%
\AgdaOperator{\AgdaFunction{∈}}\AgdaSpace{}%
\AgdaBound{b₂}\AgdaSymbol{)}\AgdaSpace{}%
\AgdaOperator{\AgdaFunction{×}}\AgdaSpace{}%
\AgdaBound{v}\AgdaSpace{}%
\AgdaOperator{\AgdaFunction{∈}}\AgdaSpace{}%
\AgdaFunction{cmdRead}\AgdaSpace{}%
\AgdaBound{ls}\AgdaSpace{}%
\AgdaBound{x₂}\AgdaSpace{}%
\AgdaOperator{\AgdaFunction{×}}\AgdaSpace{}%
\AgdaBound{v}\AgdaSpace{}%
\AgdaOperator{\AgdaFunction{∈}}\AgdaSpace{}%
\AgdaFunction{cmdWrote}\AgdaSpace{}%
\AgdaBound{ls}\AgdaSpace{}%
\AgdaBound{x₁}\AgdaSymbol{))))}\<%
\\
\>[0]\AgdaFunction{¬speculative?}\AgdaSpace{}%
\AgdaBound{b₂}\AgdaSpace{}%
\AgdaBound{ls}\AgdaSpace{}%
\AgdaSymbol{=}\AgdaSpace{}%
\AgdaFunction{g₁}\AgdaSpace{}%
\AgdaBound{b₂}\AgdaSpace{}%
\AgdaSymbol{(}\AgdaFunction{map}\AgdaSpace{}%
\AgdaField{proj₁}\AgdaSpace{}%
\AgdaBound{ls}\AgdaSymbol{)}\AgdaSpace{}%
\AgdaBound{ls}\<%
\\
\>[0][@{}l@{\AgdaIndent{0}}]%
\>[2]\AgdaKeyword{where}%
\>[2659I]\AgdaFunction{g₁}\AgdaSpace{}%
\AgdaSymbol{:}\AgdaSpace{}%
\AgdaSymbol{∀}\AgdaSpace{}%
\AgdaBound{b₂}\AgdaSpace{}%
\AgdaBound{ls₁}\AgdaSpace{}%
\AgdaBound{ls₂}\AgdaSpace{}%
\AgdaSymbol{→}\AgdaSpace{}%
\AgdaRecord{Dec}\AgdaSpace{}%
\AgdaSymbol{(}\AgdaFunction{∃[}\AgdaSpace{}%
\AgdaBound{x₁}\AgdaSpace{}%
\AgdaFunction{]}\AgdaSymbol{(}\AgdaFunction{∃[}\AgdaSpace{}%
\AgdaBound{x₂}\AgdaSpace{}%
\AgdaFunction{]}\AgdaSymbol{(}\AgdaFunction{∃[}\AgdaSpace{}%
\AgdaBound{v}\AgdaSpace{}%
\AgdaFunction{]}\AgdaSymbol{((}\AgdaBound{x₂}\AgdaSpace{}%
\AgdaOperator{\AgdaFunction{before}}\AgdaSpace{}%
\AgdaBound{x₁}\AgdaSpace{}%
\AgdaOperator{\AgdaFunction{∈}}\AgdaSpace{}%
\AgdaBound{ls₁}\AgdaSymbol{)}\AgdaSpace{}%
\AgdaOperator{\AgdaFunction{×}}\AgdaSpace{}%
\AgdaBound{x₂}\AgdaSpace{}%
\AgdaOperator{\AgdaFunction{∈}}\AgdaSpace{}%
\AgdaBound{b₂}\AgdaSpace{}%
\AgdaOperator{\AgdaFunction{×}}\AgdaSpace{}%
\AgdaOperator{\AgdaFunction{¬}}\AgdaSpace{}%
\AgdaSymbol{(}\AgdaBound{x₁}\AgdaSpace{}%
\AgdaOperator{\AgdaFunction{before}}\AgdaSpace{}%
\AgdaBound{x₂}\AgdaSpace{}%
\AgdaOperator{\AgdaFunction{∈}}\AgdaSpace{}%
\AgdaBound{b₂}\AgdaSymbol{)}\AgdaSpace{}%
\AgdaOperator{\AgdaFunction{×}}\AgdaSpace{}%
\AgdaBound{v}\AgdaSpace{}%
\AgdaOperator{\AgdaFunction{∈}}\AgdaSpace{}%
\AgdaFunction{cmdRead}\AgdaSpace{}%
\AgdaBound{ls₂}\AgdaSpace{}%
\AgdaBound{x₂}\AgdaSpace{}%
\AgdaOperator{\AgdaFunction{×}}\AgdaSpace{}%
\AgdaBound{v}\AgdaSpace{}%
\AgdaOperator{\AgdaFunction{∈}}\AgdaSpace{}%
\AgdaFunction{cmdWrote}\AgdaSpace{}%
\AgdaBound{ls₂}\AgdaSpace{}%
\AgdaBound{x₁}\AgdaSymbol{))))}\<%
\\
\>[.][@{}l@{}]\<[2659I]%
\>[8]\AgdaFunction{g₁}\AgdaSpace{}%
\AgdaBound{b₂}\AgdaSpace{}%
\AgdaInductiveConstructor{[]}\AgdaSpace{}%
\AgdaBound{ls₂}\AgdaSpace{}%
\AgdaSymbol{=}\AgdaSpace{}%
\AgdaInductiveConstructor{false}\AgdaSpace{}%
\AgdaOperator{\AgdaInductiveConstructor{Relation.Nullary.because}}\AgdaSpace{}%
\AgdaInductiveConstructor{Relation.Nullary.ofⁿ}\AgdaSpace{}%
\AgdaFunction{g₂}\<%
\\
\>[8][@{}l@{\AgdaIndent{0}}]%
\>[10]\AgdaKeyword{where}%
\>[2709I]\AgdaFunction{g₂}\AgdaSpace{}%
\AgdaSymbol{:}\AgdaSpace{}%
\AgdaOperator{\AgdaFunction{¬}}\AgdaSpace{}%
\AgdaSymbol{(}\AgdaFunction{∃[}\AgdaSpace{}%
\AgdaBound{x₁}\AgdaSpace{}%
\AgdaFunction{]}\AgdaSymbol{(}\AgdaFunction{∃[}\AgdaSpace{}%
\AgdaBound{x₂}\AgdaSpace{}%
\AgdaFunction{]}\AgdaSymbol{(}\AgdaFunction{∃[}\AgdaSpace{}%
\AgdaBound{v}\AgdaSpace{}%
\AgdaFunction{]}\AgdaSymbol{((}\AgdaBound{x₂}\AgdaSpace{}%
\AgdaOperator{\AgdaFunction{before}}\AgdaSpace{}%
\AgdaBound{x₁}\AgdaSpace{}%
\AgdaOperator{\AgdaFunction{∈}}\AgdaSpace{}%
\AgdaInductiveConstructor{[]}\AgdaSymbol{)}\AgdaSpace{}%
\AgdaOperator{\AgdaFunction{×}}\AgdaSpace{}%
\AgdaBound{x₂}\AgdaSpace{}%
\AgdaOperator{\AgdaFunction{∈}}\AgdaSpace{}%
\AgdaBound{b₂}\AgdaSpace{}%
\AgdaOperator{\AgdaFunction{×}}\AgdaSpace{}%
\AgdaOperator{\AgdaFunction{¬}}\AgdaSpace{}%
\AgdaSymbol{(}\AgdaBound{x₁}\AgdaSpace{}%
\AgdaOperator{\AgdaFunction{before}}\AgdaSpace{}%
\AgdaBound{x₂}\AgdaSpace{}%
\AgdaOperator{\AgdaFunction{∈}}\AgdaSpace{}%
\AgdaBound{b₂}\AgdaSymbol{)}\AgdaSpace{}%
\AgdaOperator{\AgdaFunction{×}}\AgdaSpace{}%
\AgdaBound{v}\AgdaSpace{}%
\AgdaOperator{\AgdaFunction{∈}}\AgdaSpace{}%
\AgdaFunction{cmdRead}\AgdaSpace{}%
\AgdaBound{ls₂}\AgdaSpace{}%
\AgdaBound{x₂}\AgdaSpace{}%
\AgdaOperator{\AgdaFunction{×}}\AgdaSpace{}%
\AgdaBound{v}\AgdaSpace{}%
\AgdaOperator{\AgdaFunction{∈}}\AgdaSpace{}%
\AgdaFunction{cmdWrote}\AgdaSpace{}%
\AgdaBound{ls₂}\AgdaSpace{}%
\AgdaBound{x₁}\AgdaSymbol{))))}\<%
\\
\>[.][@{}l@{}]\<[2709I]%
\>[16]\AgdaFunction{g₂}\AgdaSpace{}%
\AgdaSymbol{(\AgdaUnderscore{}}\AgdaSpace{}%
\AgdaOperator{\AgdaInductiveConstructor{,}}\AgdaSpace{}%
\AgdaSymbol{\AgdaUnderscore{}}\AgdaSpace{}%
\AgdaOperator{\AgdaInductiveConstructor{,}}\AgdaSpace{}%
\AgdaSymbol{\AgdaUnderscore{}}\AgdaSpace{}%
\AgdaOperator{\AgdaInductiveConstructor{,}}\AgdaSpace{}%
\AgdaSymbol{(}\AgdaBound{xs}\AgdaSpace{}%
\AgdaOperator{\AgdaInductiveConstructor{,}}\AgdaSpace{}%
\AgdaBound{ys}\AgdaSpace{}%
\AgdaOperator{\AgdaInductiveConstructor{,}}\AgdaSpace{}%
\AgdaBound{≡₁}\AgdaSpace{}%
\AgdaOperator{\AgdaInductiveConstructor{,}}\AgdaSpace{}%
\AgdaBound{x₁∈ys}\AgdaSymbol{)}\AgdaSpace{}%
\AgdaOperator{\AgdaInductiveConstructor{,}}\AgdaSpace{}%
\AgdaBound{rest}\AgdaSymbol{)}\<%
\\
\>[16][@{}l@{\AgdaIndent{0}}]%
\>[18]\AgdaSymbol{=}\AgdaSpace{}%
\AgdaFunction{contradiction}\AgdaSpace{}%
\AgdaSymbol{(}\AgdaFunction{subst}\AgdaSpace{}%
\AgdaSymbol{(λ}\AgdaSpace{}%
\AgdaBound{x}\AgdaSpace{}%
\AgdaSymbol{→}\AgdaSpace{}%
\AgdaSymbol{\AgdaUnderscore{}}\AgdaSpace{}%
\AgdaOperator{\AgdaFunction{∈}}\AgdaSpace{}%
\AgdaBound{x}\AgdaSymbol{)}\AgdaSpace{}%
\AgdaSymbol{(}\AgdaFunction{sym}\AgdaSpace{}%
\AgdaBound{≡₁}\AgdaSymbol{)}\AgdaSpace{}%
\AgdaSymbol{(}\AgdaFunction{∈-++⁺ʳ}\AgdaSpace{}%
\AgdaBound{xs}\AgdaSpace{}%
\AgdaSymbol{(}\AgdaInductiveConstructor{there}\AgdaSpace{}%
\AgdaBound{x₁∈ys}\AgdaSymbol{)))}\AgdaSpace{}%
\AgdaSymbol{(λ}\AgdaSpace{}%
\AgdaSymbol{())}\<%
\\
\>[8]\AgdaFunction{g₁}\AgdaSpace{}%
\AgdaBound{b₂}\AgdaSpace{}%
\AgdaSymbol{(}\AgdaBound{x}\AgdaSpace{}%
\AgdaOperator{\AgdaInductiveConstructor{∷}}\AgdaSpace{}%
\AgdaBound{ls₁}\AgdaSymbol{)}\AgdaSpace{}%
\AgdaBound{ls₂}\AgdaSpace{}%
\AgdaKeyword{with}\AgdaSpace{}%
\AgdaBound{x}\AgdaSpace{}%
\AgdaOperator{\AgdaPostulate{∈?}}\AgdaSpace{}%
\AgdaBound{b₂}\<%
\\
\>[8]\AgdaSymbol{...}\AgdaSpace{}%
\AgdaSymbol{|}\AgdaSpace{}%
\AgdaInductiveConstructor{yes}\AgdaSpace{}%
\AgdaBound{x∈b₂}\AgdaSpace{}%
\AgdaKeyword{with}\AgdaSpace{}%
\AgdaFunction{speculativeHazard-x?}\AgdaSpace{}%
\AgdaBound{x}\AgdaSpace{}%
\AgdaBound{b₂}\AgdaSpace{}%
\AgdaBound{ls₁}\AgdaSpace{}%
\AgdaBound{ls₂}\AgdaSpace{}%
\AgdaSymbol{(}\AgdaFunction{cmdRead}\AgdaSpace{}%
\AgdaBound{ls₂}\AgdaSpace{}%
\AgdaBound{x}\AgdaSymbol{)}\<%
\\
\>[8]\AgdaSymbol{...}\AgdaSpace{}%
\AgdaSymbol{|}\AgdaSpace{}%
\AgdaInductiveConstructor{yes}\AgdaSpace{}%
\AgdaSymbol{(}\AgdaBound{x₁}\AgdaSpace{}%
\AgdaOperator{\AgdaInductiveConstructor{,}}\AgdaSpace{}%
\AgdaBound{v}\AgdaSpace{}%
\AgdaOperator{\AgdaInductiveConstructor{,}}\AgdaSpace{}%
\AgdaBound{x₁∈ls}\AgdaSpace{}%
\AgdaOperator{\AgdaInductiveConstructor{,}}\AgdaSpace{}%
\AgdaBound{¬bf}\AgdaSpace{}%
\AgdaOperator{\AgdaInductiveConstructor{,}}\AgdaSpace{}%
\AgdaBound{v∈rs}\AgdaSpace{}%
\AgdaOperator{\AgdaInductiveConstructor{,}}\AgdaSpace{}%
\AgdaBound{v∈ws}\AgdaSymbol{)}\<%
\\
\>[8][@{}l@{\AgdaIndent{0}}]%
\>[10]\AgdaSymbol{=}\AgdaSpace{}%
\AgdaInductiveConstructor{true}\AgdaSpace{}%
\AgdaOperator{\AgdaInductiveConstructor{Relation.Nullary.because}}\AgdaSpace{}%
\AgdaInductiveConstructor{Relation.Nullary.ofʸ}\AgdaSpace{}%
\AgdaSymbol{(}\AgdaBound{x₁}\AgdaSpace{}%
\AgdaOperator{\AgdaInductiveConstructor{,}}\AgdaSpace{}%
\AgdaBound{x}\AgdaSpace{}%
\AgdaOperator{\AgdaInductiveConstructor{,}}\AgdaSpace{}%
\AgdaBound{v}\AgdaSpace{}%
\AgdaOperator{\AgdaInductiveConstructor{,}}\AgdaSpace{}%
\AgdaSymbol{(}\AgdaInductiveConstructor{[]}\AgdaSpace{}%
\AgdaOperator{\AgdaInductiveConstructor{,}}\AgdaSpace{}%
\AgdaBound{ls₁}\AgdaSpace{}%
\AgdaOperator{\AgdaInductiveConstructor{,}}\AgdaSpace{}%
\AgdaInductiveConstructor{refl}\AgdaSpace{}%
\AgdaOperator{\AgdaInductiveConstructor{,}}\AgdaSpace{}%
\AgdaBound{x₁∈ls}\AgdaSymbol{)}\AgdaSpace{}%
\AgdaOperator{\AgdaInductiveConstructor{,}}\AgdaSpace{}%
\AgdaBound{x∈b₂}\AgdaSpace{}%
\AgdaOperator{\AgdaInductiveConstructor{,}}\AgdaSpace{}%
\AgdaBound{¬bf}\AgdaSpace{}%
\AgdaOperator{\AgdaInductiveConstructor{,}}\AgdaSpace{}%
\AgdaBound{v∈rs}\AgdaSpace{}%
\AgdaOperator{\AgdaInductiveConstructor{,}}\AgdaSpace{}%
\AgdaBound{v∈ws}\AgdaSymbol{)}\<%
\\
\>[8]\AgdaSymbol{...}\AgdaSpace{}%
\AgdaSymbol{|}\AgdaSpace{}%
\AgdaInductiveConstructor{no}\AgdaSpace{}%
\AgdaBound{¬p}\AgdaSpace{}%
\AgdaKeyword{with}\AgdaSpace{}%
\AgdaFunction{g₁}\AgdaSpace{}%
\AgdaBound{b₂}\AgdaSpace{}%
\AgdaBound{ls₁}\AgdaSpace{}%
\AgdaBound{ls₂}\<%
\\
\>[8]\AgdaSymbol{...}\AgdaSpace{}%
\AgdaSymbol{|}\AgdaSpace{}%
\AgdaInductiveConstructor{yes}\AgdaSpace{}%
\AgdaSymbol{(}\AgdaBound{x₁}\AgdaSpace{}%
\AgdaOperator{\AgdaInductiveConstructor{,}}\AgdaSpace{}%
\AgdaBound{x₂}\AgdaSpace{}%
\AgdaOperator{\AgdaInductiveConstructor{,}}\AgdaSpace{}%
\AgdaBound{v}\AgdaSpace{}%
\AgdaOperator{\AgdaInductiveConstructor{,}}\AgdaSpace{}%
\AgdaSymbol{(}\AgdaBound{xs}\AgdaSpace{}%
\AgdaOperator{\AgdaInductiveConstructor{,}}\AgdaSpace{}%
\AgdaBound{ys}\AgdaSpace{}%
\AgdaOperator{\AgdaInductiveConstructor{,}}\AgdaSpace{}%
\AgdaBound{≡₁}\AgdaSpace{}%
\AgdaOperator{\AgdaInductiveConstructor{,}}\AgdaSpace{}%
\AgdaBound{∈₁}\AgdaSymbol{)}\AgdaSpace{}%
\AgdaOperator{\AgdaInductiveConstructor{,}}\AgdaSpace{}%
\AgdaBound{rest}\AgdaSymbol{)}\<%
\\
\>[8][@{}l@{\AgdaIndent{0}}]%
\>[10]\AgdaSymbol{=}\AgdaSpace{}%
\AgdaInductiveConstructor{true}\AgdaSpace{}%
\AgdaOperator{\AgdaInductiveConstructor{Relation.Nullary.because}}\AgdaSpace{}%
\AgdaInductiveConstructor{Relation.Nullary.ofʸ}\AgdaSpace{}%
\AgdaSymbol{(}\AgdaBound{x₁}\AgdaSpace{}%
\AgdaOperator{\AgdaInductiveConstructor{,}}\AgdaSpace{}%
\AgdaBound{x₂}\AgdaSpace{}%
\AgdaOperator{\AgdaInductiveConstructor{,}}\AgdaSpace{}%
\AgdaBound{v}\AgdaSpace{}%
\AgdaOperator{\AgdaInductiveConstructor{,}}\AgdaSpace{}%
\AgdaSymbol{(}\AgdaBound{x}\AgdaSpace{}%
\AgdaOperator{\AgdaInductiveConstructor{∷}}\AgdaSpace{}%
\AgdaBound{xs}\AgdaSpace{}%
\AgdaOperator{\AgdaInductiveConstructor{,}}\AgdaSpace{}%
\AgdaBound{ys}\AgdaSpace{}%
\AgdaOperator{\AgdaInductiveConstructor{,}}\AgdaSpace{}%
\AgdaFunction{cong}\AgdaSpace{}%
\AgdaSymbol{(}\AgdaBound{x}\AgdaSpace{}%
\AgdaOperator{\AgdaInductiveConstructor{∷\AgdaUnderscore{}}}\AgdaSymbol{)}\AgdaSpace{}%
\AgdaBound{≡₁}\AgdaSpace{}%
\AgdaOperator{\AgdaInductiveConstructor{,}}\AgdaSpace{}%
\AgdaBound{∈₁}\AgdaSymbol{)}\AgdaSpace{}%
\AgdaOperator{\AgdaInductiveConstructor{,}}\AgdaSpace{}%
\AgdaBound{rest}\AgdaSymbol{)}\<%
\\
\>[8]\AgdaSymbol{...}\AgdaSpace{}%
\AgdaSymbol{|}\AgdaSpace{}%
\AgdaInductiveConstructor{no}\AgdaSpace{}%
\AgdaBound{¬sh}\AgdaSpace{}%
\AgdaSymbol{=}\AgdaSpace{}%
\AgdaInductiveConstructor{false}\AgdaSpace{}%
\AgdaOperator{\AgdaInductiveConstructor{Relation.Nullary.because}}\AgdaSpace{}%
\AgdaInductiveConstructor{Relation.Nullary.ofⁿ}\AgdaSpace{}%
\AgdaFunction{g₂}\<%
\\
\>[8][@{}l@{\AgdaIndent{0}}]%
\>[10]\AgdaKeyword{where}%
\>[2891I]\AgdaFunction{g₂}\AgdaSpace{}%
\AgdaSymbol{:}\AgdaSpace{}%
\AgdaOperator{\AgdaFunction{¬}}\AgdaSpace{}%
\AgdaSymbol{(}\AgdaFunction{∃[}\AgdaSpace{}%
\AgdaBound{x₁}\AgdaSpace{}%
\AgdaFunction{]}\AgdaSymbol{(}\AgdaFunction{∃[}\AgdaSpace{}%
\AgdaBound{x₂}\AgdaSpace{}%
\AgdaFunction{]}\AgdaSymbol{(}\AgdaFunction{∃[}\AgdaSpace{}%
\AgdaBound{v}\AgdaSpace{}%
\AgdaFunction{]}\AgdaSymbol{((}\AgdaBound{x₂}\AgdaSpace{}%
\AgdaOperator{\AgdaFunction{before}}\AgdaSpace{}%
\AgdaBound{x₁}\AgdaSpace{}%
\AgdaOperator{\AgdaFunction{∈}}\AgdaSpace{}%
\AgdaSymbol{(}\AgdaBound{x}\AgdaSpace{}%
\AgdaOperator{\AgdaInductiveConstructor{∷}}\AgdaSpace{}%
\AgdaBound{ls₁}\AgdaSymbol{))}\AgdaSpace{}%
\AgdaOperator{\AgdaFunction{×}}\AgdaSpace{}%
\AgdaBound{x₂}\AgdaSpace{}%
\AgdaOperator{\AgdaFunction{∈}}\AgdaSpace{}%
\AgdaBound{b₂}\AgdaSpace{}%
\AgdaOperator{\AgdaFunction{×}}\AgdaSpace{}%
\AgdaOperator{\AgdaFunction{¬}}\AgdaSpace{}%
\AgdaSymbol{(}\AgdaBound{x₁}\AgdaSpace{}%
\AgdaOperator{\AgdaFunction{before}}\AgdaSpace{}%
\AgdaBound{x₂}\AgdaSpace{}%
\AgdaOperator{\AgdaFunction{∈}}\AgdaSpace{}%
\AgdaBound{b₂}\AgdaSymbol{)}\AgdaSpace{}%
\AgdaOperator{\AgdaFunction{×}}\AgdaSpace{}%
\AgdaBound{v}\AgdaSpace{}%
\AgdaOperator{\AgdaFunction{∈}}\AgdaSpace{}%
\AgdaFunction{cmdRead}\AgdaSpace{}%
\AgdaBound{ls₂}\AgdaSpace{}%
\AgdaBound{x₂}\AgdaSpace{}%
\AgdaOperator{\AgdaFunction{×}}\AgdaSpace{}%
\AgdaBound{v}\AgdaSpace{}%
\AgdaOperator{\AgdaFunction{∈}}\AgdaSpace{}%
\AgdaFunction{cmdWrote}\AgdaSpace{}%
\AgdaBound{ls₂}\AgdaSpace{}%
\AgdaBound{x₁}\AgdaSymbol{))))}\<%
\\
\>[.][@{}l@{}]\<[2891I]%
\>[16]\AgdaFunction{g₂}\AgdaSpace{}%
\AgdaSymbol{(}\AgdaBound{x₁}\AgdaSpace{}%
\AgdaOperator{\AgdaInductiveConstructor{,}}\AgdaSpace{}%
\AgdaBound{x₂}\AgdaSpace{}%
\AgdaOperator{\AgdaInductiveConstructor{,}}\AgdaSpace{}%
\AgdaBound{v}\AgdaSpace{}%
\AgdaOperator{\AgdaInductiveConstructor{,}}\AgdaSpace{}%
\AgdaSymbol{(}\AgdaInductiveConstructor{[]}\AgdaSpace{}%
\AgdaOperator{\AgdaInductiveConstructor{,}}\AgdaSpace{}%
\AgdaBound{ys}\AgdaSpace{}%
\AgdaOperator{\AgdaInductiveConstructor{,}}\AgdaSpace{}%
\AgdaBound{≡₁}\AgdaSpace{}%
\AgdaOperator{\AgdaInductiveConstructor{,}}\AgdaSpace{}%
\AgdaBound{∈₁}\AgdaSymbol{)}\AgdaSpace{}%
\AgdaOperator{\AgdaInductiveConstructor{,}}\AgdaSpace{}%
\AgdaBound{x₂∈b₂}\AgdaSpace{}%
\AgdaOperator{\AgdaInductiveConstructor{,}}\AgdaSpace{}%
\AgdaBound{¬bf}\AgdaSpace{}%
\AgdaOperator{\AgdaInductiveConstructor{,}}\AgdaSpace{}%
\AgdaBound{v∈rs}\AgdaSpace{}%
\AgdaOperator{\AgdaInductiveConstructor{,}}\AgdaSpace{}%
\AgdaBound{v∈ws}\AgdaSymbol{)}\<%
\\
\>[16][@{}l@{\AgdaIndent{0}}]%
\>[18]\AgdaSymbol{=}\AgdaSpace{}%
\AgdaBound{¬p}%
\>[2952I]\AgdaSymbol{(}\AgdaBound{x₁}\AgdaSpace{}%
\AgdaOperator{\AgdaInductiveConstructor{,}}\AgdaSpace{}%
\AgdaBound{v}\AgdaSpace{}%
\AgdaOperator{\AgdaInductiveConstructor{,}}\AgdaSpace{}%
\AgdaFunction{subst}\AgdaSpace{}%
\AgdaSymbol{(λ}\AgdaSpace{}%
\AgdaBound{x₃}\AgdaSpace{}%
\AgdaSymbol{→}\AgdaSpace{}%
\AgdaBound{x₁}\AgdaSpace{}%
\AgdaOperator{\AgdaFunction{∈}}\AgdaSpace{}%
\AgdaBound{x₃}\AgdaSymbol{)}\AgdaSpace{}%
\AgdaSymbol{(}\AgdaFunction{sym}\AgdaSpace{}%
\AgdaSymbol{(}\AgdaFunction{∷-injectiveʳ}\AgdaSpace{}%
\AgdaBound{≡₁}\AgdaSymbol{))}\AgdaSpace{}%
\AgdaBound{∈₁}\AgdaSpace{}%
\AgdaOperator{\AgdaInductiveConstructor{,}}\AgdaSpace{}%
\AgdaFunction{subst}\AgdaSpace{}%
\AgdaSymbol{(λ}\AgdaSpace{}%
\AgdaBound{x₃}\AgdaSpace{}%
\AgdaSymbol{→}\AgdaSpace{}%
\AgdaOperator{\AgdaFunction{¬}}\AgdaSpace{}%
\AgdaSymbol{(}\AgdaBound{x₁}\AgdaSpace{}%
\AgdaOperator{\AgdaFunction{before}}\AgdaSpace{}%
\AgdaBound{x₃}\AgdaSpace{}%
\AgdaOperator{\AgdaFunction{∈}}\AgdaSpace{}%
\AgdaBound{b₂}\AgdaSymbol{))}\AgdaSpace{}%
\AgdaSymbol{(}\AgdaFunction{sym}\AgdaSpace{}%
\AgdaSymbol{(}\AgdaFunction{∷-injectiveˡ}\AgdaSpace{}%
\AgdaBound{≡₁}\AgdaSymbol{))}\AgdaSpace{}%
\AgdaBound{¬bf}\<%
\\
\>[.][@{}l@{}]\<[2952I]%
\>[23]\AgdaOperator{\AgdaInductiveConstructor{,}}\AgdaSpace{}%
\AgdaFunction{subst}\AgdaSpace{}%
\AgdaSymbol{(λ}\AgdaSpace{}%
\AgdaBound{x₃}\AgdaSpace{}%
\AgdaSymbol{→}\AgdaSpace{}%
\AgdaBound{v}\AgdaSpace{}%
\AgdaOperator{\AgdaFunction{∈}}\AgdaSpace{}%
\AgdaFunction{cmdRead}\AgdaSpace{}%
\AgdaBound{ls₂}\AgdaSpace{}%
\AgdaBound{x₃}\AgdaSymbol{)}\AgdaSpace{}%
\AgdaSymbol{(}\AgdaFunction{sym}\AgdaSpace{}%
\AgdaSymbol{(}\AgdaFunction{∷-injectiveˡ}\AgdaSpace{}%
\AgdaBound{≡₁}\AgdaSymbol{))}\AgdaSpace{}%
\AgdaBound{v∈rs}\AgdaSpace{}%
\AgdaOperator{\AgdaInductiveConstructor{,}}\AgdaSpace{}%
\AgdaBound{v∈ws}\AgdaSymbol{)}\<%
\\
\>[16]\AgdaFunction{g₂}\AgdaSpace{}%
\AgdaSymbol{(}\AgdaBound{x₁}\AgdaSpace{}%
\AgdaOperator{\AgdaInductiveConstructor{,}}\AgdaSpace{}%
\AgdaBound{x₂}\AgdaSpace{}%
\AgdaOperator{\AgdaInductiveConstructor{,}}\AgdaSpace{}%
\AgdaBound{v}\AgdaSpace{}%
\AgdaOperator{\AgdaInductiveConstructor{,}}\AgdaSpace{}%
\AgdaSymbol{(}\AgdaBound{x₃}\AgdaSpace{}%
\AgdaOperator{\AgdaInductiveConstructor{∷}}\AgdaSpace{}%
\AgdaBound{xs}\AgdaSpace{}%
\AgdaOperator{\AgdaInductiveConstructor{,}}\AgdaSpace{}%
\AgdaBound{ys}\AgdaSpace{}%
\AgdaOperator{\AgdaInductiveConstructor{,}}\AgdaSpace{}%
\AgdaBound{≡₁}\AgdaSpace{}%
\AgdaOperator{\AgdaInductiveConstructor{,}}\AgdaSpace{}%
\AgdaBound{∈₁}\AgdaSymbol{)}\AgdaSpace{}%
\AgdaOperator{\AgdaInductiveConstructor{,}}\AgdaSpace{}%
\AgdaBound{rest}\AgdaSymbol{)}\AgdaSpace{}%
\AgdaSymbol{=}\AgdaSpace{}%
\AgdaBound{¬sh}\AgdaSpace{}%
\AgdaSymbol{(}\AgdaBound{x₁}\AgdaSpace{}%
\AgdaOperator{\AgdaInductiveConstructor{,}}\AgdaSpace{}%
\AgdaBound{x₂}\AgdaSpace{}%
\AgdaOperator{\AgdaInductiveConstructor{,}}\AgdaSpace{}%
\AgdaBound{v}\AgdaSpace{}%
\AgdaOperator{\AgdaInductiveConstructor{,}}\AgdaSpace{}%
\AgdaSymbol{(}\AgdaBound{xs}\AgdaSpace{}%
\AgdaOperator{\AgdaInductiveConstructor{,}}\AgdaSpace{}%
\AgdaBound{ys}\AgdaSpace{}%
\AgdaOperator{\AgdaInductiveConstructor{,}}\AgdaSpace{}%
\AgdaSymbol{(}\AgdaFunction{∷-injectiveʳ}\AgdaSpace{}%
\AgdaBound{≡₁}\AgdaSymbol{)}\AgdaSpace{}%
\AgdaOperator{\AgdaInductiveConstructor{,}}\AgdaSpace{}%
\AgdaBound{∈₁}\AgdaSymbol{)}\AgdaSpace{}%
\AgdaOperator{\AgdaInductiveConstructor{,}}\AgdaSpace{}%
\AgdaBound{rest}\AgdaSymbol{)}\<%
\\
\>[8]\AgdaFunction{g₁}\AgdaSpace{}%
\AgdaBound{b₂}\AgdaSpace{}%
\AgdaSymbol{(}\AgdaBound{x}\AgdaSpace{}%
\AgdaOperator{\AgdaInductiveConstructor{∷}}\AgdaSpace{}%
\AgdaBound{ls₁}\AgdaSymbol{)}\AgdaSpace{}%
\AgdaBound{ls₂}\AgdaSpace{}%
\AgdaSymbol{|}\AgdaSpace{}%
\AgdaInductiveConstructor{no}\AgdaSpace{}%
\AgdaBound{x∉b₂}\AgdaSpace{}%
\AgdaKeyword{with}\AgdaSpace{}%
\AgdaFunction{g₁}\AgdaSpace{}%
\AgdaBound{b₂}\AgdaSpace{}%
\AgdaBound{ls₁}\AgdaSpace{}%
\AgdaBound{ls₂}\<%
\\
\>[8]\AgdaSymbol{...}\AgdaSpace{}%
\AgdaSymbol{|}\AgdaSpace{}%
\AgdaInductiveConstructor{yes}\AgdaSpace{}%
\AgdaSymbol{(}\AgdaBound{x₁}\AgdaSpace{}%
\AgdaOperator{\AgdaInductiveConstructor{,}}\AgdaSpace{}%
\AgdaBound{x₂}\AgdaSpace{}%
\AgdaOperator{\AgdaInductiveConstructor{,}}\AgdaSpace{}%
\AgdaBound{v}\AgdaSpace{}%
\AgdaOperator{\AgdaInductiveConstructor{,}}\AgdaSpace{}%
\AgdaSymbol{(}\AgdaBound{xs}\AgdaSpace{}%
\AgdaOperator{\AgdaInductiveConstructor{,}}\AgdaSpace{}%
\AgdaBound{ys}\AgdaSpace{}%
\AgdaOperator{\AgdaInductiveConstructor{,}}\AgdaSpace{}%
\AgdaBound{≡₁}\AgdaSpace{}%
\AgdaOperator{\AgdaInductiveConstructor{,}}\AgdaSpace{}%
\AgdaBound{∈₁}\AgdaSymbol{)}\AgdaSpace{}%
\AgdaOperator{\AgdaInductiveConstructor{,}}\AgdaSpace{}%
\AgdaBound{rest}\AgdaSymbol{)}\<%
\\
\>[8][@{}l@{\AgdaIndent{0}}]%
\>[10]\AgdaSymbol{=}\AgdaSpace{}%
\AgdaInductiveConstructor{true}\AgdaSpace{}%
\AgdaOperator{\AgdaInductiveConstructor{Relation.Nullary.because}}\AgdaSpace{}%
\AgdaInductiveConstructor{Relation.Nullary.ofʸ}\AgdaSpace{}%
\AgdaSymbol{(}\AgdaBound{x₁}\AgdaSpace{}%
\AgdaOperator{\AgdaInductiveConstructor{,}}\AgdaSpace{}%
\AgdaBound{x₂}\AgdaSpace{}%
\AgdaOperator{\AgdaInductiveConstructor{,}}\AgdaSpace{}%
\AgdaBound{v}\AgdaSpace{}%
\AgdaOperator{\AgdaInductiveConstructor{,}}\AgdaSpace{}%
\AgdaSymbol{(}\AgdaBound{x}\AgdaSpace{}%
\AgdaOperator{\AgdaInductiveConstructor{∷}}\AgdaSpace{}%
\AgdaBound{xs}\AgdaSpace{}%
\AgdaOperator{\AgdaInductiveConstructor{,}}\AgdaSpace{}%
\AgdaBound{ys}\AgdaSpace{}%
\AgdaOperator{\AgdaInductiveConstructor{,}}\AgdaSpace{}%
\AgdaFunction{cong}\AgdaSpace{}%
\AgdaSymbol{(}\AgdaBound{x}\AgdaSpace{}%
\AgdaOperator{\AgdaInductiveConstructor{∷\AgdaUnderscore{}}}\AgdaSymbol{)}\AgdaSpace{}%
\AgdaBound{≡₁}\AgdaSpace{}%
\AgdaOperator{\AgdaInductiveConstructor{,}}\AgdaSpace{}%
\AgdaBound{∈₁}\AgdaSymbol{)}\AgdaSpace{}%
\AgdaOperator{\AgdaInductiveConstructor{,}}\AgdaSpace{}%
\AgdaBound{rest}\AgdaSymbol{)}\<%
\\
\>[8]\AgdaSymbol{...}\AgdaSpace{}%
\AgdaSymbol{|}\AgdaSpace{}%
\AgdaInductiveConstructor{no}\AgdaSpace{}%
\AgdaBound{¬sh}\AgdaSpace{}%
\AgdaSymbol{=}\AgdaSpace{}%
\AgdaInductiveConstructor{false}\AgdaSpace{}%
\AgdaOperator{\AgdaInductiveConstructor{Relation.Nullary.because}}\AgdaSpace{}%
\AgdaInductiveConstructor{Relation.Nullary.ofⁿ}\AgdaSpace{}%
\AgdaFunction{g₂}\<%
\\
\>[8][@{}l@{\AgdaIndent{0}}]%
\>[10]\AgdaKeyword{where}%
\>[3093I]\AgdaFunction{g₂}\AgdaSpace{}%
\AgdaSymbol{:}\AgdaSpace{}%
\AgdaOperator{\AgdaFunction{¬}}\AgdaSpace{}%
\AgdaSymbol{(}\AgdaFunction{∃[}\AgdaSpace{}%
\AgdaBound{x₁}\AgdaSpace{}%
\AgdaFunction{]}\AgdaSymbol{(}\AgdaFunction{∃[}\AgdaSpace{}%
\AgdaBound{x₂}\AgdaSpace{}%
\AgdaFunction{]}\AgdaSymbol{(}\AgdaFunction{∃[}\AgdaSpace{}%
\AgdaBound{v}\AgdaSpace{}%
\AgdaFunction{]}\AgdaSymbol{((}\AgdaBound{x₂}\AgdaSpace{}%
\AgdaOperator{\AgdaFunction{before}}\AgdaSpace{}%
\AgdaBound{x₁}\AgdaSpace{}%
\AgdaOperator{\AgdaFunction{∈}}\AgdaSpace{}%
\AgdaSymbol{(}\AgdaBound{x}\AgdaSpace{}%
\AgdaOperator{\AgdaInductiveConstructor{∷}}\AgdaSpace{}%
\AgdaBound{ls₁}\AgdaSymbol{))}\AgdaSpace{}%
\AgdaOperator{\AgdaFunction{×}}\AgdaSpace{}%
\AgdaBound{x₂}\AgdaSpace{}%
\AgdaOperator{\AgdaFunction{∈}}\AgdaSpace{}%
\AgdaBound{b₂}\AgdaSpace{}%
\AgdaOperator{\AgdaFunction{×}}\AgdaSpace{}%
\AgdaOperator{\AgdaFunction{¬}}\AgdaSpace{}%
\AgdaSymbol{(}\AgdaBound{x₁}\AgdaSpace{}%
\AgdaOperator{\AgdaFunction{before}}\AgdaSpace{}%
\AgdaBound{x₂}\AgdaSpace{}%
\AgdaOperator{\AgdaFunction{∈}}\AgdaSpace{}%
\AgdaBound{b₂}\AgdaSymbol{)}\AgdaSpace{}%
\AgdaOperator{\AgdaFunction{×}}\AgdaSpace{}%
\AgdaBound{v}\AgdaSpace{}%
\AgdaOperator{\AgdaFunction{∈}}\AgdaSpace{}%
\AgdaFunction{cmdRead}\AgdaSpace{}%
\AgdaBound{ls₂}\AgdaSpace{}%
\AgdaBound{x₂}\AgdaSpace{}%
\AgdaOperator{\AgdaFunction{×}}\AgdaSpace{}%
\AgdaBound{v}\AgdaSpace{}%
\AgdaOperator{\AgdaFunction{∈}}\AgdaSpace{}%
\AgdaFunction{cmdWrote}\AgdaSpace{}%
\AgdaBound{ls₂}\AgdaSpace{}%
\AgdaBound{x₁}\AgdaSymbol{))))}\<%
\\
\>[.][@{}l@{}]\<[3093I]%
\>[16]\AgdaFunction{g₂}\AgdaSpace{}%
\AgdaSymbol{(}\AgdaBound{x₁}\AgdaSpace{}%
\AgdaOperator{\AgdaInductiveConstructor{,}}\AgdaSpace{}%
\AgdaBound{x₂}\AgdaSpace{}%
\AgdaOperator{\AgdaInductiveConstructor{,}}\AgdaSpace{}%
\AgdaBound{v}\AgdaSpace{}%
\AgdaOperator{\AgdaInductiveConstructor{,}}\AgdaSpace{}%
\AgdaSymbol{(}\AgdaInductiveConstructor{[]}\AgdaSpace{}%
\AgdaOperator{\AgdaInductiveConstructor{,}}\AgdaSpace{}%
\AgdaBound{ys}\AgdaSpace{}%
\AgdaOperator{\AgdaInductiveConstructor{,}}\AgdaSpace{}%
\AgdaBound{≡₁}\AgdaSpace{}%
\AgdaOperator{\AgdaInductiveConstructor{,}}\AgdaSpace{}%
\AgdaBound{∈₁}\AgdaSymbol{)}\AgdaSpace{}%
\AgdaOperator{\AgdaInductiveConstructor{,}}\AgdaSpace{}%
\AgdaBound{x₂∈b₂}\AgdaSpace{}%
\AgdaOperator{\AgdaInductiveConstructor{,}}\AgdaSpace{}%
\AgdaBound{¬bf}\AgdaSpace{}%
\AgdaOperator{\AgdaInductiveConstructor{,}}\AgdaSpace{}%
\AgdaBound{v∈rs}\AgdaSpace{}%
\AgdaOperator{\AgdaInductiveConstructor{,}}\AgdaSpace{}%
\AgdaBound{v∈ws}\AgdaSymbol{)}\<%
\\
\>[16][@{}l@{\AgdaIndent{0}}]%
\>[18]\AgdaSymbol{=}\AgdaSpace{}%
\AgdaFunction{contradiction}\AgdaSpace{}%
\AgdaSymbol{(}\AgdaFunction{subst}\AgdaSpace{}%
\AgdaSymbol{(λ}\AgdaSpace{}%
\AgdaBound{x₃}\AgdaSpace{}%
\AgdaSymbol{→}\AgdaSpace{}%
\AgdaBound{x₃}\AgdaSpace{}%
\AgdaOperator{\AgdaFunction{∈}}\AgdaSpace{}%
\AgdaBound{b₂}\AgdaSymbol{)}\AgdaSpace{}%
\AgdaSymbol{(}\AgdaFunction{sym}\AgdaSpace{}%
\AgdaSymbol{(}\AgdaFunction{∷-injectiveˡ}\AgdaSpace{}%
\AgdaBound{≡₁}\AgdaSymbol{))}\AgdaSpace{}%
\AgdaBound{x₂∈b₂}\AgdaSymbol{)}\AgdaSpace{}%
\AgdaBound{x∉b₂}\<%
\\
\>[16]\AgdaFunction{g₂}\AgdaSpace{}%
\AgdaSymbol{(}\AgdaBound{x₁}\AgdaSpace{}%
\AgdaOperator{\AgdaInductiveConstructor{,}}\AgdaSpace{}%
\AgdaBound{x₂}\AgdaSpace{}%
\AgdaOperator{\AgdaInductiveConstructor{,}}\AgdaSpace{}%
\AgdaBound{v}\AgdaSpace{}%
\AgdaOperator{\AgdaInductiveConstructor{,}}\AgdaSpace{}%
\AgdaSymbol{(}\AgdaBound{x₃}\AgdaSpace{}%
\AgdaOperator{\AgdaInductiveConstructor{∷}}\AgdaSpace{}%
\AgdaBound{xs}\AgdaSpace{}%
\AgdaOperator{\AgdaInductiveConstructor{,}}\AgdaSpace{}%
\AgdaBound{ys}\AgdaSpace{}%
\AgdaOperator{\AgdaInductiveConstructor{,}}\AgdaSpace{}%
\AgdaBound{≡₁}\AgdaSpace{}%
\AgdaOperator{\AgdaInductiveConstructor{,}}\AgdaSpace{}%
\AgdaBound{∈₁}\AgdaSymbol{)}\AgdaSpace{}%
\AgdaOperator{\AgdaInductiveConstructor{,}}\AgdaSpace{}%
\AgdaBound{rest}\AgdaSymbol{)}\AgdaSpace{}%
\AgdaSymbol{=}\AgdaSpace{}%
\AgdaBound{¬sh}\AgdaSpace{}%
\AgdaSymbol{(}\AgdaBound{x₁}\AgdaSpace{}%
\AgdaOperator{\AgdaInductiveConstructor{,}}\AgdaSpace{}%
\AgdaBound{x₂}\AgdaSpace{}%
\AgdaOperator{\AgdaInductiveConstructor{,}}\AgdaSpace{}%
\AgdaBound{v}\AgdaSpace{}%
\AgdaOperator{\AgdaInductiveConstructor{,}}\AgdaSpace{}%
\AgdaSymbol{(}\AgdaBound{xs}\AgdaSpace{}%
\AgdaOperator{\AgdaInductiveConstructor{,}}\AgdaSpace{}%
\AgdaBound{ys}\AgdaSpace{}%
\AgdaOperator{\AgdaInductiveConstructor{,}}\AgdaSpace{}%
\AgdaSymbol{(}\AgdaFunction{∷-injectiveʳ}\AgdaSpace{}%
\AgdaBound{≡₁}\AgdaSymbol{)}\AgdaSpace{}%
\AgdaOperator{\AgdaInductiveConstructor{,}}\AgdaSpace{}%
\AgdaBound{∈₁}\AgdaSymbol{)}\AgdaSpace{}%
\AgdaOperator{\AgdaInductiveConstructor{,}}\AgdaSpace{}%
\AgdaBound{rest}\AgdaSymbol{)}\<%
\\
\\[\AgdaEmptyExtraSkip]%
\>[0]\AgdaFunction{hazard?}\AgdaSpace{}%
\AgdaSymbol{:}\AgdaSpace{}%
\AgdaSymbol{∀}\AgdaSpace{}%
\AgdaBound{s}\AgdaSpace{}%
\AgdaBound{x}\AgdaSpace{}%
\AgdaBound{b₂}\AgdaSpace{}%
\AgdaBound{ls}\AgdaSpace{}%
\AgdaSymbol{→}\AgdaSpace{}%
\AgdaRecord{Dec}\AgdaSpace{}%
\AgdaSymbol{(}\AgdaDatatype{Hazard}\AgdaSpace{}%
\AgdaBound{s}\AgdaSpace{}%
\AgdaBound{x}\AgdaSpace{}%
\AgdaBound{b₂}\AgdaSpace{}%
\AgdaBound{ls}\AgdaSymbol{)}\<%
\\
\>[0]\AgdaFunction{hazard?}\AgdaSpace{}%
\AgdaBound{s}\AgdaSpace{}%
\AgdaBound{x}\AgdaSpace{}%
\AgdaBound{b₂}\AgdaSpace{}%
\AgdaBound{ls}\AgdaSpace{}%
\AgdaKeyword{with}\AgdaSpace{}%
\AgdaFunction{intersection?2}\AgdaSpace{}%
\AgdaSymbol{(}\AgdaPostulate{cmdWriteNames}\AgdaSpace{}%
\AgdaBound{x}\AgdaSpace{}%
\AgdaBound{s}\AgdaSymbol{)}\AgdaSpace{}%
\AgdaSymbol{(}\AgdaFunction{filesRead}\AgdaSpace{}%
\AgdaBound{ls}\AgdaSymbol{)}\<%
\\
\>[0]\AgdaSymbol{...}\AgdaSpace{}%
\AgdaSymbol{|}\AgdaSpace{}%
\AgdaInductiveConstructor{yes}\AgdaSpace{}%
\AgdaSymbol{(}\AgdaBound{v}\AgdaSpace{}%
\AgdaOperator{\AgdaInductiveConstructor{,}}\AgdaSpace{}%
\AgdaBound{∈₁}\AgdaSpace{}%
\AgdaOperator{\AgdaInductiveConstructor{,}}\AgdaSpace{}%
\AgdaBound{∈₂}\AgdaSymbol{)}\AgdaSpace{}%
\AgdaSymbol{=}\AgdaSpace{}%
\AgdaInductiveConstructor{true}\AgdaSpace{}%
\AgdaOperator{\AgdaInductiveConstructor{Relation.Nullary.because}}\AgdaSpace{}%
\AgdaInductiveConstructor{Relation.Nullary.ofʸ}\AgdaSpace{}%
\AgdaSymbol{(}\AgdaInductiveConstructor{ReadWrite}\AgdaSpace{}%
\AgdaBound{∈₁}\AgdaSpace{}%
\AgdaBound{∈₂}\AgdaSymbol{)}\<%
\\
\>[0]\AgdaSymbol{...}\AgdaSpace{}%
\AgdaSymbol{|}\AgdaSpace{}%
\AgdaInductiveConstructor{no}\AgdaSpace{}%
\AgdaBound{¬x₁}\AgdaSpace{}%
\AgdaKeyword{with}\AgdaSpace{}%
\AgdaFunction{intersection?2}\AgdaSpace{}%
\AgdaSymbol{(}\AgdaPostulate{cmdWriteNames}\AgdaSpace{}%
\AgdaBound{x}\AgdaSpace{}%
\AgdaBound{s}\AgdaSymbol{)}\AgdaSpace{}%
\AgdaSymbol{(}\AgdaFunction{filesWrote}\AgdaSpace{}%
\AgdaBound{ls}\AgdaSymbol{)}\<%
\\
\>[0]\AgdaSymbol{...}\AgdaSpace{}%
\AgdaSymbol{|}\AgdaSpace{}%
\AgdaInductiveConstructor{yes}\AgdaSpace{}%
\AgdaSymbol{(}\AgdaBound{v}\AgdaSpace{}%
\AgdaOperator{\AgdaInductiveConstructor{,}}\AgdaSpace{}%
\AgdaBound{∈₁}\AgdaSpace{}%
\AgdaOperator{\AgdaInductiveConstructor{,}}\AgdaSpace{}%
\AgdaBound{∈₂}\AgdaSymbol{)}\AgdaSpace{}%
\AgdaSymbol{=}\AgdaSpace{}%
\AgdaInductiveConstructor{true}\AgdaSpace{}%
\AgdaOperator{\AgdaInductiveConstructor{Relation.Nullary.because}}\AgdaSpace{}%
\AgdaInductiveConstructor{Relation.Nullary.ofʸ}\AgdaSpace{}%
\AgdaSymbol{(}\AgdaInductiveConstructor{WriteWrite}\AgdaSpace{}%
\AgdaBound{∈₁}\AgdaSpace{}%
\AgdaBound{∈₂}\AgdaSymbol{)}\<%
\\
\>[0]\AgdaSymbol{...}\AgdaSpace{}%
\AgdaSymbol{|}\AgdaSpace{}%
\AgdaInductiveConstructor{no}\AgdaSpace{}%
\AgdaBound{¬x₂}\AgdaSpace{}%
\AgdaKeyword{with}\AgdaSpace{}%
\AgdaFunction{¬speculative?}\AgdaSpace{}%
\AgdaBound{b₂}\AgdaSpace{}%
\AgdaSymbol{(}\AgdaFunction{save}\AgdaSpace{}%
\AgdaBound{s}\AgdaSpace{}%
\AgdaBound{x}\AgdaSpace{}%
\AgdaBound{ls}\AgdaSymbol{)}\<%
\\
\>[0]\AgdaSymbol{...}\AgdaSpace{}%
\AgdaSymbol{|}\AgdaSpace{}%
\AgdaInductiveConstructor{yes}\AgdaSpace{}%
\AgdaSymbol{(}\AgdaBound{x₁}\AgdaSpace{}%
\AgdaOperator{\AgdaInductiveConstructor{,}}\AgdaSpace{}%
\AgdaBound{x₂}\AgdaSpace{}%
\AgdaOperator{\AgdaInductiveConstructor{,}}\AgdaSpace{}%
\AgdaBound{v}\AgdaSpace{}%
\AgdaOperator{\AgdaInductiveConstructor{,}}\AgdaSpace{}%
\AgdaBound{bf}\AgdaSpace{}%
\AgdaOperator{\AgdaInductiveConstructor{,}}\AgdaSpace{}%
\AgdaBound{x₂∈b₂}\AgdaSpace{}%
\AgdaOperator{\AgdaInductiveConstructor{,}}\AgdaSpace{}%
\AgdaBound{¬bf}\AgdaSpace{}%
\AgdaOperator{\AgdaInductiveConstructor{,}}\AgdaSpace{}%
\AgdaBound{v∈₁}\AgdaSpace{}%
\AgdaOperator{\AgdaInductiveConstructor{,}}\AgdaSpace{}%
\AgdaBound{v∈₂}\AgdaSymbol{)}\<%
\\
\>[0][@{}l@{\AgdaIndent{0}}]%
\>[2]\AgdaSymbol{=}\AgdaSpace{}%
\AgdaInductiveConstructor{true}\AgdaSpace{}%
\AgdaOperator{\AgdaInductiveConstructor{Relation.Nullary.because}}\AgdaSpace{}%
\AgdaInductiveConstructor{Relation.Nullary.ofʸ}\AgdaSpace{}%
\AgdaSymbol{(}\AgdaInductiveConstructor{Speculative}\AgdaSpace{}%
\AgdaBound{x₁}\AgdaSpace{}%
\AgdaBound{x₂}\AgdaSpace{}%
\AgdaBound{bf}\AgdaSpace{}%
\AgdaBound{x₂∈b₂}\AgdaSpace{}%
\AgdaBound{¬bf}\AgdaSpace{}%
\AgdaBound{v∈₁}\AgdaSpace{}%
\AgdaBound{v∈₂}\AgdaSymbol{)}\<%
\\
\>[0]\AgdaSymbol{...}\AgdaSpace{}%
\AgdaSymbol{|}\AgdaSpace{}%
\AgdaInductiveConstructor{no}\AgdaSpace{}%
\AgdaBound{¬sh}\AgdaSpace{}%
\AgdaSymbol{=}\AgdaSpace{}%
\AgdaInductiveConstructor{false}\AgdaSpace{}%
\AgdaOperator{\AgdaInductiveConstructor{Relation.Nullary.because}}\AgdaSpace{}%
\AgdaInductiveConstructor{Relation.Nullary.ofⁿ}\AgdaSpace{}%
\AgdaFunction{g₁}\<%
\\
\>[0][@{}l@{\AgdaIndent{0}}]%
\>[2]\AgdaKeyword{where}%
\>[3309I]\AgdaFunction{g₁}\AgdaSpace{}%
\AgdaSymbol{:}\AgdaSpace{}%
\AgdaOperator{\AgdaFunction{¬}}\AgdaSpace{}%
\AgdaDatatype{Hazard}\AgdaSpace{}%
\AgdaBound{s}\AgdaSpace{}%
\AgdaBound{x}\AgdaSpace{}%
\AgdaBound{b₂}\AgdaSpace{}%
\AgdaBound{ls}\<%
\\
\>[.][@{}l@{}]\<[3309I]%
\>[8]\AgdaFunction{g₁}\AgdaSpace{}%
\AgdaSymbol{(}\AgdaInductiveConstructor{ReadWrite}\AgdaSpace{}%
\AgdaBound{x}\AgdaSpace{}%
\AgdaBound{x₁}\AgdaSymbol{)}\AgdaSpace{}%
\AgdaSymbol{=}\AgdaSpace{}%
\AgdaBound{¬x₁}\AgdaSpace{}%
\AgdaSymbol{(\AgdaUnderscore{}}\AgdaSpace{}%
\AgdaOperator{\AgdaInductiveConstructor{,}}\AgdaSpace{}%
\AgdaSymbol{(}\AgdaBound{x}\AgdaSpace{}%
\AgdaOperator{\AgdaInductiveConstructor{,}}\AgdaSpace{}%
\AgdaBound{x₁}\AgdaSymbol{))}\<%
\\
\>[8]\AgdaFunction{g₁}\AgdaSpace{}%
\AgdaSymbol{(}\AgdaInductiveConstructor{WriteWrite}\AgdaSpace{}%
\AgdaBound{x}\AgdaSpace{}%
\AgdaBound{x₁}\AgdaSymbol{)}\AgdaSpace{}%
\AgdaSymbol{=}\AgdaSpace{}%
\AgdaBound{¬x₂}\AgdaSpace{}%
\AgdaSymbol{(\AgdaUnderscore{}}\AgdaSpace{}%
\AgdaOperator{\AgdaInductiveConstructor{,}}\AgdaSpace{}%
\AgdaSymbol{(}\AgdaBound{x}\AgdaSpace{}%
\AgdaOperator{\AgdaInductiveConstructor{,}}\AgdaSpace{}%
\AgdaBound{x₁}\AgdaSymbol{))}\<%
\\
\>[8]\AgdaFunction{g₁}\AgdaSpace{}%
\AgdaSymbol{(}\AgdaInductiveConstructor{Speculative}\AgdaSpace{}%
\AgdaBound{x₁}\AgdaSpace{}%
\AgdaBound{x₂}\AgdaSpace{}%
\AgdaBound{y}\AgdaSpace{}%
\AgdaBound{x₃}\AgdaSpace{}%
\AgdaBound{x₄}\AgdaSpace{}%
\AgdaBound{x₅}\AgdaSpace{}%
\AgdaBound{x₆}\AgdaSymbol{)}\<%
\\
\>[8][@{}l@{\AgdaIndent{0}}]%
\>[10]\AgdaSymbol{=}\AgdaSpace{}%
\AgdaBound{¬sh}\AgdaSpace{}%
\AgdaSymbol{(}\AgdaBound{x₁}\AgdaSpace{}%
\AgdaOperator{\AgdaInductiveConstructor{,}}\AgdaSpace{}%
\AgdaBound{x₂}\AgdaSpace{}%
\AgdaOperator{\AgdaInductiveConstructor{,}}\AgdaSpace{}%
\AgdaSymbol{\AgdaUnderscore{}}\AgdaSpace{}%
\AgdaOperator{\AgdaInductiveConstructor{,}}\AgdaSpace{}%
\AgdaBound{y}\AgdaSpace{}%
\AgdaOperator{\AgdaInductiveConstructor{,}}\AgdaSpace{}%
\AgdaBound{x₃}\AgdaSpace{}%
\AgdaOperator{\AgdaInductiveConstructor{,}}\AgdaSpace{}%
\AgdaBound{x₄}\AgdaSpace{}%
\AgdaOperator{\AgdaInductiveConstructor{,}}\AgdaSpace{}%
\AgdaBound{x₅}\AgdaSpace{}%
\AgdaOperator{\AgdaInductiveConstructor{,}}\AgdaSpace{}%
\AgdaBound{x₆}\AgdaSymbol{)}\<%
\\
\\[\AgdaEmptyExtraSkip]%
\>[0]\AgdaFunction{hazardfree?}\AgdaSpace{}%
\AgdaSymbol{:}\AgdaSpace{}%
\AgdaSymbol{∀}\AgdaSpace{}%
\AgdaBound{s}\AgdaSpace{}%
\AgdaBound{b₁}\AgdaSpace{}%
\AgdaBound{b₂}\AgdaSpace{}%
\AgdaBound{ls}\AgdaSpace{}%
\AgdaSymbol{→}\AgdaSpace{}%
\AgdaRecord{Dec}\AgdaSpace{}%
\AgdaSymbol{(}\AgdaDatatype{HazardFree}\AgdaSpace{}%
\AgdaBound{s}\AgdaSpace{}%
\AgdaBound{b₁}\AgdaSpace{}%
\AgdaBound{b₂}\AgdaSpace{}%
\AgdaBound{ls}\AgdaSymbol{)}\<%
\\
\>[0]\AgdaFunction{hazardfree?}\AgdaSpace{}%
\AgdaBound{s}\AgdaSpace{}%
\AgdaInductiveConstructor{[]}\AgdaSpace{}%
\AgdaBound{b₂}\AgdaSpace{}%
\AgdaBound{ls}\AgdaSpace{}%
\AgdaSymbol{=}\AgdaSpace{}%
\AgdaInductiveConstructor{true}\AgdaSpace{}%
\AgdaOperator{\AgdaInductiveConstructor{Relation.Nullary.because}}\AgdaSpace{}%
\AgdaInductiveConstructor{Relation.Nullary.ofʸ}\AgdaSpace{}%
\AgdaInductiveConstructor{[]}\<%
\\
\>[0]\AgdaFunction{hazardfree?}\AgdaSpace{}%
\AgdaBound{s}\AgdaSpace{}%
\AgdaSymbol{(}\AgdaBound{x}\AgdaSpace{}%
\AgdaOperator{\AgdaInductiveConstructor{∷}}\AgdaSpace{}%
\AgdaBound{b₁}\AgdaSymbol{)}\AgdaSpace{}%
\AgdaBound{b₂}\AgdaSpace{}%
\AgdaBound{ls}\AgdaSpace{}%
\AgdaKeyword{with}\AgdaSpace{}%
\AgdaFunction{hazard?}\AgdaSpace{}%
\AgdaBound{s}\AgdaSpace{}%
\AgdaBound{x}\AgdaSpace{}%
\AgdaBound{b₂}\AgdaSpace{}%
\AgdaBound{ls}\<%
\\
\>[0]\AgdaSymbol{...}\AgdaSpace{}%
\AgdaSymbol{|}\AgdaSpace{}%
\AgdaInductiveConstructor{yes}\AgdaSpace{}%
\AgdaBound{hz}\AgdaSpace{}%
\AgdaSymbol{=}\AgdaSpace{}%
\AgdaInductiveConstructor{false}\AgdaSpace{}%
\AgdaOperator{\AgdaInductiveConstructor{Relation.Nullary.because}}\AgdaSpace{}%
\AgdaInductiveConstructor{Relation.Nullary.ofⁿ}\AgdaSpace{}%
\AgdaFunction{g₁}\<%
\\
\>[0][@{}l@{\AgdaIndent{0}}]%
\>[2]\AgdaKeyword{where}%
\>[3403I]\AgdaFunction{g₁}\AgdaSpace{}%
\AgdaSymbol{:}\AgdaSpace{}%
\AgdaOperator{\AgdaFunction{¬}}\AgdaSpace{}%
\AgdaDatatype{HazardFree}\AgdaSpace{}%
\AgdaBound{s}\AgdaSpace{}%
\AgdaSymbol{(}\AgdaBound{x}\AgdaSpace{}%
\AgdaOperator{\AgdaInductiveConstructor{∷}}\AgdaSpace{}%
\AgdaBound{b₁}\AgdaSymbol{)}\AgdaSpace{}%
\AgdaBound{b₂}\AgdaSpace{}%
\AgdaBound{ls}\<%
\\
\>[.][@{}l@{}]\<[3403I]%
\>[8]\AgdaFunction{g₁}\AgdaSpace{}%
\AgdaSymbol{(}\AgdaBound{¬hz}\AgdaSpace{}%
\AgdaOperator{\AgdaInductiveConstructor{∷}}\AgdaSpace{}%
\AgdaBound{hf}\AgdaSymbol{)}\AgdaSpace{}%
\AgdaSymbol{=}\AgdaSpace{}%
\AgdaBound{¬hz}\AgdaSpace{}%
\AgdaBound{hz}\<%
\\
\>[0]\AgdaSymbol{...}\AgdaSpace{}%
\AgdaSymbol{|}\AgdaSpace{}%
\AgdaInductiveConstructor{no}\AgdaSpace{}%
\AgdaBound{¬hz}\AgdaSpace{}%
\AgdaKeyword{with}\AgdaSpace{}%
\AgdaFunction{hazardfree?}\AgdaSpace{}%
\AgdaSymbol{(}\AgdaPostulate{run}\AgdaSpace{}%
\AgdaBound{x}\AgdaSpace{}%
\AgdaBound{s}\AgdaSymbol{)}\AgdaSpace{}%
\AgdaBound{b₁}\AgdaSpace{}%
\AgdaBound{b₂}\AgdaSpace{}%
\AgdaSymbol{(}\AgdaFunction{save}\AgdaSpace{}%
\AgdaBound{s}\AgdaSpace{}%
\AgdaBound{x}\AgdaSpace{}%
\AgdaBound{ls}\AgdaSymbol{)}\<%
\\
\>[0]\AgdaSymbol{...}\AgdaSpace{}%
\AgdaSymbol{|}\AgdaSpace{}%
\AgdaInductiveConstructor{yes}\AgdaSpace{}%
\AgdaBound{hf}\AgdaSpace{}%
\AgdaSymbol{=}\AgdaSpace{}%
\AgdaInductiveConstructor{true}\AgdaSpace{}%
\AgdaOperator{\AgdaInductiveConstructor{Relation.Nullary.because}}\AgdaSpace{}%
\AgdaInductiveConstructor{Relation.Nullary.ofʸ}\AgdaSpace{}%
\AgdaSymbol{(}\AgdaBound{¬hz}\AgdaSpace{}%
\AgdaOperator{\AgdaInductiveConstructor{∷}}\AgdaSpace{}%
\AgdaBound{hf}\AgdaSymbol{)}\<%
\\
\>[0]\AgdaSymbol{...}\AgdaSpace{}%
\AgdaSymbol{|}\AgdaSpace{}%
\AgdaInductiveConstructor{no}\AgdaSpace{}%
\AgdaBound{¬hf}\AgdaSpace{}%
\AgdaSymbol{=}\AgdaSpace{}%
\AgdaInductiveConstructor{false}\AgdaSpace{}%
\AgdaOperator{\AgdaInductiveConstructor{Relation.Nullary.because}}\AgdaSpace{}%
\AgdaInductiveConstructor{Relation.Nullary.ofⁿ}\AgdaSpace{}%
\AgdaFunction{g₁}\<%
\\
\>[0][@{}l@{\AgdaIndent{0}}]%
\>[2]\AgdaKeyword{where}%
\>[3451I]\AgdaFunction{g₁}\AgdaSpace{}%
\AgdaSymbol{:}\AgdaSpace{}%
\AgdaOperator{\AgdaFunction{¬}}\AgdaSpace{}%
\AgdaDatatype{HazardFree}\AgdaSpace{}%
\AgdaBound{s}\AgdaSpace{}%
\AgdaSymbol{(}\AgdaBound{x}\AgdaSpace{}%
\AgdaOperator{\AgdaInductiveConstructor{∷}}\AgdaSpace{}%
\AgdaBound{b₁}\AgdaSymbol{)}\AgdaSpace{}%
\AgdaBound{b₂}\AgdaSpace{}%
\AgdaBound{ls}\<%
\\
\>[.][@{}l@{}]\<[3451I]%
\>[8]\AgdaFunction{g₁}\AgdaSpace{}%
\AgdaSymbol{(\AgdaUnderscore{}}\AgdaSpace{}%
\AgdaOperator{\AgdaInductiveConstructor{∷}}\AgdaSpace{}%
\AgdaBound{hf}\AgdaSymbol{)}\AgdaSpace{}%
\AgdaSymbol{=}\AgdaSpace{}%
\AgdaBound{¬hf}\AgdaSpace{}%
\AgdaBound{hf}\<%
\end{code}

\begin{code}[hide]%
\>[0]\AgdaKeyword{open}\AgdaSpace{}%
\AgdaKeyword{import}\AgdaSpace{}%
\AgdaModule{Functional.State}\AgdaSpace{}%
\AgdaSymbol{as}\AgdaSpace{}%
\AgdaModule{St}\AgdaSpace{}%
\AgdaKeyword{using}\AgdaSpace{}%
\AgdaSymbol{(}\AgdaFunction{Oracle}\AgdaSpace{}%
\AgdaSymbol{;}\AgdaSpace{}%
\AgdaFunction{FileSystem}\AgdaSpace{}%
\AgdaSymbol{;}\AgdaSpace{}%
\AgdaFunction{Cmd}\AgdaSymbol{)}\<%
\\
\\[\AgdaEmptyExtraSkip]%
\>[0]\AgdaKeyword{module}\AgdaSpace{}%
\AgdaModule{Functional.Script.Exec}\AgdaSpace{}%
\AgdaSymbol{(}\AgdaBound{oracle}\AgdaSpace{}%
\AgdaSymbol{:}\AgdaSpace{}%
\AgdaFunction{Oracle}\AgdaSymbol{)}\AgdaSpace{}%
\AgdaKeyword{where}\<%
\\
\\[\AgdaEmptyExtraSkip]%
\>[0]\AgdaKeyword{open}\AgdaSpace{}%
\AgdaKeyword{import}\AgdaSpace{}%
\AgdaModule{Common.List.Properties}\AgdaSpace{}%
\AgdaSymbol{as}\AgdaSpace{}%
\AgdaModule{CLP}\AgdaSpace{}%
\AgdaKeyword{using}\AgdaSpace{}%
\AgdaSymbol{(}\AgdaFunction{before-∷ʳ⁺}\AgdaSymbol{)}\<%
\\
\>[0]\AgdaKeyword{open}\AgdaSpace{}%
\AgdaKeyword{import}\AgdaSpace{}%
\AgdaModule{Functional.State.Helpers}\AgdaSpace{}%
\AgdaSymbol{(}\AgdaBound{oracle}\AgdaSymbol{)}\AgdaSpace{}%
\AgdaKeyword{using}\AgdaSpace{}%
\AgdaSymbol{(}run \AgdaSymbol{;} cmdReadNames \AgdaSymbol{;} cmdWriteNames \AgdaSymbol{;} cmdReadWriteNames\AgdaSymbol{)}\<%
\\
\>[0]\AgdaKeyword{open}\AgdaSpace{}%
\AgdaKeyword{import}\AgdaSpace{}%
\AgdaModule{Functional.State.Properties}\AgdaSpace{}%
\AgdaSymbol{(}\AgdaBound{oracle}\AgdaSymbol{)}\AgdaSpace{}%
\AgdaKeyword{using}\AgdaSpace{}%
\AgdaSymbol{(}lemma1-sym \AgdaSymbol{;} lemma5 \AgdaSymbol{;} run-≡\AgdaSymbol{)}\<%
\\
\>[0]\AgdaKeyword{open}\AgdaSpace{}%
\AgdaKeyword{import}\AgdaSpace{}%
\AgdaModule{Agda.Builtin.Equality}\<%
\\
\>[0]\AgdaKeyword{open}\AgdaSpace{}%
\AgdaKeyword{import}\AgdaSpace{}%
\AgdaModule{Data.Sum}\AgdaSpace{}%
\AgdaKeyword{using}\AgdaSpace{}%
\AgdaSymbol{(}\AgdaOperator{\AgdaDatatype{\AgdaUnderscore{}⊎\AgdaUnderscore{}}}\AgdaSymbol{)}\<%
\\
\>[0]\AgdaKeyword{open}\AgdaSpace{}%
\AgdaKeyword{import}\AgdaSpace{}%
\AgdaModule{Data.List}\AgdaSpace{}%
\AgdaKeyword{using}\AgdaSpace{}%
\AgdaSymbol{(}\AgdaInductiveConstructor{[]}\AgdaSpace{}%
\AgdaSymbol{;}\AgdaSpace{}%
\AgdaOperator{\AgdaInductiveConstructor{\AgdaUnderscore{}∷\AgdaUnderscore{}}}\AgdaSpace{}%
\AgdaSymbol{;}\AgdaSpace{}%
\AgdaDatatype{List}\AgdaSpace{}%
\AgdaSymbol{;}\AgdaSpace{}%
\AgdaFunction{map}\AgdaSpace{}%
\AgdaSymbol{;}\AgdaSpace{}%
\AgdaOperator{\AgdaFunction{\AgdaUnderscore{}++\AgdaUnderscore{}}}\AgdaSpace{}%
\AgdaSymbol{;}\AgdaSpace{}%
\AgdaOperator{\AgdaFunction{\AgdaUnderscore{}∷ʳ\AgdaUnderscore{}}}\AgdaSpace{}%
\AgdaSymbol{;}\AgdaSpace{}%
\AgdaOperator{\AgdaFunction{[\AgdaUnderscore{}]}}\AgdaSpace{}%
\AgdaSymbol{;}\AgdaSpace{}%
\AgdaFunction{foldr}\AgdaSpace{}%
\AgdaSymbol{;}\AgdaSpace{}%
\AgdaFunction{reverse}\AgdaSymbol{)}\<%
\\
\>[0]\AgdaKeyword{open}\AgdaSpace{}%
\AgdaKeyword{import}\AgdaSpace{}%
\AgdaModule{Data.Product}\AgdaSpace{}%
\AgdaKeyword{using}\AgdaSpace{}%
\AgdaSymbol{(}\AgdaField{proj₁}\AgdaSpace{}%
\AgdaSymbol{;}\AgdaSpace{}%
\AgdaField{proj₂}\AgdaSpace{}%
\AgdaSymbol{;}\AgdaSpace{}%
\AgdaOperator{\AgdaInductiveConstructor{\AgdaUnderscore{},\AgdaUnderscore{}}}\AgdaSpace{}%
\AgdaSymbol{;}\AgdaSpace{}%
\AgdaOperator{\AgdaFunction{\AgdaUnderscore{}×\AgdaUnderscore{}}}\AgdaSpace{}%
\AgdaSymbol{;}\AgdaSpace{}%
\AgdaFunction{∄-syntax}\AgdaSpace{}%
\AgdaSymbol{;}\AgdaSpace{}%
\AgdaFunction{∃-syntax}\AgdaSpace{}%
\AgdaSymbol{;}\AgdaSpace{}%
\AgdaFunction{Σ-syntax}\AgdaSymbol{)}\<%
\\
\\[\AgdaEmptyExtraSkip]%
\>[0]\AgdaKeyword{open}\AgdaSpace{}%
\AgdaKeyword{import}\AgdaSpace{}%
\AgdaModule{Functional.Build}\AgdaSpace{}%
\AgdaSymbol{(}\AgdaBound{oracle}\AgdaSymbol{)}\AgdaSpace{}%
\AgdaKeyword{using}\AgdaSpace{}%
\AgdaSymbol{(}Build\AgdaSymbol{)}\<%
\\
\>[0]\AgdaKeyword{open}\AgdaSpace{}%
\AgdaKeyword{import}\AgdaSpace{}%
\AgdaModule{Functional.File}\AgdaSpace{}%
\AgdaKeyword{using}\AgdaSpace{}%
\AgdaSymbol{(}\AgdaFunction{FileName}\AgdaSpace{}%
\AgdaSymbol{;}\AgdaSpace{}%
\AgdaFunction{Files}\AgdaSymbol{)}\<%
\\
\>[0]\AgdaKeyword{open}\AgdaSpace{}%
\AgdaKeyword{import}\AgdaSpace{}%
\AgdaModule{Data.List.Relation.Binary.Disjoint.Propositional}\AgdaSpace{}%
\AgdaKeyword{using}\AgdaSpace{}%
\AgdaSymbol{(}\AgdaFunction{Disjoint}\AgdaSymbol{)}\<%
\\
\>[0]\AgdaKeyword{open}\AgdaSpace{}%
\AgdaKeyword{import}\AgdaSpace{}%
\AgdaModule{Relation.Binary.PropositionalEquality}\AgdaSpace{}%
\AgdaKeyword{using}\AgdaSpace{}%
\AgdaSymbol{(}\AgdaFunction{decSetoid}\AgdaSpace{}%
\AgdaSymbol{;}\AgdaSpace{}%
\AgdaFunction{trans}\AgdaSpace{}%
\AgdaSymbol{;}\AgdaSpace{}%
\AgdaFunction{sym}\AgdaSpace{}%
\AgdaSymbol{;}\AgdaSpace{}%
\AgdaFunction{subst}\AgdaSpace{}%
\AgdaSymbol{;}\AgdaSpace{}%
\AgdaFunction{cong}\AgdaSpace{}%
\AgdaSymbol{;}\AgdaSpace{}%
\AgdaFunction{cong₂}\AgdaSymbol{)}\<%
\\
\>[0]\AgdaKeyword{open}\AgdaSpace{}%
\AgdaKeyword{import}\AgdaSpace{}%
\AgdaModule{Data.String.Properties}\AgdaSpace{}%
\AgdaKeyword{using}\AgdaSpace{}%
\AgdaSymbol{(}\AgdaOperator{\AgdaFunction{\AgdaUnderscore{}≟\AgdaUnderscore{}}}\AgdaSymbol{)}\<%
\\
\>[0]\AgdaKeyword{open}\AgdaSpace{}%
\AgdaKeyword{import}\AgdaSpace{}%
\AgdaModule{Data.List.Membership.DecSetoid}\AgdaSpace{}%
\AgdaSymbol{(}\AgdaFunction{decSetoid}\AgdaSpace{}%
\AgdaOperator{\AgdaFunction{\AgdaUnderscore{}≟\AgdaUnderscore{}}}\AgdaSymbol{)}\AgdaSpace{}%
\AgdaKeyword{using}\AgdaSpace{}%
\AgdaSymbol{(}\AgdaUnderscore{}∈\AgdaUnderscore{} \AgdaSymbol{;} \AgdaUnderscore{}∈?\AgdaUnderscore{} \AgdaSymbol{;} \AgdaUnderscore{}∉\AgdaUnderscore{}\AgdaSymbol{)}\<%
\\
\>[0]\AgdaKeyword{open}\AgdaSpace{}%
\AgdaKeyword{import}\AgdaSpace{}%
\AgdaModule{Data.List.Properties}\AgdaSpace{}%
\AgdaKeyword{using}\AgdaSpace{}%
\AgdaSymbol{(}\AgdaFunction{∷-injective}\AgdaSpace{}%
\AgdaSymbol{;}\AgdaSpace{}%
\AgdaFunction{∷-injectiveʳ}\AgdaSpace{}%
\AgdaSymbol{;}\AgdaSpace{}%
\AgdaFunction{∷-injectiveˡ}\AgdaSpace{}%
\AgdaSymbol{;}\AgdaSpace{}%
\AgdaFunction{++-identityʳ}\AgdaSymbol{)}\<%
\\
\>[0]\AgdaKeyword{open}\AgdaSpace{}%
\AgdaKeyword{import}\AgdaSpace{}%
\AgdaModule{Relation.Nullary}\AgdaSpace{}%
\AgdaKeyword{using}\AgdaSpace{}%
\AgdaSymbol{(}\AgdaInductiveConstructor{yes}\AgdaSpace{}%
\AgdaSymbol{;}\AgdaSpace{}%
\AgdaInductiveConstructor{no}\AgdaSpace{}%
\AgdaSymbol{;}\AgdaSpace{}%
\AgdaOperator{\AgdaFunction{¬\AgdaUnderscore{}}}\AgdaSymbol{)}\<%
\\
\>[0]\AgdaKeyword{open}\AgdaSpace{}%
\AgdaKeyword{import}\AgdaSpace{}%
\AgdaModule{Data.List.Relation.Unary.Any}\AgdaSpace{}%
\AgdaKeyword{using}\AgdaSpace{}%
\AgdaSymbol{(}\AgdaFunction{tail}\AgdaSpace{}%
\AgdaSymbol{;}\AgdaSpace{}%
\AgdaInductiveConstructor{here}\AgdaSpace{}%
\AgdaSymbol{;}\AgdaSpace{}%
\AgdaInductiveConstructor{there}\AgdaSymbol{)}\<%
\\
\>[0]\AgdaKeyword{open}\AgdaSpace{}%
\AgdaKeyword{import}\AgdaSpace{}%
\AgdaModule{Data.List.Relation.Unary.Any.Properties}\AgdaSpace{}%
\AgdaKeyword{using}\AgdaSpace{}%
\AgdaSymbol{(}\AgdaFunction{++⁺ˡ}\AgdaSymbol{)}\<%
\\
\>[0]\AgdaKeyword{open}\AgdaSpace{}%
\AgdaKeyword{import}\AgdaSpace{}%
\AgdaModule{Data.List.Relation.Binary.Permutation.Propositional}\AgdaSpace{}%
\AgdaKeyword{using}\AgdaSpace{}%
\AgdaSymbol{(}\AgdaOperator{\AgdaDatatype{\AgdaUnderscore{}↭\AgdaUnderscore{}}}\AgdaSymbol{)}\<%
\\
\>[0]\AgdaKeyword{open}\AgdaSpace{}%
\AgdaKeyword{import}\AgdaSpace{}%
\AgdaModule{Relation.Nullary.Negation}\AgdaSpace{}%
\AgdaKeyword{using}\AgdaSpace{}%
\AgdaSymbol{(}\AgdaFunction{contradiction}\AgdaSymbol{)}\<%
\\
\>[0]\AgdaKeyword{open}\AgdaSpace{}%
\AgdaKeyword{import}\AgdaSpace{}%
\AgdaModule{Data.List.Membership.Propositional.Properties}\AgdaSpace{}%
\AgdaKeyword{using}\AgdaSpace{}%
\AgdaSymbol{(}\AgdaFunction{∈-++⁻}\AgdaSpace{}%
\AgdaSymbol{;}\AgdaSpace{}%
\AgdaFunction{∈-++⁺ˡ}\AgdaSpace{}%
\AgdaSymbol{;}\AgdaSpace{}%
\AgdaFunction{∈-++⁺ʳ}\AgdaSpace{}%
\AgdaSymbol{;}\AgdaSpace{}%
\AgdaFunction{∈-insert}\AgdaSpace{}%
\AgdaSymbol{;}\AgdaSpace{}%
\AgdaFunction{∈-∃++}\AgdaSpace{}%
\AgdaSymbol{;}\AgdaSpace{}%
\AgdaFunction{∈-resp-≋}\AgdaSymbol{)}\<%
\\
\>[0]\AgdaKeyword{open}\AgdaSpace{}%
\AgdaKeyword{import}\AgdaSpace{}%
\AgdaModule{Data.Sum}\AgdaSpace{}%
\AgdaKeyword{using}\AgdaSpace{}%
\AgdaSymbol{(}\AgdaInductiveConstructor{inj₁}\AgdaSpace{}%
\AgdaSymbol{;}\AgdaSpace{}%
\AgdaInductiveConstructor{inj₂}\AgdaSymbol{)}\<%
\\
\>[0]\AgdaKeyword{open}\AgdaSpace{}%
\AgdaKeyword{import}\AgdaSpace{}%
\AgdaModule{Data.List.Relation.Binary.Subset.Propositional}\AgdaSpace{}%
\AgdaKeyword{using}\AgdaSpace{}%
\AgdaSymbol{(}\AgdaOperator{\AgdaFunction{\AgdaUnderscore{}⊆\AgdaUnderscore{}}}\AgdaSymbol{)}\<%
\\
\>[0]\AgdaKeyword{open}\AgdaSpace{}%
\AgdaKeyword{import}\AgdaSpace{}%
\AgdaModule{Data.Empty}\AgdaSpace{}%
\AgdaKeyword{using}\AgdaSpace{}%
\AgdaSymbol{(}\AgdaDatatype{⊥}\AgdaSymbol{)}\<%
\\
\>[0]\AgdaKeyword{open}\AgdaSpace{}%
\AgdaKeyword{import}\AgdaSpace{}%
\AgdaModule{Data.List.Relation.Unary.All}\AgdaSpace{}%
\AgdaKeyword{using}\AgdaSpace{}%
\AgdaSymbol{(}\AgdaDatatype{All}\AgdaSpace{}%
\AgdaSymbol{;}\AgdaSpace{}%
\AgdaFunction{lookup}\AgdaSymbol{)}\<%
\\
\>[0]\AgdaKeyword{open}\AgdaSpace{}%
\AgdaKeyword{import}\AgdaSpace{}%
\AgdaModule{Data.List.Relation.Unary.All.Properties}\AgdaSpace{}%
\AgdaKeyword{using}\AgdaSpace{}%
\AgdaSymbol{(}\AgdaFunction{++⁻ˡ}\AgdaSpace{}%
\AgdaSymbol{;}\AgdaSpace{}%
\AgdaFunction{++⁻ʳ}\AgdaSpace{}%
\AgdaSymbol{;}\AgdaSpace{}%
\AgdaFunction{++⁻}\AgdaSpace{}%
\AgdaSymbol{)}\<%
\\
\>[0]\AgdaKeyword{open}\AgdaSpace{}%
\AgdaKeyword{import}\AgdaSpace{}%
\AgdaModule{Function}\AgdaSpace{}%
\AgdaKeyword{using}\AgdaSpace{}%
\AgdaSymbol{(}\AgdaOperator{\AgdaFunction{\AgdaUnderscore{}∘\AgdaUnderscore{}}}\AgdaSymbol{)}\<%
\end{code}

\newcommand{\script}{%
\begin{code}%
\>[0]\AgdaFunction{script}\AgdaSpace{}%
\AgdaSymbol{:}\AgdaSpace{}%
\AgdaFunction{Build}\AgdaSpace{}%
\AgdaSymbol{→}\AgdaSpace{}%
\AgdaFunction{FileSystem}\AgdaSpace{}%
\AgdaSymbol{→}\AgdaSpace{}%
\AgdaFunction{FileSystem}\<%
\\
\>[0]\AgdaFunction{script}\AgdaSpace{}%
\AgdaInductiveConstructor{[]}\AgdaSpace{}%
\AgdaBound{sys}\AgdaSpace{}%
\AgdaSymbol{=}\AgdaSpace{}%
\AgdaBound{sys}\<%
\\
\>[0]\AgdaFunction{script}\AgdaSpace{}%
\AgdaSymbol{(}\AgdaBound{x}\AgdaSpace{}%
\AgdaOperator{\AgdaInductiveConstructor{∷}}\AgdaSpace{}%
\AgdaBound{b}\AgdaSymbol{)}\AgdaSpace{}%
\AgdaBound{sys}\AgdaSpace{}%
\AgdaSymbol{=}\AgdaSpace{}%
\AgdaFunction{script}\AgdaSpace{}%
\AgdaBound{b}\AgdaSpace{}%
\AgdaSymbol{(}\AgdaFunction{run}\AgdaSpace{}%
\AgdaBound{x}\AgdaSpace{}%
\AgdaBound{sys}\AgdaSymbol{)}\<%
\end{code}}

\begin{code}[hide]%
\>[0]\AgdaFunction{buildReadNames}\AgdaSpace{}%
\AgdaSymbol{:}\AgdaSpace{}%
\AgdaFunction{FileSystem}\AgdaSpace{}%
\AgdaSymbol{->}\AgdaSpace{}%
\AgdaFunction{Build}\AgdaSpace{}%
\AgdaSymbol{->}\AgdaSpace{}%
\AgdaDatatype{List}\AgdaSpace{}%
\AgdaFunction{FileName}\<%
\\
\>[0]\AgdaFunction{buildReadNames}\AgdaSpace{}%
\AgdaSymbol{\AgdaUnderscore{}}\AgdaSpace{}%
\AgdaInductiveConstructor{[]}\AgdaSpace{}%
\AgdaSymbol{=}\AgdaSpace{}%
\AgdaInductiveConstructor{[]}\<%
\\
\>[0]\AgdaFunction{buildReadNames}\AgdaSpace{}%
\AgdaBound{sys}\AgdaSpace{}%
\AgdaSymbol{(}\AgdaBound{x}\AgdaSpace{}%
\AgdaOperator{\AgdaInductiveConstructor{∷}}\AgdaSpace{}%
\AgdaBound{b}\AgdaSymbol{)}\AgdaSpace{}%
\AgdaSymbol{=}\AgdaSpace{}%
\AgdaSymbol{(}\AgdaFunction{cmdReadNames}\AgdaSpace{}%
\AgdaBound{x}\AgdaSpace{}%
\AgdaBound{sys}\AgdaSymbol{)}\AgdaSpace{}%
\AgdaOperator{\AgdaFunction{++}}\AgdaSpace{}%
\AgdaFunction{buildReadNames}\AgdaSpace{}%
\AgdaSymbol{(}\AgdaFunction{run}\AgdaSpace{}%
\AgdaBound{x}\AgdaSpace{}%
\AgdaBound{sys}\AgdaSymbol{)}\AgdaSpace{}%
\AgdaBound{b}\<%
\\
\\[\AgdaEmptyExtraSkip]%
\>[0]\AgdaFunction{buildWriteNames}\AgdaSpace{}%
\AgdaSymbol{:}\AgdaSpace{}%
\AgdaFunction{FileSystem}\AgdaSpace{}%
\AgdaSymbol{->}\AgdaSpace{}%
\AgdaFunction{Build}\AgdaSpace{}%
\AgdaSymbol{->}\AgdaSpace{}%
\AgdaDatatype{List}\AgdaSpace{}%
\AgdaFunction{FileName}\<%
\\
\>[0]\AgdaFunction{buildWriteNames}\AgdaSpace{}%
\AgdaSymbol{\AgdaUnderscore{}}\AgdaSpace{}%
\AgdaInductiveConstructor{[]}\AgdaSpace{}%
\AgdaSymbol{=}\AgdaSpace{}%
\AgdaInductiveConstructor{[]}\<%
\\
\>[0]\AgdaFunction{buildWriteNames}\AgdaSpace{}%
\AgdaBound{sys}\AgdaSpace{}%
\AgdaSymbol{(}\AgdaBound{x}\AgdaSpace{}%
\AgdaOperator{\AgdaInductiveConstructor{∷}}\AgdaSpace{}%
\AgdaBound{b}\AgdaSymbol{)}\AgdaSpace{}%
\AgdaSymbol{=}\AgdaSpace{}%
\AgdaSymbol{(}\AgdaFunction{cmdWriteNames}\AgdaSpace{}%
\AgdaBound{x}\AgdaSpace{}%
\AgdaBound{sys}\AgdaSymbol{)}\AgdaSpace{}%
\AgdaOperator{\AgdaFunction{++}}\AgdaSpace{}%
\AgdaFunction{buildWriteNames}\AgdaSpace{}%
\AgdaSymbol{(}\AgdaFunction{run}\AgdaSpace{}%
\AgdaBound{x}\AgdaSpace{}%
\AgdaBound{sys}\AgdaSymbol{)}\AgdaSpace{}%
\AgdaBound{b}\<%
\\
\\[\AgdaEmptyExtraSkip]%
\>[0]\AgdaFunction{build-rws}\AgdaSpace{}%
\AgdaSymbol{:}\AgdaSpace{}%
\AgdaFunction{FileSystem}\AgdaSpace{}%
\AgdaSymbol{->}\AgdaSpace{}%
\AgdaFunction{Build}\AgdaSpace{}%
\AgdaSymbol{->}\AgdaSpace{}%
\AgdaDatatype{List}\AgdaSpace{}%
\AgdaFunction{FileName}\AgdaSpace{}%
\AgdaSymbol{->}\AgdaSpace{}%
\AgdaDatatype{List}\AgdaSpace{}%
\AgdaFunction{FileName}\<%
\\
\>[0]\AgdaFunction{build-rws}\AgdaSpace{}%
\AgdaBound{sys}\AgdaSpace{}%
\AgdaInductiveConstructor{[]}\AgdaSpace{}%
\AgdaBound{ls}\AgdaSpace{}%
\AgdaSymbol{=}\AgdaSpace{}%
\AgdaBound{ls}\<%
\\
\>[0]\AgdaFunction{build-rws}\AgdaSpace{}%
\AgdaBound{sys}\AgdaSpace{}%
\AgdaSymbol{(}\AgdaBound{x}\AgdaSpace{}%
\AgdaOperator{\AgdaInductiveConstructor{∷}}\AgdaSpace{}%
\AgdaBound{b}\AgdaSymbol{)}\AgdaSpace{}%
\AgdaBound{ls}\AgdaSpace{}%
\AgdaSymbol{=}\AgdaSpace{}%
\AgdaFunction{build-rws}\AgdaSpace{}%
\AgdaSymbol{(}\AgdaFunction{run}\AgdaSpace{}%
\AgdaBound{x}\AgdaSpace{}%
\AgdaBound{sys}\AgdaSymbol{)}\AgdaSpace{}%
\AgdaBound{b}\AgdaSpace{}%
\AgdaSymbol{(}\AgdaFunction{cmdReadWriteNames}\AgdaSpace{}%
\AgdaBound{x}\AgdaSpace{}%
\AgdaBound{sys}\AgdaSpace{}%
\AgdaOperator{\AgdaFunction{++}}\AgdaSpace{}%
\AgdaBound{ls}\AgdaSymbol{)}\<%
\\
\\[\AgdaEmptyExtraSkip]%
\\[\AgdaEmptyExtraSkip]%
\>[0]\AgdaComment{-- proofs to go in own file eventually:        }\<%
\\
\\[\AgdaEmptyExtraSkip]%
\>[0]\AgdaComment{-- use this instead of h₂}\<%
\\
\>[0]\AgdaFunction{h₄}\AgdaSpace{}%
\AgdaSymbol{:}\AgdaSpace{}%
\AgdaSymbol{(}\AgdaBound{s}\AgdaSpace{}%
\AgdaBound{s₁}\AgdaSpace{}%
\AgdaSymbol{:}\AgdaSpace{}%
\AgdaFunction{FileSystem}\AgdaSymbol{)}\AgdaSpace{}%
\AgdaSymbol{(}\AgdaBound{x}\AgdaSpace{}%
\AgdaSymbol{:}\AgdaSpace{}%
\AgdaFunction{Cmd}\AgdaSymbol{)}\AgdaSpace{}%
\AgdaSymbol{->}\AgdaSpace{}%
\AgdaDatatype{All}\AgdaSpace{}%
\AgdaSymbol{(λ}\AgdaSpace{}%
\AgdaBound{f₁}\AgdaSpace{}%
\AgdaSymbol{→}\AgdaSpace{}%
\AgdaBound{s}\AgdaSpace{}%
\AgdaBound{f₁}\AgdaSpace{}%
\AgdaOperator{\AgdaDatatype{≡}}\AgdaSpace{}%
\AgdaBound{s₁}\AgdaSpace{}%
\AgdaBound{f₁}\AgdaSymbol{)}\AgdaSpace{}%
\AgdaSymbol{(}\AgdaFunction{cmdReadNames}\AgdaSpace{}%
\AgdaBound{x}\AgdaSpace{}%
\AgdaBound{s₁}\AgdaSymbol{)}\AgdaSpace{}%
\AgdaSymbol{->}\AgdaSpace{}%
\AgdaField{proj₁}\AgdaSpace{}%
\AgdaSymbol{(}\AgdaBound{oracle}\AgdaSpace{}%
\AgdaBound{x}\AgdaSymbol{)}\AgdaSpace{}%
\AgdaBound{s}\AgdaSpace{}%
\AgdaOperator{\AgdaDatatype{≡}}\AgdaSpace{}%
\AgdaField{proj₁}\AgdaSpace{}%
\AgdaSymbol{(}\AgdaBound{oracle}\AgdaSpace{}%
\AgdaBound{x}\AgdaSymbol{)}\AgdaSpace{}%
\AgdaBound{s₁}\<%
\\
\>[0]\AgdaFunction{h₄}\AgdaSpace{}%
\AgdaBound{s}\AgdaSpace{}%
\AgdaBound{s₁}\AgdaSpace{}%
\AgdaBound{x}\AgdaSpace{}%
\AgdaBound{all₁}\AgdaSpace{}%
\AgdaSymbol{=}\AgdaSpace{}%
\AgdaFunction{sym}\AgdaSpace{}%
\AgdaSymbol{(}\AgdaField{proj₂}\AgdaSpace{}%
\AgdaSymbol{(}\AgdaBound{oracle}\AgdaSpace{}%
\AgdaBound{x}\AgdaSymbol{)}\AgdaSpace{}%
\AgdaBound{s₁}\AgdaSpace{}%
\AgdaBound{s}\AgdaSpace{}%
\AgdaSymbol{λ}\AgdaSpace{}%
\AgdaBound{f₁}\AgdaSpace{}%
\AgdaBound{x₁}\AgdaSpace{}%
\AgdaSymbol{→}\AgdaSpace{}%
\AgdaFunction{sym}\AgdaSpace{}%
\AgdaSymbol{(}\AgdaFunction{lookup}\AgdaSpace{}%
\AgdaBound{all₁}\AgdaSpace{}%
\AgdaBound{x₁}\AgdaSymbol{))}\<%
\\
\\[\AgdaEmptyExtraSkip]%
\>[0]\AgdaFunction{h₃}\AgdaSpace{}%
\AgdaSymbol{:}\AgdaSpace{}%
\AgdaSymbol{\{}\AgdaBound{s}\AgdaSpace{}%
\AgdaBound{s₁}\AgdaSpace{}%
\AgdaSymbol{:}\AgdaSpace{}%
\AgdaFunction{FileSystem}\AgdaSymbol{\}}\AgdaSpace{}%
\AgdaSymbol{(}\AgdaBound{x}\AgdaSpace{}%
\AgdaSymbol{:}\AgdaSpace{}%
\AgdaFunction{Cmd}\AgdaSymbol{)}\AgdaSpace{}%
\AgdaSymbol{(}\AgdaBound{xs}\AgdaSpace{}%
\AgdaBound{ys}\AgdaSpace{}%
\AgdaSymbol{:}\AgdaSpace{}%
\AgdaFunction{Build}\AgdaSymbol{)}\AgdaSpace{}%
\AgdaSymbol{->}\AgdaSpace{}%
\AgdaDatatype{All}\AgdaSpace{}%
\AgdaSymbol{(λ}\AgdaSpace{}%
\AgdaBound{f₁}\AgdaSpace{}%
\AgdaSymbol{→}\AgdaSpace{}%
\AgdaBound{s}\AgdaSpace{}%
\AgdaBound{f₁}\AgdaSpace{}%
\AgdaOperator{\AgdaDatatype{≡}}\AgdaSpace{}%
\AgdaBound{s₁}\AgdaSpace{}%
\AgdaBound{f₁}\AgdaSymbol{)}\AgdaSpace{}%
\AgdaSymbol{(}\AgdaFunction{buildReadNames}\AgdaSpace{}%
\AgdaBound{s₁}\AgdaSpace{}%
\AgdaSymbol{(}\AgdaBound{xs}\AgdaSpace{}%
\AgdaOperator{\AgdaFunction{++}}\AgdaSpace{}%
\AgdaBound{x}\AgdaSpace{}%
\AgdaOperator{\AgdaInductiveConstructor{∷}}\AgdaSpace{}%
\AgdaBound{ys}\AgdaSymbol{))}\AgdaSpace{}%
\AgdaSymbol{->}\AgdaSpace{}%
\AgdaDatatype{All}\AgdaSpace{}%
\AgdaSymbol{(λ}\AgdaSpace{}%
\AgdaBound{f₁}\AgdaSpace{}%
\AgdaSymbol{→}\AgdaSpace{}%
\AgdaFunction{script}\AgdaSpace{}%
\AgdaBound{xs}\AgdaSpace{}%
\AgdaBound{s}\AgdaSpace{}%
\AgdaBound{f₁}\AgdaSpace{}%
\AgdaOperator{\AgdaDatatype{≡}}\AgdaSpace{}%
\AgdaFunction{script}\AgdaSpace{}%
\AgdaBound{xs}\AgdaSpace{}%
\AgdaBound{s₁}\AgdaSpace{}%
\AgdaBound{f₁}\AgdaSymbol{)}\AgdaSpace{}%
\AgdaSymbol{(}\AgdaFunction{cmdReadNames}\AgdaSpace{}%
\AgdaBound{x}\AgdaSpace{}%
\AgdaSymbol{(}\AgdaFunction{script}\AgdaSpace{}%
\AgdaBound{xs}\AgdaSpace{}%
\AgdaBound{s₁}\AgdaSymbol{))}\<%
\\
\>[0]\AgdaFunction{h₃}\AgdaSpace{}%
\AgdaSymbol{\{}\AgdaBound{s}\AgdaSymbol{\}}\AgdaSpace{}%
\AgdaSymbol{\{}\AgdaBound{s₁}\AgdaSymbol{\}}\AgdaSpace{}%
\AgdaBound{x}\AgdaSpace{}%
\AgdaInductiveConstructor{[]}\AgdaSpace{}%
\AgdaBound{ys}\AgdaSpace{}%
\AgdaBound{all₁}\AgdaSpace{}%
\AgdaSymbol{=}\AgdaSpace{}%
\AgdaFunction{++⁻ˡ}\AgdaSpace{}%
\AgdaSymbol{(}\AgdaFunction{cmdReadNames}\AgdaSpace{}%
\AgdaBound{x}\AgdaSpace{}%
\AgdaBound{s₁}\AgdaSymbol{)}\AgdaSpace{}%
\AgdaBound{all₁}\<%
\\
\>[0]\AgdaFunction{h₃}\AgdaSpace{}%
\AgdaSymbol{\{}\AgdaBound{s}\AgdaSymbol{\}}\AgdaSpace{}%
\AgdaSymbol{\{}\AgdaBound{s₁}\AgdaSymbol{\}}\AgdaSpace{}%
\AgdaBound{x}\AgdaSpace{}%
\AgdaSymbol{(}\AgdaBound{x₁}\AgdaSpace{}%
\AgdaOperator{\AgdaInductiveConstructor{∷}}\AgdaSpace{}%
\AgdaBound{xs}\AgdaSymbol{)}\AgdaSpace{}%
\AgdaBound{ys}\AgdaSpace{}%
\AgdaBound{all₁}\AgdaSpace{}%
\AgdaKeyword{with}\AgdaSpace{}%
\AgdaFunction{++⁻}\AgdaSpace{}%
\AgdaSymbol{(}\AgdaFunction{cmdReadNames}\AgdaSpace{}%
\AgdaBound{x₁}\AgdaSpace{}%
\AgdaBound{s₁}\AgdaSymbol{)}\AgdaSpace{}%
\AgdaBound{all₁}\<%
\\
\>[0]\AgdaSymbol{...}\AgdaSpace{}%
\AgdaSymbol{|}\AgdaSpace{}%
\AgdaBound{all₂}\AgdaSpace{}%
\AgdaOperator{\AgdaInductiveConstructor{,}}\AgdaSpace{}%
\AgdaBound{all₃}\AgdaSpace{}%
\AgdaSymbol{=}\AgdaSpace{}%
\AgdaFunction{h₃}\AgdaSpace{}%
\AgdaSymbol{\{}\AgdaFunction{run}\AgdaSpace{}%
\AgdaBound{x₁}\AgdaSpace{}%
\AgdaBound{s}\AgdaSymbol{\}}\AgdaSpace{}%
\AgdaSymbol{\{}\AgdaFunction{run}\AgdaSpace{}%
\AgdaBound{x₁}\AgdaSpace{}%
\AgdaBound{s₁}\AgdaSymbol{\}}\AgdaSpace{}%
\AgdaBound{x}\AgdaSpace{}%
\AgdaBound{xs}\AgdaSpace{}%
\AgdaBound{ys}\AgdaSpace{}%
\AgdaSymbol{(}\AgdaFunction{lemma1-sym}\AgdaSpace{}%
\AgdaSymbol{\{}\AgdaBound{s}\AgdaSymbol{\}}\AgdaSpace{}%
\AgdaSymbol{\{}\AgdaBound{s₁}\AgdaSymbol{\}}\AgdaSpace{}%
\AgdaSymbol{(}\AgdaFunction{buildReadNames}\AgdaSpace{}%
\AgdaSymbol{(}\AgdaFunction{run}\AgdaSpace{}%
\AgdaBound{x₁}\AgdaSpace{}%
\AgdaBound{s₁}\AgdaSymbol{)}\AgdaSpace{}%
\AgdaSymbol{(}\AgdaBound{xs}\AgdaSpace{}%
\AgdaOperator{\AgdaFunction{++}}\AgdaSpace{}%
\AgdaBound{x}\AgdaSpace{}%
\AgdaOperator{\AgdaInductiveConstructor{∷}}\AgdaSpace{}%
\AgdaBound{ys}\AgdaSymbol{))}\AgdaSpace{}%
\AgdaBound{x₁}\AgdaSpace{}%
\AgdaBound{all₂}\AgdaSpace{}%
\AgdaBound{all₃}\AgdaSymbol{)}\<%
\\
\\[\AgdaEmptyExtraSkip]%
\\[\AgdaEmptyExtraSkip]%
\>[0]\AgdaFunction{h₁}\AgdaSpace{}%
\AgdaSymbol{:}\AgdaSpace{}%
\AgdaSymbol{\{}\AgdaBound{s}\AgdaSpace{}%
\AgdaSymbol{:}\AgdaSpace{}%
\AgdaFunction{FileSystem}\AgdaSymbol{\}}\AgdaSpace{}%
\AgdaSymbol{(}\AgdaBound{f₁}\AgdaSpace{}%
\AgdaSymbol{:}\AgdaSpace{}%
\AgdaFunction{FileName}\AgdaSymbol{)}\AgdaSpace{}%
\AgdaSymbol{->}\AgdaSpace{}%
\AgdaSymbol{(}\AgdaBound{x}\AgdaSpace{}%
\AgdaBound{w}\AgdaSpace{}%
\AgdaSymbol{:}\AgdaSpace{}%
\AgdaFunction{Cmd}\AgdaSymbol{)}\AgdaSpace{}%
\AgdaSymbol{(}\AgdaBound{xs}\AgdaSpace{}%
\AgdaBound{ys}\AgdaSpace{}%
\AgdaBound{ls₁}\AgdaSpace{}%
\AgdaBound{ls₂}\AgdaSpace{}%
\AgdaSymbol{:}\AgdaSpace{}%
\AgdaFunction{Build}\AgdaSymbol{)}\AgdaSpace{}%
\AgdaSymbol{->}\AgdaSpace{}%
\AgdaBound{f₁}\AgdaSpace{}%
\AgdaOperator{\AgdaFunction{∈}}\AgdaSpace{}%
\AgdaFunction{cmdReadNames}\AgdaSpace{}%
\AgdaBound{w}\AgdaSpace{}%
\AgdaSymbol{(}\AgdaFunction{script}\AgdaSpace{}%
\AgdaBound{ls₁}\AgdaSpace{}%
\AgdaBound{s}\AgdaSymbol{)}\AgdaSpace{}%
\AgdaSymbol{->}\AgdaSpace{}%
\AgdaFunction{Disjoint}\AgdaSpace{}%
\AgdaSymbol{(}\AgdaFunction{cmdWriteNames}\AgdaSpace{}%
\AgdaBound{x}\AgdaSpace{}%
\AgdaSymbol{(}\AgdaFunction{script}\AgdaSpace{}%
\AgdaBound{xs}\AgdaSpace{}%
\AgdaBound{s}\AgdaSymbol{))}\AgdaSpace{}%
\AgdaSymbol{(}\AgdaFunction{buildReadNames}\AgdaSpace{}%
\AgdaSymbol{(}\AgdaFunction{run}\AgdaSpace{}%
\AgdaBound{x}\AgdaSpace{}%
\AgdaSymbol{(}\AgdaFunction{script}\AgdaSpace{}%
\AgdaBound{xs}\AgdaSpace{}%
\AgdaBound{s}\AgdaSymbol{))}\AgdaSpace{}%
\AgdaBound{ys}\AgdaSymbol{)}\AgdaSpace{}%
\AgdaSymbol{->}\AgdaSpace{}%
\AgdaBound{xs}\AgdaSpace{}%
\AgdaOperator{\AgdaFunction{++}}\AgdaSpace{}%
\AgdaBound{ys}\AgdaSpace{}%
\AgdaOperator{\AgdaDatatype{≡}}\AgdaSpace{}%
\AgdaBound{ls₁}\AgdaSpace{}%
\AgdaOperator{\AgdaFunction{++}}\AgdaSpace{}%
\AgdaBound{w}\AgdaSpace{}%
\AgdaOperator{\AgdaInductiveConstructor{∷}}\AgdaSpace{}%
\AgdaBound{ls₂}\AgdaSpace{}%
\AgdaSymbol{->}\AgdaSpace{}%
\AgdaFunction{∃[}\AgdaSpace{}%
\AgdaBound{ls₃}\AgdaSpace{}%
\AgdaFunction{]}\AgdaSymbol{(}\AgdaFunction{∃[}\AgdaSpace{}%
\AgdaBound{ls₄}\AgdaSpace{}%
\AgdaFunction{]}\AgdaSymbol{(}\AgdaBound{xs}\AgdaSpace{}%
\AgdaOperator{\AgdaFunction{++}}\AgdaSpace{}%
\AgdaBound{x}\AgdaSpace{}%
\AgdaOperator{\AgdaInductiveConstructor{∷}}\AgdaSpace{}%
\AgdaBound{ys}\AgdaSpace{}%
\AgdaOperator{\AgdaDatatype{≡}}\AgdaSpace{}%
\AgdaBound{ls₃}\AgdaSpace{}%
\AgdaOperator{\AgdaFunction{++}}\AgdaSpace{}%
\AgdaBound{w}\AgdaSpace{}%
\AgdaOperator{\AgdaInductiveConstructor{∷}}\AgdaSpace{}%
\AgdaBound{ls₄}\AgdaSpace{}%
\AgdaOperator{\AgdaFunction{×}}\AgdaSpace{}%
\AgdaBound{f₁}\AgdaSpace{}%
\AgdaOperator{\AgdaFunction{∈}}\AgdaSpace{}%
\AgdaFunction{cmdReadNames}\AgdaSpace{}%
\AgdaBound{w}\AgdaSpace{}%
\AgdaSymbol{(}\AgdaFunction{script}\AgdaSpace{}%
\AgdaBound{ls₃}\AgdaSpace{}%
\AgdaBound{s}\AgdaSymbol{)))}\<%
\\
\>[0]\AgdaFunction{h₁}\AgdaSpace{}%
\AgdaSymbol{\{}\AgdaBound{s}\AgdaSymbol{\}}\AgdaSpace{}%
\AgdaBound{f₁}\AgdaSpace{}%
\AgdaBound{x}\AgdaSpace{}%
\AgdaBound{w}\AgdaSpace{}%
\AgdaInductiveConstructor{[]}\AgdaSpace{}%
\AgdaBound{ys}\AgdaSpace{}%
\AgdaBound{ls₁}\AgdaSpace{}%
\AgdaBound{ls₂}\AgdaSpace{}%
\AgdaBound{f₁∈}\AgdaSpace{}%
\AgdaBound{dsj}\AgdaSpace{}%
\AgdaBound{ys≡ls₁++w∷ls₂}\AgdaSpace{}%
\AgdaKeyword{with}\AgdaSpace{}%
\AgdaFunction{h₃}\AgdaSpace{}%
\AgdaBound{w}\AgdaSpace{}%
\AgdaBound{ls₁}\AgdaSpace{}%
\AgdaBound{ls₂}\AgdaSpace{}%
\AgdaSymbol{(}\AgdaFunction{lemma5}\AgdaSpace{}%
\AgdaSymbol{(}\AgdaFunction{bs}\AgdaSpace{}%
\AgdaSymbol{(}\AgdaBound{ls₁}\AgdaSpace{}%
\AgdaOperator{\AgdaFunction{++}}\AgdaSpace{}%
\AgdaBound{w}\AgdaSpace{}%
\AgdaOperator{\AgdaInductiveConstructor{∷}}\AgdaSpace{}%
\AgdaBound{ls₂}\AgdaSymbol{))}\AgdaSpace{}%
\AgdaFunction{ws}\AgdaSpace{}%
\AgdaSymbol{(}\AgdaFunction{subst}\AgdaSpace{}%
\AgdaSymbol{(λ}\AgdaSpace{}%
\AgdaBound{x₁}\AgdaSpace{}%
\AgdaSymbol{→}\AgdaSpace{}%
\AgdaFunction{Disjoint}\AgdaSpace{}%
\AgdaSymbol{(}\AgdaFunction{map}\AgdaSpace{}%
\AgdaField{proj₁}\AgdaSpace{}%
\AgdaFunction{ws}\AgdaSymbol{)}\AgdaSpace{}%
\AgdaSymbol{(}\AgdaFunction{bs}\AgdaSpace{}%
\AgdaBound{x₁}\AgdaSymbol{))}\AgdaSpace{}%
\AgdaBound{ys≡ls₁++w∷ls₂}\AgdaSpace{}%
\AgdaBound{dsj}\AgdaSymbol{))}\<%
\\
\>[0][@{}l@{\AgdaIndent{0}}]%
\>[2]\AgdaKeyword{where}%
\>[562I]\AgdaFunction{bs}\AgdaSpace{}%
\AgdaSymbol{:}\AgdaSpace{}%
\AgdaFunction{Build}\AgdaSpace{}%
\AgdaSymbol{->}\AgdaSpace{}%
\AgdaDatatype{List}\AgdaSpace{}%
\AgdaFunction{FileName}\<%
\\
\>[.][@{}l@{}]\<[562I]%
\>[8]\AgdaFunction{bs}\AgdaSpace{}%
\AgdaBound{b}\AgdaSpace{}%
\AgdaSymbol{=}\AgdaSpace{}%
\AgdaFunction{buildReadNames}\AgdaSpace{}%
\AgdaSymbol{(}\AgdaFunction{run}\AgdaSpace{}%
\AgdaBound{x}\AgdaSpace{}%
\AgdaBound{s}\AgdaSymbol{)}\AgdaSpace{}%
\AgdaBound{b}\<%
\\
\>[8]\AgdaFunction{ws}\AgdaSpace{}%
\AgdaSymbol{:}\AgdaSpace{}%
\AgdaFunction{Files}\<%
\\
\>[8]\AgdaFunction{ws}\AgdaSpace{}%
\AgdaSymbol{=}\AgdaSpace{}%
\AgdaField{proj₂}\AgdaSpace{}%
\AgdaSymbol{(}\AgdaField{proj₁}\AgdaSpace{}%
\AgdaSymbol{(}\AgdaBound{oracle}\AgdaSpace{}%
\AgdaBound{x}\AgdaSymbol{)}\AgdaSpace{}%
\AgdaBound{s}\AgdaSymbol{)}\<%
\\
\>[0]\AgdaSymbol{...}\AgdaSpace{}%
\AgdaSymbol{|}\AgdaSpace{}%
\AgdaBound{all₂}\AgdaSpace{}%
\AgdaSymbol{=}\AgdaSpace{}%
\AgdaSymbol{(}\AgdaBound{x}\AgdaSpace{}%
\AgdaOperator{\AgdaInductiveConstructor{∷}}\AgdaSpace{}%
\AgdaBound{ls₁}\AgdaSymbol{)}\AgdaSpace{}%
\AgdaOperator{\AgdaInductiveConstructor{,}}\AgdaSpace{}%
\AgdaBound{ls₂}\AgdaSpace{}%
\AgdaOperator{\AgdaInductiveConstructor{,}}\AgdaSpace{}%
\AgdaFunction{cong}\AgdaSpace{}%
\AgdaSymbol{(}\AgdaBound{x}\AgdaSpace{}%
\AgdaOperator{\AgdaInductiveConstructor{∷\AgdaUnderscore{}}}\AgdaSymbol{)}\AgdaSpace{}%
\AgdaBound{ys≡ls₁++w∷ls₂}\AgdaSpace{}%
\AgdaOperator{\AgdaInductiveConstructor{,}}\AgdaSpace{}%
\AgdaFunction{subst}\AgdaSpace{}%
\AgdaSymbol{(λ}\AgdaSpace{}%
\AgdaBound{x₁}\AgdaSpace{}%
\AgdaSymbol{→}\AgdaSpace{}%
\AgdaBound{f₁}\AgdaSpace{}%
\AgdaOperator{\AgdaFunction{∈}}\AgdaSpace{}%
\AgdaBound{x₁}\AgdaSymbol{)}\AgdaSpace{}%
\AgdaSymbol{(}\AgdaFunction{cong}\AgdaSpace{}%
\AgdaSymbol{((}\AgdaFunction{map}\AgdaSpace{}%
\AgdaField{proj₁}\AgdaSymbol{)}\AgdaSpace{}%
\AgdaOperator{\AgdaFunction{∘}}\AgdaSpace{}%
\AgdaField{proj₁}\AgdaSymbol{)}\AgdaSpace{}%
\AgdaSymbol{(}\AgdaFunction{h₄}\AgdaSpace{}%
\AgdaSymbol{(}\AgdaFunction{script}\AgdaSpace{}%
\AgdaBound{ls₁}\AgdaSpace{}%
\AgdaBound{s}\AgdaSymbol{)}\AgdaSpace{}%
\AgdaSymbol{(}\AgdaFunction{script}\AgdaSpace{}%
\AgdaBound{ls₁}\AgdaSpace{}%
\AgdaSymbol{(}\AgdaFunction{run}\AgdaSpace{}%
\AgdaBound{x}\AgdaSpace{}%
\AgdaBound{s}\AgdaSymbol{))}\AgdaSpace{}%
\AgdaBound{w}\AgdaSpace{}%
\AgdaBound{all₂}\AgdaSymbol{))}\AgdaSpace{}%
\AgdaBound{f₁∈}\<%
\\
\>[0]\AgdaFunction{h₁}\AgdaSpace{}%
\AgdaBound{f₁}\AgdaSpace{}%
\AgdaBound{x}\AgdaSpace{}%
\AgdaBound{w}\AgdaSpace{}%
\AgdaSymbol{(}\AgdaBound{x₁}\AgdaSpace{}%
\AgdaOperator{\AgdaInductiveConstructor{∷}}\AgdaSpace{}%
\AgdaBound{xs}\AgdaSymbol{)}\AgdaSpace{}%
\AgdaBound{ys}\AgdaSpace{}%
\AgdaInductiveConstructor{[]}\AgdaSpace{}%
\AgdaBound{ls₂}\AgdaSpace{}%
\AgdaBound{f₁∈}\AgdaSpace{}%
\AgdaBound{dsj}\AgdaSpace{}%
\AgdaBound{x₁∷xs++ys≡w∷ls₂}\AgdaSpace{}%
\AgdaKeyword{with}\AgdaSpace{}%
\AgdaFunction{∷-injective}\AgdaSpace{}%
\AgdaBound{x₁∷xs++ys≡w∷ls₂}\<%
\\
\>[0]\AgdaSymbol{...}\AgdaSpace{}%
\AgdaSymbol{|}\AgdaSpace{}%
\AgdaBound{x₁≡w}\AgdaSpace{}%
\AgdaOperator{\AgdaInductiveConstructor{,}}\AgdaSpace{}%
\AgdaSymbol{\AgdaUnderscore{}}\AgdaSpace{}%
\AgdaSymbol{=}\AgdaSpace{}%
\AgdaInductiveConstructor{[]}\AgdaSpace{}%
\AgdaOperator{\AgdaInductiveConstructor{,}}\AgdaSpace{}%
\AgdaBound{xs}\AgdaSpace{}%
\AgdaOperator{\AgdaFunction{++}}\AgdaSpace{}%
\AgdaBound{x}\AgdaSpace{}%
\AgdaOperator{\AgdaInductiveConstructor{∷}}\AgdaSpace{}%
\AgdaBound{ys}\AgdaSpace{}%
\AgdaOperator{\AgdaInductiveConstructor{,}}\AgdaSpace{}%
\AgdaFunction{cong₂}\AgdaSpace{}%
\AgdaOperator{\AgdaInductiveConstructor{\AgdaUnderscore{}∷\AgdaUnderscore{}}}\AgdaSpace{}%
\AgdaBound{x₁≡w}\AgdaSpace{}%
\AgdaInductiveConstructor{refl}\AgdaSpace{}%
\AgdaOperator{\AgdaInductiveConstructor{,}}\AgdaSpace{}%
\AgdaBound{f₁∈}\<%
\\
\>[0]\AgdaFunction{h₁}\AgdaSpace{}%
\AgdaSymbol{\{}\AgdaBound{s}\AgdaSymbol{\}}\AgdaSpace{}%
\AgdaBound{f₁}\AgdaSpace{}%
\AgdaBound{x}\AgdaSpace{}%
\AgdaBound{w}\AgdaSpace{}%
\AgdaSymbol{(}\AgdaBound{x₁}\AgdaSpace{}%
\AgdaOperator{\AgdaInductiveConstructor{∷}}\AgdaSpace{}%
\AgdaBound{xs}\AgdaSymbol{)}\AgdaSpace{}%
\AgdaBound{ys}\AgdaSpace{}%
\AgdaSymbol{(}\AgdaBound{x₂}\AgdaSpace{}%
\AgdaOperator{\AgdaInductiveConstructor{∷}}\AgdaSpace{}%
\AgdaBound{ls₁}\AgdaSymbol{)}\AgdaSpace{}%
\AgdaBound{ls₂}\AgdaSpace{}%
\AgdaBound{f₁∈}\AgdaSpace{}%
\AgdaBound{dsj}\AgdaSpace{}%
\AgdaBound{x₁∷xs++ys≡x₂∷ls₁++w∷ls₂}\AgdaSpace{}%
\AgdaKeyword{with}\AgdaSpace{}%
\AgdaFunction{∷-injective}\AgdaSpace{}%
\AgdaBound{x₁∷xs++ys≡x₂∷ls₁++w∷ls₂}\<%
\\
\>[0]\AgdaSymbol{...}\AgdaSpace{}%
\AgdaSymbol{|}\AgdaSpace{}%
\AgdaBound{x₁≡x₂}\AgdaSpace{}%
\AgdaOperator{\AgdaInductiveConstructor{,}}\AgdaSpace{}%
\AgdaBound{xs++ys≡ls₁++w∷ls₂}\AgdaSpace{}%
\AgdaKeyword{with}\AgdaSpace{}%
\AgdaFunction{h₁}\AgdaSpace{}%
\AgdaSymbol{\{}\AgdaFunction{run}\AgdaSpace{}%
\AgdaBound{x₂}\AgdaSpace{}%
\AgdaBound{s}\AgdaSymbol{\}}\AgdaSpace{}%
\AgdaBound{f₁}\AgdaSpace{}%
\AgdaBound{x}\AgdaSpace{}%
\AgdaBound{w}\AgdaSpace{}%
\AgdaBound{xs}\AgdaSpace{}%
\AgdaBound{ys}\AgdaSpace{}%
\AgdaBound{ls₁}\AgdaSpace{}%
\AgdaBound{ls₂}\AgdaSpace{}%
\AgdaBound{f₁∈}\AgdaSpace{}%
\AgdaFunction{dsj₂}\AgdaSpace{}%
\AgdaBound{xs++ys≡ls₁++w∷ls₂}\<%
\\
\>[0][@{}l@{\AgdaIndent{0}}]%
\>[2]\AgdaKeyword{where}%
\>[692I]\AgdaFunction{dsj₂}\AgdaSpace{}%
\AgdaSymbol{:}\AgdaSpace{}%
\AgdaFunction{Disjoint}\AgdaSpace{}%
\AgdaSymbol{(}\AgdaFunction{cmdWriteNames}\AgdaSpace{}%
\AgdaBound{x}\AgdaSpace{}%
\AgdaSymbol{(}\AgdaFunction{script}\AgdaSpace{}%
\AgdaBound{xs}\AgdaSpace{}%
\AgdaSymbol{(}\AgdaFunction{run}\AgdaSpace{}%
\AgdaBound{x₂}\AgdaSpace{}%
\AgdaBound{s}\AgdaSymbol{)))}\AgdaSpace{}%
\AgdaSymbol{(}\AgdaFunction{buildReadNames}\AgdaSpace{}%
\AgdaSymbol{(}\AgdaFunction{run}\AgdaSpace{}%
\AgdaBound{x}\AgdaSpace{}%
\AgdaSymbol{(}\AgdaFunction{script}\AgdaSpace{}%
\AgdaBound{xs}\AgdaSpace{}%
\AgdaSymbol{(}\AgdaFunction{run}\AgdaSpace{}%
\AgdaBound{x₂}\AgdaSpace{}%
\AgdaBound{s}\AgdaSymbol{)))}\AgdaSpace{}%
\AgdaBound{ys}\AgdaSymbol{)}\<%
\\
\>[.][@{}l@{}]\<[692I]%
\>[8]\AgdaFunction{dsj₂}\AgdaSpace{}%
\AgdaSymbol{=}\AgdaSpace{}%
\AgdaSymbol{λ}\AgdaSpace{}%
\AgdaBound{x₃}\AgdaSpace{}%
\AgdaSymbol{→}\AgdaSpace{}%
\AgdaBound{dsj}%
\>[716I]\AgdaSymbol{(}\AgdaFunction{subst}\AgdaSpace{}%
\AgdaSymbol{(λ}\AgdaSpace{}%
\AgdaBound{x₄}\AgdaSpace{}%
\AgdaSymbol{→}\AgdaSpace{}%
\AgdaSymbol{\AgdaUnderscore{}}\AgdaSpace{}%
\AgdaOperator{\AgdaFunction{∈}}\AgdaSpace{}%
\AgdaFunction{cmdWriteNames}\AgdaSpace{}%
\AgdaBound{x}\AgdaSpace{}%
\AgdaSymbol{(}\AgdaFunction{script}\AgdaSpace{}%
\AgdaBound{xs}\AgdaSpace{}%
\AgdaSymbol{(}\AgdaFunction{run}\AgdaSpace{}%
\AgdaBound{x₄}\AgdaSpace{}%
\AgdaBound{s}\AgdaSymbol{)))}\AgdaSpace{}%
\AgdaSymbol{(}\AgdaFunction{sym}\AgdaSpace{}%
\AgdaBound{x₁≡x₂}\AgdaSymbol{)}\AgdaSpace{}%
\AgdaSymbol{(}\AgdaField{proj₁}\AgdaSpace{}%
\AgdaBound{x₃}\AgdaSymbol{)}\<%
\\
\>[.][@{}l@{}]\<[716I]%
\>[26]\AgdaOperator{\AgdaInductiveConstructor{,}}\AgdaSpace{}%
\AgdaFunction{subst}\AgdaSpace{}%
\AgdaSymbol{(λ}\AgdaSpace{}%
\AgdaBound{x₄}\AgdaSpace{}%
\AgdaSymbol{→}\AgdaSpace{}%
\AgdaSymbol{\AgdaUnderscore{}}\AgdaSpace{}%
\AgdaOperator{\AgdaFunction{∈}}\AgdaSpace{}%
\AgdaFunction{buildReadNames}\AgdaSpace{}%
\AgdaSymbol{(}\AgdaFunction{run}\AgdaSpace{}%
\AgdaBound{x}\AgdaSpace{}%
\AgdaSymbol{(}\AgdaFunction{script}\AgdaSpace{}%
\AgdaBound{xs}\AgdaSpace{}%
\AgdaSymbol{(}\AgdaFunction{run}\AgdaSpace{}%
\AgdaBound{x₄}\AgdaSpace{}%
\AgdaBound{s}\AgdaSymbol{)))}\AgdaSpace{}%
\AgdaBound{ys}\AgdaSymbol{)}\AgdaSpace{}%
\AgdaSymbol{(}\AgdaFunction{sym}\AgdaSpace{}%
\AgdaBound{x₁≡x₂}\AgdaSymbol{)}\AgdaSpace{}%
\AgdaSymbol{(}\AgdaField{proj₂}\AgdaSpace{}%
\AgdaBound{x₃}\AgdaSymbol{))}\<%
\\
\>[0]\AgdaSymbol{...}\AgdaSpace{}%
\AgdaSymbol{|}\AgdaSpace{}%
\AgdaBound{ls₃}\AgdaSpace{}%
\AgdaOperator{\AgdaInductiveConstructor{,}}\AgdaSpace{}%
\AgdaBound{ls₄}\AgdaSpace{}%
\AgdaOperator{\AgdaInductiveConstructor{,}}\AgdaSpace{}%
\AgdaBound{xs++x∷ys≡ls₃++w∷ls₄}\AgdaSpace{}%
\AgdaOperator{\AgdaInductiveConstructor{,}}\AgdaSpace{}%
\AgdaBound{f₁∈₂}\AgdaSpace{}%
\AgdaSymbol{=}\AgdaSpace{}%
\AgdaBound{x₂}\AgdaSpace{}%
\AgdaOperator{\AgdaInductiveConstructor{∷}}\AgdaSpace{}%
\AgdaBound{ls₃}\AgdaSpace{}%
\AgdaOperator{\AgdaInductiveConstructor{,}}\AgdaSpace{}%
\AgdaBound{ls₄}\AgdaSpace{}%
\AgdaOperator{\AgdaInductiveConstructor{,}}\AgdaSpace{}%
\AgdaFunction{cong₂}\AgdaSpace{}%
\AgdaOperator{\AgdaInductiveConstructor{\AgdaUnderscore{}∷\AgdaUnderscore{}}}\AgdaSpace{}%
\AgdaBound{x₁≡x₂}\AgdaSpace{}%
\AgdaBound{xs++x∷ys≡ls₃++w∷ls₄}\AgdaSpace{}%
\AgdaOperator{\AgdaInductiveConstructor{,}}\AgdaSpace{}%
\AgdaBound{f₁∈₂}\<%
\\
\\[\AgdaEmptyExtraSkip]%
\>[0]\AgdaFunction{script-≡}\AgdaSpace{}%
\AgdaSymbol{:}\AgdaSpace{}%
\AgdaSymbol{\{}\AgdaBound{sys}\AgdaSpace{}%
\AgdaBound{sys₁}\AgdaSpace{}%
\AgdaSymbol{:}\AgdaSpace{}%
\AgdaFunction{FileSystem}\AgdaSymbol{\}}\AgdaSpace{}%
\AgdaSymbol{(}\AgdaBound{b}\AgdaSpace{}%
\AgdaSymbol{:}\AgdaSpace{}%
\AgdaFunction{Build}\AgdaSymbol{)}\AgdaSpace{}%
\AgdaSymbol{->}\AgdaSpace{}%
\AgdaSymbol{(∀}\AgdaSpace{}%
\AgdaBound{f₁}\AgdaSpace{}%
\AgdaSymbol{→}\AgdaSpace{}%
\AgdaBound{sys}\AgdaSpace{}%
\AgdaBound{f₁}\AgdaSpace{}%
\AgdaOperator{\AgdaDatatype{≡}}\AgdaSpace{}%
\AgdaBound{sys₁}\AgdaSpace{}%
\AgdaBound{f₁}\AgdaSymbol{)}\AgdaSpace{}%
\AgdaSymbol{->}\AgdaSpace{}%
\AgdaSymbol{(∀}\AgdaSpace{}%
\AgdaBound{f₁}\AgdaSpace{}%
\AgdaSymbol{→}\AgdaSpace{}%
\AgdaFunction{script}\AgdaSpace{}%
\AgdaBound{b}\AgdaSpace{}%
\AgdaBound{sys}\AgdaSpace{}%
\AgdaBound{f₁}\AgdaSpace{}%
\AgdaOperator{\AgdaDatatype{≡}}\AgdaSpace{}%
\AgdaFunction{script}\AgdaSpace{}%
\AgdaBound{b}\AgdaSpace{}%
\AgdaBound{sys₁}\AgdaSpace{}%
\AgdaBound{f₁}\AgdaSymbol{)}\<%
\\
\>[0]\AgdaFunction{script-≡}\AgdaSpace{}%
\AgdaInductiveConstructor{[]}\AgdaSpace{}%
\AgdaBound{∀₁}\AgdaSpace{}%
\AgdaBound{f₁}\AgdaSpace{}%
\AgdaSymbol{=}\AgdaSpace{}%
\AgdaBound{∀₁}\AgdaSpace{}%
\AgdaBound{f₁}\<%
\\
\>[0]\AgdaFunction{script-≡}\AgdaSpace{}%
\AgdaSymbol{(}\AgdaBound{x}\AgdaSpace{}%
\AgdaOperator{\AgdaInductiveConstructor{∷}}\AgdaSpace{}%
\AgdaBound{b}\AgdaSymbol{)}\AgdaSpace{}%
\AgdaBound{∀₁}\AgdaSpace{}%
\AgdaSymbol{=}\AgdaSpace{}%
\AgdaFunction{script-≡}\AgdaSpace{}%
\AgdaBound{b}\AgdaSpace{}%
\AgdaSymbol{λ}\AgdaSpace{}%
\AgdaBound{f₁}\AgdaSpace{}%
\AgdaSymbol{→}\AgdaSpace{}%
\AgdaFunction{run-≡}\AgdaSpace{}%
\AgdaBound{x}\AgdaSpace{}%
\AgdaBound{∀₁}\AgdaSpace{}%
\AgdaBound{f₁}\<%
\end{code}

\begin{code}[hide]%
\>[0]\AgdaKeyword{open}\AgdaSpace{}%
\AgdaKeyword{import}\AgdaSpace{}%
\AgdaModule{Functional.State}\AgdaSpace{}%
\AgdaSymbol{as}\AgdaSpace{}%
\AgdaModule{St}\AgdaSpace{}%
\AgdaKeyword{using}\AgdaSpace{}%
\AgdaSymbol{(}\AgdaFunction{Oracle}\AgdaSpace{}%
\AgdaSymbol{;}\AgdaSpace{}%
\AgdaFunction{Cmd}\AgdaSpace{}%
\AgdaSymbol{;}\AgdaSpace{}%
\AgdaFunction{extend}\AgdaSpace{}%
\AgdaSymbol{;}\AgdaSpace{}%
\AgdaFunction{read}\AgdaSpace{}%
\AgdaSymbol{;}\AgdaSpace{}%
\AgdaFunction{Memory}\AgdaSymbol{)}\<%
\\
\\[\AgdaEmptyExtraSkip]%
\>[0]\AgdaKeyword{module}\AgdaSpace{}%
\AgdaModule{Functional.Script.Proofs}\AgdaSpace{}%
\AgdaSymbol{(}\AgdaBound{oracle}\AgdaSpace{}%
\AgdaSymbol{:}\AgdaSpace{}%
\AgdaFunction{Oracle}\AgdaSymbol{)}\AgdaSpace{}%
\AgdaKeyword{where}\<%
\\
\\[\AgdaEmptyExtraSkip]%
\>[0]\AgdaKeyword{open}\AgdaSpace{}%
\AgdaKeyword{import}\AgdaSpace{}%
\AgdaModule{Agda.Builtin.Equality}\<%
\\
\>[0]\AgdaKeyword{open}\AgdaSpace{}%
\AgdaKeyword{import}\AgdaSpace{}%
\AgdaModule{Functional.State.Helpers}\AgdaSpace{}%
\AgdaSymbol{(}\AgdaBound{oracle}\AgdaSymbol{)}\AgdaSpace{}%
\AgdaKeyword{using}\AgdaSpace{}%
\AgdaSymbol{(}cmdWriteNames \AgdaSymbol{;} run \AgdaSymbol{;} cmdReadNames \AgdaSymbol{;} cmdWrites\AgdaSymbol{)}\<%
\\
\>[0]\AgdaKeyword{open}\AgdaSpace{}%
\AgdaKeyword{import}\AgdaSpace{}%
\AgdaModule{Functional.State.Properties}\AgdaSpace{}%
\AgdaSymbol{(}\AgdaBound{oracle}\AgdaSymbol{)}\AgdaSpace{}%
\AgdaKeyword{using}\AgdaSpace{}%
\AgdaSymbol{(}lemma5 \AgdaSymbol{;} lemma3\AgdaSymbol{)}\<%
\\
\>[0]\AgdaKeyword{open}\AgdaSpace{}%
\AgdaKeyword{import}\AgdaSpace{}%
\AgdaModule{Functional.Script.Exec}\AgdaSpace{}%
\AgdaSymbol{(}\AgdaBound{oracle}\AgdaSymbol{)}\AgdaSpace{}%
\AgdaKeyword{using}\AgdaSpace{}%
\AgdaSymbol{(}script \AgdaSymbol{;} buildWriteNames \AgdaSymbol{;} buildReadNames\AgdaSymbol{)}\<%
\\
\>[0]\AgdaKeyword{open}\AgdaSpace{}%
\AgdaKeyword{import}\AgdaSpace{}%
\AgdaModule{Functional.Script.Hazard}\AgdaSpace{}%
\AgdaSymbol{(}\AgdaBound{oracle}\AgdaSymbol{)}\AgdaSpace{}%
\AgdaKeyword{using}\AgdaSpace{}%
\AgdaSymbol{(}HazardFree \AgdaSymbol{;} FileInfo\AgdaSymbol{)}\<%
\\
\>[0]\AgdaKeyword{open}\AgdaSpace{}%
\AgdaKeyword{import}\AgdaSpace{}%
\AgdaModule{Functional.Script.Hazard.Properties}\AgdaSpace{}%
\AgdaSymbol{(}\AgdaBound{oracle}\AgdaSymbol{)}\AgdaSpace{}%
\AgdaKeyword{using}\AgdaSpace{}%
\AgdaSymbol{(}hf-∷ʳ-l \AgdaSymbol{;} hf-drop-mid \AgdaSymbol{;} hf=>disjoint \AgdaSymbol{;} hf=>disjointRW \AgdaSymbol{;} hf=>disjointWW \AgdaSymbol{;} hf=>disjointWR \AgdaSymbol{;} hf-∷ʳ-r\AgdaSymbol{)}\<%
\\
\>[0]\AgdaKeyword{open}\AgdaSpace{}%
\AgdaKeyword{import}\AgdaSpace{}%
\AgdaModule{Data.Sum}\AgdaSpace{}%
\AgdaKeyword{using}\AgdaSpace{}%
\AgdaSymbol{(}\AgdaInductiveConstructor{inj₁}\AgdaSpace{}%
\AgdaSymbol{;}\AgdaSpace{}%
\AgdaInductiveConstructor{inj₂}\AgdaSymbol{)}\<%
\\
\>[0]\AgdaKeyword{open}\AgdaSpace{}%
\AgdaKeyword{import}\AgdaSpace{}%
\AgdaModule{Data.Maybe}\AgdaSpace{}%
\AgdaKeyword{using}\AgdaSpace{}%
\AgdaSymbol{(}\AgdaInductiveConstructor{just}\AgdaSymbol{)}\<%
\\
\>[0]\AgdaKeyword{open}\AgdaSpace{}%
\AgdaKeyword{import}\AgdaSpace{}%
\AgdaModule{Data.List}\AgdaSpace{}%
\AgdaKeyword{using}\AgdaSpace{}%
\AgdaSymbol{(}\AgdaOperator{\AgdaFunction{\AgdaUnderscore{}∷ʳ\AgdaUnderscore{}}}\AgdaSpace{}%
\AgdaSymbol{;}\AgdaSpace{}%
\AgdaDatatype{List}\AgdaSpace{}%
\AgdaSymbol{;}\AgdaSpace{}%
\AgdaFunction{foldr}\AgdaSymbol{)}\<%
\\
\>[0]\AgdaKeyword{open}\AgdaSpace{}%
\AgdaKeyword{import}\AgdaSpace{}%
\AgdaModule{Data.List.Properties}\AgdaSpace{}%
\AgdaKeyword{using}\AgdaSpace{}%
\AgdaSymbol{(}\AgdaFunction{unfold-reverse}\AgdaSpace{}%
\AgdaSymbol{;}\AgdaSpace{}%
\AgdaFunction{reverse-involutive}\AgdaSpace{}%
\AgdaSymbol{;}\AgdaSpace{}%
\AgdaFunction{++-identityʳ}\AgdaSpace{}%
\AgdaSymbol{;}\AgdaSpace{}%
\AgdaFunction{length-reverse}\AgdaSpace{}%
\AgdaSymbol{;}\AgdaSpace{}%
\AgdaFunction{∷-injectiveˡ}\AgdaSpace{}%
\AgdaSymbol{;}\AgdaSpace{}%
\AgdaFunction{∷-injectiveʳ}\AgdaSymbol{)}\<%
\\
\>[0]\AgdaKeyword{open}\AgdaSpace{}%
\AgdaKeyword{import}\AgdaSpace{}%
\AgdaModule{Data.List.Relation.Binary.Permutation.Propositional}\AgdaSpace{}%
\AgdaKeyword{using}\AgdaSpace{}%
\AgdaSymbol{(}\AgdaOperator{\AgdaDatatype{\AgdaUnderscore{}↭\AgdaUnderscore{}}}\AgdaSpace{}%
\AgdaSymbol{;}\AgdaSpace{}%
\AgdaFunction{↭-sym}\AgdaSpace{}%
\AgdaSymbol{;}\AgdaSpace{}%
\AgdaFunction{↭-refl}\AgdaSpace{}%
\AgdaSymbol{;}\AgdaSpace{}%
\AgdaFunction{↭-trans}\AgdaSymbol{)}\<%
\\
\>[0]\AgdaKeyword{open}\AgdaSpace{}%
\AgdaKeyword{import}\AgdaSpace{}%
\AgdaModule{Data.List.Relation.Binary.Permutation.Propositional.Properties}\AgdaSpace{}%
\AgdaKeyword{using}\AgdaSpace{}%
\AgdaSymbol{(}\AgdaFunction{Any-resp-↭}\AgdaSpace{}%
\AgdaSymbol{;}\AgdaSpace{}%
\AgdaFunction{drop-mid}\AgdaSpace{}%
\AgdaSymbol{;}\AgdaSpace{}%
\AgdaFunction{↭-length}\AgdaSpace{}%
\AgdaSymbol{;}\AgdaSpace{}%
\AgdaFunction{∈-resp-↭}\AgdaSpace{}%
\AgdaSymbol{;}\AgdaSpace{}%
\AgdaFunction{++↭ʳ++}\AgdaSymbol{)}\<%
\\
\>[0]\AgdaKeyword{open}\AgdaSpace{}%
\AgdaKeyword{import}\AgdaSpace{}%
\AgdaModule{Data.List.Relation.Unary.Unique.Propositional}\AgdaSpace{}%
\AgdaKeyword{using}\AgdaSpace{}%
\AgdaSymbol{(}\AgdaDatatype{Unique}\AgdaSpace{}%
\AgdaSymbol{;}\AgdaSpace{}%
\AgdaOperator{\AgdaInductiveConstructor{\AgdaUnderscore{}∷\AgdaUnderscore{}}}\AgdaSpace{}%
\AgdaSymbol{;}\AgdaSpace{}%
\AgdaFunction{head}\AgdaSpace{}%
\AgdaSymbol{;}\AgdaSpace{}%
\AgdaFunction{tail}\AgdaSpace{}%
\AgdaSymbol{;}\AgdaSpace{}%
\AgdaInductiveConstructor{[]}\AgdaSymbol{)}\<%
\\
\>[0]\AgdaKeyword{open}\AgdaSpace{}%
\AgdaKeyword{import}\AgdaSpace{}%
\AgdaModule{Data.List.Relation.Unary.Unique.Propositional.Properties}\AgdaSpace{}%
\AgdaKeyword{using}\AgdaSpace{}%
\AgdaSymbol{(}\AgdaFunction{++⁺}\AgdaSymbol{)}\<%
\\
\>[0]\AgdaKeyword{open}\AgdaSpace{}%
\AgdaKeyword{import}\AgdaSpace{}%
\AgdaModule{Data.List}\AgdaSpace{}%
\AgdaKeyword{using}\AgdaSpace{}%
\AgdaSymbol{(}\AgdaFunction{map}\AgdaSpace{}%
\AgdaSymbol{;}\AgdaSpace{}%
\AgdaFunction{reverse}\AgdaSpace{}%
\AgdaSymbol{;}\AgdaSpace{}%
\AgdaFunction{length}\AgdaSpace{}%
\AgdaSymbol{;}\AgdaSpace{}%
\AgdaInductiveConstructor{[]}\AgdaSpace{}%
\AgdaSymbol{;}\AgdaSpace{}%
\AgdaOperator{\AgdaInductiveConstructor{\AgdaUnderscore{}∷\AgdaUnderscore{}}}\AgdaSpace{}%
\AgdaSymbol{;}\AgdaSpace{}%
\AgdaOperator{\AgdaFunction{\AgdaUnderscore{}++\AgdaUnderscore{}}}\AgdaSpace{}%
\AgdaSymbol{;}\AgdaSpace{}%
\AgdaOperator{\AgdaFunction{[\AgdaUnderscore{}]}}\AgdaSymbol{)}\<%
\\
\>[0]\AgdaKeyword{open}\AgdaSpace{}%
\AgdaKeyword{import}\AgdaSpace{}%
\AgdaModule{Data.Product}\AgdaSpace{}%
\AgdaKeyword{using}\AgdaSpace{}%
\AgdaSymbol{(}\AgdaField{proj₁}\AgdaSpace{}%
\AgdaSymbol{;}\AgdaSpace{}%
\AgdaField{proj₂}\AgdaSpace{}%
\AgdaSymbol{;}\AgdaSpace{}%
\AgdaOperator{\AgdaInductiveConstructor{\AgdaUnderscore{},\AgdaUnderscore{}}}\AgdaSymbol{)}\<%
\\
\>[0]\AgdaKeyword{open}\AgdaSpace{}%
\AgdaKeyword{import}\AgdaSpace{}%
\AgdaModule{Data.List.Relation.Binary.Disjoint.Propositional}\AgdaSpace{}%
\AgdaKeyword{using}\AgdaSpace{}%
\AgdaSymbol{(}\AgdaFunction{Disjoint}\AgdaSymbol{)}\<%
\\
\>[0]\AgdaKeyword{open}\AgdaSpace{}%
\AgdaKeyword{import}\AgdaSpace{}%
\AgdaModule{Data.List.Membership.Propositional}\AgdaSpace{}%
\AgdaKeyword{using}\AgdaSpace{}%
\AgdaSymbol{(}\AgdaOperator{\AgdaFunction{\AgdaUnderscore{}∈\AgdaUnderscore{}}}\AgdaSpace{}%
\AgdaSymbol{;}\AgdaSpace{}%
\AgdaOperator{\AgdaFunction{\AgdaUnderscore{}∉\AgdaUnderscore{}}}\AgdaSymbol{)}\<%
\\
\>[0]\AgdaKeyword{open}\AgdaSpace{}%
\AgdaKeyword{import}\AgdaSpace{}%
\AgdaModule{Data.List.Membership.Propositional.Properties}\AgdaSpace{}%
\AgdaKeyword{using}\AgdaSpace{}%
\AgdaSymbol{(}\AgdaFunction{∈-∃++}\AgdaSpace{}%
\AgdaSymbol{;}\AgdaSpace{}%
\AgdaFunction{∈-++⁻}\AgdaSpace{}%
\AgdaSymbol{;}\AgdaSpace{}%
\AgdaFunction{∈-++⁺ˡ}\AgdaSpace{}%
\AgdaSymbol{;}\AgdaSpace{}%
\AgdaFunction{∈-++⁺ʳ}\AgdaSymbol{)}\<%
\\
\>[0]\AgdaKeyword{open}\AgdaSpace{}%
\AgdaKeyword{import}\AgdaSpace{}%
\AgdaModule{Data.List.Relation.Unary.Any}\AgdaSpace{}%
\AgdaSymbol{as}\AgdaSpace{}%
\AgdaModule{Any}\AgdaSpace{}%
\AgdaKeyword{using}\AgdaSpace{}%
\AgdaSymbol{(}\AgdaInductiveConstructor{here}\AgdaSpace{}%
\AgdaSymbol{;}\AgdaSpace{}%
\AgdaInductiveConstructor{there}\AgdaSymbol{)}\<%
\\
\>[0]\AgdaKeyword{open}\AgdaSpace{}%
\AgdaKeyword{import}\AgdaSpace{}%
\AgdaModule{Data.List.Relation.Unary.Any.Properties}\AgdaSpace{}%
\AgdaKeyword{using}\AgdaSpace{}%
\AgdaSymbol{(}\AgdaFunction{reverse⁺}\AgdaSpace{}%
\AgdaSymbol{;}\AgdaSpace{}%
\AgdaFunction{reverse⁻}\AgdaSymbol{)}\<%
\\
\>[0]\AgdaKeyword{open}\AgdaSpace{}%
\AgdaKeyword{import}\AgdaSpace{}%
\AgdaModule{Relation.Binary.PropositionalEquality}\AgdaSpace{}%
\AgdaKeyword{using}\AgdaSpace{}%
\AgdaSymbol{(}\AgdaFunction{subst}\AgdaSpace{}%
\AgdaSymbol{;}\AgdaSpace{}%
\AgdaFunction{subst₂}\AgdaSpace{}%
\AgdaSymbol{;}\AgdaSpace{}%
\AgdaFunction{sym}\AgdaSpace{}%
\AgdaSymbol{;}\AgdaSpace{}%
\AgdaFunction{trans}\AgdaSpace{}%
\AgdaSymbol{;}\AgdaSpace{}%
\AgdaFunction{decSetoid}\AgdaSpace{}%
\AgdaSymbol{;}\AgdaSpace{}%
\AgdaFunction{cong-app}\AgdaSpace{}%
\AgdaSymbol{;}\AgdaSpace{}%
\AgdaFunction{cong}\AgdaSymbol{)}\<%
\\
\>[0]\AgdaKeyword{open}\AgdaSpace{}%
\AgdaKeyword{import}\AgdaSpace{}%
\AgdaModule{Data.List.Relation.Unary.All}\AgdaSpace{}%
\AgdaSymbol{as}\AgdaSpace{}%
\AgdaModule{All}\AgdaSpace{}%
\AgdaKeyword{using}\AgdaSpace{}%
\AgdaSymbol{(}\AgdaDatatype{All}\AgdaSpace{}%
\AgdaSymbol{;}\AgdaSpace{}%
\AgdaOperator{\AgdaInductiveConstructor{\AgdaUnderscore{}∷\AgdaUnderscore{}}}\AgdaSpace{}%
\AgdaSymbol{;}\AgdaSpace{}%
\AgdaFunction{lookup}\AgdaSymbol{)}\<%
\\
\>[0]\AgdaKeyword{open}\AgdaSpace{}%
\AgdaKeyword{import}\AgdaSpace{}%
\AgdaModule{Data.List.Relation.Unary.All.Properties}\AgdaSpace{}%
\AgdaSymbol{as}\AgdaSpace{}%
\AgdaModule{AllP}\AgdaSpace{}%
\AgdaKeyword{hiding}\AgdaSpace{}%
\AgdaSymbol{(}\AgdaFunction{++⁺}\AgdaSymbol{)}\<%
\\
\>[0]\AgdaKeyword{open}\AgdaSpace{}%
\AgdaKeyword{import}\AgdaSpace{}%
\AgdaModule{Relation.Nullary}\AgdaSpace{}%
\AgdaKeyword{using}\AgdaSpace{}%
\AgdaSymbol{(}\AgdaOperator{\AgdaFunction{¬\AgdaUnderscore{}}}\AgdaSymbol{)}\<%
\\
\>[0]\AgdaKeyword{open}\AgdaSpace{}%
\AgdaKeyword{import}\AgdaSpace{}%
\AgdaModule{Functional.Build}\AgdaSpace{}%
\AgdaSymbol{(}\AgdaBound{oracle}\AgdaSymbol{)}\AgdaSpace{}%
\AgdaKeyword{using}\AgdaSpace{}%
\AgdaSymbol{(}Build \AgdaSymbol{;} UniqueEvidence \AgdaSymbol{;} PreCond\AgdaSymbol{)}\<%
\\
\>[0]\AgdaKeyword{open}\AgdaSpace{}%
\AgdaKeyword{import}\AgdaSpace{}%
\AgdaModule{Functional.Script.Properties}\AgdaSpace{}%
\AgdaSymbol{(}\AgdaBound{oracle}\AgdaSymbol{)}\AgdaSpace{}%
\AgdaKeyword{using}\AgdaSpace{}%
\AgdaSymbol{(}exec-f₁≡ \AgdaSymbol{;} exec-≡f₁ \AgdaSymbol{;} writes≡ \AgdaSymbol{;} exec-∷≡ \AgdaSymbol{;} exec≡₃\AgdaSymbol{)}\<%
\\
\>[0]\AgdaKeyword{open}\AgdaSpace{}%
\AgdaKeyword{import}\AgdaSpace{}%
\AgdaModule{Data.List.Relation.Binary.Subset.Propositional}\AgdaSpace{}%
\AgdaKeyword{using}\AgdaSpace{}%
\AgdaSymbol{(}\AgdaOperator{\AgdaFunction{\AgdaUnderscore{}⊆\AgdaUnderscore{}}}\AgdaSymbol{)}\<%
\\
\>[0]\AgdaKeyword{open}\AgdaSpace{}%
\AgdaKeyword{import}\AgdaSpace{}%
\AgdaModule{Data.String.Properties}\AgdaSpace{}%
\AgdaKeyword{using}\AgdaSpace{}%
\AgdaSymbol{(}\AgdaOperator{\AgdaFunction{\AgdaUnderscore{}≟\AgdaUnderscore{}}}\AgdaSymbol{)}\<%
\\
\>[0]\AgdaKeyword{open}\AgdaSpace{}%
\AgdaKeyword{import}\AgdaSpace{}%
\AgdaModule{Data.List.Membership.DecSetoid}\AgdaSpace{}%
\AgdaSymbol{(}\AgdaFunction{decSetoid}\AgdaSpace{}%
\AgdaOperator{\AgdaFunction{\AgdaUnderscore{}≟\AgdaUnderscore{}}}\AgdaSymbol{)}\AgdaSpace{}%
\AgdaKeyword{using}\AgdaSpace{}%
\AgdaSymbol{(}\AgdaUnderscore{}∈?\AgdaUnderscore{}\AgdaSymbol{)}\<%
\\
\>[0]\AgdaKeyword{open}\AgdaSpace{}%
\AgdaKeyword{import}\AgdaSpace{}%
\AgdaModule{Relation.Nullary}\AgdaSpace{}%
\AgdaKeyword{using}\AgdaSpace{}%
\AgdaSymbol{(}\AgdaInductiveConstructor{yes}\AgdaSpace{}%
\AgdaSymbol{;}\AgdaSpace{}%
\AgdaInductiveConstructor{no}\AgdaSymbol{)}\<%
\\
\>[0]\AgdaKeyword{open}\AgdaSpace{}%
\AgdaKeyword{import}\AgdaSpace{}%
\AgdaModule{Data.Product.Properties}\AgdaSpace{}%
\AgdaKeyword{using}\AgdaSpace{}%
\AgdaSymbol{(}\AgdaFunction{,-injectiveʳ}\AgdaSpace{}%
\AgdaSymbol{;}\AgdaSpace{}%
\AgdaFunction{,-injectiveˡ}\AgdaSymbol{)}\<%
\\
\>[0]\AgdaKeyword{open}\AgdaSpace{}%
\AgdaKeyword{import}\AgdaSpace{}%
\AgdaModule{Relation.Nullary.Negation}\AgdaSpace{}%
\AgdaKeyword{using}\AgdaSpace{}%
\AgdaSymbol{(}\AgdaFunction{contradiction}\AgdaSymbol{)}\<%
\\
\>[0]\AgdaKeyword{open}\AgdaSpace{}%
\AgdaKeyword{import}\AgdaSpace{}%
\AgdaModule{Function.Base}\AgdaSpace{}%
\AgdaKeyword{using}\AgdaSpace{}%
\AgdaSymbol{(}\AgdaOperator{\AgdaFunction{\AgdaUnderscore{}∘\AgdaUnderscore{}}}\AgdaSymbol{)}\<%
\\
\\[\AgdaEmptyExtraSkip]%
\>[0]\AgdaComment{\{- If we follow the concept of hazardfree, it doesn't make sense, for 
  a hazardfree build to contain duplicates because those commands would need to do the
same thing, which would cause a hazard. + rattle doesn't run duplicates, so ....
lets just assume builds are unique. make my life easier...-\}}\<%
\\
\>[0]\AgdaFunction{all-drop-mid}\AgdaSpace{}%
\AgdaSymbol{:}\AgdaSpace{}%
\AgdaSymbol{∀}\AgdaSpace{}%
\AgdaSymbol{(}\AgdaBound{xs}\AgdaSpace{}%
\AgdaSymbol{:}\AgdaSpace{}%
\AgdaFunction{Build}\AgdaSymbol{)}\AgdaSpace{}%
\AgdaSymbol{\{}\AgdaBound{ys}\AgdaSymbol{\}}\AgdaSpace{}%
\AgdaSymbol{\{}\AgdaBound{x}\AgdaSymbol{\}}\AgdaSpace{}%
\AgdaSymbol{\{}\AgdaBound{x₁}\AgdaSymbol{\}}\AgdaSpace{}%
\AgdaSymbol{→}\AgdaSpace{}%
\AgdaDatatype{All}\AgdaSpace{}%
\AgdaSymbol{(λ}\AgdaSpace{}%
\AgdaBound{y}\AgdaSpace{}%
\AgdaSymbol{→}\AgdaSpace{}%
\AgdaOperator{\AgdaFunction{¬}}\AgdaSpace{}%
\AgdaBound{x₁}\AgdaSpace{}%
\AgdaOperator{\AgdaDatatype{≡}}\AgdaSpace{}%
\AgdaBound{y}\AgdaSymbol{)}\AgdaSpace{}%
\AgdaSymbol{(}\AgdaBound{xs}\AgdaSpace{}%
\AgdaOperator{\AgdaFunction{++}}\AgdaSpace{}%
\AgdaBound{x}\AgdaSpace{}%
\AgdaOperator{\AgdaInductiveConstructor{∷}}\AgdaSpace{}%
\AgdaBound{ys}\AgdaSymbol{)}\AgdaSpace{}%
\AgdaSymbol{→}\AgdaSpace{}%
\AgdaDatatype{All}\AgdaSpace{}%
\AgdaSymbol{(λ}\AgdaSpace{}%
\AgdaBound{y}\AgdaSpace{}%
\AgdaSymbol{→}\AgdaSpace{}%
\AgdaOperator{\AgdaFunction{¬}}\AgdaSpace{}%
\AgdaBound{x₁}\AgdaSpace{}%
\AgdaOperator{\AgdaDatatype{≡}}\AgdaSpace{}%
\AgdaBound{y}\AgdaSymbol{)}\AgdaSpace{}%
\AgdaSymbol{(}\AgdaBound{xs}\AgdaSpace{}%
\AgdaOperator{\AgdaFunction{++}}\AgdaSpace{}%
\AgdaBound{ys}\AgdaSymbol{)}\<%
\\
\>[0]\AgdaFunction{all-drop-mid}\AgdaSpace{}%
\AgdaInductiveConstructor{[]}\AgdaSpace{}%
\AgdaBound{all₁}\AgdaSpace{}%
\AgdaSymbol{=}\AgdaSpace{}%
\AgdaFunction{All.tail}\AgdaSpace{}%
\AgdaBound{all₁}\<%
\\
\>[0]\AgdaFunction{all-drop-mid}\AgdaSpace{}%
\AgdaSymbol{(}\AgdaBound{x₂}\AgdaSpace{}%
\AgdaOperator{\AgdaInductiveConstructor{∷}}\AgdaSpace{}%
\AgdaBound{xs}\AgdaSymbol{)}\AgdaSpace{}%
\AgdaBound{all₁}\AgdaSpace{}%
\AgdaSymbol{=}\AgdaSpace{}%
\AgdaSymbol{(}\AgdaFunction{All.head}\AgdaSpace{}%
\AgdaBound{all₁}\AgdaSymbol{)}\AgdaSpace{}%
\AgdaOperator{\AgdaInductiveConstructor{∷}}\AgdaSpace{}%
\AgdaSymbol{(}\AgdaFunction{all-drop-mid}\AgdaSpace{}%
\AgdaBound{xs}\AgdaSpace{}%
\AgdaSymbol{(}\AgdaFunction{All.tail}\AgdaSpace{}%
\AgdaBound{all₁}\AgdaSymbol{))}\<%
\\
\\[\AgdaEmptyExtraSkip]%
\>[0]\AgdaFunction{unique-drop-mid}\AgdaSpace{}%
\AgdaSymbol{:}\AgdaSpace{}%
\AgdaSymbol{∀}\AgdaSpace{}%
\AgdaSymbol{(}\AgdaBound{xs}\AgdaSpace{}%
\AgdaSymbol{:}\AgdaSpace{}%
\AgdaFunction{Build}\AgdaSymbol{)}\AgdaSpace{}%
\AgdaSymbol{\{}\AgdaBound{ys}\AgdaSymbol{\}}\AgdaSpace{}%
\AgdaSymbol{\{}\AgdaBound{x}\AgdaSymbol{\}}\AgdaSpace{}%
\AgdaSymbol{→}\AgdaSpace{}%
\AgdaDatatype{Unique}\AgdaSpace{}%
\AgdaSymbol{(}\AgdaBound{xs}\AgdaSpace{}%
\AgdaOperator{\AgdaFunction{++}}\AgdaSpace{}%
\AgdaBound{x}\AgdaSpace{}%
\AgdaOperator{\AgdaInductiveConstructor{∷}}\AgdaSpace{}%
\AgdaBound{ys}\AgdaSymbol{)}\AgdaSpace{}%
\AgdaSymbol{→}\AgdaSpace{}%
\AgdaDatatype{Unique}\AgdaSpace{}%
\AgdaSymbol{(}\AgdaBound{xs}\AgdaSpace{}%
\AgdaOperator{\AgdaFunction{++}}\AgdaSpace{}%
\AgdaBound{ys}\AgdaSymbol{)}\<%
\\
\>[0]\AgdaFunction{unique-drop-mid}\AgdaSpace{}%
\AgdaInductiveConstructor{[]}\AgdaSpace{}%
\AgdaSymbol{(}\AgdaBound{x₁}\AgdaSpace{}%
\AgdaOperator{\AgdaInductiveConstructor{∷}}\AgdaSpace{}%
\AgdaBound{u}\AgdaSymbol{)}\AgdaSpace{}%
\AgdaSymbol{=}\AgdaSpace{}%
\AgdaBound{u}\<%
\\
\>[0]\AgdaFunction{unique-drop-mid}\AgdaSpace{}%
\AgdaSymbol{(}\AgdaBound{x₁}\AgdaSpace{}%
\AgdaOperator{\AgdaInductiveConstructor{∷}}\AgdaSpace{}%
\AgdaBound{xs}\AgdaSymbol{)}\AgdaSpace{}%
\AgdaBound{u}\AgdaSpace{}%
\AgdaSymbol{=}\AgdaSpace{}%
\AgdaSymbol{(}\AgdaFunction{all-drop-mid}\AgdaSpace{}%
\AgdaBound{xs}\AgdaSpace{}%
\AgdaSymbol{(}\AgdaFunction{head}\AgdaSpace{}%
\AgdaBound{u}\AgdaSymbol{))}\AgdaSpace{}%
\AgdaOperator{\AgdaInductiveConstructor{∷}}\AgdaSpace{}%
\AgdaSymbol{(}\AgdaFunction{unique-drop-mid}\AgdaSpace{}%
\AgdaBound{xs}\AgdaSpace{}%
\AgdaSymbol{(}\AgdaFunction{tail}\AgdaSpace{}%
\AgdaBound{u}\AgdaSymbol{))}\<%
\\
\\[\AgdaEmptyExtraSkip]%
\>[0]\AgdaFunction{unique→disjoint}\AgdaSpace{}%
\AgdaSymbol{:}\AgdaSpace{}%
\AgdaSymbol{∀}\AgdaSpace{}%
\AgdaSymbol{\{}\AgdaBound{x₁}\AgdaSpace{}%
\AgdaSymbol{:}\AgdaSpace{}%
\AgdaFunction{Cmd}\AgdaSymbol{\}}\AgdaSpace{}%
\AgdaBound{xs}\AgdaSpace{}%
\AgdaSymbol{→}\AgdaSpace{}%
\AgdaDatatype{All}\AgdaSpace{}%
\AgdaSymbol{(λ}\AgdaSpace{}%
\AgdaBound{y}\AgdaSpace{}%
\AgdaSymbol{→}\AgdaSpace{}%
\AgdaOperator{\AgdaFunction{¬}}\AgdaSpace{}%
\AgdaBound{x₁}\AgdaSpace{}%
\AgdaOperator{\AgdaDatatype{≡}}\AgdaSpace{}%
\AgdaBound{y}\AgdaSymbol{)}\AgdaSpace{}%
\AgdaBound{xs}\AgdaSpace{}%
\AgdaSymbol{→}\AgdaSpace{}%
\AgdaFunction{Disjoint}\AgdaSpace{}%
\AgdaBound{xs}\AgdaSpace{}%
\AgdaSymbol{(}\AgdaBound{x₁}\AgdaSpace{}%
\AgdaOperator{\AgdaInductiveConstructor{∷}}\AgdaSpace{}%
\AgdaInductiveConstructor{[]}\AgdaSymbol{)}\<%
\\
\>[0]\AgdaFunction{unique→disjoint}\AgdaSpace{}%
\AgdaInductiveConstructor{[]}\AgdaSpace{}%
\AgdaInductiveConstructor{All.[]}\AgdaSpace{}%
\AgdaSymbol{=}\AgdaSpace{}%
\AgdaSymbol{λ}\AgdaSpace{}%
\AgdaSymbol{()}\<%
\\
\>[0]\AgdaFunction{unique→disjoint}\AgdaSpace{}%
\AgdaSymbol{(}\AgdaBound{x}\AgdaSpace{}%
\AgdaOperator{\AgdaInductiveConstructor{∷}}\AgdaSpace{}%
\AgdaBound{xs}\AgdaSymbol{)}\AgdaSpace{}%
\AgdaSymbol{(}\AgdaBound{¬x₁≡x}\AgdaSpace{}%
\AgdaOperator{\AgdaInductiveConstructor{∷}}\AgdaSpace{}%
\AgdaBound{all₁}\AgdaSymbol{)}\AgdaSpace{}%
\AgdaBound{x₂}\AgdaSpace{}%
\AgdaSymbol{=}\AgdaSpace{}%
\AgdaFunction{unique→disjoint}\AgdaSpace{}%
\AgdaBound{xs}\AgdaSpace{}%
\AgdaBound{all₁}\AgdaSpace{}%
\AgdaSymbol{(}\AgdaFunction{Any.tail}\AgdaSpace{}%
\AgdaSymbol{(λ}\AgdaSpace{}%
\AgdaBound{v≡x}\AgdaSpace{}%
\AgdaSymbol{→}\AgdaSpace{}%
\AgdaBound{¬x₁≡x}\AgdaSpace{}%
\AgdaSymbol{(}\AgdaFunction{trans}\AgdaSpace{}%
\AgdaSymbol{(}\AgdaFunction{g₁}\AgdaSpace{}%
\AgdaSymbol{(}\AgdaField{proj₂}\AgdaSpace{}%
\AgdaBound{x₂}\AgdaSymbol{))}\AgdaSpace{}%
\AgdaBound{v≡x}\AgdaSymbol{))}\AgdaSpace{}%
\AgdaSymbol{(}\AgdaField{proj₁}\AgdaSpace{}%
\AgdaBound{x₂}\AgdaSymbol{)}\AgdaSpace{}%
\AgdaOperator{\AgdaInductiveConstructor{,}}\AgdaSpace{}%
\AgdaField{proj₂}\AgdaSpace{}%
\AgdaBound{x₂}\AgdaSymbol{)}\<%
\\
\>[0][@{}l@{\AgdaIndent{0}}]%
\>[2]\AgdaKeyword{where}%
\>[397I]\AgdaFunction{g₁}\AgdaSpace{}%
\AgdaSymbol{:}\AgdaSpace{}%
\AgdaSymbol{∀}\AgdaSpace{}%
\AgdaSymbol{\{}\AgdaBound{v}\AgdaSymbol{\}}\AgdaSpace{}%
\AgdaSymbol{\{}\AgdaBound{x₁}\AgdaSymbol{\}}\AgdaSpace{}%
\AgdaSymbol{→}\AgdaSpace{}%
\AgdaBound{v}\AgdaSpace{}%
\AgdaOperator{\AgdaFunction{∈}}\AgdaSpace{}%
\AgdaBound{x₁}\AgdaSpace{}%
\AgdaOperator{\AgdaInductiveConstructor{∷}}\AgdaSpace{}%
\AgdaInductiveConstructor{[]}\AgdaSpace{}%
\AgdaSymbol{→}\AgdaSpace{}%
\AgdaBound{x₁}\AgdaSpace{}%
\AgdaOperator{\AgdaDatatype{≡}}\AgdaSpace{}%
\AgdaBound{v}\<%
\\
\>[.][@{}l@{}]\<[397I]%
\>[8]\AgdaFunction{g₁}\AgdaSpace{}%
\AgdaSymbol{(}\AgdaInductiveConstructor{here}\AgdaSpace{}%
\AgdaBound{px}\AgdaSymbol{)}\AgdaSpace{}%
\AgdaSymbol{=}\AgdaSpace{}%
\AgdaFunction{sym}\AgdaSpace{}%
\AgdaBound{px}\<%
\\
\\[\AgdaEmptyExtraSkip]%
\>[0]\AgdaFunction{all-reverse}\AgdaSpace{}%
\AgdaSymbol{:}\AgdaSpace{}%
\AgdaSymbol{∀}\AgdaSpace{}%
\AgdaSymbol{\{}\AgdaBound{x₁}\AgdaSpace{}%
\AgdaSymbol{:}\AgdaSpace{}%
\AgdaFunction{Cmd}\AgdaSymbol{\}}\AgdaSpace{}%
\AgdaBound{xs}\AgdaSpace{}%
\AgdaSymbol{→}\AgdaSpace{}%
\AgdaDatatype{All}\AgdaSpace{}%
\AgdaSymbol{(λ}\AgdaSpace{}%
\AgdaBound{y}\AgdaSpace{}%
\AgdaSymbol{→}\AgdaSpace{}%
\AgdaOperator{\AgdaFunction{¬}}\AgdaSpace{}%
\AgdaBound{x₁}\AgdaSpace{}%
\AgdaOperator{\AgdaDatatype{≡}}\AgdaSpace{}%
\AgdaBound{y}\AgdaSymbol{)}\AgdaSpace{}%
\AgdaBound{xs}\AgdaSpace{}%
\AgdaSymbol{→}\AgdaSpace{}%
\AgdaDatatype{All}\AgdaSpace{}%
\AgdaSymbol{(λ}\AgdaSpace{}%
\AgdaBound{y}\AgdaSpace{}%
\AgdaSymbol{→}\AgdaSpace{}%
\AgdaOperator{\AgdaFunction{¬}}\AgdaSpace{}%
\AgdaBound{x₁}\AgdaSpace{}%
\AgdaOperator{\AgdaDatatype{≡}}\AgdaSpace{}%
\AgdaBound{y}\AgdaSymbol{)}\AgdaSpace{}%
\AgdaSymbol{(}\AgdaFunction{reverse}\AgdaSpace{}%
\AgdaBound{xs}\AgdaSymbol{)}\<%
\\
\>[0]\AgdaFunction{all-reverse}\AgdaSpace{}%
\AgdaInductiveConstructor{[]}\AgdaSpace{}%
\AgdaInductiveConstructor{All.[]}\AgdaSpace{}%
\AgdaSymbol{=}\AgdaSpace{}%
\AgdaInductiveConstructor{All.[]}\<%
\\
\>[0]\AgdaFunction{all-reverse}\AgdaSpace{}%
\AgdaSymbol{(}\AgdaBound{x}\AgdaSpace{}%
\AgdaOperator{\AgdaInductiveConstructor{∷}}\AgdaSpace{}%
\AgdaBound{xs}\AgdaSymbol{)}\AgdaSpace{}%
\AgdaSymbol{(}\AgdaBound{px}\AgdaSpace{}%
\AgdaOperator{\AgdaInductiveConstructor{∷}}\AgdaSpace{}%
\AgdaBound{all₁}\AgdaSymbol{)}\AgdaSpace{}%
\AgdaSymbol{=}\AgdaSpace{}%
\AgdaFunction{subst}\AgdaSpace{}%
\AgdaSymbol{(λ}\AgdaSpace{}%
\AgdaBound{x₂}\AgdaSpace{}%
\AgdaSymbol{→}\AgdaSpace{}%
\AgdaDatatype{All}\AgdaSpace{}%
\AgdaSymbol{(λ}\AgdaSpace{}%
\AgdaBound{y}\AgdaSpace{}%
\AgdaSymbol{→}\AgdaSpace{}%
\AgdaOperator{\AgdaFunction{¬}}\AgdaSpace{}%
\AgdaSymbol{\AgdaUnderscore{}}\AgdaSpace{}%
\AgdaOperator{\AgdaDatatype{≡}}\AgdaSpace{}%
\AgdaBound{y}\AgdaSymbol{)}\AgdaSpace{}%
\AgdaBound{x₂}\AgdaSymbol{)}\AgdaSpace{}%
\AgdaSymbol{(}\AgdaFunction{sym}\AgdaSpace{}%
\AgdaSymbol{(}\AgdaFunction{unfold-reverse}\AgdaSpace{}%
\AgdaBound{x}\AgdaSpace{}%
\AgdaBound{xs}\AgdaSymbol{))}\AgdaSpace{}%
\AgdaSymbol{(}\AgdaFunction{AllP.++⁺}\AgdaSpace{}%
\AgdaSymbol{(}\AgdaFunction{all-reverse}\AgdaSpace{}%
\AgdaBound{xs}\AgdaSpace{}%
\AgdaBound{all₁}\AgdaSymbol{)}\AgdaSpace{}%
\AgdaSymbol{(}\AgdaBound{px}\AgdaSpace{}%
\AgdaOperator{\AgdaInductiveConstructor{∷}}\AgdaSpace{}%
\AgdaInductiveConstructor{All.[]}\AgdaSymbol{))}\<%
\\
\\[\AgdaEmptyExtraSkip]%
\>[0]\AgdaFunction{unique=>∉}\AgdaSpace{}%
\AgdaSymbol{:}\AgdaSpace{}%
\AgdaSymbol{∀}\AgdaSpace{}%
\AgdaSymbol{(}\AgdaBound{x}\AgdaSpace{}%
\AgdaSymbol{:}\AgdaSpace{}%
\AgdaFunction{Cmd}\AgdaSymbol{)}\AgdaSpace{}%
\AgdaBound{xs}\AgdaSpace{}%
\AgdaSymbol{→}\AgdaSpace{}%
\AgdaDatatype{All}\AgdaSpace{}%
\AgdaSymbol{(λ}\AgdaSpace{}%
\AgdaBound{y}\AgdaSpace{}%
\AgdaSymbol{→}\AgdaSpace{}%
\AgdaOperator{\AgdaFunction{¬}}\AgdaSpace{}%
\AgdaBound{x}\AgdaSpace{}%
\AgdaOperator{\AgdaDatatype{≡}}\AgdaSpace{}%
\AgdaBound{y}\AgdaSymbol{)}\AgdaSpace{}%
\AgdaBound{xs}\AgdaSpace{}%
\AgdaSymbol{→}\AgdaSpace{}%
\AgdaBound{x}\AgdaSpace{}%
\AgdaOperator{\AgdaFunction{∉}}\AgdaSpace{}%
\AgdaBound{xs}\<%
\\
\>[0]\AgdaFunction{unique=>∉}\AgdaSpace{}%
\AgdaBound{x}\AgdaSpace{}%
\AgdaBound{xs}\AgdaSpace{}%
\AgdaBound{px}\AgdaSpace{}%
\AgdaBound{x₁}\AgdaSpace{}%
\AgdaKeyword{with}\AgdaSpace{}%
\AgdaFunction{lookup}\AgdaSpace{}%
\AgdaBound{px}\AgdaSpace{}%
\AgdaBound{x₁}\<%
\\
\>[0]\AgdaSymbol{...}\AgdaSpace{}%
\AgdaSymbol{|}\AgdaSpace{}%
\AgdaBound{a}\AgdaSpace{}%
\AgdaSymbol{=}\AgdaSpace{}%
\AgdaBound{a}\AgdaSpace{}%
\AgdaInductiveConstructor{refl}\<%
\\
\\[\AgdaEmptyExtraSkip]%
\>[0]\AgdaFunction{unique=>¬}\AgdaSpace{}%
\AgdaSymbol{:}\AgdaSpace{}%
\AgdaSymbol{∀}\AgdaSpace{}%
\AgdaSymbol{(}\AgdaBound{v}\AgdaSpace{}%
\AgdaSymbol{:}\AgdaSpace{}%
\AgdaFunction{Cmd}\AgdaSymbol{)}\AgdaSpace{}%
\AgdaBound{x}\AgdaSpace{}%
\AgdaBound{xs}\AgdaSpace{}%
\AgdaBound{ys}\AgdaSpace{}%
\AgdaSymbol{→}\AgdaSpace{}%
\AgdaBound{v}\AgdaSpace{}%
\AgdaOperator{\AgdaFunction{∈}}\AgdaSpace{}%
\AgdaBound{ys}\AgdaSpace{}%
\AgdaSymbol{→}\AgdaSpace{}%
\AgdaDatatype{Unique}\AgdaSpace{}%
\AgdaSymbol{(}\AgdaBound{xs}\AgdaSpace{}%
\AgdaOperator{\AgdaFunction{++}}\AgdaSpace{}%
\AgdaBound{x}\AgdaSpace{}%
\AgdaOperator{\AgdaInductiveConstructor{∷}}\AgdaSpace{}%
\AgdaBound{ys}\AgdaSymbol{)}\AgdaSpace{}%
\AgdaSymbol{→}\AgdaSpace{}%
\AgdaOperator{\AgdaFunction{¬}}\AgdaSpace{}%
\AgdaBound{v}\AgdaSpace{}%
\AgdaOperator{\AgdaDatatype{≡}}\AgdaSpace{}%
\AgdaBound{x}\<%
\\
\>[0]\AgdaFunction{unique=>¬}\AgdaSpace{}%
\AgdaBound{v}\AgdaSpace{}%
\AgdaBound{x}\AgdaSpace{}%
\AgdaInductiveConstructor{[]}\AgdaSpace{}%
\AgdaBound{ys}\AgdaSpace{}%
\AgdaBound{v∈ys}\AgdaSpace{}%
\AgdaSymbol{(}\AgdaBound{px}\AgdaSpace{}%
\AgdaOperator{\AgdaInductiveConstructor{∷}}\AgdaSpace{}%
\AgdaBound{u}\AgdaSymbol{)}\AgdaSpace{}%
\AgdaSymbol{=}\AgdaSpace{}%
\AgdaSymbol{λ}\AgdaSpace{}%
\AgdaBound{x₁}\AgdaSpace{}%
\AgdaSymbol{→}\AgdaSpace{}%
\AgdaSymbol{(}\AgdaFunction{lookup}\AgdaSpace{}%
\AgdaBound{px}\AgdaSpace{}%
\AgdaBound{v∈ys}\AgdaSymbol{)}\AgdaSpace{}%
\AgdaSymbol{(}\AgdaFunction{sym}\AgdaSpace{}%
\AgdaBound{x₁}\AgdaSymbol{)}\<%
\\
\>[0]\AgdaFunction{unique=>¬}\AgdaSpace{}%
\AgdaBound{v}\AgdaSpace{}%
\AgdaBound{x}\AgdaSpace{}%
\AgdaSymbol{(}\AgdaBound{x₁}\AgdaSpace{}%
\AgdaOperator{\AgdaInductiveConstructor{∷}}\AgdaSpace{}%
\AgdaBound{xs}\AgdaSymbol{)}\AgdaSpace{}%
\AgdaBound{ys}\AgdaSpace{}%
\AgdaBound{v∈ys}\AgdaSpace{}%
\AgdaSymbol{(}\AgdaBound{px}\AgdaSpace{}%
\AgdaOperator{\AgdaInductiveConstructor{∷}}\AgdaSpace{}%
\AgdaBound{u}\AgdaSymbol{)}\AgdaSpace{}%
\AgdaSymbol{=}\AgdaSpace{}%
\AgdaFunction{unique=>¬}\AgdaSpace{}%
\AgdaBound{v}\AgdaSpace{}%
\AgdaBound{x}\AgdaSpace{}%
\AgdaBound{xs}\AgdaSpace{}%
\AgdaBound{ys}\AgdaSpace{}%
\AgdaBound{v∈ys}\AgdaSpace{}%
\AgdaBound{u}\<%
\\
\\[\AgdaEmptyExtraSkip]%
\>[0]\AgdaFunction{unique-reverse}\AgdaSpace{}%
\AgdaSymbol{:}\AgdaSpace{}%
\AgdaSymbol{∀}\AgdaSpace{}%
\AgdaBound{xs}\AgdaSpace{}%
\AgdaSymbol{→}\AgdaSpace{}%
\AgdaDatatype{Unique}\AgdaSpace{}%
\AgdaBound{xs}\AgdaSpace{}%
\AgdaSymbol{→}\AgdaSpace{}%
\AgdaDatatype{Unique}\AgdaSpace{}%
\AgdaSymbol{(}\AgdaFunction{reverse}\AgdaSpace{}%
\AgdaBound{xs}\AgdaSymbol{)}\<%
\\
\>[0]\AgdaFunction{unique-reverse}\AgdaSpace{}%
\AgdaInductiveConstructor{[]}\AgdaSpace{}%
\AgdaBound{u}\AgdaSpace{}%
\AgdaSymbol{=}\AgdaSpace{}%
\AgdaInductiveConstructor{[]}\<%
\\
\>[0]\AgdaFunction{unique-reverse}\AgdaSpace{}%
\AgdaSymbol{(}\AgdaBound{x₁}\AgdaSpace{}%
\AgdaOperator{\AgdaInductiveConstructor{∷}}\AgdaSpace{}%
\AgdaBound{xs}\AgdaSymbol{)}\AgdaSpace{}%
\AgdaSymbol{(}\AgdaBound{px}\AgdaSpace{}%
\AgdaOperator{\AgdaInductiveConstructor{∷}}\AgdaSpace{}%
\AgdaBound{u}\AgdaSymbol{)}\AgdaSpace{}%
\AgdaKeyword{with}\AgdaSpace{}%
\AgdaFunction{++⁺}\AgdaSpace{}%
\AgdaSymbol{(}\AgdaFunction{unique-reverse}\AgdaSpace{}%
\AgdaBound{xs}\AgdaSpace{}%
\AgdaBound{u}\AgdaSymbol{)}\AgdaSpace{}%
\AgdaSymbol{(}\AgdaInductiveConstructor{All.[]}\AgdaSpace{}%
\AgdaOperator{\AgdaInductiveConstructor{∷}}\AgdaSpace{}%
\AgdaInductiveConstructor{[]}\AgdaSymbol{)}\AgdaSpace{}%
\AgdaSymbol{(}\AgdaFunction{unique→disjoint}\AgdaSpace{}%
\AgdaSymbol{(}\AgdaFunction{reverse}\AgdaSpace{}%
\AgdaBound{xs}\AgdaSymbol{)}\AgdaSpace{}%
\AgdaSymbol{(}\AgdaFunction{all-reverse}\AgdaSpace{}%
\AgdaBound{xs}\AgdaSpace{}%
\AgdaBound{px}\AgdaSymbol{))}\<%
\\
\>[0]\AgdaSymbol{...}\AgdaSpace{}%
\AgdaSymbol{|}\AgdaSpace{}%
\AgdaBound{u₁}\AgdaSpace{}%
\AgdaSymbol{=}\AgdaSpace{}%
\AgdaFunction{subst}\AgdaSpace{}%
\AgdaSymbol{(λ}\AgdaSpace{}%
\AgdaBound{ls}\AgdaSpace{}%
\AgdaSymbol{→}\AgdaSpace{}%
\AgdaDatatype{Unique}\AgdaSpace{}%
\AgdaBound{ls}\AgdaSymbol{)}\AgdaSpace{}%
\AgdaSymbol{(}\AgdaFunction{sym}\AgdaSpace{}%
\AgdaSymbol{(}\AgdaFunction{unfold-reverse}\AgdaSpace{}%
\AgdaBound{x₁}\AgdaSpace{}%
\AgdaBound{xs}\AgdaSymbol{))}\AgdaSpace{}%
\AgdaBound{u₁}\<%
\\
\\[\AgdaEmptyExtraSkip]%
\>[0]\AgdaFunction{↭-reverse}\AgdaSpace{}%
\AgdaSymbol{:}\AgdaSpace{}%
\AgdaSymbol{∀}\AgdaSpace{}%
\AgdaSymbol{(}\AgdaBound{xs}\AgdaSpace{}%
\AgdaSymbol{:}\AgdaSpace{}%
\AgdaFunction{Build}\AgdaSymbol{)}\AgdaSpace{}%
\AgdaSymbol{→}\AgdaSpace{}%
\AgdaBound{xs}\AgdaSpace{}%
\AgdaOperator{\AgdaDatatype{↭}}\AgdaSpace{}%
\AgdaFunction{reverse}\AgdaSpace{}%
\AgdaBound{xs}\<%
\\
\>[0]\AgdaFunction{↭-reverse}\AgdaSpace{}%
\AgdaBound{xs}\AgdaSpace{}%
\AgdaSymbol{=}\AgdaSpace{}%
\AgdaFunction{subst}\AgdaSpace{}%
\AgdaSymbol{(λ}\AgdaSpace{}%
\AgdaBound{x}\AgdaSpace{}%
\AgdaSymbol{→}\AgdaSpace{}%
\AgdaBound{x}\AgdaSpace{}%
\AgdaOperator{\AgdaDatatype{↭}}\AgdaSpace{}%
\AgdaFunction{reverse}\AgdaSpace{}%
\AgdaBound{xs}\AgdaSymbol{)}\AgdaSpace{}%
\AgdaSymbol{(}\AgdaFunction{++-identityʳ}\AgdaSpace{}%
\AgdaBound{xs}\AgdaSymbol{)}\AgdaSpace{}%
\AgdaSymbol{(}\AgdaFunction{++↭ʳ++}\AgdaSpace{}%
\AgdaBound{xs}\AgdaSpace{}%
\AgdaInductiveConstructor{[]}\AgdaSymbol{)}\<%
\\
\\[\AgdaEmptyExtraSkip]%
\>[0]\AgdaFunction{helper5}\AgdaSpace{}%
\AgdaSymbol{:}\AgdaSpace{}%
\AgdaSymbol{∀}\AgdaSpace{}%
\AgdaSymbol{\{}\AgdaBound{s}\AgdaSymbol{\}}\AgdaSpace{}%
\AgdaBound{xs}\AgdaSpace{}%
\AgdaBound{x}\AgdaSpace{}%
\AgdaSymbol{→}\AgdaSpace{}%
\AgdaFunction{Disjoint}\AgdaSpace{}%
\AgdaSymbol{(}\AgdaFunction{cmdWriteNames}\AgdaSpace{}%
\AgdaBound{x}\AgdaSpace{}%
\AgdaBound{s}\AgdaSymbol{)}\AgdaSpace{}%
\AgdaBound{xs}\AgdaSpace{}%
\AgdaSymbol{→}\AgdaSpace{}%
\AgdaDatatype{All}\AgdaSpace{}%
\AgdaSymbol{(λ}\AgdaSpace{}%
\AgdaBound{f₁}\AgdaSpace{}%
\AgdaSymbol{→}\AgdaSpace{}%
\AgdaBound{s}\AgdaSpace{}%
\AgdaBound{f₁}\AgdaSpace{}%
\AgdaOperator{\AgdaDatatype{≡}}\AgdaSpace{}%
\AgdaFunction{run}\AgdaSpace{}%
\AgdaBound{x}\AgdaSpace{}%
\AgdaBound{s}\AgdaSpace{}%
\AgdaBound{f₁}\AgdaSymbol{)}\AgdaSpace{}%
\AgdaBound{xs}\<%
\\
\>[0]\AgdaFunction{helper5}\AgdaSpace{}%
\AgdaInductiveConstructor{[]}\AgdaSpace{}%
\AgdaBound{x}\AgdaSpace{}%
\AgdaBound{dsj}\AgdaSpace{}%
\AgdaSymbol{=}\AgdaSpace{}%
\AgdaInductiveConstructor{All.[]}\<%
\\
\>[0]\AgdaFunction{helper5}\AgdaSpace{}%
\AgdaSymbol{\{}\AgdaBound{s}\AgdaSymbol{\}}\AgdaSpace{}%
\AgdaSymbol{(}\AgdaBound{x₁}\AgdaSpace{}%
\AgdaOperator{\AgdaInductiveConstructor{∷}}\AgdaSpace{}%
\AgdaBound{xs}\AgdaSymbol{)}\AgdaSpace{}%
\AgdaBound{x}\AgdaSpace{}%
\AgdaBound{dsj}\AgdaSpace{}%
\AgdaSymbol{=}\AgdaSpace{}%
\AgdaSymbol{(}\AgdaFunction{lemma3}\AgdaSpace{}%
\AgdaBound{x₁}\AgdaSpace{}%
\AgdaSymbol{(}\AgdaFunction{cmdWrites}\AgdaSpace{}%
\AgdaBound{x}\AgdaSpace{}%
\AgdaBound{s}\AgdaSymbol{)}\AgdaSpace{}%
\AgdaSymbol{λ}\AgdaSpace{}%
\AgdaBound{x₂}\AgdaSpace{}%
\AgdaSymbol{→}\AgdaSpace{}%
\AgdaBound{dsj}\AgdaSpace{}%
\AgdaSymbol{(}\AgdaBound{x₂}\AgdaSpace{}%
\AgdaOperator{\AgdaInductiveConstructor{,}}\AgdaSpace{}%
\AgdaInductiveConstructor{here}\AgdaSpace{}%
\AgdaInductiveConstructor{refl}\AgdaSymbol{))}\AgdaSpace{}%
\AgdaOperator{\AgdaInductiveConstructor{All.∷}}\AgdaSpace{}%
\AgdaSymbol{(}\AgdaFunction{helper5}\AgdaSpace{}%
\AgdaBound{xs}\AgdaSpace{}%
\AgdaBound{x}\AgdaSpace{}%
\AgdaSymbol{λ}\AgdaSpace{}%
\AgdaBound{x₂}\AgdaSpace{}%
\AgdaSymbol{→}\AgdaSpace{}%
\AgdaBound{dsj}\AgdaSpace{}%
\AgdaSymbol{((}\AgdaField{proj₁}\AgdaSpace{}%
\AgdaBound{x₂}\AgdaSymbol{)}\AgdaSpace{}%
\AgdaOperator{\AgdaInductiveConstructor{,}}\AgdaSpace{}%
\AgdaInductiveConstructor{there}\AgdaSpace{}%
\AgdaSymbol{(}\AgdaField{proj₂}\AgdaSpace{}%
\AgdaBound{x₂}\AgdaSymbol{)))}\<%
\\
\\[\AgdaEmptyExtraSkip]%
\>[0]\AgdaComment{-- kind of a dumb function that could be inlined?}\<%
\\
\>[0]\AgdaFunction{helper4}\AgdaSpace{}%
\AgdaSymbol{:}\AgdaSpace{}%
\AgdaSymbol{∀}\AgdaSpace{}%
\AgdaSymbol{\{}\AgdaBound{f₁}\AgdaSymbol{\}}\AgdaSpace{}%
\AgdaSymbol{\{}\AgdaBound{s}\AgdaSymbol{\}}\AgdaSpace{}%
\AgdaBound{bs}\AgdaSpace{}%
\AgdaBound{x}\AgdaSpace{}%
\AgdaSymbol{→}\AgdaSpace{}%
\AgdaBound{f₁}\AgdaSpace{}%
\AgdaOperator{\AgdaFunction{∈}}\AgdaSpace{}%
\AgdaFunction{cmdReadNames}\AgdaSpace{}%
\AgdaBound{x}\AgdaSpace{}%
\AgdaBound{s}\AgdaSpace{}%
\AgdaSymbol{→}\AgdaSpace{}%
\AgdaFunction{Disjoint}\AgdaSpace{}%
\AgdaSymbol{(}\AgdaFunction{cmdReadNames}\AgdaSpace{}%
\AgdaBound{x}\AgdaSpace{}%
\AgdaBound{s}\AgdaSymbol{)}\AgdaSpace{}%
\AgdaSymbol{(}\AgdaFunction{buildWriteNames}\AgdaSpace{}%
\AgdaBound{s}\AgdaSpace{}%
\AgdaBound{bs}\AgdaSymbol{)}\AgdaSpace{}%
\AgdaSymbol{→}\AgdaSpace{}%
\AgdaBound{f₁}\AgdaSpace{}%
\AgdaOperator{\AgdaFunction{∉}}\AgdaSpace{}%
\AgdaFunction{buildWriteNames}\AgdaSpace{}%
\AgdaBound{s}\AgdaSpace{}%
\AgdaBound{bs}\<%
\\
\>[0]\AgdaFunction{helper4}\AgdaSpace{}%
\AgdaBound{bs}\AgdaSpace{}%
\AgdaBound{x}\AgdaSpace{}%
\AgdaBound{f₁∈reads}\AgdaSpace{}%
\AgdaBound{dsj}\AgdaSpace{}%
\AgdaBound{f₁∈writes}\AgdaSpace{}%
\AgdaSymbol{=}\AgdaSpace{}%
\AgdaBound{dsj}\AgdaSpace{}%
\AgdaSymbol{(}\AgdaBound{f₁∈reads}\AgdaSpace{}%
\AgdaOperator{\AgdaInductiveConstructor{,}}\AgdaSpace{}%
\AgdaBound{f₁∈writes}\AgdaSymbol{)}\<%
\\
\\[\AgdaEmptyExtraSkip]%
\>[0]\AgdaFunction{helper3}\AgdaSpace{}%
\AgdaSymbol{:}\AgdaSpace{}%
\AgdaSymbol{∀}\AgdaSpace{}%
\AgdaSymbol{\{}\AgdaBound{s}\AgdaSymbol{\}}\AgdaSpace{}%
\AgdaBound{as}\AgdaSpace{}%
\AgdaBound{bs}\AgdaSpace{}%
\AgdaBound{xs}\AgdaSpace{}%
\AgdaBound{x}\AgdaSpace{}%
\AgdaSymbol{→}\AgdaSpace{}%
\AgdaFunction{Disjoint}\AgdaSpace{}%
\AgdaSymbol{(}\AgdaFunction{cmdReadNames}\AgdaSpace{}%
\AgdaBound{x}\AgdaSpace{}%
\AgdaSymbol{(}\AgdaFunction{script}\AgdaSpace{}%
\AgdaBound{as}\AgdaSpace{}%
\AgdaBound{s}\AgdaSymbol{))}\AgdaSpace{}%
\AgdaSymbol{(}\AgdaFunction{buildWriteNames}\AgdaSpace{}%
\AgdaSymbol{(}\AgdaFunction{script}\AgdaSpace{}%
\AgdaBound{as}\AgdaSpace{}%
\AgdaBound{s}\AgdaSymbol{)}\AgdaSpace{}%
\AgdaBound{bs}\AgdaSymbol{)}\AgdaSpace{}%
\AgdaSymbol{→}\AgdaSpace{}%
\AgdaSymbol{(∀}\AgdaSpace{}%
\AgdaBound{f₁}\AgdaSpace{}%
\AgdaSymbol{→}\AgdaSpace{}%
\AgdaFunction{script}\AgdaSpace{}%
\AgdaSymbol{(}\AgdaBound{as}\AgdaSpace{}%
\AgdaOperator{\AgdaFunction{++}}\AgdaSpace{}%
\AgdaBound{bs}\AgdaSymbol{)}\AgdaSpace{}%
\AgdaBound{s}\AgdaSpace{}%
\AgdaBound{f₁}\AgdaSpace{}%
\AgdaOperator{\AgdaDatatype{≡}}\AgdaSpace{}%
\AgdaFunction{script}\AgdaSpace{}%
\AgdaBound{xs}\AgdaSpace{}%
\AgdaBound{s}\AgdaSpace{}%
\AgdaBound{f₁}\AgdaSymbol{)}\AgdaSpace{}%
\AgdaSymbol{→}\AgdaSpace{}%
\AgdaField{proj₁}\AgdaSpace{}%
\AgdaSymbol{(}\AgdaBound{oracle}\AgdaSpace{}%
\AgdaBound{x}\AgdaSymbol{)}\AgdaSpace{}%
\AgdaSymbol{(}\AgdaFunction{script}\AgdaSpace{}%
\AgdaBound{as}\AgdaSpace{}%
\AgdaBound{s}\AgdaSymbol{)}\AgdaSpace{}%
\AgdaOperator{\AgdaDatatype{≡}}\AgdaSpace{}%
\AgdaField{proj₁}\AgdaSpace{}%
\AgdaSymbol{(}\AgdaBound{oracle}\AgdaSpace{}%
\AgdaBound{x}\AgdaSymbol{)}\AgdaSpace{}%
\AgdaSymbol{(}\AgdaFunction{script}\AgdaSpace{}%
\AgdaBound{xs}\AgdaSpace{}%
\AgdaBound{s}\AgdaSymbol{)}\<%
\\
\>[0]\AgdaFunction{helper3}\AgdaSpace{}%
\AgdaSymbol{\{}\AgdaBound{s}\AgdaSymbol{\}}\AgdaSpace{}%
\AgdaBound{as}\AgdaSpace{}%
\AgdaBound{bs}\AgdaSpace{}%
\AgdaBound{xs}\AgdaSpace{}%
\AgdaBound{x}\AgdaSpace{}%
\AgdaBound{dsj}\AgdaSpace{}%
\AgdaBound{∀₁}\AgdaSpace{}%
\AgdaSymbol{=}\AgdaSpace{}%
\AgdaField{proj₂}\AgdaSpace{}%
\AgdaSymbol{(}\AgdaBound{oracle}\AgdaSpace{}%
\AgdaBound{x}\AgdaSymbol{)}\AgdaSpace{}%
\AgdaSymbol{(}\AgdaFunction{script}\AgdaSpace{}%
\AgdaBound{as}\AgdaSpace{}%
\AgdaBound{s}\AgdaSymbol{)}\AgdaSpace{}%
\AgdaSymbol{(}\AgdaFunction{script}\AgdaSpace{}%
\AgdaBound{xs}\AgdaSpace{}%
\AgdaBound{s}\AgdaSymbol{)}\AgdaSpace{}%
\AgdaSymbol{λ}\AgdaSpace{}%
\AgdaBound{f₁}\AgdaSpace{}%
\AgdaBound{x₁}\AgdaSpace{}%
\AgdaSymbol{→}\AgdaSpace{}%
\AgdaFunction{trans}\AgdaSpace{}%
\AgdaSymbol{(}\AgdaFunction{exec-≡f₁}\AgdaSpace{}%
\AgdaBound{s}\AgdaSpace{}%
\AgdaBound{f₁}\AgdaSpace{}%
\AgdaBound{as}\AgdaSpace{}%
\AgdaBound{bs}\AgdaSpace{}%
\AgdaSymbol{(}\AgdaFunction{f₁∉₁}\AgdaSpace{}%
\AgdaBound{x₁}\AgdaSymbol{))}\AgdaSpace{}%
\AgdaSymbol{(}\AgdaBound{∀₁}\AgdaSpace{}%
\AgdaBound{f₁}\AgdaSymbol{)}\<%
\\
\>[0][@{}l@{\AgdaIndent{0}}]%
\>[2]\AgdaKeyword{where}%
\>[822I]\AgdaFunction{f₁∉₁}\AgdaSpace{}%
\AgdaSymbol{:}\AgdaSpace{}%
\AgdaSymbol{∀}\AgdaSpace{}%
\AgdaSymbol{\{}\AgdaBound{f₁}\AgdaSymbol{\}}\AgdaSpace{}%
\AgdaSymbol{→}\AgdaSpace{}%
\AgdaBound{f₁}\AgdaSpace{}%
\AgdaOperator{\AgdaFunction{∈}}\AgdaSpace{}%
\AgdaFunction{cmdReadNames}\AgdaSpace{}%
\AgdaBound{x}\AgdaSpace{}%
\AgdaSymbol{(}\AgdaFunction{script}\AgdaSpace{}%
\AgdaBound{as}\AgdaSpace{}%
\AgdaBound{s}\AgdaSymbol{)}\AgdaSpace{}%
\AgdaSymbol{→}\AgdaSpace{}%
\AgdaBound{f₁}\AgdaSpace{}%
\AgdaOperator{\AgdaFunction{∉}}\AgdaSpace{}%
\AgdaFunction{buildWriteNames}\AgdaSpace{}%
\AgdaSymbol{(}\AgdaFunction{script}\AgdaSpace{}%
\AgdaBound{as}\AgdaSpace{}%
\AgdaBound{s}\AgdaSymbol{)}\AgdaSpace{}%
\AgdaBound{bs}\<%
\\
\>[.][@{}l@{}]\<[822I]%
\>[8]\AgdaFunction{f₁∉₁}\AgdaSpace{}%
\AgdaBound{f₁∈reads}\AgdaSpace{}%
\AgdaSymbol{=}\AgdaSpace{}%
\AgdaFunction{helper4}\AgdaSpace{}%
\AgdaBound{bs}\AgdaSpace{}%
\AgdaBound{x}\AgdaSpace{}%
\AgdaBound{f₁∈reads}\AgdaSpace{}%
\AgdaBound{dsj}\<%
\\
\\[\AgdaEmptyExtraSkip]%
\>[0]\AgdaFunction{helper8}\AgdaSpace{}%
\AgdaSymbol{:}\AgdaSpace{}%
\AgdaSymbol{∀}\AgdaSpace{}%
\AgdaSymbol{\{}\AgdaBound{s}\AgdaSymbol{\}}\AgdaSpace{}%
\AgdaSymbol{\{}\AgdaBound{f₁}\AgdaSymbol{\}}\AgdaSpace{}%
\AgdaBound{xs}\AgdaSpace{}%
\AgdaSymbol{→}\AgdaSpace{}%
\AgdaBound{f₁}\AgdaSpace{}%
\AgdaOperator{\AgdaFunction{∉}}\AgdaSpace{}%
\AgdaSymbol{(}\AgdaFunction{buildWriteNames}\AgdaSpace{}%
\AgdaBound{s}\AgdaSpace{}%
\AgdaBound{xs}\AgdaSymbol{)}\AgdaSpace{}%
\AgdaSymbol{→}\AgdaSpace{}%
\AgdaFunction{script}\AgdaSpace{}%
\AgdaBound{xs}\AgdaSpace{}%
\AgdaBound{s}\AgdaSpace{}%
\AgdaBound{f₁}\AgdaSpace{}%
\AgdaOperator{\AgdaDatatype{≡}}\AgdaSpace{}%
\AgdaBound{s}\AgdaSpace{}%
\AgdaBound{f₁}\<%
\\
\>[0]\AgdaFunction{helper8}\AgdaSpace{}%
\AgdaInductiveConstructor{[]}\AgdaSpace{}%
\AgdaBound{f₁∉}\AgdaSpace{}%
\AgdaSymbol{=}\AgdaSpace{}%
\AgdaInductiveConstructor{refl}\<%
\\
\>[0]\AgdaFunction{helper8}\AgdaSpace{}%
\AgdaSymbol{\{}\AgdaBound{s}\AgdaSymbol{\}}\AgdaSpace{}%
\AgdaSymbol{\{}\AgdaBound{f₁}\AgdaSymbol{\}}\AgdaSpace{}%
\AgdaSymbol{(}\AgdaBound{x}\AgdaSpace{}%
\AgdaOperator{\AgdaInductiveConstructor{∷}}\AgdaSpace{}%
\AgdaBound{xs}\AgdaSymbol{)}\AgdaSpace{}%
\AgdaBound{f₁∉}\AgdaSpace{}%
\AgdaSymbol{=}\AgdaSpace{}%
\AgdaFunction{trans}%
\>[880I]\AgdaSymbol{(}\AgdaFunction{helper8}\AgdaSpace{}%
\AgdaBound{xs}\AgdaSpace{}%
\AgdaSymbol{λ}\AgdaSpace{}%
\AgdaBound{x₁}\AgdaSpace{}%
\AgdaSymbol{→}\AgdaSpace{}%
\AgdaBound{f₁∉}\AgdaSpace{}%
\AgdaSymbol{(}\AgdaFunction{∈-++⁺ʳ}\AgdaSpace{}%
\AgdaSymbol{(}\AgdaFunction{cmdWriteNames}\AgdaSpace{}%
\AgdaBound{x}\AgdaSpace{}%
\AgdaBound{s}\AgdaSymbol{)}\AgdaSpace{}%
\AgdaBound{x₁}\AgdaSymbol{))}\<%
\\
\>[.][@{}l@{}]\<[880I]%
\>[38]\AgdaSymbol{(}\AgdaFunction{sym}\AgdaSpace{}%
\AgdaSymbol{(}\AgdaFunction{lemma3}\AgdaSpace{}%
\AgdaBound{f₁}\AgdaSpace{}%
\AgdaSymbol{(}\AgdaFunction{cmdWrites}\AgdaSpace{}%
\AgdaBound{x}\AgdaSpace{}%
\AgdaBound{s}\AgdaSymbol{)}\AgdaSpace{}%
\AgdaSymbol{λ}\AgdaSpace{}%
\AgdaBound{x₁}\AgdaSpace{}%
\AgdaSymbol{→}\AgdaSpace{}%
\AgdaBound{f₁∉}\AgdaSpace{}%
\AgdaSymbol{(}\AgdaFunction{∈-++⁺ˡ}\AgdaSpace{}%
\AgdaBound{x₁}\AgdaSymbol{)))}\<%
\\
\\[\AgdaEmptyExtraSkip]%
\>[0]\AgdaFunction{helper7}\AgdaSpace{}%
\AgdaSymbol{:}\AgdaSpace{}%
\AgdaSymbol{∀}\AgdaSpace{}%
\AgdaSymbol{\{}\AgdaBound{s}\AgdaSymbol{\}}\AgdaSpace{}%
\AgdaSymbol{\{}\AgdaBound{s₁}\AgdaSymbol{\}}\AgdaSpace{}%
\AgdaBound{f₁}\AgdaSpace{}%
\AgdaBound{xs}\AgdaSpace{}%
\AgdaBound{ys}\AgdaSpace{}%
\AgdaSymbol{→}\AgdaSpace{}%
\AgdaBound{xs}\AgdaSpace{}%
\AgdaOperator{\AgdaDatatype{≡}}\AgdaSpace{}%
\AgdaBound{ys}\AgdaSpace{}%
\AgdaSymbol{→}\AgdaSpace{}%
\AgdaBound{f₁}\AgdaSpace{}%
\AgdaOperator{\AgdaFunction{∈}}\AgdaSpace{}%
\AgdaFunction{map}\AgdaSpace{}%
\AgdaField{proj₁}\AgdaSpace{}%
\AgdaBound{xs}\AgdaSpace{}%
\AgdaSymbol{→}\AgdaSpace{}%
\AgdaFunction{foldr}\AgdaSpace{}%
\AgdaFunction{extend}\AgdaSpace{}%
\AgdaBound{s}\AgdaSpace{}%
\AgdaBound{xs}\AgdaSpace{}%
\AgdaBound{f₁}\AgdaSpace{}%
\AgdaOperator{\AgdaDatatype{≡}}\AgdaSpace{}%
\AgdaFunction{foldr}\AgdaSpace{}%
\AgdaFunction{extend}\AgdaSpace{}%
\AgdaBound{s₁}\AgdaSpace{}%
\AgdaBound{ys}\AgdaSpace{}%
\AgdaBound{f₁}\<%
\\
\>[0]\AgdaFunction{helper7}\AgdaSpace{}%
\AgdaBound{g₁}\AgdaSpace{}%
\AgdaSymbol{((}\AgdaBound{f}\AgdaSpace{}%
\AgdaOperator{\AgdaInductiveConstructor{,}}\AgdaSpace{}%
\AgdaBound{v}\AgdaSymbol{)}\AgdaSpace{}%
\AgdaOperator{\AgdaInductiveConstructor{∷}}\AgdaSpace{}%
\AgdaBound{xs}\AgdaSymbol{)}\AgdaSpace{}%
\AgdaSymbol{((}\AgdaBound{f₁}\AgdaSpace{}%
\AgdaOperator{\AgdaInductiveConstructor{,}}\AgdaSpace{}%
\AgdaBound{v₁}\AgdaSymbol{)}\AgdaSpace{}%
\AgdaOperator{\AgdaInductiveConstructor{∷}}\AgdaSpace{}%
\AgdaBound{ys}\AgdaSymbol{)}\AgdaSpace{}%
\AgdaBound{≡₁}\AgdaSpace{}%
\AgdaBound{f₁∈}\AgdaSpace{}%
\AgdaKeyword{with}\AgdaSpace{}%
\AgdaBound{f}\AgdaSpace{}%
\AgdaOperator{\AgdaFunction{≟}}\AgdaSpace{}%
\AgdaBound{g₁}\AgdaSpace{}%
\AgdaSymbol{|}\AgdaSpace{}%
\AgdaBound{f₁}\AgdaSpace{}%
\AgdaOperator{\AgdaFunction{≟}}\AgdaSpace{}%
\AgdaBound{g₁}\<%
\\
\>[0]\AgdaSymbol{...}\AgdaSpace{}%
\AgdaSymbol{|}\AgdaSpace{}%
\AgdaInductiveConstructor{yes}\AgdaSpace{}%
\AgdaBound{f≡g₁}\AgdaSpace{}%
\AgdaSymbol{|}\AgdaSpace{}%
\AgdaInductiveConstructor{yes}\AgdaSpace{}%
\AgdaBound{f₁≡g₁}\AgdaSpace{}%
\AgdaSymbol{=}\AgdaSpace{}%
\AgdaFunction{cong}\AgdaSpace{}%
\AgdaInductiveConstructor{just}\AgdaSpace{}%
\AgdaSymbol{(}\AgdaFunction{,-injectiveʳ}\AgdaSpace{}%
\AgdaSymbol{(}\AgdaFunction{∷-injectiveˡ}\AgdaSpace{}%
\AgdaBound{≡₁}\AgdaSymbol{))}\<%
\\
\>[0]\AgdaSymbol{...}\AgdaSpace{}%
\AgdaSymbol{|}\AgdaSpace{}%
\AgdaInductiveConstructor{yes}\AgdaSpace{}%
\AgdaBound{f≡g₁}\AgdaSpace{}%
\AgdaSymbol{|}\AgdaSpace{}%
\AgdaInductiveConstructor{no}\AgdaSpace{}%
\AgdaBound{¬f₁≡g₁}\AgdaSpace{}%
\AgdaSymbol{=}\AgdaSpace{}%
\AgdaFunction{contradiction}\AgdaSpace{}%
\AgdaSymbol{(}\AgdaFunction{trans}\AgdaSpace{}%
\AgdaSymbol{(}\AgdaFunction{sym}\AgdaSpace{}%
\AgdaSymbol{(}\AgdaFunction{,-injectiveˡ}\AgdaSpace{}%
\AgdaSymbol{(}\AgdaFunction{∷-injectiveˡ}\AgdaSpace{}%
\AgdaBound{≡₁}\AgdaSymbol{)))}\AgdaSpace{}%
\AgdaBound{f≡g₁}\AgdaSymbol{)}\AgdaSpace{}%
\AgdaBound{¬f₁≡g₁}\<%
\\
\>[0]\AgdaSymbol{...}\AgdaSpace{}%
\AgdaSymbol{|}\AgdaSpace{}%
\AgdaInductiveConstructor{no}\AgdaSpace{}%
\AgdaBound{¬f≡g₁}\AgdaSpace{}%
\AgdaSymbol{|}\AgdaSpace{}%
\AgdaInductiveConstructor{yes}\AgdaSpace{}%
\AgdaBound{f₁≡g₁}\AgdaSpace{}%
\AgdaSymbol{=}\AgdaSpace{}%
\AgdaFunction{contradiction}\AgdaSpace{}%
\AgdaSymbol{(}\AgdaFunction{trans}\AgdaSpace{}%
\AgdaSymbol{(}\AgdaFunction{,-injectiveˡ}\AgdaSpace{}%
\AgdaSymbol{(}\AgdaFunction{∷-injectiveˡ}\AgdaSpace{}%
\AgdaBound{≡₁}\AgdaSymbol{))}\AgdaSpace{}%
\AgdaBound{f₁≡g₁}\AgdaSymbol{)}\AgdaSpace{}%
\AgdaBound{¬f≡g₁}\<%
\\
\>[0]\AgdaSymbol{...}\AgdaSpace{}%
\AgdaSymbol{|}\AgdaSpace{}%
\AgdaInductiveConstructor{no}\AgdaSpace{}%
\AgdaBound{¬f≡g₁}\AgdaSpace{}%
\AgdaSymbol{|}\AgdaSpace{}%
\AgdaInductiveConstructor{no}\AgdaSpace{}%
\AgdaBound{¬f₁≡g₁}\AgdaSpace{}%
\AgdaSymbol{=}\AgdaSpace{}%
\AgdaFunction{helper7}\AgdaSpace{}%
\AgdaBound{g₁}\AgdaSpace{}%
\AgdaBound{xs}\AgdaSpace{}%
\AgdaBound{ys}\AgdaSpace{}%
\AgdaSymbol{(}\AgdaFunction{∷-injectiveʳ}\AgdaSpace{}%
\AgdaBound{≡₁}\AgdaSymbol{)}\AgdaSpace{}%
\AgdaSymbol{(}\AgdaFunction{Any.tail}\AgdaSpace{}%
\AgdaSymbol{(λ}\AgdaSpace{}%
\AgdaBound{x}\AgdaSpace{}%
\AgdaSymbol{→}\AgdaSpace{}%
\AgdaBound{¬f≡g₁}\AgdaSpace{}%
\AgdaSymbol{(}\AgdaFunction{sym}\AgdaSpace{}%
\AgdaBound{x}\AgdaSymbol{))}\AgdaSpace{}%
\AgdaBound{f₁∈}\AgdaSymbol{)}\<%
\\
\\[\AgdaEmptyExtraSkip]%
\>[0]\AgdaFunction{script≡}\AgdaSpace{}%
\AgdaSymbol{:}\AgdaSpace{}%
\AgdaSymbol{∀}\AgdaSpace{}%
\AgdaSymbol{\{}\AgdaBound{s}\AgdaSymbol{\}}\AgdaSpace{}%
\AgdaBound{xs}\AgdaSpace{}%
\AgdaBound{ys}\AgdaSpace{}%
\AgdaSymbol{→}\AgdaSpace{}%
\AgdaFunction{script}\AgdaSpace{}%
\AgdaSymbol{(}\AgdaBound{xs}\AgdaSpace{}%
\AgdaOperator{\AgdaFunction{++}}\AgdaSpace{}%
\AgdaBound{ys}\AgdaSymbol{)}\AgdaSpace{}%
\AgdaBound{s}\AgdaSpace{}%
\AgdaOperator{\AgdaDatatype{≡}}\AgdaSpace{}%
\AgdaFunction{script}\AgdaSpace{}%
\AgdaBound{ys}\AgdaSpace{}%
\AgdaSymbol{(}\AgdaFunction{script}\AgdaSpace{}%
\AgdaBound{xs}\AgdaSpace{}%
\AgdaBound{s}\AgdaSymbol{)}\<%
\\
\>[0]\AgdaFunction{script≡}\AgdaSpace{}%
\AgdaInductiveConstructor{[]}\AgdaSpace{}%
\AgdaBound{ys}\AgdaSpace{}%
\AgdaSymbol{=}\AgdaSpace{}%
\AgdaInductiveConstructor{refl}\<%
\\
\>[0]\AgdaFunction{script≡}\AgdaSpace{}%
\AgdaSymbol{(}\AgdaBound{x}\AgdaSpace{}%
\AgdaOperator{\AgdaInductiveConstructor{∷}}\AgdaSpace{}%
\AgdaBound{xs}\AgdaSymbol{)}\AgdaSpace{}%
\AgdaBound{ys}\AgdaSpace{}%
\AgdaSymbol{=}\AgdaSpace{}%
\AgdaFunction{script≡}\AgdaSpace{}%
\AgdaBound{xs}\AgdaSpace{}%
\AgdaBound{ys}\<%
\\
\\[\AgdaEmptyExtraSkip]%
\>[0]\AgdaComment{-- putting x in middle doesnt change result if x doesnt write to file. }\<%
\\
\>[0]\AgdaFunction{helper6}\AgdaSpace{}%
\AgdaSymbol{:}\AgdaSpace{}%
\AgdaSymbol{∀}\AgdaSpace{}%
\AgdaSymbol{\{}\AgdaBound{s}\AgdaSymbol{\}}\AgdaSpace{}%
\AgdaSymbol{\{}\AgdaBound{f₁}\AgdaSymbol{\}}\AgdaSpace{}%
\AgdaBound{xs}\AgdaSpace{}%
\AgdaBound{x}\AgdaSpace{}%
\AgdaSymbol{→}\AgdaSpace{}%
\AgdaBound{f₁}\AgdaSpace{}%
\AgdaOperator{\AgdaFunction{∉}}\AgdaSpace{}%
\AgdaFunction{cmdWriteNames}\AgdaSpace{}%
\AgdaBound{x}\AgdaSpace{}%
\AgdaBound{s}\AgdaSpace{}%
\AgdaSymbol{→}\AgdaSpace{}%
\AgdaDatatype{All}\AgdaSpace{}%
\AgdaSymbol{(λ}\AgdaSpace{}%
\AgdaBound{f₁}\AgdaSpace{}%
\AgdaSymbol{→}\AgdaSpace{}%
\AgdaBound{s}\AgdaSpace{}%
\AgdaBound{f₁}\AgdaSpace{}%
\AgdaOperator{\AgdaDatatype{≡}}\AgdaSpace{}%
\AgdaFunction{run}\AgdaSpace{}%
\AgdaBound{x}\AgdaSpace{}%
\AgdaBound{s}\AgdaSpace{}%
\AgdaBound{f₁}\AgdaSymbol{)}\AgdaSpace{}%
\AgdaSymbol{(}\AgdaFunction{buildReadNames}\AgdaSpace{}%
\AgdaSymbol{(}\AgdaFunction{run}\AgdaSpace{}%
\AgdaBound{x}\AgdaSpace{}%
\AgdaBound{s}\AgdaSymbol{)}\AgdaSpace{}%
\AgdaBound{xs}\AgdaSymbol{)}\AgdaSpace{}%
\AgdaSymbol{→}\AgdaSpace{}%
\AgdaFunction{script}\AgdaSpace{}%
\AgdaSymbol{(}\AgdaBound{x}\AgdaSpace{}%
\AgdaOperator{\AgdaInductiveConstructor{∷}}\AgdaSpace{}%
\AgdaBound{xs}\AgdaSymbol{)}\AgdaSpace{}%
\AgdaBound{s}\AgdaSpace{}%
\AgdaBound{f₁}\AgdaSpace{}%
\AgdaOperator{\AgdaDatatype{≡}}\AgdaSpace{}%
\AgdaFunction{script}\AgdaSpace{}%
\AgdaBound{xs}\AgdaSpace{}%
\AgdaBound{s}\AgdaSpace{}%
\AgdaBound{f₁}\<%
\\
\>[0]\AgdaFunction{helper6}\AgdaSpace{}%
\AgdaSymbol{\{}\AgdaBound{s}\AgdaSymbol{\}}\AgdaSpace{}%
\AgdaBound{xs}\AgdaSpace{}%
\AgdaBound{x}\AgdaSpace{}%
\AgdaBound{f₁∉}\AgdaSpace{}%
\AgdaBound{all₁}\AgdaSpace{}%
\AgdaSymbol{=}\AgdaSpace{}%
\AgdaFunction{sym}\AgdaSpace{}%
\AgdaSymbol{(}\AgdaFunction{exec-∷≡}\AgdaSpace{}%
\AgdaSymbol{\AgdaUnderscore{}}\AgdaSpace{}%
\AgdaBound{s}\AgdaSpace{}%
\AgdaSymbol{(}\AgdaFunction{run}\AgdaSpace{}%
\AgdaBound{x}\AgdaSpace{}%
\AgdaBound{s}\AgdaSymbol{)}\AgdaSpace{}%
\AgdaBound{xs}\AgdaSpace{}%
\AgdaBound{all₁}\AgdaSpace{}%
\AgdaSymbol{(}\AgdaFunction{lemma3}\AgdaSpace{}%
\AgdaSymbol{\AgdaUnderscore{}}\AgdaSpace{}%
\AgdaSymbol{(}\AgdaFunction{cmdWrites}\AgdaSpace{}%
\AgdaBound{x}\AgdaSpace{}%
\AgdaBound{s}\AgdaSymbol{)}\AgdaSpace{}%
\AgdaBound{f₁∉}\AgdaSymbol{))}\<%
\\
\\[\AgdaEmptyExtraSkip]%
\\[\AgdaEmptyExtraSkip]%
\>[0]\AgdaFunction{add-back}\AgdaSpace{}%
\AgdaSymbol{:}\AgdaSpace{}%
\AgdaSymbol{∀}\AgdaSpace{}%
\AgdaSymbol{\{}\AgdaBound{s}\AgdaSymbol{\}}\AgdaSpace{}%
\AgdaBound{as}\AgdaSpace{}%
\AgdaBound{bs}\AgdaSpace{}%
\AgdaBound{xs}\AgdaSpace{}%
\AgdaBound{x}\AgdaSpace{}%
\AgdaSymbol{→}\AgdaSpace{}%
\AgdaFunction{Disjoint}\AgdaSpace{}%
\AgdaSymbol{(}\AgdaFunction{cmdWriteNames}\AgdaSpace{}%
\AgdaBound{x}\AgdaSpace{}%
\AgdaSymbol{(}\AgdaFunction{script}\AgdaSpace{}%
\AgdaBound{as}\AgdaSpace{}%
\AgdaBound{s}\AgdaSymbol{))}\AgdaSpace{}%
\AgdaSymbol{(}\AgdaFunction{buildWriteNames}\AgdaSpace{}%
\AgdaSymbol{(}\AgdaFunction{run}\AgdaSpace{}%
\AgdaBound{x}\AgdaSpace{}%
\AgdaSymbol{(}\AgdaFunction{script}\AgdaSpace{}%
\AgdaBound{as}\AgdaSpace{}%
\AgdaBound{s}\AgdaSymbol{))}\AgdaSpace{}%
\AgdaBound{bs}\AgdaSymbol{)}\AgdaSpace{}%
\AgdaSymbol{→}\AgdaSpace{}%
\AgdaFunction{Disjoint}\AgdaSpace{}%
\AgdaSymbol{(}\AgdaFunction{cmdReadNames}\AgdaSpace{}%
\AgdaBound{x}\AgdaSpace{}%
\AgdaSymbol{(}\AgdaFunction{script}\AgdaSpace{}%
\AgdaBound{as}\AgdaSpace{}%
\AgdaBound{s}\AgdaSymbol{))}\AgdaSpace{}%
\AgdaSymbol{(}\AgdaFunction{buildWriteNames}\AgdaSpace{}%
\AgdaSymbol{(}\AgdaFunction{run}\AgdaSpace{}%
\AgdaBound{x}\AgdaSpace{}%
\AgdaSymbol{(}\AgdaFunction{script}\AgdaSpace{}%
\AgdaBound{as}\AgdaSpace{}%
\AgdaBound{s}\AgdaSymbol{))}\AgdaSpace{}%
\AgdaBound{bs}\AgdaSymbol{)}\AgdaSpace{}%
\AgdaSymbol{→}\AgdaSpace{}%
\AgdaFunction{Disjoint}\AgdaSpace{}%
\AgdaSymbol{(}\AgdaFunction{cmdWriteNames}\AgdaSpace{}%
\AgdaBound{x}\AgdaSpace{}%
\AgdaSymbol{(}\AgdaFunction{script}\AgdaSpace{}%
\AgdaBound{as}\AgdaSpace{}%
\AgdaBound{s}\AgdaSymbol{))}\AgdaSpace{}%
\AgdaSymbol{(}\AgdaFunction{buildReadNames}\AgdaSpace{}%
\AgdaSymbol{(}\AgdaFunction{run}\AgdaSpace{}%
\AgdaBound{x}\AgdaSpace{}%
\AgdaSymbol{(}\AgdaFunction{script}\AgdaSpace{}%
\AgdaBound{as}\AgdaSpace{}%
\AgdaBound{s}\AgdaSymbol{))}\AgdaSpace{}%
\AgdaBound{bs}\AgdaSymbol{)}\AgdaSpace{}%
\AgdaSymbol{→}\AgdaSpace{}%
\AgdaSymbol{(∀}\AgdaSpace{}%
\AgdaBound{f₁}\AgdaSpace{}%
\AgdaSymbol{→}\AgdaSpace{}%
\AgdaFunction{script}\AgdaSpace{}%
\AgdaSymbol{(}\AgdaBound{as}\AgdaSpace{}%
\AgdaOperator{\AgdaFunction{++}}\AgdaSpace{}%
\AgdaBound{bs}\AgdaSymbol{)}\AgdaSpace{}%
\AgdaBound{s}\AgdaSpace{}%
\AgdaBound{f₁}\AgdaSpace{}%
\AgdaOperator{\AgdaDatatype{≡}}\AgdaSpace{}%
\AgdaFunction{script}\AgdaSpace{}%
\AgdaBound{xs}\AgdaSpace{}%
\AgdaBound{s}\AgdaSpace{}%
\AgdaBound{f₁}\AgdaSymbol{)}\AgdaSpace{}%
\AgdaSymbol{→}\AgdaSpace{}%
\AgdaSymbol{(∀}\AgdaSpace{}%
\AgdaBound{f₁}\AgdaSpace{}%
\AgdaSymbol{→}\AgdaSpace{}%
\AgdaFunction{script}\AgdaSpace{}%
\AgdaSymbol{(}\AgdaBound{as}\AgdaSpace{}%
\AgdaOperator{\AgdaFunction{++}}\AgdaSpace{}%
\AgdaBound{x}\AgdaSpace{}%
\AgdaOperator{\AgdaInductiveConstructor{∷}}\AgdaSpace{}%
\AgdaBound{bs}\AgdaSymbol{)}\AgdaSpace{}%
\AgdaBound{s}\AgdaSpace{}%
\AgdaBound{f₁}\AgdaSpace{}%
\AgdaOperator{\AgdaDatatype{≡}}\AgdaSpace{}%
\AgdaFunction{script}\AgdaSpace{}%
\AgdaSymbol{(}\AgdaBound{xs}\AgdaSpace{}%
\AgdaOperator{\AgdaFunction{++}}\AgdaSpace{}%
\AgdaBound{x}\AgdaSpace{}%
\AgdaOperator{\AgdaInductiveConstructor{∷}}\AgdaSpace{}%
\AgdaInductiveConstructor{[]}\AgdaSymbol{)}\AgdaSpace{}%
\AgdaBound{s}\AgdaSpace{}%
\AgdaBound{f₁}\AgdaSymbol{)}\<%
\\
\>[0]\AgdaFunction{add-back}\AgdaSpace{}%
\AgdaSymbol{\{}\AgdaBound{s}\AgdaSymbol{\}}\AgdaSpace{}%
\AgdaBound{as}\AgdaSpace{}%
\AgdaBound{bs}\AgdaSpace{}%
\AgdaBound{xs}\AgdaSpace{}%
\AgdaBound{x}\AgdaSpace{}%
\AgdaBound{dsj₀}\AgdaSpace{}%
\AgdaBound{dsj}\AgdaSpace{}%
\AgdaBound{dsj₁}\AgdaSpace{}%
\AgdaBound{∀₁}\AgdaSpace{}%
\AgdaBound{f₁}\AgdaSpace{}%
\AgdaKeyword{with}\AgdaSpace{}%
\AgdaFunction{helper5}\AgdaSpace{}%
\AgdaSymbol{(}\AgdaFunction{buildReadNames}\AgdaSpace{}%
\AgdaSymbol{(}\AgdaFunction{run}\AgdaSpace{}%
\AgdaBound{x}\AgdaSpace{}%
\AgdaSymbol{(}\AgdaFunction{script}\AgdaSpace{}%
\AgdaBound{as}\AgdaSpace{}%
\AgdaBound{s}\AgdaSymbol{))}\AgdaSpace{}%
\AgdaBound{bs}\AgdaSymbol{)}\AgdaSpace{}%
\AgdaBound{x}\AgdaSpace{}%
\AgdaBound{dsj₁}\<%
\\
\>[0]\AgdaSymbol{...}\AgdaSpace{}%
\AgdaSymbol{|}\AgdaSpace{}%
\AgdaBound{all₁}\AgdaSpace{}%
\AgdaKeyword{with}\AgdaSpace{}%
\AgdaFunction{writes≡}\AgdaSpace{}%
\AgdaSymbol{(}\AgdaFunction{script}\AgdaSpace{}%
\AgdaBound{as}\AgdaSpace{}%
\AgdaBound{s}\AgdaSymbol{)}\AgdaSpace{}%
\AgdaSymbol{(}\AgdaFunction{run}\AgdaSpace{}%
\AgdaBound{x}\AgdaSpace{}%
\AgdaSymbol{(}\AgdaFunction{script}\AgdaSpace{}%
\AgdaBound{as}\AgdaSpace{}%
\AgdaBound{s}\AgdaSymbol{))}\AgdaSpace{}%
\AgdaBound{bs}\AgdaSpace{}%
\AgdaBound{all₁}\<%
\\
\>[0]\AgdaSymbol{...}\AgdaSpace{}%
\AgdaSymbol{|}\AgdaSpace{}%
\AgdaBound{≡₁}%
\>[11]\AgdaKeyword{with}\AgdaSpace{}%
\AgdaFunction{helper3}\AgdaSpace{}%
\AgdaBound{as}\AgdaSpace{}%
\AgdaBound{bs}\AgdaSpace{}%
\AgdaBound{xs}\AgdaSpace{}%
\AgdaBound{x}\AgdaSpace{}%
\AgdaSymbol{(}\AgdaFunction{subst}\AgdaSpace{}%
\AgdaSymbol{(λ}\AgdaSpace{}%
\AgdaBound{x₁}\AgdaSpace{}%
\AgdaSymbol{→}\AgdaSpace{}%
\AgdaFunction{Disjoint}\AgdaSpace{}%
\AgdaSymbol{\AgdaUnderscore{}}\AgdaSpace{}%
\AgdaBound{x₁}\AgdaSymbol{)}\AgdaSpace{}%
\AgdaSymbol{(}\AgdaFunction{sym}\AgdaSpace{}%
\AgdaBound{≡₁}\AgdaSymbol{)}\AgdaSpace{}%
\AgdaBound{dsj}\AgdaSymbol{)}\AgdaSpace{}%
\AgdaBound{∀₁}\<%
\\
\>[0]\AgdaSymbol{...}\AgdaSpace{}%
\AgdaSymbol{|}\AgdaSpace{}%
\AgdaBound{x≡}%
\>[11]\AgdaKeyword{with}\AgdaSpace{}%
\AgdaBound{f₁}\AgdaSpace{}%
\AgdaOperator{\AgdaFunction{∈?}}\AgdaSpace{}%
\AgdaFunction{cmdWriteNames}\AgdaSpace{}%
\AgdaBound{x}\AgdaSpace{}%
\AgdaSymbol{(}\AgdaFunction{script}\AgdaSpace{}%
\AgdaBound{as}\AgdaSpace{}%
\AgdaBound{s}\AgdaSymbol{)}\<%
\\
\>[0]\AgdaComment{-- we know as ++ x ≡ xs ++ x. just show bs doesnt change what x wrote to.}\<%
\\
\>[0]\AgdaSymbol{...}\AgdaSpace{}%
\AgdaSymbol{|}\AgdaSpace{}%
\AgdaInductiveConstructor{yes}\AgdaSpace{}%
\AgdaBound{f₁∈}\AgdaSpace{}%
\AgdaSymbol{=}\AgdaSpace{}%
\AgdaFunction{trans}\AgdaSpace{}%
\AgdaFunction{≡₂}\AgdaSpace{}%
\AgdaFunction{≡₃}\<%
\\
\>[0][@{}l@{\AgdaIndent{0}}]%
\>[2]\AgdaKeyword{where}%
\>[1259I]\AgdaFunction{≡₂}\AgdaSpace{}%
\AgdaSymbol{:}\AgdaSpace{}%
\AgdaFunction{script}\AgdaSpace{}%
\AgdaSymbol{(}\AgdaBound{as}\AgdaSpace{}%
\AgdaOperator{\AgdaFunction{++}}\AgdaSpace{}%
\AgdaBound{x}\AgdaSpace{}%
\AgdaOperator{\AgdaInductiveConstructor{∷}}\AgdaSpace{}%
\AgdaBound{bs}\AgdaSymbol{)}\AgdaSpace{}%
\AgdaBound{s}\AgdaSpace{}%
\AgdaBound{f₁}\AgdaSpace{}%
\AgdaOperator{\AgdaDatatype{≡}}\AgdaSpace{}%
\AgdaFunction{run}\AgdaSpace{}%
\AgdaBound{x}\AgdaSpace{}%
\AgdaSymbol{(}\AgdaFunction{script}\AgdaSpace{}%
\AgdaBound{as}\AgdaSpace{}%
\AgdaBound{s}\AgdaSymbol{)}\AgdaSpace{}%
\AgdaBound{f₁}\<%
\\
\>[.][@{}l@{}]\<[1259I]%
\>[8]\AgdaFunction{≡₂}\AgdaSpace{}%
\AgdaSymbol{=}\AgdaSpace{}%
\AgdaFunction{trans}\AgdaSpace{}%
\AgdaSymbol{(}\AgdaFunction{cong-app}\AgdaSpace{}%
\AgdaSymbol{(}\AgdaFunction{script≡}\AgdaSpace{}%
\AgdaBound{as}\AgdaSpace{}%
\AgdaSymbol{(}\AgdaBound{x}\AgdaSpace{}%
\AgdaOperator{\AgdaInductiveConstructor{∷}}\AgdaSpace{}%
\AgdaBound{bs}\AgdaSymbol{))}\AgdaSpace{}%
\AgdaBound{f₁}\AgdaSymbol{)}\AgdaSpace{}%
\AgdaSymbol{(}\AgdaFunction{helper8}\AgdaSpace{}%
\AgdaBound{bs}\AgdaSpace{}%
\AgdaSymbol{λ}\AgdaSpace{}%
\AgdaBound{x₁}\AgdaSpace{}%
\AgdaSymbol{→}\AgdaSpace{}%
\AgdaBound{dsj₀}\AgdaSpace{}%
\AgdaSymbol{(}\AgdaBound{f₁∈}\AgdaSpace{}%
\AgdaOperator{\AgdaInductiveConstructor{,}}\AgdaSpace{}%
\AgdaBound{x₁}\AgdaSymbol{))}\<%
\\
\>[8]\AgdaFunction{≡₃}\AgdaSpace{}%
\AgdaSymbol{:}\AgdaSpace{}%
\AgdaFunction{run}\AgdaSpace{}%
\AgdaBound{x}\AgdaSpace{}%
\AgdaSymbol{(}\AgdaFunction{script}\AgdaSpace{}%
\AgdaBound{as}\AgdaSpace{}%
\AgdaBound{s}\AgdaSymbol{)}\AgdaSpace{}%
\AgdaBound{f₁}\AgdaSpace{}%
\AgdaOperator{\AgdaDatatype{≡}}\AgdaSpace{}%
\AgdaFunction{script}\AgdaSpace{}%
\AgdaSymbol{(}\AgdaBound{xs}\AgdaSpace{}%
\AgdaOperator{\AgdaFunction{++}}\AgdaSpace{}%
\AgdaBound{x}\AgdaSpace{}%
\AgdaOperator{\AgdaInductiveConstructor{∷}}\AgdaSpace{}%
\AgdaInductiveConstructor{[]}\AgdaSymbol{)}\AgdaSpace{}%
\AgdaBound{s}\AgdaSpace{}%
\AgdaBound{f₁}\<%
\\
\>[8]\AgdaFunction{≡₃}\AgdaSpace{}%
\AgdaSymbol{=}\AgdaSpace{}%
\AgdaFunction{trans}%
\>[1312I]\AgdaSymbol{(}\AgdaFunction{helper7}\AgdaSpace{}%
\AgdaSymbol{\{}\AgdaFunction{script}\AgdaSpace{}%
\AgdaBound{as}\AgdaSpace{}%
\AgdaBound{s}\AgdaSymbol{\}}\AgdaSpace{}%
\AgdaSymbol{\{}\AgdaFunction{script}\AgdaSpace{}%
\AgdaBound{xs}\AgdaSpace{}%
\AgdaBound{s}\AgdaSymbol{\}}\AgdaSpace{}%
\AgdaBound{f₁}\AgdaSpace{}%
\AgdaSymbol{(}\AgdaFunction{cmdWrites}\AgdaSpace{}%
\AgdaBound{x}\AgdaSpace{}%
\AgdaSymbol{(}\AgdaFunction{script}\AgdaSpace{}%
\AgdaBound{as}\AgdaSpace{}%
\AgdaBound{s}\AgdaSymbol{))}\AgdaSpace{}%
\AgdaSymbol{(}\AgdaFunction{cmdWrites}\AgdaSpace{}%
\AgdaBound{x}\AgdaSpace{}%
\AgdaSymbol{(}\AgdaFunction{script}\AgdaSpace{}%
\AgdaBound{xs}\AgdaSpace{}%
\AgdaBound{s}\AgdaSymbol{))}\AgdaSpace{}%
\AgdaSymbol{(}\AgdaFunction{cong}\AgdaSpace{}%
\AgdaField{proj₂}\AgdaSpace{}%
\AgdaBound{x≡}\AgdaSymbol{)}\AgdaSpace{}%
\AgdaBound{f₁∈}\AgdaSymbol{)}\<%
\\
\>[.][@{}l@{}]\<[1312I]%
\>[19]\AgdaSymbol{(}\AgdaFunction{sym}\AgdaSpace{}%
\AgdaSymbol{(}\AgdaFunction{cong-app}\AgdaSpace{}%
\AgdaSymbol{(}\AgdaFunction{script≡}\AgdaSpace{}%
\AgdaBound{xs}\AgdaSpace{}%
\AgdaSymbol{(}\AgdaBound{x}\AgdaSpace{}%
\AgdaOperator{\AgdaInductiveConstructor{∷}}\AgdaSpace{}%
\AgdaInductiveConstructor{[]}\AgdaSymbol{))}\AgdaSpace{}%
\AgdaBound{f₁}\AgdaSymbol{))}\<%
\\
\>[0]\<%
\\
\>[0]\AgdaSymbol{...}\AgdaSpace{}%
\AgdaSymbol{|}\AgdaSpace{}%
\AgdaInductiveConstructor{no}\AgdaSpace{}%
\AgdaBound{f₁∉}\AgdaSpace{}%
\AgdaKeyword{with}%
\>[1345I]\AgdaFunction{trans}\AgdaSpace{}%
\AgdaSymbol{(}\AgdaFunction{cong-app}\AgdaSpace{}%
\AgdaSymbol{(}\AgdaFunction{script≡}\AgdaSpace{}%
\AgdaBound{as}\AgdaSpace{}%
\AgdaSymbol{(}\AgdaBound{x}\AgdaSpace{}%
\AgdaOperator{\AgdaInductiveConstructor{∷}}\AgdaSpace{}%
\AgdaBound{bs}\AgdaSymbol{))}\AgdaSpace{}%
\AgdaBound{f₁}\AgdaSymbol{)}\AgdaSpace{}%
\AgdaSymbol{(}\AgdaFunction{trans}\AgdaSpace{}%
\AgdaSymbol{(}\AgdaFunction{helper6}\AgdaSpace{}%
\AgdaBound{bs}\AgdaSpace{}%
\AgdaBound{x}\AgdaSpace{}%
\AgdaBound{f₁∉}\AgdaSpace{}%
\AgdaBound{all₁}\AgdaSymbol{)}\AgdaSpace{}%
\AgdaSymbol{(}\AgdaFunction{sym}\AgdaSpace{}%
\AgdaSymbol{(}\AgdaFunction{cong-app}\AgdaSpace{}%
\AgdaSymbol{(}\AgdaFunction{script≡}\AgdaSpace{}%
\AgdaBound{as}\AgdaSpace{}%
\AgdaBound{bs}\AgdaSymbol{)}\AgdaSpace{}%
\AgdaBound{f₁}\AgdaSymbol{)))}\<%
\\
\>[.][@{}l@{}]\<[1345I]%
\>[18]\AgdaComment{-- script xs s f₁ ≡ script (xs ++ x ∷ []) s f₁}\<%
\\
\>[0]\AgdaSymbol{...}\AgdaSpace{}%
\AgdaSymbol{|}\AgdaSpace{}%
\AgdaBound{≡₂}%
\>[13]\AgdaKeyword{with}\AgdaSpace{}%
\AgdaFunction{trans}\AgdaSpace{}%
\AgdaSymbol{(}\AgdaFunction{lemma3}\AgdaSpace{}%
\AgdaSymbol{\AgdaUnderscore{}}\AgdaSpace{}%
\AgdaSymbol{(}\AgdaFunction{cmdWrites}\AgdaSpace{}%
\AgdaBound{x}\AgdaSpace{}%
\AgdaSymbol{(}\AgdaFunction{script}\AgdaSpace{}%
\AgdaBound{xs}\AgdaSpace{}%
\AgdaBound{s}\AgdaSymbol{))}\AgdaSpace{}%
\AgdaSymbol{(}\AgdaFunction{subst}\AgdaSpace{}%
\AgdaSymbol{(λ}\AgdaSpace{}%
\AgdaBound{x₁}\AgdaSpace{}%
\AgdaSymbol{→}\AgdaSpace{}%
\AgdaSymbol{\AgdaUnderscore{}}\AgdaSpace{}%
\AgdaOperator{\AgdaFunction{∉}}\AgdaSpace{}%
\AgdaBound{x₁}\AgdaSymbol{)}\AgdaSpace{}%
\AgdaSymbol{(}\AgdaFunction{cong}\AgdaSpace{}%
\AgdaSymbol{(}\AgdaFunction{map}\AgdaSpace{}%
\AgdaField{proj₁}\AgdaSpace{}%
\AgdaOperator{\AgdaFunction{∘}}\AgdaSpace{}%
\AgdaField{proj₂}\AgdaSymbol{)}\AgdaSpace{}%
\AgdaBound{x≡}\AgdaSymbol{)}\AgdaSpace{}%
\AgdaBound{f₁∉}\AgdaSymbol{))}\AgdaSpace{}%
\AgdaSymbol{(}\AgdaFunction{cong-app}\AgdaSpace{}%
\AgdaSymbol{(}\AgdaFunction{exec≡₃}\AgdaSpace{}%
\AgdaBound{x}\AgdaSpace{}%
\AgdaBound{xs}\AgdaSymbol{)}\AgdaSpace{}%
\AgdaBound{f₁}\AgdaSymbol{)}\<%
\\
\>[0]\AgdaSymbol{...}\AgdaSpace{}%
\AgdaSymbol{|}\AgdaSpace{}%
\AgdaBound{≡₃}\AgdaSpace{}%
\AgdaSymbol{=}\AgdaSpace{}%
\AgdaFunction{trans}\AgdaSpace{}%
\AgdaBound{≡₂}\AgdaSpace{}%
\AgdaSymbol{(}\AgdaFunction{trans}\AgdaSpace{}%
\AgdaSymbol{(}\AgdaBound{∀₁}\AgdaSpace{}%
\AgdaBound{f₁}\AgdaSymbol{)}\AgdaSpace{}%
\AgdaBound{≡₃}\AgdaSymbol{)}\<%
\\
\\[\AgdaEmptyExtraSkip]%
\\[\AgdaEmptyExtraSkip]%
\>[0]\AgdaFunction{reordered}%
\>[1403I]\AgdaSymbol{:}\AgdaSpace{}%
\AgdaSymbol{∀}\AgdaSpace{}%
\AgdaSymbol{\{}\AgdaBound{s}\AgdaSymbol{\}}\AgdaSpace{}%
\AgdaSymbol{\{}\AgdaBound{ls}\AgdaSymbol{\}}\AgdaSpace{}%
\AgdaBound{xs}\AgdaSpace{}%
\AgdaBound{ys}\AgdaSpace{}%
\AgdaSymbol{→}\AgdaSpace{}%
\AgdaFunction{length}\AgdaSpace{}%
\AgdaBound{xs}\AgdaSpace{}%
\AgdaOperator{\AgdaDatatype{≡}}\AgdaSpace{}%
\AgdaFunction{length}\AgdaSpace{}%
\AgdaBound{ys}\AgdaSpace{}%
\AgdaSymbol{→}\AgdaSpace{}%
\AgdaBound{xs}\AgdaSpace{}%
\AgdaOperator{\AgdaDatatype{↭}}\AgdaSpace{}%
\AgdaBound{ys}\AgdaSpace{}%
\AgdaSymbol{→}\AgdaSpace{}%
\AgdaFunction{UniqueEvidence}\AgdaSpace{}%
\AgdaBound{xs}\AgdaSpace{}%
\AgdaBound{ys}\AgdaSpace{}%
\AgdaSymbol{(}\AgdaFunction{map}\AgdaSpace{}%
\AgdaField{proj₁}\AgdaSpace{}%
\AgdaBound{ls}\AgdaSymbol{)}\<%
\\
\>[1403I][@{}l@{\AgdaIndent{0}}]%
\>[11]\AgdaSymbol{→}\AgdaSpace{}%
\AgdaDatatype{HazardFree}\AgdaSpace{}%
\AgdaBound{s}\AgdaSpace{}%
\AgdaBound{xs}\AgdaSpace{}%
\AgdaSymbol{(}\AgdaFunction{reverse}\AgdaSpace{}%
\AgdaBound{ys}\AgdaSymbol{)}\AgdaSpace{}%
\AgdaBound{ls}\AgdaSpace{}%
\AgdaSymbol{→}\AgdaSpace{}%
\AgdaSymbol{(∀}\AgdaSpace{}%
\AgdaBound{f₁}\AgdaSpace{}%
\AgdaSymbol{→}\AgdaSpace{}%
\AgdaFunction{script}\AgdaSpace{}%
\AgdaBound{xs}\AgdaSpace{}%
\AgdaBound{s}\AgdaSpace{}%
\AgdaBound{f₁}\AgdaSpace{}%
\AgdaOperator{\AgdaDatatype{≡}}\AgdaSpace{}%
\AgdaFunction{script}\AgdaSpace{}%
\AgdaSymbol{(}\AgdaFunction{reverse}\AgdaSpace{}%
\AgdaBound{ys}\AgdaSymbol{)}\AgdaSpace{}%
\AgdaBound{s}\AgdaSpace{}%
\AgdaBound{f₁}\AgdaSymbol{)}\<%
\\
\>[0]\AgdaFunction{reordered}\AgdaSpace{}%
\AgdaInductiveConstructor{[]}\AgdaSpace{}%
\AgdaInductiveConstructor{[]}\AgdaSpace{}%
\AgdaSymbol{\AgdaUnderscore{}}\AgdaSpace{}%
\AgdaBound{p}\AgdaSpace{}%
\AgdaSymbol{\AgdaUnderscore{}}\AgdaSpace{}%
\AgdaBound{hf}\AgdaSpace{}%
\AgdaBound{f₁}\AgdaSpace{}%
\AgdaSymbol{=}\AgdaSpace{}%
\AgdaInductiveConstructor{refl}\<%
\\
\>[0]\AgdaCatchallClause{\AgdaFunction{reordered}}\AgdaSpace{}%
\AgdaCatchallClause{\AgdaSymbol{\{}}\AgdaCatchallClause{\AgdaBound{s}}\AgdaCatchallClause{\AgdaSymbol{\}}}\AgdaSpace{}%
\AgdaCatchallClause{\AgdaSymbol{\{}}\AgdaCatchallClause{\AgdaBound{ls}}\AgdaCatchallClause{\AgdaSymbol{\}}}\AgdaSpace{}%
\AgdaCatchallClause{\AgdaBound{xs}}\AgdaSpace{}%
\AgdaCatchallClause{\AgdaSymbol{(}}\AgdaCatchallClause{\AgdaBound{x}}\AgdaSpace{}%
\AgdaCatchallClause{\AgdaOperator{\AgdaInductiveConstructor{∷}}}\AgdaSpace{}%
\AgdaCatchallClause{\AgdaBound{ys}}\AgdaCatchallClause{\AgdaSymbol{)}}\AgdaSpace{}%
\AgdaCatchallClause{\AgdaSymbol{\AgdaUnderscore{}}}\AgdaSpace{}%
\AgdaCatchallClause{\AgdaBound{p}}\AgdaSpace{}%
\AgdaCatchallClause{\AgdaSymbol{(}}\AgdaCatchallClause{\AgdaBound{uxs}}\AgdaSpace{}%
\AgdaCatchallClause{\AgdaOperator{\AgdaInductiveConstructor{,}}}\AgdaSpace{}%
\AgdaCatchallClause{\AgdaBound{ux}}\AgdaSpace{}%
\AgdaCatchallClause{\AgdaOperator{\AgdaInductiveConstructor{∷}}}\AgdaSpace{}%
\AgdaCatchallClause{\AgdaBound{uys}}\AgdaSpace{}%
\AgdaCatchallClause{\AgdaOperator{\AgdaInductiveConstructor{,}}}\AgdaSpace{}%
\AgdaCatchallClause{\AgdaBound{uls}}\AgdaSpace{}%
\AgdaCatchallClause{\AgdaOperator{\AgdaInductiveConstructor{,}}}\AgdaSpace{}%
\AgdaCatchallClause{\AgdaBound{dsj}}\AgdaCatchallClause{\AgdaSymbol{)}}\AgdaSpace{}%
\AgdaCatchallClause{\AgdaBound{hf}}\AgdaSpace{}%
\AgdaCatchallClause{\AgdaBound{f₁}}\AgdaSpace{}%
\AgdaKeyword{with}\AgdaSpace{}%
\AgdaFunction{∈-∃++}\AgdaSpace{}%
\AgdaSymbol{(}\AgdaFunction{∈-resp-↭}\AgdaSpace{}%
\AgdaSymbol{(}\AgdaFunction{↭-sym}\AgdaSpace{}%
\AgdaBound{p}\AgdaSymbol{)}\AgdaSpace{}%
\AgdaSymbol{(}\AgdaInductiveConstructor{here}\AgdaSpace{}%
\AgdaInductiveConstructor{refl}\AgdaSymbol{))}\<%
\\
\>[0]\AgdaSymbol{...}\AgdaSpace{}%
\AgdaSymbol{|}\AgdaSpace{}%
\AgdaSymbol{(}\AgdaBound{as}\AgdaSpace{}%
\AgdaOperator{\AgdaInductiveConstructor{,}}\AgdaSpace{}%
\AgdaBound{bs}\AgdaSpace{}%
\AgdaOperator{\AgdaInductiveConstructor{,}}\AgdaSpace{}%
\AgdaBound{≡₁}\AgdaSymbol{)}\AgdaSpace{}%
\AgdaKeyword{with}\AgdaSpace{}%
\AgdaFunction{add-back}\AgdaSpace{}%
\AgdaBound{as}\AgdaSpace{}%
\AgdaBound{bs}\AgdaSpace{}%
\AgdaSymbol{(}\AgdaFunction{reverse}\AgdaSpace{}%
\AgdaBound{ys}\AgdaSymbol{)}\AgdaSpace{}%
\AgdaBound{x}\AgdaSpace{}%
\AgdaFunction{dsj₁}\AgdaSpace{}%
\AgdaFunction{dsj₂}\AgdaSpace{}%
\AgdaFunction{dsj₃}\AgdaSpace{}%
\AgdaSymbol{(}\AgdaFunction{reordered}\AgdaSpace{}%
\AgdaSymbol{(}\AgdaBound{as}\AgdaSpace{}%
\AgdaOperator{\AgdaFunction{++}}\AgdaSpace{}%
\AgdaBound{bs}\AgdaSymbol{)}\AgdaSpace{}%
\AgdaBound{ys}\AgdaSpace{}%
\AgdaSymbol{(}\AgdaFunction{↭-length}\AgdaSpace{}%
\AgdaFunction{↭₂}\AgdaSymbol{)}\AgdaSpace{}%
\AgdaFunction{↭₂}\AgdaSpace{}%
\AgdaSymbol{(}\AgdaFunction{uas++bs}\AgdaSpace{}%
\AgdaOperator{\AgdaInductiveConstructor{,}}\AgdaSpace{}%
\AgdaBound{uys}\AgdaSpace{}%
\AgdaOperator{\AgdaInductiveConstructor{,}}\AgdaSpace{}%
\AgdaBound{uls}\AgdaSpace{}%
\AgdaOperator{\AgdaInductiveConstructor{,}}\AgdaSpace{}%
\AgdaFunction{dsj₀}\AgdaSymbol{)}\AgdaSpace{}%
\AgdaFunction{hf₁}\AgdaSymbol{)}\<%
\\
\>[0][@{}l@{\AgdaIndent{0}}]%
\>[2]\AgdaKeyword{where}%
\>[1513I]\AgdaFunction{↭₂}\AgdaSpace{}%
\AgdaSymbol{:}\AgdaSpace{}%
\AgdaBound{as}\AgdaSpace{}%
\AgdaOperator{\AgdaFunction{++}}\AgdaSpace{}%
\AgdaBound{bs}\AgdaSpace{}%
\AgdaOperator{\AgdaDatatype{↭}}\AgdaSpace{}%
\AgdaBound{ys}\<%
\\
\>[.][@{}l@{}]\<[1513I]%
\>[8]\AgdaFunction{↭₂}\AgdaSpace{}%
\AgdaSymbol{=}\AgdaSpace{}%
\AgdaFunction{drop-mid}\AgdaSpace{}%
\AgdaBound{as}\AgdaSpace{}%
\AgdaInductiveConstructor{[]}\AgdaSpace{}%
\AgdaSymbol{(}\AgdaFunction{subst}\AgdaSpace{}%
\AgdaSymbol{(λ}\AgdaSpace{}%
\AgdaBound{x₁}\AgdaSpace{}%
\AgdaSymbol{→}\AgdaSpace{}%
\AgdaBound{x₁}\AgdaSpace{}%
\AgdaOperator{\AgdaDatatype{↭}}\AgdaSpace{}%
\AgdaBound{x}\AgdaSpace{}%
\AgdaOperator{\AgdaInductiveConstructor{∷}}\AgdaSpace{}%
\AgdaBound{ys}\AgdaSymbol{)}\AgdaSpace{}%
\AgdaBound{≡₁}\AgdaSpace{}%
\AgdaBound{p}\AgdaSymbol{)}\<%
\\
\>[8]\AgdaFunction{uas++bs}\AgdaSpace{}%
\AgdaSymbol{:}\AgdaSpace{}%
\AgdaDatatype{Unique}\AgdaSpace{}%
\AgdaSymbol{(}\AgdaBound{as}\AgdaSpace{}%
\AgdaOperator{\AgdaFunction{++}}\AgdaSpace{}%
\AgdaBound{bs}\AgdaSymbol{)}\<%
\\
\>[8]\AgdaFunction{uas++bs}\AgdaSpace{}%
\AgdaSymbol{=}\AgdaSpace{}%
\AgdaFunction{unique-drop-mid}\AgdaSpace{}%
\AgdaBound{as}\AgdaSpace{}%
\AgdaSymbol{(}\AgdaFunction{subst}\AgdaSpace{}%
\AgdaSymbol{(λ}\AgdaSpace{}%
\AgdaBound{x₁}\AgdaSpace{}%
\AgdaSymbol{→}\AgdaSpace{}%
\AgdaDatatype{Unique}\AgdaSpace{}%
\AgdaBound{x₁}\AgdaSymbol{)}\AgdaSpace{}%
\AgdaBound{≡₁}\AgdaSpace{}%
\AgdaBound{uxs}\AgdaSymbol{)}\<%
\\
\>[8]\AgdaFunction{dsj₀}\AgdaSpace{}%
\AgdaSymbol{:}\AgdaSpace{}%
\AgdaFunction{Disjoint}\AgdaSpace{}%
\AgdaSymbol{(}\AgdaBound{as}\AgdaSpace{}%
\AgdaOperator{\AgdaFunction{++}}\AgdaSpace{}%
\AgdaBound{bs}\AgdaSymbol{)}\AgdaSpace{}%
\AgdaSymbol{(}\AgdaFunction{map}\AgdaSpace{}%
\AgdaField{proj₁}\AgdaSpace{}%
\AgdaBound{ls}\AgdaSymbol{)}\<%
\\
\>[8]\AgdaFunction{dsj₀}\AgdaSpace{}%
\AgdaBound{x₂}\AgdaSpace{}%
\AgdaKeyword{with}\AgdaSpace{}%
\AgdaFunction{∈-++⁻}\AgdaSpace{}%
\AgdaBound{as}\AgdaSpace{}%
\AgdaSymbol{(}\AgdaField{proj₁}\AgdaSpace{}%
\AgdaBound{x₂}\AgdaSymbol{)}\<%
\\
\>[8]\AgdaSymbol{...}\AgdaSpace{}%
\AgdaSymbol{|}\AgdaSpace{}%
\AgdaInductiveConstructor{inj₁}\AgdaSpace{}%
\AgdaOperator{\AgdaBound{\AgdaUnderscore{}∈as}}\AgdaSpace{}%
\AgdaSymbol{=}\AgdaSpace{}%
\AgdaBound{dsj}\AgdaSpace{}%
\AgdaSymbol{(}\AgdaFunction{subst}\AgdaSpace{}%
\AgdaSymbol{(λ}\AgdaSpace{}%
\AgdaBound{x₁}\AgdaSpace{}%
\AgdaSymbol{→}\AgdaSpace{}%
\AgdaSymbol{\AgdaUnderscore{}}\AgdaSpace{}%
\AgdaOperator{\AgdaFunction{∈}}\AgdaSpace{}%
\AgdaBound{x₁}\AgdaSymbol{)}\AgdaSpace{}%
\AgdaSymbol{(}\AgdaFunction{sym}\AgdaSpace{}%
\AgdaBound{≡₁}\AgdaSymbol{)}\AgdaSpace{}%
\AgdaSymbol{(}\AgdaFunction{∈-++⁺ˡ}\AgdaSpace{}%
\AgdaOperator{\AgdaBound{\AgdaUnderscore{}∈as}}\AgdaSymbol{)}\AgdaSpace{}%
\AgdaOperator{\AgdaInductiveConstructor{,}}\AgdaSpace{}%
\AgdaSymbol{(}\AgdaField{proj₂}\AgdaSpace{}%
\AgdaBound{x₂}\AgdaSymbol{))}\<%
\\
\>[8]\AgdaSymbol{...}\AgdaSpace{}%
\AgdaSymbol{|}\AgdaSpace{}%
\AgdaInductiveConstructor{inj₂}\AgdaSpace{}%
\AgdaOperator{\AgdaBound{\AgdaUnderscore{}∈bs}}\AgdaSpace{}%
\AgdaSymbol{=}\AgdaSpace{}%
\AgdaBound{dsj}\AgdaSpace{}%
\AgdaSymbol{(}\AgdaFunction{subst}\AgdaSpace{}%
\AgdaSymbol{(λ}\AgdaSpace{}%
\AgdaBound{x₁}\AgdaSpace{}%
\AgdaSymbol{→}\AgdaSpace{}%
\AgdaSymbol{\AgdaUnderscore{}}\AgdaSpace{}%
\AgdaOperator{\AgdaFunction{∈}}\AgdaSpace{}%
\AgdaBound{x₁}\AgdaSymbol{)}\AgdaSpace{}%
\AgdaSymbol{(}\AgdaFunction{sym}\AgdaSpace{}%
\AgdaBound{≡₁}\AgdaSymbol{)}\AgdaSpace{}%
\AgdaSymbol{(}\AgdaFunction{∈-++⁺ʳ}\AgdaSpace{}%
\AgdaBound{as}\AgdaSpace{}%
\AgdaSymbol{(}\AgdaInductiveConstructor{there}\AgdaSpace{}%
\AgdaOperator{\AgdaBound{\AgdaUnderscore{}∈bs}}\AgdaSymbol{))}\AgdaSpace{}%
\AgdaOperator{\AgdaInductiveConstructor{,}}\AgdaSpace{}%
\AgdaSymbol{(}\AgdaField{proj₂}\AgdaSpace{}%
\AgdaBound{x₂}\AgdaSymbol{))}\<%
\\
\>[8]\AgdaFunction{⊆₁}\AgdaSpace{}%
\AgdaSymbol{:}\AgdaSpace{}%
\AgdaBound{as}\AgdaSpace{}%
\AgdaOperator{\AgdaFunction{++}}\AgdaSpace{}%
\AgdaBound{x}\AgdaSpace{}%
\AgdaOperator{\AgdaInductiveConstructor{∷}}\AgdaSpace{}%
\AgdaBound{bs}\AgdaSpace{}%
\AgdaOperator{\AgdaFunction{⊆}}\AgdaSpace{}%
\AgdaFunction{reverse}\AgdaSpace{}%
\AgdaBound{ys}\AgdaSpace{}%
\AgdaOperator{\AgdaFunction{++}}\AgdaSpace{}%
\AgdaBound{x}\AgdaSpace{}%
\AgdaOperator{\AgdaInductiveConstructor{∷}}\AgdaSpace{}%
\AgdaInductiveConstructor{[]}\<%
\\
\>[8]\AgdaFunction{⊆₁}\AgdaSpace{}%
\AgdaBound{x₁∈as++x∷bs}\AgdaSpace{}%
\AgdaSymbol{=}\AgdaSpace{}%
\AgdaFunction{subst}%
\>[1621I]\AgdaSymbol{(λ}\AgdaSpace{}%
\AgdaBound{x₂}\AgdaSpace{}%
\AgdaSymbol{→}\AgdaSpace{}%
\AgdaSymbol{\AgdaUnderscore{}}\AgdaSpace{}%
\AgdaOperator{\AgdaFunction{∈}}\AgdaSpace{}%
\AgdaBound{x₂}\AgdaSymbol{)}\<%
\\
\>[.][@{}l@{}]\<[1621I]%
\>[31]\AgdaSymbol{(}\AgdaFunction{unfold-reverse}\AgdaSpace{}%
\AgdaBound{x}\AgdaSpace{}%
\AgdaBound{ys}\AgdaSymbol{)}\<%
\\
\>[31]\AgdaSymbol{(}\AgdaFunction{reverse⁺}\AgdaSpace{}%
\AgdaSymbol{(}\AgdaFunction{∈-resp-↭}\AgdaSpace{}%
\AgdaBound{p}\AgdaSpace{}%
\AgdaSymbol{(}\AgdaFunction{subst}\AgdaSpace{}%
\AgdaSymbol{(λ}\AgdaSpace{}%
\AgdaBound{x₂}\AgdaSpace{}%
\AgdaSymbol{→}\AgdaSpace{}%
\AgdaSymbol{\AgdaUnderscore{}}\AgdaSpace{}%
\AgdaOperator{\AgdaFunction{∈}}\AgdaSpace{}%
\AgdaBound{x₂}\AgdaSymbol{)}\AgdaSpace{}%
\AgdaSymbol{(}\AgdaFunction{sym}\AgdaSpace{}%
\AgdaBound{≡₁}\AgdaSymbol{)}\AgdaSpace{}%
\AgdaBound{x₁∈as++x∷bs}\AgdaSymbol{)))}\<%
\\
\>[8]\AgdaFunction{hf₀}\AgdaSpace{}%
\AgdaSymbol{:}\AgdaSpace{}%
\AgdaDatatype{HazardFree}\AgdaSpace{}%
\AgdaBound{s}\AgdaSpace{}%
\AgdaSymbol{(}\AgdaBound{as}\AgdaSpace{}%
\AgdaOperator{\AgdaFunction{++}}\AgdaSpace{}%
\AgdaBound{x}\AgdaSpace{}%
\AgdaOperator{\AgdaInductiveConstructor{∷}}\AgdaSpace{}%
\AgdaBound{bs}\AgdaSymbol{)}\AgdaSpace{}%
\AgdaSymbol{((}\AgdaFunction{reverse}\AgdaSpace{}%
\AgdaBound{ys}\AgdaSymbol{)}\AgdaSpace{}%
\AgdaOperator{\AgdaFunction{++}}\AgdaSpace{}%
\AgdaBound{x}\AgdaSpace{}%
\AgdaOperator{\AgdaInductiveConstructor{∷}}\AgdaSpace{}%
\AgdaInductiveConstructor{[]}\AgdaSymbol{)}\AgdaSpace{}%
\AgdaSymbol{\AgdaUnderscore{}}\<%
\\
\>[8]\AgdaFunction{hf₀}\AgdaSpace{}%
\AgdaSymbol{=}\AgdaSpace{}%
\AgdaFunction{subst₂}\AgdaSpace{}%
\AgdaSymbol{(λ}\AgdaSpace{}%
\AgdaBound{x₂}\AgdaSpace{}%
\AgdaBound{x₃}\AgdaSpace{}%
\AgdaSymbol{→}\AgdaSpace{}%
\AgdaDatatype{HazardFree}\AgdaSpace{}%
\AgdaBound{s}\AgdaSpace{}%
\AgdaBound{x₂}\AgdaSpace{}%
\AgdaBound{x₃}\AgdaSpace{}%
\AgdaSymbol{\AgdaUnderscore{})}\AgdaSpace{}%
\AgdaBound{≡₁}\AgdaSpace{}%
\AgdaSymbol{(}\AgdaFunction{unfold-reverse}\AgdaSpace{}%
\AgdaBound{x}\AgdaSpace{}%
\AgdaBound{ys}\AgdaSymbol{)}\AgdaSpace{}%
\AgdaBound{hf}\<%
\\
\>[8]\AgdaFunction{hf₁}\AgdaSpace{}%
\AgdaSymbol{:}\AgdaSpace{}%
\AgdaDatatype{HazardFree}\AgdaSpace{}%
\AgdaBound{s}\AgdaSpace{}%
\AgdaSymbol{(}\AgdaBound{as}\AgdaSpace{}%
\AgdaOperator{\AgdaFunction{++}}\AgdaSpace{}%
\AgdaBound{bs}\AgdaSymbol{)}\AgdaSpace{}%
\AgdaSymbol{(}\AgdaFunction{reverse}\AgdaSpace{}%
\AgdaBound{ys}\AgdaSymbol{)}\AgdaSpace{}%
\AgdaSymbol{\AgdaUnderscore{}}\<%
\\
\>[8]\AgdaFunction{hf₁}\AgdaSpace{}%
\AgdaSymbol{=}\AgdaSpace{}%
\AgdaFunction{hf-drop-mid}%
\>[1683I]\AgdaBound{as}\AgdaSpace{}%
\AgdaBound{bs}\AgdaSpace{}%
\AgdaSymbol{(}\AgdaFunction{reverse}\AgdaSpace{}%
\AgdaBound{ys}\AgdaSymbol{)}\AgdaSpace{}%
\AgdaFunction{⊆₁}\<%
\\
\>[.][@{}l@{}]\<[1683I]%
\>[26]\AgdaSymbol{(}\AgdaFunction{subst}\AgdaSpace{}%
\AgdaSymbol{(λ}\AgdaSpace{}%
\AgdaBound{x₁}\AgdaSpace{}%
\AgdaSymbol{→}\AgdaSpace{}%
\AgdaDatatype{Unique}\AgdaSpace{}%
\AgdaBound{x₁}\AgdaSymbol{)}\AgdaSpace{}%
\AgdaBound{≡₁}\AgdaSpace{}%
\AgdaBound{uxs}\AgdaSymbol{)}\<%
\\
\>[26]\AgdaSymbol{(}\AgdaFunction{subst}\AgdaSpace{}%
\AgdaSymbol{(λ}\AgdaSpace{}%
\AgdaBound{x₁}\AgdaSpace{}%
\AgdaSymbol{→}\AgdaSpace{}%
\AgdaDatatype{Unique}\AgdaSpace{}%
\AgdaBound{x₁}\AgdaSymbol{)}\AgdaSpace{}%
\AgdaSymbol{(}\AgdaFunction{unfold-reverse}\AgdaSpace{}%
\AgdaBound{x}\AgdaSpace{}%
\AgdaBound{ys}\AgdaSymbol{)}\AgdaSpace{}%
\AgdaSymbol{(}\AgdaFunction{unique-reverse}\AgdaSpace{}%
\AgdaSymbol{(}\AgdaBound{x}\AgdaSpace{}%
\AgdaOperator{\AgdaInductiveConstructor{∷}}\AgdaSpace{}%
\AgdaBound{ys}\AgdaSymbol{)}\AgdaSpace{}%
\AgdaSymbol{(}\AgdaBound{ux}\AgdaSpace{}%
\AgdaOperator{\AgdaInductiveConstructor{∷}}\AgdaSpace{}%
\AgdaBound{uys}\AgdaSymbol{)))}\<%
\\
\>[26]\AgdaBound{uls}\<%
\\
\>[26]\AgdaSymbol{(}\AgdaFunction{subst}\AgdaSpace{}%
\AgdaSymbol{(λ}\AgdaSpace{}%
\AgdaBound{x₁}\AgdaSpace{}%
\AgdaSymbol{→}\AgdaSpace{}%
\AgdaFunction{Disjoint}\AgdaSpace{}%
\AgdaBound{x₁}\AgdaSpace{}%
\AgdaSymbol{\AgdaUnderscore{})}\AgdaSpace{}%
\AgdaBound{≡₁}\AgdaSpace{}%
\AgdaBound{dsj}\AgdaSymbol{)}\<%
\\
\>[26]\AgdaFunction{hf₀}\<%
\\
\>[8]\AgdaFunction{bs⊆reverse-ys}\AgdaSpace{}%
\AgdaSymbol{:}\AgdaSpace{}%
\AgdaBound{bs}\AgdaSpace{}%
\AgdaOperator{\AgdaFunction{⊆}}\AgdaSpace{}%
\AgdaFunction{reverse}\AgdaSpace{}%
\AgdaBound{ys}\<%
\\
\>[8]\AgdaFunction{bs⊆reverse-ys}\AgdaSpace{}%
\AgdaBound{x₃}\AgdaSpace{}%
\AgdaSymbol{=}\AgdaSpace{}%
\AgdaFunction{reverse⁺}\AgdaSpace{}%
\AgdaSymbol{(}\AgdaFunction{∈-resp-↭}\AgdaSpace{}%
\AgdaFunction{↭₂}\AgdaSpace{}%
\AgdaSymbol{(}\AgdaFunction{∈-++⁺ʳ}\AgdaSpace{}%
\AgdaBound{as}\AgdaSpace{}%
\AgdaBound{x₃}\AgdaSymbol{))}\<%
\\
\>[8]\AgdaFunction{x∉reverse-ys}\AgdaSpace{}%
\AgdaSymbol{:}\AgdaSpace{}%
\AgdaBound{x}\AgdaSpace{}%
\AgdaOperator{\AgdaFunction{∉}}\AgdaSpace{}%
\AgdaFunction{reverse}\AgdaSpace{}%
\AgdaBound{ys}\<%
\\
\>[8]\AgdaFunction{x∉reverse-ys}\AgdaSpace{}%
\AgdaBound{x∈reverse-ys}\AgdaSpace{}%
\AgdaSymbol{=}\AgdaSpace{}%
\AgdaFunction{lookup}\AgdaSpace{}%
\AgdaBound{ux}\AgdaSpace{}%
\AgdaSymbol{(}\AgdaFunction{reverse⁻}\AgdaSpace{}%
\AgdaBound{x∈reverse-ys}\AgdaSymbol{)}\AgdaSpace{}%
\AgdaInductiveConstructor{refl}\<%
\\
\>[8]\AgdaFunction{dsj₁}\AgdaSpace{}%
\AgdaSymbol{:}\AgdaSpace{}%
\AgdaFunction{Disjoint}\AgdaSpace{}%
\AgdaSymbol{(}\AgdaFunction{cmdWriteNames}\AgdaSpace{}%
\AgdaBound{x}\AgdaSpace{}%
\AgdaSymbol{(}\AgdaFunction{script}\AgdaSpace{}%
\AgdaBound{as}\AgdaSpace{}%
\AgdaSymbol{\AgdaUnderscore{}))}\AgdaSpace{}%
\AgdaSymbol{(}\AgdaFunction{buildWriteNames}\AgdaSpace{}%
\AgdaSymbol{(}\AgdaFunction{run}\AgdaSpace{}%
\AgdaBound{x}\AgdaSpace{}%
\AgdaSymbol{(}\AgdaFunction{script}\AgdaSpace{}%
\AgdaBound{as}\AgdaSpace{}%
\AgdaSymbol{\AgdaUnderscore{}))}\AgdaSpace{}%
\AgdaBound{bs}\AgdaSymbol{)}\<%
\\
\>[8]\AgdaFunction{dsj₁}\AgdaSpace{}%
\AgdaSymbol{=}\AgdaSpace{}%
\AgdaFunction{hf=>disjointWW}\AgdaSpace{}%
\AgdaBound{s}\AgdaSpace{}%
\AgdaBound{x}\AgdaSpace{}%
\AgdaBound{as}\AgdaSpace{}%
\AgdaBound{bs}\AgdaSpace{}%
\AgdaSymbol{(}\AgdaFunction{reverse}\AgdaSpace{}%
\AgdaBound{ys}\AgdaSymbol{)}\AgdaSpace{}%
\AgdaSymbol{\AgdaUnderscore{}}%
\>[56]\AgdaFunction{bs⊆reverse-ys}\AgdaSpace{}%
\AgdaFunction{x∉reverse-ys}\AgdaSpace{}%
\AgdaFunction{hf₀}\<%
\\
\>[8]\AgdaFunction{dsj₂}\AgdaSpace{}%
\AgdaSymbol{:}\AgdaSpace{}%
\AgdaFunction{Disjoint}\AgdaSpace{}%
\AgdaSymbol{(}\AgdaFunction{cmdReadNames}\AgdaSpace{}%
\AgdaBound{x}\AgdaSpace{}%
\AgdaSymbol{(}\AgdaFunction{script}\AgdaSpace{}%
\AgdaBound{as}\AgdaSpace{}%
\AgdaSymbol{\AgdaUnderscore{}))}\AgdaSpace{}%
\AgdaSymbol{(}\AgdaFunction{buildWriteNames}\AgdaSpace{}%
\AgdaSymbol{(}\AgdaFunction{run}\AgdaSpace{}%
\AgdaBound{x}\AgdaSpace{}%
\AgdaSymbol{(}\AgdaFunction{script}\AgdaSpace{}%
\AgdaBound{as}\AgdaSpace{}%
\AgdaSymbol{\AgdaUnderscore{}))}\AgdaSpace{}%
\AgdaBound{bs}\AgdaSymbol{)}\<%
\\
\>[8]\AgdaFunction{dsj₂}\AgdaSpace{}%
\AgdaSymbol{=}\AgdaSpace{}%
\AgdaFunction{hf=>disjointRW}\AgdaSpace{}%
\AgdaBound{s}\AgdaSpace{}%
\AgdaBound{x}\AgdaSpace{}%
\AgdaBound{as}\AgdaSpace{}%
\AgdaBound{bs}\AgdaSpace{}%
\AgdaSymbol{(}\AgdaFunction{reverse}\AgdaSpace{}%
\AgdaBound{ys}\AgdaSymbol{)}\AgdaSpace{}%
\AgdaSymbol{\AgdaUnderscore{}}\AgdaSpace{}%
\AgdaFunction{bs⊆reverse-ys}\AgdaSpace{}%
\AgdaFunction{x∉reverse-ys}\AgdaSpace{}%
\AgdaFunction{hf₀}\<%
\\
\>[8]\AgdaFunction{dsj₃}\AgdaSpace{}%
\AgdaSymbol{:}\AgdaSpace{}%
\AgdaFunction{Disjoint}\AgdaSpace{}%
\AgdaSymbol{(}\AgdaFunction{cmdWriteNames}\AgdaSpace{}%
\AgdaBound{x}\AgdaSpace{}%
\AgdaSymbol{(}\AgdaFunction{script}\AgdaSpace{}%
\AgdaBound{as}\AgdaSpace{}%
\AgdaSymbol{\AgdaUnderscore{}))}\AgdaSpace{}%
\AgdaSymbol{(}\AgdaFunction{buildReadNames}\AgdaSpace{}%
\AgdaSymbol{(}\AgdaFunction{run}\AgdaSpace{}%
\AgdaBound{x}\AgdaSpace{}%
\AgdaSymbol{(}\AgdaFunction{script}\AgdaSpace{}%
\AgdaBound{as}\AgdaSpace{}%
\AgdaSymbol{\AgdaUnderscore{}))}\AgdaSpace{}%
\AgdaBound{bs}\AgdaSymbol{)}\<%
\\
\>[8]\AgdaFunction{dsj₃}\AgdaSpace{}%
\AgdaSymbol{=}\AgdaSpace{}%
\AgdaFunction{hf=>disjointWR}\AgdaSpace{}%
\AgdaBound{s}\AgdaSpace{}%
\AgdaBound{x}\AgdaSpace{}%
\AgdaBound{as}\AgdaSpace{}%
\AgdaBound{bs}\AgdaSpace{}%
\AgdaSymbol{(}\AgdaFunction{reverse}\AgdaSpace{}%
\AgdaBound{ys}\AgdaSymbol{)}\AgdaSpace{}%
\AgdaSymbol{\AgdaUnderscore{}}\AgdaSpace{}%
\AgdaFunction{bs⊆reverse-ys}\AgdaSpace{}%
\AgdaFunction{x∉reverse-ys}\AgdaSpace{}%
\AgdaFunction{hf₀}\<%
\\
\>[0]\AgdaSymbol{...}\AgdaSpace{}%
\AgdaSymbol{|}\AgdaSpace{}%
\AgdaBound{∀₂}\AgdaSpace{}%
\AgdaSymbol{=}\AgdaSpace{}%
\AgdaFunction{subst₂}\AgdaSpace{}%
\AgdaSymbol{(λ}\AgdaSpace{}%
\AgdaBound{x₂}\AgdaSpace{}%
\AgdaBound{x₃}\AgdaSpace{}%
\AgdaSymbol{→}\AgdaSpace{}%
\AgdaFunction{script}\AgdaSpace{}%
\AgdaBound{x₂}\AgdaSpace{}%
\AgdaSymbol{\AgdaUnderscore{}}\AgdaSpace{}%
\AgdaBound{f₁}\AgdaSpace{}%
\AgdaOperator{\AgdaDatatype{≡}}\AgdaSpace{}%
\AgdaFunction{script}\AgdaSpace{}%
\AgdaBound{x₃}\AgdaSpace{}%
\AgdaSymbol{\AgdaUnderscore{}}\AgdaSpace{}%
\AgdaBound{f₁}\AgdaSymbol{)}\AgdaSpace{}%
\AgdaSymbol{(}\AgdaFunction{sym}\AgdaSpace{}%
\AgdaBound{≡₁}\AgdaSymbol{)}\AgdaSpace{}%
\AgdaSymbol{(}\AgdaFunction{sym}\AgdaSpace{}%
\AgdaSymbol{(}\AgdaFunction{unfold-reverse}\AgdaSpace{}%
\AgdaBound{x}\AgdaSpace{}%
\AgdaBound{ys}\AgdaSymbol{))}\AgdaSpace{}%
\AgdaSymbol{(}\AgdaBound{∀₂}\AgdaSpace{}%
\AgdaBound{f₁}\AgdaSymbol{)}\<%
\end{code}

\newcommand{\reordered}{%
\begin{code}%
\>[0]\AgdaFunction{reordered≡}\AgdaSpace{}%
\AgdaSymbol{:}\AgdaSpace{}%
\AgdaSymbol{∀}\AgdaSpace{}%
\AgdaBound{s}\AgdaSpace{}%
\AgdaBound{br}\AgdaSpace{}%
\AgdaBound{bc}\AgdaSpace{}%
\AgdaSymbol{→}\AgdaSpace{}%
\AgdaFunction{PreCond}\AgdaSpace{}%
\AgdaBound{s}\AgdaSpace{}%
\AgdaBound{br}\AgdaSpace{}%
\AgdaBound{bc}\AgdaSpace{}%
\AgdaSymbol{→}\AgdaSpace{}%
\AgdaDatatype{HazardFree}\AgdaSpace{}%
\AgdaBound{s}\AgdaSpace{}%
\AgdaBound{br}\AgdaSpace{}%
\AgdaBound{bc}\AgdaSpace{}%
\AgdaInductiveConstructor{[]}\AgdaSpace{}%
\AgdaSymbol{→}\AgdaSpace{}%
\AgdaSymbol{(∀}\AgdaSpace{}%
\AgdaBound{f₁}\AgdaSpace{}%
\AgdaSymbol{→}\AgdaSpace{}%
\AgdaFunction{script}\AgdaSpace{}%
\AgdaBound{bc}\AgdaSpace{}%
\AgdaBound{s}\AgdaSpace{}%
\AgdaBound{f₁}\AgdaSpace{}%
\AgdaOperator{\AgdaDatatype{≡}}\AgdaSpace{}%
\AgdaFunction{script}\AgdaSpace{}%
\AgdaBound{br}\AgdaSpace{}%
\AgdaBound{s}\AgdaSpace{}%
\AgdaBound{f₁}\AgdaSymbol{)}\<%
\end{code}}
\begin{code}[hide]%
\>[0]\AgdaFunction{reordered≡}\AgdaSpace{}%
\AgdaBound{s}\AgdaSpace{}%
\AgdaBound{br}\AgdaSpace{}%
\AgdaBound{bc}\AgdaSpace{}%
\AgdaSymbol{(}\AgdaBound{dsb}\AgdaSpace{}%
\AgdaOperator{\AgdaInductiveConstructor{,}}\AgdaSpace{}%
\AgdaBound{ubr}\AgdaSpace{}%
\AgdaOperator{\AgdaInductiveConstructor{,}}\AgdaSpace{}%
\AgdaBound{ubc}\AgdaSpace{}%
\AgdaOperator{\AgdaInductiveConstructor{,}}\AgdaSpace{}%
\AgdaBound{pm}\AgdaSymbol{)}\AgdaSpace{}%
\AgdaBound{hf₁}\AgdaSpace{}%
\AgdaKeyword{with}\AgdaSpace{}%
\AgdaFunction{reordered}\AgdaSpace{}%
\AgdaSymbol{\{}\AgdaBound{s}\AgdaSymbol{\}}\AgdaSpace{}%
\AgdaBound{br}\AgdaSpace{}%
\AgdaSymbol{(}\AgdaFunction{reverse}\AgdaSpace{}%
\AgdaBound{bc}\AgdaSymbol{)}\AgdaSpace{}%
\AgdaSymbol{(}\AgdaFunction{↭-length}\AgdaSpace{}%
\AgdaFunction{br↭reverse-bc}\AgdaSymbol{)}\AgdaSpace{}%
\AgdaFunction{br↭reverse-bc}\AgdaSpace{}%
\AgdaFunction{ue}\AgdaSpace{}%
\AgdaSymbol{(}\AgdaFunction{subst}\AgdaSpace{}%
\AgdaSymbol{(λ}\AgdaSpace{}%
\AgdaBound{x}\AgdaSpace{}%
\AgdaSymbol{→}\AgdaSpace{}%
\AgdaDatatype{HazardFree}\AgdaSpace{}%
\AgdaBound{s}\AgdaSpace{}%
\AgdaBound{br}\AgdaSpace{}%
\AgdaBound{x}\AgdaSpace{}%
\AgdaSymbol{\AgdaUnderscore{})}\AgdaSpace{}%
\AgdaSymbol{(}\AgdaFunction{sym}\AgdaSpace{}%
\AgdaSymbol{(}\AgdaFunction{reverse-involutive}\AgdaSpace{}%
\AgdaBound{bc}\AgdaSymbol{))}\AgdaSpace{}%
\AgdaBound{hf₁}\AgdaSymbol{)}\<%
\\
\>[0][@{}l@{\AgdaIndent{0}}]%
\>[2]\AgdaKeyword{where}%
\>[1908I]\AgdaFunction{br↭reverse-bc}\AgdaSpace{}%
\AgdaSymbol{:}\AgdaSpace{}%
\AgdaBound{br}\AgdaSpace{}%
\AgdaOperator{\AgdaDatatype{↭}}\AgdaSpace{}%
\AgdaFunction{reverse}\AgdaSpace{}%
\AgdaBound{bc}\<%
\\
\>[.][@{}l@{}]\<[1908I]%
\>[8]\AgdaFunction{br↭reverse-bc}\AgdaSpace{}%
\AgdaSymbol{=}\AgdaSpace{}%
\AgdaFunction{↭-trans}\AgdaSpace{}%
\AgdaSymbol{(}\AgdaFunction{↭-sym}\AgdaSpace{}%
\AgdaBound{pm}\AgdaSymbol{)}\AgdaSpace{}%
\AgdaSymbol{(}\AgdaFunction{↭-reverse}\AgdaSpace{}%
\AgdaBound{bc}\AgdaSymbol{)}\<%
\\
\>[8]\AgdaFunction{g₁}\AgdaSpace{}%
\AgdaSymbol{:}\AgdaSpace{}%
\AgdaFunction{Disjoint}\AgdaSpace{}%
\AgdaBound{br}\AgdaSpace{}%
\AgdaInductiveConstructor{[]}\<%
\\
\>[8]\AgdaFunction{g₁}\AgdaSpace{}%
\AgdaSymbol{()}\<%
\\
\>[8]\AgdaFunction{ue}\AgdaSpace{}%
\AgdaSymbol{:}\AgdaSpace{}%
\AgdaFunction{UniqueEvidence}\AgdaSpace{}%
\AgdaBound{br}\AgdaSpace{}%
\AgdaSymbol{(}\AgdaFunction{reverse}\AgdaSpace{}%
\AgdaBound{bc}\AgdaSymbol{)}\AgdaSpace{}%
\AgdaInductiveConstructor{[]}\<%
\\
\>[8]\AgdaFunction{ue}\AgdaSpace{}%
\AgdaSymbol{=}\AgdaSpace{}%
\AgdaBound{ubr}\AgdaSpace{}%
\AgdaOperator{\AgdaInductiveConstructor{,}}\AgdaSpace{}%
\AgdaSymbol{(}\AgdaFunction{unique-reverse}\AgdaSpace{}%
\AgdaBound{bc}\AgdaSpace{}%
\AgdaBound{ubc}\AgdaSymbol{)}\AgdaSpace{}%
\AgdaOperator{\AgdaInductiveConstructor{,}}\AgdaSpace{}%
\AgdaInductiveConstructor{[]}\AgdaSpace{}%
\AgdaOperator{\AgdaInductiveConstructor{,}}\AgdaSpace{}%
\AgdaFunction{g₁}\<%
\\
\>[0]\AgdaSymbol{...}\AgdaSpace{}%
\AgdaSymbol{|}\AgdaSpace{}%
\AgdaBound{∀₁}\AgdaSpace{}%
\AgdaSymbol{=}\AgdaSpace{}%
\AgdaSymbol{λ}\AgdaSpace{}%
\AgdaBound{f₁}\AgdaSpace{}%
\AgdaSymbol{→}\AgdaSpace{}%
\AgdaFunction{subst}\AgdaSpace{}%
\AgdaSymbol{(λ}\AgdaSpace{}%
\AgdaBound{x}\AgdaSpace{}%
\AgdaSymbol{→}\AgdaSpace{}%
\AgdaFunction{script}\AgdaSpace{}%
\AgdaBound{x}\AgdaSpace{}%
\AgdaBound{s}\AgdaSpace{}%
\AgdaBound{f₁}\AgdaSpace{}%
\AgdaOperator{\AgdaDatatype{≡}}\AgdaSpace{}%
\AgdaFunction{script}\AgdaSpace{}%
\AgdaBound{br}\AgdaSpace{}%
\AgdaBound{s}\AgdaSpace{}%
\AgdaBound{f₁}\AgdaSymbol{)}\AgdaSpace{}%
\AgdaSymbol{(}\AgdaFunction{reverse-involutive}\AgdaSpace{}%
\AgdaBound{bc}\AgdaSymbol{)}\AgdaSpace{}%
\AgdaSymbol{(}\AgdaFunction{sym}\AgdaSpace{}%
\AgdaSymbol{(}\AgdaBound{∀₁}\AgdaSpace{}%
\AgdaBound{f₁}\AgdaSymbol{))}\<%
\end{code}

\begin{code}[hide]%
\>[0]\AgdaKeyword{open}\AgdaSpace{}%
\AgdaKeyword{import}\AgdaSpace{}%
\AgdaModule{Functional.State}\AgdaSpace{}%
\AgdaKeyword{using}\AgdaSpace{}%
\AgdaSymbol{(}\AgdaFunction{State}\AgdaSpace{}%
\AgdaSymbol{;}\AgdaSpace{}%
\AgdaFunction{Oracle}\AgdaSpace{}%
\AgdaSymbol{;}\AgdaSpace{}%
\AgdaFunction{Cmd}\AgdaSpace{}%
\AgdaSymbol{;}\AgdaSpace{}%
\AgdaFunction{FileSystem}\AgdaSpace{}%
\AgdaSymbol{;}\AgdaSpace{}%
\AgdaFunction{Memory}\AgdaSymbol{)}\AgdaSpace{}%
\AgdaKeyword{renaming}\AgdaSpace{}%
\AgdaSymbol{(}\AgdaFunction{save}\AgdaSpace{}%
\AgdaSymbol{to}\AgdaSpace{}%
\AgdaFunction{store}\AgdaSymbol{)}\<%
\\
\\[\AgdaEmptyExtraSkip]%
\>[0]\AgdaKeyword{module}\AgdaSpace{}%
\AgdaModule{Functional.Rattle.Exec}\AgdaSpace{}%
\AgdaSymbol{(}\AgdaBound{oracle}\AgdaSpace{}%
\AgdaSymbol{:}\AgdaSpace{}%
\AgdaFunction{Oracle}\AgdaSymbol{)}\AgdaSpace{}%
\AgdaKeyword{where}\<%
\\
\\[\AgdaEmptyExtraSkip]%
\>[0]\AgdaKeyword{open}\AgdaSpace{}%
\AgdaKeyword{import}\AgdaSpace{}%
\AgdaModule{Functional.State.Helpers}\AgdaSpace{}%
\AgdaSymbol{(}\AgdaBound{oracle}\AgdaSymbol{)}\AgdaSpace{}%
\AgdaSymbol{as}\AgdaSpace{}%
\AgdaModule{St}\AgdaSpace{}%
\AgdaKeyword{hiding}\AgdaSpace{}%
\AgdaSymbol{(}\AgdaFunction{run}\AgdaSymbol{)}\<%
\\
\>[0]\AgdaKeyword{open}\AgdaSpace{}%
\AgdaKeyword{import}\AgdaSpace{}%
\AgdaModule{Agda.Builtin.Equality}\<%
\\
\>[0]\AgdaKeyword{open}\AgdaSpace{}%
\AgdaKeyword{import}\AgdaSpace{}%
\AgdaModule{Agda.Builtin.Nat}\AgdaSpace{}%
\AgdaKeyword{renaming}\AgdaSpace{}%
\AgdaSymbol{(}\AgdaDatatype{Nat}\AgdaSpace{}%
\AgdaSymbol{to}\AgdaSpace{}%
\AgdaDatatype{ℕ}\AgdaSymbol{)}\<%
\\
\>[0]\AgdaKeyword{open}\AgdaSpace{}%
\AgdaKeyword{import}\AgdaSpace{}%
\AgdaModule{Data.Nat}\AgdaSpace{}%
\AgdaKeyword{using}\AgdaSpace{}%
\AgdaSymbol{(}\AgdaOperator{\AgdaDatatype{\AgdaUnderscore{}≤\AgdaUnderscore{}}}\AgdaSymbol{)}\<%
\\
\>[0]\AgdaKeyword{open}\AgdaSpace{}%
\AgdaKeyword{import}\AgdaSpace{}%
\AgdaModule{Data.Nat.Properties}\AgdaSpace{}%
\AgdaSymbol{as}\AgdaSpace{}%
\AgdaModule{N}\AgdaSpace{}%
\AgdaKeyword{using}\AgdaSpace{}%
\AgdaSymbol{(}\AgdaFunction{1+n≰n}\AgdaSymbol{)}\<%
\\
\>[0]\AgdaKeyword{open}\AgdaSpace{}%
\AgdaKeyword{import}\AgdaSpace{}%
\AgdaModule{Data.Bool}\AgdaSpace{}%
\AgdaKeyword{using}\AgdaSpace{}%
\AgdaSymbol{(}\AgdaDatatype{Bool}\AgdaSpace{}%
\AgdaSymbol{;}\AgdaSpace{}%
\AgdaOperator{\AgdaFunction{if\AgdaUnderscore{}then\AgdaUnderscore{}else\AgdaUnderscore{}}}\AgdaSpace{}%
\AgdaSymbol{;}\AgdaSpace{}%
\AgdaInductiveConstructor{false}\AgdaSpace{}%
\AgdaSymbol{;}\AgdaSpace{}%
\AgdaInductiveConstructor{true}\AgdaSpace{}%
\AgdaSymbol{;}\AgdaSpace{}%
\AgdaFunction{not}\AgdaSymbol{)}\<%
\\
\>[0]\AgdaKeyword{open}\AgdaSpace{}%
\AgdaKeyword{import}\AgdaSpace{}%
\AgdaModule{Data.List}\AgdaSpace{}%
\AgdaKeyword{using}\AgdaSpace{}%
\AgdaSymbol{(}\AgdaDatatype{List}\AgdaSpace{}%
\AgdaSymbol{;}\AgdaSpace{}%
\AgdaInductiveConstructor{[]}\AgdaSpace{}%
\AgdaSymbol{;}\AgdaSpace{}%
\AgdaOperator{\AgdaInductiveConstructor{\AgdaUnderscore{}∷\AgdaUnderscore{}}}\AgdaSpace{}%
\AgdaSymbol{;}\AgdaSpace{}%
\AgdaFunction{map}\AgdaSpace{}%
\AgdaSymbol{;}\AgdaSpace{}%
\AgdaFunction{filter}\AgdaSpace{}%
\AgdaSymbol{;}\AgdaSpace{}%
\AgdaOperator{\AgdaFunction{\AgdaUnderscore{}++\AgdaUnderscore{}}}\AgdaSpace{}%
\AgdaSymbol{;}\AgdaSpace{}%
\AgdaFunction{foldr}\AgdaSpace{}%
\AgdaSymbol{;}\AgdaSpace{}%
\AgdaFunction{all}\AgdaSpace{}%
\AgdaSymbol{;}\AgdaSpace{}%
\AgdaFunction{concatMap}\AgdaSpace{}%
\AgdaSymbol{;}\AgdaSpace{}%
\AgdaFunction{reverse}\AgdaSymbol{;}\AgdaSpace{}%
\AgdaOperator{\AgdaFunction{\AgdaUnderscore{}∷ʳ\AgdaUnderscore{}}}\AgdaSymbol{)}\<%
\\
\>[0]\AgdaKeyword{open}\AgdaSpace{}%
\AgdaKeyword{import}\AgdaSpace{}%
\AgdaModule{Data.List.Properties}\AgdaSpace{}%
\AgdaKeyword{using}\AgdaSpace{}%
\AgdaSymbol{(}\AgdaFunction{∷-injectiveʳ}\AgdaSpace{}%
\AgdaSymbol{;}\AgdaSpace{}%
\AgdaFunction{∷-injectiveˡ}\AgdaSpace{}%
\AgdaSymbol{;}\AgdaSpace{}%
\AgdaFunction{∷-injective}\AgdaSymbol{)}\<%
\\
\>[0]\AgdaKeyword{open}\AgdaSpace{}%
\AgdaKeyword{import}\AgdaSpace{}%
\AgdaModule{Data.String.Properties}\AgdaSpace{}%
\AgdaKeyword{using}\AgdaSpace{}%
\AgdaSymbol{(}\AgdaOperator{\AgdaFunction{\AgdaUnderscore{}≟\AgdaUnderscore{}}}\AgdaSpace{}%
\AgdaSymbol{;}\AgdaSpace{}%
\AgdaOperator{\AgdaFunction{\AgdaUnderscore{}==\AgdaUnderscore{}}}\AgdaSymbol{)}\<%
\\
\>[0]\AgdaKeyword{open}\AgdaSpace{}%
\AgdaKeyword{import}\AgdaSpace{}%
\AgdaModule{Data.List.Relation.Binary.Equality.DecPropositional}\AgdaSpace{}%
\AgdaOperator{\AgdaFunction{\AgdaUnderscore{}≟\AgdaUnderscore{}}}\AgdaSpace{}%
\AgdaKeyword{using}\AgdaSpace{}%
\AgdaSymbol{(}\AgdaUnderscore{}≡?\AgdaUnderscore{}\AgdaSymbol{)}\<%
\\
\>[0]\AgdaKeyword{open}\AgdaSpace{}%
\AgdaKeyword{import}\AgdaSpace{}%
\AgdaModule{Data.Maybe}\AgdaSpace{}%
\AgdaSymbol{as}\AgdaSpace{}%
\AgdaModule{Maybe}\AgdaSpace{}%
\AgdaKeyword{using}\AgdaSpace{}%
\AgdaSymbol{(}\AgdaDatatype{Maybe}\AgdaSpace{}%
\AgdaSymbol{;}\AgdaSpace{}%
\AgdaInductiveConstructor{just}\AgdaSpace{}%
\AgdaSymbol{;}\AgdaSpace{}%
\AgdaInductiveConstructor{nothing}\AgdaSymbol{)}\<%
\\
\>[0]\AgdaKeyword{open}\AgdaSpace{}%
\AgdaKeyword{import}\AgdaSpace{}%
\AgdaModule{Data.Maybe.Relation.Binary.Pointwise}\AgdaSpace{}%
\AgdaKeyword{using}\AgdaSpace{}%
\AgdaSymbol{(}\AgdaFunction{dec}\AgdaSymbol{)}\<%
\\
\>[0]\AgdaKeyword{open}\AgdaSpace{}%
\AgdaKeyword{import}\AgdaSpace{}%
\AgdaModule{Data.Product}\AgdaSpace{}%
\AgdaKeyword{using}\AgdaSpace{}%
\AgdaSymbol{(}\AgdaField{proj₁}\AgdaSpace{}%
\AgdaSymbol{;}\AgdaSpace{}%
\AgdaField{proj₂}\AgdaSpace{}%
\AgdaSymbol{;}\AgdaSpace{}%
\AgdaOperator{\AgdaInductiveConstructor{\AgdaUnderscore{},\AgdaUnderscore{}}}\AgdaSpace{}%
\AgdaSymbol{;}\AgdaSpace{}%
\AgdaOperator{\AgdaFunction{\AgdaUnderscore{}×\AgdaUnderscore{}}}\AgdaSpace{}%
\AgdaSymbol{;}\AgdaSpace{}%
\AgdaFunction{∃-syntax}\AgdaSpace{}%
\AgdaSymbol{;}\AgdaSpace{}%
\AgdaFunction{Σ-syntax}\AgdaSymbol{)}\<%
\\
\>[0]\AgdaKeyword{open}\AgdaSpace{}%
\AgdaKeyword{import}\AgdaSpace{}%
\AgdaModule{Data.Product.Properties}\AgdaSpace{}%
\AgdaKeyword{using}\AgdaSpace{}%
\AgdaSymbol{(}\AgdaFunction{≡-dec}\AgdaSpace{}%
\AgdaSymbol{;}\AgdaSpace{}%
\AgdaFunction{,-injectiveˡ}\AgdaSpace{}%
\AgdaSymbol{;}\AgdaSpace{}%
\AgdaFunction{,-injectiveʳ}\AgdaSymbol{)}\<%
\\
\>[0]\AgdaKeyword{open}\AgdaSpace{}%
\AgdaKeyword{import}\AgdaSpace{}%
\AgdaModule{Data.Product.Relation.Binary.Pointwise.NonDependent}\AgdaSpace{}%
\AgdaKeyword{using}\AgdaSpace{}%
\AgdaSymbol{(}\AgdaFunction{×-decidable}\AgdaSpace{}%
\AgdaSymbol{;}\AgdaSpace{}%
\AgdaFunction{≡×≡⇒≡}\AgdaSpace{}%
\AgdaSymbol{;}\AgdaSpace{}%
\AgdaFunction{≡⇒≡×≡}\AgdaSpace{}%
\AgdaSymbol{)}\<%
\\
\>[0]\AgdaKeyword{open}\AgdaSpace{}%
\AgdaKeyword{import}\AgdaSpace{}%
\AgdaModule{Function.Base}\AgdaSpace{}%
\AgdaKeyword{using}\AgdaSpace{}%
\AgdaSymbol{(}\AgdaOperator{\AgdaFunction{\AgdaUnderscore{}∘\AgdaUnderscore{}}}\AgdaSymbol{)}\<%
\\
\>[0]\AgdaKeyword{open}\AgdaSpace{}%
\AgdaKeyword{import}\AgdaSpace{}%
\AgdaModule{Functional.File}\AgdaSpace{}%
\AgdaKeyword{using}\AgdaSpace{}%
\AgdaSymbol{(}\AgdaFunction{FileName}\AgdaSpace{}%
\AgdaSymbol{;}\AgdaSpace{}%
\AgdaFunction{FileContent}\AgdaSymbol{)}\<%
\\
\>[0]\AgdaKeyword{open}\AgdaSpace{}%
\AgdaKeyword{import}\AgdaSpace{}%
\AgdaModule{Functional.Build}\AgdaSpace{}%
\AgdaSymbol{(}\AgdaBound{oracle}\AgdaSymbol{)}\AgdaSpace{}%
\AgdaKeyword{using}\AgdaSpace{}%
\AgdaSymbol{(}Build \AgdaSymbol{;} UniqueEvidence\AgdaSymbol{)}\<%
\\
\>[0]\AgdaKeyword{open}\AgdaSpace{}%
\AgdaKeyword{import}\AgdaSpace{}%
\AgdaModule{Relation.Binary}\AgdaSpace{}%
\AgdaKeyword{using}\AgdaSpace{}%
\AgdaSymbol{(}\AgdaFunction{Decidable}\AgdaSymbol{)}\<%
\\
\>[0]\AgdaKeyword{open}\AgdaSpace{}%
\AgdaKeyword{import}\AgdaSpace{}%
\AgdaModule{Relation.Nullary.Decidable}\AgdaSpace{}%
\AgdaSymbol{as}\AgdaSpace{}%
\AgdaModule{Dec}\AgdaSpace{}%
\AgdaKeyword{using}\AgdaSpace{}%
\AgdaSymbol{(}\AgdaFunction{isYes}\AgdaSpace{}%
\AgdaSymbol{;}\AgdaSpace{}%
\AgdaFunction{map′}\AgdaSymbol{)}\<%
\\
\>[0]\AgdaKeyword{open}\AgdaSpace{}%
\AgdaKeyword{import}\AgdaSpace{}%
\AgdaModule{Relation.Nullary.Negation}\AgdaSpace{}%
\AgdaKeyword{using}\AgdaSpace{}%
\AgdaSymbol{(}\AgdaFunction{¬?}\AgdaSpace{}%
\AgdaSymbol{;}\AgdaSpace{}%
\AgdaFunction{contradiction}\AgdaSymbol{)}\<%
\\
\>[0]\AgdaKeyword{open}\AgdaSpace{}%
\AgdaKeyword{import}\AgdaSpace{}%
\AgdaModule{Functional.Forward.Exec}\AgdaSpace{}%
\AgdaSymbol{(}\AgdaBound{oracle}\AgdaSymbol{)}\AgdaSpace{}%
\AgdaKeyword{using}\AgdaSpace{}%
\AgdaSymbol{(}run?\AgdaSymbol{)}\<%
\\
\>[0]\AgdaKeyword{open}\AgdaSpace{}%
\AgdaKeyword{import}\AgdaSpace{}%
\AgdaModule{Data.Sum}\AgdaSpace{}%
\AgdaKeyword{using}\AgdaSpace{}%
\AgdaSymbol{(}\AgdaOperator{\AgdaDatatype{\AgdaUnderscore{}⊎\AgdaUnderscore{}}}\AgdaSpace{}%
\AgdaSymbol{;}\AgdaSpace{}%
\AgdaInductiveConstructor{inj₁}\AgdaSpace{}%
\AgdaSymbol{;}\AgdaSpace{}%
\AgdaInductiveConstructor{inj₂}\AgdaSpace{}%
\AgdaSymbol{;}\AgdaSpace{}%
\AgdaFunction{map₂}\AgdaSymbol{)}\<%
\\
\>[0]\AgdaKeyword{open}\AgdaSpace{}%
\AgdaKeyword{import}\AgdaSpace{}%
\AgdaModule{Data.List.Membership.DecPropositional}\AgdaSpace{}%
\AgdaOperator{\AgdaFunction{\AgdaUnderscore{}≟\AgdaUnderscore{}}}\AgdaSpace{}%
\AgdaKeyword{using}\AgdaSpace{}%
\AgdaSymbol{(}\AgdaUnderscore{}∈?\AgdaUnderscore{} \AgdaSymbol{;} \AgdaUnderscore{}∈\AgdaUnderscore{} \AgdaSymbol{;} \AgdaUnderscore{}∉\AgdaUnderscore{}\AgdaSymbol{)}\<%
\\
\>[0]\AgdaKeyword{open}\AgdaSpace{}%
\AgdaKeyword{import}\AgdaSpace{}%
\AgdaModule{Data.List.Membership.Propositional.Properties}\AgdaSpace{}%
\AgdaKeyword{using}\AgdaSpace{}%
\AgdaSymbol{(}\AgdaFunction{∈-++⁺ʳ}\AgdaSymbol{)}\<%
\\
\>[0]\AgdaKeyword{open}\AgdaSpace{}%
\AgdaKeyword{import}\AgdaSpace{}%
\AgdaModule{Data.List.Relation.Unary.Unique.Propositional}\AgdaSpace{}%
\AgdaKeyword{using}\AgdaSpace{}%
\AgdaSymbol{(}\AgdaDatatype{Unique}\AgdaSymbol{)}\<%
\\
\>[0]\AgdaKeyword{open}\AgdaSpace{}%
\AgdaKeyword{import}\AgdaSpace{}%
\AgdaModule{Data.Empty}\AgdaSpace{}%
\AgdaKeyword{using}\AgdaSpace{}%
\AgdaSymbol{(}\AgdaDatatype{⊥}\AgdaSymbol{)}\<%
\\
\>[0]\AgdaKeyword{open}\AgdaSpace{}%
\AgdaKeyword{import}\AgdaSpace{}%
\AgdaModule{Relation.Nullary}\AgdaSpace{}%
\AgdaKeyword{using}\AgdaSpace{}%
\AgdaSymbol{(}\AgdaInductiveConstructor{yes}\AgdaSpace{}%
\AgdaSymbol{;}\AgdaSpace{}%
\AgdaInductiveConstructor{no}\AgdaSpace{}%
\AgdaSymbol{;}\AgdaSpace{}%
\AgdaOperator{\AgdaFunction{¬\AgdaUnderscore{}}}\AgdaSymbol{)}\<%
\\
\>[0]\AgdaKeyword{open}\AgdaSpace{}%
\AgdaKeyword{import}\AgdaSpace{}%
\AgdaModule{Data.List.Relation.Unary.Any}\AgdaSpace{}%
\AgdaKeyword{using}\AgdaSpace{}%
\AgdaSymbol{(}\AgdaFunction{tail}\AgdaSpace{}%
\AgdaSymbol{;}\AgdaSpace{}%
\AgdaInductiveConstructor{here}\AgdaSpace{}%
\AgdaSymbol{;}\AgdaSpace{}%
\AgdaInductiveConstructor{there}\AgdaSymbol{)}\<%
\\
\>[0]\AgdaKeyword{open}\AgdaSpace{}%
\AgdaKeyword{import}\AgdaSpace{}%
\AgdaModule{Common.List.Properties}\AgdaSpace{}%
\AgdaKeyword{using}\AgdaSpace{}%
\AgdaSymbol{(}\AgdaOperator{\AgdaFunction{\AgdaUnderscore{}before\AgdaUnderscore{}∈\AgdaUnderscore{}}}\AgdaSpace{}%
\AgdaSymbol{;}\AgdaSpace{}%
\AgdaFunction{before-∷}\AgdaSymbol{)}\<%
\\
\>[0]\AgdaKeyword{open}\AgdaSpace{}%
\AgdaKeyword{import}\AgdaSpace{}%
\AgdaModule{Functional.Script.Hazard}\AgdaSpace{}%
\AgdaSymbol{(}\AgdaBound{oracle}\AgdaSymbol{)}\AgdaSpace{}%
\AgdaKeyword{using}\AgdaSpace{}%
\AgdaSymbol{(}Hazard \AgdaSymbol{;} WriteWrite \AgdaSymbol{;} ReadWrite \AgdaSymbol{;} Speculative \AgdaSymbol{;} filesRead \AgdaSymbol{;} filesWrote \AgdaSymbol{;} cmdRead \AgdaSymbol{;} cmdWrote \AgdaSymbol{;} cmdsRun \AgdaSymbol{;} FileInfo \AgdaSymbol{;} ∈-cmdWrote∷ \AgdaSymbol{;} ∈-cmdRead∷ \AgdaSymbol{;} ∃Hazard \AgdaSymbol{;} before?\AgdaSymbol{)}\AgdaSpace{}%
\AgdaKeyword{renaming}\AgdaSpace{}%
\AgdaSymbol{(}save \AgdaSymbol{to} rec\AgdaSymbol{)}\<%
\\
\>[0]\AgdaKeyword{open}\AgdaSpace{}%
\AgdaKeyword{import}\AgdaSpace{}%
\AgdaModule{Data.List.Relation.Unary.AllPairs}\AgdaSpace{}%
\AgdaKeyword{using}\AgdaSpace{}%
\AgdaSymbol{(}\AgdaOperator{\AgdaInductiveConstructor{\AgdaUnderscore{}∷\AgdaUnderscore{}}}\AgdaSymbol{)}\<%
\\
\>[0]\AgdaKeyword{open}\AgdaSpace{}%
\AgdaKeyword{import}\AgdaSpace{}%
\AgdaModule{Relation.Binary.PropositionalEquality}\AgdaSpace{}%
\AgdaKeyword{using}\AgdaSpace{}%
\AgdaSymbol{(}\AgdaFunction{cong}\AgdaSpace{}%
\AgdaSymbol{;}\AgdaSpace{}%
\AgdaFunction{trans}\AgdaSpace{}%
\AgdaSymbol{;}\AgdaSpace{}%
\AgdaFunction{sym}\AgdaSpace{}%
\AgdaSymbol{;}\AgdaSpace{}%
\AgdaFunction{subst}\AgdaSpace{}%
\AgdaSymbol{;}\AgdaSpace{}%
\AgdaOperator{\AgdaFunction{\AgdaUnderscore{}≢\AgdaUnderscore{}}}\AgdaSymbol{)}\<%
\\
\>[0]\AgdaKeyword{open}\AgdaSpace{}%
\AgdaKeyword{import}\AgdaSpace{}%
\AgdaModule{Data.List.Relation.Unary.All}\AgdaSpace{}%
\AgdaKeyword{using}\AgdaSpace{}%
\AgdaSymbol{(}\AgdaDatatype{All}\AgdaSpace{}%
\AgdaSymbol{;}\AgdaSpace{}%
\AgdaFunction{lookup}\AgdaSymbol{)}\<%
\\
\>[0]\AgdaKeyword{open}\AgdaSpace{}%
\AgdaKeyword{import}\AgdaSpace{}%
\AgdaModule{Data.List.Relation.Binary.Disjoint.Propositional}\AgdaSpace{}%
\AgdaKeyword{using}\AgdaSpace{}%
\AgdaSymbol{(}\AgdaFunction{Disjoint}\AgdaSymbol{)}\<%
\\
\\[\AgdaEmptyExtraSkip]%
\>[0]\AgdaFunction{doRunR}\AgdaSpace{}%
\AgdaSymbol{:}\AgdaSpace{}%
\AgdaFunction{State}\AgdaSpace{}%
\AgdaSymbol{->}\AgdaSpace{}%
\AgdaFunction{Cmd}\AgdaSpace{}%
\AgdaSymbol{->}\AgdaSpace{}%
\AgdaFunction{State}\<%
\\
\>[0]\AgdaFunction{doRunR}\AgdaSpace{}%
\AgdaSymbol{(}\AgdaBound{s}\AgdaSpace{}%
\AgdaOperator{\AgdaInductiveConstructor{,}}\AgdaSpace{}%
\AgdaBound{m}\AgdaSymbol{)}\AgdaSpace{}%
\AgdaBound{x}\AgdaSpace{}%
\AgdaSymbol{=}\AgdaSpace{}%
\AgdaKeyword{let}\AgdaSpace{}%
\AgdaBound{s₂}%
\>[278I]\AgdaSymbol{=}\AgdaSpace{}%
\AgdaFunction{St.run}\AgdaSpace{}%
\AgdaBound{x}\AgdaSpace{}%
\AgdaBound{s}\AgdaSpace{}%
\AgdaKeyword{in}\<%
\\
\>[278I][@{}l@{\AgdaIndent{0}}]%
\>[27]\AgdaSymbol{(}\AgdaBound{s₂}\AgdaSpace{}%
\AgdaOperator{\AgdaInductiveConstructor{,}}\AgdaSpace{}%
\AgdaFunction{store}\AgdaSpace{}%
\AgdaBound{x}\AgdaSpace{}%
\AgdaSymbol{((}\AgdaFunction{cmdReadNames}\AgdaSpace{}%
\AgdaBound{x}\AgdaSpace{}%
\AgdaBound{s}\AgdaSymbol{)}\AgdaSpace{}%
\AgdaOperator{\AgdaFunction{++}}\AgdaSpace{}%
\AgdaSymbol{(}\AgdaFunction{cmdWriteNames}\AgdaSpace{}%
\AgdaBound{x}\AgdaSpace{}%
\AgdaBound{s}\AgdaSymbol{))}\AgdaSpace{}%
\AgdaBound{s₂}\AgdaSpace{}%
\AgdaBound{m}\AgdaSymbol{)}\<%
\\
\\[\AgdaEmptyExtraSkip]%
\>[0]\AgdaComment{\{- check for hazards

speculative write read
write write
read write

-\}}\<%
\\
\\[\AgdaEmptyExtraSkip]%
\>[0]\AgdaComment{\{- check for speculative hazard

   check for writewrite hazard
   check for readwrite hazard

 1. speculative
 -- 
2. write write
3. read write

-\}}\<%
\\
\\[\AgdaEmptyExtraSkip]%
\>[0]\AgdaFunction{∃Intersection}\AgdaSpace{}%
\AgdaSymbol{:}\AgdaSpace{}%
\AgdaSymbol{∀}\AgdaSpace{}%
\AgdaBound{ls₁}\AgdaSpace{}%
\AgdaBound{ls₂}\AgdaSpace{}%
\AgdaSymbol{→}\AgdaSpace{}%
\AgdaDatatype{Maybe}\AgdaSpace{}%
\AgdaSymbol{(}\AgdaFunction{∃[}\AgdaSpace{}%
\AgdaBound{v}\AgdaSpace{}%
\AgdaFunction{]}\AgdaSymbol{(}\AgdaBound{v}\AgdaSpace{}%
\AgdaOperator{\AgdaFunction{∈}}\AgdaSpace{}%
\AgdaBound{ls₁}\AgdaSpace{}%
\AgdaOperator{\AgdaFunction{×}}\AgdaSpace{}%
\AgdaBound{v}\AgdaSpace{}%
\AgdaOperator{\AgdaFunction{∈}}\AgdaSpace{}%
\AgdaBound{ls₂}\AgdaSymbol{))}\<%
\\
\>[0]\AgdaFunction{∃Intersection}\AgdaSpace{}%
\AgdaInductiveConstructor{[]}\AgdaSpace{}%
\AgdaBound{ls₂}\AgdaSpace{}%
\AgdaSymbol{=}\AgdaSpace{}%
\AgdaInductiveConstructor{nothing}\<%
\\
\>[0]\AgdaFunction{∃Intersection}\AgdaSpace{}%
\AgdaSymbol{(}\AgdaBound{x}\AgdaSpace{}%
\AgdaOperator{\AgdaInductiveConstructor{∷}}\AgdaSpace{}%
\AgdaBound{ls₁}\AgdaSymbol{)}\AgdaSpace{}%
\AgdaBound{ls₂}\AgdaSpace{}%
\AgdaKeyword{with}\AgdaSpace{}%
\AgdaBound{x}\AgdaSpace{}%
\AgdaOperator{\AgdaFunction{∈?}}\AgdaSpace{}%
\AgdaBound{ls₂}\<%
\\
\>[0]\AgdaSymbol{...}\AgdaSpace{}%
\AgdaSymbol{|}\AgdaSpace{}%
\AgdaInductiveConstructor{yes}\AgdaSpace{}%
\AgdaBound{x∈ls₂}\AgdaSpace{}%
\AgdaSymbol{=}\AgdaSpace{}%
\AgdaInductiveConstructor{just}\AgdaSpace{}%
\AgdaSymbol{(}\AgdaBound{x}\AgdaSpace{}%
\AgdaOperator{\AgdaInductiveConstructor{,}}\AgdaSpace{}%
\AgdaInductiveConstructor{here}\AgdaSpace{}%
\AgdaInductiveConstructor{refl}\AgdaSpace{}%
\AgdaOperator{\AgdaInductiveConstructor{,}}\AgdaSpace{}%
\AgdaBound{x∈ls₂}\AgdaSymbol{)}\<%
\\
\>[0]\AgdaSymbol{...}\AgdaSpace{}%
\AgdaSymbol{|}\AgdaSpace{}%
\AgdaInductiveConstructor{no}\AgdaSpace{}%
\AgdaBound{x∉ls₂}\AgdaSpace{}%
\AgdaKeyword{with}\AgdaSpace{}%
\AgdaFunction{∃Intersection}\AgdaSpace{}%
\AgdaBound{ls₁}\AgdaSpace{}%
\AgdaBound{ls₂}\<%
\\
\>[0]\AgdaSymbol{...}\AgdaSpace{}%
\AgdaSymbol{|}\AgdaSpace{}%
\AgdaInductiveConstructor{nothing}\AgdaSpace{}%
\AgdaSymbol{=}\AgdaSpace{}%
\AgdaInductiveConstructor{nothing}\<%
\\
\>[0]\AgdaSymbol{...}\AgdaSpace{}%
\AgdaSymbol{|}\AgdaSpace{}%
\AgdaInductiveConstructor{just}\AgdaSpace{}%
\AgdaSymbol{(}\AgdaBound{v}\AgdaSpace{}%
\AgdaOperator{\AgdaInductiveConstructor{,}}\AgdaSpace{}%
\AgdaBound{v∈ls₁}\AgdaSpace{}%
\AgdaOperator{\AgdaInductiveConstructor{,}}\AgdaSpace{}%
\AgdaBound{v∈ls₂}\AgdaSymbol{)}\AgdaSpace{}%
\AgdaSymbol{=}\AgdaSpace{}%
\AgdaInductiveConstructor{just}\AgdaSpace{}%
\AgdaSymbol{(}\AgdaBound{v}\AgdaSpace{}%
\AgdaOperator{\AgdaInductiveConstructor{,}}\AgdaSpace{}%
\AgdaInductiveConstructor{there}\AgdaSpace{}%
\AgdaBound{v∈ls₁}\AgdaSpace{}%
\AgdaOperator{\AgdaInductiveConstructor{,}}\AgdaSpace{}%
\AgdaBound{v∈ls₂}\AgdaSymbol{)}\<%
\\
\\[\AgdaEmptyExtraSkip]%
\>[0]\AgdaComment{\{- is the command required? 
   is x in b?
   are all commands before x in b in ls?
-\}}\<%
\\
\>[0]\AgdaFunction{required?}\AgdaSpace{}%
\AgdaSymbol{:}\AgdaSpace{}%
\AgdaSymbol{(}\AgdaBound{x}\AgdaSpace{}%
\AgdaSymbol{:}\AgdaSpace{}%
\AgdaFunction{Cmd}\AgdaSymbol{)}\AgdaSpace{}%
\AgdaSymbol{(}\AgdaBound{b}\AgdaSpace{}%
\AgdaBound{ls}\AgdaSpace{}%
\AgdaBound{bf}\AgdaSpace{}%
\AgdaSymbol{:}\AgdaSpace{}%
\AgdaFunction{Build}\AgdaSymbol{)}\AgdaSpace{}%
\AgdaSymbol{→}\AgdaSpace{}%
\AgdaDatatype{Maybe}\AgdaSpace{}%
\AgdaSymbol{(}\AgdaBound{x}\AgdaSpace{}%
\AgdaOperator{\AgdaFunction{∈}}\AgdaSpace{}%
\AgdaBound{b}\AgdaSymbol{)}\<%
\\
\>[0]\AgdaFunction{required?}\AgdaSpace{}%
\AgdaBound{x}\AgdaSpace{}%
\AgdaInductiveConstructor{[]}\AgdaSpace{}%
\AgdaBound{ls}\AgdaSpace{}%
\AgdaSymbol{\AgdaUnderscore{}}\AgdaSpace{}%
\AgdaSymbol{=}\AgdaSpace{}%
\AgdaInductiveConstructor{nothing}\<%
\\
\>[0]\AgdaFunction{required?}\AgdaSpace{}%
\AgdaBound{x}\AgdaSpace{}%
\AgdaSymbol{(}\AgdaBound{x₁}\AgdaSpace{}%
\AgdaOperator{\AgdaInductiveConstructor{∷}}\AgdaSpace{}%
\AgdaBound{b}\AgdaSymbol{)}\AgdaSpace{}%
\AgdaBound{ls}\AgdaSpace{}%
\AgdaBound{bf}\AgdaSpace{}%
\AgdaKeyword{with}\AgdaSpace{}%
\AgdaBound{x}\AgdaSpace{}%
\AgdaOperator{\AgdaFunction{≟}}\AgdaSpace{}%
\AgdaBound{x₁}\<%
\\
\>[0]\AgdaSymbol{...}\AgdaSpace{}%
\AgdaSymbol{|}\AgdaSpace{}%
\AgdaInductiveConstructor{no}\AgdaSpace{}%
\AgdaBound{¬x≡x₁}\AgdaSpace{}%
\AgdaKeyword{with}\AgdaSpace{}%
\AgdaFunction{required?}\AgdaSpace{}%
\AgdaBound{x}\AgdaSpace{}%
\AgdaBound{b}\AgdaSpace{}%
\AgdaBound{ls}\AgdaSpace{}%
\AgdaSymbol{(}\AgdaBound{x₁}\AgdaSpace{}%
\AgdaOperator{\AgdaInductiveConstructor{∷}}\AgdaSpace{}%
\AgdaBound{bf}\AgdaSymbol{)}\<%
\\
\>[0]\AgdaSymbol{...}\AgdaSpace{}%
\AgdaSymbol{|}\AgdaSpace{}%
\AgdaInductiveConstructor{nothing}\AgdaSpace{}%
\AgdaSymbol{=}\AgdaSpace{}%
\AgdaInductiveConstructor{nothing}\<%
\\
\>[0]\AgdaSymbol{...}\AgdaSpace{}%
\AgdaSymbol{|}\AgdaSpace{}%
\AgdaInductiveConstructor{just}\AgdaSpace{}%
\AgdaBound{x∈b}\AgdaSpace{}%
\AgdaSymbol{=}\AgdaSpace{}%
\AgdaInductiveConstructor{just}\AgdaSpace{}%
\AgdaSymbol{(}\AgdaInductiveConstructor{there}\AgdaSpace{}%
\AgdaBound{x∈b}\AgdaSymbol{)}\<%
\\
\>[0]\AgdaFunction{required?}\AgdaSpace{}%
\AgdaBound{x}\AgdaSpace{}%
\AgdaSymbol{(}\AgdaBound{x₁}\AgdaSpace{}%
\AgdaOperator{\AgdaInductiveConstructor{∷}}\AgdaSpace{}%
\AgdaBound{b}\AgdaSymbol{)}\AgdaSpace{}%
\AgdaBound{ls}\AgdaSpace{}%
\AgdaBound{bf}\AgdaSpace{}%
\AgdaSymbol{|}\AgdaSpace{}%
\AgdaInductiveConstructor{yes}\AgdaSpace{}%
\AgdaBound{x≡x₁}\AgdaSpace{}%
\AgdaKeyword{with}\AgdaSpace{}%
\AgdaFunction{subset?}\AgdaSpace{}%
\AgdaBound{bf}\AgdaSpace{}%
\AgdaBound{ls}\<%
\\
\>[0][@{}l@{\AgdaIndent{0}}]%
\>[2]\AgdaKeyword{where}%
\>[424I]\AgdaFunction{subset?}\AgdaSpace{}%
\AgdaSymbol{:}\AgdaSpace{}%
\AgdaSymbol{(}\AgdaBound{ls₁}\AgdaSpace{}%
\AgdaBound{ls₂}\AgdaSpace{}%
\AgdaSymbol{:}\AgdaSpace{}%
\AgdaFunction{Build}\AgdaSymbol{)}\AgdaSpace{}%
\AgdaSymbol{→}\AgdaSpace{}%
\AgdaDatatype{Bool}\<%
\\
\>[.][@{}l@{}]\<[424I]%
\>[8]\AgdaFunction{subset?}\AgdaSpace{}%
\AgdaInductiveConstructor{[]}\AgdaSpace{}%
\AgdaBound{ls₂}\AgdaSpace{}%
\AgdaSymbol{=}\AgdaSpace{}%
\AgdaInductiveConstructor{true}\<%
\\
\>[8]\AgdaFunction{subset?}\AgdaSpace{}%
\AgdaSymbol{(}\AgdaBound{x}\AgdaSpace{}%
\AgdaOperator{\AgdaInductiveConstructor{∷}}\AgdaSpace{}%
\AgdaBound{ls₁}\AgdaSymbol{)}\AgdaSpace{}%
\AgdaBound{ls₂}\AgdaSpace{}%
\AgdaKeyword{with}\AgdaSpace{}%
\AgdaBound{x}\AgdaSpace{}%
\AgdaOperator{\AgdaFunction{∈?}}\AgdaSpace{}%
\AgdaBound{ls₁}\<%
\\
\>[8]\AgdaSymbol{...}\AgdaSpace{}%
\AgdaSymbol{|}\AgdaSpace{}%
\AgdaInductiveConstructor{yes}\AgdaSpace{}%
\AgdaBound{x∈}\AgdaSpace{}%
\AgdaSymbol{=}\AgdaSpace{}%
\AgdaFunction{subset?}\AgdaSpace{}%
\AgdaBound{ls₁}\AgdaSpace{}%
\AgdaBound{ls₂}\<%
\\
\>[8]\AgdaSymbol{...}\AgdaSpace{}%
\AgdaSymbol{|}\AgdaSpace{}%
\AgdaInductiveConstructor{no}\AgdaSpace{}%
\AgdaBound{x∉}\AgdaSpace{}%
\AgdaSymbol{=}\AgdaSpace{}%
\AgdaInductiveConstructor{false}\<%
\\
\>[0]\AgdaSymbol{...}\AgdaSpace{}%
\AgdaSymbol{|}\AgdaSpace{}%
\AgdaInductiveConstructor{true}\AgdaSpace{}%
\AgdaSymbol{=}\AgdaSpace{}%
\AgdaInductiveConstructor{just}\AgdaSpace{}%
\AgdaSymbol{(}\AgdaInductiveConstructor{here}\AgdaSpace{}%
\AgdaBound{x≡x₁}\AgdaSymbol{)}\<%
\\
\>[0]\AgdaSymbol{...}\AgdaSpace{}%
\AgdaSymbol{|}\AgdaSpace{}%
\AgdaInductiveConstructor{false}\AgdaSpace{}%
\AgdaSymbol{=}\AgdaSpace{}%
\AgdaInductiveConstructor{nothing}\<%
\\
\\[\AgdaEmptyExtraSkip]%
\>[0]\AgdaFunction{∃Speculative2}\AgdaSpace{}%
\AgdaSymbol{:}\AgdaSpace{}%
\AgdaSymbol{∀}\AgdaSpace{}%
\AgdaBound{x₂}\AgdaSpace{}%
\AgdaBound{b}\AgdaSpace{}%
\AgdaBound{ls₁}\AgdaSpace{}%
\AgdaBound{ls}\AgdaSpace{}%
\AgdaBound{rs}\AgdaSpace{}%
\AgdaSymbol{→}\AgdaSpace{}%
\AgdaDatatype{Maybe}\AgdaSpace{}%
\AgdaSymbol{(}\AgdaFunction{∃[}\AgdaSpace{}%
\AgdaBound{x₁}\AgdaSpace{}%
\AgdaFunction{]}\AgdaSymbol{(}\AgdaFunction{∃[}\AgdaSpace{}%
\AgdaBound{v}\AgdaSpace{}%
\AgdaFunction{]}\AgdaSymbol{(}\AgdaBound{x₁}\AgdaSpace{}%
\AgdaOperator{\AgdaFunction{∈}}\AgdaSpace{}%
\AgdaBound{ls₁}\AgdaSpace{}%
\AgdaOperator{\AgdaFunction{×}}\AgdaSpace{}%
\AgdaOperator{\AgdaFunction{¬}}\AgdaSpace{}%
\AgdaSymbol{(}\AgdaBound{x₁}\AgdaSpace{}%
\AgdaOperator{\AgdaFunction{before}}\AgdaSpace{}%
\AgdaBound{x₂}\AgdaSpace{}%
\AgdaOperator{\AgdaFunction{∈}}\AgdaSpace{}%
\AgdaBound{b}\AgdaSymbol{)}\AgdaSpace{}%
\AgdaOperator{\AgdaFunction{×}}\AgdaSpace{}%
\AgdaBound{v}\AgdaSpace{}%
\AgdaOperator{\AgdaFunction{∈}}\AgdaSpace{}%
\AgdaBound{rs}\AgdaSpace{}%
\AgdaOperator{\AgdaFunction{×}}\AgdaSpace{}%
\AgdaBound{v}\AgdaSpace{}%
\AgdaOperator{\AgdaFunction{∈}}\AgdaSpace{}%
\AgdaFunction{cmdWrote}\AgdaSpace{}%
\AgdaBound{ls}\AgdaSpace{}%
\AgdaBound{x₁}\AgdaSymbol{)))}\<%
\\
\>[0]\AgdaFunction{∃Speculative2}\AgdaSpace{}%
\AgdaBound{x₂}\AgdaSpace{}%
\AgdaBound{b}\AgdaSpace{}%
\AgdaInductiveConstructor{[]}\AgdaSpace{}%
\AgdaBound{ls}\AgdaSpace{}%
\AgdaBound{rs}\AgdaSpace{}%
\AgdaSymbol{=}\AgdaSpace{}%
\AgdaInductiveConstructor{nothing}\<%
\\
\>[0]\AgdaFunction{∃Speculative2}\AgdaSpace{}%
\AgdaBound{x₂}\AgdaSpace{}%
\AgdaBound{b}\AgdaSpace{}%
\AgdaSymbol{(}\AgdaBound{x}\AgdaSpace{}%
\AgdaOperator{\AgdaInductiveConstructor{∷}}\AgdaSpace{}%
\AgdaBound{ls₁}\AgdaSymbol{)}\AgdaSpace{}%
\AgdaBound{ls}\AgdaSpace{}%
\AgdaBound{rs}%
\>[36]\AgdaKeyword{with}\AgdaSpace{}%
\AgdaFunction{before?}\AgdaSpace{}%
\AgdaBound{x}\AgdaSpace{}%
\AgdaBound{x₂}\AgdaSpace{}%
\AgdaBound{b}\<%
\\
\>[0]\AgdaSymbol{...}\AgdaSpace{}%
\AgdaSymbol{|}\AgdaSpace{}%
\AgdaInductiveConstructor{yes}\AgdaSpace{}%
\AgdaBound{bf}\AgdaSpace{}%
\AgdaKeyword{with}\AgdaSpace{}%
\AgdaFunction{∃Speculative2}\AgdaSpace{}%
\AgdaBound{x₂}\AgdaSpace{}%
\AgdaBound{b}\AgdaSpace{}%
\AgdaBound{ls₁}\AgdaSpace{}%
\AgdaBound{ls}\AgdaSpace{}%
\AgdaBound{rs}\<%
\\
\>[0]\AgdaSymbol{...}\AgdaSpace{}%
\AgdaSymbol{|}\AgdaSpace{}%
\AgdaInductiveConstructor{nothing}\AgdaSpace{}%
\AgdaSymbol{=}\AgdaSpace{}%
\AgdaInductiveConstructor{nothing}\<%
\\
\>[0]\AgdaSymbol{...}\AgdaSpace{}%
\AgdaSymbol{|}\AgdaSpace{}%
\AgdaInductiveConstructor{just}\AgdaSpace{}%
\AgdaSymbol{(}\AgdaBound{x₁}\AgdaSpace{}%
\AgdaOperator{\AgdaInductiveConstructor{,}}\AgdaSpace{}%
\AgdaBound{v}\AgdaSpace{}%
\AgdaOperator{\AgdaInductiveConstructor{,}}\AgdaSpace{}%
\AgdaBound{x₁∈ls}\AgdaSpace{}%
\AgdaOperator{\AgdaInductiveConstructor{,}}\AgdaSpace{}%
\AgdaBound{¬bf}\AgdaSpace{}%
\AgdaOperator{\AgdaInductiveConstructor{,}}\AgdaSpace{}%
\AgdaBound{v∈rs}\AgdaSpace{}%
\AgdaOperator{\AgdaInductiveConstructor{,}}\AgdaSpace{}%
\AgdaBound{v∈ws}\AgdaSymbol{)}\<%
\\
\>[0][@{}l@{\AgdaIndent{0}}]%
\>[2]\AgdaSymbol{=}\AgdaSpace{}%
\AgdaInductiveConstructor{just}\AgdaSpace{}%
\AgdaSymbol{(}\AgdaBound{x₁}\AgdaSpace{}%
\AgdaOperator{\AgdaInductiveConstructor{,}}\AgdaSpace{}%
\AgdaBound{v}\AgdaSpace{}%
\AgdaOperator{\AgdaInductiveConstructor{,}}\AgdaSpace{}%
\AgdaInductiveConstructor{there}\AgdaSpace{}%
\AgdaBound{x₁∈ls}\AgdaSpace{}%
\AgdaOperator{\AgdaInductiveConstructor{,}}\AgdaSpace{}%
\AgdaBound{¬bf}\AgdaSpace{}%
\AgdaOperator{\AgdaInductiveConstructor{,}}\AgdaSpace{}%
\AgdaBound{v∈rs}\AgdaSpace{}%
\AgdaOperator{\AgdaInductiveConstructor{,}}\AgdaSpace{}%
\AgdaBound{v∈ws}\AgdaSymbol{)}\AgdaSpace{}%
\AgdaComment{-- ∈-cmdWrote∷ x x₁ ls v∈ws (lookup px x₁∈ls))}\<%
\\
\>[0]\AgdaFunction{∃Speculative2}\AgdaSpace{}%
\AgdaBound{x₂}\AgdaSpace{}%
\AgdaBound{b}\AgdaSpace{}%
\AgdaSymbol{(}\AgdaBound{x}\AgdaSpace{}%
\AgdaOperator{\AgdaInductiveConstructor{∷}}\AgdaSpace{}%
\AgdaBound{ls₁}\AgdaSymbol{)}\AgdaSpace{}%
\AgdaBound{ls}\AgdaSpace{}%
\AgdaBound{rs}\AgdaSpace{}%
\AgdaSymbol{|}\AgdaSpace{}%
\AgdaInductiveConstructor{no}\AgdaSpace{}%
\AgdaBound{¬bf}\AgdaSpace{}%
\AgdaKeyword{with}\AgdaSpace{}%
\AgdaFunction{∃Intersection}\AgdaSpace{}%
\AgdaBound{rs}\AgdaSpace{}%
\AgdaSymbol{(}\AgdaFunction{cmdWrote}\AgdaSpace{}%
\AgdaBound{ls}\AgdaSpace{}%
\AgdaBound{x}\AgdaSymbol{)}\<%
\\
\>[0]\AgdaSymbol{...}\AgdaSpace{}%
\AgdaSymbol{|}\AgdaSpace{}%
\AgdaInductiveConstructor{just}\AgdaSpace{}%
\AgdaSymbol{(}\AgdaBound{v}\AgdaSpace{}%
\AgdaOperator{\AgdaInductiveConstructor{,}}\AgdaSpace{}%
\AgdaBound{v∈rs}\AgdaSpace{}%
\AgdaOperator{\AgdaInductiveConstructor{,}}\AgdaSpace{}%
\AgdaBound{v∈ws}\AgdaSymbol{)}\AgdaSpace{}%
\AgdaSymbol{=}\AgdaSpace{}%
\AgdaInductiveConstructor{just}\AgdaSpace{}%
\AgdaSymbol{(}\AgdaBound{x}\AgdaSpace{}%
\AgdaOperator{\AgdaInductiveConstructor{,}}\AgdaSpace{}%
\AgdaBound{v}\AgdaSpace{}%
\AgdaOperator{\AgdaInductiveConstructor{,}}\AgdaSpace{}%
\AgdaInductiveConstructor{here}\AgdaSpace{}%
\AgdaInductiveConstructor{refl}\AgdaSpace{}%
\AgdaOperator{\AgdaInductiveConstructor{,}}\AgdaSpace{}%
\AgdaBound{¬bf}\AgdaSpace{}%
\AgdaOperator{\AgdaInductiveConstructor{,}}\AgdaSpace{}%
\AgdaBound{v∈rs}\AgdaSpace{}%
\AgdaOperator{\AgdaInductiveConstructor{,}}\AgdaSpace{}%
\AgdaBound{v∈ws}\AgdaSymbol{)}\<%
\\
\>[0]\AgdaSymbol{...}\AgdaSpace{}%
\AgdaSymbol{|}\AgdaSpace{}%
\AgdaInductiveConstructor{nothing}\AgdaSpace{}%
\AgdaKeyword{with}\AgdaSpace{}%
\AgdaFunction{∃Speculative2}\AgdaSpace{}%
\AgdaBound{x₂}\AgdaSpace{}%
\AgdaBound{b}\AgdaSpace{}%
\AgdaBound{ls₁}\AgdaSpace{}%
\AgdaBound{ls}\AgdaSpace{}%
\AgdaBound{rs}\<%
\\
\>[0]\AgdaSymbol{...}\AgdaSpace{}%
\AgdaSymbol{|}\AgdaSpace{}%
\AgdaInductiveConstructor{nothing}\AgdaSpace{}%
\AgdaSymbol{=}\AgdaSpace{}%
\AgdaInductiveConstructor{nothing}\<%
\\
\>[0]\AgdaSymbol{...}\AgdaSpace{}%
\AgdaSymbol{|}\AgdaSpace{}%
\AgdaInductiveConstructor{just}\AgdaSpace{}%
\AgdaSymbol{(}\AgdaBound{x₁}\AgdaSpace{}%
\AgdaOperator{\AgdaInductiveConstructor{,}}\AgdaSpace{}%
\AgdaBound{v}\AgdaSpace{}%
\AgdaOperator{\AgdaInductiveConstructor{,}}\AgdaSpace{}%
\AgdaBound{x₁∈ls}\AgdaSpace{}%
\AgdaOperator{\AgdaInductiveConstructor{,}}\AgdaSpace{}%
\AgdaBound{¬bf₁}\AgdaSpace{}%
\AgdaOperator{\AgdaInductiveConstructor{,}}\AgdaSpace{}%
\AgdaBound{v∈rs}\AgdaSpace{}%
\AgdaOperator{\AgdaInductiveConstructor{,}}\AgdaSpace{}%
\AgdaBound{v∈ws}\AgdaSymbol{)}\<%
\\
\>[0][@{}l@{\AgdaIndent{0}}]%
\>[2]\AgdaSymbol{=}\AgdaSpace{}%
\AgdaInductiveConstructor{just}\AgdaSpace{}%
\AgdaSymbol{(}\AgdaBound{x₁}\AgdaSpace{}%
\AgdaOperator{\AgdaInductiveConstructor{,}}\AgdaSpace{}%
\AgdaBound{v}\AgdaSpace{}%
\AgdaOperator{\AgdaInductiveConstructor{,}}\AgdaSpace{}%
\AgdaInductiveConstructor{there}\AgdaSpace{}%
\AgdaBound{x₁∈ls}\AgdaSpace{}%
\AgdaOperator{\AgdaInductiveConstructor{,}}\AgdaSpace{}%
\AgdaBound{¬bf₁}\AgdaSpace{}%
\AgdaOperator{\AgdaInductiveConstructor{,}}\AgdaSpace{}%
\AgdaBound{v∈rs}\AgdaSpace{}%
\AgdaOperator{\AgdaInductiveConstructor{,}}\AgdaSpace{}%
\AgdaBound{v∈ws}\AgdaSymbol{)}\AgdaSpace{}%
\AgdaComment{-- ∈-cmdWrote∷ x x₁ ls v∈ws (lookup px x₁∈ls))}\<%
\\
\\[\AgdaEmptyExtraSkip]%
\>[0]\AgdaFunction{∃Speculative1}\AgdaSpace{}%
\AgdaSymbol{:}\AgdaSpace{}%
\AgdaSymbol{∀}\AgdaSpace{}%
\AgdaBound{ls₁}\AgdaSpace{}%
\AgdaBound{ls}\AgdaSpace{}%
\AgdaSymbol{\{}\AgdaBound{b}\AgdaSymbol{\}}\AgdaSpace{}%
\AgdaSymbol{(}\AgdaBound{ran}\AgdaSpace{}%
\AgdaSymbol{:}\AgdaSpace{}%
\AgdaFunction{Build}\AgdaSymbol{)}\AgdaSpace{}%
\AgdaSymbol{→}\AgdaSpace{}%
\AgdaDatatype{Maybe}\AgdaSpace{}%
\AgdaSymbol{(}\AgdaFunction{∃[}\AgdaSpace{}%
\AgdaBound{x₁}\AgdaSpace{}%
\AgdaFunction{]}\AgdaSymbol{(}\AgdaFunction{∃[}\AgdaSpace{}%
\AgdaBound{x₂}\AgdaSpace{}%
\AgdaFunction{]}\AgdaSymbol{(}\AgdaFunction{∃[}\AgdaSpace{}%
\AgdaBound{v}\AgdaSpace{}%
\AgdaFunction{]}\AgdaSymbol{((}\AgdaBound{x₂}\AgdaSpace{}%
\AgdaOperator{\AgdaFunction{before}}\AgdaSpace{}%
\AgdaBound{x₁}\AgdaSpace{}%
\AgdaOperator{\AgdaFunction{∈}}\AgdaSpace{}%
\AgdaBound{ls₁}\AgdaSymbol{)}\AgdaSpace{}%
\AgdaOperator{\AgdaFunction{×}}\AgdaSpace{}%
\AgdaBound{x₂}\AgdaSpace{}%
\AgdaOperator{\AgdaFunction{∈}}\AgdaSpace{}%
\AgdaBound{b}\AgdaSpace{}%
\AgdaOperator{\AgdaFunction{×}}\AgdaSpace{}%
\AgdaOperator{\AgdaFunction{¬}}\AgdaSpace{}%
\AgdaSymbol{(}\AgdaBound{x₁}\AgdaSpace{}%
\AgdaOperator{\AgdaFunction{before}}\AgdaSpace{}%
\AgdaBound{x₂}\AgdaSpace{}%
\AgdaOperator{\AgdaFunction{∈}}\AgdaSpace{}%
\AgdaBound{b}\AgdaSymbol{)}\AgdaSpace{}%
\AgdaOperator{\AgdaFunction{×}}\AgdaSpace{}%
\AgdaBound{v}\AgdaSpace{}%
\AgdaOperator{\AgdaFunction{∈}}\AgdaSpace{}%
\AgdaFunction{cmdRead}\AgdaSpace{}%
\AgdaBound{ls}\AgdaSpace{}%
\AgdaBound{x₂}\AgdaSpace{}%
\AgdaOperator{\AgdaFunction{×}}\AgdaSpace{}%
\AgdaBound{v}\AgdaSpace{}%
\AgdaOperator{\AgdaFunction{∈}}\AgdaSpace{}%
\AgdaFunction{cmdWrote}\AgdaSpace{}%
\AgdaBound{ls}\AgdaSpace{}%
\AgdaBound{x₁}\AgdaSymbol{))))}\<%
\\
\>[0]\AgdaFunction{∃Speculative1}\AgdaSpace{}%
\AgdaInductiveConstructor{[]}\AgdaSpace{}%
\AgdaBound{ls}\AgdaSpace{}%
\AgdaBound{ran}\AgdaSpace{}%
\AgdaSymbol{=}\AgdaSpace{}%
\AgdaInductiveConstructor{nothing}\<%
\\
\>[0]\AgdaFunction{∃Speculative1}\AgdaSpace{}%
\AgdaSymbol{(}\AgdaBound{x}\AgdaSpace{}%
\AgdaOperator{\AgdaInductiveConstructor{∷}}\AgdaSpace{}%
\AgdaBound{ls₁}\AgdaSymbol{)}\AgdaSpace{}%
\AgdaBound{ls}\AgdaSpace{}%
\AgdaSymbol{\{}\AgdaBound{b}\AgdaSymbol{\}}\AgdaSpace{}%
\AgdaBound{ran}\AgdaSpace{}%
\AgdaKeyword{with}\AgdaSpace{}%
\AgdaFunction{required?}\AgdaSpace{}%
\AgdaBound{x}\AgdaSpace{}%
\AgdaBound{b}\AgdaSpace{}%
\AgdaSymbol{(}\AgdaBound{ls₁}\AgdaSpace{}%
\AgdaOperator{\AgdaFunction{++}}\AgdaSpace{}%
\AgdaBound{ran}\AgdaSymbol{)}\AgdaSpace{}%
\AgdaInductiveConstructor{[]}\<%
\\
\>[0]\AgdaSymbol{...}\AgdaSpace{}%
\AgdaSymbol{|}\AgdaSpace{}%
\AgdaInductiveConstructor{just}\AgdaSpace{}%
\AgdaBound{x∈b}\AgdaSpace{}%
\AgdaKeyword{with}\AgdaSpace{}%
\AgdaFunction{∃Speculative2}\AgdaSpace{}%
\AgdaBound{x}\AgdaSpace{}%
\AgdaBound{b}\AgdaSpace{}%
\AgdaBound{ls₁}\AgdaSpace{}%
\AgdaBound{ls}\AgdaSpace{}%
\AgdaSymbol{(}\AgdaFunction{cmdRead}\AgdaSpace{}%
\AgdaBound{ls}\AgdaSpace{}%
\AgdaBound{x}\AgdaSymbol{)}\<%
\\
\>[0]\AgdaSymbol{...}\AgdaSpace{}%
\AgdaSymbol{|}\AgdaSpace{}%
\AgdaInductiveConstructor{just}\AgdaSpace{}%
\AgdaSymbol{(}\AgdaBound{x₁}\AgdaSpace{}%
\AgdaOperator{\AgdaInductiveConstructor{,}}\AgdaSpace{}%
\AgdaBound{v}\AgdaSpace{}%
\AgdaOperator{\AgdaInductiveConstructor{,}}\AgdaSpace{}%
\AgdaBound{x₁∈ls}\AgdaSpace{}%
\AgdaOperator{\AgdaInductiveConstructor{,}}\AgdaSpace{}%
\AgdaBound{¬bf}\AgdaSpace{}%
\AgdaOperator{\AgdaInductiveConstructor{,}}\AgdaSpace{}%
\AgdaBound{v∈rs}\AgdaSpace{}%
\AgdaOperator{\AgdaInductiveConstructor{,}}\AgdaSpace{}%
\AgdaBound{v∈ws}\AgdaSymbol{)}\<%
\\
\>[0][@{}l@{\AgdaIndent{0}}]%
\>[2]\AgdaSymbol{=}\AgdaSpace{}%
\AgdaInductiveConstructor{just}%
\>[724I]\AgdaSymbol{(}\AgdaBound{x₁}\AgdaSpace{}%
\AgdaOperator{\AgdaInductiveConstructor{,}}\AgdaSpace{}%
\AgdaBound{x}\AgdaSpace{}%
\AgdaOperator{\AgdaInductiveConstructor{,}}\AgdaSpace{}%
\AgdaBound{v}\AgdaSpace{}%
\AgdaOperator{\AgdaInductiveConstructor{,}}\AgdaSpace{}%
\AgdaSymbol{(}\AgdaInductiveConstructor{[]}\AgdaSpace{}%
\AgdaOperator{\AgdaInductiveConstructor{,}}\AgdaSpace{}%
\AgdaBound{ls₁}\AgdaSpace{}%
\AgdaOperator{\AgdaInductiveConstructor{,}}\AgdaSpace{}%
\AgdaInductiveConstructor{refl}\AgdaSpace{}%
\AgdaOperator{\AgdaInductiveConstructor{,}}\AgdaSpace{}%
\AgdaBound{x₁∈ls}\AgdaSymbol{)}\AgdaSpace{}%
\AgdaOperator{\AgdaInductiveConstructor{,}}\AgdaSpace{}%
\AgdaBound{x∈b}\AgdaSpace{}%
\AgdaOperator{\AgdaInductiveConstructor{,}}\AgdaSpace{}%
\AgdaBound{¬bf}\AgdaSpace{}%
\AgdaOperator{\AgdaInductiveConstructor{,}}\AgdaSpace{}%
\AgdaBound{v∈rs}\<%
\\
\>[.][@{}l@{}]\<[724I]%
\>[9]\AgdaOperator{\AgdaInductiveConstructor{,}}\AgdaSpace{}%
\AgdaBound{v∈ws}\AgdaSymbol{)}\<%
\\
\>[0]\AgdaSymbol{...}\AgdaSpace{}%
\AgdaSymbol{|}\AgdaSpace{}%
\AgdaInductiveConstructor{nothing}\AgdaSpace{}%
\AgdaKeyword{with}\AgdaSpace{}%
\AgdaFunction{∃Speculative1}\AgdaSpace{}%
\AgdaBound{ls₁}\AgdaSpace{}%
\AgdaBound{ls}\AgdaSpace{}%
\AgdaSymbol{(}\AgdaBound{x}\AgdaSpace{}%
\AgdaOperator{\AgdaInductiveConstructor{∷}}\AgdaSpace{}%
\AgdaBound{ran}\AgdaSymbol{)}\<%
\\
\>[0]\AgdaSymbol{...}\AgdaSpace{}%
\AgdaSymbol{|}\AgdaSpace{}%
\AgdaInductiveConstructor{nothing}\AgdaSpace{}%
\AgdaSymbol{=}\AgdaSpace{}%
\AgdaInductiveConstructor{nothing}\<%
\\
\>[0]\AgdaSymbol{...}\AgdaSpace{}%
\AgdaSymbol{|}\AgdaSpace{}%
\AgdaInductiveConstructor{just}\AgdaSpace{}%
\AgdaSymbol{(}\AgdaBound{x₁}\AgdaSpace{}%
\AgdaOperator{\AgdaInductiveConstructor{,}}\AgdaSpace{}%
\AgdaBound{x₂}\AgdaSpace{}%
\AgdaOperator{\AgdaInductiveConstructor{,}}\AgdaSpace{}%
\AgdaBound{v}\AgdaSpace{}%
\AgdaOperator{\AgdaInductiveConstructor{,}}\AgdaSpace{}%
\AgdaBound{bf}\AgdaSymbol{@(}\AgdaBound{xs}\AgdaSpace{}%
\AgdaOperator{\AgdaInductiveConstructor{,}}\AgdaSpace{}%
\AgdaBound{ys}\AgdaSpace{}%
\AgdaOperator{\AgdaInductiveConstructor{,}}\AgdaSpace{}%
\AgdaBound{≡₁}\AgdaSpace{}%
\AgdaOperator{\AgdaInductiveConstructor{,}}\AgdaSpace{}%
\AgdaBound{x₁∈ys}\AgdaSymbol{)}\AgdaSpace{}%
\AgdaOperator{\AgdaInductiveConstructor{,}}\AgdaSpace{}%
\AgdaBound{x₂∈b}\AgdaSpace{}%
\AgdaOperator{\AgdaInductiveConstructor{,}}\AgdaSpace{}%
\AgdaBound{¬bf}\AgdaSpace{}%
\AgdaOperator{\AgdaInductiveConstructor{,}}\AgdaSpace{}%
\AgdaBound{v∈cr}\AgdaSpace{}%
\AgdaOperator{\AgdaInductiveConstructor{,}}\AgdaSpace{}%
\AgdaBound{v∈cw}\AgdaSymbol{)}\<%
\\
\>[0][@{}l@{\AgdaIndent{0}}]%
\>[2]\AgdaSymbol{=}\AgdaSpace{}%
\AgdaInductiveConstructor{just}%
\>[781I]\AgdaSymbol{(}\AgdaBound{x₁}\AgdaSpace{}%
\AgdaOperator{\AgdaInductiveConstructor{,}}\AgdaSpace{}%
\AgdaBound{x₂}\AgdaSpace{}%
\AgdaOperator{\AgdaInductiveConstructor{,}}\AgdaSpace{}%
\AgdaBound{v}\AgdaSpace{}%
\AgdaOperator{\AgdaInductiveConstructor{,}}\AgdaSpace{}%
\AgdaFunction{before-∷}\AgdaSpace{}%
\AgdaBound{x₂}\AgdaSpace{}%
\AgdaBound{x₁}\AgdaSpace{}%
\AgdaBound{x}\AgdaSpace{}%
\AgdaBound{ls₁}\AgdaSpace{}%
\AgdaBound{bf}\AgdaSpace{}%
\AgdaOperator{\AgdaInductiveConstructor{,}}\AgdaSpace{}%
\AgdaBound{x₂∈b}\AgdaSpace{}%
\AgdaOperator{\AgdaInductiveConstructor{,}}\AgdaSpace{}%
\AgdaBound{¬bf}\<%
\\
\>[.][@{}l@{}]\<[781I]%
\>[9]\AgdaOperator{\AgdaInductiveConstructor{,}}\AgdaSpace{}%
\AgdaBound{v∈cr}\<%
\\
\>[9]\AgdaOperator{\AgdaInductiveConstructor{,}}\AgdaSpace{}%
\AgdaBound{v∈cw}\AgdaSymbol{)}\<%
\\
\>[0]\AgdaFunction{∃Speculative1}\AgdaSpace{}%
\AgdaSymbol{(}\AgdaBound{x}\AgdaSpace{}%
\AgdaOperator{\AgdaInductiveConstructor{∷}}\AgdaSpace{}%
\AgdaBound{ls₁}\AgdaSymbol{)}\AgdaSpace{}%
\AgdaBound{ls}\AgdaSpace{}%
\AgdaBound{ran}\AgdaSpace{}%
\AgdaSymbol{|}\AgdaSpace{}%
\AgdaInductiveConstructor{nothing}\AgdaSpace{}%
\AgdaKeyword{with}\AgdaSpace{}%
\AgdaFunction{∃Speculative1}\AgdaSpace{}%
\AgdaBound{ls₁}\AgdaSpace{}%
\AgdaBound{ls}\AgdaSpace{}%
\AgdaSymbol{(}\AgdaBound{x}\AgdaSpace{}%
\AgdaOperator{\AgdaInductiveConstructor{∷}}\AgdaSpace{}%
\AgdaBound{ran}\AgdaSymbol{)}\<%
\\
\>[0]\AgdaSymbol{...}\AgdaSpace{}%
\AgdaSymbol{|}\AgdaSpace{}%
\AgdaInductiveConstructor{nothing}\AgdaSpace{}%
\AgdaSymbol{=}\AgdaSpace{}%
\AgdaInductiveConstructor{nothing}\<%
\\
\>[0]\AgdaSymbol{...}\AgdaSpace{}%
\AgdaSymbol{|}\AgdaSpace{}%
\AgdaInductiveConstructor{just}\AgdaSpace{}%
\AgdaSymbol{(}\AgdaBound{x₁}\AgdaSpace{}%
\AgdaOperator{\AgdaInductiveConstructor{,}}\AgdaSpace{}%
\AgdaBound{x₂}\AgdaSpace{}%
\AgdaOperator{\AgdaInductiveConstructor{,}}\AgdaSpace{}%
\AgdaBound{v}\AgdaSpace{}%
\AgdaOperator{\AgdaInductiveConstructor{,}}\AgdaSpace{}%
\AgdaBound{bf}\AgdaSymbol{@(}\AgdaBound{xs}\AgdaSpace{}%
\AgdaOperator{\AgdaInductiveConstructor{,}}\AgdaSpace{}%
\AgdaBound{ys}\AgdaSpace{}%
\AgdaOperator{\AgdaInductiveConstructor{,}}\AgdaSpace{}%
\AgdaBound{≡₁}\AgdaSpace{}%
\AgdaOperator{\AgdaInductiveConstructor{,}}\AgdaSpace{}%
\AgdaBound{x₁∈ys}\AgdaSymbol{)}\AgdaSpace{}%
\AgdaOperator{\AgdaInductiveConstructor{,}}\AgdaSpace{}%
\AgdaBound{x₂∈b}\AgdaSpace{}%
\AgdaOperator{\AgdaInductiveConstructor{,}}\AgdaSpace{}%
\AgdaBound{¬bf}\AgdaSpace{}%
\AgdaOperator{\AgdaInductiveConstructor{,}}\AgdaSpace{}%
\AgdaBound{v∈cr}\AgdaSpace{}%
\AgdaOperator{\AgdaInductiveConstructor{,}}\AgdaSpace{}%
\AgdaBound{v∈cw}\AgdaSymbol{)}\<%
\\
\>[0][@{}l@{\AgdaIndent{0}}]%
\>[2]\AgdaSymbol{=}\AgdaSpace{}%
\AgdaInductiveConstructor{just}%
\>[841I]\AgdaSymbol{(}\AgdaBound{x₁}\AgdaSpace{}%
\AgdaOperator{\AgdaInductiveConstructor{,}}\AgdaSpace{}%
\AgdaBound{x₂}\AgdaSpace{}%
\AgdaOperator{\AgdaInductiveConstructor{,}}\AgdaSpace{}%
\AgdaBound{v}\AgdaSpace{}%
\AgdaOperator{\AgdaInductiveConstructor{,}}\AgdaSpace{}%
\AgdaFunction{before-∷}\AgdaSpace{}%
\AgdaBound{x₂}\AgdaSpace{}%
\AgdaBound{x₁}\AgdaSpace{}%
\AgdaBound{x}\AgdaSpace{}%
\AgdaBound{ls₁}\AgdaSpace{}%
\AgdaBound{bf}\AgdaSpace{}%
\AgdaOperator{\AgdaInductiveConstructor{,}}\AgdaSpace{}%
\AgdaBound{x₂∈b}\AgdaSpace{}%
\AgdaOperator{\AgdaInductiveConstructor{,}}\AgdaSpace{}%
\AgdaBound{¬bf}\<%
\\
\>[.][@{}l@{}]\<[841I]%
\>[9]\AgdaOperator{\AgdaInductiveConstructor{,}}\AgdaSpace{}%
\AgdaBound{v∈cr}\AgdaSpace{}%
\AgdaOperator{\AgdaInductiveConstructor{,}}\AgdaSpace{}%
\AgdaBound{v∈cw}\AgdaSymbol{)}\<%
\\
\\[\AgdaEmptyExtraSkip]%
\>[0]\AgdaFunction{∃Speculative}\AgdaSpace{}%
\AgdaSymbol{:}\AgdaSpace{}%
\AgdaSymbol{∀}\AgdaSpace{}%
\AgdaBound{s}\AgdaSpace{}%
\AgdaBound{x}\AgdaSpace{}%
\AgdaSymbol{\{}\AgdaBound{b}\AgdaSymbol{\}}\AgdaSpace{}%
\AgdaBound{ls}\AgdaSpace{}%
\AgdaSymbol{→}\AgdaSpace{}%
\AgdaDatatype{Maybe}\AgdaSpace{}%
\AgdaSymbol{(}\AgdaDatatype{Hazard}\AgdaSpace{}%
\AgdaBound{s}\AgdaSpace{}%
\AgdaBound{x}\AgdaSpace{}%
\AgdaBound{b}\AgdaSpace{}%
\AgdaBound{ls}\AgdaSymbol{)}\<%
\\
\>[0]\AgdaFunction{∃Speculative}\AgdaSpace{}%
\AgdaBound{s}\AgdaSpace{}%
\AgdaBound{x}\AgdaSpace{}%
\AgdaSymbol{\{}\AgdaBound{b}\AgdaSymbol{\}}\AgdaSpace{}%
\AgdaBound{ls}\AgdaSpace{}%
\AgdaKeyword{with}\AgdaSpace{}%
\AgdaFunction{∃Speculative1}\AgdaSpace{}%
\AgdaSymbol{(}\AgdaBound{x}\AgdaSpace{}%
\AgdaOperator{\AgdaInductiveConstructor{∷}}\AgdaSpace{}%
\AgdaFunction{cmdsRun}\AgdaSpace{}%
\AgdaBound{ls}\AgdaSymbol{)}\AgdaSpace{}%
\AgdaSymbol{(}\AgdaFunction{rec}\AgdaSpace{}%
\AgdaBound{s}\AgdaSpace{}%
\AgdaBound{x}\AgdaSpace{}%
\AgdaBound{ls}\AgdaSymbol{)}\AgdaSpace{}%
\AgdaInductiveConstructor{[]}\<%
\\
\>[0]\AgdaSymbol{...}\AgdaSpace{}%
\AgdaSymbol{|}\AgdaSpace{}%
\AgdaInductiveConstructor{nothing}\AgdaSpace{}%
\AgdaSymbol{=}\AgdaSpace{}%
\AgdaInductiveConstructor{nothing}\<%
\\
\>[0]\AgdaSymbol{...}\AgdaSpace{}%
\AgdaSymbol{|}\AgdaSpace{}%
\AgdaInductiveConstructor{just}\AgdaSpace{}%
\AgdaSymbol{(}\AgdaBound{x₁}\AgdaSpace{}%
\AgdaOperator{\AgdaInductiveConstructor{,}}\AgdaSpace{}%
\AgdaBound{x₂}\AgdaSpace{}%
\AgdaOperator{\AgdaInductiveConstructor{,}}\AgdaSpace{}%
\AgdaBound{v}\AgdaSpace{}%
\AgdaOperator{\AgdaInductiveConstructor{,}}\AgdaSpace{}%
\AgdaBound{bf}\AgdaSpace{}%
\AgdaOperator{\AgdaInductiveConstructor{,}}\AgdaSpace{}%
\AgdaBound{x₁∈b}\AgdaSpace{}%
\AgdaOperator{\AgdaInductiveConstructor{,}}\AgdaSpace{}%
\AgdaBound{¬bf}\AgdaSpace{}%
\AgdaOperator{\AgdaInductiveConstructor{,}}\AgdaSpace{}%
\AgdaBound{v∈cr}\AgdaSpace{}%
\AgdaOperator{\AgdaInductiveConstructor{,}}\AgdaSpace{}%
\AgdaBound{v∈cw}\AgdaSymbol{)}\<%
\\
\>[0][@{}l@{\AgdaIndent{0}}]%
\>[2]\AgdaSymbol{=}\AgdaSpace{}%
\AgdaInductiveConstructor{just}\AgdaSpace{}%
\AgdaSymbol{(}\AgdaInductiveConstructor{Speculative}\AgdaSpace{}%
\AgdaBound{x₁}\AgdaSpace{}%
\AgdaBound{x₂}\AgdaSpace{}%
\AgdaBound{bf}\AgdaSpace{}%
\AgdaBound{x₁∈b}\AgdaSpace{}%
\AgdaBound{¬bf}\AgdaSpace{}%
\AgdaBound{v∈cr}\AgdaSpace{}%
\AgdaBound{v∈cw}\AgdaSymbol{)}\<%
\\
\\[\AgdaEmptyExtraSkip]%
\\[\AgdaEmptyExtraSkip]%
\>[0]\AgdaComment{\{- is there a read/write or write/write hazard? -\}}\<%
\\
\>[0]\AgdaFunction{∃WriteWrite}\AgdaSpace{}%
\AgdaSymbol{:}\AgdaSpace{}%
\AgdaSymbol{∀}\AgdaSpace{}%
\AgdaBound{s}\AgdaSpace{}%
\AgdaBound{x}\AgdaSpace{}%
\AgdaSymbol{\{}\AgdaBound{b}\AgdaSymbol{\}}\AgdaSpace{}%
\AgdaBound{ls}\AgdaSpace{}%
\AgdaSymbol{→}\AgdaSpace{}%
\AgdaDatatype{Maybe}\AgdaSpace{}%
\AgdaSymbol{(}\AgdaDatatype{Hazard}\AgdaSpace{}%
\AgdaBound{s}\AgdaSpace{}%
\AgdaBound{x}\AgdaSpace{}%
\AgdaBound{b}\AgdaSpace{}%
\AgdaBound{ls}\AgdaSymbol{)}\<%
\\
\>[0]\AgdaFunction{∃WriteWrite}\AgdaSpace{}%
\AgdaBound{s}\AgdaSpace{}%
\AgdaBound{x}\AgdaSpace{}%
\AgdaBound{ls}\AgdaSpace{}%
\AgdaKeyword{with}\AgdaSpace{}%
\AgdaFunction{∃Intersection}\AgdaSpace{}%
\AgdaSymbol{(}\AgdaFunction{cmdWriteNames}\AgdaSpace{}%
\AgdaBound{x}\AgdaSpace{}%
\AgdaBound{s}\AgdaSymbol{)}\AgdaSpace{}%
\AgdaSymbol{(}\AgdaFunction{filesWrote}\AgdaSpace{}%
\AgdaBound{ls}\AgdaSymbol{)}\<%
\\
\>[0]\AgdaSymbol{...}\AgdaSpace{}%
\AgdaSymbol{|}\AgdaSpace{}%
\AgdaInductiveConstructor{nothing}\AgdaSpace{}%
\AgdaSymbol{=}\AgdaSpace{}%
\AgdaInductiveConstructor{nothing}\<%
\\
\>[0]\AgdaSymbol{...}\AgdaSpace{}%
\AgdaSymbol{|}\AgdaSpace{}%
\AgdaInductiveConstructor{just}\AgdaSpace{}%
\AgdaSymbol{(}\AgdaBound{v}\AgdaSpace{}%
\AgdaOperator{\AgdaInductiveConstructor{,}}\AgdaSpace{}%
\AgdaBound{v∈ws}\AgdaSpace{}%
\AgdaOperator{\AgdaInductiveConstructor{,}}\AgdaSpace{}%
\AgdaBound{v∈wrotes}\AgdaSymbol{)}\AgdaSpace{}%
\AgdaSymbol{=}\AgdaSpace{}%
\AgdaInductiveConstructor{just}\AgdaSpace{}%
\AgdaSymbol{(}\AgdaInductiveConstructor{WriteWrite}\AgdaSpace{}%
\AgdaBound{v∈ws}\AgdaSpace{}%
\AgdaBound{v∈wrotes}\AgdaSymbol{)}\<%
\\
\\[\AgdaEmptyExtraSkip]%
\>[0]\AgdaFunction{∃ReadWrite}\AgdaSpace{}%
\AgdaSymbol{:}\AgdaSpace{}%
\AgdaSymbol{∀}\AgdaSpace{}%
\AgdaBound{s}\AgdaSpace{}%
\AgdaBound{x}\AgdaSpace{}%
\AgdaSymbol{\{}\AgdaBound{b}\AgdaSymbol{\}}\AgdaSpace{}%
\AgdaBound{ls}\AgdaSpace{}%
\AgdaSymbol{→}\AgdaSpace{}%
\AgdaDatatype{Maybe}\AgdaSpace{}%
\AgdaSymbol{(}\AgdaDatatype{Hazard}\AgdaSpace{}%
\AgdaBound{s}\AgdaSpace{}%
\AgdaBound{x}\AgdaSpace{}%
\AgdaBound{b}\AgdaSpace{}%
\AgdaBound{ls}\AgdaSymbol{)}\<%
\\
\>[0]\AgdaFunction{∃ReadWrite}\AgdaSpace{}%
\AgdaBound{s}\AgdaSpace{}%
\AgdaBound{x}\AgdaSpace{}%
\AgdaBound{ls}\AgdaSpace{}%
\AgdaKeyword{with}\AgdaSpace{}%
\AgdaFunction{∃Intersection}\AgdaSpace{}%
\AgdaSymbol{(}\AgdaFunction{cmdWriteNames}\AgdaSpace{}%
\AgdaBound{x}\AgdaSpace{}%
\AgdaBound{s}\AgdaSymbol{)}\AgdaSpace{}%
\AgdaSymbol{(}\AgdaFunction{filesRead}\AgdaSpace{}%
\AgdaBound{ls}\AgdaSymbol{)}\<%
\\
\>[0]\AgdaSymbol{...}\AgdaSpace{}%
\AgdaSymbol{|}\AgdaSpace{}%
\AgdaInductiveConstructor{nothing}\AgdaSpace{}%
\AgdaSymbol{=}\AgdaSpace{}%
\AgdaInductiveConstructor{nothing}\<%
\\
\>[0]\AgdaSymbol{...}\AgdaSpace{}%
\AgdaSymbol{|}\AgdaSpace{}%
\AgdaInductiveConstructor{just}\AgdaSpace{}%
\AgdaSymbol{(}\AgdaBound{v}\AgdaSpace{}%
\AgdaOperator{\AgdaInductiveConstructor{,}}\AgdaSpace{}%
\AgdaBound{v∈ws}\AgdaSpace{}%
\AgdaOperator{\AgdaInductiveConstructor{,}}\AgdaSpace{}%
\AgdaBound{v∈reads}\AgdaSymbol{)}\AgdaSpace{}%
\AgdaSymbol{=}\AgdaSpace{}%
\AgdaInductiveConstructor{just}\AgdaSpace{}%
\AgdaSymbol{(}\AgdaInductiveConstructor{ReadWrite}\AgdaSpace{}%
\AgdaBound{v∈ws}\AgdaSpace{}%
\AgdaBound{v∈reads}\AgdaSymbol{)}\<%
\end{code}

\newcommand{\checkHazard}{%
\begin{code}%
\>[0]\AgdaFunction{checkHazard}\AgdaSpace{}%
\AgdaSymbol{:}\AgdaSpace{}%
\AgdaSymbol{∀}\AgdaSpace{}%
\AgdaBound{s}\AgdaSpace{}%
\AgdaBound{x}\AgdaSpace{}%
\AgdaSymbol{\{}\AgdaBound{b}\AgdaSymbol{\}}\AgdaSpace{}%
\AgdaBound{ls}\AgdaSpace{}%
\AgdaSymbol{→}\AgdaSpace{}%
\AgdaDatatype{Maybe}\AgdaSpace{}%
\AgdaSymbol{(}\AgdaDatatype{Hazard}\AgdaSpace{}%
\AgdaBound{s}\AgdaSpace{}%
\AgdaBound{x}\AgdaSpace{}%
\AgdaBound{b}\AgdaSpace{}%
\AgdaBound{ls}\AgdaSymbol{)}\<%
\end{code}}
\begin{code}[hide]%
\>[0]\AgdaFunction{checkHazard}\AgdaSpace{}%
\AgdaBound{s}\AgdaSpace{}%
\AgdaBound{x}\AgdaSpace{}%
\AgdaBound{ls}\AgdaSpace{}%
\AgdaKeyword{with}\AgdaSpace{}%
\AgdaFunction{∃WriteWrite}\AgdaSpace{}%
\AgdaBound{s}\AgdaSpace{}%
\AgdaBound{x}\AgdaSpace{}%
\AgdaBound{ls}\<%
\\
\>[0]\AgdaSymbol{...}\AgdaSpace{}%
\AgdaSymbol{|}\AgdaSpace{}%
\AgdaInductiveConstructor{just}\AgdaSpace{}%
\AgdaBound{hz}\AgdaSpace{}%
\AgdaSymbol{=}\AgdaSpace{}%
\AgdaInductiveConstructor{just}\AgdaSpace{}%
\AgdaBound{hz}\<%
\\
\>[0]\AgdaSymbol{...}\AgdaSpace{}%
\AgdaSymbol{|}\AgdaSpace{}%
\AgdaInductiveConstructor{nothing}\AgdaSpace{}%
\AgdaKeyword{with}\AgdaSpace{}%
\AgdaFunction{∃Speculative}\AgdaSpace{}%
\AgdaBound{s}\AgdaSpace{}%
\AgdaBound{x}\AgdaSpace{}%
\AgdaBound{ls}\<%
\\
\>[0]\AgdaSymbol{...}\AgdaSpace{}%
\AgdaSymbol{|}\AgdaSpace{}%
\AgdaInductiveConstructor{just}\AgdaSpace{}%
\AgdaBound{hz}\AgdaSpace{}%
\AgdaSymbol{=}\AgdaSpace{}%
\AgdaInductiveConstructor{just}\AgdaSpace{}%
\AgdaBound{hz}\<%
\\
\>[0]\AgdaSymbol{...}\AgdaSpace{}%
\AgdaSymbol{|}\AgdaSpace{}%
\AgdaInductiveConstructor{nothing}\AgdaSpace{}%
\AgdaSymbol{=}\AgdaSpace{}%
\AgdaFunction{∃ReadWrite}\AgdaSpace{}%
\AgdaBound{s}\AgdaSpace{}%
\AgdaBound{x}\AgdaSpace{}%
\AgdaBound{ls}\<%
\\
\\[\AgdaEmptyExtraSkip]%
\>[0]\AgdaFunction{doRunWError}\AgdaSpace{}%
\AgdaSymbol{:}\AgdaSpace{}%
\AgdaSymbol{∀}\AgdaSpace{}%
\AgdaSymbol{\{}\AgdaBound{b}\AgdaSymbol{\}}\AgdaSpace{}%
\AgdaSymbol{(}\AgdaBound{((s}\AgdaSpace{}%
\AgdaBound{,}\AgdaSpace{}%
\AgdaBound{m)}\AgdaSpace{}%
\AgdaBound{,}\AgdaSpace{}%
\AgdaBound{ls)}\AgdaSpace{}%
\AgdaSymbol{:}\AgdaSpace{}%
\AgdaFunction{State}\AgdaSpace{}%
\AgdaOperator{\AgdaFunction{×}}\AgdaSpace{}%
\AgdaFunction{FileInfo}\AgdaSymbol{)}\AgdaSpace{}%
\AgdaSymbol{→}\AgdaSpace{}%
\AgdaSymbol{(}\AgdaBound{x}\AgdaSpace{}%
\AgdaSymbol{:}\AgdaSpace{}%
\AgdaFunction{Cmd}\AgdaSymbol{)}\AgdaSpace{}%
\AgdaSymbol{→}\AgdaSpace{}%
\AgdaDatatype{Hazard}\AgdaSpace{}%
\AgdaBound{s}\AgdaSpace{}%
\AgdaBound{x}\AgdaSpace{}%
\AgdaBound{b}\AgdaSpace{}%
\AgdaBound{ls}\AgdaSpace{}%
\AgdaOperator{\AgdaDatatype{⊎}}\AgdaSpace{}%
\AgdaFunction{State}\AgdaSpace{}%
\AgdaOperator{\AgdaFunction{×}}\AgdaSpace{}%
\AgdaFunction{FileInfo}\<%
\\
\>[0]\AgdaFunction{doRunWError}\AgdaSpace{}%
\AgdaSymbol{((}\AgdaBound{s}\AgdaSpace{}%
\AgdaOperator{\AgdaInductiveConstructor{,}}\AgdaSpace{}%
\AgdaBound{mm}\AgdaSymbol{)}\AgdaSpace{}%
\AgdaOperator{\AgdaInductiveConstructor{,}}\AgdaSpace{}%
\AgdaBound{ls}\AgdaSymbol{)}\AgdaSpace{}%
\AgdaBound{x}\AgdaSpace{}%
\AgdaKeyword{with}\AgdaSpace{}%
\AgdaFunction{checkHazard}\AgdaSpace{}%
\AgdaBound{s}\AgdaSpace{}%
\AgdaBound{x}\AgdaSpace{}%
\AgdaBound{ls}\<%
\\
\>[0]\AgdaSymbol{...}\AgdaSpace{}%
\AgdaSymbol{|}\AgdaSpace{}%
\AgdaInductiveConstructor{just}\AgdaSpace{}%
\AgdaBound{hz}\AgdaSpace{}%
\AgdaSymbol{=}\AgdaSpace{}%
\AgdaInductiveConstructor{inj₁}\AgdaSpace{}%
\AgdaBound{hz}\<%
\\
\>[0]\AgdaSymbol{...}\AgdaSpace{}%
\AgdaSymbol{|}\AgdaSpace{}%
\AgdaInductiveConstructor{nothing}\AgdaSpace{}%
\AgdaSymbol{=}\AgdaSpace{}%
\AgdaInductiveConstructor{inj₂}\AgdaSpace{}%
\AgdaSymbol{(}\AgdaFunction{doRunR}\AgdaSpace{}%
\AgdaSymbol{(}\AgdaBound{s}\AgdaSpace{}%
\AgdaOperator{\AgdaInductiveConstructor{,}}\AgdaSpace{}%
\AgdaBound{mm}\AgdaSymbol{)}\AgdaSpace{}%
\AgdaBound{x}\AgdaSpace{}%
\AgdaOperator{\AgdaInductiveConstructor{,}}\AgdaSpace{}%
\AgdaFunction{rec}\AgdaSpace{}%
\AgdaBound{s}\AgdaSpace{}%
\AgdaBound{x}\AgdaSpace{}%
\AgdaBound{ls}\AgdaSymbol{)}\<%
\end{code}

\newcommand{\runR}{%
\begin{code}%
\>[0]\AgdaFunction{runR}%
\>[1100I]\AgdaSymbol{:}\AgdaSpace{}%
\AgdaFunction{Cmd}\AgdaSpace{}%
\AgdaSymbol{→}\AgdaSpace{}%
\AgdaSymbol{(}\AgdaFunction{FileSystem}\AgdaSpace{}%
\AgdaOperator{\AgdaFunction{×}}\AgdaSpace{}%
\AgdaFunction{Memory}\AgdaSymbol{)}\<%
\\
\>[.][@{}l@{}]\<[1100I]%
\>[5]\AgdaSymbol{→}\AgdaSpace{}%
\AgdaSymbol{(}\AgdaFunction{FileSystem}\AgdaSpace{}%
\AgdaOperator{\AgdaFunction{×}}\AgdaSpace{}%
\AgdaFunction{Memory}\AgdaSymbol{)}\<%
\\
\>[0]\AgdaFunction{runR}\AgdaSpace{}%
\AgdaBound{cmd}\AgdaSpace{}%
\AgdaBound{st}%
\>[1111I]\AgdaSymbol{=}\AgdaSpace{}%
\AgdaOperator{\AgdaFunction{if}}\AgdaSpace{}%
\AgdaSymbol{(}\AgdaFunction{run?}\AgdaSpace{}%
\AgdaBound{cmd}\AgdaSpace{}%
\AgdaBound{st}\AgdaSymbol{)}\<%
\\
\>[1111I][@{}l@{\AgdaIndent{0}}]%
\>[13]\AgdaOperator{\AgdaFunction{then}}\AgdaSpace{}%
\AgdaFunction{doRunR}\AgdaSpace{}%
\AgdaBound{st}\AgdaSpace{}%
\AgdaBound{cmd}\<%
\\
\>[13]\AgdaOperator{\AgdaFunction{else}}\AgdaSpace{}%
\AgdaBound{st}\<%
\end{code}}

\begin{code}[hide]%
\>[0]\AgdaFunction{g₂}\AgdaSpace{}%
\AgdaSymbol{:}\AgdaSpace{}%
\AgdaSymbol{∀}\AgdaSpace{}%
\AgdaSymbol{\{}\AgdaBound{x}\AgdaSymbol{\}}\AgdaSpace{}%
\AgdaBound{xs}\AgdaSpace{}%
\AgdaSymbol{→}\AgdaSpace{}%
\AgdaBound{x}\AgdaSpace{}%
\AgdaOperator{\AgdaFunction{∉}}\AgdaSpace{}%
\AgdaBound{xs}\AgdaSpace{}%
\AgdaSymbol{→}\AgdaSpace{}%
\AgdaDatatype{All}\AgdaSpace{}%
\AgdaSymbol{(λ}\AgdaSpace{}%
\AgdaBound{y}\AgdaSpace{}%
\AgdaSymbol{→}\AgdaSpace{}%
\AgdaOperator{\AgdaFunction{¬}}\AgdaSpace{}%
\AgdaBound{x}\AgdaSpace{}%
\AgdaOperator{\AgdaDatatype{≡}}\AgdaSpace{}%
\AgdaBound{y}\AgdaSymbol{)}\AgdaSpace{}%
\AgdaBound{xs}\<%
\\
\>[0]\AgdaFunction{g₂}\AgdaSpace{}%
\AgdaInductiveConstructor{[]}\AgdaSpace{}%
\AgdaBound{x∉xs}\AgdaSpace{}%
\AgdaSymbol{=}\AgdaSpace{}%
\AgdaInductiveConstructor{All.[]}\<%
\\
\>[0]\AgdaFunction{g₂}\AgdaSpace{}%
\AgdaSymbol{(}\AgdaBound{x}\AgdaSpace{}%
\AgdaOperator{\AgdaInductiveConstructor{∷}}\AgdaSpace{}%
\AgdaBound{xs}\AgdaSymbol{)}\AgdaSpace{}%
\AgdaBound{x∉xs}\AgdaSpace{}%
\AgdaSymbol{=}\AgdaSpace{}%
\AgdaSymbol{(λ}\AgdaSpace{}%
\AgdaBound{x₃}\AgdaSpace{}%
\AgdaSymbol{→}\AgdaSpace{}%
\AgdaBound{x∉xs}\AgdaSpace{}%
\AgdaSymbol{(}\AgdaInductiveConstructor{here}\AgdaSpace{}%
\AgdaBound{x₃}\AgdaSymbol{))}\AgdaSpace{}%
\AgdaOperator{\AgdaInductiveConstructor{All.∷}}\AgdaSpace{}%
\AgdaSymbol{(}\AgdaFunction{g₂}\AgdaSpace{}%
\AgdaBound{xs}\AgdaSpace{}%
\AgdaSymbol{λ}\AgdaSpace{}%
\AgdaBound{x₃}\AgdaSpace{}%
\AgdaSymbol{→}\AgdaSpace{}%
\AgdaBound{x∉xs}\AgdaSpace{}%
\AgdaSymbol{(}\AgdaInductiveConstructor{there}\AgdaSpace{}%
\AgdaBound{x₃}\AgdaSymbol{))}\<%
\end{code}

\newcommand{\runWError}{%
\begin{code}%
\>[0]\AgdaFunction{runWError}\AgdaSpace{}%
\AgdaSymbol{:}\AgdaSpace{}%
\AgdaSymbol{∀}\AgdaSpace{}%
\AgdaSymbol{\{}\AgdaBound{b}\AgdaSymbol{\}}\AgdaSpace{}%
\AgdaBound{x}\AgdaSpace{}%
\AgdaBound{s}\AgdaSpace{}%
\AgdaBound{m}\AgdaSpace{}%
\AgdaBound{ls}\<%
\\
\>[0][@{}l@{\AgdaIndent{0}}]%
\>[2]\AgdaSymbol{→}\AgdaSpace{}%
\AgdaDatatype{Hazard}\AgdaSpace{}%
\AgdaBound{s}\AgdaSpace{}%
\AgdaBound{x}\AgdaSpace{}%
\AgdaBound{b}\AgdaSpace{}%
\AgdaBound{ls}\AgdaSpace{}%
\AgdaOperator{\AgdaDatatype{⊎}}\AgdaSpace{}%
\AgdaSymbol{(}\AgdaFunction{FileSystem}\AgdaSpace{}%
\AgdaOperator{\AgdaFunction{×}}\AgdaSpace{}%
\AgdaFunction{Memory}\AgdaSymbol{)}\AgdaSpace{}%
\AgdaOperator{\AgdaFunction{×}}\AgdaSpace{}%
\AgdaFunction{FileInfo}\<%
\\
\>[0]\AgdaFunction{runWError}\AgdaSpace{}%
\AgdaBound{x}\AgdaSpace{}%
\AgdaBound{s}\AgdaSpace{}%
\AgdaBound{m}\AgdaSpace{}%
\AgdaBound{ls}\AgdaSpace{}%
\AgdaKeyword{with}\AgdaSpace{}%
\AgdaSymbol{(}\AgdaFunction{run?}\AgdaSpace{}%
\AgdaBound{x}\AgdaSpace{}%
\AgdaSymbol{(}\AgdaBound{s}\AgdaSpace{}%
\AgdaOperator{\AgdaInductiveConstructor{,}}\AgdaSpace{}%
\AgdaBound{m}\AgdaSymbol{))}\<%
\\
\>[0]\AgdaSymbol{...}\AgdaSpace{}%
\AgdaSymbol{|}\AgdaSpace{}%
\AgdaInductiveConstructor{false}\AgdaSpace{}%
\AgdaSymbol{=}\AgdaSpace{}%
\AgdaInductiveConstructor{inj₂}\AgdaSpace{}%
\AgdaSymbol{((}\AgdaBound{s}\AgdaSpace{}%
\AgdaOperator{\AgdaInductiveConstructor{,}}\AgdaSpace{}%
\AgdaBound{m}\AgdaSymbol{)}\AgdaSpace{}%
\AgdaOperator{\AgdaInductiveConstructor{,}}\AgdaSpace{}%
\AgdaBound{ls}\AgdaSymbol{)}\<%
\\
\>[0]\AgdaSymbol{...}\AgdaSpace{}%
\AgdaSymbol{|}\AgdaSpace{}%
\AgdaInductiveConstructor{true}\AgdaSpace{}%
\AgdaKeyword{with}\AgdaSpace{}%
\AgdaFunction{checkHazard}\AgdaSpace{}%
\AgdaBound{s}\AgdaSpace{}%
\AgdaBound{x}\AgdaSpace{}%
\AgdaBound{ls}\<%
\\
\>[0]\AgdaSymbol{...}\AgdaSpace{}%
\AgdaSymbol{|}\AgdaSpace{}%
\AgdaInductiveConstructor{just}\AgdaSpace{}%
\AgdaBound{hz}\AgdaSpace{}%
\AgdaSymbol{=}\AgdaSpace{}%
\AgdaInductiveConstructor{inj₁}\AgdaSpace{}%
\AgdaBound{hz}\<%
\\
\>[0]\AgdaSymbol{...}\AgdaSpace{}%
\AgdaSymbol{|}\AgdaSpace{}%
\AgdaInductiveConstructor{nothing}\AgdaSpace{}%
\AgdaSymbol{=}\AgdaSpace{}%
\AgdaInductiveConstructor{inj₂}\AgdaSpace{}%
\AgdaSymbol{(}\AgdaFunction{doRunR}\AgdaSpace{}%
\AgdaSymbol{(}\AgdaBound{s}\AgdaSpace{}%
\AgdaOperator{\AgdaInductiveConstructor{,}}\AgdaSpace{}%
\AgdaBound{m}\AgdaSymbol{)}\AgdaSpace{}%
\AgdaBound{x}\AgdaSpace{}%
\AgdaOperator{\AgdaInductiveConstructor{,}}\AgdaSpace{}%
\AgdaFunction{rec}\AgdaSpace{}%
\AgdaBound{s}\AgdaSpace{}%
\AgdaBound{x}\AgdaSpace{}%
\AgdaBound{ls}\AgdaSymbol{)}\<%
\end{code}}

\newcommand{\Rexec}{%
\begin{code}%
\>[0]\AgdaFunction{rattle-unchecked}%
\>[1226I]\AgdaSymbol{:}\AgdaSpace{}%
\AgdaFunction{Build}\AgdaSpace{}%
\AgdaSymbol{→}\AgdaSpace{}%
\AgdaSymbol{(}\AgdaFunction{FileSystem}\AgdaSpace{}%
\AgdaOperator{\AgdaFunction{×}}\AgdaSpace{}%
\AgdaFunction{Memory}\AgdaSymbol{)}\<%
\\
\>[.][@{}l@{}]\<[1226I]%
\>[17]\AgdaSymbol{→}\AgdaSpace{}%
\AgdaSymbol{(}\AgdaFunction{FileSystem}\AgdaSpace{}%
\AgdaOperator{\AgdaFunction{×}}\AgdaSpace{}%
\AgdaFunction{Memory}\AgdaSymbol{)}\<%
\\
\>[0]\AgdaFunction{rattle-unchecked}\AgdaSpace{}%
\AgdaInductiveConstructor{[]}\AgdaSpace{}%
\AgdaBound{st}\AgdaSpace{}%
\AgdaSymbol{=}\AgdaSpace{}%
\AgdaBound{st}\<%
\\
\>[0]\AgdaFunction{rattle-unchecked}\AgdaSpace{}%
\AgdaSymbol{(}\AgdaBound{x}\AgdaSpace{}%
\AgdaOperator{\AgdaInductiveConstructor{∷}}\AgdaSpace{}%
\AgdaBound{b}\AgdaSymbol{)}\AgdaSpace{}%
\AgdaBound{st}\AgdaSpace{}%
\AgdaSymbol{=}\AgdaSpace{}%
\AgdaFunction{rattle-unchecked}\AgdaSpace{}%
\AgdaBound{b}\AgdaSpace{}%
\AgdaSymbol{(}\AgdaFunction{runR}\AgdaSpace{}%
\AgdaBound{x}\AgdaSpace{}%
\AgdaBound{st}\AgdaSymbol{)}\<%
\end{code}}

\newcommand{\rattle}{%
\begin{code}%
\>[0]\AgdaFunction{rattle}%
\>[1249I]\AgdaSymbol{:}\AgdaSpace{}%
\AgdaSymbol{(}\AgdaBound{br}\AgdaSpace{}%
\AgdaBound{bs}\AgdaSpace{}%
\AgdaSymbol{:}\AgdaSpace{}%
\AgdaFunction{Build}\AgdaSymbol{)}\AgdaSpace{}%
\AgdaSymbol{→}\AgdaSpace{}%
\AgdaSymbol{(}\AgdaFunction{FileSystem}\AgdaSpace{}%
\AgdaOperator{\AgdaFunction{×}}\AgdaSpace{}%
\AgdaFunction{Memory}\AgdaSymbol{)}\AgdaSpace{}%
\AgdaOperator{\AgdaFunction{×}}\AgdaSpace{}%
\AgdaFunction{FileInfo}\<%
\\
\>[.][@{}l@{}]\<[1249I]%
\>[7]\AgdaSymbol{→}\AgdaSpace{}%
\AgdaFunction{∃Hazard}\AgdaSpace{}%
\AgdaBound{bs}\AgdaSpace{}%
\AgdaOperator{\AgdaDatatype{⊎}}\AgdaSpace{}%
\AgdaSymbol{(}\AgdaFunction{FileSystem}\AgdaSpace{}%
\AgdaOperator{\AgdaFunction{×}}\AgdaSpace{}%
\AgdaFunction{Memory}\AgdaSymbol{)}\AgdaSpace{}%
\AgdaOperator{\AgdaFunction{×}}\AgdaSpace{}%
\AgdaFunction{FileInfo}\<%
\\
\>[0]\AgdaFunction{rattle}\AgdaSpace{}%
\AgdaInductiveConstructor{[]}\AgdaSpace{}%
\AgdaBound{bs}\AgdaSpace{}%
\AgdaBound{st}\AgdaSpace{}%
\AgdaSymbol{=}\AgdaSpace{}%
\AgdaInductiveConstructor{inj₂}\AgdaSpace{}%
\AgdaBound{st}\<%
\\
\>[0]\AgdaFunction{rattle}\AgdaSpace{}%
\AgdaSymbol{(}\AgdaBound{x}\AgdaSpace{}%
\AgdaOperator{\AgdaInductiveConstructor{∷}}\AgdaSpace{}%
\AgdaBound{b₁}\AgdaSymbol{)}\AgdaSpace{}%
\AgdaBound{bs}\AgdaSpace{}%
\AgdaBound{st}\AgdaSymbol{@((}\AgdaBound{s}\AgdaSpace{}%
\AgdaOperator{\AgdaInductiveConstructor{,}}\AgdaSpace{}%
\AgdaBound{m}\AgdaSymbol{)}\AgdaSpace{}%
\AgdaOperator{\AgdaInductiveConstructor{,}}\AgdaSpace{}%
\AgdaBound{ls}\AgdaSymbol{)}\AgdaSpace{}%
\AgdaKeyword{with}\AgdaSpace{}%
\AgdaFunction{runWError}\AgdaSpace{}%
\AgdaBound{x}\AgdaSpace{}%
\AgdaBound{s}\AgdaSpace{}%
\AgdaBound{m}\AgdaSpace{}%
\AgdaBound{ls}\<%
\\
\>[0]\AgdaSymbol{...}\AgdaSpace{}%
\AgdaSymbol{|}\AgdaSpace{}%
\AgdaInductiveConstructor{inj₁}\AgdaSpace{}%
\AgdaBound{hz}\AgdaSpace{}%
\AgdaSymbol{=}\AgdaSpace{}%
\AgdaInductiveConstructor{inj₁}\AgdaSpace{}%
\AgdaSymbol{(}\AgdaField{proj₁}\AgdaSpace{}%
\AgdaSymbol{(}\AgdaField{proj₁}\AgdaSpace{}%
\AgdaBound{st}\AgdaSymbol{)}\AgdaSpace{}%
\AgdaOperator{\AgdaInductiveConstructor{,}}\AgdaSpace{}%
\AgdaBound{x}\AgdaSpace{}%
\AgdaOperator{\AgdaInductiveConstructor{,}}\AgdaSpace{}%
\AgdaField{proj₂}\AgdaSpace{}%
\AgdaBound{st}\AgdaSpace{}%
\AgdaOperator{\AgdaInductiveConstructor{,}}\AgdaSpace{}%
\AgdaBound{hz}\AgdaSymbol{)}\<%
\\
\>[0]\AgdaSymbol{...}\AgdaSpace{}%
\AgdaSymbol{|}\AgdaSpace{}%
\AgdaInductiveConstructor{inj₂}\AgdaSpace{}%
\AgdaSymbol{(}\AgdaBound{st₁}\AgdaSpace{}%
\AgdaOperator{\AgdaInductiveConstructor{,}}\AgdaSpace{}%
\AgdaBound{ls₁}\AgdaSymbol{)}\AgdaSpace{}%
\AgdaSymbol{=}\AgdaSpace{}%
\AgdaFunction{rattle}\AgdaSpace{}%
\AgdaBound{b₁}\AgdaSpace{}%
\AgdaBound{bs}\AgdaSpace{}%
\AgdaSymbol{(}\AgdaBound{st₁}\AgdaSpace{}%
\AgdaOperator{\AgdaInductiveConstructor{,}}\AgdaSpace{}%
\AgdaBound{ls₁}\AgdaSymbol{)}\<%
\end{code}}

\begin{code}[hide]%
\>[0]\AgdaKeyword{open}\AgdaSpace{}%
\AgdaKeyword{import}\AgdaSpace{}%
\AgdaModule{Functional.State}\AgdaSpace{}%
\AgdaKeyword{using}\AgdaSpace{}%
\AgdaSymbol{(}\AgdaFunction{Oracle}\AgdaSpace{}%
\AgdaSymbol{;}\AgdaSpace{}%
\AgdaFunction{FileSystem}\AgdaSpace{}%
\AgdaSymbol{;}\AgdaSpace{}%
\AgdaFunction{Memory}\AgdaSpace{}%
\AgdaSymbol{;}\AgdaSpace{}%
\AgdaFunction{Cmd}\AgdaSpace{}%
\AgdaSymbol{;}\AgdaSpace{}%
\AgdaFunction{extend}\AgdaSpace{}%
\AgdaSymbol{;}\AgdaSpace{}%
\AgdaFunction{State}\AgdaSpace{}%
\AgdaSymbol{;}\AgdaSpace{}%
\AgdaFunction{save}\AgdaSpace{}%
\AgdaSymbol{;}\AgdaSpace{}%
\AgdaFunction{reads}\AgdaSymbol{)}\<%
\\
\\[\AgdaEmptyExtraSkip]%
\>[0]\AgdaKeyword{module}\AgdaSpace{}%
\AgdaModule{Functional.Rattle.Properties}\AgdaSpace{}%
\AgdaSymbol{(}\AgdaBound{oracle}\AgdaSpace{}%
\AgdaSymbol{:}\AgdaSpace{}%
\AgdaFunction{Oracle}\AgdaSymbol{)}\AgdaSpace{}%
\AgdaKeyword{where}\<%
\\
\\[\AgdaEmptyExtraSkip]%
\>[0]\AgdaKeyword{open}\AgdaSpace{}%
\AgdaKeyword{import}\AgdaSpace{}%
\AgdaModule{Agda.Builtin.Equality}\<%
\\
\\[\AgdaEmptyExtraSkip]%
\>[0]\AgdaKeyword{open}\AgdaSpace{}%
\AgdaKeyword{import}\AgdaSpace{}%
\AgdaModule{Functional.State.Helpers}\AgdaSpace{}%
\AgdaSymbol{(}\AgdaBound{oracle}\AgdaSymbol{)}\AgdaSpace{}%
\AgdaSymbol{as}\AgdaSpace{}%
\AgdaModule{St}\AgdaSpace{}%
\AgdaKeyword{using}\AgdaSpace{}%
\AgdaSymbol{(}\AgdaPostulate{cmdReadNames}\AgdaSpace{}%
\AgdaSymbol{;}\AgdaSpace{}%
\AgdaPostulate{cmdWriteNames}\AgdaSpace{}%
\AgdaSymbol{;}\AgdaSpace{}%
\AgdaPostulate{cmdWrites}\AgdaSymbol{)}\<%
\\
\>[0]\AgdaKeyword{open}\AgdaSpace{}%
\AgdaKeyword{import}\AgdaSpace{}%
\AgdaModule{Functional.State.Properties}\AgdaSpace{}%
\AgdaSymbol{(}\AgdaBound{oracle}\AgdaSymbol{)}\AgdaSpace{}%
\AgdaKeyword{using}\AgdaSpace{}%
\AgdaSymbol{(}lemma3 \AgdaSymbol{;} lemma4 \AgdaSymbol{;} run-≡\AgdaSymbol{)}\<%
\\
\>[0]\AgdaKeyword{open}\AgdaSpace{}%
\AgdaKeyword{import}\AgdaSpace{}%
\AgdaModule{Common.List.Properties}\AgdaSpace{}%
\AgdaKeyword{using}\AgdaSpace{}%
\AgdaSymbol{(}\AgdaFunction{∈-resp-≡}\AgdaSpace{}%
\AgdaSymbol{;}\AgdaSpace{}%
\AgdaFunction{l11}\AgdaSpace{}%
\AgdaSymbol{;}\AgdaSpace{}%
\AgdaOperator{\AgdaFunction{\AgdaUnderscore{}before\AgdaUnderscore{}∈\AgdaUnderscore{}}}\AgdaSpace{}%
\AgdaSymbol{;}\AgdaSpace{}%
\AgdaFunction{l4}\AgdaSpace{}%
\AgdaSymbol{;}\AgdaSpace{}%
\AgdaFunction{concatMap-++-commute}\AgdaSymbol{)}\<%
\\
\>[0]\AgdaKeyword{open}\AgdaSpace{}%
\AgdaKeyword{import}\AgdaSpace{}%
\AgdaModule{Functional.Forward.Properties}\AgdaSpace{}%
\AgdaSymbol{(}\AgdaBound{oracle}\AgdaSymbol{)}\AgdaSpace{}%
\AgdaKeyword{using}\AgdaSpace{}%
\AgdaSymbol{(}cmdReadWrites \AgdaSymbol{;} IdempotentState \AgdaSymbol{;} IdempotentCmd \AgdaSymbol{;} cmdIdempotent\AgdaSymbol{)}\<%
\\
\>[0]\AgdaKeyword{open}\AgdaSpace{}%
\AgdaKeyword{import}\AgdaSpace{}%
\AgdaModule{Functional.Forward.Exec}\AgdaSpace{}%
\AgdaSymbol{(}\AgdaBound{oracle}\AgdaSymbol{)}\AgdaSpace{}%
\AgdaKeyword{using}\AgdaSpace{}%
\AgdaSymbol{(}get \AgdaSymbol{;} maybeAll \AgdaSymbol{;} run?\AgdaSymbol{)}\<%
\\
\>[0]\AgdaKeyword{open}\AgdaSpace{}%
\AgdaKeyword{import}\AgdaSpace{}%
\AgdaModule{Functional.File}\AgdaSpace{}%
\AgdaKeyword{using}\AgdaSpace{}%
\AgdaSymbol{(}\AgdaFunction{FileName}\AgdaSpace{}%
\AgdaSymbol{;}\AgdaSpace{}%
\AgdaFunction{FileContent}\AgdaSymbol{)}\<%
\\
\>[0]\AgdaKeyword{open}\AgdaSpace{}%
\AgdaKeyword{import}\AgdaSpace{}%
\AgdaModule{Data.Bool}\AgdaSpace{}%
\AgdaKeyword{using}\AgdaSpace{}%
\AgdaSymbol{(}\AgdaInductiveConstructor{true}\AgdaSpace{}%
\AgdaSymbol{;}\AgdaSpace{}%
\AgdaInductiveConstructor{false}\AgdaSpace{}%
\AgdaSymbol{;}\AgdaSpace{}%
\AgdaOperator{\AgdaFunction{if\AgdaUnderscore{}then\AgdaUnderscore{}else\AgdaUnderscore{}}}\AgdaSymbol{)}\<%
\\
\\[\AgdaEmptyExtraSkip]%
\>[0]\AgdaKeyword{open}\AgdaSpace{}%
\AgdaKeyword{import}\AgdaSpace{}%
\AgdaModule{Data.List.Relation.Binary.Disjoint.Propositional}\AgdaSpace{}%
\AgdaKeyword{using}\AgdaSpace{}%
\AgdaSymbol{(}\AgdaFunction{Disjoint}\AgdaSymbol{)}\<%
\\
\>[0]\AgdaKeyword{open}\AgdaSpace{}%
\AgdaKeyword{import}\AgdaSpace{}%
\AgdaModule{Data.Product}\AgdaSpace{}%
\AgdaKeyword{using}\AgdaSpace{}%
\AgdaSymbol{(}\AgdaField{proj₁}\AgdaSpace{}%
\AgdaSymbol{;}\AgdaSpace{}%
\AgdaField{proj₂}\AgdaSpace{}%
\AgdaSymbol{;}\AgdaSpace{}%
\AgdaOperator{\AgdaInductiveConstructor{\AgdaUnderscore{},\AgdaUnderscore{}}}\AgdaSpace{}%
\AgdaSymbol{;}\AgdaSpace{}%
\AgdaOperator{\AgdaFunction{\AgdaUnderscore{}×\AgdaUnderscore{}}}\AgdaSpace{}%
\AgdaSymbol{;}\AgdaSpace{}%
\AgdaFunction{∃-syntax}\AgdaSymbol{)}\<%
\\
\>[0]\AgdaKeyword{open}\AgdaSpace{}%
\AgdaKeyword{import}\AgdaSpace{}%
\AgdaModule{Data.Product.Properties}\AgdaSpace{}%
\AgdaKeyword{using}\AgdaSpace{}%
\AgdaSymbol{(}\AgdaFunction{,-injectiveʳ}\AgdaSpace{}%
\AgdaSymbol{;}\AgdaSpace{}%
\AgdaFunction{,-injectiveˡ}\AgdaSymbol{)}\<%
\\
\>[0]\AgdaKeyword{open}\AgdaSpace{}%
\AgdaKeyword{import}\AgdaSpace{}%
\AgdaModule{Data.List.Relation.Unary.All}\AgdaSpace{}%
\AgdaKeyword{using}\AgdaSpace{}%
\AgdaSymbol{(}\AgdaDatatype{All}\AgdaSpace{}%
\AgdaSymbol{;}\AgdaSpace{}%
\AgdaFunction{lookup}\AgdaSymbol{)}\<%
\\
\>[0]\AgdaKeyword{open}\AgdaSpace{}%
\AgdaKeyword{import}\AgdaSpace{}%
\AgdaModule{Data.List}\AgdaSpace{}%
\AgdaKeyword{using}\AgdaSpace{}%
\AgdaSymbol{(}\AgdaDatatype{List}\AgdaSpace{}%
\AgdaSymbol{;}\AgdaSpace{}%
\AgdaFunction{map}\AgdaSpace{}%
\AgdaSymbol{;}\AgdaSpace{}%
\AgdaOperator{\AgdaFunction{\AgdaUnderscore{}++\AgdaUnderscore{}}}\AgdaSpace{}%
\AgdaSymbol{;}\AgdaSpace{}%
\AgdaFunction{foldr}\AgdaSpace{}%
\AgdaSymbol{;}\AgdaSpace{}%
\AgdaOperator{\AgdaInductiveConstructor{\AgdaUnderscore{}∷\AgdaUnderscore{}}}\AgdaSpace{}%
\AgdaSymbol{;}\AgdaSpace{}%
\AgdaInductiveConstructor{[]}\AgdaSpace{}%
\AgdaSymbol{;}\AgdaSpace{}%
\AgdaFunction{concatMap}\AgdaSpace{}%
\AgdaSymbol{;}\AgdaSpace{}%
\AgdaFunction{reverse}\AgdaSpace{}%
\AgdaSymbol{;}\AgdaSpace{}%
\AgdaOperator{\AgdaFunction{\AgdaUnderscore{}∷ʳ\AgdaUnderscore{}}}\AgdaSpace{}%
\AgdaSymbol{;}\AgdaSpace{}%
\AgdaFunction{length}\AgdaSymbol{)}\<%
\\
\>[0]\AgdaKeyword{open}\AgdaSpace{}%
\AgdaKeyword{import}\AgdaSpace{}%
\AgdaModule{Data.List.Properties}\AgdaSpace{}%
\AgdaKeyword{using}\AgdaSpace{}%
\AgdaSymbol{(}\AgdaFunction{∷-injective}\AgdaSpace{}%
\AgdaSymbol{;}\AgdaSpace{}%
\AgdaFunction{unfold-reverse}\AgdaSpace{}%
\AgdaSymbol{;}\AgdaSpace{}%
\AgdaFunction{++-assoc}\AgdaSpace{}%
\AgdaSymbol{;}\AgdaSpace{}%
\AgdaFunction{reverse-++-commute}\AgdaSpace{}%
\AgdaSymbol{;}\AgdaSpace{}%
\AgdaFunction{reverse-involutive}\AgdaSpace{}%
\AgdaSymbol{;}\AgdaSpace{}%
\AgdaFunction{∷-injectiveˡ}\AgdaSpace{}%
\AgdaSymbol{;}\AgdaSpace{}%
\AgdaFunction{∷-injectiveʳ}\AgdaSymbol{)}\<%
\\
\>[0]\AgdaKeyword{open}\AgdaSpace{}%
\AgdaKeyword{import}\AgdaSpace{}%
\AgdaModule{Data.Maybe}\AgdaSpace{}%
\AgdaKeyword{using}\AgdaSpace{}%
\AgdaSymbol{(}\AgdaDatatype{Maybe}\AgdaSpace{}%
\AgdaSymbol{;}\AgdaSpace{}%
\AgdaInductiveConstructor{nothing}\AgdaSpace{}%
\AgdaSymbol{;}\AgdaSpace{}%
\AgdaInductiveConstructor{just}\AgdaSymbol{)}\<%
\\
\>[0]\AgdaKeyword{open}\AgdaSpace{}%
\AgdaKeyword{import}\AgdaSpace{}%
\AgdaModule{Relation.Binary.PropositionalEquality}\AgdaSpace{}%
\AgdaKeyword{using}\AgdaSpace{}%
\AgdaSymbol{(}\AgdaFunction{trans}\AgdaSpace{}%
\AgdaSymbol{;}\AgdaSpace{}%
\AgdaFunction{cong}\AgdaSpace{}%
\AgdaSymbol{;}\AgdaSpace{}%
\AgdaFunction{cong-app}\AgdaSpace{}%
\AgdaSymbol{;}\AgdaSpace{}%
\AgdaFunction{sym}\AgdaSpace{}%
\AgdaSymbol{;}\AgdaSpace{}%
\AgdaFunction{decSetoid}\AgdaSpace{}%
\AgdaSymbol{;}\AgdaSpace{}%
\AgdaFunction{cong₂}\AgdaSpace{}%
\AgdaSymbol{;}\AgdaSpace{}%
\AgdaFunction{subst}\AgdaSpace{}%
\AgdaSymbol{;}\AgdaSpace{}%
\AgdaFunction{subst₂}\AgdaSpace{}%
\AgdaSymbol{;}\AgdaSpace{}%
\AgdaFunction{inspect}\AgdaSpace{}%
\AgdaSymbol{;}\AgdaSpace{}%
\AgdaOperator{\AgdaInductiveConstructor{[\AgdaUnderscore{}]}}\AgdaSymbol{)}\<%
\\
\>[0]\AgdaKeyword{open}\AgdaSpace{}%
\AgdaKeyword{import}\AgdaSpace{}%
\AgdaModule{Data.List.Membership.Propositional.Properties}\AgdaSpace{}%
\AgdaKeyword{using}\AgdaSpace{}%
\AgdaSymbol{(}\AgdaFunction{∈-++⁺ˡ}\AgdaSpace{}%
\AgdaSymbol{;}\AgdaSpace{}%
\AgdaFunction{∈-++⁺ʳ}\AgdaSpace{}%
\AgdaSymbol{;}\AgdaSpace{}%
\AgdaFunction{∈-++⁻}\AgdaSpace{}%
\AgdaSymbol{;}\AgdaSpace{}%
\AgdaFunction{∈-concat⁻}\AgdaSpace{}%
\AgdaSymbol{;}\AgdaSpace{}%
\AgdaFunction{∈-∃++}\AgdaSymbol{)}\<%
\\
\>[0]\AgdaKeyword{open}\AgdaSpace{}%
\AgdaKeyword{import}\AgdaSpace{}%
\AgdaModule{Function.Base}\AgdaSpace{}%
\AgdaKeyword{using}\AgdaSpace{}%
\AgdaSymbol{(}\AgdaOperator{\AgdaFunction{\AgdaUnderscore{}∘\AgdaUnderscore{}}}\AgdaSymbol{)}\<%
\\
\>[0]\AgdaKeyword{open}\AgdaSpace{}%
\AgdaKeyword{import}\AgdaSpace{}%
\AgdaModule{Data.String.Properties}\AgdaSpace{}%
\AgdaKeyword{using}\AgdaSpace{}%
\AgdaSymbol{(}\AgdaOperator{\AgdaFunction{\AgdaUnderscore{}≟\AgdaUnderscore{}}}\AgdaSpace{}%
\AgdaSymbol{;}\AgdaSpace{}%
\AgdaOperator{\AgdaFunction{\AgdaUnderscore{}==\AgdaUnderscore{}}}\AgdaSymbol{)}\<%
\\
\>[0]\AgdaKeyword{open}\AgdaSpace{}%
\AgdaKeyword{import}\AgdaSpace{}%
\AgdaModule{Data.List.Membership.DecSetoid}\AgdaSpace{}%
\AgdaSymbol{(}\AgdaFunction{decSetoid}\AgdaSpace{}%
\AgdaOperator{\AgdaFunction{\AgdaUnderscore{}≟\AgdaUnderscore{}}}\AgdaSymbol{)}\AgdaSpace{}%
\AgdaKeyword{using}\AgdaSpace{}%
\AgdaSymbol{(}\AgdaUnderscore{}∈\AgdaUnderscore{} \AgdaSymbol{;} \AgdaUnderscore{}∈?\AgdaUnderscore{} \AgdaSymbol{;} \AgdaUnderscore{}∉\AgdaUnderscore{}\AgdaSymbol{)}\<%
\\
\>[0]\AgdaKeyword{open}\AgdaSpace{}%
\AgdaKeyword{import}\AgdaSpace{}%
\AgdaModule{Relation.Nullary}\AgdaSpace{}%
\AgdaKeyword{using}\AgdaSpace{}%
\AgdaSymbol{(}\AgdaInductiveConstructor{yes}\AgdaSpace{}%
\AgdaSymbol{;}\AgdaSpace{}%
\AgdaInductiveConstructor{no}\AgdaSpace{}%
\AgdaSymbol{;}\AgdaSpace{}%
\AgdaOperator{\AgdaFunction{¬\AgdaUnderscore{}}}\AgdaSymbol{)}\<%
\\
\>[0]\AgdaKeyword{open}\AgdaSpace{}%
\AgdaKeyword{import}\AgdaSpace{}%
\AgdaModule{Relation.Nullary.Negation}\AgdaSpace{}%
\AgdaKeyword{using}\AgdaSpace{}%
\AgdaSymbol{(}\AgdaFunction{contradiction}\AgdaSymbol{)}\<%
\\
\>[0]\AgdaKeyword{open}\AgdaSpace{}%
\AgdaKeyword{import}\AgdaSpace{}%
\AgdaModule{Data.List.Relation.Unary.Any}\AgdaSpace{}%
\AgdaKeyword{using}\AgdaSpace{}%
\AgdaSymbol{(}\AgdaFunction{tail}\AgdaSpace{}%
\AgdaSymbol{;}\AgdaSpace{}%
\AgdaInductiveConstructor{here}\AgdaSpace{}%
\AgdaSymbol{;}\AgdaSpace{}%
\AgdaInductiveConstructor{there}\AgdaSymbol{)}\<%
\\
\>[0]\AgdaKeyword{open}\AgdaSpace{}%
\AgdaKeyword{import}\AgdaSpace{}%
\AgdaModule{Data.List.Relation.Unary.Any.Properties}\AgdaSpace{}%
\AgdaKeyword{using}\AgdaSpace{}%
\AgdaSymbol{(}\AgdaFunction{reverse⁺}\AgdaSpace{}%
\AgdaSymbol{;}\AgdaSpace{}%
\AgdaFunction{reverse⁻}\AgdaSymbol{)}\<%
\\
\>[0]\AgdaKeyword{open}\AgdaSpace{}%
\AgdaKeyword{import}\AgdaSpace{}%
\AgdaModule{Functional.Rattle.Exec}\AgdaSpace{}%
\AgdaSymbol{(}\AgdaBound{oracle}\AgdaSymbol{)}\AgdaSpace{}%
\AgdaKeyword{using}\AgdaSpace{}%
\AgdaSymbol{(}rattle \AgdaSymbol{;} runWError \AgdaSymbol{;} runR \AgdaSymbol{;} doRunR \AgdaSymbol{;} rattle-unchecked \AgdaSymbol{;} doRunWError \AgdaSymbol{;} checkHazard \AgdaSymbol{;} g₂\AgdaSymbol{)}\<%
\\
\>[0]\AgdaKeyword{open}\AgdaSpace{}%
\AgdaKeyword{import}\AgdaSpace{}%
\AgdaModule{Functional.Build}\AgdaSpace{}%
\AgdaSymbol{(}\AgdaBound{oracle}\AgdaSymbol{)}\AgdaSpace{}%
\AgdaKeyword{using}\AgdaSpace{}%
\AgdaSymbol{(}Build \AgdaSymbol{;} UniqueEvidence \AgdaSymbol{;} PreCond \AgdaSymbol{;} DisjointBuild \AgdaSymbol{;} Cons\AgdaSymbol{)}\<%
\\
\>[0]\AgdaKeyword{open}\AgdaSpace{}%
\AgdaKeyword{import}\AgdaSpace{}%
\AgdaModule{Data.Sum}\AgdaSpace{}%
\AgdaKeyword{using}\AgdaSpace{}%
\AgdaSymbol{(}\AgdaInductiveConstructor{inj₂}\AgdaSpace{}%
\AgdaSymbol{;}\AgdaSpace{}%
\AgdaFunction{from-inj₂}\AgdaSpace{}%
\AgdaSymbol{;}\AgdaSpace{}%
\AgdaInductiveConstructor{inj₁}\AgdaSpace{}%
\AgdaSymbol{;}\AgdaSpace{}%
\AgdaOperator{\AgdaDatatype{\AgdaUnderscore{}⊎\AgdaUnderscore{}}}\AgdaSymbol{)}\<%
\\
\>[0]\AgdaKeyword{open}\AgdaSpace{}%
\AgdaKeyword{import}\AgdaSpace{}%
\AgdaModule{Data.Sum.Properties}\AgdaSpace{}%
\AgdaKeyword{using}\AgdaSpace{}%
\AgdaSymbol{(}\AgdaFunction{inj₂-injective}\AgdaSymbol{)}\<%
\\
\>[0]\AgdaKeyword{open}\AgdaSpace{}%
\AgdaKeyword{import}\AgdaSpace{}%
\AgdaModule{Functional.Script.Exec}\AgdaSpace{}%
\AgdaSymbol{(}\AgdaBound{oracle}\AgdaSymbol{)}\AgdaSpace{}%
\AgdaSymbol{as}\AgdaSpace{}%
\AgdaModule{Script}\<%
\\
\>[0]\AgdaKeyword{open}\AgdaSpace{}%
\AgdaKeyword{import}\AgdaSpace{}%
\AgdaModule{Functional.Script.Properties}\AgdaSpace{}%
\AgdaSymbol{(}\AgdaBound{oracle}\AgdaSymbol{)}\AgdaSpace{}%
\AgdaKeyword{using}\AgdaSpace{}%
\AgdaSymbol{(}dsj-≡ \AgdaSymbol{;} exec-≡f₁ \AgdaSymbol{;} writes≡ \AgdaSymbol{;} exec-∷≡ \AgdaSymbol{;} exec≡₃\AgdaSymbol{)}\<%
\\
\>[0]\AgdaKeyword{open}\AgdaSpace{}%
\AgdaKeyword{import}\AgdaSpace{}%
\AgdaModule{Functional.Script.Proofs}\AgdaSpace{}%
\AgdaSymbol{(}\AgdaBound{oracle}\AgdaSymbol{)}\AgdaSpace{}%
\AgdaKeyword{using}\AgdaSpace{}%
\AgdaSymbol{(}reordered \AgdaSymbol{;} reordered≡ \AgdaSymbol{;} unique-reverse \AgdaSymbol{;} unique-drop-mid \AgdaSymbol{;} helper3 \AgdaSymbol{;} helper5\AgdaSymbol{)}\<%
\\
\\[\AgdaEmptyExtraSkip]%
\>[0]\AgdaKeyword{open}\AgdaSpace{}%
\AgdaKeyword{import}\AgdaSpace{}%
\AgdaModule{Functional.Script.Hazard}\AgdaSpace{}%
\AgdaSymbol{(}\AgdaBound{oracle}\AgdaSymbol{)}\AgdaSpace{}%
\AgdaKeyword{using}\AgdaSpace{}%
\AgdaSymbol{(}Hazard \AgdaSymbol{;} HazardFree \AgdaSymbol{;} FileInfo \AgdaSymbol{;} \AgdaUnderscore{}∷\AgdaUnderscore{} \AgdaSymbol{;} ∃Hazard \AgdaSymbol{;} [] \AgdaSymbol{;} hazardfree? \AgdaSymbol{;} ReadWrite \AgdaSymbol{;} WriteWrite \AgdaSymbol{;} Speculative \AgdaSymbol{;} filesRead \AgdaSymbol{;} filesWrote \AgdaSymbol{;} files \AgdaSymbol{;} cmdsRun \AgdaSymbol{;} cmdWrote \AgdaSymbol{;} ∈-cmdWrote∷ \AgdaSymbol{;} ∈-cmdRead∷l \AgdaSymbol{;} ∈-filesWrote\AgdaSymbol{)}\AgdaSpace{}%
\AgdaKeyword{renaming}\AgdaSpace{}%
\AgdaSymbol{(}save \AgdaSymbol{to} rec\AgdaSymbol{)}\<%
\\
\\[\AgdaEmptyExtraSkip]%
\>[0]\AgdaKeyword{open}\AgdaSpace{}%
\AgdaKeyword{import}\AgdaSpace{}%
\AgdaModule{Functional.Script.Hazard.Properties}\AgdaSpace{}%
\AgdaSymbol{(}\AgdaBound{oracle}\AgdaSymbol{)}\AgdaSpace{}%
\AgdaKeyword{using}\AgdaSpace{}%
\AgdaSymbol{(}hf-still \AgdaSymbol{;} hf=>disjointWR \AgdaSymbol{;} hf=>disjointRW \AgdaSymbol{;} hf=>disjointWW \AgdaSymbol{;} hf-drop-mid\AgdaSymbol{)}\<%
\\
\>[0]\AgdaKeyword{open}\AgdaSpace{}%
\AgdaKeyword{import}\AgdaSpace{}%
\AgdaModule{Data.Empty}\AgdaSpace{}%
\AgdaKeyword{using}\AgdaSpace{}%
\AgdaSymbol{(}\AgdaDatatype{⊥}\AgdaSpace{}%
\AgdaSymbol{;}\AgdaSpace{}%
\AgdaFunction{⊥-elim}\AgdaSymbol{)}\<%
\\
\\[\AgdaEmptyExtraSkip]%
\>[0]\AgdaKeyword{open}\AgdaSpace{}%
\AgdaKeyword{import}\AgdaSpace{}%
\AgdaModule{Data.List.Relation.Unary.Unique.Propositional}\AgdaSpace{}%
\AgdaKeyword{using}\AgdaSpace{}%
\AgdaSymbol{(}\AgdaDatatype{Unique}\AgdaSymbol{)}\<%
\\
\>[0]\AgdaKeyword{open}\AgdaSpace{}%
\AgdaKeyword{import}\AgdaSpace{}%
\AgdaModule{Data.List.Relation.Unary.AllPairs}\AgdaSpace{}%
\AgdaKeyword{using}\AgdaSpace{}%
\AgdaSymbol{(}\AgdaOperator{\AgdaInductiveConstructor{\AgdaUnderscore{}∷\AgdaUnderscore{}}}\AgdaSymbol{)}\<%
\\
\>[0]\AgdaKeyword{open}\AgdaSpace{}%
\AgdaKeyword{import}\AgdaSpace{}%
\AgdaModule{Data.List.Relation.Binary.Subset.Propositional}\AgdaSpace{}%
\AgdaKeyword{using}\AgdaSpace{}%
\AgdaSymbol{(}\AgdaOperator{\AgdaFunction{\AgdaUnderscore{}⊆\AgdaUnderscore{}}}\AgdaSymbol{)}\<%
\\
\>[0]\AgdaKeyword{open}\AgdaSpace{}%
\AgdaKeyword{import}\AgdaSpace{}%
\AgdaModule{Data.List.Relation.Binary.Permutation.Propositional}\AgdaSpace{}%
\AgdaKeyword{using}\AgdaSpace{}%
\AgdaSymbol{(}\AgdaOperator{\AgdaDatatype{\AgdaUnderscore{}↭\AgdaUnderscore{}}}\AgdaSpace{}%
\AgdaSymbol{;}\AgdaSpace{}%
\AgdaFunction{↭-sym}\AgdaSymbol{)}\<%
\\
\>[0]\AgdaKeyword{open}\AgdaSpace{}%
\AgdaKeyword{import}\AgdaSpace{}%
\AgdaModule{Data.List.Relation.Binary.Permutation.Propositional.Properties}\AgdaSpace{}%
\AgdaKeyword{using}\AgdaSpace{}%
\AgdaSymbol{(}\AgdaFunction{∈-resp-↭}\AgdaSpace{}%
\AgdaSymbol{;}\AgdaSpace{}%
\AgdaFunction{↭-length}\AgdaSpace{}%
\AgdaSymbol{;}\AgdaSpace{}%
\AgdaFunction{drop-mid}\AgdaSymbol{)}\<%
\\
\\[\AgdaEmptyExtraSkip]%
\\[\AgdaEmptyExtraSkip]%
\\[\AgdaEmptyExtraSkip]%
\>[0]\AgdaKeyword{data}\AgdaSpace{}%
\AgdaDatatype{MemoryProperty}\AgdaSpace{}%
\AgdaSymbol{:}\AgdaSpace{}%
\AgdaFunction{Memory}\AgdaSpace{}%
\AgdaSymbol{->}\AgdaSpace{}%
\AgdaPrimitiveType{Set}\AgdaSpace{}%
\AgdaKeyword{where}\<%
\\
\>[0][@{}l@{\AgdaIndent{0}}]%
\>[1]\AgdaInductiveConstructor{[]}%
\>[6]\AgdaSymbol{:}\AgdaSpace{}%
\AgdaDatatype{MemoryProperty}\AgdaSpace{}%
\AgdaInductiveConstructor{[]}\<%
\\
\>[1]\AgdaInductiveConstructor{Cons}\AgdaSpace{}%
\AgdaSymbol{:}\AgdaSpace{}%
\AgdaSymbol{\{}\AgdaBound{mm}\AgdaSpace{}%
\AgdaSymbol{:}\AgdaSpace{}%
\AgdaFunction{Memory}\AgdaSymbol{\}}\AgdaSpace{}%
\AgdaSymbol{(}\AgdaBound{x}\AgdaSpace{}%
\AgdaSymbol{:}\AgdaSpace{}%
\AgdaFunction{Cmd}\AgdaSymbol{)}\AgdaSpace{}%
\AgdaSymbol{->}\AgdaSpace{}%
\AgdaSymbol{(}\AgdaBound{sys}\AgdaSpace{}%
\AgdaSymbol{:}\AgdaSpace{}%
\AgdaFunction{FileSystem}\AgdaSymbol{)}\AgdaSpace{}%
\AgdaSymbol{->}\AgdaSpace{}%
\AgdaSymbol{(∀}\AgdaSpace{}%
\AgdaBound{f₁}\AgdaSpace{}%
\AgdaSymbol{→}\AgdaSpace{}%
\AgdaBound{f₁}\AgdaSpace{}%
\AgdaOperator{\AgdaPostulate{∈}}\AgdaSpace{}%
\AgdaPostulate{cmdReadNames}\AgdaSpace{}%
\AgdaBound{x}\AgdaSpace{}%
\AgdaBound{sys}\AgdaSpace{}%
\AgdaSymbol{->}\AgdaSpace{}%
\AgdaBound{sys}\AgdaSpace{}%
\AgdaBound{f₁}\AgdaSpace{}%
\AgdaOperator{\AgdaDatatype{≡}}\AgdaSpace{}%
\AgdaPostulate{St.run}\AgdaSpace{}%
\AgdaBound{x}\AgdaSpace{}%
\AgdaBound{sys}\AgdaSpace{}%
\AgdaBound{f₁}\AgdaSymbol{)}\AgdaSpace{}%
\AgdaSymbol{->}\AgdaSpace{}%
\AgdaDatatype{MemoryProperty}\AgdaSpace{}%
\AgdaBound{mm}\AgdaSpace{}%
\AgdaSymbol{->}\AgdaSpace{}%
\AgdaDatatype{MemoryProperty}\AgdaSpace{}%
\AgdaSymbol{((}\AgdaBound{x}\AgdaSpace{}%
\AgdaOperator{\AgdaInductiveConstructor{,}}\AgdaSpace{}%
\AgdaFunction{map}\AgdaSpace{}%
\AgdaSymbol{(λ}\AgdaSpace{}%
\AgdaBound{f₁}\AgdaSpace{}%
\AgdaSymbol{→}\AgdaSpace{}%
\AgdaBound{f₁}\AgdaSpace{}%
\AgdaOperator{\AgdaInductiveConstructor{,}}\AgdaSpace{}%
\AgdaSymbol{(}\AgdaPostulate{St.run}\AgdaSpace{}%
\AgdaBound{x}\AgdaSpace{}%
\AgdaBound{sys}\AgdaSymbol{)}\AgdaSpace{}%
\AgdaBound{f₁}\AgdaSymbol{)}\AgdaSpace{}%
\AgdaSymbol{(}\AgdaPostulate{cmdReadWrites}\AgdaSpace{}%
\AgdaBound{x}\AgdaSpace{}%
\AgdaBound{sys}\AgdaSymbol{))}\AgdaSpace{}%
\AgdaOperator{\AgdaInductiveConstructor{∷}}\AgdaSpace{}%
\AgdaBound{mm}\AgdaSymbol{)}\<%
\\
\>[0]\<%
\\
\\[\AgdaEmptyExtraSkip]%
\>[0]\AgdaFunction{getProperty}\AgdaSpace{}%
\AgdaSymbol{:}\AgdaSpace{}%
\AgdaSymbol{\{}\AgdaBound{mm}\AgdaSpace{}%
\AgdaSymbol{:}\AgdaSpace{}%
\AgdaFunction{Memory}\AgdaSymbol{\}}\AgdaSpace{}%
\AgdaSymbol{(}\AgdaBound{x}\AgdaSpace{}%
\AgdaSymbol{:}\AgdaSpace{}%
\AgdaFunction{Cmd}\AgdaSymbol{)}\AgdaSpace{}%
\AgdaSymbol{->}\AgdaSpace{}%
\AgdaDatatype{MemoryProperty}\AgdaSpace{}%
\AgdaBound{mm}\AgdaSpace{}%
\AgdaSymbol{->}\AgdaSpace{}%
\AgdaSymbol{(}\AgdaBound{x∈}\AgdaSpace{}%
\AgdaSymbol{:}\AgdaSpace{}%
\AgdaBound{x}\AgdaSpace{}%
\AgdaOperator{\AgdaPostulate{∈}}\AgdaSpace{}%
\AgdaFunction{map}\AgdaSpace{}%
\AgdaField{proj₁}\AgdaSpace{}%
\AgdaBound{mm}\AgdaSymbol{)}\AgdaSpace{}%
\AgdaSymbol{->}\AgdaSpace{}%
\AgdaFunction{∃[}\AgdaSpace{}%
\AgdaBound{sys}\AgdaSpace{}%
\AgdaFunction{]}\AgdaSymbol{(}\AgdaPostulate{get}\AgdaSpace{}%
\AgdaBound{x}\AgdaSpace{}%
\AgdaBound{mm}\AgdaSpace{}%
\AgdaBound{x∈}\AgdaSpace{}%
\AgdaOperator{\AgdaDatatype{≡}}\AgdaSpace{}%
\AgdaFunction{map}\AgdaSpace{}%
\AgdaSymbol{(λ}\AgdaSpace{}%
\AgdaBound{f₁}\AgdaSpace{}%
\AgdaSymbol{→}\AgdaSpace{}%
\AgdaBound{f₁}\AgdaSpace{}%
\AgdaOperator{\AgdaInductiveConstructor{,}}\AgdaSpace{}%
\AgdaSymbol{(}\AgdaPostulate{St.run}\AgdaSpace{}%
\AgdaBound{x}\AgdaSpace{}%
\AgdaBound{sys}\AgdaSymbol{)}\AgdaSpace{}%
\AgdaBound{f₁}\AgdaSymbol{)}\AgdaSpace{}%
\AgdaSymbol{(}\AgdaPostulate{cmdReadWrites}\AgdaSpace{}%
\AgdaBound{x}\AgdaSpace{}%
\AgdaBound{sys}\AgdaSymbol{)}\AgdaSpace{}%
\AgdaOperator{\AgdaFunction{×}}\AgdaSpace{}%
\AgdaSymbol{∀}\AgdaSpace{}%
\AgdaBound{f₁}\AgdaSpace{}%
\AgdaSymbol{→}\AgdaSpace{}%
\AgdaBound{f₁}\AgdaSpace{}%
\AgdaOperator{\AgdaPostulate{∈}}\AgdaSpace{}%
\AgdaPostulate{cmdReadNames}\AgdaSpace{}%
\AgdaBound{x}\AgdaSpace{}%
\AgdaBound{sys}\AgdaSpace{}%
\AgdaSymbol{->}\AgdaSpace{}%
\AgdaBound{sys}\AgdaSpace{}%
\AgdaBound{f₁}\AgdaSpace{}%
\AgdaOperator{\AgdaDatatype{≡}}\AgdaSpace{}%
\AgdaPostulate{St.run}\AgdaSpace{}%
\AgdaBound{x}\AgdaSpace{}%
\AgdaBound{sys}\AgdaSpace{}%
\AgdaBound{f₁}\AgdaSymbol{)}\<%
\\
\>[0]\AgdaFunction{getProperty}\AgdaSpace{}%
\AgdaBound{x}\AgdaSpace{}%
\AgdaSymbol{(}\AgdaInductiveConstructor{Cons}\AgdaSpace{}%
\AgdaBound{x₁}\AgdaSpace{}%
\AgdaBound{sys}\AgdaSpace{}%
\AgdaBound{∀₁}\AgdaSpace{}%
\AgdaBound{mp}\AgdaSymbol{)}\AgdaSpace{}%
\AgdaBound{x∈}\AgdaSpace{}%
\AgdaKeyword{with}\AgdaSpace{}%
\AgdaBound{x}\AgdaSpace{}%
\AgdaOperator{\AgdaFunction{≟}}\AgdaSpace{}%
\AgdaBound{x₁}\<%
\\
\>[0]\AgdaSymbol{...}\AgdaSpace{}%
\AgdaSymbol{|}\AgdaSpace{}%
\AgdaInductiveConstructor{yes}\AgdaSpace{}%
\AgdaBound{x≡x₁}\AgdaSpace{}%
\AgdaSymbol{=}\AgdaSpace{}%
\AgdaBound{sys}\AgdaSpace{}%
\AgdaOperator{\AgdaInductiveConstructor{,}}\AgdaSpace{}%
\AgdaFunction{cong}\AgdaSpace{}%
\AgdaSymbol{(λ}\AgdaSpace{}%
\AgdaBound{x₂}\AgdaSpace{}%
\AgdaSymbol{→}\AgdaSpace{}%
\AgdaFunction{map}\AgdaSpace{}%
\AgdaSymbol{(λ}\AgdaSpace{}%
\AgdaBound{f₁}\AgdaSpace{}%
\AgdaSymbol{→}\AgdaSpace{}%
\AgdaBound{f₁}\AgdaSpace{}%
\AgdaOperator{\AgdaInductiveConstructor{,}}\AgdaSpace{}%
\AgdaFunction{foldr}\AgdaSpace{}%
\AgdaFunction{extend}\AgdaSpace{}%
\AgdaBound{sys}\AgdaSpace{}%
\AgdaSymbol{(}\AgdaField{proj₂}\AgdaSpace{}%
\AgdaSymbol{(}\AgdaField{proj₁}\AgdaSpace{}%
\AgdaSymbol{(}\AgdaBound{oracle}\AgdaSpace{}%
\AgdaBound{x₂}\AgdaSymbol{)}\AgdaSpace{}%
\AgdaBound{sys}\AgdaSymbol{))}\AgdaSpace{}%
\AgdaBound{f₁}\AgdaSymbol{)}\AgdaSpace{}%
\AgdaSymbol{(}\AgdaPostulate{cmdReadWrites}\AgdaSpace{}%
\AgdaBound{x₂}\AgdaSpace{}%
\AgdaBound{sys}\AgdaSymbol{))}\AgdaSpace{}%
\AgdaSymbol{(}\AgdaFunction{sym}\AgdaSpace{}%
\AgdaBound{x≡x₁}\AgdaSymbol{)}\AgdaSpace{}%
\AgdaOperator{\AgdaInductiveConstructor{,}}\AgdaSpace{}%
\AgdaSymbol{λ}\AgdaSpace{}%
\AgdaBound{f₁}\AgdaSpace{}%
\AgdaBound{x₂}\AgdaSpace{}%
\AgdaSymbol{→}\AgdaSpace{}%
\AgdaFunction{subst}\AgdaSpace{}%
\AgdaSymbol{(λ}\AgdaSpace{}%
\AgdaBound{x₃}\AgdaSpace{}%
\AgdaSymbol{→}\AgdaSpace{}%
\AgdaBound{sys}\AgdaSpace{}%
\AgdaBound{f₁}\AgdaSpace{}%
\AgdaOperator{\AgdaDatatype{≡}}\AgdaSpace{}%
\AgdaPostulate{St.run}\AgdaSpace{}%
\AgdaBound{x₃}\AgdaSpace{}%
\AgdaBound{sys}\AgdaSpace{}%
\AgdaBound{f₁}\AgdaSymbol{)}\AgdaSpace{}%
\AgdaSymbol{(}\AgdaFunction{sym}\AgdaSpace{}%
\AgdaBound{x≡x₁}\AgdaSymbol{)}\AgdaSpace{}%
\AgdaSymbol{(}\AgdaBound{∀₁}\AgdaSpace{}%
\AgdaBound{f₁}\AgdaSpace{}%
\AgdaSymbol{(}\AgdaFunction{subst}\AgdaSpace{}%
\AgdaSymbol{(λ}\AgdaSpace{}%
\AgdaBound{x₃}\AgdaSpace{}%
\AgdaSymbol{→}\AgdaSpace{}%
\AgdaBound{f₁}\AgdaSpace{}%
\AgdaOperator{\AgdaPostulate{∈}}\AgdaSpace{}%
\AgdaFunction{map}\AgdaSpace{}%
\AgdaField{proj₁}\AgdaSpace{}%
\AgdaSymbol{(}\AgdaField{proj₁}\AgdaSpace{}%
\AgdaSymbol{(}\AgdaField{proj₁}\AgdaSpace{}%
\AgdaSymbol{(}\AgdaBound{oracle}\AgdaSpace{}%
\AgdaBound{x₃}\AgdaSymbol{)}\AgdaSpace{}%
\AgdaBound{sys}\AgdaSymbol{)))}\AgdaSpace{}%
\AgdaBound{x≡x₁}\AgdaSpace{}%
\AgdaBound{x₂}\AgdaSymbol{))}\<%
\\
\>[0]\AgdaSymbol{...}\AgdaSpace{}%
\AgdaSymbol{|}\AgdaSpace{}%
\AgdaInductiveConstructor{no}\AgdaSpace{}%
\AgdaBound{¬x≡x₁}\AgdaSpace{}%
\AgdaSymbol{=}\AgdaSpace{}%
\AgdaFunction{getProperty}\AgdaSpace{}%
\AgdaBound{x}\AgdaSpace{}%
\AgdaBound{mp}\AgdaSpace{}%
\AgdaSymbol{(}\AgdaFunction{tail}\AgdaSpace{}%
\AgdaBound{¬x≡x₁}\AgdaSpace{}%
\AgdaBound{x∈}\AgdaSymbol{)}\<%
\\
\\[\AgdaEmptyExtraSkip]%
\>[0]\AgdaFunction{lemma1}\AgdaSpace{}%
\AgdaSymbol{:}\AgdaSpace{}%
\AgdaSymbol{∀}\AgdaSpace{}%
\AgdaSymbol{(}\AgdaBound{s}\AgdaSpace{}%
\AgdaSymbol{:}\AgdaSpace{}%
\AgdaFunction{FileSystem}\AgdaSymbol{)}\AgdaSpace{}%
\AgdaSymbol{\{}\AgdaBound{s₁}\AgdaSymbol{\}}\AgdaSpace{}%
\AgdaSymbol{\{}\AgdaBound{x}\AgdaSymbol{\}}\AgdaSpace{}%
\AgdaBound{ls}\AgdaSpace{}%
\AgdaBound{ls₁}\AgdaSpace{}%
\AgdaSymbol{→}\AgdaSpace{}%
\AgdaDatatype{All}\AgdaSpace{}%
\AgdaSymbol{(λ}\AgdaSpace{}%
\AgdaSymbol{(}\AgdaBound{f₁}\AgdaSpace{}%
\AgdaOperator{\AgdaInductiveConstructor{,}}\AgdaSpace{}%
\AgdaBound{v₁}\AgdaSymbol{)}\AgdaSpace{}%
\AgdaSymbol{→}\AgdaSpace{}%
\AgdaBound{s}\AgdaSpace{}%
\AgdaBound{f₁}\AgdaSpace{}%
\AgdaOperator{\AgdaDatatype{≡}}\AgdaSpace{}%
\AgdaBound{v₁}\AgdaSymbol{)}\AgdaSpace{}%
\AgdaBound{ls}\AgdaSpace{}%
\AgdaSymbol{→}\AgdaSpace{}%
\AgdaBound{ls}\AgdaSpace{}%
\AgdaOperator{\AgdaDatatype{≡}}\AgdaSpace{}%
\AgdaFunction{map}\AgdaSpace{}%
\AgdaSymbol{(λ}\AgdaSpace{}%
\AgdaBound{f₁}\AgdaSpace{}%
\AgdaSymbol{→}\AgdaSpace{}%
\AgdaBound{f₁}\AgdaSpace{}%
\AgdaOperator{\AgdaInductiveConstructor{,}}\AgdaSpace{}%
\AgdaSymbol{(}\AgdaPostulate{St.run}\AgdaSpace{}%
\AgdaBound{x}\AgdaSpace{}%
\AgdaBound{s₁}\AgdaSymbol{)}\AgdaSpace{}%
\AgdaBound{f₁}\AgdaSymbol{)}\AgdaSpace{}%
\AgdaBound{ls₁}\AgdaSpace{}%
\AgdaSymbol{→}\AgdaSpace{}%
\AgdaDatatype{All}\AgdaSpace{}%
\AgdaSymbol{(λ}\AgdaSpace{}%
\AgdaBound{f₁}\AgdaSpace{}%
\AgdaSymbol{→}\AgdaSpace{}%
\AgdaBound{s}\AgdaSpace{}%
\AgdaBound{f₁}\AgdaSpace{}%
\AgdaOperator{\AgdaDatatype{≡}}\AgdaSpace{}%
\AgdaPostulate{St.run}\AgdaSpace{}%
\AgdaBound{x}\AgdaSpace{}%
\AgdaBound{s₁}\AgdaSpace{}%
\AgdaBound{f₁}\AgdaSymbol{)}\AgdaSpace{}%
\AgdaBound{ls₁}\<%
\\
\>[0]\AgdaFunction{lemma1}\AgdaSpace{}%
\AgdaBound{s}\AgdaSpace{}%
\AgdaInductiveConstructor{[]}\AgdaSpace{}%
\AgdaInductiveConstructor{[]}\AgdaSpace{}%
\AgdaBound{all₁}\AgdaSpace{}%
\AgdaBound{≡₁}\AgdaSpace{}%
\AgdaSymbol{=}\AgdaSpace{}%
\AgdaInductiveConstructor{All.[]}\<%
\\
\>[0]\AgdaFunction{lemma1}\AgdaSpace{}%
\AgdaBound{s}\AgdaSpace{}%
\AgdaSymbol{(}\AgdaBound{x₁}\AgdaSpace{}%
\AgdaOperator{\AgdaInductiveConstructor{∷}}\AgdaSpace{}%
\AgdaBound{ls}\AgdaSymbol{)}\AgdaSpace{}%
\AgdaSymbol{(}\AgdaBound{x}\AgdaSpace{}%
\AgdaOperator{\AgdaInductiveConstructor{∷}}\AgdaSpace{}%
\AgdaBound{ls₁}\AgdaSymbol{)}\AgdaSpace{}%
\AgdaSymbol{(}\AgdaBound{px}\AgdaSpace{}%
\AgdaOperator{\AgdaInductiveConstructor{All.∷}}\AgdaSpace{}%
\AgdaBound{all₁}\AgdaSymbol{)}\AgdaSpace{}%
\AgdaBound{≡₁}\AgdaSpace{}%
\AgdaKeyword{with}\AgdaSpace{}%
\AgdaFunction{∷-injective}\AgdaSpace{}%
\AgdaBound{≡₁}\<%
\\
\>[0]\AgdaSymbol{...}\AgdaSpace{}%
\AgdaSymbol{|}\AgdaSpace{}%
\AgdaBound{x₁≡x}\AgdaSpace{}%
\AgdaOperator{\AgdaInductiveConstructor{,}}\AgdaSpace{}%
\AgdaBound{≡₂}\AgdaSpace{}%
\AgdaSymbol{=}\AgdaSpace{}%
\AgdaSymbol{(}\AgdaFunction{trans}\AgdaSpace{}%
\AgdaSymbol{(}\AgdaFunction{subst}\AgdaSpace{}%
\AgdaSymbol{(λ}\AgdaSpace{}%
\AgdaBound{x₂}\AgdaSpace{}%
\AgdaSymbol{→}\AgdaSpace{}%
\AgdaBound{s}\AgdaSpace{}%
\AgdaBound{x₂}\AgdaSpace{}%
\AgdaOperator{\AgdaDatatype{≡}}\AgdaSpace{}%
\AgdaField{proj₂}\AgdaSpace{}%
\AgdaBound{x₁}\AgdaSymbol{)}\AgdaSpace{}%
\AgdaSymbol{(}\AgdaFunction{,-injectiveˡ}\AgdaSpace{}%
\AgdaBound{x₁≡x}\AgdaSymbol{)}\AgdaSpace{}%
\AgdaBound{px}\AgdaSymbol{)}\AgdaSpace{}%
\AgdaSymbol{(}\AgdaFunction{,-injectiveʳ}\AgdaSpace{}%
\AgdaBound{x₁≡x}\AgdaSymbol{))}\AgdaSpace{}%
\AgdaOperator{\AgdaInductiveConstructor{All.∷}}\AgdaSpace{}%
\AgdaSymbol{(}\AgdaFunction{lemma1}\AgdaSpace{}%
\AgdaBound{s}\AgdaSpace{}%
\AgdaBound{ls}\AgdaSpace{}%
\AgdaBound{ls₁}\AgdaSpace{}%
\AgdaBound{all₁}\AgdaSpace{}%
\AgdaBound{≡₂}\AgdaSymbol{)}\<%
\\
\\[\AgdaEmptyExtraSkip]%
\>[0]\AgdaComment{\{- Want to prove that system will be the same whether or not we run the command; -\}}\<%
\\
\>[0]\AgdaComment{\{- If a command's inputs and outputs are unchanged from when it was last run,
 then running it will have no effect. -\}}\<%
\\
\>[0]\AgdaFunction{noEffect}\AgdaSpace{}%
\AgdaSymbol{:}\AgdaSpace{}%
\AgdaSymbol{∀}\AgdaSpace{}%
\AgdaSymbol{\{}\AgdaBound{s₁}\AgdaSymbol{\}}\AgdaSpace{}%
\AgdaSymbol{\{}\AgdaBound{s₂}\AgdaSpace{}%
\AgdaSymbol{:}\AgdaSpace{}%
\AgdaFunction{FileSystem}\AgdaSymbol{\}}\AgdaSpace{}%
\AgdaSymbol{\{}\AgdaBound{mm}\AgdaSymbol{\}}\AgdaSpace{}%
\AgdaBound{x}\AgdaSpace{}%
\AgdaSymbol{→}\AgdaSpace{}%
\AgdaSymbol{(∀}\AgdaSpace{}%
\AgdaBound{f₁}\AgdaSpace{}%
\AgdaSymbol{→}\AgdaSpace{}%
\AgdaBound{s₁}\AgdaSpace{}%
\AgdaBound{f₁}\AgdaSpace{}%
\AgdaOperator{\AgdaDatatype{≡}}\AgdaSpace{}%
\AgdaBound{s₂}\AgdaSpace{}%
\AgdaBound{f₁}\AgdaSymbol{)}\AgdaSpace{}%
\AgdaSymbol{→}\AgdaSpace{}%
\AgdaDatatype{MemoryProperty}\AgdaSpace{}%
\AgdaBound{mm}\AgdaSpace{}%
\AgdaSymbol{→}\AgdaSpace{}%
\AgdaSymbol{(}\AgdaBound{x∈}\AgdaSpace{}%
\AgdaSymbol{:}\AgdaSpace{}%
\AgdaBound{x}\AgdaSpace{}%
\AgdaOperator{\AgdaPostulate{∈}}\AgdaSpace{}%
\AgdaFunction{map}\AgdaSpace{}%
\AgdaField{proj₁}\AgdaSpace{}%
\AgdaBound{mm}\AgdaSymbol{)}\AgdaSpace{}%
\AgdaSymbol{→}\AgdaSpace{}%
\AgdaDatatype{All}\AgdaSpace{}%
\AgdaSymbol{(λ}\AgdaSpace{}%
\AgdaSymbol{(}\AgdaBound{f₁}\AgdaSpace{}%
\AgdaOperator{\AgdaInductiveConstructor{,}}\AgdaSpace{}%
\AgdaBound{v₁}\AgdaSymbol{)}\AgdaSpace{}%
\AgdaSymbol{→}\AgdaSpace{}%
\AgdaBound{s₂}\AgdaSpace{}%
\AgdaBound{f₁}\AgdaSpace{}%
\AgdaOperator{\AgdaDatatype{≡}}\AgdaSpace{}%
\AgdaBound{v₁}\AgdaSymbol{)}\AgdaSpace{}%
\AgdaSymbol{(}\AgdaPostulate{get}\AgdaSpace{}%
\AgdaBound{x}\AgdaSpace{}%
\AgdaBound{mm}\AgdaSpace{}%
\AgdaBound{x∈}\AgdaSymbol{)}\AgdaSpace{}%
\AgdaSymbol{→}\AgdaSpace{}%
\AgdaSymbol{∀}\AgdaSpace{}%
\AgdaBound{f₁}\AgdaSpace{}%
\AgdaSymbol{→}\AgdaSpace{}%
\AgdaBound{s₂}\AgdaSpace{}%
\AgdaBound{f₁}\AgdaSpace{}%
\AgdaOperator{\AgdaDatatype{≡}}\AgdaSpace{}%
\AgdaPostulate{St.run}\AgdaSpace{}%
\AgdaBound{x}\AgdaSpace{}%
\AgdaBound{s₁}\AgdaSpace{}%
\AgdaBound{f₁}\<%
\\
\>[0]\AgdaFunction{noEffect}\AgdaSpace{}%
\AgdaSymbol{\{}\AgdaBound{s₁}\AgdaSymbol{\}}\AgdaSpace{}%
\AgdaSymbol{\{}\AgdaBound{s₂}\AgdaSymbol{\}}\AgdaSpace{}%
\AgdaSymbol{\{}\AgdaBound{mm}\AgdaSymbol{\}}\AgdaSpace{}%
\AgdaBound{x}\AgdaSpace{}%
\AgdaBound{∀₁}\AgdaSpace{}%
\AgdaBound{mp}\AgdaSpace{}%
\AgdaBound{x∈}\AgdaSpace{}%
\AgdaBound{all₁}\AgdaSpace{}%
\AgdaBound{f₁}\AgdaSpace{}%
\AgdaKeyword{with}\AgdaSpace{}%
\AgdaFunction{getProperty}\AgdaSpace{}%
\AgdaBound{x}\AgdaSpace{}%
\AgdaBound{mp}\AgdaSpace{}%
\AgdaBound{x∈}\<%
\\
\>[0]\AgdaSymbol{...}\AgdaSpace{}%
\AgdaSymbol{|}\AgdaSpace{}%
\AgdaBound{s₃}\AgdaSpace{}%
\AgdaOperator{\AgdaInductiveConstructor{,}}\AgdaSpace{}%
\AgdaBound{get≡}\AgdaSpace{}%
\AgdaOperator{\AgdaInductiveConstructor{,}}\AgdaSpace{}%
\AgdaBound{∀₂}\AgdaSpace{}%
\AgdaKeyword{with}\AgdaSpace{}%
\AgdaBound{f₁}\AgdaSpace{}%
\AgdaOperator{\AgdaPostulate{∈?}}\AgdaSpace{}%
\AgdaPostulate{cmdWriteNames}\AgdaSpace{}%
\AgdaBound{x}\AgdaSpace{}%
\AgdaBound{s₁}\<%
\\
\>[0]\AgdaSymbol{...}\AgdaSpace{}%
\AgdaSymbol{|}\AgdaSpace{}%
\AgdaInductiveConstructor{no}\AgdaSpace{}%
\AgdaBound{f₁∉}\AgdaSpace{}%
\AgdaSymbol{=}\AgdaSpace{}%
\AgdaFunction{trans}\AgdaSpace{}%
\AgdaSymbol{(}\AgdaFunction{sym}\AgdaSpace{}%
\AgdaSymbol{(}\AgdaBound{∀₁}\AgdaSpace{}%
\AgdaBound{f₁}\AgdaSymbol{))}\AgdaSpace{}%
\AgdaSymbol{(}\AgdaPostulate{lemma3}\AgdaSpace{}%
\AgdaBound{f₁}\AgdaSpace{}%
\AgdaSymbol{(}\AgdaField{proj₂}\AgdaSpace{}%
\AgdaSymbol{(}\AgdaField{proj₁}\AgdaSpace{}%
\AgdaSymbol{(}\AgdaBound{oracle}\AgdaSpace{}%
\AgdaBound{x}\AgdaSymbol{)}\AgdaSpace{}%
\AgdaBound{s₁}\AgdaSymbol{))}\AgdaSpace{}%
\AgdaBound{f₁∉}\AgdaSymbol{)}\<%
\\
\>[0]\AgdaSymbol{...}\AgdaSpace{}%
\AgdaSymbol{|}\AgdaSpace{}%
\AgdaInductiveConstructor{yes}\AgdaSpace{}%
\AgdaBound{f₁∈}\AgdaSpace{}%
\AgdaKeyword{with}\AgdaSpace{}%
\AgdaFunction{lemma1}\AgdaSpace{}%
\AgdaBound{s₂}\AgdaSpace{}%
\AgdaSymbol{(}\AgdaPostulate{get}\AgdaSpace{}%
\AgdaBound{x}\AgdaSpace{}%
\AgdaBound{mm}\AgdaSpace{}%
\AgdaBound{x∈}\AgdaSymbol{)}\AgdaSpace{}%
\AgdaSymbol{(}\AgdaPostulate{cmdReadWrites}\AgdaSpace{}%
\AgdaBound{x}\AgdaSpace{}%
\AgdaBound{s₃}\AgdaSymbol{)}\AgdaSpace{}%
\AgdaBound{all₁}\AgdaSpace{}%
\AgdaBound{get≡}\<%
\\
\>[0]\AgdaSymbol{...}\AgdaSpace{}%
\AgdaSymbol{|}\AgdaSpace{}%
\AgdaBound{all₂}\AgdaSpace{}%
\AgdaKeyword{with}\AgdaSpace{}%
\AgdaField{proj₂}\AgdaSpace{}%
\AgdaSymbol{(}\AgdaBound{oracle}\AgdaSpace{}%
\AgdaBound{x}\AgdaSymbol{)}\AgdaSpace{}%
\AgdaBound{s₃}\AgdaSpace{}%
\AgdaBound{s₁}\AgdaSpace{}%
\AgdaSymbol{(λ}\AgdaSpace{}%
\AgdaBound{f₂}\AgdaSpace{}%
\AgdaBound{x₁}\AgdaSpace{}%
\AgdaSymbol{→}\AgdaSpace{}%
\AgdaFunction{sym}\AgdaSpace{}%
\AgdaSymbol{(}\AgdaFunction{trans}\AgdaSpace{}%
\AgdaSymbol{(}\AgdaBound{∀₁}\AgdaSpace{}%
\AgdaBound{f₂}\AgdaSymbol{)}\AgdaSpace{}%
\AgdaSymbol{(}\AgdaFunction{trans}\AgdaSpace{}%
\AgdaSymbol{(}\AgdaFunction{lookup}\AgdaSpace{}%
\AgdaBound{all₂}\AgdaSpace{}%
\AgdaSymbol{(}\AgdaFunction{∈-++⁺ˡ}\AgdaSpace{}%
\AgdaBound{x₁}\AgdaSymbol{))}\AgdaSpace{}%
\AgdaSymbol{(}\AgdaFunction{sym}\AgdaSpace{}%
\AgdaSymbol{(}\AgdaBound{∀₂}\AgdaSpace{}%
\AgdaBound{f₂}\AgdaSpace{}%
\AgdaBound{x₁}\AgdaSymbol{)))))}\<%
\\
\>[0]\AgdaSymbol{...}\AgdaSpace{}%
\AgdaSymbol{|}\AgdaSpace{}%
\AgdaBound{≡₁}\AgdaSpace{}%
\AgdaSymbol{=}\AgdaSpace{}%
\AgdaFunction{trans}%
\>[746I]\AgdaSymbol{(}\AgdaFunction{lookup}\AgdaSpace{}%
\AgdaBound{all₂}\AgdaSpace{}%
\AgdaSymbol{(}\AgdaFunction{∈-++⁺ʳ}\AgdaSpace{}%
\AgdaSymbol{(}\AgdaPostulate{cmdReadNames}\AgdaSpace{}%
\AgdaBound{x}\AgdaSpace{}%
\AgdaBound{s₃}\AgdaSymbol{)}\AgdaSpace{}%
\AgdaFunction{f₁∈₁}\AgdaSymbol{))}\<%
\\
\>[.][@{}l@{}]\<[746I]%
\>[17]\AgdaSymbol{(}\AgdaFunction{subst}%
\>[753I]\AgdaSymbol{(λ}\AgdaSpace{}%
\AgdaBound{x₁}\AgdaSpace{}%
\AgdaSymbol{→}\AgdaSpace{}%
\AgdaFunction{foldr}\AgdaSpace{}%
\AgdaFunction{extend}\AgdaSpace{}%
\AgdaBound{s₃}\AgdaSpace{}%
\AgdaBound{x₁}\AgdaSpace{}%
\AgdaBound{f₁}\AgdaSpace{}%
\AgdaOperator{\AgdaDatatype{≡}}\AgdaSpace{}%
\AgdaFunction{foldr}\AgdaSpace{}%
\AgdaFunction{extend}\AgdaSpace{}%
\AgdaBound{s₁}\AgdaSpace{}%
\AgdaSymbol{(}\AgdaField{proj₂}\AgdaSpace{}%
\AgdaSymbol{(}\AgdaField{proj₁}\AgdaSpace{}%
\AgdaSymbol{(}\AgdaBound{oracle}\AgdaSpace{}%
\AgdaBound{x}\AgdaSymbol{)}\AgdaSpace{}%
\AgdaBound{s₁}\AgdaSymbol{))}\AgdaSpace{}%
\AgdaBound{f₁}\AgdaSymbol{)}\<%
\\
\>[.][@{}l@{}]\<[753I]%
\>[24]\AgdaSymbol{(}\AgdaFunction{sym}\AgdaSpace{}%
\AgdaSymbol{(}\AgdaFunction{cong}\AgdaSpace{}%
\AgdaField{proj₂}\AgdaSpace{}%
\AgdaBound{≡₁}\AgdaSymbol{))}\<%
\\
\>[24]\AgdaSymbol{(}\AgdaPostulate{lemma4}\AgdaSpace{}%
\AgdaSymbol{(}\AgdaField{proj₂}\AgdaSpace{}%
\AgdaSymbol{(}\AgdaField{proj₁}\AgdaSpace{}%
\AgdaSymbol{(}\AgdaBound{oracle}\AgdaSpace{}%
\AgdaBound{x}\AgdaSymbol{)}\AgdaSpace{}%
\AgdaBound{s₁}\AgdaSymbol{))}\AgdaSpace{}%
\AgdaBound{f₁}\AgdaSpace{}%
\AgdaBound{f₁∈}\AgdaSymbol{))}\<%
\\
\>[0][@{}l@{\AgdaIndent{0}}]%
\>[2]\AgdaKeyword{where}%
\>[781I]\AgdaFunction{f₁∈₁}\AgdaSpace{}%
\AgdaSymbol{:}\AgdaSpace{}%
\AgdaBound{f₁}\AgdaSpace{}%
\AgdaOperator{\AgdaPostulate{∈}}\AgdaSpace{}%
\AgdaPostulate{cmdWriteNames}\AgdaSpace{}%
\AgdaBound{x}\AgdaSpace{}%
\AgdaBound{s₃}\<%
\\
\>[.][@{}l@{}]\<[781I]%
\>[8]\AgdaFunction{f₁∈₁}\AgdaSpace{}%
\AgdaSymbol{=}\AgdaSpace{}%
\AgdaFunction{subst}\AgdaSpace{}%
\AgdaSymbol{(λ}\AgdaSpace{}%
\AgdaBound{ls}\AgdaSpace{}%
\AgdaSymbol{→}\AgdaSpace{}%
\AgdaBound{f₁}\AgdaSpace{}%
\AgdaOperator{\AgdaPostulate{∈}}\AgdaSpace{}%
\AgdaBound{ls}\AgdaSymbol{)}\AgdaSpace{}%
\AgdaSymbol{(}\AgdaFunction{sym}\AgdaSpace{}%
\AgdaSymbol{(}\AgdaFunction{cong}\AgdaSpace{}%
\AgdaSymbol{(}\AgdaFunction{map}\AgdaSpace{}%
\AgdaField{proj₁}\AgdaSpace{}%
\AgdaOperator{\AgdaFunction{∘}}\AgdaSpace{}%
\AgdaField{proj₂}\AgdaSymbol{)}\AgdaSpace{}%
\AgdaBound{≡₁}\AgdaSymbol{))}\AgdaSpace{}%
\AgdaBound{f₁∈}\<%
\\
\\[\AgdaEmptyExtraSkip]%
\\[\AgdaEmptyExtraSkip]%
\>[0]\AgdaFunction{doRunSoundness}\AgdaSpace{}%
\AgdaSymbol{:}\AgdaSpace{}%
\AgdaSymbol{∀}\AgdaSpace{}%
\AgdaBound{st}\AgdaSpace{}%
\AgdaBound{ls}\AgdaSpace{}%
\AgdaSymbol{\{}\AgdaBound{st₁}\AgdaSymbol{\}}\AgdaSpace{}%
\AgdaSymbol{\{}\AgdaBound{ls₁}\AgdaSymbol{\}}\AgdaSpace{}%
\AgdaBound{b}\AgdaSpace{}%
\AgdaBound{x}\AgdaSpace{}%
\AgdaSymbol{→}\AgdaSpace{}%
\AgdaPostulate{doRunWError}\AgdaSpace{}%
\AgdaSymbol{\{}\AgdaBound{b}\AgdaSymbol{\}}\AgdaSpace{}%
\AgdaSymbol{(}\AgdaBound{st}\AgdaSpace{}%
\AgdaOperator{\AgdaInductiveConstructor{,}}\AgdaSpace{}%
\AgdaBound{ls}\AgdaSymbol{)}\AgdaSpace{}%
\AgdaBound{x}\AgdaSpace{}%
\AgdaOperator{\AgdaDatatype{≡}}\AgdaSpace{}%
\AgdaInductiveConstructor{inj₂}\AgdaSpace{}%
\AgdaSymbol{(}\AgdaBound{st₁}\AgdaSpace{}%
\AgdaOperator{\AgdaInductiveConstructor{,}}\AgdaSpace{}%
\AgdaBound{ls₁}\AgdaSymbol{)}\AgdaSpace{}%
\AgdaSymbol{→}\AgdaSpace{}%
\AgdaPostulate{doRunR}\AgdaSpace{}%
\AgdaBound{st}\AgdaSpace{}%
\AgdaBound{x}\AgdaSpace{}%
\AgdaOperator{\AgdaDatatype{≡}}\AgdaSpace{}%
\AgdaBound{st₁}\<%
\\
\>[0]\AgdaFunction{doRunSoundness}\AgdaSpace{}%
\AgdaBound{st}\AgdaSpace{}%
\AgdaBound{ls}\AgdaSpace{}%
\AgdaBound{b}\AgdaSpace{}%
\AgdaBound{x}\AgdaSpace{}%
\AgdaBound{≡₁}\AgdaSpace{}%
\AgdaKeyword{with}\AgdaSpace{}%
\AgdaPostulate{checkHazard}\AgdaSpace{}%
\AgdaSymbol{(}\AgdaField{proj₁}\AgdaSpace{}%
\AgdaBound{st}\AgdaSymbol{)}\AgdaSpace{}%
\AgdaBound{x}\AgdaSpace{}%
\AgdaSymbol{\{}\AgdaBound{b}\AgdaSymbol{\}}\AgdaSpace{}%
\AgdaBound{ls}\<%
\\
\>[0]\AgdaSymbol{...}\AgdaSpace{}%
\AgdaSymbol{|}\AgdaSpace{}%
\AgdaInductiveConstructor{nothing}\AgdaSpace{}%
\AgdaSymbol{=}\AgdaSpace{}%
\AgdaFunction{cong}\AgdaSpace{}%
\AgdaField{proj₁}\AgdaSpace{}%
\AgdaSymbol{(}\AgdaFunction{inj₂-injective}\AgdaSpace{}%
\AgdaBound{≡₁}\AgdaSymbol{)}\<%
\\
\\[\AgdaEmptyExtraSkip]%
\>[0]\AgdaFunction{runSoundness}\AgdaSpace{}%
\AgdaSymbol{:}\AgdaSpace{}%
\AgdaSymbol{∀}\AgdaSpace{}%
\AgdaBound{s}\AgdaSpace{}%
\AgdaBound{m}\AgdaSpace{}%
\AgdaBound{ls}\AgdaSpace{}%
\AgdaBound{st₁}\AgdaSpace{}%
\AgdaBound{ls₁}\AgdaSpace{}%
\AgdaBound{b}\AgdaSpace{}%
\AgdaBound{x}\AgdaSpace{}%
\AgdaSymbol{→}\AgdaSpace{}%
\AgdaPostulate{runWError}\AgdaSpace{}%
\AgdaSymbol{\{}\AgdaBound{b}\AgdaSymbol{\}}\AgdaSpace{}%
\AgdaBound{x}\AgdaSpace{}%
\AgdaBound{s}\AgdaSpace{}%
\AgdaBound{m}\AgdaSpace{}%
\AgdaBound{ls}\AgdaSpace{}%
\AgdaOperator{\AgdaDatatype{≡}}\AgdaSpace{}%
\AgdaInductiveConstructor{inj₂}\AgdaSpace{}%
\AgdaSymbol{(}\AgdaBound{st₁}\AgdaSpace{}%
\AgdaOperator{\AgdaInductiveConstructor{,}}\AgdaSpace{}%
\AgdaBound{ls₁}\AgdaSymbol{)}\AgdaSpace{}%
\AgdaSymbol{→}\AgdaSpace{}%
\AgdaPostulate{runR}\AgdaSpace{}%
\AgdaBound{x}\AgdaSpace{}%
\AgdaSymbol{(}\AgdaBound{s}\AgdaSpace{}%
\AgdaOperator{\AgdaInductiveConstructor{,}}\AgdaSpace{}%
\AgdaBound{m}\AgdaSymbol{)}\AgdaSpace{}%
\AgdaOperator{\AgdaDatatype{≡}}\AgdaSpace{}%
\AgdaBound{st₁}\<%
\\
\>[0]\AgdaFunction{runSoundness}\AgdaSpace{}%
\AgdaBound{s}\AgdaSpace{}%
\AgdaBound{m}\AgdaSpace{}%
\AgdaBound{ls}\AgdaSpace{}%
\AgdaBound{st₁}\AgdaSpace{}%
\AgdaBound{ls₁}\AgdaSpace{}%
\AgdaBound{b}\AgdaSpace{}%
\AgdaBound{x}\AgdaSpace{}%
\AgdaBound{≡₁}\AgdaSpace{}%
\AgdaKeyword{with}\AgdaSpace{}%
\AgdaPostulate{run?}\AgdaSpace{}%
\AgdaBound{x}\AgdaSpace{}%
\AgdaSymbol{(}\AgdaBound{s}\AgdaSpace{}%
\AgdaOperator{\AgdaInductiveConstructor{,}}\AgdaSpace{}%
\AgdaBound{m}\AgdaSymbol{)}\<%
\\
\>[0]\AgdaSymbol{...}\AgdaSpace{}%
\AgdaSymbol{|}\AgdaSpace{}%
\AgdaInductiveConstructor{false}\AgdaSpace{}%
\AgdaSymbol{=}\AgdaSpace{}%
\AgdaFunction{cong}\AgdaSpace{}%
\AgdaField{proj₁}\AgdaSpace{}%
\AgdaSymbol{(}\AgdaFunction{inj₂-injective}\AgdaSpace{}%
\AgdaBound{≡₁}\AgdaSymbol{)}\<%
\\
\>[0]\AgdaSymbol{...}\AgdaSpace{}%
\AgdaSymbol{|}\AgdaSpace{}%
\AgdaInductiveConstructor{true}\AgdaSpace{}%
\AgdaKeyword{with}\AgdaSpace{}%
\AgdaPostulate{checkHazard}\AgdaSpace{}%
\AgdaBound{s}\AgdaSpace{}%
\AgdaBound{x}\AgdaSpace{}%
\AgdaSymbol{\{}\AgdaBound{b}\AgdaSymbol{\}}\AgdaSpace{}%
\AgdaBound{ls}\<%
\\
\>[0]\AgdaSymbol{...}\AgdaSpace{}%
\AgdaSymbol{|}\AgdaSpace{}%
\AgdaInductiveConstructor{nothing}\AgdaSpace{}%
\AgdaSymbol{=}\AgdaSpace{}%
\AgdaFunction{cong}\AgdaSpace{}%
\AgdaField{proj₁}\AgdaSpace{}%
\AgdaSymbol{(}\AgdaFunction{inj₂-injective}\AgdaSpace{}%
\AgdaBound{≡₁}\AgdaSymbol{)}\<%
\\
\>[0]\AgdaComment{-- doRunSoundness (s , m) ls b x ≡₁}\<%
\end{code}

\begin{code}[hide]%
\>[0]\AgdaFunction{soundness-inner}\AgdaSpace{}%
\AgdaSymbol{:}\AgdaSpace{}%
\AgdaSymbol{∀}\AgdaSpace{}%
\AgdaSymbol{\{}\AgdaBound{st₁}\AgdaSymbol{\}}\AgdaSpace{}%
\AgdaSymbol{\{}\AgdaBound{ls₁}\AgdaSymbol{\}}\AgdaSpace{}%
\AgdaBound{st}\AgdaSpace{}%
\AgdaBound{ls}\AgdaSpace{}%
\AgdaBound{b₁}\AgdaSpace{}%
\AgdaBound{b₂}\AgdaSpace{}%
\AgdaSymbol{→}\AgdaSpace{}%
\AgdaPostulate{rattle}\AgdaSpace{}%
\AgdaBound{b₁}\AgdaSpace{}%
\AgdaBound{b₂}\AgdaSpace{}%
\AgdaSymbol{(}\AgdaBound{st}\AgdaSpace{}%
\AgdaOperator{\AgdaInductiveConstructor{,}}\AgdaSpace{}%
\AgdaBound{ls}\AgdaSymbol{)}\AgdaSpace{}%
\AgdaOperator{\AgdaDatatype{≡}}\AgdaSpace{}%
\AgdaInductiveConstructor{inj₂}\AgdaSpace{}%
\AgdaSymbol{(}\AgdaBound{st₁}\AgdaSpace{}%
\AgdaOperator{\AgdaInductiveConstructor{,}}\AgdaSpace{}%
\AgdaBound{ls₁}\AgdaSymbol{)}\AgdaSpace{}%
\AgdaSymbol{→}\AgdaSpace{}%
\AgdaPostulate{rattle-unchecked}\AgdaSpace{}%
\AgdaBound{b₁}\AgdaSpace{}%
\AgdaBound{st}\AgdaSpace{}%
\AgdaOperator{\AgdaDatatype{≡}}\AgdaSpace{}%
\AgdaBound{st₁}\<%
\\
\>[0]\AgdaFunction{soundness-inner}\AgdaSpace{}%
\AgdaBound{st}\AgdaSpace{}%
\AgdaBound{ls}\AgdaSpace{}%
\AgdaInductiveConstructor{[]}\AgdaSpace{}%
\AgdaBound{b₂}\AgdaSpace{}%
\AgdaBound{≡₁}\AgdaSpace{}%
\AgdaSymbol{=}\AgdaSpace{}%
\AgdaFunction{cong}\AgdaSpace{}%
\AgdaField{proj₁}\AgdaSpace{}%
\AgdaSymbol{(}\AgdaFunction{inj₂-injective}\AgdaSpace{}%
\AgdaBound{≡₁}\AgdaSymbol{)}\<%
\\
\>[0]\AgdaFunction{soundness-inner}\AgdaSpace{}%
\AgdaSymbol{(}\AgdaBound{s}\AgdaSpace{}%
\AgdaOperator{\AgdaInductiveConstructor{,}}\AgdaSpace{}%
\AgdaBound{m}\AgdaSymbol{)}\AgdaSpace{}%
\AgdaBound{ls}\AgdaSpace{}%
\AgdaSymbol{(}\AgdaBound{x}\AgdaSpace{}%
\AgdaOperator{\AgdaInductiveConstructor{∷}}\AgdaSpace{}%
\AgdaBound{b₁}\AgdaSymbol{)}\AgdaSpace{}%
\AgdaBound{b₂}%
\>[40]\AgdaBound{≡₁}\AgdaSpace{}%
\AgdaKeyword{with}\AgdaSpace{}%
\AgdaPostulate{runWError}\AgdaSpace{}%
\AgdaSymbol{\{}\AgdaBound{b₂}\AgdaSymbol{\}}\AgdaSpace{}%
\AgdaBound{x}\AgdaSpace{}%
\AgdaBound{s}\AgdaSpace{}%
\AgdaBound{m}\AgdaSpace{}%
\AgdaBound{ls}\AgdaSpace{}%
\AgdaSymbol{|}\AgdaSpace{}%
\AgdaFunction{inspect}\AgdaSpace{}%
\AgdaSymbol{(}\AgdaPostulate{runWError}\AgdaSpace{}%
\AgdaSymbol{\{}\AgdaBound{b₂}\AgdaSymbol{\}}\AgdaSpace{}%
\AgdaBound{x}\AgdaSpace{}%
\AgdaBound{s}\AgdaSpace{}%
\AgdaBound{m}\AgdaSymbol{)}\AgdaSpace{}%
\AgdaBound{ls}\<%
\\
\>[0]\AgdaSymbol{...}\AgdaSpace{}%
\AgdaSymbol{|}\AgdaSpace{}%
\AgdaInductiveConstructor{inj₂}\AgdaSpace{}%
\AgdaSymbol{(}\AgdaBound{st₂}\AgdaSpace{}%
\AgdaOperator{\AgdaInductiveConstructor{,}}\AgdaSpace{}%
\AgdaBound{ls₂}\AgdaSymbol{)}\AgdaSpace{}%
\AgdaSymbol{|}\AgdaSpace{}%
\AgdaOperator{\AgdaInductiveConstructor{[}}\AgdaSpace{}%
\AgdaBound{≡₂}\AgdaSpace{}%
\AgdaOperator{\AgdaInductiveConstructor{]}}\AgdaSpace{}%
\AgdaKeyword{with}\AgdaSpace{}%
\AgdaFunction{runSoundness}\AgdaSpace{}%
\AgdaBound{s}\AgdaSpace{}%
\AgdaBound{m}\AgdaSpace{}%
\AgdaBound{ls}\AgdaSpace{}%
\AgdaBound{st₂}\AgdaSpace{}%
\AgdaBound{ls₂}\AgdaSpace{}%
\AgdaBound{b₂}\AgdaSpace{}%
\AgdaBound{x}\AgdaSpace{}%
\AgdaBound{≡₂}\<%
\\
\>[0]\AgdaSymbol{...}\AgdaSpace{}%
\AgdaSymbol{|}\AgdaSpace{}%
\AgdaBound{≡st₂}\AgdaSpace{}%
\AgdaSymbol{=}\AgdaSpace{}%
\AgdaFunction{subst}\AgdaSpace{}%
\AgdaSymbol{(λ}\AgdaSpace{}%
\AgdaBound{x₁}\AgdaSpace{}%
\AgdaSymbol{→}\AgdaSpace{}%
\AgdaPostulate{rattle-unchecked}\AgdaSpace{}%
\AgdaBound{b₁}\AgdaSpace{}%
\AgdaBound{x₁}\AgdaSpace{}%
\AgdaOperator{\AgdaDatatype{≡}}\AgdaSpace{}%
\AgdaSymbol{\AgdaUnderscore{})}\AgdaSpace{}%
\AgdaSymbol{(}\AgdaFunction{sym}\AgdaSpace{}%
\AgdaBound{≡st₂}\AgdaSymbol{)}\AgdaSpace{}%
\AgdaSymbol{(}\AgdaFunction{soundness-inner}\AgdaSpace{}%
\AgdaBound{st₂}\AgdaSpace{}%
\AgdaBound{ls₂}\AgdaSpace{}%
\AgdaBound{b₁}\AgdaSpace{}%
\AgdaBound{b₂}\AgdaSpace{}%
\AgdaBound{≡₁}\AgdaSymbol{)}\<%
\\
\\[\AgdaEmptyExtraSkip]%
\>[0]\AgdaFunction{OKBuild}\AgdaSpace{}%
\AgdaSymbol{:}\AgdaSpace{}%
\AgdaFunction{State}\AgdaSpace{}%
\AgdaSymbol{→}\AgdaSpace{}%
\AgdaPostulate{FileInfo}\AgdaSpace{}%
\AgdaSymbol{→}\AgdaSpace{}%
\AgdaPostulate{Build}\AgdaSpace{}%
\AgdaSymbol{→}\AgdaSpace{}%
\AgdaPostulate{Build}\AgdaSpace{}%
\AgdaSymbol{→}\AgdaSpace{}%
\AgdaPrimitiveType{Set}\<%
\\
\>[0]\AgdaFunction{OKBuild}\AgdaSpace{}%
\AgdaSymbol{(}\AgdaBound{s}\AgdaSpace{}%
\AgdaOperator{\AgdaInductiveConstructor{,}}\AgdaSpace{}%
\AgdaBound{mm}\AgdaSymbol{)}\AgdaSpace{}%
\AgdaBound{ls}\AgdaSpace{}%
\AgdaBound{b₁}\AgdaSpace{}%
\AgdaBound{b₂}\AgdaSpace{}%
\AgdaSymbol{=}\AgdaSpace{}%
\AgdaPostulate{DisjointBuild}\AgdaSpace{}%
\AgdaBound{s}\AgdaSpace{}%
\AgdaBound{b₁}\AgdaSpace{}%
\AgdaOperator{\AgdaFunction{×}}\AgdaSpace{}%
\AgdaDatatype{MemoryProperty}\AgdaSpace{}%
\AgdaBound{mm}\AgdaSpace{}%
\AgdaOperator{\AgdaFunction{×}}\AgdaSpace{}%
\AgdaPostulate{UniqueEvidence}\AgdaSpace{}%
\AgdaBound{b₁}\AgdaSpace{}%
\AgdaBound{b₂}\AgdaSpace{}%
\AgdaSymbol{(}\AgdaFunction{map}\AgdaSpace{}%
\AgdaField{proj₁}\AgdaSpace{}%
\AgdaBound{ls}\AgdaSymbol{)}\<%
\end{code}

\begin{code}[hide]%
\>[0]\AgdaFunction{completeness-inner}\AgdaSpace{}%
\AgdaSymbol{:}\AgdaSpace{}%
\AgdaSymbol{∀}\AgdaSpace{}%
\AgdaBound{st}\AgdaSpace{}%
\AgdaBound{ls}\AgdaSpace{}%
\AgdaBound{b₁}\AgdaSpace{}%
\AgdaBound{b₂}\AgdaSpace{}%
\AgdaSymbol{→}\AgdaSpace{}%
\AgdaFunction{OKBuild}\AgdaSpace{}%
\AgdaBound{st}\AgdaSpace{}%
\AgdaBound{ls}\AgdaSpace{}%
\AgdaBound{b₁}\AgdaSpace{}%
\AgdaBound{b₂}\AgdaSpace{}%
\AgdaSymbol{→}\AgdaSpace{}%
\AgdaPostulate{HazardFree}\AgdaSpace{}%
\AgdaSymbol{(}\AgdaField{proj₁}\AgdaSpace{}%
\AgdaBound{st}\AgdaSymbol{)}\AgdaSpace{}%
\AgdaBound{b₁}\AgdaSpace{}%
\AgdaBound{b₂}\AgdaSpace{}%
\AgdaBound{ls}\<%
\\
\>[0][@{}l@{\AgdaIndent{0}}]%
\>[13]\AgdaSymbol{→}\AgdaSpace{}%
\AgdaFunction{∃[}\AgdaSpace{}%
\AgdaBound{st₁}\AgdaSpace{}%
\AgdaFunction{]}\AgdaSymbol{(}\AgdaFunction{∃[}\AgdaSpace{}%
\AgdaBound{ls₁}\AgdaSpace{}%
\AgdaFunction{]}\AgdaSymbol{(}\AgdaPostulate{rattle}\AgdaSpace{}%
\AgdaBound{b₁}\AgdaSpace{}%
\AgdaBound{b₂}\AgdaSpace{}%
\AgdaSymbol{(}\AgdaBound{st}\AgdaSpace{}%
\AgdaOperator{\AgdaInductiveConstructor{,}}\AgdaSpace{}%
\AgdaBound{ls}\AgdaSymbol{)}\AgdaSpace{}%
\AgdaOperator{\AgdaDatatype{≡}}\AgdaSpace{}%
\AgdaInductiveConstructor{inj₂}\AgdaSpace{}%
\AgdaSymbol{(}\AgdaBound{st₁}\AgdaSpace{}%
\AgdaOperator{\AgdaInductiveConstructor{,}}\AgdaSpace{}%
\AgdaBound{ls₁}\AgdaSymbol{)))}\<%
\\
\>[0]\AgdaFunction{completeness-inner}\AgdaSpace{}%
\AgdaBound{st}\AgdaSpace{}%
\AgdaBound{ls}\AgdaSpace{}%
\AgdaInductiveConstructor{[]}\AgdaSpace{}%
\AgdaSymbol{\AgdaUnderscore{}}\AgdaSpace{}%
\AgdaSymbol{(}\AgdaBound{dsb}\AgdaSpace{}%
\AgdaOperator{\AgdaInductiveConstructor{,}}\AgdaSpace{}%
\AgdaBound{mp}\AgdaSpace{}%
\AgdaOperator{\AgdaInductiveConstructor{,}}\AgdaSpace{}%
\AgdaSymbol{(}\AgdaBound{ub₁}\AgdaSpace{}%
\AgdaOperator{\AgdaInductiveConstructor{,}}\AgdaSpace{}%
\AgdaBound{ub₂}\AgdaSpace{}%
\AgdaOperator{\AgdaInductiveConstructor{,}}\AgdaSpace{}%
\AgdaBound{uls}\AgdaSpace{}%
\AgdaOperator{\AgdaInductiveConstructor{,}}\AgdaSpace{}%
\AgdaBound{dsj}\AgdaSymbol{))}\AgdaSpace{}%
\AgdaBound{hf}\AgdaSpace{}%
\AgdaSymbol{=}\AgdaSpace{}%
\AgdaBound{st}\AgdaSpace{}%
\AgdaOperator{\AgdaInductiveConstructor{,}}\AgdaSpace{}%
\AgdaBound{ls}\AgdaSpace{}%
\AgdaOperator{\AgdaInductiveConstructor{,}}\AgdaSpace{}%
\AgdaInductiveConstructor{refl}\<%
\\
\>[0]\AgdaFunction{completeness-inner}\AgdaSpace{}%
\AgdaBound{st}\AgdaSymbol{@(}\AgdaBound{s}\AgdaSpace{}%
\AgdaOperator{\AgdaInductiveConstructor{,}}\AgdaSpace{}%
\AgdaBound{mm}\AgdaSymbol{)}\AgdaSpace{}%
\AgdaBound{ls}\AgdaSpace{}%
\AgdaSymbol{(}\AgdaBound{x}\AgdaSpace{}%
\AgdaOperator{\AgdaInductiveConstructor{∷}}\AgdaSpace{}%
\AgdaBound{b₁}\AgdaSymbol{)}\AgdaSpace{}%
\AgdaBound{b₂}\AgdaSpace{}%
\AgdaSymbol{((}\AgdaInductiveConstructor{Cons}\AgdaSpace{}%
\AgdaDottedPattern{\AgdaSymbol{.}}\AgdaDottedPattern{\AgdaBound{x}}\AgdaSpace{}%
\AgdaBound{ds}\AgdaSpace{}%
\AgdaDottedPattern{\AgdaSymbol{.}}\AgdaDottedPattern{\AgdaBound{b₁}}\AgdaSpace{}%
\AgdaBound{dsb}\AgdaSymbol{)}\AgdaSpace{}%
\AgdaOperator{\AgdaInductiveConstructor{,}}\AgdaSpace{}%
\AgdaBound{mp}\AgdaSpace{}%
\AgdaOperator{\AgdaInductiveConstructor{,}}\AgdaSpace{}%
\AgdaSymbol{((}\AgdaBound{px}\AgdaSpace{}%
\AgdaOperator{\AgdaInductiveConstructor{∷}}\AgdaSpace{}%
\AgdaBound{ub₁}\AgdaSymbol{)}\AgdaSpace{}%
\AgdaOperator{\AgdaInductiveConstructor{,}}\AgdaSpace{}%
\AgdaBound{ub₂}\AgdaSpace{}%
\AgdaOperator{\AgdaInductiveConstructor{,}}\AgdaSpace{}%
\AgdaBound{uls}\AgdaSpace{}%
\AgdaOperator{\AgdaInductiveConstructor{,}}\AgdaSpace{}%
\AgdaBound{dsj₁}\AgdaSymbol{))}%
\>[109]\AgdaSymbol{(}\AgdaBound{¬hz}\AgdaSpace{}%
\AgdaOperator{\AgdaInductiveConstructor{∷}}\AgdaSpace{}%
\AgdaBound{hf}\AgdaSymbol{)}\AgdaSpace{}%
\AgdaKeyword{with}\AgdaSpace{}%
\AgdaBound{x}\AgdaSpace{}%
\AgdaOperator{\AgdaPostulate{∈?}}\AgdaSpace{}%
\AgdaFunction{map}\AgdaSpace{}%
\AgdaField{proj₁}\AgdaSpace{}%
\AgdaBound{mm}\<%
\\
\>[0]\AgdaSymbol{...}\AgdaSpace{}%
\AgdaSymbol{|}\AgdaSpace{}%
\AgdaInductiveConstructor{yes}\AgdaSpace{}%
\AgdaBound{x∈}\AgdaSpace{}%
\AgdaKeyword{with}\AgdaSpace{}%
\AgdaPostulate{maybeAll}\AgdaSpace{}%
\AgdaSymbol{\{}\AgdaBound{s}\AgdaSymbol{\}}\AgdaSpace{}%
\AgdaSymbol{(}\AgdaPostulate{get}\AgdaSpace{}%
\AgdaBound{x}\AgdaSpace{}%
\AgdaBound{mm}\AgdaSpace{}%
\AgdaBound{x∈}\AgdaSymbol{)}\<%
\\
\>[0]\AgdaSymbol{...}\AgdaSpace{}%
\AgdaSymbol{|}\AgdaSpace{}%
\AgdaInductiveConstructor{nothing}\AgdaSpace{}%
\AgdaKeyword{with}\AgdaSpace{}%
\AgdaPostulate{checkHazard}\AgdaSpace{}%
\AgdaBound{s}\AgdaSpace{}%
\AgdaBound{x}\AgdaSpace{}%
\AgdaSymbol{\{}\AgdaBound{b₂}\AgdaSymbol{\}}\AgdaSpace{}%
\AgdaBound{ls}\<%
\\
\>[0]\AgdaSymbol{...}\AgdaSpace{}%
\AgdaSymbol{|}\AgdaSpace{}%
\AgdaInductiveConstructor{just}\AgdaSpace{}%
\AgdaBound{hz}\AgdaSpace{}%
\AgdaSymbol{=}\AgdaSpace{}%
\AgdaFunction{⊥-elim}\AgdaSpace{}%
\AgdaSymbol{(}\AgdaBound{¬hz}\AgdaSpace{}%
\AgdaBound{hz}\AgdaSymbol{)}\<%
\\
\>[0]\AgdaSymbol{...}\AgdaSpace{}%
\AgdaSymbol{|}\AgdaSpace{}%
\AgdaInductiveConstructor{nothing}\AgdaSpace{}%
\AgdaSymbol{=}\AgdaSpace{}%
\AgdaFunction{completeness-inner}\AgdaSpace{}%
\AgdaFunction{st₂}\AgdaSpace{}%
\AgdaFunction{ls₂}\AgdaSpace{}%
\AgdaBound{b₁}\AgdaSpace{}%
\AgdaBound{b₂}\AgdaSpace{}%
\AgdaSymbol{(}\AgdaBound{dsb}\AgdaSpace{}%
\AgdaOperator{\AgdaInductiveConstructor{,}}\AgdaSpace{}%
\AgdaFunction{mp₂}\AgdaSpace{}%
\AgdaOperator{\AgdaInductiveConstructor{,}}\AgdaSpace{}%
\AgdaSymbol{(}\AgdaBound{ub₁}\AgdaSpace{}%
\AgdaOperator{\AgdaInductiveConstructor{,}}\AgdaSpace{}%
\AgdaBound{ub₂}\AgdaSpace{}%
\AgdaOperator{\AgdaInductiveConstructor{,}}\AgdaSpace{}%
\AgdaFunction{uls₂}\AgdaSpace{}%
\AgdaOperator{\AgdaInductiveConstructor{,}}\AgdaSpace{}%
\AgdaFunction{dsj₂}\AgdaSymbol{))}\AgdaSpace{}%
\AgdaBound{hf}\<%
\\
\>[0][@{}l@{\AgdaIndent{0}}]%
\>[2]\AgdaKeyword{where}%
\>[1176I]\AgdaFunction{dsj₂}\AgdaSpace{}%
\AgdaSymbol{:}\AgdaSpace{}%
\AgdaFunction{Disjoint}\AgdaSpace{}%
\AgdaBound{b₁}\AgdaSpace{}%
\AgdaSymbol{(}\AgdaBound{x}\AgdaSpace{}%
\AgdaOperator{\AgdaInductiveConstructor{∷}}\AgdaSpace{}%
\AgdaFunction{map}\AgdaSpace{}%
\AgdaField{proj₁}\AgdaSpace{}%
\AgdaBound{ls}\AgdaSymbol{)}\<%
\\
\>[.][@{}l@{}]\<[1176I]%
\>[8]\AgdaFunction{dsj₂}\AgdaSpace{}%
\AgdaSymbol{=}\AgdaSpace{}%
\AgdaSymbol{λ}\AgdaSpace{}%
\AgdaBound{x₁}\AgdaSpace{}%
\AgdaSymbol{→}\AgdaSpace{}%
\AgdaBound{dsj₁}\AgdaSpace{}%
\AgdaSymbol{(}\AgdaInductiveConstructor{there}\AgdaSpace{}%
\AgdaSymbol{(}\AgdaField{proj₁}\AgdaSpace{}%
\AgdaBound{x₁}\AgdaSymbol{)}\AgdaSpace{}%
\AgdaOperator{\AgdaInductiveConstructor{,}}\AgdaSpace{}%
\AgdaFunction{tail}\AgdaSpace{}%
\AgdaSymbol{(λ}\AgdaSpace{}%
\AgdaBound{v≡x}\AgdaSpace{}%
\AgdaSymbol{→}\AgdaSpace{}%
\AgdaFunction{lookup}\AgdaSpace{}%
\AgdaBound{px}\AgdaSpace{}%
\AgdaSymbol{(}\AgdaField{proj₁}\AgdaSpace{}%
\AgdaBound{x₁}\AgdaSymbol{)}\AgdaSpace{}%
\AgdaSymbol{(}\AgdaFunction{sym}\AgdaSpace{}%
\AgdaBound{v≡x}\AgdaSymbol{))}\AgdaSpace{}%
\AgdaSymbol{(}\AgdaField{proj₂}\AgdaSpace{}%
\AgdaBound{x₁}\AgdaSymbol{))}\<%
\\
\>[8]\AgdaFunction{st₂}\AgdaSpace{}%
\AgdaSymbol{:}\AgdaSpace{}%
\AgdaFunction{State}\<%
\\
\>[8]\AgdaFunction{st₂}\AgdaSpace{}%
\AgdaSymbol{=}\AgdaSpace{}%
\AgdaPostulate{St.run}\AgdaSpace{}%
\AgdaBound{x}\AgdaSpace{}%
\AgdaBound{s}\AgdaSpace{}%
\AgdaOperator{\AgdaInductiveConstructor{,}}\AgdaSpace{}%
\AgdaFunction{save}\AgdaSpace{}%
\AgdaBound{x}\AgdaSpace{}%
\AgdaSymbol{((}\AgdaPostulate{cmdReadNames}\AgdaSpace{}%
\AgdaBound{x}\AgdaSpace{}%
\AgdaBound{s}\AgdaSymbol{)}\AgdaSpace{}%
\AgdaOperator{\AgdaFunction{++}}\AgdaSpace{}%
\AgdaSymbol{(}\AgdaPostulate{cmdWriteNames}\AgdaSpace{}%
\AgdaBound{x}\AgdaSpace{}%
\AgdaBound{s}\AgdaSymbol{))}\AgdaSpace{}%
\AgdaSymbol{(}\AgdaPostulate{St.run}\AgdaSpace{}%
\AgdaBound{x}\AgdaSpace{}%
\AgdaBound{s}\AgdaSymbol{)}\AgdaSpace{}%
\AgdaBound{mm}\<%
\\
\>[8]\AgdaFunction{ls₂}\AgdaSpace{}%
\AgdaSymbol{:}\AgdaSpace{}%
\AgdaPostulate{FileInfo}\<%
\\
\>[8]\AgdaFunction{ls₂}\AgdaSpace{}%
\AgdaSymbol{=}\AgdaSpace{}%
\AgdaPostulate{rec}\AgdaSpace{}%
\AgdaBound{s}\AgdaSpace{}%
\AgdaBound{x}\AgdaSpace{}%
\AgdaBound{ls}\<%
\\
\>[8]\AgdaFunction{uls₂}\AgdaSpace{}%
\AgdaSymbol{:}\AgdaSpace{}%
\AgdaDatatype{Unique}\AgdaSpace{}%
\AgdaSymbol{(}\AgdaBound{x}\AgdaSpace{}%
\AgdaOperator{\AgdaInductiveConstructor{∷}}\AgdaSpace{}%
\AgdaFunction{map}\AgdaSpace{}%
\AgdaField{proj₁}\AgdaSpace{}%
\AgdaBound{ls}\AgdaSymbol{)}\<%
\\
\>[8]\AgdaFunction{uls₂}\AgdaSpace{}%
\AgdaSymbol{=}\AgdaSpace{}%
\AgdaPostulate{g₂}\AgdaSpace{}%
\AgdaSymbol{(}\AgdaFunction{map}\AgdaSpace{}%
\AgdaField{proj₁}\AgdaSpace{}%
\AgdaBound{ls}\AgdaSymbol{)}\AgdaSpace{}%
\AgdaSymbol{(λ}\AgdaSpace{}%
\AgdaBound{x₁}\AgdaSpace{}%
\AgdaSymbol{→}\AgdaSpace{}%
\AgdaBound{dsj₁}\AgdaSpace{}%
\AgdaSymbol{(}\AgdaInductiveConstructor{here}\AgdaSpace{}%
\AgdaInductiveConstructor{refl}\AgdaSpace{}%
\AgdaOperator{\AgdaInductiveConstructor{,}}\AgdaSpace{}%
\AgdaBound{x₁}\AgdaSymbol{))}\AgdaSpace{}%
\AgdaOperator{\AgdaInductiveConstructor{∷}}\AgdaSpace{}%
\AgdaBound{uls}\<%
\\
\>[8]\AgdaFunction{mp₂}\AgdaSpace{}%
\AgdaSymbol{:}\AgdaSpace{}%
\AgdaDatatype{MemoryProperty}\AgdaSpace{}%
\AgdaSymbol{(}\AgdaField{proj₂}\AgdaSpace{}%
\AgdaFunction{st₂}\AgdaSymbol{)}\<%
\\
\>[8]\AgdaFunction{mp₂}\AgdaSpace{}%
\AgdaSymbol{=}\AgdaSpace{}%
\AgdaInductiveConstructor{MemoryProperty.Cons}\AgdaSpace{}%
\AgdaBound{x}\AgdaSpace{}%
\AgdaBound{s}\AgdaSpace{}%
\AgdaSymbol{(λ}\AgdaSpace{}%
\AgdaBound{f₁}\AgdaSpace{}%
\AgdaBound{x₂}\AgdaSpace{}%
\AgdaSymbol{→}\AgdaSpace{}%
\AgdaPostulate{lemma3}\AgdaSpace{}%
\AgdaBound{f₁}\AgdaSpace{}%
\AgdaSymbol{(}\AgdaField{proj₂}\AgdaSpace{}%
\AgdaSymbol{(}\AgdaField{proj₁}\AgdaSpace{}%
\AgdaSymbol{(}\AgdaBound{oracle}\AgdaSpace{}%
\AgdaBound{x}\AgdaSymbol{)}\AgdaSpace{}%
\AgdaBound{s}\AgdaSymbol{))}\AgdaSpace{}%
\AgdaSymbol{λ}\AgdaSpace{}%
\AgdaBound{x₃}\AgdaSpace{}%
\AgdaSymbol{→}\AgdaSpace{}%
\AgdaBound{ds}\AgdaSpace{}%
\AgdaSymbol{(}\AgdaBound{x₂}\AgdaSpace{}%
\AgdaOperator{\AgdaInductiveConstructor{,}}\AgdaSpace{}%
\AgdaBound{x₃}\AgdaSymbol{))}\AgdaSpace{}%
\AgdaBound{mp}\<%
\\
\\[\AgdaEmptyExtraSkip]%
\>[0]\AgdaFunction{completeness-inner}\AgdaSpace{}%
\AgdaBound{st}\AgdaSymbol{@(}\AgdaBound{s}\AgdaSpace{}%
\AgdaOperator{\AgdaInductiveConstructor{,}}\AgdaSpace{}%
\AgdaBound{mm}\AgdaSymbol{)}\AgdaSpace{}%
\AgdaBound{ls}\AgdaSpace{}%
\AgdaSymbol{(}\AgdaBound{x}\AgdaSpace{}%
\AgdaOperator{\AgdaInductiveConstructor{∷}}\AgdaSpace{}%
\AgdaBound{b₁}\AgdaSymbol{)}\AgdaSpace{}%
\AgdaBound{b₂}\AgdaSpace{}%
\AgdaSymbol{((}\AgdaInductiveConstructor{Cons}\AgdaSpace{}%
\AgdaDottedPattern{\AgdaSymbol{.}}\AgdaDottedPattern{\AgdaBound{x}}\AgdaSpace{}%
\AgdaBound{x₁}\AgdaSpace{}%
\AgdaDottedPattern{\AgdaSymbol{.}}\AgdaDottedPattern{\AgdaBound{b₁}}\AgdaSpace{}%
\AgdaBound{dsb}\AgdaSymbol{)}\AgdaSpace{}%
\AgdaOperator{\AgdaInductiveConstructor{,}}\AgdaSpace{}%
\AgdaBound{mp}\AgdaSpace{}%
\AgdaOperator{\AgdaInductiveConstructor{,}}\AgdaSpace{}%
\AgdaSymbol{((}\AgdaBound{px}\AgdaSpace{}%
\AgdaOperator{\AgdaInductiveConstructor{∷}}\AgdaSpace{}%
\AgdaBound{ub₁}\AgdaSymbol{)}\AgdaSpace{}%
\AgdaOperator{\AgdaInductiveConstructor{,}}\AgdaSpace{}%
\AgdaBound{ub₂}\AgdaSpace{}%
\AgdaOperator{\AgdaInductiveConstructor{,}}\AgdaSpace{}%
\AgdaBound{uls}\AgdaSpace{}%
\AgdaOperator{\AgdaInductiveConstructor{,}}\AgdaSpace{}%
\AgdaBound{dsj₁}\AgdaSymbol{))}%
\>[109]\AgdaSymbol{(}\AgdaBound{¬hz}\AgdaSpace{}%
\AgdaOperator{\AgdaInductiveConstructor{∷}}\AgdaSpace{}%
\AgdaBound{hf}\AgdaSymbol{)}\AgdaSpace{}%
\AgdaSymbol{|}\AgdaSpace{}%
\AgdaInductiveConstructor{yes}\AgdaSpace{}%
\AgdaBound{x∈}\AgdaSpace{}%
\AgdaSymbol{|}\AgdaSpace{}%
\AgdaInductiveConstructor{just}\AgdaSpace{}%
\AgdaBound{all₁}\AgdaSpace{}%
\AgdaKeyword{with}\AgdaSpace{}%
\AgdaFunction{noEffect}\AgdaSpace{}%
\AgdaBound{x}\AgdaSpace{}%
\AgdaSymbol{(λ}\AgdaSpace{}%
\AgdaBound{f₁}\AgdaSpace{}%
\AgdaSymbol{→}\AgdaSpace{}%
\AgdaInductiveConstructor{refl}\AgdaSymbol{)}\AgdaSpace{}%
\AgdaBound{mp}\AgdaSpace{}%
\AgdaBound{x∈}\AgdaSpace{}%
\AgdaBound{all₁}\<%
\\
\>[0]\AgdaSymbol{...}\AgdaSpace{}%
\AgdaSymbol{|}\AgdaSpace{}%
\AgdaBound{∀₁}\AgdaSpace{}%
\AgdaSymbol{=}\AgdaSpace{}%
\AgdaFunction{completeness-inner}\AgdaSpace{}%
\AgdaBound{st}\AgdaSpace{}%
\AgdaBound{ls}\AgdaSpace{}%
\AgdaBound{b₁}\AgdaSpace{}%
\AgdaBound{b₂}\AgdaSpace{}%
\AgdaSymbol{((}\AgdaPostulate{dsj-≡}\AgdaSpace{}%
\AgdaSymbol{(}\AgdaPostulate{St.run}\AgdaSpace{}%
\AgdaBound{x}\AgdaSpace{}%
\AgdaBound{s}\AgdaSymbol{)}\AgdaSpace{}%
\AgdaBound{s}\AgdaSpace{}%
\AgdaBound{b₁}\AgdaSpace{}%
\AgdaBound{∀₁}\AgdaSpace{}%
\AgdaBound{dsb}\AgdaSymbol{)}\AgdaSpace{}%
\AgdaOperator{\AgdaInductiveConstructor{,}}\AgdaSpace{}%
\AgdaBound{mp}\AgdaSpace{}%
\AgdaOperator{\AgdaInductiveConstructor{,}}\AgdaSpace{}%
\AgdaSymbol{(}\AgdaBound{ub₁}\AgdaSpace{}%
\AgdaOperator{\AgdaInductiveConstructor{,}}\AgdaSpace{}%
\AgdaBound{ub₂}\AgdaSpace{}%
\AgdaOperator{\AgdaInductiveConstructor{,}}\AgdaSpace{}%
\AgdaBound{uls}\AgdaSpace{}%
\AgdaOperator{\AgdaInductiveConstructor{,}}\AgdaSpace{}%
\AgdaFunction{dsj₃}\AgdaSymbol{))}%
\>[110]\AgdaFunction{hf₂}\<%
\\
\>[0][@{}l@{\AgdaIndent{0}}]%
\>[2]\AgdaKeyword{where}%
\>[1351I]\AgdaFunction{dsj₃}\AgdaSpace{}%
\AgdaSymbol{:}\AgdaSpace{}%
\AgdaFunction{Disjoint}\AgdaSpace{}%
\AgdaBound{b₁}\AgdaSpace{}%
\AgdaSymbol{(}\AgdaFunction{map}\AgdaSpace{}%
\AgdaField{proj₁}\AgdaSpace{}%
\AgdaBound{ls}\AgdaSymbol{)}\<%
\\
\>[.][@{}l@{}]\<[1351I]%
\>[8]\AgdaFunction{dsj₃}\AgdaSpace{}%
\AgdaSymbol{=}\AgdaSpace{}%
\AgdaSymbol{λ}\AgdaSpace{}%
\AgdaBound{x₂}\AgdaSpace{}%
\AgdaSymbol{→}\AgdaSpace{}%
\AgdaBound{dsj₁}\AgdaSpace{}%
\AgdaSymbol{(}\AgdaInductiveConstructor{there}\AgdaSpace{}%
\AgdaSymbol{(}\AgdaField{proj₁}\AgdaSpace{}%
\AgdaBound{x₂}\AgdaSymbol{)}\AgdaSpace{}%
\AgdaOperator{\AgdaInductiveConstructor{,}}\AgdaSpace{}%
\AgdaField{proj₂}\AgdaSpace{}%
\AgdaBound{x₂}\AgdaSymbol{)}\<%
\\
\>[8]\AgdaFunction{dsj₂}\AgdaSpace{}%
\AgdaSymbol{:}\AgdaSpace{}%
\AgdaFunction{Disjoint}\AgdaSpace{}%
\AgdaBound{b₁}\AgdaSpace{}%
\AgdaSymbol{(}\AgdaBound{x}\AgdaSpace{}%
\AgdaOperator{\AgdaInductiveConstructor{∷}}\AgdaSpace{}%
\AgdaFunction{map}\AgdaSpace{}%
\AgdaField{proj₁}\AgdaSpace{}%
\AgdaBound{ls}\AgdaSymbol{)}\<%
\\
\>[8]\AgdaFunction{dsj₂}\AgdaSpace{}%
\AgdaSymbol{=}\AgdaSpace{}%
\AgdaSymbol{λ}\AgdaSpace{}%
\AgdaBound{x₁}\AgdaSpace{}%
\AgdaSymbol{→}\AgdaSpace{}%
\AgdaBound{dsj₁}\AgdaSpace{}%
\AgdaSymbol{(}\AgdaInductiveConstructor{there}\AgdaSpace{}%
\AgdaSymbol{(}\AgdaField{proj₁}\AgdaSpace{}%
\AgdaBound{x₁}\AgdaSymbol{)}\AgdaSpace{}%
\AgdaOperator{\AgdaInductiveConstructor{,}}\AgdaSpace{}%
\AgdaFunction{tail}\AgdaSpace{}%
\AgdaSymbol{(λ}\AgdaSpace{}%
\AgdaBound{v≡x}\AgdaSpace{}%
\AgdaSymbol{→}\AgdaSpace{}%
\AgdaFunction{lookup}\AgdaSpace{}%
\AgdaBound{px}\AgdaSpace{}%
\AgdaSymbol{(}\AgdaField{proj₁}\AgdaSpace{}%
\AgdaBound{x₁}\AgdaSymbol{)}\AgdaSpace{}%
\AgdaSymbol{(}\AgdaFunction{sym}\AgdaSpace{}%
\AgdaBound{v≡x}\AgdaSymbol{))}\AgdaSpace{}%
\AgdaSymbol{(}\AgdaField{proj₂}\AgdaSpace{}%
\AgdaBound{x₁}\AgdaSymbol{))}\<%
\\
\>[8]\AgdaFunction{uls₂}\AgdaSpace{}%
\AgdaSymbol{:}\AgdaSpace{}%
\AgdaDatatype{Unique}\AgdaSpace{}%
\AgdaSymbol{(}\AgdaBound{x}\AgdaSpace{}%
\AgdaOperator{\AgdaInductiveConstructor{∷}}\AgdaSpace{}%
\AgdaFunction{map}\AgdaSpace{}%
\AgdaField{proj₁}\AgdaSpace{}%
\AgdaBound{ls}\AgdaSymbol{)}\<%
\\
\>[8]\AgdaFunction{uls₂}\AgdaSpace{}%
\AgdaSymbol{=}\AgdaSpace{}%
\AgdaPostulate{g₂}\AgdaSpace{}%
\AgdaSymbol{(}\AgdaFunction{map}\AgdaSpace{}%
\AgdaField{proj₁}\AgdaSpace{}%
\AgdaBound{ls}\AgdaSymbol{)}\AgdaSpace{}%
\AgdaSymbol{(λ}\AgdaSpace{}%
\AgdaBound{x₁}\AgdaSpace{}%
\AgdaSymbol{→}\AgdaSpace{}%
\AgdaBound{dsj₁}\AgdaSpace{}%
\AgdaSymbol{(}\AgdaInductiveConstructor{here}\AgdaSpace{}%
\AgdaInductiveConstructor{refl}\AgdaSpace{}%
\AgdaOperator{\AgdaInductiveConstructor{,}}\AgdaSpace{}%
\AgdaBound{x₁}\AgdaSymbol{))}\AgdaSpace{}%
\AgdaOperator{\AgdaInductiveConstructor{∷}}\AgdaSpace{}%
\AgdaBound{uls}\<%
\\
\>[8]\AgdaFunction{hf₂}\AgdaSpace{}%
\AgdaSymbol{:}\AgdaSpace{}%
\AgdaPostulate{HazardFree}\AgdaSpace{}%
\AgdaBound{s}\AgdaSpace{}%
\AgdaBound{b₁}\AgdaSpace{}%
\AgdaBound{b₂}\AgdaSpace{}%
\AgdaBound{ls}\<%
\\
\>[8]\AgdaFunction{hf₂}\AgdaSpace{}%
\AgdaSymbol{=}\AgdaSpace{}%
\AgdaPostulate{hf-still}\AgdaSpace{}%
\AgdaBound{b₁}\AgdaSpace{}%
\AgdaInductiveConstructor{[]}\AgdaSpace{}%
\AgdaSymbol{((}\AgdaBound{x}\AgdaSpace{}%
\AgdaOperator{\AgdaInductiveConstructor{,}}\AgdaSpace{}%
\AgdaPostulate{cmdReadNames}\AgdaSpace{}%
\AgdaBound{x}\AgdaSpace{}%
\AgdaBound{s}\AgdaSpace{}%
\AgdaOperator{\AgdaInductiveConstructor{,}}\AgdaSpace{}%
\AgdaPostulate{cmdWriteNames}\AgdaSpace{}%
\AgdaBound{x}\AgdaSpace{}%
\AgdaBound{s}\AgdaSymbol{)}\AgdaSpace{}%
\AgdaOperator{\AgdaInductiveConstructor{∷}}\AgdaSpace{}%
\AgdaInductiveConstructor{[]}\AgdaSymbol{)}\AgdaSpace{}%
\AgdaBound{ls}\AgdaSpace{}%
\AgdaSymbol{(λ}\AgdaSpace{}%
\AgdaBound{f₁}\AgdaSpace{}%
\AgdaBound{x₂}\AgdaSpace{}%
\AgdaSymbol{→}\AgdaSpace{}%
\AgdaFunction{sym}\AgdaSpace{}%
\AgdaSymbol{(}\AgdaBound{∀₁}\AgdaSpace{}%
\AgdaBound{f₁}\AgdaSymbol{))}\AgdaSpace{}%
\AgdaBound{ub₁}\AgdaSpace{}%
\AgdaFunction{uls₂}\AgdaSpace{}%
\AgdaFunction{dsj₂}\AgdaSpace{}%
\AgdaBound{hf}\<%
\\
\>[0]\<%
\\
\>[0]\AgdaFunction{completeness-inner}\AgdaSpace{}%
\AgdaBound{st}\AgdaSymbol{@(}\AgdaBound{s}\AgdaSpace{}%
\AgdaOperator{\AgdaInductiveConstructor{,}}\AgdaSpace{}%
\AgdaBound{mm}\AgdaSymbol{)}\AgdaSpace{}%
\AgdaBound{ls}\AgdaSpace{}%
\AgdaSymbol{(}\AgdaBound{x}\AgdaSpace{}%
\AgdaOperator{\AgdaInductiveConstructor{∷}}\AgdaSpace{}%
\AgdaBound{b₁}\AgdaSymbol{)}\AgdaSpace{}%
\AgdaBound{b₂}\AgdaSpace{}%
\AgdaSymbol{((}\AgdaInductiveConstructor{Cons}\AgdaSpace{}%
\AgdaDottedPattern{\AgdaSymbol{.}}\AgdaDottedPattern{\AgdaBound{x}}\AgdaSpace{}%
\AgdaBound{ds}\AgdaSpace{}%
\AgdaDottedPattern{\AgdaSymbol{.}}\AgdaDottedPattern{\AgdaBound{b₁}}\AgdaSpace{}%
\AgdaBound{dsb}\AgdaSymbol{)}\AgdaSpace{}%
\AgdaOperator{\AgdaInductiveConstructor{,}}\AgdaSpace{}%
\AgdaBound{mp}\AgdaSpace{}%
\AgdaOperator{\AgdaInductiveConstructor{,}}\AgdaSpace{}%
\AgdaSymbol{((}\AgdaBound{px}\AgdaSpace{}%
\AgdaOperator{\AgdaInductiveConstructor{∷}}\AgdaSpace{}%
\AgdaBound{ub₁}\AgdaSymbol{)}\AgdaSpace{}%
\AgdaOperator{\AgdaInductiveConstructor{,}}\AgdaSpace{}%
\AgdaBound{ub₂}\AgdaSpace{}%
\AgdaOperator{\AgdaInductiveConstructor{,}}\AgdaSpace{}%
\AgdaBound{uls}\AgdaSpace{}%
\AgdaOperator{\AgdaInductiveConstructor{,}}\AgdaSpace{}%
\AgdaBound{dsj₁}\AgdaSymbol{))}%
\>[109]\AgdaSymbol{(}\AgdaBound{¬hz}\AgdaSpace{}%
\AgdaOperator{\AgdaInductiveConstructor{∷}}\AgdaSpace{}%
\AgdaBound{hf}\AgdaSymbol{)}\AgdaSpace{}%
\AgdaSymbol{|}%
\>[123]\AgdaInductiveConstructor{no}\AgdaSpace{}%
\AgdaBound{x∉}\AgdaSpace{}%
\AgdaKeyword{with}\AgdaSpace{}%
\AgdaPostulate{checkHazard}\AgdaSpace{}%
\AgdaBound{s}\AgdaSpace{}%
\AgdaBound{x}\AgdaSpace{}%
\AgdaSymbol{\{}\AgdaBound{b₂}\AgdaSymbol{\}}\AgdaSpace{}%
\AgdaBound{ls}\<%
\\
\>[0]\AgdaSymbol{...}\AgdaSpace{}%
\AgdaSymbol{|}\AgdaSpace{}%
\AgdaInductiveConstructor{just}\AgdaSpace{}%
\AgdaBound{hz}\AgdaSpace{}%
\AgdaSymbol{=}\AgdaSpace{}%
\AgdaFunction{⊥-elim}\AgdaSpace{}%
\AgdaSymbol{(}\AgdaBound{¬hz}\AgdaSpace{}%
\AgdaBound{hz}\AgdaSymbol{)}\<%
\\
\>[0]\AgdaSymbol{...}\AgdaSpace{}%
\AgdaSymbol{|}\AgdaSpace{}%
\AgdaInductiveConstructor{nothing}\AgdaSpace{}%
\AgdaSymbol{=}\AgdaSpace{}%
\AgdaFunction{completeness-inner}\AgdaSpace{}%
\AgdaSymbol{(}\AgdaPostulate{St.run}\AgdaSpace{}%
\AgdaBound{x}\AgdaSpace{}%
\AgdaBound{s}\AgdaSpace{}%
\AgdaOperator{\AgdaInductiveConstructor{,}}\AgdaSpace{}%
\AgdaFunction{save}\AgdaSpace{}%
\AgdaBound{x}\AgdaSpace{}%
\AgdaSymbol{((}\AgdaPostulate{cmdReadNames}\AgdaSpace{}%
\AgdaBound{x}\AgdaSpace{}%
\AgdaBound{s}\AgdaSymbol{)}\AgdaSpace{}%
\AgdaOperator{\AgdaFunction{++}}\AgdaSpace{}%
\AgdaSymbol{(}\AgdaPostulate{cmdWriteNames}\AgdaSpace{}%
\AgdaBound{x}\AgdaSpace{}%
\AgdaBound{s}\AgdaSymbol{))}\AgdaSpace{}%
\AgdaSymbol{(}\AgdaPostulate{St.run}\AgdaSpace{}%
\AgdaBound{x}\AgdaSpace{}%
\AgdaBound{s}\AgdaSymbol{)}\AgdaSpace{}%
\AgdaBound{mm}\AgdaSymbol{)}\AgdaSpace{}%
\AgdaSymbol{(}\AgdaPostulate{rec}\AgdaSpace{}%
\AgdaBound{s}\AgdaSpace{}%
\AgdaBound{x}\AgdaSpace{}%
\AgdaBound{ls}\AgdaSymbol{)}\AgdaSpace{}%
\AgdaBound{b₁}\AgdaSpace{}%
\AgdaBound{b₂}\AgdaSpace{}%
\AgdaSymbol{(}\AgdaBound{dsb}\AgdaSpace{}%
\AgdaOperator{\AgdaInductiveConstructor{,}}\AgdaSpace{}%
\AgdaSymbol{(}\AgdaInductiveConstructor{MemoryProperty.Cons}\AgdaSpace{}%
\AgdaBound{x}\AgdaSpace{}%
\AgdaBound{s}\AgdaSpace{}%
\AgdaSymbol{(λ}\AgdaSpace{}%
\AgdaBound{f₁}\AgdaSpace{}%
\AgdaBound{x₂}\AgdaSpace{}%
\AgdaSymbol{→}\AgdaSpace{}%
\AgdaPostulate{lemma3}\AgdaSpace{}%
\AgdaBound{f₁}\AgdaSpace{}%
\AgdaSymbol{(}\AgdaField{proj₂}\AgdaSpace{}%
\AgdaSymbol{(}\AgdaField{proj₁}\AgdaSpace{}%
\AgdaSymbol{(}\AgdaBound{oracle}\AgdaSpace{}%
\AgdaBound{x}\AgdaSymbol{)}\AgdaSpace{}%
\AgdaBound{s}\AgdaSymbol{))}\AgdaSpace{}%
\AgdaSymbol{λ}\AgdaSpace{}%
\AgdaBound{x₃}\AgdaSpace{}%
\AgdaSymbol{→}\AgdaSpace{}%
\AgdaBound{ds}\AgdaSpace{}%
\AgdaSymbol{(}\AgdaBound{x₂}\AgdaSpace{}%
\AgdaOperator{\AgdaInductiveConstructor{,}}\AgdaSpace{}%
\AgdaBound{x₃}\AgdaSymbol{))}\AgdaSpace{}%
\AgdaBound{mp}\AgdaSymbol{)}\AgdaSpace{}%
\AgdaOperator{\AgdaInductiveConstructor{,}}\AgdaSpace{}%
\AgdaSymbol{(}\AgdaBound{ub₁}\AgdaSpace{}%
\AgdaOperator{\AgdaInductiveConstructor{,}}\AgdaSpace{}%
\AgdaBound{ub₂}\AgdaSpace{}%
\AgdaOperator{\AgdaInductiveConstructor{,}}\AgdaSpace{}%
\AgdaSymbol{(}\AgdaPostulate{g₂}\AgdaSpace{}%
\AgdaSymbol{(}\AgdaFunction{map}\AgdaSpace{}%
\AgdaField{proj₁}\AgdaSpace{}%
\AgdaBound{ls}\AgdaSymbol{)}\AgdaSpace{}%
\AgdaSymbol{(λ}\AgdaSpace{}%
\AgdaBound{x₁}\AgdaSpace{}%
\AgdaSymbol{→}\AgdaSpace{}%
\AgdaBound{dsj₁}\AgdaSpace{}%
\AgdaSymbol{(}\AgdaInductiveConstructor{here}\AgdaSpace{}%
\AgdaInductiveConstructor{refl}\AgdaSpace{}%
\AgdaOperator{\AgdaInductiveConstructor{,}}\AgdaSpace{}%
\AgdaBound{x₁}\AgdaSymbol{))}\AgdaSpace{}%
\AgdaOperator{\AgdaInductiveConstructor{∷}}\AgdaSpace{}%
\AgdaBound{uls}\AgdaSymbol{)}\AgdaSpace{}%
\AgdaOperator{\AgdaInductiveConstructor{,}}\AgdaSpace{}%
\AgdaFunction{dsj₂}\AgdaSymbol{))}\AgdaSpace{}%
\AgdaBound{hf}\<%
\\
\>[0][@{}l@{\AgdaIndent{0}}]%
\>[2]\AgdaKeyword{where}%
\>[1568I]\AgdaFunction{dsj₂}\AgdaSpace{}%
\AgdaSymbol{:}\AgdaSpace{}%
\AgdaFunction{Disjoint}\AgdaSpace{}%
\AgdaBound{b₁}\AgdaSpace{}%
\AgdaSymbol{(}\AgdaBound{x}\AgdaSpace{}%
\AgdaOperator{\AgdaInductiveConstructor{∷}}\AgdaSpace{}%
\AgdaFunction{map}\AgdaSpace{}%
\AgdaField{proj₁}\AgdaSpace{}%
\AgdaBound{ls}\AgdaSymbol{)}\<%
\\
\>[.][@{}l@{}]\<[1568I]%
\>[8]\AgdaFunction{dsj₂}\AgdaSpace{}%
\AgdaSymbol{=}\AgdaSpace{}%
\AgdaSymbol{λ}\AgdaSpace{}%
\AgdaBound{x₁}\AgdaSpace{}%
\AgdaSymbol{→}\AgdaSpace{}%
\AgdaBound{dsj₁}\AgdaSpace{}%
\AgdaSymbol{(}\AgdaInductiveConstructor{there}\AgdaSpace{}%
\AgdaSymbol{(}\AgdaField{proj₁}\AgdaSpace{}%
\AgdaBound{x₁}\AgdaSymbol{)}\AgdaSpace{}%
\AgdaOperator{\AgdaInductiveConstructor{,}}\AgdaSpace{}%
\AgdaFunction{tail}\AgdaSpace{}%
\AgdaSymbol{(λ}\AgdaSpace{}%
\AgdaBound{v≡x}\AgdaSpace{}%
\AgdaSymbol{→}\AgdaSpace{}%
\AgdaFunction{lookup}\AgdaSpace{}%
\AgdaBound{px}\AgdaSpace{}%
\AgdaSymbol{(}\AgdaField{proj₁}\AgdaSpace{}%
\AgdaBound{x₁}\AgdaSymbol{)}\AgdaSpace{}%
\AgdaSymbol{(}\AgdaFunction{sym}\AgdaSpace{}%
\AgdaBound{v≡x}\AgdaSymbol{))}\AgdaSpace{}%
\AgdaSymbol{(}\AgdaField{proj₂}\AgdaSpace{}%
\AgdaBound{x₁}\AgdaSymbol{))}\<%
\end{code}

\newcommand{\completeness}{%
\begin{code}%
\>[0]\AgdaFunction{completeness}\AgdaSpace{}%
\AgdaSymbol{:}\AgdaSpace{}%
\AgdaSymbol{∀}\AgdaSpace{}%
\AgdaBound{s}\AgdaSpace{}%
\AgdaBound{br}\AgdaSpace{}%
\AgdaBound{bs}\AgdaSpace{}%
\AgdaSymbol{→}\AgdaSpace{}%
\AgdaPostulate{PreCond}\AgdaSpace{}%
\AgdaBound{s}\AgdaSpace{}%
\AgdaBound{br}\AgdaSpace{}%
\AgdaBound{bs}\AgdaSpace{}%
\AgdaSymbol{→}\AgdaSpace{}%
\AgdaPostulate{HazardFree}\AgdaSpace{}%
\AgdaBound{s}\AgdaSpace{}%
\AgdaBound{br}\AgdaSpace{}%
\AgdaBound{bs}\AgdaSpace{}%
\AgdaInductiveConstructor{[]}\AgdaSpace{}%
\AgdaSymbol{→}\AgdaSpace{}%
\AgdaFunction{∃[}\AgdaSpace{}%
\AgdaBound{st}\AgdaSpace{}%
\AgdaFunction{]}\AgdaSymbol{(}\AgdaFunction{∃[}\AgdaSpace{}%
\AgdaBound{ls}\AgdaSpace{}%
\AgdaFunction{]}\AgdaSymbol{(}\AgdaPostulate{rattle}\AgdaSpace{}%
\AgdaBound{br}\AgdaSpace{}%
\AgdaBound{bs}\AgdaSpace{}%
\AgdaSymbol{((}\AgdaBound{s}\AgdaSpace{}%
\AgdaOperator{\AgdaInductiveConstructor{,}}\AgdaSpace{}%
\AgdaInductiveConstructor{[]}\AgdaSymbol{)}\AgdaSpace{}%
\AgdaOperator{\AgdaInductiveConstructor{,}}\AgdaSpace{}%
\AgdaInductiveConstructor{[]}\AgdaSymbol{)}\AgdaSpace{}%
\AgdaOperator{\AgdaDatatype{≡}}\AgdaSpace{}%
\AgdaInductiveConstructor{inj₂}\AgdaSpace{}%
\AgdaSymbol{(}\AgdaBound{st}\AgdaSpace{}%
\AgdaOperator{\AgdaInductiveConstructor{,}}\AgdaSpace{}%
\AgdaBound{ls}\AgdaSymbol{)))}\<%
\end{code}}
\begin{code}[hide]%
\>[0]\AgdaFunction{completeness}\AgdaSpace{}%
\AgdaBound{s}\AgdaSpace{}%
\AgdaBound{br}\AgdaSpace{}%
\AgdaBound{bc}\AgdaSpace{}%
\AgdaSymbol{(}\AgdaBound{dsb}\AgdaSpace{}%
\AgdaOperator{\AgdaInductiveConstructor{,}}\AgdaSpace{}%
\AgdaBound{ubr}\AgdaSpace{}%
\AgdaOperator{\AgdaInductiveConstructor{,}}\AgdaSpace{}%
\AgdaBound{ubc}\AgdaSpace{}%
\AgdaOperator{\AgdaInductiveConstructor{,}}\AgdaSpace{}%
\AgdaSymbol{\AgdaUnderscore{})}\AgdaSpace{}%
\AgdaBound{hf}\AgdaSpace{}%
\AgdaSymbol{=}\AgdaSpace{}%
\AgdaFunction{completeness-inner}\AgdaSpace{}%
\AgdaSymbol{(}\AgdaBound{s}\AgdaSpace{}%
\AgdaOperator{\AgdaInductiveConstructor{,}}\AgdaSpace{}%
\AgdaInductiveConstructor{[]}\AgdaSymbol{)}\AgdaSpace{}%
\AgdaInductiveConstructor{[]}\AgdaSpace{}%
\AgdaBound{br}\AgdaSpace{}%
\AgdaBound{bc}\AgdaSpace{}%
\AgdaSymbol{(}\AgdaBound{dsb}\AgdaSpace{}%
\AgdaOperator{\AgdaInductiveConstructor{,}}\AgdaSpace{}%
\AgdaSymbol{(}\AgdaInductiveConstructor{[]}\AgdaSpace{}%
\AgdaOperator{\AgdaInductiveConstructor{,}}\AgdaSpace{}%
\AgdaBound{ubr}\AgdaSpace{}%
\AgdaOperator{\AgdaInductiveConstructor{,}}\AgdaSpace{}%
\AgdaSymbol{(}\AgdaBound{ubc}\AgdaSpace{}%
\AgdaOperator{\AgdaInductiveConstructor{,}}\AgdaSpace{}%
\AgdaSymbol{(}\AgdaInductiveConstructor{Data.List.Relation.Unary.AllPairs.[]}\AgdaSpace{}%
\AgdaOperator{\AgdaInductiveConstructor{,}}\AgdaSpace{}%
\AgdaFunction{g₁}\AgdaSymbol{))))}\AgdaSpace{}%
\AgdaBound{hf}\<%
\\
\>[0][@{}l@{\AgdaIndent{0}}]%
\>[2]\AgdaKeyword{where}%
\>[1663I]\AgdaFunction{g₁}\AgdaSpace{}%
\AgdaSymbol{:}\AgdaSpace{}%
\AgdaFunction{Disjoint}\AgdaSpace{}%
\AgdaBound{br}\AgdaSpace{}%
\AgdaInductiveConstructor{[]}\<%
\\
\>[.][@{}l@{}]\<[1663I]%
\>[8]\AgdaFunction{g₁}\AgdaSpace{}%
\AgdaSymbol{()}\<%
\end{code}

\begin{code}[hide]%
\>[0]\AgdaFunction{script≡rattle-inner}\AgdaSpace{}%
\AgdaSymbol{:}\AgdaSpace{}%
\AgdaSymbol{∀}\AgdaSpace{}%
\AgdaSymbol{\{}\AgdaBound{s₁}\AgdaSymbol{\}}\AgdaSpace{}%
\AgdaSymbol{\{}\AgdaBound{s₂}\AgdaSymbol{\}}\AgdaSpace{}%
\AgdaBound{m}\AgdaSpace{}%
\AgdaBound{b₁}\AgdaSpace{}%
\AgdaSymbol{→}\AgdaSpace{}%
\AgdaSymbol{(∀}\AgdaSpace{}%
\AgdaBound{f₁}\AgdaSpace{}%
\AgdaSymbol{→}\AgdaSpace{}%
\AgdaBound{s₁}\AgdaSpace{}%
\AgdaBound{f₁}\AgdaSpace{}%
\AgdaOperator{\AgdaDatatype{≡}}\AgdaSpace{}%
\AgdaBound{s₂}\AgdaSpace{}%
\AgdaBound{f₁}\AgdaSymbol{)}\AgdaSpace{}%
\AgdaSymbol{→}\AgdaSpace{}%
\AgdaPostulate{DisjointBuild}\AgdaSpace{}%
\AgdaBound{s₂}\AgdaSpace{}%
\AgdaBound{b₁}\AgdaSpace{}%
\AgdaSymbol{→}\AgdaSpace{}%
\AgdaDatatype{MemoryProperty}\AgdaSpace{}%
\AgdaBound{m}\<%
\\
\>[0][@{}l@{\AgdaIndent{0}}]%
\>[14]\AgdaSymbol{→}\AgdaSpace{}%
\AgdaSymbol{(∀}\AgdaSpace{}%
\AgdaBound{f₁}\AgdaSpace{}%
\AgdaSymbol{→}\AgdaSpace{}%
\AgdaPostulate{script}\AgdaSpace{}%
\AgdaBound{b₁}\AgdaSpace{}%
\AgdaBound{s₁}\AgdaSpace{}%
\AgdaBound{f₁}\AgdaSpace{}%
\AgdaOperator{\AgdaDatatype{≡}}\AgdaSpace{}%
\AgdaField{proj₁}\AgdaSpace{}%
\AgdaSymbol{(}\AgdaPostulate{rattle-unchecked}\AgdaSpace{}%
\AgdaBound{b₁}\AgdaSpace{}%
\AgdaSymbol{(}\AgdaBound{s₂}\AgdaSpace{}%
\AgdaOperator{\AgdaInductiveConstructor{,}}\AgdaSpace{}%
\AgdaBound{m}\AgdaSymbol{))}\AgdaSpace{}%
\AgdaBound{f₁}\AgdaSymbol{)}\<%
\\
\>[0]\AgdaFunction{script≡rattle-inner}\AgdaSpace{}%
\AgdaBound{mm}\AgdaSpace{}%
\AgdaInductiveConstructor{[]}\AgdaSpace{}%
\AgdaBound{∀₁}\AgdaSpace{}%
\AgdaBound{dsb}\AgdaSpace{}%
\AgdaBound{mp}\AgdaSpace{}%
\AgdaSymbol{=}\AgdaSpace{}%
\AgdaBound{∀₁}\<%
\\
\>[0]\AgdaFunction{script≡rattle-inner}\AgdaSpace{}%
\AgdaSymbol{\{}\AgdaBound{s₁}\AgdaSymbol{\}}\AgdaSpace{}%
\AgdaSymbol{\{}\AgdaBound{s₂}\AgdaSymbol{\}}\AgdaSpace{}%
\AgdaBound{mm}\AgdaSpace{}%
\AgdaSymbol{(}\AgdaBound{x}\AgdaSpace{}%
\AgdaOperator{\AgdaInductiveConstructor{∷}}\AgdaSpace{}%
\AgdaBound{b₁}\AgdaSymbol{)}\AgdaSpace{}%
\AgdaBound{∀₁}\AgdaSpace{}%
\AgdaSymbol{(}\AgdaInductiveConstructor{Cons}\AgdaSpace{}%
\AgdaDottedPattern{\AgdaSymbol{.}}\AgdaDottedPattern{\AgdaBound{x}}\AgdaSpace{}%
\AgdaBound{dsj}\AgdaSpace{}%
\AgdaDottedPattern{\AgdaSymbol{.}}\AgdaDottedPattern{\AgdaBound{b₁}}\AgdaSpace{}%
\AgdaBound{dsb}\AgdaSymbol{)}\AgdaSpace{}%
\AgdaBound{mp}\AgdaSpace{}%
\AgdaBound{f₁}\AgdaSpace{}%
\AgdaKeyword{with}\AgdaSpace{}%
\AgdaBound{x}\AgdaSpace{}%
\AgdaOperator{\AgdaPostulate{∈?}}\AgdaSpace{}%
\AgdaFunction{map}\AgdaSpace{}%
\AgdaField{proj₁}\AgdaSpace{}%
\AgdaBound{mm}\<%
\\
\>[0]\AgdaSymbol{...}\AgdaSpace{}%
\AgdaSymbol{|}\AgdaSpace{}%
\AgdaInductiveConstructor{no}\AgdaSpace{}%
\AgdaBound{x∉}\AgdaSpace{}%
\AgdaSymbol{=}\AgdaSpace{}%
\AgdaFunction{script≡rattle-inner}\AgdaSpace{}%
\AgdaSymbol{\{}\AgdaPostulate{St.run}\AgdaSpace{}%
\AgdaBound{x}\AgdaSpace{}%
\AgdaBound{s₁}\AgdaSymbol{\}}\AgdaSpace{}%
\AgdaSymbol{\{}\AgdaPostulate{St.run}\AgdaSpace{}%
\AgdaBound{x}\AgdaSpace{}%
\AgdaBound{s₂}\AgdaSymbol{\}}\AgdaSpace{}%
\AgdaSymbol{(}\AgdaFunction{save}\AgdaSpace{}%
\AgdaBound{x}\AgdaSpace{}%
\AgdaSymbol{(}\AgdaPostulate{cmdReadNames}\AgdaSpace{}%
\AgdaBound{x}\AgdaSpace{}%
\AgdaBound{s₂}\AgdaSpace{}%
\AgdaOperator{\AgdaFunction{++}}\AgdaSpace{}%
\AgdaPostulate{cmdWriteNames}\AgdaSpace{}%
\AgdaBound{x}\AgdaSpace{}%
\AgdaBound{s₂}\AgdaSymbol{)}\AgdaSpace{}%
\AgdaSymbol{(}\AgdaPostulate{St.run}\AgdaSpace{}%
\AgdaBound{x}\AgdaSpace{}%
\AgdaBound{s₂}\AgdaSymbol{)}\AgdaSpace{}%
\AgdaBound{mm}\AgdaSymbol{)}\AgdaSpace{}%
\AgdaBound{b₁}\AgdaSpace{}%
\AgdaSymbol{(}\AgdaPostulate{run-≡}\AgdaSpace{}%
\AgdaBound{x}\AgdaSpace{}%
\AgdaBound{∀₁}\AgdaSymbol{)}\AgdaSpace{}%
\AgdaBound{dsb}\AgdaSpace{}%
\AgdaFunction{mp₁}\AgdaSpace{}%
\AgdaBound{f₁}\<%
\\
\>[0][@{}l@{\AgdaIndent{0}}]%
\>[2]\AgdaKeyword{where}%
\>[1764I]\AgdaFunction{mp₁}\AgdaSpace{}%
\AgdaSymbol{:}\AgdaSpace{}%
\AgdaDatatype{MemoryProperty}\AgdaSpace{}%
\AgdaSymbol{(}\AgdaFunction{save}\AgdaSpace{}%
\AgdaBound{x}\AgdaSpace{}%
\AgdaSymbol{(}\AgdaPostulate{cmdReadNames}\AgdaSpace{}%
\AgdaBound{x}\AgdaSpace{}%
\AgdaBound{s₂}\AgdaSpace{}%
\AgdaOperator{\AgdaFunction{++}}\AgdaSpace{}%
\AgdaPostulate{cmdWriteNames}\AgdaSpace{}%
\AgdaBound{x}\AgdaSpace{}%
\AgdaBound{s₂}\AgdaSymbol{)}\AgdaSpace{}%
\AgdaSymbol{(}\AgdaPostulate{St.run}\AgdaSpace{}%
\AgdaBound{x}\AgdaSpace{}%
\AgdaBound{s₂}\AgdaSymbol{)}\AgdaSpace{}%
\AgdaBound{mm}\AgdaSymbol{)}\<%
\\
\>[.][@{}l@{}]\<[1764I]%
\>[8]\AgdaFunction{mp₁}\AgdaSpace{}%
\AgdaSymbol{=}\AgdaSpace{}%
\AgdaInductiveConstructor{Cons}\AgdaSpace{}%
\AgdaBound{x}\AgdaSpace{}%
\AgdaBound{s₂}\AgdaSpace{}%
\AgdaSymbol{(λ}\AgdaSpace{}%
\AgdaBound{f₂}\AgdaSpace{}%
\AgdaBound{x₁}\AgdaSpace{}%
\AgdaSymbol{→}\AgdaSpace{}%
\AgdaPostulate{lemma3}\AgdaSpace{}%
\AgdaBound{f₂}\AgdaSpace{}%
\AgdaSymbol{(}\AgdaField{proj₂}\AgdaSpace{}%
\AgdaSymbol{(}\AgdaField{proj₁}\AgdaSpace{}%
\AgdaSymbol{(}\AgdaBound{oracle}\AgdaSpace{}%
\AgdaBound{x}\AgdaSymbol{)}\AgdaSpace{}%
\AgdaBound{s₂}\AgdaSymbol{))}\AgdaSpace{}%
\AgdaSymbol{λ}\AgdaSpace{}%
\AgdaBound{x₂}\AgdaSpace{}%
\AgdaSymbol{→}\AgdaSpace{}%
\AgdaBound{dsj}\AgdaSpace{}%
\AgdaSymbol{(}\AgdaBound{x₁}\AgdaSpace{}%
\AgdaOperator{\AgdaInductiveConstructor{,}}\AgdaSpace{}%
\AgdaBound{x₂}\AgdaSymbol{))}\AgdaSpace{}%
\AgdaBound{mp}\<%
\\
\>[0]\AgdaSymbol{...}\AgdaSpace{}%
\AgdaSymbol{|}\AgdaSpace{}%
\AgdaInductiveConstructor{yes}\AgdaSpace{}%
\AgdaBound{x∈}\AgdaSpace{}%
\AgdaKeyword{with}\AgdaSpace{}%
\AgdaPostulate{maybeAll}\AgdaSpace{}%
\AgdaSymbol{\{}\AgdaBound{s₂}\AgdaSymbol{\}}\AgdaSpace{}%
\AgdaSymbol{(}\AgdaPostulate{get}\AgdaSpace{}%
\AgdaBound{x}\AgdaSpace{}%
\AgdaBound{mm}\AgdaSpace{}%
\AgdaBound{x∈}\AgdaSymbol{)}\<%
\\
\>[0]\AgdaSymbol{...}\AgdaSpace{}%
\AgdaSymbol{|}\AgdaSpace{}%
\AgdaInductiveConstructor{nothing}\AgdaSpace{}%
\AgdaSymbol{=}\AgdaSpace{}%
\AgdaFunction{script≡rattle-inner}\AgdaSpace{}%
\AgdaSymbol{\{}\AgdaPostulate{St.run}\AgdaSpace{}%
\AgdaBound{x}\AgdaSpace{}%
\AgdaBound{s₁}\AgdaSymbol{\}}\AgdaSpace{}%
\AgdaSymbol{\{}\AgdaPostulate{St.run}\AgdaSpace{}%
\AgdaBound{x}\AgdaSpace{}%
\AgdaBound{s₂}\AgdaSymbol{\}}\AgdaSpace{}%
\AgdaSymbol{(}\AgdaFunction{save}\AgdaSpace{}%
\AgdaBound{x}\AgdaSpace{}%
\AgdaSymbol{(}\AgdaPostulate{cmdReadNames}\AgdaSpace{}%
\AgdaBound{x}\AgdaSpace{}%
\AgdaBound{s₂}\AgdaSpace{}%
\AgdaOperator{\AgdaFunction{++}}\AgdaSpace{}%
\AgdaPostulate{cmdWriteNames}\AgdaSpace{}%
\AgdaBound{x}\AgdaSpace{}%
\AgdaBound{s₂}\AgdaSymbol{)}\AgdaSpace{}%
\AgdaSymbol{(}\AgdaPostulate{St.run}\AgdaSpace{}%
\AgdaBound{x}\AgdaSpace{}%
\AgdaBound{s₂}\AgdaSymbol{)}\AgdaSpace{}%
\AgdaBound{mm}\AgdaSymbol{)}\AgdaSpace{}%
\AgdaBound{b₁}\AgdaSpace{}%
\AgdaSymbol{(}\AgdaPostulate{run-≡}\AgdaSpace{}%
\AgdaBound{x}\AgdaSpace{}%
\AgdaBound{∀₁}\AgdaSymbol{)}\AgdaSpace{}%
\AgdaBound{dsb}\AgdaSpace{}%
\AgdaFunction{mp₁}\AgdaSpace{}%
\AgdaBound{f₁}\<%
\\
\>[0][@{}l@{\AgdaIndent{0}}]%
\>[2]\AgdaKeyword{where}%
\>[1843I]\AgdaFunction{mp₁}\AgdaSpace{}%
\AgdaSymbol{:}\AgdaSpace{}%
\AgdaDatatype{MemoryProperty}\AgdaSpace{}%
\AgdaSymbol{(}\AgdaFunction{save}\AgdaSpace{}%
\AgdaBound{x}\AgdaSpace{}%
\AgdaSymbol{(}\AgdaPostulate{cmdReadNames}\AgdaSpace{}%
\AgdaBound{x}\AgdaSpace{}%
\AgdaBound{s₂}\AgdaSpace{}%
\AgdaOperator{\AgdaFunction{++}}\AgdaSpace{}%
\AgdaPostulate{cmdWriteNames}\AgdaSpace{}%
\AgdaBound{x}\AgdaSpace{}%
\AgdaBound{s₂}\AgdaSymbol{)}\AgdaSpace{}%
\AgdaSymbol{(}\AgdaPostulate{St.run}\AgdaSpace{}%
\AgdaBound{x}\AgdaSpace{}%
\AgdaBound{s₂}\AgdaSymbol{)}\AgdaSpace{}%
\AgdaBound{mm}\AgdaSymbol{)}\<%
\\
\>[.][@{}l@{}]\<[1843I]%
\>[8]\AgdaFunction{mp₁}\AgdaSpace{}%
\AgdaSymbol{=}\AgdaSpace{}%
\AgdaInductiveConstructor{Cons}\AgdaSpace{}%
\AgdaBound{x}\AgdaSpace{}%
\AgdaBound{s₂}\AgdaSpace{}%
\AgdaSymbol{(λ}\AgdaSpace{}%
\AgdaBound{f₂}\AgdaSpace{}%
\AgdaBound{x₁}\AgdaSpace{}%
\AgdaSymbol{→}\AgdaSpace{}%
\AgdaPostulate{lemma3}\AgdaSpace{}%
\AgdaBound{f₂}\AgdaSpace{}%
\AgdaSymbol{(}\AgdaField{proj₂}\AgdaSpace{}%
\AgdaSymbol{(}\AgdaField{proj₁}\AgdaSpace{}%
\AgdaSymbol{(}\AgdaBound{oracle}\AgdaSpace{}%
\AgdaBound{x}\AgdaSymbol{)}\AgdaSpace{}%
\AgdaBound{s₂}\AgdaSymbol{))}\AgdaSpace{}%
\AgdaSymbol{λ}\AgdaSpace{}%
\AgdaBound{x₂}\AgdaSpace{}%
\AgdaSymbol{→}\AgdaSpace{}%
\AgdaBound{dsj}\AgdaSpace{}%
\AgdaSymbol{(}\AgdaBound{x₁}\AgdaSpace{}%
\AgdaOperator{\AgdaInductiveConstructor{,}}\AgdaSpace{}%
\AgdaBound{x₂}\AgdaSymbol{))}\AgdaSpace{}%
\AgdaBound{mp}\<%
\\
\>[0]\AgdaSymbol{...}\AgdaSpace{}%
\AgdaSymbol{|}\AgdaSpace{}%
\AgdaInductiveConstructor{just}\AgdaSpace{}%
\AgdaBound{all₁}\AgdaSpace{}%
\AgdaSymbol{=}\AgdaSpace{}%
\AgdaFunction{script≡rattle-inner}\AgdaSpace{}%
\AgdaSymbol{\{}\AgdaPostulate{St.run}\AgdaSpace{}%
\AgdaBound{x}\AgdaSpace{}%
\AgdaBound{s₁}\AgdaSymbol{\}}\AgdaSpace{}%
\AgdaSymbol{\{}\AgdaBound{s₂}\AgdaSymbol{\}}\AgdaSpace{}%
\AgdaBound{mm}\AgdaSpace{}%
\AgdaBound{b₁}\AgdaSpace{}%
\AgdaFunction{∀₂}\AgdaSpace{}%
\AgdaSymbol{(}\AgdaPostulate{dsj-≡}\AgdaSpace{}%
\AgdaSymbol{(}\AgdaPostulate{St.run}\AgdaSpace{}%
\AgdaBound{x}\AgdaSpace{}%
\AgdaBound{s₂}\AgdaSymbol{)}\AgdaSpace{}%
\AgdaBound{s₂}\AgdaSpace{}%
\AgdaBound{b₁}\AgdaSpace{}%
\AgdaFunction{∀₃}\AgdaSpace{}%
\AgdaBound{dsb}\AgdaSymbol{)}\AgdaSpace{}%
\AgdaBound{mp}\AgdaSpace{}%
\AgdaBound{f₁}\<%
\\
\>[0][@{}l@{\AgdaIndent{0}}]%
\>[2]\AgdaKeyword{where}%
\>[1904I]\AgdaFunction{∀₂}\AgdaSpace{}%
\AgdaSymbol{:}\AgdaSpace{}%
\AgdaSymbol{∀}\AgdaSpace{}%
\AgdaBound{f₁}\AgdaSpace{}%
\AgdaSymbol{→}\AgdaSpace{}%
\AgdaPostulate{St.run}\AgdaSpace{}%
\AgdaBound{x}\AgdaSpace{}%
\AgdaBound{s₁}\AgdaSpace{}%
\AgdaBound{f₁}\AgdaSpace{}%
\AgdaOperator{\AgdaDatatype{≡}}\AgdaSpace{}%
\AgdaBound{s₂}\AgdaSpace{}%
\AgdaBound{f₁}\<%
\\
\>[.][@{}l@{}]\<[1904I]%
\>[8]\AgdaFunction{∀₂}\AgdaSpace{}%
\AgdaBound{f₁}\AgdaSpace{}%
\AgdaSymbol{=}\AgdaSpace{}%
\AgdaFunction{sym}\AgdaSpace{}%
\AgdaSymbol{(}\AgdaFunction{noEffect}\AgdaSpace{}%
\AgdaBound{x}\AgdaSpace{}%
\AgdaBound{∀₁}\AgdaSpace{}%
\AgdaBound{mp}\AgdaSpace{}%
\AgdaBound{x∈}\AgdaSpace{}%
\AgdaBound{all₁}\AgdaSpace{}%
\AgdaBound{f₁}\AgdaSymbol{)}\<%
\\
\>[8]\AgdaFunction{∀₃}\AgdaSpace{}%
\AgdaSymbol{:}\AgdaSpace{}%
\AgdaSymbol{∀}\AgdaSpace{}%
\AgdaBound{f₁}\AgdaSpace{}%
\AgdaSymbol{→}\AgdaSpace{}%
\AgdaBound{s₂}\AgdaSpace{}%
\AgdaBound{f₁}\AgdaSpace{}%
\AgdaOperator{\AgdaDatatype{≡}}\AgdaSpace{}%
\AgdaPostulate{St.run}\AgdaSpace{}%
\AgdaBound{x}\AgdaSpace{}%
\AgdaBound{s₂}\AgdaSpace{}%
\AgdaBound{f₁}\<%
\\
\>[8]\AgdaFunction{∀₃}\AgdaSpace{}%
\AgdaSymbol{=}\AgdaSpace{}%
\AgdaFunction{noEffect}\AgdaSpace{}%
\AgdaBound{x}\AgdaSpace{}%
\AgdaSymbol{(λ}\AgdaSpace{}%
\AgdaBound{f₂}\AgdaSpace{}%
\AgdaSymbol{→}\AgdaSpace{}%
\AgdaInductiveConstructor{refl}\AgdaSymbol{)}\AgdaSpace{}%
\AgdaBound{mp}\AgdaSpace{}%
\AgdaBound{x∈}\AgdaSpace{}%
\AgdaBound{all₁}\<%
\\
\\[\AgdaEmptyExtraSkip]%
\>[0]\AgdaComment{-- rattle produces a State and the System in that state is equivalent to the one produced by script}\<%
\end{code}

\newcommand{\eqtoscript}{%
\begin{code}%
\>[0]\AgdaFunction{≡toScript}\AgdaSpace{}%
\AgdaSymbol{:}\AgdaSpace{}%
\AgdaFunction{FileSystem}\AgdaSpace{}%
\AgdaSymbol{→}\AgdaSpace{}%
\AgdaPostulate{Build}\AgdaSpace{}%
\AgdaSymbol{→}\AgdaSpace{}%
\AgdaPostulate{Build}\AgdaSpace{}%
\AgdaSymbol{→}\AgdaSpace{}%
\AgdaPrimitiveType{Set}\<%
\\
\>[0]\AgdaFunction{≡toScript}\AgdaSpace{}%
\AgdaBound{s}\AgdaSpace{}%
\AgdaBound{br}\AgdaSpace{}%
\AgdaBound{bs}\AgdaSpace{}%
\AgdaSymbol{=}\AgdaSpace{}%
\AgdaFunction{∃[}\AgdaSpace{}%
\AgdaBound{s₁}\AgdaSpace{}%
\AgdaFunction{]}\AgdaSymbol{(}\AgdaFunction{∃[}\AgdaSpace{}%
\AgdaBound{m}\AgdaSpace{}%
\AgdaFunction{]}\AgdaSymbol{(}\AgdaFunction{∃[}\AgdaSpace{}%
\AgdaBound{ls}\AgdaSpace{}%
\AgdaFunction{]}\AgdaSymbol{(}\AgdaPostulate{rattle}\AgdaSpace{}%
\AgdaBound{br}\AgdaSpace{}%
\AgdaBound{bs}\AgdaSpace{}%
\AgdaSymbol{((}\AgdaBound{s}\AgdaSpace{}%
\AgdaOperator{\AgdaInductiveConstructor{,}}\AgdaSpace{}%
\AgdaInductiveConstructor{[]}\AgdaSymbol{)}\AgdaSpace{}%
\AgdaOperator{\AgdaInductiveConstructor{,}}\AgdaSpace{}%
\AgdaInductiveConstructor{[]}\AgdaSymbol{)}\AgdaSpace{}%
\AgdaOperator{\AgdaDatatype{≡}}\AgdaSpace{}%
\AgdaInductiveConstructor{inj₂}\AgdaSpace{}%
\AgdaSymbol{((}\AgdaBound{s₁}\AgdaSpace{}%
\AgdaOperator{\AgdaInductiveConstructor{,}}\AgdaSpace{}%
\AgdaBound{m}\AgdaSymbol{)}\AgdaSpace{}%
\AgdaOperator{\AgdaInductiveConstructor{,}}\AgdaSpace{}%
\AgdaBound{ls}\AgdaSymbol{)}\AgdaSpace{}%
\AgdaOperator{\AgdaFunction{×}}\AgdaSpace{}%
\AgdaSymbol{∀}\AgdaSpace{}%
\AgdaBound{f₁}\AgdaSpace{}%
\AgdaSymbol{→}\AgdaSpace{}%
\AgdaBound{s₁}\AgdaSpace{}%
\AgdaBound{f₁}\AgdaSpace{}%
\AgdaOperator{\AgdaDatatype{≡}}\AgdaSpace{}%
\AgdaPostulate{script}\AgdaSpace{}%
\AgdaBound{bs}\AgdaSpace{}%
\AgdaBound{s}\AgdaSpace{}%
\AgdaBound{f₁}\AgdaSymbol{)))}\<%
\end{code}}

\newcommand{\lemmasr}{%
\begin{code}%
\>[0]\AgdaFunction{script≡rattle-unchecked}\AgdaSpace{}%
\AgdaSymbol{:}\AgdaSpace{}%
\AgdaSymbol{∀}\AgdaSpace{}%
\AgdaBound{s}\AgdaSpace{}%
\AgdaBound{b}\AgdaSpace{}%
\AgdaSymbol{→}\AgdaSpace{}%
\AgdaPostulate{DisjointBuild}\AgdaSpace{}%
\AgdaBound{s}\AgdaSpace{}%
\AgdaBound{b}\AgdaSpace{}%
\AgdaSymbol{→}\AgdaSpace{}%
\AgdaSymbol{(∀}\AgdaSpace{}%
\AgdaBound{f₁}\AgdaSpace{}%
\AgdaSymbol{→}\AgdaSpace{}%
\AgdaPostulate{script}\AgdaSpace{}%
\AgdaBound{b}\AgdaSpace{}%
\AgdaBound{s}\AgdaSpace{}%
\AgdaBound{f₁}\AgdaSpace{}%
\AgdaOperator{\AgdaDatatype{≡}}\AgdaSpace{}%
\AgdaField{proj₁}\AgdaSpace{}%
\AgdaSymbol{(}\AgdaPostulate{rattle-unchecked}\AgdaSpace{}%
\AgdaBound{b}\AgdaSpace{}%
\AgdaSymbol{(}\AgdaBound{s}\AgdaSpace{}%
\AgdaOperator{\AgdaInductiveConstructor{,}}\AgdaSpace{}%
\AgdaInductiveConstructor{[]}\AgdaSymbol{))}\AgdaSpace{}%
\AgdaBound{f₁}\AgdaSymbol{)}\<%
\end{code}}
\begin{code}[hide]%
\>[0]\AgdaFunction{script≡rattle-unchecked}\AgdaSpace{}%
\AgdaBound{s}\AgdaSpace{}%
\AgdaBound{b}\AgdaSpace{}%
\AgdaBound{dsb}\AgdaSpace{}%
\AgdaSymbol{=}\AgdaSpace{}%
\AgdaFunction{script≡rattle-inner}\AgdaSpace{}%
\AgdaInductiveConstructor{[]}\AgdaSpace{}%
\AgdaBound{b}\AgdaSpace{}%
\AgdaSymbol{(λ}\AgdaSpace{}%
\AgdaBound{f₁}\AgdaSpace{}%
\AgdaSymbol{→}\AgdaSpace{}%
\AgdaInductiveConstructor{refl}\AgdaSymbol{)}\AgdaSpace{}%
\AgdaBound{dsb}\AgdaSpace{}%
\AgdaInductiveConstructor{[]}\<%
\end{code}

\newcommand{\soundness}{%
\begin{code}%
\>[0]\AgdaFunction{soundness}%
\>[2028I]\AgdaSymbol{:}\AgdaSpace{}%
\AgdaSymbol{∀}\AgdaSpace{}%
\AgdaSymbol{\{}\AgdaBound{s₁}\AgdaSymbol{\}}\AgdaSpace{}%
\AgdaSymbol{\{}\AgdaBound{m₁}\AgdaSymbol{\}}\AgdaSpace{}%
\AgdaSymbol{\{}\AgdaBound{ls}\AgdaSymbol{\}}\AgdaSpace{}%
\AgdaBound{s}\AgdaSpace{}%
\AgdaBound{br}\AgdaSpace{}%
\AgdaBound{bs}\AgdaSpace{}%
\AgdaSymbol{→}\AgdaSpace{}%
\AgdaPostulate{DisjointBuild}\AgdaSpace{}%
\AgdaBound{s}\AgdaSpace{}%
\AgdaBound{br}\AgdaSpace{}%
\AgdaSymbol{→}\AgdaSpace{}%
\AgdaPostulate{rattle}\AgdaSpace{}%
\AgdaBound{br}\AgdaSpace{}%
\AgdaBound{bs}\AgdaSpace{}%
\AgdaSymbol{((}\AgdaBound{s}\AgdaSpace{}%
\AgdaOperator{\AgdaInductiveConstructor{,}}\AgdaSpace{}%
\AgdaInductiveConstructor{[]}\AgdaSymbol{)}\AgdaSpace{}%
\AgdaOperator{\AgdaInductiveConstructor{,}}\AgdaSpace{}%
\AgdaInductiveConstructor{[]}\AgdaSymbol{)}\AgdaSpace{}%
\AgdaOperator{\AgdaDatatype{≡}}\AgdaSpace{}%
\AgdaInductiveConstructor{inj₂}\AgdaSpace{}%
\AgdaSymbol{((}\AgdaBound{s₁}\AgdaSpace{}%
\AgdaOperator{\AgdaInductiveConstructor{,}}\AgdaSpace{}%
\AgdaBound{m₁}\AgdaSymbol{)}\AgdaSpace{}%
\AgdaOperator{\AgdaInductiveConstructor{,}}\AgdaSpace{}%
\AgdaBound{ls}\AgdaSymbol{)}\<%
\\
\>[.][@{}l@{}]\<[2028I]%
\>[10]\AgdaSymbol{→}\AgdaSpace{}%
\AgdaSymbol{(∀}\AgdaSpace{}%
\AgdaBound{f₁}\AgdaSpace{}%
\AgdaSymbol{→}\AgdaSpace{}%
\AgdaPostulate{script}\AgdaSpace{}%
\AgdaBound{br}\AgdaSpace{}%
\AgdaBound{s}\AgdaSpace{}%
\AgdaBound{f₁}\AgdaSpace{}%
\AgdaOperator{\AgdaDatatype{≡}}\AgdaSpace{}%
\AgdaBound{s₁}\AgdaSpace{}%
\AgdaBound{f₁}\AgdaSymbol{)}\<%
\end{code}}
\begin{code}[hide]%
\>[0]\AgdaFunction{soundness}\AgdaSpace{}%
\AgdaBound{s}\AgdaSpace{}%
\AgdaBound{br}\AgdaSpace{}%
\AgdaBound{bc}\AgdaSpace{}%
\AgdaBound{dsb}\AgdaSpace{}%
\AgdaBound{≡₁}\AgdaSpace{}%
\AgdaBound{f₁}\AgdaSpace{}%
\AgdaSymbol{=}\AgdaSpace{}%
\AgdaFunction{trans}%
\>[2074I]\AgdaSymbol{(}\AgdaFunction{script≡rattle-unchecked}\AgdaSpace{}%
\AgdaBound{s}\AgdaSpace{}%
\AgdaBound{br}\AgdaSpace{}%
\AgdaBound{dsb}\AgdaSpace{}%
\AgdaBound{f₁}\AgdaSymbol{)}\<%
\\
\>[.][@{}l@{}]\<[2074I]%
\>[36]\AgdaSymbol{(}\AgdaFunction{cong-app}\AgdaSpace{}%
\AgdaSymbol{(}\AgdaFunction{,-injectiveˡ}\AgdaSpace{}%
\AgdaSymbol{(}\AgdaFunction{soundness-inner}\AgdaSpace{}%
\AgdaSymbol{(}\AgdaBound{s}\AgdaSpace{}%
\AgdaOperator{\AgdaInductiveConstructor{,}}\AgdaSpace{}%
\AgdaInductiveConstructor{[]}\AgdaSymbol{)}\AgdaSpace{}%
\AgdaInductiveConstructor{[]}\AgdaSpace{}%
\AgdaBound{br}\AgdaSpace{}%
\AgdaBound{bc}\AgdaSpace{}%
\AgdaBound{≡₁}\AgdaSymbol{))}\AgdaSpace{}%
\AgdaBound{f₁}\AgdaSymbol{)}\<%
\end{code}

\begin{code}[hide]%
\>[0]\AgdaComment{-- correctness is if you have any build then either you get the right answer (the one the script gave) or you get an error and there was a hazard.}\<%
\end{code}
\newcommand{\correct}{%
\begin{code}%
\>[0]\AgdaFunction{correct-rattle}\AgdaSpace{}%
\AgdaSymbol{:}\AgdaSpace{}%
\AgdaSymbol{∀}\AgdaSpace{}%
\AgdaBound{s}\AgdaSpace{}%
\AgdaBound{b}\AgdaSpace{}%
\AgdaSymbol{→}\AgdaSpace{}%
\AgdaPostulate{PreCond}\AgdaSpace{}%
\AgdaBound{s}\AgdaSpace{}%
\AgdaBound{b}\AgdaSpace{}%
\AgdaBound{b}\AgdaSpace{}%
\AgdaSymbol{→}\AgdaSpace{}%
\AgdaOperator{\AgdaFunction{¬}}\AgdaSpace{}%
\AgdaPostulate{HazardFree}\AgdaSpace{}%
\AgdaBound{s}\AgdaSpace{}%
\AgdaBound{b}\AgdaSpace{}%
\AgdaBound{b}\AgdaSpace{}%
\AgdaInductiveConstructor{[]}\AgdaSpace{}%
\AgdaOperator{\AgdaDatatype{⊎}}\AgdaSpace{}%
\AgdaFunction{≡toScript}\AgdaSpace{}%
\AgdaBound{s}\AgdaSpace{}%
\AgdaBound{b}\AgdaSpace{}%
\AgdaBound{b}\<%
\end{code}}
\begin{code}[hide]%
\>[0]\AgdaFunction{correct-rattle}\AgdaSpace{}%
\AgdaBound{s}\AgdaSpace{}%
\AgdaBound{b}\AgdaSpace{}%
\AgdaBound{pc}\AgdaSpace{}%
\AgdaKeyword{with}\AgdaSpace{}%
\AgdaPostulate{rattle}\AgdaSpace{}%
\AgdaBound{b}\AgdaSpace{}%
\AgdaBound{b}\AgdaSpace{}%
\AgdaSymbol{((}\AgdaBound{s}\AgdaSpace{}%
\AgdaOperator{\AgdaInductiveConstructor{,}}\AgdaSpace{}%
\AgdaInductiveConstructor{[]}\AgdaSymbol{)}\AgdaSpace{}%
\AgdaOperator{\AgdaInductiveConstructor{,}}\AgdaSpace{}%
\AgdaInductiveConstructor{[]}\AgdaSymbol{)}\AgdaSpace{}%
\AgdaSymbol{|}\AgdaSpace{}%
\AgdaFunction{inspect}\AgdaSpace{}%
\AgdaSymbol{(}\AgdaPostulate{rattle}\AgdaSpace{}%
\AgdaBound{b}\AgdaSpace{}%
\AgdaBound{b}\AgdaSymbol{)}\AgdaSpace{}%
\AgdaSymbol{((}\AgdaBound{s}\AgdaSpace{}%
\AgdaOperator{\AgdaInductiveConstructor{,}}\AgdaSpace{}%
\AgdaInductiveConstructor{[]}\AgdaSymbol{)}\AgdaSpace{}%
\AgdaOperator{\AgdaInductiveConstructor{,}}\AgdaSpace{}%
\AgdaInductiveConstructor{[]}\AgdaSymbol{)}\<%
\\
\>[0]\AgdaSymbol{...}\AgdaSpace{}%
\AgdaSymbol{|}\AgdaSpace{}%
\AgdaInductiveConstructor{inj₁}\AgdaSpace{}%
\AgdaBound{hz}\AgdaSpace{}%
\AgdaSymbol{|}\AgdaSpace{}%
\AgdaOperator{\AgdaInductiveConstructor{[}}\AgdaSpace{}%
\AgdaBound{≡₁}\AgdaSpace{}%
\AgdaOperator{\AgdaInductiveConstructor{]}}\AgdaSpace{}%
\AgdaSymbol{=}\AgdaSpace{}%
\AgdaInductiveConstructor{inj₁}\AgdaSpace{}%
\AgdaFunction{g₁}\<%
\\
\>[0][@{}l@{\AgdaIndent{0}}]%
\>[2]\AgdaKeyword{where}%
\>[2142I]\AgdaFunction{g₁}\AgdaSpace{}%
\AgdaSymbol{:}\AgdaSpace{}%
\AgdaOperator{\AgdaFunction{¬}}\AgdaSpace{}%
\AgdaPostulate{HazardFree}\AgdaSpace{}%
\AgdaBound{s}\AgdaSpace{}%
\AgdaBound{b}\AgdaSpace{}%
\AgdaBound{b}\AgdaSpace{}%
\AgdaInductiveConstructor{[]}\<%
\\
\>[.][@{}l@{}]\<[2142I]%
\>[8]\AgdaFunction{g₁}\AgdaSpace{}%
\AgdaBound{hf}\AgdaSpace{}%
\AgdaKeyword{with}\AgdaSpace{}%
\AgdaFunction{completeness}\AgdaSpace{}%
\AgdaBound{s}\AgdaSpace{}%
\AgdaBound{b}\AgdaSpace{}%
\AgdaBound{b}\AgdaSpace{}%
\AgdaBound{pc}\AgdaSpace{}%
\AgdaBound{hf}\<%
\\
\>[8]\AgdaSymbol{...}\AgdaSpace{}%
\AgdaSymbol{|}\AgdaSpace{}%
\AgdaBound{a}\AgdaSpace{}%
\AgdaOperator{\AgdaInductiveConstructor{,}}\AgdaSpace{}%
\AgdaBound{fst}\AgdaSpace{}%
\AgdaOperator{\AgdaInductiveConstructor{,}}\AgdaSpace{}%
\AgdaBound{≡₂}\AgdaSpace{}%
\AgdaSymbol{=}\AgdaSpace{}%
\AgdaFunction{contradiction}\AgdaSpace{}%
\AgdaSymbol{(}\AgdaFunction{trans}\AgdaSpace{}%
\AgdaSymbol{(}\AgdaFunction{sym}\AgdaSpace{}%
\AgdaBound{≡₁}\AgdaSymbol{)}\AgdaSpace{}%
\AgdaBound{≡₂}\AgdaSymbol{)}\AgdaSpace{}%
\AgdaSymbol{λ}\AgdaSpace{}%
\AgdaSymbol{()}\<%
\\
\>[0]\AgdaSymbol{...}\AgdaSpace{}%
\AgdaSymbol{|}\AgdaSpace{}%
\AgdaInductiveConstructor{inj₂}\AgdaSpace{}%
\AgdaSymbol{((}\AgdaBound{s₁}\AgdaSpace{}%
\AgdaOperator{\AgdaInductiveConstructor{,}}\AgdaSpace{}%
\AgdaBound{mm₁}\AgdaSymbol{)}\AgdaSpace{}%
\AgdaOperator{\AgdaInductiveConstructor{,}}\AgdaSpace{}%
\AgdaBound{ls₁}\AgdaSymbol{)}\AgdaSpace{}%
\AgdaSymbol{|}\AgdaSpace{}%
\AgdaOperator{\AgdaInductiveConstructor{[}}\AgdaSpace{}%
\AgdaBound{≡₁}\AgdaSpace{}%
\AgdaOperator{\AgdaInductiveConstructor{]}}\AgdaSpace{}%
\AgdaSymbol{=}\AgdaSpace{}%
\AgdaInductiveConstructor{inj₂}\AgdaSpace{}%
\AgdaSymbol{(}\AgdaBound{s₁}\AgdaSpace{}%
\AgdaOperator{\AgdaInductiveConstructor{,}}\AgdaSpace{}%
\AgdaBound{mm₁}\AgdaSpace{}%
\AgdaOperator{\AgdaInductiveConstructor{,}}\AgdaSpace{}%
\AgdaBound{ls₁}\AgdaSpace{}%
\AgdaOperator{\AgdaInductiveConstructor{,}}\AgdaSpace{}%
\AgdaInductiveConstructor{refl}\AgdaSpace{}%
\AgdaOperator{\AgdaInductiveConstructor{,}}\AgdaSpace{}%
\AgdaFunction{∀≡}\AgdaSymbol{)}\<%
\\
\>[0][@{}l@{\AgdaIndent{0}}]%
\>[2]\AgdaKeyword{where}%
\>[2194I]\AgdaFunction{∀≡}\AgdaSpace{}%
\AgdaSymbol{:}\AgdaSpace{}%
\AgdaSymbol{∀}\AgdaSpace{}%
\AgdaBound{f₁}\AgdaSpace{}%
\AgdaSymbol{→}\AgdaSpace{}%
\AgdaBound{s₁}\AgdaSpace{}%
\AgdaBound{f₁}\AgdaSpace{}%
\AgdaOperator{\AgdaDatatype{≡}}\AgdaSpace{}%
\AgdaPostulate{script}\AgdaSpace{}%
\AgdaBound{b}\AgdaSpace{}%
\AgdaBound{s}\AgdaSpace{}%
\AgdaBound{f₁}\<%
\\
\>[.][@{}l@{}]\<[2194I]%
\>[8]\AgdaFunction{∀≡}\AgdaSpace{}%
\AgdaBound{f₁}\AgdaSpace{}%
\AgdaSymbol{=}\AgdaSpace{}%
\AgdaFunction{sym}\AgdaSpace{}%
\AgdaSymbol{(}\AgdaFunction{soundness}\AgdaSpace{}%
\AgdaBound{s}\AgdaSpace{}%
\AgdaBound{b}\AgdaSpace{}%
\AgdaBound{b}\AgdaSpace{}%
\AgdaSymbol{(}\AgdaField{proj₁}\AgdaSpace{}%
\AgdaBound{pc}\AgdaSymbol{)}\AgdaSpace{}%
\AgdaBound{≡₁}\AgdaSpace{}%
\AgdaBound{f₁}\AgdaSymbol{)}\<%
\end{code}

\begin{code}[hide]%
\>[0]\AgdaComment{-- want to prove if execWError original build produces a hazard then execWError of the reordered build will produce a hazard too.}\<%
\\
\>[0]\AgdaComment{-- this would also mean if the reordered build doesnt produce a hazard, then the original build doesn't produce a hazard.}\<%
\\
\>[0]\AgdaComment{-- which means it produces a state.}\<%
\\
\>[0]\AgdaComment{-- then we use soundness to prove theyre equal to their non hazard versions ; then we use the reordering proof to show they're equivalent??}\<%
\\
\\[\AgdaEmptyExtraSkip]%
\>[0]\AgdaComment{-- for every pair of commands in b₂ (x , y) where x is before y; if x writes to something y reads; then (x , y) in b₁ too.}\<%
\\
\\[\AgdaEmptyExtraSkip]%
\>[0]\AgdaFunction{before=>∈}\AgdaSpace{}%
\AgdaSymbol{:}\AgdaSpace{}%
\AgdaSymbol{∀}\AgdaSpace{}%
\AgdaSymbol{\{}\AgdaBound{x₁}\AgdaSymbol{\}}\AgdaSpace{}%
\AgdaSymbol{\{}\AgdaBound{x₂}\AgdaSymbol{\}}\AgdaSpace{}%
\AgdaSymbol{\{}\AgdaBound{xs}\AgdaSymbol{\}}\AgdaSpace{}%
\AgdaSymbol{→}\AgdaSpace{}%
\AgdaBound{x₁}\AgdaSpace{}%
\AgdaOperator{\AgdaFunction{before}}\AgdaSpace{}%
\AgdaBound{x₂}\AgdaSpace{}%
\AgdaOperator{\AgdaFunction{∈}}\AgdaSpace{}%
\AgdaBound{xs}\AgdaSpace{}%
\AgdaSymbol{→}\AgdaSpace{}%
\AgdaSymbol{(}\AgdaBound{x₁}\AgdaSpace{}%
\AgdaOperator{\AgdaPostulate{∈}}\AgdaSpace{}%
\AgdaBound{xs}\AgdaSpace{}%
\AgdaOperator{\AgdaFunction{×}}\AgdaSpace{}%
\AgdaBound{x₂}\AgdaSpace{}%
\AgdaOperator{\AgdaPostulate{∈}}\AgdaSpace{}%
\AgdaBound{xs}\AgdaSymbol{)}\<%
\\
\>[0]\AgdaFunction{before=>∈}\AgdaSpace{}%
\AgdaSymbol{(}\AgdaBound{ys}\AgdaSpace{}%
\AgdaOperator{\AgdaInductiveConstructor{,}}\AgdaSpace{}%
\AgdaBound{zs}\AgdaSpace{}%
\AgdaOperator{\AgdaInductiveConstructor{,}}\AgdaSpace{}%
\AgdaBound{≡₁}\AgdaSpace{}%
\AgdaOperator{\AgdaInductiveConstructor{,}}\AgdaSpace{}%
\AgdaBound{x₂∈zs}\AgdaSymbol{)}\AgdaSpace{}%
\AgdaSymbol{=}%
\>[2244I]\AgdaSymbol{(}\AgdaFunction{subst}\AgdaSpace{}%
\AgdaSymbol{(λ}\AgdaSpace{}%
\AgdaBound{x}\AgdaSpace{}%
\AgdaSymbol{→}\AgdaSpace{}%
\AgdaSymbol{\AgdaUnderscore{}}\AgdaSpace{}%
\AgdaOperator{\AgdaPostulate{∈}}\AgdaSpace{}%
\AgdaBound{x}\AgdaSymbol{)}\AgdaSpace{}%
\AgdaSymbol{(}\AgdaFunction{sym}\AgdaSpace{}%
\AgdaBound{≡₁}\AgdaSymbol{)}\AgdaSpace{}%
\AgdaSymbol{(}\AgdaFunction{∈-++⁺ʳ}\AgdaSpace{}%
\AgdaBound{ys}\AgdaSpace{}%
\AgdaSymbol{(}\AgdaInductiveConstructor{here}\AgdaSpace{}%
\AgdaInductiveConstructor{refl}\AgdaSymbol{)))}\<%
\\
\>[.][@{}l@{}]\<[2244I]%
\>[35]\AgdaOperator{\AgdaInductiveConstructor{,}}\AgdaSpace{}%
\AgdaSymbol{(}\AgdaFunction{subst}\AgdaSpace{}%
\AgdaSymbol{(λ}\AgdaSpace{}%
\AgdaBound{x}\AgdaSpace{}%
\AgdaSymbol{→}\AgdaSpace{}%
\AgdaSymbol{\AgdaUnderscore{}}\AgdaSpace{}%
\AgdaOperator{\AgdaPostulate{∈}}\AgdaSpace{}%
\AgdaBound{x}\AgdaSymbol{)}\AgdaSpace{}%
\AgdaSymbol{(}\AgdaFunction{sym}\AgdaSpace{}%
\AgdaBound{≡₁}\AgdaSymbol{)}\AgdaSpace{}%
\AgdaSymbol{(}\AgdaFunction{∈-++⁺ʳ}\AgdaSpace{}%
\AgdaBound{ys}\AgdaSpace{}%
\AgdaSymbol{(}\AgdaInductiveConstructor{there}\AgdaSpace{}%
\AgdaBound{x₂∈zs}\AgdaSymbol{)))}\<%
\\
\\[\AgdaEmptyExtraSkip]%
\>[0]\AgdaFunction{still-hazard}\AgdaSpace{}%
\AgdaSymbol{:}\AgdaSpace{}%
\AgdaSymbol{∀}\AgdaSpace{}%
\AgdaSymbol{\{}\AgdaBound{s}\AgdaSymbol{\}}\AgdaSpace{}%
\AgdaSymbol{\{}\AgdaBound{ls}\AgdaSymbol{\}}\AgdaSpace{}%
\AgdaBound{x₁}\AgdaSpace{}%
\AgdaBound{x}\AgdaSpace{}%
\AgdaBound{b}\AgdaSpace{}%
\AgdaSymbol{→}\AgdaSpace{}%
\AgdaOperator{\AgdaFunction{¬}}\AgdaSpace{}%
\AgdaBound{x₁}\AgdaSpace{}%
\AgdaOperator{\AgdaDatatype{≡}}\AgdaSpace{}%
\AgdaBound{x}\AgdaSpace{}%
\AgdaSymbol{→}\AgdaSpace{}%
\AgdaBound{x}\AgdaSpace{}%
\AgdaOperator{\AgdaPostulate{∉}}\AgdaSpace{}%
\AgdaSymbol{(}\AgdaFunction{map}\AgdaSpace{}%
\AgdaField{proj₁}\AgdaSpace{}%
\AgdaBound{ls}\AgdaSymbol{)}\AgdaSpace{}%
\AgdaSymbol{→}\AgdaSpace{}%
\AgdaPostulate{Hazard}\AgdaSpace{}%
\AgdaBound{s}\AgdaSpace{}%
\AgdaBound{x₁}\AgdaSpace{}%
\AgdaSymbol{(}\AgdaBound{b}\AgdaSpace{}%
\AgdaOperator{\AgdaFunction{++}}\AgdaSpace{}%
\AgdaBound{x}\AgdaSpace{}%
\AgdaOperator{\AgdaInductiveConstructor{∷}}\AgdaSpace{}%
\AgdaInductiveConstructor{[]}\AgdaSymbol{)}\AgdaSpace{}%
\AgdaBound{ls}\AgdaSpace{}%
\AgdaSymbol{→}\AgdaSpace{}%
\AgdaPostulate{Hazard}\AgdaSpace{}%
\AgdaBound{s}\AgdaSpace{}%
\AgdaBound{x₁}\AgdaSpace{}%
\AgdaBound{b}\AgdaSpace{}%
\AgdaBound{ls}\<%
\\
\>[0]\AgdaFunction{still-hazard}\AgdaSpace{}%
\AgdaBound{x₁}\AgdaSpace{}%
\AgdaBound{x}\AgdaSpace{}%
\AgdaBound{b}\AgdaSpace{}%
\AgdaSymbol{\AgdaUnderscore{}}\AgdaSpace{}%
\AgdaBound{x∉ls}\AgdaSpace{}%
\AgdaSymbol{(}\AgdaPostulate{ReadWrite}\AgdaSpace{}%
\AgdaBound{x₂}\AgdaSpace{}%
\AgdaBound{x₃}\AgdaSymbol{)}\AgdaSpace{}%
\AgdaSymbol{=}\AgdaSpace{}%
\AgdaSymbol{(}\AgdaPostulate{ReadWrite}\AgdaSpace{}%
\AgdaBound{x₂}\AgdaSpace{}%
\AgdaBound{x₃}\AgdaSymbol{)}\<%
\\
\>[0]\AgdaFunction{still-hazard}\AgdaSpace{}%
\AgdaBound{x₁}\AgdaSpace{}%
\AgdaBound{x}\AgdaSpace{}%
\AgdaBound{b}\AgdaSpace{}%
\AgdaSymbol{\AgdaUnderscore{}}\AgdaSpace{}%
\AgdaBound{x∉ls}\AgdaSpace{}%
\AgdaSymbol{(}\AgdaPostulate{WriteWrite}\AgdaSpace{}%
\AgdaBound{x₂}\AgdaSpace{}%
\AgdaBound{x₃}\AgdaSymbol{)}\AgdaSpace{}%
\AgdaSymbol{=}\AgdaSpace{}%
\AgdaSymbol{(}\AgdaPostulate{WriteWrite}\AgdaSpace{}%
\AgdaBound{x₂}\AgdaSpace{}%
\AgdaBound{x₃}\AgdaSymbol{)}\<%
\\
\>[0]\AgdaFunction{still-hazard}\AgdaSpace{}%
\AgdaBound{x₁}\AgdaSpace{}%
\AgdaBound{x}\AgdaSpace{}%
\AgdaBound{b}\AgdaSpace{}%
\AgdaBound{¬x₁≡x}\AgdaSpace{}%
\AgdaBound{x∉ls}\AgdaSpace{}%
\AgdaSymbol{(}\AgdaPostulate{Speculative}\AgdaSpace{}%
\AgdaBound{x₂}\AgdaSpace{}%
\AgdaBound{x₃}\AgdaSpace{}%
\AgdaBound{x₄}\AgdaSpace{}%
\AgdaBound{x₅}\AgdaSpace{}%
\AgdaBound{x₆}\AgdaSpace{}%
\AgdaBound{x₇}\AgdaSpace{}%
\AgdaBound{x₈}\AgdaSymbol{)}\<%
\\
\>[0][@{}l@{\AgdaIndent{0}}]%
\>[2]\AgdaSymbol{=}%
\>[2341I]\AgdaPostulate{Speculative}\AgdaSpace{}%
\AgdaBound{x₂}\AgdaSpace{}%
\AgdaBound{x₃}\AgdaSpace{}%
\AgdaBound{x₄}\AgdaSpace{}%
\AgdaFunction{x₃∈b}\AgdaSpace{}%
\AgdaFunction{¬bf}\AgdaSpace{}%
\AgdaBound{x₇}\AgdaSpace{}%
\AgdaBound{x₈}\<%
\\
\>[.][@{}l@{}]\<[2341I]%
\>[4]\AgdaKeyword{where}%
\>[2349I]\AgdaFunction{x₃∈b}\AgdaSpace{}%
\AgdaSymbol{:}\AgdaSpace{}%
\AgdaBound{x₃}\AgdaSpace{}%
\AgdaOperator{\AgdaPostulate{∈}}\AgdaSpace{}%
\AgdaBound{b}\<%
\\
\>[.][@{}l@{}]\<[2349I]%
\>[10]\AgdaFunction{x₃∈b}\AgdaSpace{}%
\AgdaKeyword{with}\AgdaSpace{}%
\AgdaFunction{∈-++⁻}\AgdaSpace{}%
\AgdaBound{b}\AgdaSpace{}%
\AgdaBound{x₅}\<%
\\
\>[10]\AgdaSymbol{...}\AgdaSpace{}%
\AgdaSymbol{|}\AgdaSpace{}%
\AgdaInductiveConstructor{inj₁}\AgdaSpace{}%
\AgdaBound{x₃∈₁}\AgdaSpace{}%
\AgdaSymbol{=}\AgdaSpace{}%
\AgdaBound{x₃∈₁}\<%
\\
\>[10]\AgdaSymbol{...}\AgdaSpace{}%
\AgdaSymbol{|}\AgdaSpace{}%
\AgdaInductiveConstructor{inj₂}\AgdaSpace{}%
\AgdaSymbol{(}\AgdaInductiveConstructor{here}\AgdaSpace{}%
\AgdaBound{px₁}\AgdaSymbol{)}\AgdaSpace{}%
\AgdaKeyword{with}\AgdaSpace{}%
\AgdaFunction{before=>∈}\AgdaSpace{}%
\AgdaBound{x₄}\<%
\\
\>[10]\AgdaSymbol{...}\AgdaSpace{}%
\AgdaSymbol{|}\AgdaSpace{}%
\AgdaInductiveConstructor{here}\AgdaSpace{}%
\AgdaBound{px}\AgdaSpace{}%
\AgdaOperator{\AgdaInductiveConstructor{,}}\AgdaSpace{}%
\AgdaBound{∈₂}\AgdaSpace{}%
\AgdaSymbol{=}\AgdaSpace{}%
\AgdaFunction{contradiction}\AgdaSpace{}%
\AgdaSymbol{(}\AgdaFunction{sym}\AgdaSpace{}%
\AgdaSymbol{(}\AgdaFunction{trans}\AgdaSpace{}%
\AgdaSymbol{(}\AgdaFunction{sym}\AgdaSpace{}%
\AgdaBound{px₁}\AgdaSymbol{)}\AgdaSpace{}%
\AgdaBound{px}\AgdaSymbol{))}\AgdaSpace{}%
\AgdaBound{¬x₁≡x}\<%
\\
\>[10]\AgdaSymbol{...}\AgdaSpace{}%
\AgdaSymbol{|}\AgdaSpace{}%
\AgdaInductiveConstructor{there}\AgdaSpace{}%
\AgdaBound{∈₁}\AgdaSpace{}%
\AgdaOperator{\AgdaInductiveConstructor{,}}\AgdaSpace{}%
\AgdaBound{∈₂}\AgdaSpace{}%
\AgdaSymbol{=}\AgdaSpace{}%
\AgdaFunction{contradiction}\AgdaSpace{}%
\AgdaSymbol{(}\AgdaFunction{subst}\AgdaSpace{}%
\AgdaSymbol{(λ}\AgdaSpace{}%
\AgdaBound{x₉}\AgdaSpace{}%
\AgdaSymbol{→}\AgdaSpace{}%
\AgdaBound{x₉}\AgdaSpace{}%
\AgdaOperator{\AgdaPostulate{∈}}\AgdaSpace{}%
\AgdaSymbol{\AgdaUnderscore{})}\AgdaSpace{}%
\AgdaBound{px₁}\AgdaSpace{}%
\AgdaBound{∈₁}\AgdaSymbol{)}\AgdaSpace{}%
\AgdaBound{x∉ls}\<%
\\
\>[10]\AgdaFunction{¬bf}\AgdaSpace{}%
\AgdaSymbol{:}\AgdaSpace{}%
\AgdaOperator{\AgdaFunction{¬}}\AgdaSpace{}%
\AgdaSymbol{(}\AgdaBound{x₂}\AgdaSpace{}%
\AgdaOperator{\AgdaFunction{before}}\AgdaSpace{}%
\AgdaBound{x₃}\AgdaSpace{}%
\AgdaOperator{\AgdaFunction{∈}}\AgdaSpace{}%
\AgdaBound{b}\AgdaSymbol{)}\<%
\\
\>[10]\AgdaFunction{¬bf}\AgdaSpace{}%
\AgdaSymbol{(}\AgdaBound{xs}\AgdaSpace{}%
\AgdaOperator{\AgdaInductiveConstructor{,}}\AgdaSpace{}%
\AgdaBound{ys}\AgdaSpace{}%
\AgdaOperator{\AgdaInductiveConstructor{,}}\AgdaSpace{}%
\AgdaBound{≡₁}\AgdaSpace{}%
\AgdaOperator{\AgdaInductiveConstructor{,}}\AgdaSpace{}%
\AgdaBound{∈₁}\AgdaSymbol{)}\AgdaSpace{}%
\AgdaSymbol{=}\AgdaSpace{}%
\AgdaBound{x₆}\AgdaSpace{}%
\AgdaSymbol{(}\AgdaBound{xs}\AgdaSpace{}%
\AgdaOperator{\AgdaInductiveConstructor{,}}\AgdaSpace{}%
\AgdaBound{ys}\AgdaSpace{}%
\AgdaOperator{\AgdaFunction{++}}\AgdaSpace{}%
\AgdaBound{x}\AgdaSpace{}%
\AgdaOperator{\AgdaInductiveConstructor{∷}}\AgdaSpace{}%
\AgdaInductiveConstructor{[]}\AgdaSpace{}%
\AgdaOperator{\AgdaInductiveConstructor{,}}\AgdaSpace{}%
\AgdaFunction{trans}\AgdaSpace{}%
\AgdaSymbol{(}\AgdaFunction{cong}\AgdaSpace{}%
\AgdaSymbol{(}\AgdaOperator{\AgdaFunction{\AgdaUnderscore{}++}}\AgdaSpace{}%
\AgdaBound{x}\AgdaSpace{}%
\AgdaOperator{\AgdaInductiveConstructor{∷}}\AgdaSpace{}%
\AgdaInductiveConstructor{[]}\AgdaSymbol{)}\AgdaSpace{}%
\AgdaBound{≡₁}\AgdaSymbol{)}\AgdaSpace{}%
\AgdaSymbol{(}\AgdaFunction{++-assoc}\AgdaSpace{}%
\AgdaBound{xs}\AgdaSpace{}%
\AgdaSymbol{(}\AgdaBound{x₂}\AgdaSpace{}%
\AgdaOperator{\AgdaInductiveConstructor{∷}}\AgdaSpace{}%
\AgdaBound{ys}\AgdaSymbol{)}\AgdaSpace{}%
\AgdaSymbol{(}\AgdaBound{x}\AgdaSpace{}%
\AgdaOperator{\AgdaInductiveConstructor{∷}}\AgdaSpace{}%
\AgdaInductiveConstructor{[]}\AgdaSymbol{))}\AgdaSpace{}%
\AgdaOperator{\AgdaInductiveConstructor{,}}\AgdaSpace{}%
\AgdaFunction{∈-++⁺ˡ}\AgdaSpace{}%
\AgdaBound{∈₁}\AgdaSymbol{)}\<%
\\
\\[\AgdaEmptyExtraSkip]%
\>[0]\AgdaFunction{script-rec}\AgdaSpace{}%
\AgdaSymbol{:}\AgdaSpace{}%
\AgdaSymbol{∀}\AgdaSpace{}%
\AgdaSymbol{(}\AgdaBound{b}\AgdaSpace{}%
\AgdaSymbol{:}\AgdaSpace{}%
\AgdaPostulate{Build}\AgdaSymbol{)}\AgdaSpace{}%
\AgdaBound{s}\AgdaSpace{}%
\AgdaBound{ls}\AgdaSpace{}%
\AgdaSymbol{→}\AgdaSpace{}%
\AgdaPostulate{FileInfo}\<%
\\
\>[0]\AgdaFunction{script-rec}\AgdaSpace{}%
\AgdaInductiveConstructor{[]}\AgdaSpace{}%
\AgdaBound{s}\AgdaSpace{}%
\AgdaBound{ls}\AgdaSpace{}%
\AgdaSymbol{=}\AgdaSpace{}%
\AgdaBound{ls}\<%
\\
\>[0]\AgdaFunction{script-rec}\AgdaSpace{}%
\AgdaSymbol{(}\AgdaBound{x}\AgdaSpace{}%
\AgdaOperator{\AgdaInductiveConstructor{∷}}\AgdaSpace{}%
\AgdaBound{b}\AgdaSymbol{)}\AgdaSpace{}%
\AgdaBound{s}\AgdaSpace{}%
\AgdaBound{ls}\AgdaSpace{}%
\AgdaSymbol{=}\AgdaSpace{}%
\AgdaFunction{script-rec}\AgdaSpace{}%
\AgdaBound{b}\AgdaSpace{}%
\AgdaSymbol{(}\AgdaPostulate{St.run}\AgdaSpace{}%
\AgdaBound{x}\AgdaSpace{}%
\AgdaBound{s}\AgdaSymbol{)}\AgdaSpace{}%
\AgdaSymbol{(}\AgdaPostulate{rec}\AgdaSpace{}%
\AgdaBound{s}\AgdaSpace{}%
\AgdaBound{x}\AgdaSpace{}%
\AgdaBound{ls}\AgdaSymbol{)}\<%
\\
\\[\AgdaEmptyExtraSkip]%
\>[0]\AgdaFunction{script-rec-++}\AgdaSpace{}%
\AgdaSymbol{:}\AgdaSpace{}%
\AgdaSymbol{∀}\AgdaSpace{}%
\AgdaSymbol{\{}\AgdaBound{s}\AgdaSymbol{\}}\AgdaSpace{}%
\AgdaSymbol{\{}\AgdaBound{as}\AgdaSymbol{\}}\AgdaSpace{}%
\AgdaSymbol{\{}\AgdaBound{bs}\AgdaSymbol{\}}\AgdaSpace{}%
\AgdaBound{xs}\AgdaSpace{}%
\AgdaSymbol{→}\AgdaSpace{}%
\AgdaFunction{script-rec}\AgdaSpace{}%
\AgdaBound{xs}\AgdaSpace{}%
\AgdaBound{s}\AgdaSpace{}%
\AgdaSymbol{(}\AgdaBound{as}\AgdaSpace{}%
\AgdaOperator{\AgdaFunction{++}}\AgdaSpace{}%
\AgdaBound{bs}\AgdaSymbol{)}\AgdaSpace{}%
\AgdaOperator{\AgdaDatatype{≡}}\AgdaSpace{}%
\AgdaFunction{script-rec}\AgdaSpace{}%
\AgdaBound{xs}\AgdaSpace{}%
\AgdaBound{s}\AgdaSpace{}%
\AgdaBound{as}\AgdaSpace{}%
\AgdaOperator{\AgdaFunction{++}}\AgdaSpace{}%
\AgdaBound{bs}\<%
\\
\>[0]\AgdaFunction{script-rec-++}\AgdaSpace{}%
\AgdaInductiveConstructor{[]}\AgdaSpace{}%
\AgdaSymbol{=}\AgdaSpace{}%
\AgdaInductiveConstructor{refl}\<%
\\
\>[0]\AgdaFunction{script-rec-++}\AgdaSpace{}%
\AgdaSymbol{(}\AgdaBound{x}\AgdaSpace{}%
\AgdaOperator{\AgdaInductiveConstructor{∷}}\AgdaSpace{}%
\AgdaBound{xs}\AgdaSymbol{)}\AgdaSpace{}%
\AgdaSymbol{=}\AgdaSpace{}%
\AgdaFunction{script-rec-++}\AgdaSpace{}%
\AgdaBound{xs}\<%
\\
\\[\AgdaEmptyExtraSkip]%
\>[0]\AgdaFunction{before-reverse}\AgdaSpace{}%
\AgdaSymbol{:}\AgdaSpace{}%
\AgdaSymbol{∀}\AgdaSpace{}%
\AgdaBound{x₁}\AgdaSpace{}%
\AgdaBound{x₂}\AgdaSpace{}%
\AgdaSymbol{(}\AgdaBound{xs}\AgdaSpace{}%
\AgdaSymbol{:}\AgdaSpace{}%
\AgdaPostulate{Build}\AgdaSymbol{)}\AgdaSpace{}%
\AgdaSymbol{→}\AgdaSpace{}%
\AgdaBound{x₁}\AgdaSpace{}%
\AgdaOperator{\AgdaFunction{before}}\AgdaSpace{}%
\AgdaBound{x₂}\AgdaSpace{}%
\AgdaOperator{\AgdaFunction{∈}}\AgdaSpace{}%
\AgdaBound{xs}\AgdaSpace{}%
\AgdaSymbol{→}\AgdaSpace{}%
\AgdaBound{x₂}\AgdaSpace{}%
\AgdaOperator{\AgdaFunction{before}}\AgdaSpace{}%
\AgdaBound{x₁}\AgdaSpace{}%
\AgdaOperator{\AgdaFunction{∈}}\AgdaSpace{}%
\AgdaSymbol{(}\AgdaFunction{reverse}\AgdaSpace{}%
\AgdaBound{xs}\AgdaSymbol{)}\<%
\\
\>[0]\AgdaFunction{before-reverse}\AgdaSpace{}%
\AgdaBound{x₁}\AgdaSpace{}%
\AgdaBound{x₂}\AgdaSpace{}%
\AgdaBound{xs}\AgdaSpace{}%
\AgdaSymbol{(}\AgdaBound{ys}\AgdaSpace{}%
\AgdaOperator{\AgdaInductiveConstructor{,}}\AgdaSpace{}%
\AgdaBound{zs}\AgdaSpace{}%
\AgdaOperator{\AgdaInductiveConstructor{,}}\AgdaSpace{}%
\AgdaBound{≡₁}\AgdaSpace{}%
\AgdaOperator{\AgdaInductiveConstructor{,}}\AgdaSpace{}%
\AgdaBound{x₂∈zs}\AgdaSymbol{)}\AgdaSpace{}%
\AgdaKeyword{with}\AgdaSpace{}%
\AgdaFunction{∈-∃++}\AgdaSpace{}%
\AgdaBound{x₂∈zs}\<%
\\
\>[0]\AgdaSymbol{...}\AgdaSpace{}%
\AgdaSymbol{|}\AgdaSpace{}%
\AgdaSymbol{(}\AgdaBound{as}\AgdaSpace{}%
\AgdaOperator{\AgdaInductiveConstructor{,}}\AgdaSpace{}%
\AgdaBound{bs}\AgdaSpace{}%
\AgdaOperator{\AgdaInductiveConstructor{,}}\AgdaSpace{}%
\AgdaBound{≡₂}\AgdaSymbol{)}\AgdaSpace{}%
\AgdaSymbol{=}\AgdaSpace{}%
\AgdaFunction{reverse}\AgdaSpace{}%
\AgdaBound{bs}\AgdaSpace{}%
\AgdaOperator{\AgdaInductiveConstructor{,}}\AgdaSpace{}%
\AgdaFunction{reverse}\AgdaSpace{}%
\AgdaSymbol{(}\AgdaBound{ys}\AgdaSpace{}%
\AgdaOperator{\AgdaFunction{++}}\AgdaSpace{}%
\AgdaBound{x₁}\AgdaSpace{}%
\AgdaOperator{\AgdaInductiveConstructor{∷}}\AgdaSpace{}%
\AgdaBound{as}\AgdaSymbol{)}\AgdaSpace{}%
\AgdaOperator{\AgdaInductiveConstructor{,}}\AgdaSpace{}%
\AgdaFunction{≡₃}\AgdaSpace{}%
\AgdaOperator{\AgdaInductiveConstructor{,}}\AgdaSpace{}%
\AgdaFunction{reverse⁺}\AgdaSpace{}%
\AgdaSymbol{(}\AgdaFunction{∈-++⁺ʳ}\AgdaSpace{}%
\AgdaBound{ys}\AgdaSpace{}%
\AgdaSymbol{(}\AgdaInductiveConstructor{here}\AgdaSpace{}%
\AgdaInductiveConstructor{refl}\AgdaSymbol{))}\<%
\\
\>[0][@{}l@{\AgdaIndent{0}}]%
\>[2]\AgdaKeyword{where}%
\>[2557I]\AgdaFunction{≡₃}\AgdaSpace{}%
\AgdaSymbol{:}\AgdaSpace{}%
\AgdaFunction{reverse}\AgdaSpace{}%
\AgdaBound{xs}\AgdaSpace{}%
\AgdaOperator{\AgdaDatatype{≡}}\AgdaSpace{}%
\AgdaFunction{reverse}\AgdaSpace{}%
\AgdaBound{bs}\AgdaSpace{}%
\AgdaOperator{\AgdaFunction{++}}\AgdaSpace{}%
\AgdaBound{x₂}\AgdaSpace{}%
\AgdaOperator{\AgdaInductiveConstructor{∷}}\AgdaSpace{}%
\AgdaFunction{reverse}\AgdaSpace{}%
\AgdaSymbol{(}\AgdaBound{ys}\AgdaSpace{}%
\AgdaOperator{\AgdaFunction{++}}\AgdaSpace{}%
\AgdaBound{x₁}\AgdaSpace{}%
\AgdaOperator{\AgdaInductiveConstructor{∷}}\AgdaSpace{}%
\AgdaBound{as}\AgdaSymbol{)}\<%
\\
\>[.][@{}l@{}]\<[2557I]%
\>[8]\AgdaFunction{≡₃}\AgdaSpace{}%
\AgdaKeyword{with}\AgdaSpace{}%
\AgdaFunction{trans}\AgdaSpace{}%
\AgdaSymbol{(}\AgdaFunction{subst}\AgdaSpace{}%
\AgdaSymbol{(λ}\AgdaSpace{}%
\AgdaBound{x}\AgdaSpace{}%
\AgdaSymbol{→}\AgdaSpace{}%
\AgdaBound{xs}\AgdaSpace{}%
\AgdaOperator{\AgdaDatatype{≡}}\AgdaSpace{}%
\AgdaBound{ys}\AgdaSpace{}%
\AgdaOperator{\AgdaFunction{++}}\AgdaSpace{}%
\AgdaBound{x₁}\AgdaSpace{}%
\AgdaOperator{\AgdaInductiveConstructor{∷}}\AgdaSpace{}%
\AgdaBound{x}\AgdaSymbol{)}\AgdaSpace{}%
\AgdaBound{≡₂}\AgdaSpace{}%
\AgdaBound{≡₁}\AgdaSymbol{)}\AgdaSpace{}%
\AgdaSymbol{(}\AgdaFunction{sym}\AgdaSpace{}%
\AgdaSymbol{(}\AgdaFunction{++-assoc}\AgdaSpace{}%
\AgdaBound{ys}\AgdaSpace{}%
\AgdaSymbol{(}\AgdaBound{x₁}\AgdaSpace{}%
\AgdaOperator{\AgdaInductiveConstructor{∷}}\AgdaSpace{}%
\AgdaBound{as}\AgdaSymbol{)}\AgdaSpace{}%
\AgdaSymbol{(}\AgdaBound{x₂}\AgdaSpace{}%
\AgdaOperator{\AgdaInductiveConstructor{∷}}\AgdaSpace{}%
\AgdaBound{bs}\AgdaSymbol{)))}\<%
\\
\>[8]\AgdaSymbol{...}\AgdaSpace{}%
\AgdaSymbol{|}\AgdaSpace{}%
\AgdaBound{a}\AgdaSpace{}%
\AgdaSymbol{=}\AgdaSpace{}%
\AgdaFunction{trans}%
\>[2601I]\AgdaSymbol{(}\AgdaFunction{cong}\AgdaSpace{}%
\AgdaFunction{reverse}\AgdaSpace{}%
\AgdaBound{a}\AgdaSymbol{)}\<%
\\
\>[.][@{}l@{}]\<[2601I]%
\>[24]\AgdaSymbol{(}\AgdaFunction{trans}%
\>[2604I]\AgdaSymbol{(}\AgdaFunction{reverse-++-commute}\AgdaSpace{}%
\AgdaSymbol{(}\AgdaBound{ys}\AgdaSpace{}%
\AgdaOperator{\AgdaFunction{++}}\AgdaSpace{}%
\AgdaBound{x₁}\AgdaSpace{}%
\AgdaOperator{\AgdaInductiveConstructor{∷}}\AgdaSpace{}%
\AgdaBound{as}\AgdaSymbol{)}\AgdaSpace{}%
\AgdaSymbol{(}\AgdaBound{x₂}\AgdaSpace{}%
\AgdaOperator{\AgdaInductiveConstructor{∷}}\AgdaSpace{}%
\AgdaBound{bs}\AgdaSymbol{))}\<%
\\
\>[.][@{}l@{}]\<[2604I]%
\>[31]\AgdaSymbol{(}\AgdaFunction{trans}%
\>[2613I]\AgdaSymbol{(}\AgdaFunction{cong}\AgdaSpace{}%
\AgdaSymbol{(}\AgdaOperator{\AgdaFunction{\AgdaUnderscore{}++}}\AgdaSpace{}%
\AgdaFunction{reverse}\AgdaSpace{}%
\AgdaSymbol{(}\AgdaBound{ys}\AgdaSpace{}%
\AgdaOperator{\AgdaFunction{++}}\AgdaSpace{}%
\AgdaBound{x₁}\AgdaSpace{}%
\AgdaOperator{\AgdaInductiveConstructor{∷}}\AgdaSpace{}%
\AgdaBound{as}\AgdaSymbol{))}\AgdaSpace{}%
\AgdaSymbol{(}\AgdaFunction{unfold-reverse}\AgdaSpace{}%
\AgdaBound{x₂}\AgdaSpace{}%
\AgdaBound{bs}\AgdaSymbol{))}\<%
\\
\>[.][@{}l@{}]\<[2613I]%
\>[38]\AgdaSymbol{(}\AgdaFunction{l4}\AgdaSpace{}%
\AgdaBound{x₂}\AgdaSpace{}%
\AgdaSymbol{(}\AgdaFunction{reverse}\AgdaSpace{}%
\AgdaBound{bs}\AgdaSymbol{))))}\<%
\\
\\[\AgdaEmptyExtraSkip]%
\\[\AgdaEmptyExtraSkip]%
\>[0]\AgdaFunction{preserves-++}\AgdaSpace{}%
\AgdaSymbol{:}\AgdaSpace{}%
\AgdaSymbol{∀}\AgdaSpace{}%
\AgdaSymbol{\{}\AgdaBound{s}\AgdaSymbol{\}}\AgdaSpace{}%
\AgdaSymbol{\{}\AgdaBound{ls}\AgdaSymbol{\}}\AgdaSpace{}%
\AgdaBound{x}\AgdaSpace{}%
\AgdaBound{b₁}\AgdaSpace{}%
\AgdaBound{b₂}\AgdaSpace{}%
\AgdaSymbol{→}\AgdaSpace{}%
\AgdaSymbol{(}\AgdaPostulate{cmdsRun}\AgdaSpace{}%
\AgdaSymbol{(}\AgdaFunction{script-rec}\AgdaSpace{}%
\AgdaBound{b₁}\AgdaSpace{}%
\AgdaBound{s}\AgdaSpace{}%
\AgdaBound{ls}\AgdaSymbol{))}\AgdaSpace{}%
\AgdaOperator{\AgdaDatatype{≡}}\AgdaSpace{}%
\AgdaFunction{reverse}\AgdaSpace{}%
\AgdaBound{b₂}\AgdaSpace{}%
\AgdaSymbol{→}\AgdaSpace{}%
\AgdaFunction{Disjoint}\AgdaSpace{}%
\AgdaSymbol{(}\AgdaPostulate{files}\AgdaSpace{}%
\AgdaSymbol{(}\AgdaFunction{script-rec}\AgdaSpace{}%
\AgdaBound{b₁}\AgdaSpace{}%
\AgdaBound{s}\AgdaSpace{}%
\AgdaBound{ls}\AgdaSymbol{))}\AgdaSpace{}%
\AgdaSymbol{(}\AgdaPostulate{cmdWriteNames}\AgdaSpace{}%
\AgdaBound{x}\AgdaSpace{}%
\AgdaSymbol{(}\AgdaPostulate{script}\AgdaSpace{}%
\AgdaBound{b₁}\AgdaSpace{}%
\AgdaBound{s}\AgdaSymbol{))}\AgdaSpace{}%
\AgdaSymbol{→}\AgdaSpace{}%
\AgdaBound{x}\AgdaSpace{}%
\AgdaOperator{\AgdaPostulate{∉}}\AgdaSpace{}%
\AgdaSymbol{(}\AgdaFunction{map}\AgdaSpace{}%
\AgdaField{proj₁}\AgdaSpace{}%
\AgdaBound{ls}\AgdaSymbol{)}\AgdaSpace{}%
\AgdaSymbol{→}\AgdaSpace{}%
\AgdaBound{x}\AgdaSpace{}%
\AgdaOperator{\AgdaPostulate{∉}}\AgdaSpace{}%
\AgdaBound{b₁}\AgdaSpace{}%
\AgdaSymbol{→}\AgdaSpace{}%
\AgdaPostulate{HazardFree}\AgdaSpace{}%
\AgdaBound{s}\AgdaSpace{}%
\AgdaBound{b₁}\AgdaSpace{}%
\AgdaBound{b₂}\AgdaSpace{}%
\AgdaBound{ls}\AgdaSpace{}%
\AgdaSymbol{→}\AgdaSpace{}%
\AgdaPostulate{HazardFree}\AgdaSpace{}%
\AgdaBound{s}\AgdaSpace{}%
\AgdaSymbol{(}\AgdaBound{b₁}\AgdaSpace{}%
\AgdaOperator{\AgdaFunction{++}}\AgdaSpace{}%
\AgdaBound{x}\AgdaSpace{}%
\AgdaOperator{\AgdaInductiveConstructor{∷}}\AgdaSpace{}%
\AgdaInductiveConstructor{[]}\AgdaSymbol{)}\AgdaSpace{}%
\AgdaSymbol{(}\AgdaBound{b₂}\AgdaSpace{}%
\AgdaOperator{\AgdaFunction{++}}\AgdaSpace{}%
\AgdaBound{x}\AgdaSpace{}%
\AgdaOperator{\AgdaInductiveConstructor{∷}}\AgdaSpace{}%
\AgdaInductiveConstructor{[]}\AgdaSymbol{)}\AgdaSpace{}%
\AgdaBound{ls}\<%
\\
\>[0]\AgdaFunction{preserves-++}\AgdaSpace{}%
\AgdaBound{x}\AgdaSpace{}%
\AgdaInductiveConstructor{[]}\AgdaSpace{}%
\AgdaBound{b₂}\AgdaSpace{}%
\AgdaBound{≡₁}\AgdaSpace{}%
\AgdaBound{dsj}\AgdaSpace{}%
\AgdaSymbol{\AgdaUnderscore{}}\AgdaSpace{}%
\AgdaSymbol{\AgdaUnderscore{}}\AgdaSpace{}%
\AgdaInductiveConstructor{[]}\AgdaSpace{}%
\AgdaSymbol{=}\AgdaSpace{}%
\AgdaFunction{g₁}\AgdaSpace{}%
\AgdaOperator{\AgdaInductiveConstructor{∷}}\AgdaSpace{}%
\AgdaInductiveConstructor{[]}\<%
\\
\>[0][@{}l@{\AgdaIndent{0}}]%
\>[2]\AgdaKeyword{where}%
\>[2697I]\AgdaFunction{g₁}\AgdaSpace{}%
\AgdaSymbol{:}\AgdaSpace{}%
\AgdaOperator{\AgdaFunction{¬}}\AgdaSpace{}%
\AgdaPostulate{Hazard}\AgdaSpace{}%
\AgdaSymbol{\AgdaUnderscore{}}\AgdaSpace{}%
\AgdaBound{x}\AgdaSpace{}%
\AgdaSymbol{(}\AgdaBound{b₂}\AgdaSpace{}%
\AgdaOperator{\AgdaFunction{++}}\AgdaSpace{}%
\AgdaBound{x}\AgdaSpace{}%
\AgdaOperator{\AgdaInductiveConstructor{∷}}\AgdaSpace{}%
\AgdaInductiveConstructor{[]}\AgdaSymbol{)}\AgdaSpace{}%
\AgdaSymbol{\AgdaUnderscore{}}\<%
\\
\>[.][@{}l@{}]\<[2697I]%
\>[8]\AgdaFunction{g₁}\AgdaSpace{}%
\AgdaSymbol{(}\AgdaPostulate{ReadWrite}\AgdaSpace{}%
\AgdaBound{x}\AgdaSpace{}%
\AgdaBound{x₁}\AgdaSymbol{)}\AgdaSpace{}%
\AgdaSymbol{=}\AgdaSpace{}%
\AgdaBound{dsj}\AgdaSpace{}%
\AgdaSymbol{(}\AgdaFunction{∈-++⁺ˡ}\AgdaSpace{}%
\AgdaBound{x₁}\AgdaSpace{}%
\AgdaOperator{\AgdaInductiveConstructor{,}}\AgdaSpace{}%
\AgdaBound{x}\AgdaSymbol{)}\<%
\\
\>[8]\AgdaFunction{g₁}\AgdaSpace{}%
\AgdaSymbol{(}\AgdaPostulate{WriteWrite}\AgdaSpace{}%
\AgdaBound{x}\AgdaSpace{}%
\AgdaBound{x₁}\AgdaSymbol{)}\AgdaSpace{}%
\AgdaSymbol{=}\AgdaSpace{}%
\AgdaBound{dsj}\AgdaSpace{}%
\AgdaSymbol{(}\AgdaFunction{∈-++⁺ʳ}\AgdaSpace{}%
\AgdaSymbol{\AgdaUnderscore{}}\AgdaSpace{}%
\AgdaBound{x₁}\AgdaSpace{}%
\AgdaOperator{\AgdaInductiveConstructor{,}}\AgdaSpace{}%
\AgdaBound{x}\AgdaSymbol{)}\<%
\\
\>[8]\AgdaFunction{g₁}\AgdaSpace{}%
\AgdaSymbol{(}\AgdaPostulate{Speculative}\AgdaSpace{}%
\AgdaBound{x₁}\AgdaSpace{}%
\AgdaBound{x₂}\AgdaSpace{}%
\AgdaBound{y}\AgdaSpace{}%
\AgdaBound{x₃}\AgdaSpace{}%
\AgdaBound{x₄}\AgdaSpace{}%
\AgdaBound{x₅}\AgdaSpace{}%
\AgdaBound{x₆}\AgdaSymbol{)}\AgdaSpace{}%
\AgdaKeyword{with}\AgdaSpace{}%
\AgdaFunction{before-reverse}\AgdaSpace{}%
\AgdaBound{x₂}\AgdaSpace{}%
\AgdaBound{x₁}\AgdaSpace{}%
\AgdaSymbol{(}\AgdaBound{x}\AgdaSpace{}%
\AgdaOperator{\AgdaInductiveConstructor{∷}}\AgdaSpace{}%
\AgdaFunction{reverse}\AgdaSpace{}%
\AgdaBound{b₂}\AgdaSymbol{)}\AgdaSpace{}%
\AgdaSymbol{(}\AgdaFunction{subst}\AgdaSpace{}%
\AgdaSymbol{(λ}\AgdaSpace{}%
\AgdaBound{x₇}\AgdaSpace{}%
\AgdaSymbol{→}\AgdaSpace{}%
\AgdaBound{x₂}\AgdaSpace{}%
\AgdaOperator{\AgdaFunction{before}}\AgdaSpace{}%
\AgdaBound{x₁}\AgdaSpace{}%
\AgdaOperator{\AgdaFunction{∈}}\AgdaSpace{}%
\AgdaBound{x₇}\AgdaSymbol{)}\AgdaSpace{}%
\AgdaSymbol{(}\AgdaFunction{cong}\AgdaSpace{}%
\AgdaSymbol{(}\AgdaBound{x}\AgdaSpace{}%
\AgdaOperator{\AgdaInductiveConstructor{∷\AgdaUnderscore{}}}\AgdaSymbol{)}\AgdaSpace{}%
\AgdaBound{≡₁}\AgdaSymbol{)}\AgdaSpace{}%
\AgdaBound{y}\AgdaSymbol{)}\<%
\\
\>[8]\AgdaSymbol{...}\AgdaSpace{}%
\AgdaSymbol{|}\AgdaSpace{}%
\AgdaBound{bf}\AgdaSpace{}%
\AgdaSymbol{=}\AgdaSpace{}%
\AgdaBound{x₄}\AgdaSpace{}%
\AgdaSymbol{(}\AgdaFunction{subst}\AgdaSpace{}%
\AgdaSymbol{(λ}\AgdaSpace{}%
\AgdaBound{x₁₀}\AgdaSpace{}%
\AgdaSymbol{→}\AgdaSpace{}%
\AgdaBound{x₁}\AgdaSpace{}%
\AgdaOperator{\AgdaFunction{before}}\AgdaSpace{}%
\AgdaBound{x₂}\AgdaSpace{}%
\AgdaOperator{\AgdaFunction{∈}}\AgdaSpace{}%
\AgdaBound{x₁₀}\AgdaSymbol{)}\AgdaSpace{}%
\AgdaSymbol{(}\AgdaFunction{trans}\AgdaSpace{}%
\AgdaSymbol{(}\AgdaFunction{unfold-reverse}\AgdaSpace{}%
\AgdaBound{x}\AgdaSpace{}%
\AgdaSymbol{(}\AgdaFunction{reverse}\AgdaSpace{}%
\AgdaBound{b₂}\AgdaSymbol{))}\AgdaSpace{}%
\AgdaSymbol{(}\AgdaFunction{cong}\AgdaSpace{}%
\AgdaSymbol{(}\AgdaOperator{\AgdaFunction{\AgdaUnderscore{}++}}\AgdaSpace{}%
\AgdaBound{x}\AgdaSpace{}%
\AgdaOperator{\AgdaInductiveConstructor{∷}}\AgdaSpace{}%
\AgdaInductiveConstructor{[]}\AgdaSymbol{)}\AgdaSpace{}%
\AgdaSymbol{(}\AgdaFunction{reverse-involutive}\AgdaSpace{}%
\AgdaBound{b₂}\AgdaSymbol{)))}\AgdaSpace{}%
\AgdaBound{bf}\AgdaSymbol{)}\<%
\\
\>[0]\AgdaFunction{preserves-++}\AgdaSpace{}%
\AgdaSymbol{\{}\AgdaBound{s}\AgdaSymbol{\}}\AgdaSpace{}%
\AgdaSymbol{\{}\AgdaBound{ls}\AgdaSymbol{\}}\AgdaSpace{}%
\AgdaBound{x}\AgdaSpace{}%
\AgdaSymbol{(}\AgdaBound{x₁}\AgdaSpace{}%
\AgdaOperator{\AgdaInductiveConstructor{∷}}\AgdaSpace{}%
\AgdaBound{b₁}\AgdaSymbol{)}\AgdaSpace{}%
\AgdaBound{b₂}\AgdaSpace{}%
\AgdaBound{≡₁}\AgdaSpace{}%
\AgdaBound{dsj}\AgdaSpace{}%
\AgdaBound{x∉ls}\AgdaSpace{}%
\AgdaBound{x∉b₁}\AgdaSpace{}%
\AgdaSymbol{(}\AgdaBound{x₂}\AgdaSpace{}%
\AgdaOperator{\AgdaInductiveConstructor{∷}}\AgdaSpace{}%
\AgdaBound{hf}\AgdaSymbol{)}\AgdaSpace{}%
\AgdaSymbol{=}\AgdaSpace{}%
\AgdaSymbol{(}\AgdaFunction{g₁}\AgdaSpace{}%
\AgdaBound{x∉ls}\AgdaSpace{}%
\AgdaBound{x₂}\AgdaSymbol{)}\AgdaSpace{}%
\AgdaOperator{\AgdaInductiveConstructor{∷}}\AgdaSpace{}%
\AgdaSymbol{(}\AgdaFunction{preserves-++}\AgdaSpace{}%
\AgdaBound{x}\AgdaSpace{}%
\AgdaBound{b₁}\AgdaSpace{}%
\AgdaBound{b₂}\AgdaSpace{}%
\AgdaBound{≡₁}\AgdaSpace{}%
\AgdaBound{dsj}\AgdaSpace{}%
\AgdaFunction{x∉₁}\AgdaSpace{}%
\AgdaSymbol{(λ}\AgdaSpace{}%
\AgdaBound{x₃}\AgdaSpace{}%
\AgdaSymbol{→}\AgdaSpace{}%
\AgdaBound{x∉b₁}\AgdaSpace{}%
\AgdaSymbol{(}\AgdaInductiveConstructor{there}\AgdaSpace{}%
\AgdaBound{x₃}\AgdaSymbol{))}\AgdaSpace{}%
\AgdaBound{hf}\AgdaSymbol{)}\<%
\\
\>[0][@{}l@{\AgdaIndent{0}}]%
\>[2]\AgdaKeyword{where}%
\>[2817I]\AgdaFunction{g₁}\AgdaSpace{}%
\AgdaSymbol{:}\AgdaSpace{}%
\AgdaSymbol{∀}\AgdaSpace{}%
\AgdaSymbol{\{}\AgdaBound{s}\AgdaSymbol{\}}\AgdaSpace{}%
\AgdaSymbol{\{}\AgdaBound{ls}\AgdaSymbol{\}}\AgdaSpace{}%
\AgdaSymbol{→}\AgdaSpace{}%
\AgdaBound{x}\AgdaSpace{}%
\AgdaOperator{\AgdaPostulate{∉}}\AgdaSpace{}%
\AgdaSymbol{(}\AgdaFunction{map}\AgdaSpace{}%
\AgdaField{proj₁}\AgdaSpace{}%
\AgdaBound{ls}\AgdaSymbol{)}\AgdaSpace{}%
\AgdaSymbol{→}\AgdaSpace{}%
\AgdaOperator{\AgdaFunction{¬}}\AgdaSpace{}%
\AgdaPostulate{Hazard}\AgdaSpace{}%
\AgdaBound{s}\AgdaSpace{}%
\AgdaBound{x₁}\AgdaSpace{}%
\AgdaBound{b₂}\AgdaSpace{}%
\AgdaBound{ls}\AgdaSpace{}%
\AgdaSymbol{→}\AgdaSpace{}%
\AgdaOperator{\AgdaFunction{¬}}\AgdaSpace{}%
\AgdaPostulate{Hazard}\AgdaSpace{}%
\AgdaBound{s}\AgdaSpace{}%
\AgdaBound{x₁}\AgdaSpace{}%
\AgdaSymbol{(}\AgdaBound{b₂}\AgdaSpace{}%
\AgdaOperator{\AgdaFunction{++}}\AgdaSpace{}%
\AgdaBound{x}\AgdaSpace{}%
\AgdaOperator{\AgdaInductiveConstructor{∷}}\AgdaSpace{}%
\AgdaInductiveConstructor{[]}\AgdaSymbol{)}\AgdaSpace{}%
\AgdaBound{ls}\<%
\\
\>[.][@{}l@{}]\<[2817I]%
\>[8]\AgdaFunction{g₁}\AgdaSpace{}%
\AgdaBound{x∉ls}\AgdaSpace{}%
\AgdaBound{¬hz}\AgdaSpace{}%
\AgdaBound{hz}\AgdaSpace{}%
\AgdaSymbol{=}\AgdaSpace{}%
\AgdaBound{¬hz}\AgdaSpace{}%
\AgdaSymbol{(}\AgdaFunction{still-hazard}\AgdaSpace{}%
\AgdaSymbol{\AgdaUnderscore{}}\AgdaSpace{}%
\AgdaBound{x}\AgdaSpace{}%
\AgdaBound{b₂}\AgdaSpace{}%
\AgdaSymbol{(λ}\AgdaSpace{}%
\AgdaBound{x₄}\AgdaSpace{}%
\AgdaSymbol{→}\AgdaSpace{}%
\AgdaBound{x∉b₁}\AgdaSpace{}%
\AgdaSymbol{(}\AgdaInductiveConstructor{here}\AgdaSpace{}%
\AgdaSymbol{(}\AgdaFunction{sym}\AgdaSpace{}%
\AgdaBound{x₄}\AgdaSymbol{)))}\AgdaSpace{}%
\AgdaBound{x∉ls}\AgdaSpace{}%
\AgdaBound{hz}\AgdaSymbol{)}\<%
\\
\>[8]\AgdaFunction{x∉₁}\AgdaSpace{}%
\AgdaSymbol{:}\AgdaSpace{}%
\AgdaBound{x}\AgdaSpace{}%
\AgdaOperator{\AgdaPostulate{∉}}\AgdaSpace{}%
\AgdaBound{x₁}\AgdaSpace{}%
\AgdaOperator{\AgdaInductiveConstructor{∷}}\AgdaSpace{}%
\AgdaFunction{map}\AgdaSpace{}%
\AgdaField{proj₁}\AgdaSpace{}%
\AgdaBound{ls}\<%
\\
\>[8]\AgdaFunction{x∉₁}\AgdaSpace{}%
\AgdaSymbol{(}\AgdaInductiveConstructor{here}\AgdaSpace{}%
\AgdaBound{px}\AgdaSymbol{)}\AgdaSpace{}%
\AgdaSymbol{=}\AgdaSpace{}%
\AgdaBound{x∉b₁}\AgdaSpace{}%
\AgdaSymbol{(}\AgdaInductiveConstructor{here}\AgdaSpace{}%
\AgdaBound{px}\AgdaSymbol{)}\<%
\\
\>[8]\AgdaFunction{x∉₁}\AgdaSpace{}%
\AgdaSymbol{(}\AgdaInductiveConstructor{there}\AgdaSpace{}%
\AgdaBound{x∈ls}\AgdaSymbol{)}\AgdaSpace{}%
\AgdaSymbol{=}\AgdaSpace{}%
\AgdaBound{x∉ls}\AgdaSpace{}%
\AgdaBound{x∈ls}\<%
\\
\\[\AgdaEmptyExtraSkip]%
\\[\AgdaEmptyExtraSkip]%
\>[0]\AgdaFunction{add-back-read}%
\>[2883I]\AgdaSymbol{:}\AgdaSpace{}%
\AgdaSymbol{∀}\AgdaSpace{}%
\AgdaSymbol{\{}\AgdaBound{s}\AgdaSymbol{\}}\AgdaSpace{}%
\AgdaSymbol{\{}\AgdaBound{ls}\AgdaSymbol{\}}\AgdaSpace{}%
\AgdaBound{as}\AgdaSpace{}%
\AgdaBound{bs}\AgdaSpace{}%
\AgdaBound{xs}\AgdaSpace{}%
\AgdaBound{x}\AgdaSpace{}%
\AgdaSymbol{→}\AgdaSpace{}%
\AgdaPostulate{filesRead}\AgdaSpace{}%
\AgdaSymbol{(}\AgdaFunction{script-rec}\AgdaSpace{}%
\AgdaBound{xs}\AgdaSpace{}%
\AgdaBound{s}\AgdaSpace{}%
\AgdaBound{ls}\AgdaSymbol{)}\AgdaSpace{}%
\AgdaOperator{\AgdaFunction{⊆}}\AgdaSpace{}%
\AgdaPostulate{filesRead}\AgdaSpace{}%
\AgdaSymbol{(}\AgdaFunction{script-rec}\AgdaSpace{}%
\AgdaSymbol{(}\AgdaBound{as}\AgdaSpace{}%
\AgdaOperator{\AgdaFunction{++}}\AgdaSpace{}%
\AgdaBound{bs}\AgdaSymbol{)}\AgdaSpace{}%
\AgdaBound{s}\AgdaSpace{}%
\AgdaBound{ls}\AgdaSymbol{)}\<%
\\
\>[.][@{}l@{}]\<[2883I]%
\>[14]\AgdaSymbol{→}\AgdaSpace{}%
\AgdaPostulate{filesRead}\AgdaSpace{}%
\AgdaSymbol{(}\AgdaFunction{script-rec}\AgdaSpace{}%
\AgdaSymbol{(}\AgdaBound{xs}\AgdaSpace{}%
\AgdaOperator{\AgdaFunction{++}}\AgdaSpace{}%
\AgdaBound{x}\AgdaSpace{}%
\AgdaOperator{\AgdaInductiveConstructor{∷}}\AgdaSpace{}%
\AgdaInductiveConstructor{[]}\AgdaSymbol{)}\AgdaSpace{}%
\AgdaBound{s}\AgdaSpace{}%
\AgdaBound{ls}\AgdaSymbol{)}\AgdaSpace{}%
\AgdaOperator{\AgdaFunction{⊆}}\AgdaSpace{}%
\AgdaPostulate{filesRead}\AgdaSpace{}%
\AgdaSymbol{(}\AgdaFunction{script-rec}\AgdaSpace{}%
\AgdaSymbol{(}\AgdaBound{as}\AgdaSpace{}%
\AgdaOperator{\AgdaFunction{++}}\AgdaSpace{}%
\AgdaBound{x}\AgdaSpace{}%
\AgdaOperator{\AgdaInductiveConstructor{∷}}\AgdaSpace{}%
\AgdaBound{bs}\AgdaSymbol{)}\AgdaSpace{}%
\AgdaBound{s}\AgdaSpace{}%
\AgdaBound{ls}\AgdaSymbol{)}\<%
\\
\>[0]\AgdaFunction{add-back-read}\AgdaSpace{}%
\AgdaSymbol{\{}\AgdaBound{s}\AgdaSymbol{\}}\AgdaSpace{}%
\AgdaSymbol{\{}\AgdaBound{ls}\AgdaSymbol{\}}\AgdaSpace{}%
\AgdaBound{as}\AgdaSpace{}%
\AgdaBound{bs}\AgdaSpace{}%
\AgdaBound{xs}\AgdaSpace{}%
\AgdaBound{x}\AgdaSpace{}%
\AgdaBound{⊆₁}\AgdaSpace{}%
\AgdaBound{x₂}\AgdaSpace{}%
\AgdaKeyword{with}\AgdaSpace{}%
\AgdaFunction{∈-++⁻}\AgdaSpace{}%
\AgdaSymbol{(}\AgdaPostulate{filesRead}\AgdaSpace{}%
\AgdaSymbol{(}\AgdaFunction{script-rec}\AgdaSpace{}%
\AgdaBound{xs}\AgdaSpace{}%
\AgdaBound{s}\AgdaSpace{}%
\AgdaBound{ls}\AgdaSymbol{))}\AgdaSpace{}%
\AgdaSymbol{\{!!\}}\<%
\\
\>[0]\AgdaComment{\{- proves filesRead (script-rec (xs ++ x ∷ []) s ls ≡ filesRead script-rec xs ++ cmdwriteNames x -\}}\<%
\\
\>[0]\AgdaSymbol{...}\AgdaSpace{}%
\AgdaSymbol{|}\AgdaSpace{}%
\AgdaInductiveConstructor{inj₁}\AgdaSpace{}%
\AgdaBound{∈₁}\AgdaSpace{}%
\AgdaKeyword{with}\AgdaSpace{}%
\AgdaBound{⊆₁}\AgdaSpace{}%
\AgdaBound{∈₁}\<%
\\
\>[0]\AgdaSymbol{...}\AgdaSpace{}%
\AgdaSymbol{|}\AgdaSpace{}%
\AgdaBound{a}\AgdaSpace{}%
\AgdaSymbol{=}\AgdaSpace{}%
\AgdaSymbol{\{!!\}}\<%
\\
\>[0]\AgdaComment{\{- prove (filesRead (script-rec (as ++ bs) s ls subset of (filesRead (script-rec (as ++ x ∷ bs) s ls
-\}}\<%
\\
\>[0]\AgdaFunction{add-back-read}\AgdaSpace{}%
\AgdaSymbol{\{}\AgdaBound{s}\AgdaSymbol{\}}\AgdaSpace{}%
\AgdaSymbol{\{}\AgdaBound{ls}\AgdaSymbol{\}}\AgdaSpace{}%
\AgdaBound{xs}\AgdaSpace{}%
\AgdaBound{as}\AgdaSpace{}%
\AgdaBound{bs}\AgdaSpace{}%
\AgdaBound{x}\AgdaSpace{}%
\AgdaBound{⊆₁}\AgdaSpace{}%
\AgdaBound{x₂}\AgdaSpace{}%
\AgdaSymbol{|}\AgdaSpace{}%
\AgdaInductiveConstructor{inj₂}\AgdaSpace{}%
\AgdaBound{∈₁}\AgdaSpace{}%
\AgdaSymbol{=}\AgdaSpace{}%
\AgdaSymbol{\{!!\}}\<%
\\
\>[0]\AgdaComment{\{- prove (cmdWriteNames x (script xs s)) subset of filesRead (script-rec (as ++ x ∷ bs))... -\}}\<%
\\
\\[\AgdaEmptyExtraSkip]%
\\[\AgdaEmptyExtraSkip]%
\>[0]\AgdaFunction{filesRead-sub}\AgdaSpace{}%
\AgdaSymbol{:}%
\>[17]\AgdaSymbol{∀}\AgdaSpace{}%
\AgdaSymbol{\{}\AgdaBound{s}\AgdaSymbol{\}}\AgdaSpace{}%
\AgdaSymbol{\{}\AgdaBound{ls}\AgdaSymbol{\}}\AgdaSpace{}%
\AgdaBound{xs}\AgdaSpace{}%
\AgdaBound{ys}\AgdaSpace{}%
\AgdaSymbol{→}\AgdaSpace{}%
\AgdaFunction{length}\AgdaSpace{}%
\AgdaBound{xs}\AgdaSpace{}%
\AgdaOperator{\AgdaDatatype{≡}}\AgdaSpace{}%
\AgdaFunction{length}\AgdaSpace{}%
\AgdaBound{ys}\AgdaSpace{}%
\AgdaSymbol{→}\AgdaSpace{}%
\AgdaBound{xs}\AgdaSpace{}%
\AgdaOperator{\AgdaDatatype{↭}}\AgdaSpace{}%
\AgdaBound{ys}\AgdaSpace{}%
\AgdaSymbol{→}\AgdaSpace{}%
\AgdaPostulate{HazardFree}\AgdaSpace{}%
\AgdaBound{s}\AgdaSpace{}%
\AgdaBound{xs}\AgdaSpace{}%
\AgdaSymbol{(}\AgdaFunction{reverse}\AgdaSpace{}%
\AgdaBound{ys}\AgdaSymbol{)}\AgdaSpace{}%
\AgdaBound{ls}\AgdaSpace{}%
\AgdaSymbol{→}\AgdaSpace{}%
\AgdaPostulate{filesRead}\AgdaSpace{}%
\AgdaSymbol{(}\AgdaFunction{script-rec}\AgdaSpace{}%
\AgdaSymbol{(}\AgdaFunction{reverse}\AgdaSpace{}%
\AgdaBound{ys}\AgdaSymbol{)}\AgdaSpace{}%
\AgdaBound{s}\AgdaSpace{}%
\AgdaBound{ls}\AgdaSymbol{)}\AgdaSpace{}%
\AgdaOperator{\AgdaFunction{⊆}}\AgdaSpace{}%
\AgdaPostulate{filesRead}\AgdaSpace{}%
\AgdaSymbol{(}\AgdaFunction{script-rec}\AgdaSpace{}%
\AgdaBound{xs}\AgdaSpace{}%
\AgdaBound{s}\AgdaSpace{}%
\AgdaBound{ls}\AgdaSymbol{)}\<%
\\
\>[0]\AgdaFunction{filesRead-sub}\AgdaSpace{}%
\AgdaInductiveConstructor{[]}\AgdaSpace{}%
\AgdaInductiveConstructor{[]}\AgdaSpace{}%
\AgdaSymbol{\AgdaUnderscore{}}\AgdaSpace{}%
\AgdaBound{p}\AgdaSpace{}%
\AgdaBound{hf}\AgdaSpace{}%
\AgdaSymbol{=}\AgdaSpace{}%
\AgdaSymbol{λ}\AgdaSpace{}%
\AgdaBound{x₂}\AgdaSpace{}%
\AgdaSymbol{→}\AgdaSpace{}%
\AgdaBound{x₂}\<%
\\
\>[0]\AgdaFunction{filesRead-sub}\AgdaSpace{}%
\AgdaSymbol{\{}\AgdaBound{s}\AgdaSymbol{\}}\AgdaSpace{}%
\AgdaBound{xs}\AgdaSpace{}%
\AgdaSymbol{(}\AgdaBound{x}\AgdaSpace{}%
\AgdaOperator{\AgdaInductiveConstructor{∷}}\AgdaSpace{}%
\AgdaBound{ys}\AgdaSymbol{)}\AgdaSpace{}%
\AgdaSymbol{\AgdaUnderscore{}}\AgdaSpace{}%
\AgdaBound{p}\AgdaSpace{}%
\AgdaBound{hf}\AgdaSpace{}%
\AgdaKeyword{with}\AgdaSpace{}%
\AgdaFunction{∈-∃++}\AgdaSpace{}%
\AgdaSymbol{(}\AgdaFunction{∈-resp-↭}\AgdaSpace{}%
\AgdaSymbol{(}\AgdaFunction{↭-sym}\AgdaSpace{}%
\AgdaBound{p}\AgdaSymbol{)}\AgdaSpace{}%
\AgdaSymbol{(}\AgdaInductiveConstructor{here}\AgdaSpace{}%
\AgdaInductiveConstructor{refl}\AgdaSymbol{))}\<%
\\
\>[0]\AgdaSymbol{...}\AgdaSpace{}%
\AgdaSymbol{|}\AgdaSpace{}%
\AgdaSymbol{(}\AgdaBound{as}\AgdaSpace{}%
\AgdaOperator{\AgdaInductiveConstructor{,}}\AgdaSpace{}%
\AgdaBound{bs}\AgdaSpace{}%
\AgdaOperator{\AgdaInductiveConstructor{,}}\AgdaSpace{}%
\AgdaBound{≡₁}\AgdaSymbol{)}\AgdaSpace{}%
\AgdaKeyword{with}\AgdaSpace{}%
\AgdaFunction{add-back-read}\AgdaSpace{}%
\AgdaBound{as}\AgdaSpace{}%
\AgdaBound{bs}\AgdaSpace{}%
\AgdaSymbol{(}\AgdaFunction{reverse}\AgdaSpace{}%
\AgdaBound{ys}\AgdaSymbol{)}\AgdaSpace{}%
\AgdaBound{x}\AgdaSpace{}%
\AgdaSymbol{(}\AgdaFunction{filesRead-sub}\AgdaSpace{}%
\AgdaSymbol{(}\AgdaBound{as}\AgdaSpace{}%
\AgdaOperator{\AgdaFunction{++}}\AgdaSpace{}%
\AgdaBound{bs}\AgdaSymbol{)}\AgdaSpace{}%
\AgdaBound{ys}\AgdaSpace{}%
\AgdaSymbol{(}\AgdaFunction{↭-length}\AgdaSpace{}%
\AgdaFunction{↭₂}\AgdaSymbol{)}\AgdaSpace{}%
\AgdaFunction{↭₂}\AgdaSpace{}%
\AgdaFunction{hf₁}\AgdaSymbol{)}\<%
\\
\>[0][@{}l@{\AgdaIndent{0}}]%
\>[2]\AgdaKeyword{where}%
\>[3045I]\AgdaFunction{↭₂}\AgdaSpace{}%
\AgdaSymbol{:}\AgdaSpace{}%
\AgdaBound{as}\AgdaSpace{}%
\AgdaOperator{\AgdaFunction{++}}\AgdaSpace{}%
\AgdaBound{bs}\AgdaSpace{}%
\AgdaOperator{\AgdaDatatype{↭}}\AgdaSpace{}%
\AgdaBound{ys}\<%
\\
\>[.][@{}l@{}]\<[3045I]%
\>[8]\AgdaFunction{↭₂}\AgdaSpace{}%
\AgdaSymbol{=}\AgdaSpace{}%
\AgdaFunction{drop-mid}\AgdaSpace{}%
\AgdaBound{as}\AgdaSpace{}%
\AgdaInductiveConstructor{[]}\AgdaSpace{}%
\AgdaSymbol{(}\AgdaFunction{subst}\AgdaSpace{}%
\AgdaSymbol{(λ}\AgdaSpace{}%
\AgdaBound{x₁}\AgdaSpace{}%
\AgdaSymbol{→}\AgdaSpace{}%
\AgdaBound{x₁}\AgdaSpace{}%
\AgdaOperator{\AgdaDatatype{↭}}\AgdaSpace{}%
\AgdaBound{x}\AgdaSpace{}%
\AgdaOperator{\AgdaInductiveConstructor{∷}}\AgdaSpace{}%
\AgdaBound{ys}\AgdaSymbol{)}\AgdaSpace{}%
\AgdaBound{≡₁}\AgdaSpace{}%
\AgdaBound{p}\AgdaSymbol{)}\<%
\\
\>[8]\AgdaFunction{hf₀}\AgdaSpace{}%
\AgdaSymbol{:}\AgdaSpace{}%
\AgdaPostulate{HazardFree}\AgdaSpace{}%
\AgdaBound{s}\AgdaSpace{}%
\AgdaSymbol{(}\AgdaBound{as}\AgdaSpace{}%
\AgdaOperator{\AgdaFunction{++}}\AgdaSpace{}%
\AgdaBound{x}\AgdaSpace{}%
\AgdaOperator{\AgdaInductiveConstructor{∷}}\AgdaSpace{}%
\AgdaBound{bs}\AgdaSymbol{)}\AgdaSpace{}%
\AgdaSymbol{((}\AgdaFunction{reverse}\AgdaSpace{}%
\AgdaBound{ys}\AgdaSymbol{)}\AgdaSpace{}%
\AgdaOperator{\AgdaFunction{++}}\AgdaSpace{}%
\AgdaBound{x}\AgdaSpace{}%
\AgdaOperator{\AgdaInductiveConstructor{∷}}\AgdaSpace{}%
\AgdaInductiveConstructor{[]}\AgdaSymbol{)}\AgdaSpace{}%
\AgdaSymbol{\AgdaUnderscore{}}\<%
\\
\>[8]\AgdaFunction{hf₀}\AgdaSpace{}%
\AgdaSymbol{=}\AgdaSpace{}%
\AgdaFunction{subst₂}\AgdaSpace{}%
\AgdaSymbol{(λ}\AgdaSpace{}%
\AgdaBound{x₂}\AgdaSpace{}%
\AgdaBound{x₃}\AgdaSpace{}%
\AgdaSymbol{→}\AgdaSpace{}%
\AgdaPostulate{HazardFree}\AgdaSpace{}%
\AgdaBound{s}\AgdaSpace{}%
\AgdaBound{x₂}\AgdaSpace{}%
\AgdaBound{x₃}\AgdaSpace{}%
\AgdaSymbol{\AgdaUnderscore{})}\AgdaSpace{}%
\AgdaBound{≡₁}\AgdaSpace{}%
\AgdaSymbol{(}\AgdaFunction{unfold-reverse}\AgdaSpace{}%
\AgdaBound{x}\AgdaSpace{}%
\AgdaBound{ys}\AgdaSymbol{)}\AgdaSpace{}%
\AgdaBound{hf}\<%
\\
\>[8]\AgdaFunction{hf₁}\AgdaSpace{}%
\AgdaSymbol{:}\AgdaSpace{}%
\AgdaPostulate{HazardFree}\AgdaSpace{}%
\AgdaBound{s}\AgdaSpace{}%
\AgdaSymbol{(}\AgdaBound{as}\AgdaSpace{}%
\AgdaOperator{\AgdaFunction{++}}\AgdaSpace{}%
\AgdaBound{bs}\AgdaSymbol{)}\AgdaSpace{}%
\AgdaSymbol{(}\AgdaFunction{reverse}\AgdaSpace{}%
\AgdaBound{ys}\AgdaSymbol{)}\AgdaSpace{}%
\AgdaSymbol{\AgdaUnderscore{}}\<%
\\
\>[8]\AgdaFunction{hf₁}\AgdaSpace{}%
\AgdaSymbol{=}\AgdaSpace{}%
\AgdaPostulate{hf-drop-mid}%
\>[3109I]\AgdaBound{as}\AgdaSpace{}%
\AgdaBound{bs}\AgdaSpace{}%
\AgdaSymbol{(}\AgdaFunction{reverse}\AgdaSpace{}%
\AgdaBound{ys}\AgdaSymbol{)}\AgdaSpace{}%
\AgdaSymbol{\{!!\}}\AgdaSpace{}%
\AgdaComment{-- (λ x₂ → ∈-resp-↭ (subst₂ (λ x₀ x₁ → x₀ ↭ x₁) ≡₁ (unfold-reverse x (reverse ys)) p) x₂)}\<%
\\
\>[.][@{}l@{}]\<[3109I]%
\>[26]\AgdaSymbol{\{!!\}}\AgdaSpace{}%
\AgdaSymbol{\{!!\}}\AgdaSpace{}%
\AgdaSymbol{\{!!\}}\AgdaSpace{}%
\AgdaSymbol{\{!!\}}\AgdaSpace{}%
\AgdaFunction{hf₀}\<%
\\
\>[0]\AgdaSymbol{...}\AgdaSpace{}%
\AgdaSymbol{|}\AgdaSpace{}%
\AgdaBound{a}\AgdaSpace{}%
\AgdaSymbol{=}\AgdaSpace{}%
\AgdaFunction{subst₂}\AgdaSpace{}%
\AgdaSymbol{(λ}\AgdaSpace{}%
\AgdaBound{x₂}\AgdaSpace{}%
\AgdaBound{x₃}\AgdaSpace{}%
\AgdaSymbol{→}\AgdaSpace{}%
\AgdaPostulate{filesRead}\AgdaSpace{}%
\AgdaSymbol{(}\AgdaFunction{script-rec}\AgdaSpace{}%
\AgdaBound{x₂}\AgdaSpace{}%
\AgdaBound{s}\AgdaSpace{}%
\AgdaSymbol{\AgdaUnderscore{})}\AgdaSpace{}%
\AgdaOperator{\AgdaFunction{⊆}}\AgdaSpace{}%
\AgdaPostulate{filesRead}\AgdaSpace{}%
\AgdaSymbol{(}\AgdaFunction{script-rec}\AgdaSpace{}%
\AgdaBound{x₃}\AgdaSpace{}%
\AgdaBound{s}\AgdaSpace{}%
\AgdaSymbol{\AgdaUnderscore{}))}\AgdaSpace{}%
\AgdaSymbol{(}\AgdaFunction{sym}\AgdaSpace{}%
\AgdaSymbol{(}\AgdaFunction{unfold-reverse}\AgdaSpace{}%
\AgdaBound{x}\AgdaSpace{}%
\AgdaBound{ys}\AgdaSymbol{))}\AgdaSpace{}%
\AgdaSymbol{(}\AgdaFunction{sym}\AgdaSpace{}%
\AgdaBound{≡₁}\AgdaSymbol{)}\AgdaSpace{}%
\AgdaBound{a}\<%
\\
\\[\AgdaEmptyExtraSkip]%
\\[\AgdaEmptyExtraSkip]%
\>[0]\AgdaFunction{extra-lemma3-reads}\AgdaSpace{}%
\AgdaSymbol{:}\AgdaSpace{}%
\AgdaSymbol{∀}\AgdaSpace{}%
\AgdaSymbol{\{}\AgdaBound{s}\AgdaSymbol{\}}\AgdaSpace{}%
\AgdaSymbol{\{}\AgdaBound{ls}\AgdaSymbol{\}}\AgdaSpace{}%
\AgdaBound{bs}\AgdaSpace{}%
\AgdaBound{xs}\AgdaSpace{}%
\AgdaBound{x}\AgdaSpace{}%
\AgdaSymbol{→}\AgdaSpace{}%
\AgdaBound{x}\AgdaSpace{}%
\AgdaOperator{\AgdaPostulate{∉}}\AgdaSpace{}%
\AgdaBound{bs}\AgdaSpace{}%
\AgdaSymbol{→}\AgdaSpace{}%
\AgdaBound{x}\AgdaSpace{}%
\AgdaOperator{\AgdaPostulate{∉}}\AgdaSpace{}%
\AgdaBound{xs}\AgdaSpace{}%
\AgdaSymbol{→}\AgdaSpace{}%
\AgdaBound{bs}\AgdaSpace{}%
\AgdaOperator{\AgdaFunction{⊆}}\AgdaSpace{}%
\AgdaBound{xs}\AgdaSpace{}%
\AgdaSymbol{→}\AgdaSpace{}%
\AgdaPostulate{HazardFree}\AgdaSpace{}%
\AgdaBound{s}\AgdaSpace{}%
\AgdaBound{bs}\AgdaSpace{}%
\AgdaSymbol{(}\AgdaBound{xs}\AgdaSpace{}%
\AgdaOperator{\AgdaFunction{++}}\AgdaSpace{}%
\AgdaBound{x}\AgdaSpace{}%
\AgdaOperator{\AgdaInductiveConstructor{∷}}\AgdaSpace{}%
\AgdaInductiveConstructor{[]}\AgdaSymbol{)}\AgdaSpace{}%
\AgdaBound{ls}\AgdaSpace{}%
\AgdaSymbol{→}\AgdaSpace{}%
\AgdaFunction{Disjoint}\AgdaSpace{}%
\AgdaSymbol{(}\AgdaPostulate{filesRead}\AgdaSpace{}%
\AgdaSymbol{(}\AgdaFunction{script-rec}\AgdaSpace{}%
\AgdaBound{bs}\AgdaSpace{}%
\AgdaBound{s}\AgdaSpace{}%
\AgdaInductiveConstructor{[]}\AgdaSymbol{))}\AgdaSpace{}%
\AgdaSymbol{(}\AgdaPostulate{cmdWrote}\AgdaSpace{}%
\AgdaBound{ls}\AgdaSpace{}%
\AgdaBound{x}\AgdaSymbol{)}\<%
\\
\>[0]\AgdaFunction{extra-lemma3-reads}\AgdaSpace{}%
\AgdaSymbol{\{}\AgdaBound{s}\AgdaSymbol{\}}\AgdaSpace{}%
\AgdaSymbol{\{}\AgdaBound{ls}\AgdaSymbol{\}}\AgdaSpace{}%
\AgdaSymbol{(}\AgdaBound{x₁}\AgdaSpace{}%
\AgdaOperator{\AgdaInductiveConstructor{∷}}\AgdaSpace{}%
\AgdaBound{bs}\AgdaSymbol{)}\AgdaSpace{}%
\AgdaBound{xs}\AgdaSpace{}%
\AgdaBound{x}\AgdaSpace{}%
\AgdaBound{x∉bs}\AgdaSpace{}%
\AgdaBound{x∉xs}\AgdaSpace{}%
\AgdaBound{bs⊆xs}\AgdaSpace{}%
\AgdaSymbol{(}\AgdaBound{¬hz}\AgdaSpace{}%
\AgdaOperator{\AgdaInductiveConstructor{∷}}\AgdaSpace{}%
\AgdaBound{hf}\AgdaSymbol{)}\AgdaSpace{}%
\AgdaSymbol{(}\AgdaBound{fst}\AgdaSpace{}%
\AgdaOperator{\AgdaInductiveConstructor{,}}\AgdaSpace{}%
\AgdaBound{v∈ls}\AgdaSymbol{)}\AgdaSpace{}%
\AgdaKeyword{with}\AgdaSpace{}%
\AgdaFunction{∈-++⁻}\AgdaSpace{}%
\AgdaSymbol{(}\AgdaPostulate{filesRead}\AgdaSpace{}%
\AgdaSymbol{(}\AgdaFunction{script-rec}\AgdaSpace{}%
\AgdaBound{bs}\AgdaSpace{}%
\AgdaSymbol{(}\AgdaPostulate{St.run}\AgdaSpace{}%
\AgdaBound{x₁}\AgdaSpace{}%
\AgdaBound{s}\AgdaSymbol{)}\AgdaSpace{}%
\AgdaInductiveConstructor{[]}\AgdaSymbol{))}\AgdaSpace{}%
\AgdaSymbol{(}\AgdaFunction{subst}\AgdaSpace{}%
\AgdaSymbol{(λ}\AgdaSpace{}%
\AgdaBound{x₂}\AgdaSpace{}%
\AgdaSymbol{→}\AgdaSpace{}%
\AgdaSymbol{\AgdaUnderscore{}}\AgdaSpace{}%
\AgdaOperator{\AgdaPostulate{∈}}\AgdaSpace{}%
\AgdaBound{x₂}\AgdaSymbol{)}\AgdaSpace{}%
\AgdaFunction{≡₁}\AgdaSpace{}%
\AgdaBound{fst}\AgdaSymbol{)}\<%
\\
\>[0][@{}l@{\AgdaIndent{0}}]%
\>[2]\AgdaKeyword{where}%
\>[3218I]\AgdaFunction{≡₁}\AgdaSpace{}%
\AgdaSymbol{:}\AgdaSpace{}%
\AgdaPostulate{filesRead}\AgdaSpace{}%
\AgdaSymbol{(}\AgdaFunction{script-rec}\AgdaSpace{}%
\AgdaBound{bs}\AgdaSpace{}%
\AgdaSymbol{(}\AgdaPostulate{St.run}\AgdaSpace{}%
\AgdaBound{x₁}\AgdaSpace{}%
\AgdaBound{s}\AgdaSymbol{)}\AgdaSpace{}%
\AgdaSymbol{(}\AgdaPostulate{rec}\AgdaSpace{}%
\AgdaBound{s}\AgdaSpace{}%
\AgdaBound{x₁}\AgdaSpace{}%
\AgdaInductiveConstructor{[]}\AgdaSymbol{))}\AgdaSpace{}%
\AgdaOperator{\AgdaDatatype{≡}}\AgdaSpace{}%
\AgdaPostulate{filesRead}\AgdaSpace{}%
\AgdaSymbol{(}\AgdaFunction{script-rec}\AgdaSpace{}%
\AgdaBound{bs}\AgdaSpace{}%
\AgdaSymbol{(}\AgdaPostulate{St.run}\AgdaSpace{}%
\AgdaBound{x₁}\AgdaSpace{}%
\AgdaBound{s}\AgdaSymbol{)}\AgdaSpace{}%
\AgdaInductiveConstructor{[]}\AgdaSymbol{)}\AgdaSpace{}%
\AgdaOperator{\AgdaFunction{++}}\AgdaSpace{}%
\AgdaSymbol{(}\AgdaPostulate{filesRead}\AgdaSpace{}%
\AgdaSymbol{(}\AgdaPostulate{rec}\AgdaSpace{}%
\AgdaBound{s}\AgdaSpace{}%
\AgdaBound{x₁}\AgdaSpace{}%
\AgdaInductiveConstructor{[]}\AgdaSymbol{))}\<%
\\
\>[.][@{}l@{}]\<[3218I]%
\>[8]\AgdaFunction{≡₁}\AgdaSpace{}%
\AgdaSymbol{=}\AgdaSpace{}%
\AgdaFunction{trans}\AgdaSpace{}%
\AgdaSymbol{(}\AgdaFunction{cong}\AgdaSpace{}%
\AgdaPostulate{filesRead}\AgdaSpace{}%
\AgdaSymbol{(}\AgdaFunction{script-rec-++}\AgdaSpace{}%
\AgdaBound{bs}\AgdaSymbol{))}\AgdaSpace{}%
\AgdaSymbol{(}\AgdaFunction{concatMap-++-commute}\AgdaSpace{}%
\AgdaSymbol{(}\AgdaField{proj₁}\AgdaSpace{}%
\AgdaOperator{\AgdaFunction{∘}}\AgdaSpace{}%
\AgdaField{proj₂}\AgdaSymbol{)}\AgdaSpace{}%
\AgdaSymbol{(}\AgdaFunction{script-rec}\AgdaSpace{}%
\AgdaBound{bs}\AgdaSpace{}%
\AgdaSymbol{(}\AgdaPostulate{St.run}\AgdaSpace{}%
\AgdaBound{x₁}\AgdaSpace{}%
\AgdaBound{s}\AgdaSymbol{)}\AgdaSpace{}%
\AgdaInductiveConstructor{[]}\AgdaSymbol{)}\AgdaSpace{}%
\AgdaSymbol{(}\AgdaPostulate{rec}\AgdaSpace{}%
\AgdaBound{s}\AgdaSpace{}%
\AgdaBound{x₁}\AgdaSpace{}%
\AgdaInductiveConstructor{[]}\AgdaSymbol{))}\<%
\\
\>[0]\AgdaSymbol{...}\AgdaSpace{}%
\AgdaSymbol{|}\AgdaSpace{}%
\AgdaInductiveConstructor{inj₁}\AgdaSpace{}%
\AgdaBound{∈₁}\AgdaSpace{}%
\AgdaSymbol{=}\AgdaSpace{}%
\AgdaSymbol{(}\AgdaFunction{extra-lemma3-reads}\AgdaSpace{}%
\AgdaBound{bs}\AgdaSpace{}%
\AgdaBound{xs}\AgdaSpace{}%
\AgdaBound{x}\AgdaSpace{}%
\AgdaSymbol{(λ}\AgdaSpace{}%
\AgdaBound{x₂}\AgdaSpace{}%
\AgdaSymbol{→}\AgdaSpace{}%
\AgdaBound{x∉bs}\AgdaSpace{}%
\AgdaSymbol{(}\AgdaInductiveConstructor{there}\AgdaSpace{}%
\AgdaBound{x₂}\AgdaSymbol{))}\AgdaSpace{}%
\AgdaBound{x∉xs}\AgdaSpace{}%
\AgdaSymbol{(λ}\AgdaSpace{}%
\AgdaBound{x₂}\AgdaSpace{}%
\AgdaSymbol{→}\AgdaSpace{}%
\AgdaBound{bs⊆xs}\AgdaSpace{}%
\AgdaSymbol{(}\AgdaInductiveConstructor{there}\AgdaSpace{}%
\AgdaBound{x₂}\AgdaSymbol{))}\AgdaSpace{}%
\AgdaBound{hf}\AgdaSymbol{)}\AgdaSpace{}%
\AgdaSymbol{(}\AgdaBound{∈₁}\AgdaSpace{}%
\AgdaOperator{\AgdaInductiveConstructor{,}}\AgdaSpace{}%
\AgdaPostulate{∈-cmdWrote∷}\AgdaSpace{}%
\AgdaSymbol{(}\AgdaBound{x₁}\AgdaSpace{}%
\AgdaOperator{\AgdaInductiveConstructor{,}}\AgdaSpace{}%
\AgdaSymbol{\AgdaUnderscore{}}\AgdaSpace{}%
\AgdaOperator{\AgdaInductiveConstructor{,}}\AgdaSpace{}%
\AgdaSymbol{\AgdaUnderscore{})}\AgdaSpace{}%
\AgdaBound{x}\AgdaSpace{}%
\AgdaBound{ls}\AgdaSpace{}%
\AgdaBound{v∈ls}\AgdaSpace{}%
\AgdaSymbol{λ}\AgdaSpace{}%
\AgdaBound{x₁≡x}\AgdaSpace{}%
\AgdaSymbol{→}\AgdaSpace{}%
\AgdaBound{x∉bs}\AgdaSpace{}%
\AgdaSymbol{(}\AgdaInductiveConstructor{here}\AgdaSpace{}%
\AgdaSymbol{(}\AgdaFunction{sym}\AgdaSpace{}%
\AgdaBound{x₁≡x}\AgdaSymbol{)))}\<%
\\
\>[0]\AgdaSymbol{...}\AgdaSpace{}%
\AgdaSymbol{|}\AgdaSpace{}%
\AgdaInductiveConstructor{inj₂}\AgdaSpace{}%
\AgdaBound{∈₁}\AgdaSpace{}%
\AgdaSymbol{=}\AgdaSpace{}%
\AgdaBound{¬hz}\AgdaSpace{}%
\AgdaSymbol{(}\AgdaPostulate{Speculative}\AgdaSpace{}%
\AgdaBound{x}\AgdaSpace{}%
\AgdaBound{x₁}\AgdaSpace{}%
\AgdaFunction{bf}\AgdaSpace{}%
\AgdaSymbol{(}\AgdaFunction{∈-++⁺ˡ}\AgdaSpace{}%
\AgdaSymbol{(}\AgdaBound{bs⊆xs}\AgdaSpace{}%
\AgdaSymbol{(}\AgdaInductiveConstructor{here}\AgdaSpace{}%
\AgdaInductiveConstructor{refl}\AgdaSymbol{)))}\AgdaSpace{}%
\AgdaSymbol{(}\AgdaFunction{¬bf}\AgdaSpace{}%
\AgdaBound{xs}\AgdaSpace{}%
\AgdaSymbol{(}\AgdaBound{bs⊆xs}\AgdaSpace{}%
\AgdaSymbol{(}\AgdaInductiveConstructor{here}\AgdaSpace{}%
\AgdaInductiveConstructor{refl}\AgdaSymbol{))}\AgdaSpace{}%
\AgdaBound{x∉xs}\AgdaSymbol{)}\AgdaSpace{}%
\AgdaSymbol{(}\AgdaPostulate{∈-cmdRead∷l}\AgdaSpace{}%
\AgdaSymbol{(}\AgdaBound{x₁}\AgdaSpace{}%
\AgdaOperator{\AgdaInductiveConstructor{,}}\AgdaSpace{}%
\AgdaSymbol{\AgdaUnderscore{}}\AgdaSpace{}%
\AgdaOperator{\AgdaInductiveConstructor{,}}\AgdaSpace{}%
\AgdaSymbol{\AgdaUnderscore{})}\AgdaSpace{}%
\AgdaBound{ls}\AgdaSpace{}%
\AgdaSymbol{(}\AgdaFunction{∈₂}\AgdaSpace{}%
\AgdaBound{∈₁}\AgdaSymbol{))}\AgdaSpace{}%
\AgdaSymbol{(}\AgdaPostulate{∈-cmdWrote∷}\AgdaSpace{}%
\AgdaSymbol{(}\AgdaBound{x₁}\AgdaSpace{}%
\AgdaOperator{\AgdaInductiveConstructor{,}}\AgdaSpace{}%
\AgdaSymbol{\AgdaUnderscore{}}\AgdaSpace{}%
\AgdaOperator{\AgdaInductiveConstructor{,}}\AgdaSpace{}%
\AgdaSymbol{\AgdaUnderscore{})}\AgdaSpace{}%
\AgdaBound{x}\AgdaSpace{}%
\AgdaBound{ls}\AgdaSpace{}%
\AgdaBound{v∈ls}\AgdaSpace{}%
\AgdaSymbol{λ}\AgdaSpace{}%
\AgdaBound{x₁≡x}\AgdaSpace{}%
\AgdaSymbol{→}\AgdaSpace{}%
\AgdaBound{x∉bs}\AgdaSpace{}%
\AgdaSymbol{(}\AgdaInductiveConstructor{here}\AgdaSpace{}%
\AgdaSymbol{(}\AgdaFunction{sym}\AgdaSpace{}%
\AgdaBound{x₁≡x}\AgdaSymbol{))))}\<%
\\
\>[0][@{}l@{\AgdaIndent{0}}]%
\>[2]\AgdaKeyword{where}%
\>[3348I]\AgdaFunction{∈₂}\AgdaSpace{}%
\AgdaSymbol{:}\AgdaSpace{}%
\AgdaSymbol{∀}\AgdaSpace{}%
\AgdaSymbol{\{}\AgdaBound{v}\AgdaSymbol{\}}\AgdaSpace{}%
\AgdaSymbol{→}\AgdaSpace{}%
\AgdaBound{v}\AgdaSpace{}%
\AgdaOperator{\AgdaPostulate{∈}}\AgdaSpace{}%
\AgdaPostulate{cmdReadNames}\AgdaSpace{}%
\AgdaBound{x₁}\AgdaSpace{}%
\AgdaBound{s}\AgdaSpace{}%
\AgdaOperator{\AgdaFunction{++}}\AgdaSpace{}%
\AgdaInductiveConstructor{[]}\AgdaSpace{}%
\AgdaSymbol{→}\AgdaSpace{}%
\AgdaBound{v}\AgdaSpace{}%
\AgdaOperator{\AgdaPostulate{∈}}\AgdaSpace{}%
\AgdaPostulate{cmdReadNames}\AgdaSpace{}%
\AgdaBound{x₁}\AgdaSpace{}%
\AgdaBound{s}\<%
\\
\>[.][@{}l@{}]\<[3348I]%
\>[8]\AgdaFunction{∈₂}\AgdaSpace{}%
\AgdaBound{v∈}\AgdaSpace{}%
\AgdaKeyword{with}\AgdaSpace{}%
\AgdaFunction{∈-++⁻}\AgdaSpace{}%
\AgdaSymbol{(}\AgdaPostulate{cmdReadNames}\AgdaSpace{}%
\AgdaBound{x₁}\AgdaSpace{}%
\AgdaBound{s}\AgdaSymbol{)}\AgdaSpace{}%
\AgdaBound{v∈}\<%
\\
\>[8]\AgdaSymbol{...}\AgdaSpace{}%
\AgdaSymbol{|}\AgdaSpace{}%
\AgdaInductiveConstructor{inj₁}\AgdaSpace{}%
\AgdaBound{v∈₁}\AgdaSpace{}%
\AgdaSymbol{=}\AgdaSpace{}%
\AgdaBound{v∈₁}\<%
\\
\>[8]\AgdaSymbol{...}\AgdaSpace{}%
\AgdaSymbol{|}\AgdaSpace{}%
\AgdaInductiveConstructor{inj₂}\AgdaSpace{}%
\AgdaSymbol{()}\<%
\\
\>[8]\AgdaFunction{bf}\AgdaSpace{}%
\AgdaSymbol{:}\AgdaSpace{}%
\AgdaBound{x₁}\AgdaSpace{}%
\AgdaOperator{\AgdaFunction{before}}\AgdaSpace{}%
\AgdaBound{x}\AgdaSpace{}%
\AgdaOperator{\AgdaFunction{∈}}\AgdaSpace{}%
\AgdaSymbol{(}\AgdaBound{x₁}\AgdaSpace{}%
\AgdaOperator{\AgdaInductiveConstructor{∷}}\AgdaSpace{}%
\AgdaPostulate{cmdsRun}\AgdaSpace{}%
\AgdaBound{ls}\AgdaSymbol{)}\<%
\\
\>[8]\AgdaFunction{bf}\AgdaSpace{}%
\AgdaSymbol{=}\AgdaSpace{}%
\AgdaInductiveConstructor{[]}\AgdaSpace{}%
\AgdaOperator{\AgdaInductiveConstructor{,}}\AgdaSpace{}%
\AgdaPostulate{cmdsRun}\AgdaSpace{}%
\AgdaBound{ls}\AgdaSpace{}%
\AgdaOperator{\AgdaInductiveConstructor{,}}\AgdaSpace{}%
\AgdaInductiveConstructor{refl}\AgdaSpace{}%
\AgdaOperator{\AgdaInductiveConstructor{,}}\AgdaSpace{}%
\AgdaSymbol{\{!!\}}\<%
\\
\>[8]\AgdaFunction{¬bf}\AgdaSpace{}%
\AgdaSymbol{:}\AgdaSpace{}%
\AgdaSymbol{∀}\AgdaSpace{}%
\AgdaBound{xs}\AgdaSpace{}%
\AgdaSymbol{→}\AgdaSpace{}%
\AgdaBound{x₁}\AgdaSpace{}%
\AgdaOperator{\AgdaPostulate{∈}}\AgdaSpace{}%
\AgdaBound{xs}\AgdaSpace{}%
\AgdaSymbol{→}\AgdaSpace{}%
\AgdaBound{x}\AgdaSpace{}%
\AgdaOperator{\AgdaPostulate{∉}}\AgdaSpace{}%
\AgdaBound{xs}\AgdaSpace{}%
\AgdaSymbol{→}\AgdaSpace{}%
\AgdaOperator{\AgdaFunction{¬}}\AgdaSpace{}%
\AgdaSymbol{(}\AgdaBound{x}\AgdaSpace{}%
\AgdaOperator{\AgdaFunction{before}}\AgdaSpace{}%
\AgdaBound{x₁}\AgdaSpace{}%
\AgdaOperator{\AgdaFunction{∈}}\AgdaSpace{}%
\AgdaSymbol{(}\AgdaBound{xs}\AgdaSpace{}%
\AgdaOperator{\AgdaFunction{++}}\AgdaSpace{}%
\AgdaBound{x}\AgdaSpace{}%
\AgdaOperator{\AgdaInductiveConstructor{∷}}\AgdaSpace{}%
\AgdaInductiveConstructor{[]}\AgdaSymbol{))}\<%
\\
\>[8]\AgdaFunction{¬bf}\AgdaSpace{}%
\AgdaSymbol{(}\AgdaBound{x₂}\AgdaSpace{}%
\AgdaOperator{\AgdaInductiveConstructor{∷}}\AgdaSpace{}%
\AgdaBound{xs}\AgdaSymbol{)}\AgdaSpace{}%
\AgdaSymbol{\AgdaUnderscore{}}%
\>[25]\AgdaBound{x∉x₂∷xs}\AgdaSpace{}%
\AgdaSymbol{(}\AgdaInductiveConstructor{[]}\AgdaSpace{}%
\AgdaOperator{\AgdaInductiveConstructor{,}}\AgdaSpace{}%
\AgdaBound{zs}\AgdaSpace{}%
\AgdaOperator{\AgdaInductiveConstructor{,}}\AgdaSpace{}%
\AgdaBound{≡₁}\AgdaSpace{}%
\AgdaOperator{\AgdaInductiveConstructor{,}}\AgdaSpace{}%
\AgdaBound{∈₁}\AgdaSymbol{)}\AgdaSpace{}%
\AgdaSymbol{=}\AgdaSpace{}%
\AgdaFunction{contradiction}\AgdaSpace{}%
\AgdaSymbol{(}\AgdaInductiveConstructor{here}\AgdaSpace{}%
\AgdaSymbol{(}\AgdaFunction{sym}\AgdaSpace{}%
\AgdaSymbol{(}\AgdaFunction{∷-injectiveˡ}\AgdaSpace{}%
\AgdaBound{≡₁}\AgdaSymbol{)))}\AgdaSpace{}%
\AgdaBound{x∉x₂∷xs}\<%
\\
\>[8]\AgdaFunction{¬bf}\AgdaSpace{}%
\AgdaSymbol{(}\AgdaBound{x₃}\AgdaSpace{}%
\AgdaOperator{\AgdaInductiveConstructor{∷}}\AgdaSpace{}%
\AgdaBound{xs}\AgdaSymbol{)}\AgdaSpace{}%
\AgdaBound{x₁∈xs}\AgdaSpace{}%
\AgdaBound{x∉xs}\AgdaSpace{}%
\AgdaSymbol{(}\AgdaBound{x₂}\AgdaSpace{}%
\AgdaOperator{\AgdaInductiveConstructor{∷}}\AgdaSpace{}%
\AgdaBound{ys}\AgdaSpace{}%
\AgdaOperator{\AgdaInductiveConstructor{,}}\AgdaSpace{}%
\AgdaBound{zs}\AgdaSpace{}%
\AgdaOperator{\AgdaInductiveConstructor{,}}\AgdaSpace{}%
\AgdaBound{≡₁}\AgdaSpace{}%
\AgdaOperator{\AgdaInductiveConstructor{,}}\AgdaSpace{}%
\AgdaBound{∈₁}\AgdaSymbol{)}\AgdaSpace{}%
\AgdaSymbol{=}\AgdaSpace{}%
\AgdaFunction{¬bf}\AgdaSpace{}%
\AgdaBound{xs}\AgdaSpace{}%
\AgdaSymbol{(}\AgdaFunction{tail}\AgdaSpace{}%
\AgdaSymbol{\{!!\}}\AgdaSpace{}%
\AgdaBound{x₁∈xs}\AgdaSymbol{)}\AgdaSpace{}%
\AgdaSymbol{(λ}\AgdaSpace{}%
\AgdaBound{x₄}\AgdaSpace{}%
\AgdaSymbol{→}\AgdaSpace{}%
\AgdaBound{x∉xs}\AgdaSpace{}%
\AgdaSymbol{(}\AgdaInductiveConstructor{there}\AgdaSpace{}%
\AgdaBound{x₄}\AgdaSymbol{))}\AgdaSpace{}%
\AgdaSymbol{\{!!\}}\<%
\\
\\[\AgdaEmptyExtraSkip]%
\\[\AgdaEmptyExtraSkip]%
\>[0]\AgdaFunction{extra-lemma3-writes}\AgdaSpace{}%
\AgdaSymbol{:}\AgdaSpace{}%
\AgdaSymbol{∀}\AgdaSpace{}%
\AgdaSymbol{\{}\AgdaBound{s}\AgdaSymbol{\}}\AgdaSpace{}%
\AgdaSymbol{\{}\AgdaBound{ls}\AgdaSymbol{\}}\AgdaSpace{}%
\AgdaBound{bs}\AgdaSpace{}%
\AgdaBound{xs}\AgdaSpace{}%
\AgdaBound{x}\AgdaSpace{}%
\AgdaSymbol{→}\AgdaSpace{}%
\AgdaBound{x}\AgdaSpace{}%
\AgdaOperator{\AgdaPostulate{∉}}\AgdaSpace{}%
\AgdaBound{bs}\AgdaSpace{}%
\AgdaSymbol{→}\AgdaSpace{}%
\AgdaBound{x}\AgdaSpace{}%
\AgdaOperator{\AgdaPostulate{∉}}\AgdaSpace{}%
\AgdaBound{xs}\AgdaSpace{}%
\AgdaSymbol{→}\AgdaSpace{}%
\AgdaBound{bs}\AgdaSpace{}%
\AgdaOperator{\AgdaFunction{⊆}}\AgdaSpace{}%
\AgdaBound{xs}\AgdaSpace{}%
\AgdaSymbol{→}\AgdaSpace{}%
\AgdaPostulate{HazardFree}\AgdaSpace{}%
\AgdaBound{s}\AgdaSpace{}%
\AgdaBound{bs}\AgdaSpace{}%
\AgdaSymbol{(}\AgdaBound{xs}\AgdaSpace{}%
\AgdaOperator{\AgdaFunction{++}}\AgdaSpace{}%
\AgdaBound{x}\AgdaSpace{}%
\AgdaOperator{\AgdaInductiveConstructor{∷}}\AgdaSpace{}%
\AgdaInductiveConstructor{[]}\AgdaSymbol{)}\AgdaSpace{}%
\AgdaBound{ls}\AgdaSpace{}%
\AgdaSymbol{→}\AgdaSpace{}%
\AgdaFunction{Disjoint}\AgdaSpace{}%
\AgdaSymbol{(}\AgdaPostulate{filesWrote}\AgdaSpace{}%
\AgdaSymbol{(}\AgdaFunction{script-rec}\AgdaSpace{}%
\AgdaBound{bs}\AgdaSpace{}%
\AgdaBound{s}\AgdaSpace{}%
\AgdaInductiveConstructor{[]}\AgdaSymbol{))}\AgdaSpace{}%
\AgdaSymbol{(}\AgdaPostulate{cmdWrote}\AgdaSpace{}%
\AgdaBound{ls}\AgdaSpace{}%
\AgdaBound{x}\AgdaSymbol{)}\<%
\\
\>[0]\AgdaFunction{extra-lemma3-writes}\AgdaSpace{}%
\AgdaSymbol{\{}\AgdaBound{s}\AgdaSymbol{\}}\AgdaSpace{}%
\AgdaSymbol{\{}\AgdaBound{ls}\AgdaSymbol{\}}\AgdaSpace{}%
\AgdaSymbol{(}\AgdaBound{x₁}\AgdaSpace{}%
\AgdaOperator{\AgdaInductiveConstructor{∷}}\AgdaSpace{}%
\AgdaBound{bs}\AgdaSymbol{)}\AgdaSpace{}%
\AgdaBound{xs}\AgdaSpace{}%
\AgdaBound{x}\AgdaSpace{}%
\AgdaBound{x∉bs}\AgdaSpace{}%
\AgdaBound{x∉xs}\AgdaSpace{}%
\AgdaBound{bs⊆xs}\AgdaSpace{}%
\AgdaSymbol{(}\AgdaBound{¬hz}\AgdaSpace{}%
\AgdaOperator{\AgdaInductiveConstructor{∷}}\AgdaSpace{}%
\AgdaBound{hf}\AgdaSymbol{)}\AgdaSpace{}%
\AgdaSymbol{(}\AgdaBound{fst}\AgdaSpace{}%
\AgdaOperator{\AgdaInductiveConstructor{,}}\AgdaSpace{}%
\AgdaBound{v∈ls}\AgdaSymbol{)}\AgdaSpace{}%
\AgdaKeyword{with}\AgdaSpace{}%
\AgdaFunction{∈-++⁻}\AgdaSpace{}%
\AgdaSymbol{(}\AgdaPostulate{filesWrote}\AgdaSpace{}%
\AgdaSymbol{(}\AgdaFunction{script-rec}\AgdaSpace{}%
\AgdaBound{bs}\AgdaSpace{}%
\AgdaSymbol{(}\AgdaPostulate{St.run}\AgdaSpace{}%
\AgdaBound{x₁}\AgdaSpace{}%
\AgdaBound{s}\AgdaSymbol{)}\AgdaSpace{}%
\AgdaInductiveConstructor{[]}\AgdaSymbol{))}\AgdaSpace{}%
\AgdaSymbol{(}\AgdaFunction{subst}\AgdaSpace{}%
\AgdaSymbol{(λ}\AgdaSpace{}%
\AgdaBound{x₂}\AgdaSpace{}%
\AgdaSymbol{→}\AgdaSpace{}%
\AgdaSymbol{\AgdaUnderscore{}}\AgdaSpace{}%
\AgdaOperator{\AgdaPostulate{∈}}\AgdaSpace{}%
\AgdaBound{x₂}\AgdaSymbol{)}\AgdaSpace{}%
\AgdaFunction{≡₁}\AgdaSpace{}%
\AgdaBound{fst}\AgdaSymbol{)}\<%
\\
\>[0][@{}l@{\AgdaIndent{0}}]%
\>[2]\AgdaKeyword{where}%
\>[3539I]\AgdaFunction{≡₁}\AgdaSpace{}%
\AgdaSymbol{:}\AgdaSpace{}%
\AgdaPostulate{filesWrote}\AgdaSpace{}%
\AgdaSymbol{(}\AgdaFunction{script-rec}\AgdaSpace{}%
\AgdaBound{bs}\AgdaSpace{}%
\AgdaSymbol{(}\AgdaPostulate{St.run}\AgdaSpace{}%
\AgdaBound{x₁}\AgdaSpace{}%
\AgdaBound{s}\AgdaSymbol{)}\AgdaSpace{}%
\AgdaSymbol{(}\AgdaPostulate{rec}\AgdaSpace{}%
\AgdaBound{s}\AgdaSpace{}%
\AgdaBound{x₁}\AgdaSpace{}%
\AgdaInductiveConstructor{[]}\AgdaSymbol{))}\AgdaSpace{}%
\AgdaOperator{\AgdaDatatype{≡}}\AgdaSpace{}%
\AgdaPostulate{filesWrote}\AgdaSpace{}%
\AgdaSymbol{(}\AgdaFunction{script-rec}\AgdaSpace{}%
\AgdaBound{bs}\AgdaSpace{}%
\AgdaSymbol{(}\AgdaPostulate{St.run}\AgdaSpace{}%
\AgdaBound{x₁}\AgdaSpace{}%
\AgdaBound{s}\AgdaSymbol{)}\AgdaSpace{}%
\AgdaInductiveConstructor{[]}\AgdaSymbol{)}\AgdaSpace{}%
\AgdaOperator{\AgdaFunction{++}}\AgdaSpace{}%
\AgdaSymbol{(}\AgdaPostulate{filesWrote}\AgdaSpace{}%
\AgdaSymbol{(}\AgdaPostulate{rec}\AgdaSpace{}%
\AgdaBound{s}\AgdaSpace{}%
\AgdaBound{x₁}\AgdaSpace{}%
\AgdaInductiveConstructor{[]}\AgdaSymbol{))}\<%
\\
\>[.][@{}l@{}]\<[3539I]%
\>[8]\AgdaFunction{≡₁}\AgdaSpace{}%
\AgdaSymbol{=}\AgdaSpace{}%
\AgdaFunction{trans}\AgdaSpace{}%
\AgdaSymbol{(}\AgdaFunction{cong}\AgdaSpace{}%
\AgdaPostulate{filesWrote}\AgdaSpace{}%
\AgdaSymbol{(}\AgdaFunction{script-rec-++}\AgdaSpace{}%
\AgdaBound{bs}\AgdaSymbol{))}\AgdaSpace{}%
\AgdaSymbol{(}\AgdaFunction{concatMap-++-commute}\AgdaSpace{}%
\AgdaSymbol{(}\AgdaField{proj₂}\AgdaSpace{}%
\AgdaOperator{\AgdaFunction{∘}}\AgdaSpace{}%
\AgdaField{proj₂}\AgdaSymbol{)}\AgdaSpace{}%
\AgdaSymbol{(}\AgdaFunction{script-rec}\AgdaSpace{}%
\AgdaBound{bs}\AgdaSpace{}%
\AgdaSymbol{(}\AgdaPostulate{St.run}\AgdaSpace{}%
\AgdaBound{x₁}\AgdaSpace{}%
\AgdaBound{s}\AgdaSymbol{)}\AgdaSpace{}%
\AgdaInductiveConstructor{[]}\AgdaSymbol{)}\AgdaSpace{}%
\AgdaSymbol{(}\AgdaPostulate{rec}\AgdaSpace{}%
\AgdaBound{s}\AgdaSpace{}%
\AgdaBound{x₁}\AgdaSpace{}%
\AgdaInductiveConstructor{[]}\AgdaSymbol{))}\<%
\\
\>[0]\AgdaSymbol{...}\AgdaSpace{}%
\AgdaSymbol{|}\AgdaSpace{}%
\AgdaInductiveConstructor{inj₁}\AgdaSpace{}%
\AgdaBound{∈₁}\AgdaSpace{}%
\AgdaSymbol{=}%
\>[3589I]\AgdaSymbol{(}\AgdaFunction{extra-lemma3-writes}\AgdaSpace{}%
\AgdaBound{bs}\AgdaSpace{}%
\AgdaBound{xs}\AgdaSpace{}%
\AgdaBound{x}\AgdaSpace{}%
\AgdaSymbol{(λ}\AgdaSpace{}%
\AgdaBound{x₂}\AgdaSpace{}%
\AgdaSymbol{→}\AgdaSpace{}%
\AgdaBound{x∉bs}\AgdaSpace{}%
\AgdaSymbol{(}\AgdaInductiveConstructor{there}\AgdaSpace{}%
\AgdaBound{x₂}\AgdaSymbol{))}\AgdaSpace{}%
\AgdaBound{x∉xs}\AgdaSpace{}%
\AgdaSymbol{(λ}\AgdaSpace{}%
\AgdaBound{x₂}\AgdaSpace{}%
\AgdaSymbol{→}\AgdaSpace{}%
\AgdaBound{bs⊆xs}\AgdaSpace{}%
\AgdaSymbol{(}\AgdaInductiveConstructor{there}\AgdaSpace{}%
\AgdaBound{x₂}\AgdaSymbol{))}\AgdaSpace{}%
\AgdaBound{hf}\AgdaSymbol{)}\<%
\\
\>[.][@{}l@{}]\<[3589I]%
\>[16]\AgdaSymbol{(}\AgdaBound{∈₁}\AgdaSpace{}%
\AgdaOperator{\AgdaInductiveConstructor{,}}\AgdaSpace{}%
\AgdaPostulate{∈-cmdWrote∷}\AgdaSpace{}%
\AgdaSymbol{(}\AgdaBound{x₁}\AgdaSpace{}%
\AgdaOperator{\AgdaInductiveConstructor{,}}\AgdaSpace{}%
\AgdaSymbol{\AgdaUnderscore{}}\AgdaSpace{}%
\AgdaOperator{\AgdaInductiveConstructor{,}}\AgdaSpace{}%
\AgdaSymbol{\AgdaUnderscore{})}\AgdaSpace{}%
\AgdaBound{x}\AgdaSpace{}%
\AgdaBound{ls}\AgdaSpace{}%
\AgdaBound{v∈ls}\AgdaSpace{}%
\AgdaSymbol{λ}\AgdaSpace{}%
\AgdaBound{x₁≡x}\AgdaSpace{}%
\AgdaSymbol{→}\AgdaSpace{}%
\AgdaBound{x∉bs}\AgdaSpace{}%
\AgdaSymbol{(}\AgdaInductiveConstructor{here}\AgdaSpace{}%
\AgdaSymbol{(}\AgdaFunction{sym}\AgdaSpace{}%
\AgdaBound{x₁≡x}\AgdaSymbol{)))}\<%
\\
\>[0]\AgdaSymbol{...}\AgdaSpace{}%
\AgdaSymbol{|}\AgdaSpace{}%
\AgdaInductiveConstructor{inj₂}\AgdaSpace{}%
\AgdaBound{∈₁}\AgdaSpace{}%
\AgdaSymbol{=}\AgdaSpace{}%
\AgdaBound{¬hz}\AgdaSpace{}%
\AgdaSymbol{(}\AgdaPostulate{WriteWrite}\AgdaSpace{}%
\AgdaSymbol{(}\AgdaFunction{∈₂}\AgdaSpace{}%
\AgdaBound{∈₁}\AgdaSymbol{)}\AgdaSpace{}%
\AgdaSymbol{(}\AgdaPostulate{∈-filesWrote}\AgdaSpace{}%
\AgdaBound{ls}\AgdaSpace{}%
\AgdaBound{v∈ls}\AgdaSymbol{))}\<%
\\
\>[0][@{}l@{\AgdaIndent{0}}]%
\>[2]\AgdaKeyword{where}%
\>[3635I]\AgdaFunction{∈₂}\AgdaSpace{}%
\AgdaSymbol{:}\AgdaSpace{}%
\AgdaSymbol{∀}\AgdaSpace{}%
\AgdaSymbol{\{}\AgdaBound{v}\AgdaSymbol{\}}\AgdaSpace{}%
\AgdaSymbol{→}\AgdaSpace{}%
\AgdaBound{v}\AgdaSpace{}%
\AgdaOperator{\AgdaPostulate{∈}}\AgdaSpace{}%
\AgdaPostulate{cmdWriteNames}\AgdaSpace{}%
\AgdaBound{x₁}\AgdaSpace{}%
\AgdaBound{s}\AgdaSpace{}%
\AgdaOperator{\AgdaFunction{++}}\AgdaSpace{}%
\AgdaInductiveConstructor{[]}\AgdaSpace{}%
\AgdaSymbol{→}\AgdaSpace{}%
\AgdaBound{v}\AgdaSpace{}%
\AgdaOperator{\AgdaPostulate{∈}}\AgdaSpace{}%
\AgdaPostulate{cmdWriteNames}\AgdaSpace{}%
\AgdaBound{x₁}\AgdaSpace{}%
\AgdaBound{s}\<%
\\
\>[.][@{}l@{}]\<[3635I]%
\>[8]\AgdaFunction{∈₂}\AgdaSpace{}%
\AgdaBound{v∈}\AgdaSpace{}%
\AgdaKeyword{with}\AgdaSpace{}%
\AgdaFunction{∈-++⁻}\AgdaSpace{}%
\AgdaSymbol{(}\AgdaPostulate{cmdWriteNames}\AgdaSpace{}%
\AgdaBound{x₁}\AgdaSpace{}%
\AgdaBound{s}\AgdaSymbol{)}\AgdaSpace{}%
\AgdaBound{v∈}\<%
\\
\>[8]\AgdaSymbol{...}\AgdaSpace{}%
\AgdaSymbol{|}\AgdaSpace{}%
\AgdaInductiveConstructor{inj₁}\AgdaSpace{}%
\AgdaBound{v∈₁}\AgdaSpace{}%
\AgdaSymbol{=}\AgdaSpace{}%
\AgdaBound{v∈₁}\<%
\\
\>[8]\AgdaSymbol{...}\AgdaSpace{}%
\AgdaSymbol{|}\AgdaSpace{}%
\AgdaInductiveConstructor{inj₂}\AgdaSpace{}%
\AgdaSymbol{()}\<%
\\
\\[\AgdaEmptyExtraSkip]%
\\[\AgdaEmptyExtraSkip]%
\>[0]\AgdaFunction{extra-lemma3}\AgdaSpace{}%
\AgdaSymbol{:}\AgdaSpace{}%
\AgdaSymbol{∀}\AgdaSpace{}%
\AgdaSymbol{\{}\AgdaBound{s}\AgdaSymbol{\}}\AgdaSpace{}%
\AgdaSymbol{\{}\AgdaBound{ls}\AgdaSymbol{\}}\AgdaSpace{}%
\AgdaBound{bs}\AgdaSpace{}%
\AgdaBound{xs}\AgdaSpace{}%
\AgdaBound{x}\AgdaSpace{}%
\AgdaSymbol{→}\AgdaSpace{}%
\AgdaBound{x}\AgdaSpace{}%
\AgdaOperator{\AgdaPostulate{∉}}\AgdaSpace{}%
\AgdaBound{bs}\AgdaSpace{}%
\AgdaSymbol{→}\AgdaSpace{}%
\AgdaBound{x}\AgdaSpace{}%
\AgdaOperator{\AgdaPostulate{∉}}\AgdaSpace{}%
\AgdaBound{xs}\AgdaSpace{}%
\AgdaSymbol{→}\AgdaSpace{}%
\AgdaBound{bs}\AgdaSpace{}%
\AgdaOperator{\AgdaFunction{⊆}}\AgdaSpace{}%
\AgdaBound{xs}\AgdaSpace{}%
\AgdaSymbol{→}\AgdaSpace{}%
\AgdaPostulate{HazardFree}\AgdaSpace{}%
\AgdaBound{s}\AgdaSpace{}%
\AgdaBound{bs}\AgdaSpace{}%
\AgdaSymbol{(}\AgdaBound{xs}\AgdaSpace{}%
\AgdaOperator{\AgdaFunction{++}}\AgdaSpace{}%
\AgdaBound{x}\AgdaSpace{}%
\AgdaOperator{\AgdaInductiveConstructor{∷}}\AgdaSpace{}%
\AgdaInductiveConstructor{[]}\AgdaSymbol{)}\AgdaSpace{}%
\AgdaBound{ls}\AgdaSpace{}%
\AgdaSymbol{→}\AgdaSpace{}%
\AgdaFunction{Disjoint}\AgdaSpace{}%
\AgdaSymbol{(}\AgdaPostulate{files}\AgdaSpace{}%
\AgdaSymbol{(}\AgdaFunction{script-rec}\AgdaSpace{}%
\AgdaBound{bs}\AgdaSpace{}%
\AgdaBound{s}\AgdaSpace{}%
\AgdaInductiveConstructor{[]}\AgdaSymbol{))}\AgdaSpace{}%
\AgdaSymbol{(}\AgdaPostulate{cmdWrote}\AgdaSpace{}%
\AgdaBound{ls}\AgdaSpace{}%
\AgdaBound{x}\AgdaSymbol{)}\<%
\\
\>[0]\AgdaFunction{extra-lemma3}\AgdaSpace{}%
\AgdaSymbol{\{}\AgdaBound{s}\AgdaSymbol{\}}\AgdaSpace{}%
\AgdaBound{bs}\AgdaSpace{}%
\AgdaBound{xs}\AgdaSpace{}%
\AgdaBound{x}\AgdaSpace{}%
\AgdaBound{x∉bs}\AgdaSpace{}%
\AgdaBound{x∉xs}\AgdaSpace{}%
\AgdaBound{bs⊆xs}\AgdaSpace{}%
\AgdaBound{hf}\AgdaSpace{}%
\AgdaBound{x₂}\AgdaSpace{}%
\AgdaKeyword{with}\AgdaSpace{}%
\AgdaFunction{∈-++⁻}\AgdaSpace{}%
\AgdaSymbol{(}\AgdaPostulate{filesRead}\AgdaSpace{}%
\AgdaSymbol{(}\AgdaFunction{script-rec}\AgdaSpace{}%
\AgdaBound{bs}\AgdaSpace{}%
\AgdaBound{s}\AgdaSpace{}%
\AgdaInductiveConstructor{[]}\AgdaSymbol{))}\AgdaSpace{}%
\AgdaSymbol{(}\AgdaField{proj₁}\AgdaSpace{}%
\AgdaBound{x₂}\AgdaSymbol{)}\<%
\\
\>[0]\AgdaSymbol{...}\AgdaSpace{}%
\AgdaSymbol{|}\AgdaSpace{}%
\AgdaInductiveConstructor{inj₁}\AgdaSpace{}%
\AgdaBound{∈₁}\AgdaSpace{}%
\AgdaSymbol{=}\AgdaSpace{}%
\AgdaFunction{contradiction}\AgdaSpace{}%
\AgdaSymbol{(}\AgdaBound{∈₁}\AgdaSpace{}%
\AgdaOperator{\AgdaInductiveConstructor{,}}\AgdaSpace{}%
\AgdaSymbol{(}\AgdaField{proj₂}\AgdaSpace{}%
\AgdaBound{x₂}\AgdaSymbol{))}\AgdaSpace{}%
\AgdaSymbol{(}\AgdaFunction{extra-lemma3-reads}\AgdaSpace{}%
\AgdaBound{bs}\AgdaSpace{}%
\AgdaBound{xs}\AgdaSpace{}%
\AgdaBound{x}\AgdaSpace{}%
\AgdaBound{x∉bs}\AgdaSpace{}%
\AgdaBound{x∉xs}\AgdaSpace{}%
\AgdaBound{bs⊆xs}\AgdaSpace{}%
\AgdaBound{hf}\AgdaSymbol{)}\<%
\\
\>[0]\AgdaSymbol{...}\AgdaSpace{}%
\AgdaSymbol{|}\AgdaSpace{}%
\AgdaInductiveConstructor{inj₂}\AgdaSpace{}%
\AgdaBound{∈₁}\AgdaSpace{}%
\AgdaSymbol{=}\AgdaSpace{}%
\AgdaFunction{contradiction}\AgdaSpace{}%
\AgdaSymbol{(}\AgdaBound{∈₁}\AgdaSpace{}%
\AgdaOperator{\AgdaInductiveConstructor{,}}\AgdaSpace{}%
\AgdaSymbol{(}\AgdaField{proj₂}\AgdaSpace{}%
\AgdaBound{x₂}\AgdaSymbol{))}\AgdaSpace{}%
\AgdaSymbol{(}\AgdaFunction{extra-lemma3-writes}\AgdaSpace{}%
\AgdaBound{bs}\AgdaSpace{}%
\AgdaBound{xs}\AgdaSpace{}%
\AgdaBound{x}\AgdaSpace{}%
\AgdaBound{x∉bs}\AgdaSpace{}%
\AgdaBound{x∉xs}\AgdaSpace{}%
\AgdaBound{bs⊆xs}\AgdaSpace{}%
\AgdaBound{hf}\AgdaSymbol{)}\<%
\\
\\[\AgdaEmptyExtraSkip]%
\\[\AgdaEmptyExtraSkip]%
\>[0]\AgdaFunction{extra-lemma}\AgdaSpace{}%
\AgdaSymbol{:}\AgdaSpace{}%
\AgdaSymbol{∀}\AgdaSpace{}%
\AgdaSymbol{\{}\AgdaBound{s}\AgdaSymbol{\}}\AgdaSpace{}%
\AgdaSymbol{\{}\AgdaBound{ls}\AgdaSymbol{\}}\AgdaSpace{}%
\AgdaBound{as}\AgdaSpace{}%
\AgdaBound{bs}\AgdaSpace{}%
\AgdaBound{xs}\AgdaSpace{}%
\AgdaBound{x}\AgdaSpace{}%
\AgdaSymbol{→}\AgdaSpace{}%
\AgdaPostulate{HazardFree}\AgdaSpace{}%
\AgdaBound{s}\AgdaSpace{}%
\AgdaSymbol{(}\AgdaBound{as}\AgdaSpace{}%
\AgdaOperator{\AgdaFunction{++}}\AgdaSpace{}%
\AgdaBound{x}\AgdaSpace{}%
\AgdaOperator{\AgdaInductiveConstructor{∷}}\AgdaSpace{}%
\AgdaBound{bs}\AgdaSymbol{)}\AgdaSpace{}%
\AgdaSymbol{(}\AgdaBound{xs}\AgdaSpace{}%
\AgdaOperator{\AgdaFunction{++}}\AgdaSpace{}%
\AgdaBound{x}\AgdaSpace{}%
\AgdaOperator{\AgdaInductiveConstructor{∷}}\AgdaSpace{}%
\AgdaInductiveConstructor{[]}\AgdaSymbol{)}\AgdaSpace{}%
\AgdaBound{ls}\AgdaSpace{}%
\AgdaSymbol{→}\AgdaSpace{}%
\AgdaFunction{Disjoint}\AgdaSpace{}%
\AgdaSymbol{(}\AgdaPostulate{files}\AgdaSpace{}%
\AgdaSymbol{(}\AgdaFunction{script-rec}\AgdaSpace{}%
\AgdaBound{as}\AgdaSpace{}%
\AgdaBound{s}\AgdaSpace{}%
\AgdaBound{ls}\AgdaSymbol{))}\AgdaSpace{}%
\AgdaSymbol{(}\AgdaPostulate{cmdWriteNames}\AgdaSpace{}%
\AgdaBound{x}\AgdaSpace{}%
\AgdaSymbol{(}\AgdaPostulate{script}\AgdaSpace{}%
\AgdaBound{as}\AgdaSpace{}%
\AgdaBound{s}\AgdaSymbol{))}\AgdaSpace{}%
\AgdaOperator{\AgdaFunction{×}}\AgdaSpace{}%
\AgdaFunction{Disjoint}\AgdaSpace{}%
\AgdaSymbol{(}\AgdaPostulate{files}\AgdaSpace{}%
\AgdaSymbol{(}\AgdaFunction{script-rec}\AgdaSpace{}%
\AgdaBound{bs}\AgdaSpace{}%
\AgdaSymbol{(}\AgdaPostulate{St.run}\AgdaSpace{}%
\AgdaBound{x}\AgdaSpace{}%
\AgdaSymbol{(}\AgdaPostulate{script}\AgdaSpace{}%
\AgdaBound{as}\AgdaSpace{}%
\AgdaBound{s}\AgdaSymbol{))}\AgdaSpace{}%
\AgdaInductiveConstructor{[]}\AgdaSymbol{))}\AgdaSpace{}%
\AgdaSymbol{(}\AgdaPostulate{cmdWriteNames}\AgdaSpace{}%
\AgdaBound{x}\AgdaSpace{}%
\AgdaSymbol{(}\AgdaPostulate{script}\AgdaSpace{}%
\AgdaBound{as}\AgdaSpace{}%
\AgdaBound{s}\AgdaSymbol{))}\<%
\\
\>[0]\AgdaFunction{extra-lemma}\AgdaSpace{}%
\AgdaSymbol{\{}\AgdaBound{s}\AgdaSymbol{\}}\AgdaSpace{}%
\AgdaSymbol{\{}\AgdaBound{ls}\AgdaSymbol{\}}\AgdaSpace{}%
\AgdaInductiveConstructor{[]}\AgdaSpace{}%
\AgdaBound{bs}\AgdaSpace{}%
\AgdaBound{xs}\AgdaSpace{}%
\AgdaBound{x}\AgdaSpace{}%
\AgdaSymbol{(}\AgdaBound{¬hz}\AgdaSpace{}%
\AgdaOperator{\AgdaInductiveConstructor{∷}}\AgdaSpace{}%
\AgdaBound{hf}\AgdaSymbol{)}\AgdaSpace{}%
\AgdaSymbol{=}\AgdaSpace{}%
\AgdaFunction{dsj}\AgdaSpace{}%
\AgdaOperator{\AgdaInductiveConstructor{,}}\AgdaSpace{}%
\AgdaSymbol{(}\AgdaFunction{extra-lemma3}\AgdaSpace{}%
\AgdaBound{bs}\AgdaSpace{}%
\AgdaBound{xs}\AgdaSpace{}%
\AgdaBound{x}\AgdaSpace{}%
\AgdaSymbol{\{!!\}}\AgdaSpace{}%
\AgdaSymbol{\{!!\}}\AgdaSpace{}%
\AgdaSymbol{\{!!\}}\AgdaSpace{}%
\AgdaSymbol{(}\AgdaFunction{subst}\AgdaSpace{}%
\AgdaSymbol{(λ}\AgdaSpace{}%
\AgdaBound{x₁}\AgdaSpace{}%
\AgdaSymbol{→}\AgdaSpace{}%
\AgdaPostulate{HazardFree}\AgdaSpace{}%
\AgdaBound{x₁}\AgdaSpace{}%
\AgdaSymbol{\AgdaUnderscore{}}\AgdaSpace{}%
\AgdaSymbol{\AgdaUnderscore{}}\AgdaSpace{}%
\AgdaSymbol{\AgdaUnderscore{})}\AgdaSpace{}%
\AgdaSymbol{(}\AgdaFunction{cong}\AgdaSpace{}%
\AgdaSymbol{(}\AgdaPostulate{St.run}\AgdaSpace{}%
\AgdaBound{x}\AgdaSymbol{)}\AgdaSpace{}%
\AgdaInductiveConstructor{refl}\AgdaSymbol{)}\AgdaSpace{}%
\AgdaBound{hf}\AgdaSymbol{))}\<%
\\
\>[0][@{}l@{\AgdaIndent{0}}]%
\>[2]\AgdaKeyword{where}%
\>[3842I]\AgdaFunction{dsj}\AgdaSpace{}%
\AgdaSymbol{:}\AgdaSpace{}%
\AgdaFunction{Disjoint}\AgdaSpace{}%
\AgdaSymbol{(}\AgdaPostulate{files}\AgdaSpace{}%
\AgdaBound{ls}\AgdaSymbol{)}\AgdaSpace{}%
\AgdaSymbol{(}\AgdaPostulate{cmdWriteNames}\AgdaSpace{}%
\AgdaBound{x}\AgdaSpace{}%
\AgdaSymbol{(}\AgdaPostulate{script}\AgdaSpace{}%
\AgdaInductiveConstructor{[]}\AgdaSpace{}%
\AgdaBound{s}\AgdaSymbol{))}\<%
\\
\>[.][@{}l@{}]\<[3842I]%
\>[8]\AgdaFunction{dsj}\AgdaSpace{}%
\AgdaBound{x₂}\AgdaSpace{}%
\AgdaKeyword{with}\AgdaSpace{}%
\AgdaFunction{∈-++⁻}\AgdaSpace{}%
\AgdaSymbol{(}\AgdaPostulate{filesRead}\AgdaSpace{}%
\AgdaBound{ls}\AgdaSymbol{)}\AgdaSpace{}%
\AgdaSymbol{(}\AgdaField{proj₁}\AgdaSpace{}%
\AgdaBound{x₂}\AgdaSymbol{)}\<%
\\
\>[8]\AgdaSymbol{...}\AgdaSpace{}%
\AgdaSymbol{|}\AgdaSpace{}%
\AgdaInductiveConstructor{inj₁}\AgdaSpace{}%
\AgdaBound{∈₁}\AgdaSpace{}%
\AgdaSymbol{=}\AgdaSpace{}%
\AgdaBound{¬hz}\AgdaSpace{}%
\AgdaSymbol{(}\AgdaPostulate{ReadWrite}\AgdaSpace{}%
\AgdaSymbol{(}\AgdaField{proj₂}\AgdaSpace{}%
\AgdaBound{x₂}\AgdaSymbol{)}\AgdaSpace{}%
\AgdaBound{∈₁}\AgdaSymbol{)}\<%
\\
\>[8]\AgdaSymbol{...}\AgdaSpace{}%
\AgdaSymbol{|}\AgdaSpace{}%
\AgdaInductiveConstructor{inj₂}\AgdaSpace{}%
\AgdaBound{∈₁}\AgdaSpace{}%
\AgdaSymbol{=}\AgdaSpace{}%
\AgdaBound{¬hz}\AgdaSpace{}%
\AgdaSymbol{(}\AgdaPostulate{WriteWrite}\AgdaSpace{}%
\AgdaSymbol{(}\AgdaField{proj₂}\AgdaSpace{}%
\AgdaBound{x₂}\AgdaSymbol{)}\AgdaSpace{}%
\AgdaBound{∈₁}\AgdaSymbol{)}\<%
\\
\>[0]\AgdaFunction{extra-lemma}\AgdaSpace{}%
\AgdaSymbol{(}\AgdaBound{x₂}\AgdaSpace{}%
\AgdaOperator{\AgdaInductiveConstructor{∷}}\AgdaSpace{}%
\AgdaBound{as}\AgdaSymbol{)}\AgdaSpace{}%
\AgdaBound{bs}\AgdaSpace{}%
\AgdaBound{xs}\AgdaSpace{}%
\AgdaBound{x}\AgdaSpace{}%
\AgdaSymbol{(\AgdaUnderscore{}}\AgdaSpace{}%
\AgdaOperator{\AgdaInductiveConstructor{∷}}\AgdaSpace{}%
\AgdaBound{hf}\AgdaSymbol{)}\AgdaSpace{}%
\AgdaSymbol{=}\AgdaSpace{}%
\AgdaFunction{extra-lemma}\AgdaSpace{}%
\AgdaBound{as}\AgdaSpace{}%
\AgdaBound{bs}\AgdaSpace{}%
\AgdaBound{xs}\AgdaSpace{}%
\AgdaBound{x}\AgdaSpace{}%
\AgdaBound{hf}\<%
\\
\\[\AgdaEmptyExtraSkip]%
\>[0]\AgdaFunction{preserves-help2}\AgdaSpace{}%
\AgdaSymbol{:}\AgdaSpace{}%
\AgdaSymbol{∀}\AgdaSpace{}%
\AgdaSymbol{\{}\AgdaBound{s}\AgdaSymbol{\}}\AgdaSpace{}%
\AgdaSymbol{\{}\AgdaBound{ls}\AgdaSymbol{\}}\AgdaSpace{}%
\AgdaBound{as}\AgdaSpace{}%
\AgdaBound{bs}\AgdaSpace{}%
\AgdaBound{ys}\AgdaSpace{}%
\AgdaBound{x}\AgdaSpace{}%
\AgdaSymbol{→}\AgdaSpace{}%
\AgdaPostulate{filesWrote}\AgdaSpace{}%
\AgdaSymbol{(}\AgdaFunction{script-rec}\AgdaSpace{}%
\AgdaSymbol{(}\AgdaFunction{reverse}\AgdaSpace{}%
\AgdaBound{ys}\AgdaSymbol{)}\AgdaSpace{}%
\AgdaBound{s}\AgdaSpace{}%
\AgdaBound{ls}\AgdaSymbol{)}\AgdaSpace{}%
\AgdaOperator{\AgdaFunction{⊆}}\AgdaSpace{}%
\AgdaPostulate{filesWrote}\AgdaSpace{}%
\AgdaSymbol{(}\AgdaFunction{script-rec}\AgdaSpace{}%
\AgdaBound{as}\AgdaSpace{}%
\AgdaBound{s}\AgdaSpace{}%
\AgdaBound{ls}\AgdaSymbol{)}\AgdaSpace{}%
\AgdaOperator{\AgdaFunction{++}}\AgdaSpace{}%
\AgdaPostulate{filesWrote}\AgdaSpace{}%
\AgdaSymbol{(}\AgdaFunction{script-rec}\AgdaSpace{}%
\AgdaBound{bs}\AgdaSpace{}%
\AgdaSymbol{(}\AgdaPostulate{St.run}\AgdaSpace{}%
\AgdaBound{x}\AgdaSpace{}%
\AgdaSymbol{(}\AgdaPostulate{script}\AgdaSpace{}%
\AgdaBound{as}\AgdaSpace{}%
\AgdaBound{s}\AgdaSymbol{))}\AgdaSpace{}%
\AgdaInductiveConstructor{[]}\AgdaSymbol{)}\<%
\\
\>[0]\AgdaFunction{preserves-help2}\AgdaSpace{}%
\AgdaSymbol{=}\AgdaSpace{}%
\AgdaSymbol{\{!!\}}\<%
\\
\>[0]\AgdaFunction{preserves-help1}\AgdaSpace{}%
\AgdaSymbol{:}\AgdaSpace{}%
\AgdaSymbol{∀}\AgdaSpace{}%
\AgdaSymbol{\{}\AgdaBound{s}\AgdaSymbol{\}}\AgdaSpace{}%
\AgdaSymbol{\{}\AgdaBound{ls}\AgdaSymbol{\}}\AgdaSpace{}%
\AgdaBound{as}\AgdaSpace{}%
\AgdaBound{bs}\AgdaSpace{}%
\AgdaBound{ys}\AgdaSpace{}%
\AgdaBound{x}\AgdaSpace{}%
\AgdaSymbol{→}\AgdaSpace{}%
\AgdaPostulate{filesRead}\AgdaSpace{}%
\AgdaSymbol{(}\AgdaFunction{script-rec}\AgdaSpace{}%
\AgdaSymbol{(}\AgdaFunction{reverse}\AgdaSpace{}%
\AgdaBound{ys}\AgdaSymbol{)}\AgdaSpace{}%
\AgdaBound{s}\AgdaSpace{}%
\AgdaBound{ls}\AgdaSymbol{)}\AgdaSpace{}%
\AgdaOperator{\AgdaFunction{⊆}}\AgdaSpace{}%
\AgdaPostulate{filesRead}\AgdaSpace{}%
\AgdaSymbol{(}\AgdaFunction{script-rec}\AgdaSpace{}%
\AgdaBound{as}\AgdaSpace{}%
\AgdaBound{s}\AgdaSpace{}%
\AgdaBound{ls}\AgdaSymbol{)}\AgdaSpace{}%
\AgdaOperator{\AgdaFunction{++}}\AgdaSpace{}%
\AgdaPostulate{filesRead}\AgdaSpace{}%
\AgdaSymbol{(}\AgdaFunction{script-rec}\AgdaSpace{}%
\AgdaBound{bs}\AgdaSpace{}%
\AgdaSymbol{(}\AgdaPostulate{St.run}\AgdaSpace{}%
\AgdaBound{x}\AgdaSpace{}%
\AgdaSymbol{(}\AgdaPostulate{script}\AgdaSpace{}%
\AgdaBound{as}\AgdaSpace{}%
\AgdaBound{s}\AgdaSymbol{))}\AgdaSpace{}%
\AgdaInductiveConstructor{[]}\AgdaSymbol{)}\<%
\\
\>[0]\AgdaFunction{preserves-help1}\AgdaSpace{}%
\AgdaSymbol{=}\AgdaSpace{}%
\AgdaSymbol{\{!!\}}\<%
\\
\\[\AgdaEmptyExtraSkip]%
\\[\AgdaEmptyExtraSkip]%
\\[\AgdaEmptyExtraSkip]%
\\[\AgdaEmptyExtraSkip]%
\\[\AgdaEmptyExtraSkip]%
\\[\AgdaEmptyExtraSkip]%
\\[\AgdaEmptyExtraSkip]%
\>[0]\AgdaFunction{dsj-helper}\AgdaSpace{}%
\AgdaSymbol{:}\AgdaSpace{}%
\AgdaSymbol{∀}\AgdaSpace{}%
\AgdaSymbol{\{}\AgdaBound{s}\AgdaSymbol{\}}\AgdaSpace{}%
\AgdaSymbol{\{}\AgdaBound{ls}\AgdaSymbol{\}}\AgdaSpace{}%
\AgdaBound{xs}\AgdaSpace{}%
\AgdaBound{x}\AgdaSpace{}%
\AgdaSymbol{→}\AgdaSpace{}%
\AgdaFunction{Disjoint}\AgdaSpace{}%
\AgdaSymbol{(}\AgdaPostulate{files}\AgdaSpace{}%
\AgdaSymbol{(}\AgdaFunction{script-rec}\AgdaSpace{}%
\AgdaBound{xs}\AgdaSpace{}%
\AgdaBound{s}\AgdaSpace{}%
\AgdaBound{ls}\AgdaSymbol{))}\AgdaSpace{}%
\AgdaSymbol{(}\AgdaPostulate{cmdWriteNames}\AgdaSpace{}%
\AgdaBound{x}\AgdaSpace{}%
\AgdaSymbol{(}\AgdaPostulate{script}\AgdaSpace{}%
\AgdaBound{xs}\AgdaSpace{}%
\AgdaBound{s}\AgdaSymbol{))}\<%
\\
\>[0]\AgdaFunction{dsj-helper}\AgdaSpace{}%
\AgdaBound{xs}\AgdaSpace{}%
\AgdaBound{x}\AgdaSpace{}%
\AgdaBound{x₁}\AgdaSpace{}%
\AgdaSymbol{=}\AgdaSpace{}%
\AgdaSymbol{\{!!\}}\<%
\\
\\[\AgdaEmptyExtraSkip]%
\>[0]\AgdaComment{\{- with extra-lemma as bs (reverse ys) x (subst₂ (λ x₂ x₃ → HazardFree s x₂ x₃ ls) ≡₁ (unfold-reverse x ys) hf)
... | a₁ , a₂ with ∈-++⁻ (filesRead (script-rec (reverse ys) \AgdaUnderscore{} ls)) (proj₁ x₁)
... | inj₁ ∈₁ with ∈-++⁻ (filesRead (script-rec as s ls)) (⊆₁ ∈₁) 
  where ⊆₁ : filesRead (script-rec (reverse ys) s ls) ⊆ filesRead (script-rec as s ls) ++ filesRead (script-rec bs (St.run x (script as s)) [])
                ⊆₁ = \{!!\}
        ... | inj₁ ∈₂ = a₁ (∈-++⁺ˡ ∈₂ , ∈as (proj₂ x₁))
        ... | inj₂ ∈₂ = a₂ (∈-++⁺ˡ ∈₂ , ∈as (proj₂ x₁))
        -- in filesWrite (script-rec (reverse ys) ...
        dsj x₁ | (a₁ , a₂) | inj₂ ∈₁ with ∈-++⁻ (filesWrote (script-rec as s ls)) (⊆₁ ∈₁)
          where ⊆₁ : filesWrote (script-rec (reverse ys) s ls) ⊆ filesWrote (script-rec as s ls) ++ filesWrote (script-rec bs (St.run x (script as s)) [])
                ⊆₁ = \{!!\}
        ... | inj₁ ∈₂ = a₁ (∈-++⁺ʳ (filesRead (script-rec as s ls)) ∈₂ , ∈as (proj₂ x₁))
        ... | inj₂ ∈₂ = a₂ (∈-++⁺ʳ (filesRead (script-rec bs (St.run x (script as s)) [])) ∈₂ , ∈as (proj₂ x₁)) -\}}\<%
\\
\\[\AgdaEmptyExtraSkip]%
\>[0]\AgdaComment{\{- ∈as : ∀ \{v\} → v ∈ cmdWriteNames x (script (reverse ys) s) → v ∈ cmdWriteNames x (script as s)
        ∈as v∈ with helper5 (buildReadNames (St.run x (script as s)) bs) x \{!!\}
        ... | all₁ with writes≡ (script as s) (St.run x (script as s)) bs all₁
        ... | ≡₁ with helper3 as bs (reverse ys) x (subst (λ x₁ → Disjoint \AgdaUnderscore{} x₁) (sym ≡₁) \{!!\}) \{!!\}
        ... | x≡ = \{!!\}

        -\}}\<%
\\
\\[\AgdaEmptyExtraSkip]%
\\[\AgdaEmptyExtraSkip]%
\\[\AgdaEmptyExtraSkip]%
\>[0]\AgdaFunction{preservesHazardFree}\AgdaSpace{}%
\AgdaSymbol{:}\AgdaSpace{}%
\AgdaSymbol{∀}\AgdaSpace{}%
\AgdaSymbol{\{}\AgdaBound{s}\AgdaSymbol{\}}\AgdaSpace{}%
\AgdaSymbol{\{}\AgdaBound{ls}\AgdaSymbol{\}}\AgdaSpace{}%
\AgdaBound{xs}\AgdaSpace{}%
\AgdaBound{ys}\AgdaSpace{}%
\AgdaSymbol{→}\AgdaSpace{}%
\AgdaFunction{length}\AgdaSpace{}%
\AgdaBound{xs}\AgdaSpace{}%
\AgdaOperator{\AgdaDatatype{≡}}\AgdaSpace{}%
\AgdaFunction{length}\AgdaSpace{}%
\AgdaBound{ys}\AgdaSpace{}%
\AgdaSymbol{→}\AgdaSpace{}%
\AgdaBound{xs}\AgdaSpace{}%
\AgdaOperator{\AgdaDatatype{↭}}\AgdaSpace{}%
\AgdaBound{ys}\AgdaSpace{}%
\AgdaSymbol{→}\AgdaSpace{}%
\AgdaPostulate{UniqueEvidence}\AgdaSpace{}%
\AgdaBound{xs}\AgdaSpace{}%
\AgdaBound{ys}\AgdaSpace{}%
\AgdaSymbol{(}\AgdaFunction{map}\AgdaSpace{}%
\AgdaField{proj₁}\AgdaSpace{}%
\AgdaBound{ls}\AgdaSymbol{)}\AgdaSpace{}%
\AgdaSymbol{→}\AgdaSpace{}%
\AgdaPostulate{HazardFree}\AgdaSpace{}%
\AgdaBound{s}\AgdaSpace{}%
\AgdaBound{xs}\AgdaSpace{}%
\AgdaSymbol{(}\AgdaFunction{reverse}\AgdaSpace{}%
\AgdaBound{ys}\AgdaSymbol{)}\AgdaSpace{}%
\AgdaBound{ls}\AgdaSpace{}%
\AgdaSymbol{→}\AgdaSpace{}%
\AgdaPostulate{HazardFree}\AgdaSpace{}%
\AgdaBound{s}\AgdaSpace{}%
\AgdaSymbol{(}\AgdaFunction{reverse}\AgdaSpace{}%
\AgdaBound{ys}\AgdaSymbol{)}\AgdaSpace{}%
\AgdaSymbol{(}\AgdaFunction{reverse}\AgdaSpace{}%
\AgdaBound{ys}\AgdaSymbol{)}\AgdaSpace{}%
\AgdaBound{ls}\<%
\\
\>[0]\AgdaFunction{preservesHazardFree}\AgdaSpace{}%
\AgdaInductiveConstructor{[]}\AgdaSpace{}%
\AgdaInductiveConstructor{[]}\AgdaSpace{}%
\AgdaSymbol{\AgdaUnderscore{}}\AgdaSpace{}%
\AgdaBound{p}\AgdaSpace{}%
\AgdaSymbol{\AgdaUnderscore{}}\AgdaSpace{}%
\AgdaBound{hf}\AgdaSpace{}%
\AgdaSymbol{=}\AgdaSpace{}%
\AgdaInductiveConstructor{[]}\<%
\\
\>[0]\AgdaFunction{preservesHazardFree}\AgdaSpace{}%
\AgdaSymbol{\{}\AgdaBound{s}\AgdaSymbol{\}}\AgdaSpace{}%
\AgdaSymbol{\{}\AgdaBound{ls}\AgdaSymbol{\}}\AgdaSpace{}%
\AgdaBound{xs}\AgdaSpace{}%
\AgdaSymbol{(}\AgdaBound{x}\AgdaSpace{}%
\AgdaOperator{\AgdaInductiveConstructor{∷}}\AgdaSpace{}%
\AgdaBound{ys}\AgdaSymbol{)}\AgdaSpace{}%
\AgdaSymbol{\AgdaUnderscore{}}\AgdaSpace{}%
\AgdaBound{p}\AgdaSpace{}%
\AgdaSymbol{(}\AgdaBound{uxs}\AgdaSpace{}%
\AgdaOperator{\AgdaInductiveConstructor{,}}\AgdaSpace{}%
\AgdaBound{ux}\AgdaSpace{}%
\AgdaOperator{\AgdaInductiveConstructor{∷}}\AgdaSpace{}%
\AgdaBound{uys}\AgdaSpace{}%
\AgdaOperator{\AgdaInductiveConstructor{,}}\AgdaSpace{}%
\AgdaBound{uls}\AgdaSpace{}%
\AgdaOperator{\AgdaInductiveConstructor{,}}\AgdaSpace{}%
\AgdaBound{dsj}\AgdaSymbol{)}\AgdaSpace{}%
\AgdaBound{hf}\AgdaSpace{}%
\AgdaKeyword{with}\AgdaSpace{}%
\AgdaFunction{∈-∃++}\AgdaSpace{}%
\AgdaSymbol{(}\AgdaFunction{∈-resp-↭}\AgdaSpace{}%
\AgdaSymbol{(}\AgdaFunction{↭-sym}\AgdaSpace{}%
\AgdaBound{p}\AgdaSymbol{)}\AgdaSpace{}%
\AgdaSymbol{(}\AgdaInductiveConstructor{here}\AgdaSpace{}%
\AgdaInductiveConstructor{refl}\AgdaSymbol{))}\<%
\\
\>[0]\AgdaSymbol{...}%
\>[4053I]\AgdaSymbol{|}\AgdaSpace{}%
\AgdaSymbol{(}\AgdaBound{as}\AgdaSpace{}%
\AgdaOperator{\AgdaInductiveConstructor{,}}\AgdaSpace{}%
\AgdaBound{bs}\AgdaSpace{}%
\AgdaOperator{\AgdaInductiveConstructor{,}}\AgdaSpace{}%
\AgdaBound{≡₁}\AgdaSymbol{)}\AgdaSpace{}%
\AgdaKeyword{with}\AgdaSpace{}%
\AgdaFunction{preserves-++}\AgdaSpace{}%
\AgdaBound{x}\AgdaSpace{}%
\AgdaSymbol{(}\AgdaFunction{reverse}\AgdaSpace{}%
\AgdaBound{ys}\AgdaSymbol{)}\AgdaSpace{}%
\AgdaSymbol{(}\AgdaFunction{reverse}\AgdaSpace{}%
\AgdaBound{ys}\AgdaSymbol{)}\AgdaSpace{}%
\AgdaSymbol{\{!!\}}\AgdaSpace{}%
\AgdaFunction{dsj₄}\AgdaSpace{}%
\AgdaSymbol{\{!!\}}\AgdaSpace{}%
\AgdaSymbol{\{!!\}}\<%
\\
\>[.][@{}l@{}]\<[4053I]%
\>[4]\AgdaSymbol{(}\AgdaFunction{preservesHazardFree}\AgdaSpace{}%
\AgdaSymbol{(}\AgdaBound{as}\AgdaSpace{}%
\AgdaOperator{\AgdaFunction{++}}\AgdaSpace{}%
\AgdaBound{bs}\AgdaSymbol{)}\AgdaSpace{}%
\AgdaBound{ys}\AgdaSpace{}%
\AgdaSymbol{(}\AgdaFunction{↭-length}\AgdaSpace{}%
\AgdaFunction{↭₂}\AgdaSymbol{)}\AgdaSpace{}%
\AgdaFunction{↭₂}\AgdaSpace{}%
\AgdaSymbol{(}\AgdaFunction{uas++bs}\AgdaSpace{}%
\AgdaOperator{\AgdaInductiveConstructor{,}}\AgdaSpace{}%
\AgdaBound{uys}\AgdaSpace{}%
\AgdaOperator{\AgdaInductiveConstructor{,}}\AgdaSpace{}%
\AgdaBound{uls}\AgdaSpace{}%
\AgdaOperator{\AgdaInductiveConstructor{,}}\AgdaSpace{}%
\AgdaFunction{dsj₀}\AgdaSymbol{)}\AgdaSpace{}%
\AgdaFunction{hf₁}\AgdaSymbol{)}\<%
\\
\>[0][@{}l@{\AgdaIndent{0}}]%
\>[2]\AgdaKeyword{where}%
\>[4085I]\AgdaFunction{↭₂}\AgdaSpace{}%
\AgdaSymbol{:}\AgdaSpace{}%
\AgdaBound{as}\AgdaSpace{}%
\AgdaOperator{\AgdaFunction{++}}\AgdaSpace{}%
\AgdaBound{bs}\AgdaSpace{}%
\AgdaOperator{\AgdaDatatype{↭}}\AgdaSpace{}%
\AgdaBound{ys}\<%
\\
\>[.][@{}l@{}]\<[4085I]%
\>[8]\AgdaFunction{↭₂}\AgdaSpace{}%
\AgdaSymbol{=}\AgdaSpace{}%
\AgdaFunction{drop-mid}\AgdaSpace{}%
\AgdaBound{as}\AgdaSpace{}%
\AgdaInductiveConstructor{[]}\AgdaSpace{}%
\AgdaSymbol{(}\AgdaFunction{subst}\AgdaSpace{}%
\AgdaSymbol{(λ}\AgdaSpace{}%
\AgdaBound{x₁}\AgdaSpace{}%
\AgdaSymbol{→}\AgdaSpace{}%
\AgdaBound{x₁}\AgdaSpace{}%
\AgdaOperator{\AgdaDatatype{↭}}\AgdaSpace{}%
\AgdaBound{x}\AgdaSpace{}%
\AgdaOperator{\AgdaInductiveConstructor{∷}}\AgdaSpace{}%
\AgdaBound{ys}\AgdaSymbol{)}\AgdaSpace{}%
\AgdaBound{≡₁}\AgdaSpace{}%
\AgdaBound{p}\AgdaSymbol{)}\<%
\\
\>[8]\AgdaFunction{uas++bs}\AgdaSpace{}%
\AgdaSymbol{:}\AgdaSpace{}%
\AgdaDatatype{Unique}\AgdaSpace{}%
\AgdaSymbol{(}\AgdaBound{as}\AgdaSpace{}%
\AgdaOperator{\AgdaFunction{++}}\AgdaSpace{}%
\AgdaBound{bs}\AgdaSymbol{)}\<%
\\
\>[8]\AgdaFunction{uas++bs}\AgdaSpace{}%
\AgdaSymbol{=}\AgdaSpace{}%
\AgdaPostulate{unique-drop-mid}\AgdaSpace{}%
\AgdaBound{as}\AgdaSpace{}%
\AgdaSymbol{(}\AgdaFunction{subst}\AgdaSpace{}%
\AgdaSymbol{(λ}\AgdaSpace{}%
\AgdaBound{x₁}\AgdaSpace{}%
\AgdaSymbol{→}\AgdaSpace{}%
\AgdaDatatype{Unique}\AgdaSpace{}%
\AgdaBound{x₁}\AgdaSymbol{)}\AgdaSpace{}%
\AgdaBound{≡₁}\AgdaSpace{}%
\AgdaBound{uxs}\AgdaSymbol{)}\<%
\\
\>[8]\AgdaFunction{dsj₀}\AgdaSpace{}%
\AgdaSymbol{:}\AgdaSpace{}%
\AgdaFunction{Disjoint}\AgdaSpace{}%
\AgdaSymbol{(}\AgdaBound{as}\AgdaSpace{}%
\AgdaOperator{\AgdaFunction{++}}\AgdaSpace{}%
\AgdaBound{bs}\AgdaSymbol{)}\AgdaSpace{}%
\AgdaSymbol{(}\AgdaFunction{map}\AgdaSpace{}%
\AgdaField{proj₁}\AgdaSpace{}%
\AgdaBound{ls}\AgdaSymbol{)}\<%
\\
\>[8]\AgdaFunction{dsj₀}\AgdaSpace{}%
\AgdaBound{x₂}\AgdaSpace{}%
\AgdaKeyword{with}\AgdaSpace{}%
\AgdaFunction{∈-++⁻}\AgdaSpace{}%
\AgdaBound{as}\AgdaSpace{}%
\AgdaSymbol{(}\AgdaField{proj₁}\AgdaSpace{}%
\AgdaBound{x₂}\AgdaSymbol{)}\<%
\\
\>[8]\AgdaSymbol{...}\AgdaSpace{}%
\AgdaSymbol{|}\AgdaSpace{}%
\AgdaInductiveConstructor{inj₁}\AgdaSpace{}%
\AgdaOperator{\AgdaBound{\AgdaUnderscore{}∈as}}\AgdaSpace{}%
\AgdaSymbol{=}\AgdaSpace{}%
\AgdaBound{dsj}\AgdaSpace{}%
\AgdaSymbol{(}\AgdaFunction{subst}\AgdaSpace{}%
\AgdaSymbol{(λ}\AgdaSpace{}%
\AgdaBound{x₁}\AgdaSpace{}%
\AgdaSymbol{→}\AgdaSpace{}%
\AgdaSymbol{\AgdaUnderscore{}}\AgdaSpace{}%
\AgdaOperator{\AgdaPostulate{∈}}\AgdaSpace{}%
\AgdaBound{x₁}\AgdaSymbol{)}\AgdaSpace{}%
\AgdaSymbol{(}\AgdaFunction{sym}\AgdaSpace{}%
\AgdaBound{≡₁}\AgdaSymbol{)}\AgdaSpace{}%
\AgdaSymbol{(}\AgdaFunction{∈-++⁺ˡ}\AgdaSpace{}%
\AgdaOperator{\AgdaBound{\AgdaUnderscore{}∈as}}\AgdaSymbol{)}\AgdaSpace{}%
\AgdaOperator{\AgdaInductiveConstructor{,}}\AgdaSpace{}%
\AgdaSymbol{(}\AgdaField{proj₂}\AgdaSpace{}%
\AgdaBound{x₂}\AgdaSymbol{))}\<%
\\
\>[8]\AgdaSymbol{...}\AgdaSpace{}%
\AgdaSymbol{|}\AgdaSpace{}%
\AgdaInductiveConstructor{inj₂}\AgdaSpace{}%
\AgdaOperator{\AgdaBound{\AgdaUnderscore{}∈bs}}\AgdaSpace{}%
\AgdaSymbol{=}\AgdaSpace{}%
\AgdaBound{dsj}\AgdaSpace{}%
\AgdaSymbol{(}\AgdaFunction{subst}\AgdaSpace{}%
\AgdaSymbol{(λ}\AgdaSpace{}%
\AgdaBound{x₁}\AgdaSpace{}%
\AgdaSymbol{→}\AgdaSpace{}%
\AgdaSymbol{\AgdaUnderscore{}}\AgdaSpace{}%
\AgdaOperator{\AgdaPostulate{∈}}\AgdaSpace{}%
\AgdaBound{x₁}\AgdaSymbol{)}\AgdaSpace{}%
\AgdaSymbol{(}\AgdaFunction{sym}\AgdaSpace{}%
\AgdaBound{≡₁}\AgdaSymbol{)}\AgdaSpace{}%
\AgdaSymbol{(}\AgdaFunction{∈-++⁺ʳ}\AgdaSpace{}%
\AgdaBound{as}\AgdaSpace{}%
\AgdaSymbol{(}\AgdaInductiveConstructor{there}\AgdaSpace{}%
\AgdaOperator{\AgdaBound{\AgdaUnderscore{}∈bs}}\AgdaSymbol{))}\AgdaSpace{}%
\AgdaOperator{\AgdaInductiveConstructor{,}}\AgdaSpace{}%
\AgdaSymbol{(}\AgdaField{proj₂}\AgdaSpace{}%
\AgdaBound{x₂}\AgdaSymbol{))}\<%
\\
\>[8]\AgdaFunction{⊆₁}\AgdaSpace{}%
\AgdaSymbol{:}\AgdaSpace{}%
\AgdaBound{as}\AgdaSpace{}%
\AgdaOperator{\AgdaFunction{++}}\AgdaSpace{}%
\AgdaBound{x}\AgdaSpace{}%
\AgdaOperator{\AgdaInductiveConstructor{∷}}\AgdaSpace{}%
\AgdaBound{bs}\AgdaSpace{}%
\AgdaOperator{\AgdaFunction{⊆}}\AgdaSpace{}%
\AgdaFunction{reverse}\AgdaSpace{}%
\AgdaBound{ys}\AgdaSpace{}%
\AgdaOperator{\AgdaFunction{++}}\AgdaSpace{}%
\AgdaBound{x}\AgdaSpace{}%
\AgdaOperator{\AgdaInductiveConstructor{∷}}\AgdaSpace{}%
\AgdaInductiveConstructor{[]}\<%
\\
\>[8]\AgdaFunction{⊆₁}\AgdaSpace{}%
\AgdaBound{x₁∈as++x∷bs}\AgdaSpace{}%
\AgdaSymbol{=}\AgdaSpace{}%
\AgdaFunction{subst}%
\>[4193I]\AgdaSymbol{(λ}\AgdaSpace{}%
\AgdaBound{x₂}\AgdaSpace{}%
\AgdaSymbol{→}\AgdaSpace{}%
\AgdaSymbol{\AgdaUnderscore{}}\AgdaSpace{}%
\AgdaOperator{\AgdaPostulate{∈}}\AgdaSpace{}%
\AgdaBound{x₂}\AgdaSymbol{)}\<%
\\
\>[.][@{}l@{}]\<[4193I]%
\>[31]\AgdaSymbol{(}\AgdaFunction{unfold-reverse}\AgdaSpace{}%
\AgdaBound{x}\AgdaSpace{}%
\AgdaBound{ys}\AgdaSymbol{)}\<%
\\
\>[31]\AgdaSymbol{(}\AgdaFunction{reverse⁺}\AgdaSpace{}%
\AgdaSymbol{(}\AgdaFunction{∈-resp-↭}\AgdaSpace{}%
\AgdaBound{p}\AgdaSpace{}%
\AgdaSymbol{(}\AgdaFunction{subst}\AgdaSpace{}%
\AgdaSymbol{(λ}\AgdaSpace{}%
\AgdaBound{x₂}\AgdaSpace{}%
\AgdaSymbol{→}\AgdaSpace{}%
\AgdaSymbol{\AgdaUnderscore{}}\AgdaSpace{}%
\AgdaOperator{\AgdaPostulate{∈}}\AgdaSpace{}%
\AgdaBound{x₂}\AgdaSymbol{)}\AgdaSpace{}%
\AgdaSymbol{(}\AgdaFunction{sym}\AgdaSpace{}%
\AgdaBound{≡₁}\AgdaSymbol{)}\AgdaSpace{}%
\AgdaBound{x₁∈as++x∷bs}\AgdaSymbol{)))}\<%
\\
\>[8]\AgdaFunction{hf₀}\AgdaSpace{}%
\AgdaSymbol{:}\AgdaSpace{}%
\AgdaPostulate{HazardFree}\AgdaSpace{}%
\AgdaBound{s}\AgdaSpace{}%
\AgdaSymbol{(}\AgdaBound{as}\AgdaSpace{}%
\AgdaOperator{\AgdaFunction{++}}\AgdaSpace{}%
\AgdaBound{x}\AgdaSpace{}%
\AgdaOperator{\AgdaInductiveConstructor{∷}}\AgdaSpace{}%
\AgdaBound{bs}\AgdaSymbol{)}\AgdaSpace{}%
\AgdaSymbol{((}\AgdaFunction{reverse}\AgdaSpace{}%
\AgdaBound{ys}\AgdaSymbol{)}\AgdaSpace{}%
\AgdaOperator{\AgdaFunction{++}}\AgdaSpace{}%
\AgdaBound{x}\AgdaSpace{}%
\AgdaOperator{\AgdaInductiveConstructor{∷}}\AgdaSpace{}%
\AgdaInductiveConstructor{[]}\AgdaSymbol{)}\AgdaSpace{}%
\AgdaSymbol{\AgdaUnderscore{}}\<%
\\
\>[8]\AgdaFunction{hf₀}\AgdaSpace{}%
\AgdaSymbol{=}\AgdaSpace{}%
\AgdaFunction{subst₂}\AgdaSpace{}%
\AgdaSymbol{(λ}\AgdaSpace{}%
\AgdaBound{x₂}\AgdaSpace{}%
\AgdaBound{x₃}\AgdaSpace{}%
\AgdaSymbol{→}\AgdaSpace{}%
\AgdaPostulate{HazardFree}\AgdaSpace{}%
\AgdaBound{s}\AgdaSpace{}%
\AgdaBound{x₂}\AgdaSpace{}%
\AgdaBound{x₃}\AgdaSpace{}%
\AgdaSymbol{\AgdaUnderscore{})}\AgdaSpace{}%
\AgdaBound{≡₁}\AgdaSpace{}%
\AgdaSymbol{(}\AgdaFunction{unfold-reverse}\AgdaSpace{}%
\AgdaBound{x}\AgdaSpace{}%
\AgdaBound{ys}\AgdaSymbol{)}\AgdaSpace{}%
\AgdaBound{hf}\<%
\\
\>[8]\AgdaFunction{hf₁}\AgdaSpace{}%
\AgdaSymbol{:}\AgdaSpace{}%
\AgdaPostulate{HazardFree}\AgdaSpace{}%
\AgdaBound{s}\AgdaSpace{}%
\AgdaSymbol{(}\AgdaBound{as}\AgdaSpace{}%
\AgdaOperator{\AgdaFunction{++}}\AgdaSpace{}%
\AgdaBound{bs}\AgdaSymbol{)}\AgdaSpace{}%
\AgdaSymbol{(}\AgdaFunction{reverse}\AgdaSpace{}%
\AgdaBound{ys}\AgdaSymbol{)}\AgdaSpace{}%
\AgdaSymbol{\AgdaUnderscore{}}\<%
\\
\>[8]\AgdaFunction{hf₁}\AgdaSpace{}%
\AgdaSymbol{=}\AgdaSpace{}%
\AgdaPostulate{hf-drop-mid}%
\>[4255I]\AgdaBound{as}\AgdaSpace{}%
\AgdaBound{bs}\AgdaSpace{}%
\AgdaSymbol{(}\AgdaFunction{reverse}\AgdaSpace{}%
\AgdaBound{ys}\AgdaSymbol{)}\AgdaSpace{}%
\AgdaFunction{⊆₁}\<%
\\
\>[.][@{}l@{}]\<[4255I]%
\>[26]\AgdaSymbol{(}\AgdaFunction{subst}\AgdaSpace{}%
\AgdaSymbol{(λ}\AgdaSpace{}%
\AgdaBound{x₁}\AgdaSpace{}%
\AgdaSymbol{→}\AgdaSpace{}%
\AgdaDatatype{Unique}\AgdaSpace{}%
\AgdaBound{x₁}\AgdaSymbol{)}\AgdaSpace{}%
\AgdaBound{≡₁}\AgdaSpace{}%
\AgdaBound{uxs}\AgdaSymbol{)}\<%
\\
\>[26]\AgdaSymbol{(}\AgdaFunction{subst}\AgdaSpace{}%
\AgdaSymbol{(λ}\AgdaSpace{}%
\AgdaBound{x₁}\AgdaSpace{}%
\AgdaSymbol{→}\AgdaSpace{}%
\AgdaDatatype{Unique}\AgdaSpace{}%
\AgdaBound{x₁}\AgdaSymbol{)}\AgdaSpace{}%
\AgdaSymbol{(}\AgdaFunction{unfold-reverse}\AgdaSpace{}%
\AgdaBound{x}\AgdaSpace{}%
\AgdaBound{ys}\AgdaSymbol{)}\AgdaSpace{}%
\AgdaSymbol{(}\AgdaPostulate{unique-reverse}\AgdaSpace{}%
\AgdaSymbol{(}\AgdaBound{x}\AgdaSpace{}%
\AgdaOperator{\AgdaInductiveConstructor{∷}}\AgdaSpace{}%
\AgdaBound{ys}\AgdaSymbol{)}\AgdaSpace{}%
\AgdaSymbol{(}\AgdaBound{ux}\AgdaSpace{}%
\AgdaOperator{\AgdaInductiveConstructor{∷}}\AgdaSpace{}%
\AgdaBound{uys}\AgdaSymbol{)))}\<%
\\
\>[26]\AgdaBound{uls}\<%
\\
\>[26]\AgdaSymbol{(}\AgdaFunction{subst}\AgdaSpace{}%
\AgdaSymbol{(λ}\AgdaSpace{}%
\AgdaBound{x₁}\AgdaSpace{}%
\AgdaSymbol{→}\AgdaSpace{}%
\AgdaFunction{Disjoint}\AgdaSpace{}%
\AgdaBound{x₁}\AgdaSpace{}%
\AgdaSymbol{\AgdaUnderscore{})}\AgdaSpace{}%
\AgdaBound{≡₁}\AgdaSpace{}%
\AgdaBound{dsj}\AgdaSymbol{)}\<%
\\
\>[26]\AgdaFunction{hf₀}\<%
\\
\>[8]\AgdaFunction{bs⊆reverse-ys}\AgdaSpace{}%
\AgdaSymbol{:}\AgdaSpace{}%
\AgdaBound{bs}\AgdaSpace{}%
\AgdaOperator{\AgdaFunction{⊆}}\AgdaSpace{}%
\AgdaFunction{reverse}\AgdaSpace{}%
\AgdaBound{ys}\<%
\\
\>[8]\AgdaFunction{bs⊆reverse-ys}\AgdaSpace{}%
\AgdaBound{x₃}\AgdaSpace{}%
\AgdaSymbol{=}\AgdaSpace{}%
\AgdaFunction{reverse⁺}\AgdaSpace{}%
\AgdaSymbol{(}\AgdaFunction{∈-resp-↭}\AgdaSpace{}%
\AgdaFunction{↭₂}\AgdaSpace{}%
\AgdaSymbol{(}\AgdaFunction{∈-++⁺ʳ}\AgdaSpace{}%
\AgdaBound{as}\AgdaSpace{}%
\AgdaBound{x₃}\AgdaSymbol{))}\<%
\\
\>[8]\AgdaFunction{x∉reverse-ys}\AgdaSpace{}%
\AgdaSymbol{:}\AgdaSpace{}%
\AgdaBound{x}\AgdaSpace{}%
\AgdaOperator{\AgdaPostulate{∉}}\AgdaSpace{}%
\AgdaFunction{reverse}\AgdaSpace{}%
\AgdaBound{ys}\<%
\\
\>[8]\AgdaFunction{x∉reverse-ys}\AgdaSpace{}%
\AgdaBound{x∈reverse-ys}\AgdaSpace{}%
\AgdaSymbol{=}\AgdaSpace{}%
\AgdaFunction{lookup}\AgdaSpace{}%
\AgdaBound{ux}\AgdaSpace{}%
\AgdaSymbol{(}\AgdaFunction{reverse⁻}\AgdaSpace{}%
\AgdaBound{x∈reverse-ys}\AgdaSymbol{)}\AgdaSpace{}%
\AgdaInductiveConstructor{refl}\<%
\\
\>[8]\AgdaFunction{dsj₁}\AgdaSpace{}%
\AgdaSymbol{:}\AgdaSpace{}%
\AgdaFunction{Disjoint}\AgdaSpace{}%
\AgdaSymbol{(}\AgdaPostulate{cmdWriteNames}\AgdaSpace{}%
\AgdaBound{x}\AgdaSpace{}%
\AgdaSymbol{(}\AgdaPostulate{script}\AgdaSpace{}%
\AgdaBound{as}\AgdaSpace{}%
\AgdaSymbol{\AgdaUnderscore{}))}\AgdaSpace{}%
\AgdaSymbol{(}\AgdaPostulate{buildWriteNames}\AgdaSpace{}%
\AgdaSymbol{(}\AgdaPostulate{St.run}\AgdaSpace{}%
\AgdaBound{x}\AgdaSpace{}%
\AgdaSymbol{(}\AgdaPostulate{script}\AgdaSpace{}%
\AgdaBound{as}\AgdaSpace{}%
\AgdaSymbol{\AgdaUnderscore{}))}\AgdaSpace{}%
\AgdaBound{bs}\AgdaSymbol{)}\<%
\\
\>[8]\AgdaFunction{dsj₁}\AgdaSpace{}%
\AgdaSymbol{=}\AgdaSpace{}%
\AgdaPostulate{hf=>disjointWW}\AgdaSpace{}%
\AgdaBound{s}\AgdaSpace{}%
\AgdaBound{x}\AgdaSpace{}%
\AgdaBound{as}\AgdaSpace{}%
\AgdaBound{bs}\AgdaSpace{}%
\AgdaSymbol{(}\AgdaFunction{reverse}\AgdaSpace{}%
\AgdaBound{ys}\AgdaSymbol{)}\AgdaSpace{}%
\AgdaSymbol{\AgdaUnderscore{}}%
\>[56]\AgdaFunction{bs⊆reverse-ys}\AgdaSpace{}%
\AgdaFunction{x∉reverse-ys}\AgdaSpace{}%
\AgdaFunction{hf₀}\<%
\\
\>[8]\AgdaFunction{dsj₂}\AgdaSpace{}%
\AgdaSymbol{:}\AgdaSpace{}%
\AgdaFunction{Disjoint}\AgdaSpace{}%
\AgdaSymbol{(}\AgdaPostulate{cmdReadNames}\AgdaSpace{}%
\AgdaBound{x}\AgdaSpace{}%
\AgdaSymbol{(}\AgdaPostulate{script}\AgdaSpace{}%
\AgdaBound{as}\AgdaSpace{}%
\AgdaSymbol{\AgdaUnderscore{}))}\AgdaSpace{}%
\AgdaSymbol{(}\AgdaPostulate{buildWriteNames}\AgdaSpace{}%
\AgdaSymbol{(}\AgdaPostulate{St.run}\AgdaSpace{}%
\AgdaBound{x}\AgdaSpace{}%
\AgdaSymbol{(}\AgdaPostulate{script}\AgdaSpace{}%
\AgdaBound{as}\AgdaSpace{}%
\AgdaSymbol{\AgdaUnderscore{}))}\AgdaSpace{}%
\AgdaBound{bs}\AgdaSymbol{)}\<%
\\
\>[8]\AgdaFunction{dsj₂}\AgdaSpace{}%
\AgdaSymbol{=}\AgdaSpace{}%
\AgdaPostulate{hf=>disjointRW}\AgdaSpace{}%
\AgdaBound{s}\AgdaSpace{}%
\AgdaBound{x}\AgdaSpace{}%
\AgdaBound{as}\AgdaSpace{}%
\AgdaBound{bs}\AgdaSpace{}%
\AgdaSymbol{(}\AgdaFunction{reverse}\AgdaSpace{}%
\AgdaBound{ys}\AgdaSymbol{)}\AgdaSpace{}%
\AgdaSymbol{\AgdaUnderscore{}}\AgdaSpace{}%
\AgdaFunction{bs⊆reverse-ys}\AgdaSpace{}%
\AgdaFunction{x∉reverse-ys}\AgdaSpace{}%
\AgdaFunction{hf₀}\<%
\\
\>[8]\AgdaFunction{dsj₃}\AgdaSpace{}%
\AgdaSymbol{:}\AgdaSpace{}%
\AgdaFunction{Disjoint}\AgdaSpace{}%
\AgdaSymbol{(}\AgdaPostulate{cmdWriteNames}\AgdaSpace{}%
\AgdaBound{x}\AgdaSpace{}%
\AgdaSymbol{(}\AgdaPostulate{script}\AgdaSpace{}%
\AgdaBound{as}\AgdaSpace{}%
\AgdaSymbol{\AgdaUnderscore{}))}\AgdaSpace{}%
\AgdaSymbol{(}\AgdaPostulate{buildReadNames}\AgdaSpace{}%
\AgdaSymbol{(}\AgdaPostulate{St.run}\AgdaSpace{}%
\AgdaBound{x}\AgdaSpace{}%
\AgdaSymbol{(}\AgdaPostulate{script}\AgdaSpace{}%
\AgdaBound{as}\AgdaSpace{}%
\AgdaSymbol{\AgdaUnderscore{}))}\AgdaSpace{}%
\AgdaBound{bs}\AgdaSymbol{)}\<%
\\
\>[8]\AgdaFunction{dsj₃}\AgdaSpace{}%
\AgdaSymbol{=}\AgdaSpace{}%
\AgdaPostulate{hf=>disjointWR}\AgdaSpace{}%
\AgdaBound{s}\AgdaSpace{}%
\AgdaBound{x}\AgdaSpace{}%
\AgdaBound{as}\AgdaSpace{}%
\AgdaBound{bs}\AgdaSpace{}%
\AgdaSymbol{(}\AgdaFunction{reverse}\AgdaSpace{}%
\AgdaBound{ys}\AgdaSymbol{)}\AgdaSpace{}%
\AgdaSymbol{\AgdaUnderscore{}}\AgdaSpace{}%
\AgdaFunction{bs⊆reverse-ys}\AgdaSpace{}%
\AgdaFunction{x∉reverse-ys}\AgdaSpace{}%
\AgdaFunction{hf₀}\<%
\\
\\[\AgdaEmptyExtraSkip]%
\>[8]\AgdaFunction{dsj₄}\AgdaSpace{}%
\AgdaSymbol{:}\AgdaSpace{}%
\AgdaFunction{Disjoint}\AgdaSpace{}%
\AgdaSymbol{(}\AgdaPostulate{files}\AgdaSpace{}%
\AgdaSymbol{(}\AgdaFunction{script-rec}\AgdaSpace{}%
\AgdaSymbol{(}\AgdaFunction{reverse}\AgdaSpace{}%
\AgdaBound{ys}\AgdaSymbol{)}\AgdaSpace{}%
\AgdaSymbol{\AgdaUnderscore{}}\AgdaSpace{}%
\AgdaBound{ls}\AgdaSymbol{))}\AgdaSpace{}%
\AgdaSymbol{(}\AgdaPostulate{cmdWriteNames}\AgdaSpace{}%
\AgdaBound{x}\AgdaSpace{}%
\AgdaSymbol{(}\AgdaPostulate{script}\AgdaSpace{}%
\AgdaSymbol{(}\AgdaFunction{reverse}\AgdaSpace{}%
\AgdaBound{ys}\AgdaSymbol{)}\AgdaSpace{}%
\AgdaSymbol{\AgdaUnderscore{}))}\<%
\\
\>[8]\AgdaFunction{dsj₄}\AgdaSpace{}%
\AgdaSymbol{=}\AgdaSpace{}%
\AgdaSymbol{\{!!\}}\<%
\\
\>[0]\AgdaSymbol{...}\AgdaSpace{}%
\AgdaSymbol{|}\AgdaSpace{}%
\AgdaBound{a₂}\AgdaSpace{}%
\AgdaSymbol{=}\AgdaSpace{}%
\AgdaFunction{subst}\AgdaSpace{}%
\AgdaSymbol{(λ}\AgdaSpace{}%
\AgdaBound{x₁}\AgdaSpace{}%
\AgdaSymbol{→}\AgdaSpace{}%
\AgdaPostulate{HazardFree}\AgdaSpace{}%
\AgdaSymbol{\AgdaUnderscore{}}\AgdaSpace{}%
\AgdaBound{x₁}\AgdaSpace{}%
\AgdaBound{x₁}\AgdaSpace{}%
\AgdaSymbol{\AgdaUnderscore{})}\AgdaSpace{}%
\AgdaSymbol{(}\AgdaFunction{sym}\AgdaSpace{}%
\AgdaSymbol{(}\AgdaFunction{unfold-reverse}\AgdaSpace{}%
\AgdaBound{x}\AgdaSpace{}%
\AgdaBound{ys}\AgdaSymbol{))}\AgdaSpace{}%
\AgdaBound{a₂}\<%
\\
\\[\AgdaEmptyExtraSkip]%
\>[0]\AgdaFunction{preservesHazards}\AgdaSpace{}%
\AgdaSymbol{:}\AgdaSpace{}%
\AgdaSymbol{∀}\AgdaSpace{}%
\AgdaBound{s}\AgdaSpace{}%
\AgdaBound{ls}\AgdaSpace{}%
\AgdaBound{b₁}\AgdaSpace{}%
\AgdaBound{b₂}\AgdaSpace{}%
\AgdaSymbol{→}\AgdaSpace{}%
\AgdaBound{b₁}\AgdaSpace{}%
\AgdaOperator{\AgdaFunction{⊆}}\AgdaSpace{}%
\AgdaBound{b₂}\AgdaSpace{}%
\AgdaSymbol{→}\AgdaSpace{}%
\AgdaDatatype{Unique}\AgdaSpace{}%
\AgdaBound{b₁}\AgdaSpace{}%
\AgdaSymbol{→}\AgdaSpace{}%
\AgdaDatatype{Unique}\AgdaSpace{}%
\AgdaBound{b₂}\AgdaSpace{}%
\AgdaSymbol{→}\AgdaSpace{}%
\AgdaOperator{\AgdaFunction{¬}}\AgdaSpace{}%
\AgdaPostulate{HazardFree}\AgdaSpace{}%
\AgdaBound{s}\AgdaSpace{}%
\AgdaBound{b₁}\AgdaSpace{}%
\AgdaBound{b₁}\AgdaSpace{}%
\AgdaBound{ls}\AgdaSpace{}%
\AgdaSymbol{→}\AgdaSpace{}%
\AgdaOperator{\AgdaFunction{¬}}\AgdaSpace{}%
\AgdaPostulate{HazardFree}\AgdaSpace{}%
\AgdaBound{s}\AgdaSpace{}%
\AgdaBound{b₂}\AgdaSpace{}%
\AgdaBound{b₁}\AgdaSpace{}%
\AgdaBound{ls}\<%
\\
\>[0]\AgdaFunction{preservesHazards}\AgdaSpace{}%
\AgdaBound{s}\AgdaSpace{}%
\AgdaBound{ls}\AgdaSpace{}%
\AgdaBound{b₁}\AgdaSpace{}%
\AgdaBound{b₂}\AgdaSpace{}%
\AgdaBound{⊆₁}\AgdaSpace{}%
\AgdaBound{ue}\AgdaSpace{}%
\AgdaBound{ue₁}\AgdaSpace{}%
\AgdaBound{¬hf}\AgdaSpace{}%
\AgdaBound{hf}\AgdaSpace{}%
\AgdaSymbol{=}\AgdaSpace{}%
\AgdaBound{¬hf}\AgdaSpace{}%
\AgdaSymbol{\{!!\}}\AgdaSpace{}%
\AgdaComment{-- (preservesHazardFree s ls b₁ b₂ hf)}\<%
\end{code}

\begin{code}[hide]%
\>[0]\AgdaFunction{correct2}\AgdaSpace{}%
\AgdaSymbol{:}\AgdaSpace{}%
\AgdaSymbol{∀}\AgdaSpace{}%
\AgdaBound{br}\AgdaSpace{}%
\AgdaBound{bc}\AgdaSpace{}%
\AgdaBound{s}\AgdaSpace{}%
\AgdaBound{m}\AgdaSpace{}%
\AgdaBound{ls}\AgdaSpace{}%
\AgdaSymbol{→}\AgdaSpace{}%
\AgdaFunction{OKBuild}\AgdaSpace{}%
\AgdaSymbol{(}\AgdaBound{s}\AgdaSpace{}%
\AgdaOperator{\AgdaInductiveConstructor{,}}\AgdaSpace{}%
\AgdaBound{m}\AgdaSymbol{)}\AgdaSpace{}%
\AgdaBound{ls}\AgdaSpace{}%
\AgdaBound{br}\AgdaSpace{}%
\AgdaBound{bc}\AgdaSpace{}%
\AgdaSymbol{→}\AgdaSpace{}%
\AgdaBound{bc}\AgdaSpace{}%
\AgdaOperator{\AgdaDatatype{↭}}\AgdaSpace{}%
\AgdaBound{br}\AgdaSpace{}%
\AgdaSymbol{→}\AgdaSpace{}%
\AgdaOperator{\AgdaFunction{¬}}\AgdaSpace{}%
\AgdaPostulate{HazardFree}\AgdaSpace{}%
\AgdaBound{s}\AgdaSpace{}%
\AgdaBound{bc}\AgdaSpace{}%
\AgdaBound{bc}\AgdaSpace{}%
\AgdaBound{ls}\AgdaSpace{}%
\AgdaOperator{\AgdaDatatype{⊎}}\AgdaSpace{}%
\AgdaFunction{∃[}\AgdaSpace{}%
\AgdaBound{s₁}\AgdaSpace{}%
\AgdaFunction{]}\AgdaSymbol{(}\AgdaFunction{∃[}\AgdaSpace{}%
\AgdaBound{m₁}\AgdaSpace{}%
\AgdaFunction{]}\AgdaSymbol{(}\AgdaFunction{∃[}\AgdaSpace{}%
\AgdaBound{ls₁}\AgdaSpace{}%
\AgdaFunction{]}\AgdaSymbol{(}\AgdaPostulate{rattle}\AgdaSpace{}%
\AgdaBound{br}\AgdaSpace{}%
\AgdaBound{bc}\AgdaSpace{}%
\AgdaSymbol{((}\AgdaBound{s}\AgdaSpace{}%
\AgdaOperator{\AgdaInductiveConstructor{,}}\AgdaSpace{}%
\AgdaBound{m}\AgdaSymbol{)}\AgdaSpace{}%
\AgdaOperator{\AgdaInductiveConstructor{,}}\AgdaSpace{}%
\AgdaBound{ls}\AgdaSymbol{)}\AgdaSpace{}%
\AgdaOperator{\AgdaDatatype{≡}}\AgdaSpace{}%
\AgdaInductiveConstructor{inj₂}\AgdaSpace{}%
\AgdaSymbol{((}\AgdaBound{s₁}\AgdaSpace{}%
\AgdaOperator{\AgdaInductiveConstructor{,}}\AgdaSpace{}%
\AgdaBound{m₁}\AgdaSymbol{)}\AgdaSpace{}%
\AgdaOperator{\AgdaInductiveConstructor{,}}\AgdaSpace{}%
\AgdaBound{ls₁}\AgdaSymbol{)}\AgdaSpace{}%
\AgdaOperator{\AgdaFunction{×}}\AgdaSpace{}%
\AgdaSymbol{∀}\AgdaSpace{}%
\AgdaBound{f₁}\AgdaSpace{}%
\AgdaSymbol{→}\AgdaSpace{}%
\AgdaBound{s₁}\AgdaSpace{}%
\AgdaBound{f₁}\AgdaSpace{}%
\AgdaOperator{\AgdaDatatype{≡}}\AgdaSpace{}%
\AgdaPostulate{script}\AgdaSpace{}%
\AgdaBound{bc}\AgdaSpace{}%
\AgdaBound{s}\AgdaSpace{}%
\AgdaBound{f₁}\AgdaSymbol{)))}\<%
\\
\>[0]\AgdaFunction{correct2}\AgdaSpace{}%
\AgdaBound{b₁}\AgdaSpace{}%
\AgdaBound{b₂}\AgdaSpace{}%
\AgdaBound{s}\AgdaSpace{}%
\AgdaBound{mm}\AgdaSpace{}%
\AgdaBound{ls}\AgdaSpace{}%
\AgdaSymbol{(}\AgdaBound{dsb}\AgdaSpace{}%
\AgdaOperator{\AgdaInductiveConstructor{,}}\AgdaSpace{}%
\AgdaBound{mp}\AgdaSpace{}%
\AgdaOperator{\AgdaInductiveConstructor{,}}\AgdaSpace{}%
\AgdaBound{ue}\AgdaSymbol{)}\AgdaSpace{}%
\AgdaBound{p}\AgdaSpace{}%
\AgdaKeyword{with}\AgdaSpace{}%
\AgdaPostulate{rattle}\AgdaSpace{}%
\AgdaBound{b₁}\AgdaSpace{}%
\AgdaBound{b₂}\AgdaSpace{}%
\AgdaSymbol{((}\AgdaBound{s}\AgdaSpace{}%
\AgdaOperator{\AgdaInductiveConstructor{,}}\AgdaSpace{}%
\AgdaBound{mm}\AgdaSymbol{)}\AgdaSpace{}%
\AgdaOperator{\AgdaInductiveConstructor{,}}\AgdaSpace{}%
\AgdaBound{ls}\AgdaSymbol{)}\AgdaSpace{}%
\AgdaSymbol{|}\AgdaSpace{}%
\AgdaFunction{inspect}\AgdaSpace{}%
\AgdaSymbol{(}\AgdaPostulate{rattle}\AgdaSpace{}%
\AgdaBound{b₁}\AgdaSpace{}%
\AgdaBound{b₂}\AgdaSymbol{)}\AgdaSpace{}%
\AgdaSymbol{((}\AgdaBound{s}\AgdaSpace{}%
\AgdaOperator{\AgdaInductiveConstructor{,}}\AgdaSpace{}%
\AgdaBound{mm}\AgdaSymbol{)}\AgdaSpace{}%
\AgdaOperator{\AgdaInductiveConstructor{,}}\AgdaSpace{}%
\AgdaBound{ls}\AgdaSymbol{)}\<%
\\
\>[0]\AgdaSymbol{...}\AgdaSpace{}%
\AgdaSymbol{|}\AgdaSpace{}%
\AgdaInductiveConstructor{inj₁}\AgdaSpace{}%
\AgdaBound{hz}\AgdaSpace{}%
\AgdaSymbol{|}\AgdaSpace{}%
\AgdaOperator{\AgdaInductiveConstructor{[}}\AgdaSpace{}%
\AgdaBound{≡₁}\AgdaSpace{}%
\AgdaOperator{\AgdaInductiveConstructor{]}}\AgdaSpace{}%
\AgdaSymbol{=}\AgdaSpace{}%
\AgdaSymbol{\{!!\}}\<%
\\
\\[\AgdaEmptyExtraSkip]%
\>[0]\AgdaComment{\{- proof plan:
  we know the reorderd build produced a hazard. this doesnt mean the original build has a hazard.  
  we could prove this case if we know the incorrect build doesn't corrupt the "inputs" of the build. 
  then we can run the correct build sequentially... and we would show that all the outputs of the build are still correct; so
  this says the systems are not exactly equivalent, but are equivalent for the outputs the build was MEANT to write.  
  for us to do both of these cases; we should change this lemma to say it only produces the same outputs for a certain set of files
-\}}\<%
\\
\>[0]\AgdaSymbol{...}\AgdaSpace{}%
\AgdaSymbol{|}\AgdaSpace{}%
\AgdaInductiveConstructor{inj₂}\AgdaSpace{}%
\AgdaSymbol{(}\AgdaBound{st}\AgdaSpace{}%
\AgdaOperator{\AgdaInductiveConstructor{,}}\AgdaSpace{}%
\AgdaSymbol{\AgdaUnderscore{})}\AgdaSpace{}%
\AgdaSymbol{|}\AgdaSpace{}%
\AgdaOperator{\AgdaInductiveConstructor{[}}\AgdaSpace{}%
\AgdaBound{≡₁}\AgdaSpace{}%
\AgdaOperator{\AgdaInductiveConstructor{]}}\AgdaSpace{}%
\AgdaSymbol{=}\AgdaSpace{}%
\AgdaSymbol{\{!!\}}\AgdaSpace{}%
\AgdaComment{-- inj₂ ((st , \AgdaUnderscore{}) , refl , λ f₁ → sym (trans \{!!\} \{!!\}))}\<%
\\
\\[\AgdaEmptyExtraSkip]%
\>[0]\AgdaComment{\{-soundness : execwError b₁ = exec b₁ = script.exec b₁
                            exec b₂ = script.exec b₂ -\}}\<%
\\
\\[\AgdaEmptyExtraSkip]%
\>[0]\AgdaComment{\{- proof plan: 
  we know the reordered build doesn't produce a hazard.  which via some math should mean the original build doesnt produce a hazard. 
  1. get evidnce running the original build returns inj₂ 
  2. apply soundness to both.  
  3. use reordering proof. to show theyre the same for all files??  Of course we can only show that for two builds which are permutations where there are no extra commands.
     if we want to support the extra command thing the reodering proof needs to be expanded.
-\}}\<%
\\
\\[\AgdaEmptyExtraSkip]%
\\[\AgdaEmptyExtraSkip]%
\>[0]\AgdaComment{\{- maybe in the paper we could just prove this for the case where speculation doesnt cause hazards. and explain future work proving the other case?
  so prove a subset of the correct2 lemma?
-\}}\<%
\end{code}

\newcommand{\correctS}{%
\begin{code}%
\>[0]\AgdaFunction{correct-speculation}\AgdaSpace{}%
\AgdaSymbol{:}\AgdaSpace{}%
\AgdaSymbol{∀}\AgdaSpace{}%
\AgdaBound{s}\AgdaSpace{}%
\AgdaBound{br}\AgdaSpace{}%
\AgdaBound{bc}\AgdaSpace{}%
\AgdaSymbol{→}\AgdaSpace{}%
\AgdaPostulate{PreCond}\AgdaSpace{}%
\AgdaBound{s}\AgdaSpace{}%
\AgdaBound{br}\AgdaSpace{}%
\AgdaBound{bc}\AgdaSpace{}%
\AgdaSymbol{→}\AgdaSpace{}%
\AgdaOperator{\AgdaFunction{¬}}\AgdaSpace{}%
\AgdaPostulate{HazardFree}\AgdaSpace{}%
\AgdaBound{s}\AgdaSpace{}%
\AgdaBound{bc}\AgdaSpace{}%
\AgdaBound{bc}\AgdaSpace{}%
\AgdaInductiveConstructor{[]}\AgdaSpace{}%
\AgdaOperator{\AgdaDatatype{⊎}}\AgdaSpace{}%
\AgdaFunction{≡toScript}\AgdaSpace{}%
\AgdaBound{s}\AgdaSpace{}%
\AgdaBound{br}\AgdaSpace{}%
\AgdaBound{bc}\<%
\end{code}}
\begin{code}[hide]%
\>[0]\AgdaFunction{correct-speculation}\AgdaSpace{}%
\AgdaBound{s}\AgdaSpace{}%
\AgdaBound{br}\AgdaSpace{}%
\AgdaBound{bc}\AgdaSpace{}%
\AgdaBound{pc}\AgdaSpace{}%
\AgdaSymbol{=}\AgdaSpace{}%
\AgdaSymbol{\{!!\}}\<%
\end{code}

\begin{code}[hide]%
\>[0]\AgdaFunction{semi-correct2}\AgdaSpace{}%
\AgdaSymbol{:}\AgdaSpace{}%
\AgdaSymbol{∀}\AgdaSpace{}%
\AgdaBound{s}\AgdaSpace{}%
\AgdaBound{m}\AgdaSpace{}%
\AgdaBound{ls}\AgdaSpace{}%
\AgdaBound{b₁}\AgdaSpace{}%
\AgdaBound{b₂}\AgdaSpace{}%
\AgdaSymbol{→}\AgdaSpace{}%
\AgdaPostulate{DisjointBuild}\AgdaSpace{}%
\AgdaBound{s}\AgdaSpace{}%
\AgdaBound{b₁}\AgdaSpace{}%
\AgdaSymbol{→}\AgdaSpace{}%
\AgdaDatatype{MemoryProperty}\AgdaSpace{}%
\AgdaBound{m}\AgdaSpace{}%
\AgdaSymbol{→}\AgdaSpace{}%
\AgdaPostulate{UniqueEvidence}\AgdaSpace{}%
\AgdaBound{b₁}\AgdaSpace{}%
\AgdaBound{b₂}\AgdaSpace{}%
\AgdaSymbol{(}\AgdaFunction{map}\AgdaSpace{}%
\AgdaField{proj₁}\AgdaSpace{}%
\AgdaBound{ls}\AgdaSymbol{)}\AgdaSpace{}%
\AgdaSymbol{→}\AgdaSpace{}%
\AgdaBound{b₂}\AgdaSpace{}%
\AgdaOperator{\AgdaDatatype{↭}}\AgdaSpace{}%
\AgdaBound{b₁}\AgdaSpace{}%
\AgdaSymbol{→}\AgdaSpace{}%
\AgdaOperator{\AgdaFunction{¬}}\AgdaSpace{}%
\AgdaPostulate{HazardFree}\AgdaSpace{}%
\AgdaBound{s}\AgdaSpace{}%
\AgdaBound{b₁}\AgdaSpace{}%
\AgdaBound{b₂}\AgdaSpace{}%
\AgdaBound{ls}\AgdaSpace{}%
\AgdaOperator{\AgdaDatatype{⊎}}\AgdaSpace{}%
\AgdaOperator{\AgdaFunction{¬}}\AgdaSpace{}%
\AgdaPostulate{HazardFree}\AgdaSpace{}%
\AgdaBound{s}\AgdaSpace{}%
\AgdaBound{b₂}\AgdaSpace{}%
\AgdaBound{b₂}\AgdaSpace{}%
\AgdaBound{ls}\AgdaSpace{}%
\AgdaOperator{\AgdaDatatype{⊎}}\AgdaSpace{}%
\AgdaFunction{∃[}\AgdaSpace{}%
\AgdaBound{s₁}\AgdaSpace{}%
\AgdaFunction{]}\AgdaSymbol{(}\AgdaFunction{∃[}\AgdaSpace{}%
\AgdaBound{m₁}\AgdaSpace{}%
\AgdaFunction{]}\AgdaSymbol{(}\AgdaFunction{∃[}\AgdaSpace{}%
\AgdaBound{ls₁}\AgdaSpace{}%
\AgdaFunction{]}\AgdaSymbol{(}\AgdaPostulate{rattle}\AgdaSpace{}%
\AgdaBound{b₁}\AgdaSpace{}%
\AgdaBound{b₂}\AgdaSpace{}%
\AgdaSymbol{((}\AgdaBound{s}\AgdaSpace{}%
\AgdaOperator{\AgdaInductiveConstructor{,}}\AgdaSpace{}%
\AgdaBound{m}\AgdaSymbol{)}\AgdaSpace{}%
\AgdaOperator{\AgdaInductiveConstructor{,}}\AgdaSpace{}%
\AgdaBound{ls}\AgdaSymbol{)}\AgdaSpace{}%
\AgdaOperator{\AgdaDatatype{≡}}\AgdaSpace{}%
\AgdaInductiveConstructor{inj₂}\AgdaSpace{}%
\AgdaSymbol{((}\AgdaBound{s₁}\AgdaSpace{}%
\AgdaOperator{\AgdaInductiveConstructor{,}}\AgdaSpace{}%
\AgdaBound{m₁}\AgdaSymbol{)}\AgdaSpace{}%
\AgdaOperator{\AgdaInductiveConstructor{,}}\AgdaSpace{}%
\AgdaBound{ls₁}\AgdaSymbol{)}\AgdaSpace{}%
\AgdaOperator{\AgdaFunction{×}}\AgdaSpace{}%
\AgdaSymbol{∀}\AgdaSpace{}%
\AgdaBound{f₁}\AgdaSpace{}%
\AgdaSymbol{→}\AgdaSpace{}%
\AgdaBound{s₁}\AgdaSpace{}%
\AgdaBound{f₁}\AgdaSpace{}%
\AgdaOperator{\AgdaDatatype{≡}}\AgdaSpace{}%
\AgdaPostulate{script}\AgdaSpace{}%
\AgdaBound{b₂}\AgdaSpace{}%
\AgdaBound{s}\AgdaSpace{}%
\AgdaBound{f₁}\AgdaSymbol{)))}\<%
\\
\>[0][@{}l@{\AgdaIndent{0}}]%
\>[13]\AgdaComment{-- ≡toScript (s , m) ls b₁ b₂ b₂}\<%
\\
\>[0]\AgdaFunction{semi-correct2}\AgdaSpace{}%
\AgdaBound{s}\AgdaSpace{}%
\AgdaBound{mm}\AgdaSpace{}%
\AgdaBound{ls}\AgdaSpace{}%
\AgdaBound{b₁}\AgdaSpace{}%
\AgdaBound{b₂}\AgdaSpace{}%
\AgdaBound{dsb}\AgdaSpace{}%
\AgdaBound{mp}\AgdaSpace{}%
\AgdaBound{ue}\AgdaSpace{}%
\AgdaBound{b₂↭b₁}\AgdaSpace{}%
\AgdaKeyword{with}\AgdaSpace{}%
\AgdaPostulate{hazardfree?}\AgdaSpace{}%
\AgdaBound{s}\AgdaSpace{}%
\AgdaBound{b₁}\AgdaSpace{}%
\AgdaBound{b₂}\AgdaSpace{}%
\AgdaBound{ls}\<%
\\
\>[0]\AgdaSymbol{...}\AgdaSpace{}%
\AgdaSymbol{|}\AgdaSpace{}%
\AgdaInductiveConstructor{no}\AgdaSpace{}%
\AgdaBound{hz}\AgdaSpace{}%
\AgdaSymbol{=}\AgdaSpace{}%
\AgdaInductiveConstructor{inj₁}\AgdaSpace{}%
\AgdaBound{hz}\<%
\\
\>[0]\AgdaSymbol{...}\AgdaSpace{}%
\AgdaSymbol{|}\AgdaSpace{}%
\AgdaInductiveConstructor{yes}\AgdaSpace{}%
\AgdaBound{hf}\AgdaSpace{}%
\AgdaKeyword{with}\AgdaSpace{}%
\AgdaPostulate{hazardfree?}\AgdaSpace{}%
\AgdaBound{s}\AgdaSpace{}%
\AgdaBound{b₂}\AgdaSpace{}%
\AgdaBound{b₂}\AgdaSpace{}%
\AgdaBound{ls}\<%
\\
\>[0]\AgdaSymbol{...}\AgdaSpace{}%
\AgdaSymbol{|}\AgdaSpace{}%
\AgdaInductiveConstructor{no}\AgdaSpace{}%
\AgdaBound{hz}\AgdaSpace{}%
\AgdaSymbol{=}\AgdaSpace{}%
\AgdaInductiveConstructor{inj₂}\AgdaSpace{}%
\AgdaSymbol{(}\AgdaInductiveConstructor{inj₁}\AgdaSpace{}%
\AgdaBound{hz}\AgdaSymbol{)}\<%
\\
\>[0]\AgdaSymbol{...}\AgdaSpace{}%
\AgdaSymbol{|}\AgdaSpace{}%
\AgdaInductiveConstructor{yes}\AgdaSpace{}%
\AgdaBound{hf₁}\AgdaSpace{}%
\AgdaKeyword{with}\AgdaSpace{}%
\AgdaFunction{completeness-inner}\AgdaSpace{}%
\AgdaSymbol{(}\AgdaBound{s}\AgdaSpace{}%
\AgdaOperator{\AgdaInductiveConstructor{,}}\AgdaSpace{}%
\AgdaBound{mm}\AgdaSymbol{)}\AgdaSpace{}%
\AgdaBound{ls}\AgdaSpace{}%
\AgdaBound{b₁}\AgdaSpace{}%
\AgdaBound{b₂}\AgdaSpace{}%
\AgdaSymbol{(}\AgdaBound{dsb}\AgdaSpace{}%
\AgdaOperator{\AgdaInductiveConstructor{,}}\AgdaSpace{}%
\AgdaBound{mp}\AgdaSpace{}%
\AgdaOperator{\AgdaInductiveConstructor{,}}\AgdaSpace{}%
\AgdaBound{ue}\AgdaSymbol{)}\AgdaSpace{}%
\AgdaBound{hf}\<%
\\
\>[0]\AgdaSymbol{...}\AgdaSpace{}%
\AgdaSymbol{|}\AgdaSpace{}%
\AgdaSymbol{(}\AgdaBound{s₁}\AgdaSpace{}%
\AgdaOperator{\AgdaInductiveConstructor{,}}\AgdaSpace{}%
\AgdaBound{mm₁}\AgdaSymbol{)}\AgdaSpace{}%
\AgdaOperator{\AgdaInductiveConstructor{,}}\AgdaSpace{}%
\AgdaBound{ls₁}\AgdaSpace{}%
\AgdaOperator{\AgdaInductiveConstructor{,}}\AgdaSpace{}%
\AgdaBound{≡₁}\AgdaSpace{}%
\AgdaSymbol{=}\AgdaSpace{}%
\AgdaInductiveConstructor{inj₂}\AgdaSpace{}%
\AgdaSymbol{(}\AgdaInductiveConstructor{inj₂}\AgdaSpace{}%
\AgdaSymbol{(}\AgdaBound{s₁}\AgdaSpace{}%
\AgdaOperator{\AgdaInductiveConstructor{,}}\AgdaSpace{}%
\AgdaBound{mm₁}\AgdaSpace{}%
\AgdaOperator{\AgdaInductiveConstructor{,}}\AgdaSpace{}%
\AgdaBound{ls₁}\AgdaSpace{}%
\AgdaOperator{\AgdaInductiveConstructor{,}}\AgdaSpace{}%
\AgdaBound{≡₁}\AgdaSpace{}%
\AgdaOperator{\AgdaInductiveConstructor{,}}\AgdaSpace{}%
\AgdaSymbol{λ}\AgdaSpace{}%
\AgdaBound{f₁}\AgdaSpace{}%
\AgdaSymbol{→}\AgdaSpace{}%
\AgdaSymbol{\{!!\}))}\<%
\\
\>[0]\AgdaComment{-- sym (trans (reordered b₂ b₁ ls b₂↭b₁ ue \{!!\} hf f₁) (trans (script≡rattle-inner mm b₁ (λ f₂ → refl) dsb mp f₁) (cong-app (cong proj₁ (soundness-inner (s , mm) ls b₁ b₂ ≡₁)) f₁)))))}\<%
\\
\>[0]\AgdaComment{-- }\<%
\end{code}

\newcommand{\correctP}{%
\begin{code}%
\>[0]\AgdaFunction{semi-correct}\AgdaSpace{}%
\AgdaSymbol{:}\AgdaSpace{}%
\AgdaSymbol{∀}\AgdaSpace{}%
\AgdaBound{s}\AgdaSpace{}%
\AgdaBound{br}\AgdaSpace{}%
\AgdaBound{bs}\AgdaSpace{}%
\AgdaSymbol{→}\AgdaSpace{}%
\AgdaPostulate{PreCond}\AgdaSpace{}%
\AgdaBound{s}\AgdaSpace{}%
\AgdaBound{br}\AgdaSpace{}%
\AgdaBound{bs}\AgdaSpace{}%
\AgdaSymbol{→}\AgdaSpace{}%
\AgdaOperator{\AgdaFunction{¬}}\AgdaSpace{}%
\AgdaPostulate{HazardFree}\AgdaSpace{}%
\AgdaBound{s}\AgdaSpace{}%
\AgdaBound{br}\AgdaSpace{}%
\AgdaBound{bs}\AgdaSpace{}%
\AgdaInductiveConstructor{[]}\AgdaSpace{}%
\AgdaOperator{\AgdaDatatype{⊎}}\AgdaSpace{}%
\AgdaOperator{\AgdaFunction{¬}}\AgdaSpace{}%
\AgdaPostulate{HazardFree}\AgdaSpace{}%
\AgdaBound{s}\AgdaSpace{}%
\AgdaBound{bs}\AgdaSpace{}%
\AgdaBound{bs}\AgdaSpace{}%
\AgdaInductiveConstructor{[]}\AgdaSpace{}%
\AgdaOperator{\AgdaDatatype{⊎}}\AgdaSpace{}%
\AgdaFunction{≡toScript}\AgdaSpace{}%
\AgdaBound{s}\AgdaSpace{}%
\AgdaBound{br}\AgdaSpace{}%
\AgdaBound{bs}\<%
\\
\>[0]\AgdaFunction{semi-correct}\AgdaSpace{}%
\AgdaBound{s}\AgdaSpace{}%
\AgdaBound{br}\AgdaSpace{}%
\AgdaBound{bs}\AgdaSpace{}%
\AgdaBound{pc}\AgdaSpace{}%
\AgdaKeyword{with}\AgdaSpace{}%
\AgdaPostulate{hazardfree?}\AgdaSpace{}%
\AgdaBound{s}\AgdaSpace{}%
\AgdaBound{br}\AgdaSpace{}%
\AgdaBound{bs}\AgdaSpace{}%
\AgdaInductiveConstructor{[]}\<%
\\
\>[0]\AgdaSymbol{...}\AgdaSpace{}%
\AgdaSymbol{|}\AgdaSpace{}%
\AgdaInductiveConstructor{no}\AgdaSpace{}%
\AgdaBound{hz}\AgdaSpace{}%
\AgdaSymbol{=}\AgdaSpace{}%
\AgdaInductiveConstructor{inj₁}\AgdaSpace{}%
\AgdaBound{hz}\<%
\\
\>[0]\AgdaSymbol{...}\AgdaSpace{}%
\AgdaSymbol{|}\AgdaSpace{}%
\AgdaInductiveConstructor{yes}\AgdaSpace{}%
\AgdaBound{hf₁}\AgdaSpace{}%
\AgdaKeyword{with}\AgdaSpace{}%
\AgdaFunction{completeness}\AgdaSpace{}%
\AgdaBound{s}\AgdaSpace{}%
\AgdaBound{br}\AgdaSpace{}%
\AgdaBound{bs}\AgdaSpace{}%
\AgdaBound{pc}\AgdaSpace{}%
\AgdaBound{hf₁}\<%
\\
\>[0]\AgdaSymbol{...}\AgdaSpace{}%
\AgdaSymbol{|}\AgdaSpace{}%
\AgdaSymbol{(}\AgdaBound{s₁}\AgdaSpace{}%
\AgdaOperator{\AgdaInductiveConstructor{,}}\AgdaSpace{}%
\AgdaBound{m₁}\AgdaSymbol{)}\AgdaSpace{}%
\AgdaOperator{\AgdaInductiveConstructor{,}}\AgdaSpace{}%
\AgdaBound{ls}\AgdaSpace{}%
\AgdaOperator{\AgdaInductiveConstructor{,}}\AgdaSpace{}%
\AgdaBound{≡₁}\AgdaSpace{}%
\AgdaSymbol{=}\AgdaSpace{}%
\AgdaInductiveConstructor{inj₂}\AgdaSpace{}%
\AgdaSymbol{(}\AgdaInductiveConstructor{inj₂}\AgdaSpace{}%
\AgdaSymbol{(}\AgdaBound{s₁}\AgdaSpace{}%
\AgdaOperator{\AgdaInductiveConstructor{,}}\AgdaSpace{}%
\AgdaBound{m₁}\AgdaSpace{}%
\AgdaOperator{\AgdaInductiveConstructor{,}}\AgdaSpace{}%
\AgdaBound{ls}\AgdaSpace{}%
\AgdaOperator{\AgdaInductiveConstructor{,}}\AgdaSpace{}%
\AgdaBound{≡₁}\AgdaSpace{}%
\AgdaOperator{\AgdaInductiveConstructor{,}}\AgdaSpace{}%
\AgdaFunction{∀≡}\AgdaSymbol{))}\<%
\\
\>[0][@{}l@{\AgdaIndent{0}}]%
\>[2]\AgdaKeyword{where}%
\>[4830I]\AgdaFunction{∀≡}\AgdaSpace{}%
\AgdaSymbol{:}\AgdaSpace{}%
\AgdaSymbol{∀}\AgdaSpace{}%
\AgdaBound{f₁}\AgdaSpace{}%
\AgdaSymbol{→}\AgdaSpace{}%
\AgdaBound{s₁}\AgdaSpace{}%
\AgdaBound{f₁}\AgdaSpace{}%
\AgdaOperator{\AgdaDatatype{≡}}\AgdaSpace{}%
\AgdaPostulate{script}\AgdaSpace{}%
\AgdaBound{bs}\AgdaSpace{}%
\AgdaBound{s}\AgdaSpace{}%
\AgdaBound{f₁}\<%
\\
\>[.][@{}l@{}]\<[4830I]%
\>[8]\AgdaFunction{∀≡}\AgdaSpace{}%
\AgdaBound{f₁}\AgdaSpace{}%
\AgdaSymbol{=}\AgdaSpace{}%
\AgdaFunction{sym}\AgdaSpace{}%
\AgdaSymbol{(}\AgdaFunction{trans}%
\>[4846I]\AgdaSymbol{(}\AgdaPostulate{reordered≡}\AgdaSpace{}%
\AgdaBound{s}\AgdaSpace{}%
\AgdaBound{br}\AgdaSpace{}%
\AgdaBound{bs}\AgdaSpace{}%
\AgdaBound{pc}\AgdaSpace{}%
\AgdaBound{hf₁}\AgdaSpace{}%
\AgdaBound{f₁}\AgdaSymbol{)}\<%
\\
\>[.][@{}l@{}]\<[4846I]%
\>[27]\AgdaSymbol{(}\AgdaFunction{soundness}\AgdaSpace{}%
\AgdaBound{s}\AgdaSpace{}%
\AgdaBound{br}\AgdaSpace{}%
\AgdaBound{bs}\AgdaSpace{}%
\AgdaSymbol{(}\AgdaField{proj₁}\AgdaSpace{}%
\AgdaBound{pc}\AgdaSymbol{)}\AgdaSpace{}%
\AgdaBound{≡₁}\AgdaSpace{}%
\AgdaBound{f₁}\AgdaSymbol{))}\<%
\end{code}}

\begin{code}[hide]%
\>[0]\AgdaKeyword{open}\AgdaSpace{}%
\AgdaKeyword{import}\AgdaSpace{}%
\AgdaModule{Functional.State}\AgdaSpace{}%
\AgdaKeyword{using}\AgdaSpace{}%
\AgdaSymbol{(}\AgdaFunction{State}\AgdaSpace{}%
\AgdaSymbol{;}\AgdaSpace{}%
\AgdaFunction{Oracle}\AgdaSpace{}%
\AgdaSymbol{;}\AgdaSpace{}%
\AgdaFunction{Cmd}\AgdaSpace{}%
\AgdaSymbol{;}\AgdaSpace{}%
\AgdaFunction{FileSystem}\AgdaSpace{}%
\AgdaSymbol{;}\AgdaSpace{}%
\AgdaFunction{Memory}\AgdaSpace{}%
\AgdaSymbol{;}\AgdaSpace{}%
\AgdaFunction{extend}\AgdaSymbol{)}\AgdaSpace{}%
\AgdaKeyword{renaming}\AgdaSpace{}%
\AgdaSymbol{(}\AgdaFunction{save}\AgdaSpace{}%
\AgdaSymbol{to}\AgdaSpace{}%
\AgdaFunction{store}\AgdaSymbol{)}\<%
\\
\\[\AgdaEmptyExtraSkip]%
\>[0]\AgdaKeyword{module}\AgdaSpace{}%
\AgdaModule{Functional.Forward.Exec}\AgdaSpace{}%
\AgdaSymbol{(}\AgdaBound{oracle}\AgdaSpace{}%
\AgdaSymbol{:}\AgdaSpace{}%
\AgdaFunction{Oracle}\AgdaSymbol{)}\AgdaSpace{}%
\AgdaKeyword{where}\<%
\\
\\[\AgdaEmptyExtraSkip]%
\>[0]\AgdaKeyword{open}\AgdaSpace{}%
\AgdaKeyword{import}\AgdaSpace{}%
\AgdaModule{Agda.Builtin.Equality}\<%
\\
\>[0]\AgdaKeyword{open}\AgdaSpace{}%
\AgdaKeyword{import}\AgdaSpace{}%
\AgdaModule{Functional.State.Helpers}\AgdaSpace{}%
\AgdaSymbol{(}\AgdaBound{oracle}\AgdaSymbol{)}\AgdaSpace{}%
\AgdaSymbol{as}\AgdaSpace{}%
\AgdaModule{St}\AgdaSpace{}%
\AgdaKeyword{using}\AgdaSpace{}%
\AgdaSymbol{(}\AgdaFunction{cmdReadNames}\AgdaSymbol{)}\<%
\\
\>[0]\AgdaKeyword{open}\AgdaSpace{}%
\AgdaKeyword{import}\AgdaSpace{}%
\AgdaModule{Data.Bool}\AgdaSpace{}%
\AgdaKeyword{using}\AgdaSpace{}%
\AgdaSymbol{(}\AgdaDatatype{Bool}\AgdaSpace{}%
\AgdaSymbol{;}\AgdaSpace{}%
\AgdaOperator{\AgdaFunction{if\AgdaUnderscore{}then\AgdaUnderscore{}else\AgdaUnderscore{}}}\AgdaSymbol{)}\<%
\\
\>[0]\AgdaKeyword{open}\AgdaSpace{}%
\AgdaKeyword{import}\AgdaSpace{}%
\AgdaModule{Data.List}\AgdaSpace{}%
\AgdaKeyword{using}\AgdaSpace{}%
\AgdaSymbol{(}\AgdaDatatype{List}\AgdaSpace{}%
\AgdaSymbol{;}\AgdaSpace{}%
\AgdaInductiveConstructor{[]}\AgdaSpace{}%
\AgdaSymbol{;}\AgdaSpace{}%
\AgdaOperator{\AgdaInductiveConstructor{\AgdaUnderscore{}∷\AgdaUnderscore{}}}\AgdaSpace{}%
\AgdaSymbol{;}\AgdaSpace{}%
\AgdaFunction{map}\AgdaSpace{}%
\AgdaSymbol{;}\AgdaSpace{}%
\AgdaFunction{filter}\AgdaSpace{}%
\AgdaSymbol{;}\AgdaSpace{}%
\AgdaFunction{foldr}\AgdaSpace{}%
\AgdaSymbol{;}\AgdaSpace{}%
\AgdaOperator{\AgdaFunction{\AgdaUnderscore{}++\AgdaUnderscore{}}}\AgdaSymbol{)}\<%
\\
\>[0]\AgdaKeyword{open}\AgdaSpace{}%
\AgdaKeyword{import}\AgdaSpace{}%
\AgdaModule{Data.String.Properties}\AgdaSpace{}%
\AgdaKeyword{using}\AgdaSpace{}%
\AgdaSymbol{(}\AgdaOperator{\AgdaFunction{\AgdaUnderscore{}≟\AgdaUnderscore{}}}\AgdaSpace{}%
\AgdaSymbol{;}\AgdaSpace{}%
\AgdaOperator{\AgdaFunction{\AgdaUnderscore{}==\AgdaUnderscore{}}}\AgdaSymbol{)}\<%
\\
\>[0]\AgdaKeyword{open}\AgdaSpace{}%
\AgdaKeyword{import}\AgdaSpace{}%
\AgdaModule{Data.List.Relation.Binary.Equality.DecPropositional}\AgdaSpace{}%
\AgdaOperator{\AgdaFunction{\AgdaUnderscore{}≟\AgdaUnderscore{}}}\AgdaSpace{}%
\AgdaKeyword{using}\AgdaSpace{}%
\AgdaSymbol{(}\AgdaUnderscore{}≡?\AgdaUnderscore{}\AgdaSymbol{)}\<%
\\
\>[0]\AgdaKeyword{open}\AgdaSpace{}%
\AgdaKeyword{import}\AgdaSpace{}%
\AgdaModule{Data.Maybe}\AgdaSpace{}%
\AgdaSymbol{as}\AgdaSpace{}%
\AgdaModule{Maybe}\AgdaSpace{}%
\AgdaKeyword{using}\AgdaSpace{}%
\AgdaSymbol{(}\AgdaDatatype{Maybe}\AgdaSpace{}%
\AgdaSymbol{;}\AgdaSpace{}%
\AgdaInductiveConstructor{just}\AgdaSpace{}%
\AgdaSymbol{;}\AgdaSpace{}%
\AgdaFunction{decToMaybe}\AgdaSpace{}%
\AgdaSymbol{;}\AgdaSpace{}%
\AgdaFunction{is-nothing}\AgdaSpace{}%
\AgdaSymbol{;}\AgdaSpace{}%
\AgdaFunction{maybe}\AgdaSpace{}%
\AgdaSymbol{;}\AgdaSpace{}%
\AgdaInductiveConstructor{nothing}\AgdaSymbol{)}\<%
\\
\>[0]\AgdaKeyword{open}\AgdaSpace{}%
\AgdaKeyword{import}\AgdaSpace{}%
\AgdaModule{Data.Maybe.Properties}\AgdaSpace{}%
\AgdaKeyword{using}\AgdaSpace{}%
\AgdaSymbol{(}\AgdaFunction{≡-dec}\AgdaSymbol{)}\<%
\\
\>[0]\AgdaKeyword{open}\AgdaSpace{}%
\AgdaKeyword{import}\AgdaSpace{}%
\AgdaModule{Data.Product}\AgdaSpace{}%
\AgdaKeyword{using}\AgdaSpace{}%
\AgdaSymbol{(}\AgdaField{proj₁}\AgdaSpace{}%
\AgdaSymbol{;}\AgdaSpace{}%
\AgdaField{proj₂}\AgdaSpace{}%
\AgdaSymbol{;}\AgdaSpace{}%
\AgdaOperator{\AgdaInductiveConstructor{\AgdaUnderscore{},\AgdaUnderscore{}}}\AgdaSpace{}%
\AgdaSymbol{;}\AgdaSpace{}%
\AgdaOperator{\AgdaFunction{\AgdaUnderscore{}×\AgdaUnderscore{}}}\AgdaSpace{}%
\AgdaSymbol{;}\AgdaSpace{}%
\AgdaFunction{∃-syntax}\AgdaSymbol{)}\<%
\\
\>[0]\AgdaKeyword{open}\AgdaSpace{}%
\AgdaKeyword{import}\AgdaSpace{}%
\AgdaModule{Function.Base}\AgdaSpace{}%
\AgdaKeyword{using}\AgdaSpace{}%
\AgdaSymbol{(}\AgdaOperator{\AgdaFunction{\AgdaUnderscore{}∘\AgdaUnderscore{}}}\AgdaSymbol{)}\<%
\\
\>[0]\AgdaKeyword{open}\AgdaSpace{}%
\AgdaKeyword{import}\AgdaSpace{}%
\AgdaModule{Functional.File}\AgdaSpace{}%
\AgdaKeyword{using}\AgdaSpace{}%
\AgdaSymbol{(}\AgdaFunction{FileName}\AgdaSpace{}%
\AgdaSymbol{;}\AgdaSpace{}%
\AgdaFunction{FileContent}\AgdaSpace{}%
\AgdaSymbol{;}\AgdaSpace{}%
\AgdaFunction{Files}\AgdaSymbol{)}\<%
\\
\>[0]\AgdaKeyword{open}\AgdaSpace{}%
\AgdaKeyword{import}\AgdaSpace{}%
\AgdaModule{Functional.Build}\AgdaSpace{}%
\AgdaSymbol{(}\AgdaBound{oracle}\AgdaSymbol{)}\AgdaSpace{}%
\AgdaKeyword{using}\AgdaSpace{}%
\AgdaSymbol{(}Build\AgdaSymbol{)}\<%
\\
\>[0]\AgdaKeyword{open}\AgdaSpace{}%
\AgdaKeyword{import}\AgdaSpace{}%
\AgdaModule{Relation.Nullary.Decidable.Core}\AgdaSpace{}%
\AgdaKeyword{using}\AgdaSpace{}%
\AgdaSymbol{(}\AgdaFunction{isNo}\AgdaSymbol{)}\<%
\\
\>[0]\AgdaKeyword{open}\AgdaSpace{}%
\AgdaKeyword{import}\AgdaSpace{}%
\AgdaModule{Relation.Nullary.Negation}\AgdaSpace{}%
\AgdaKeyword{using}\AgdaSpace{}%
\AgdaSymbol{(}\AgdaFunction{¬?}\AgdaSymbol{)}\<%
\\
\>[0]\AgdaKeyword{open}\AgdaSpace{}%
\AgdaKeyword{import}\AgdaSpace{}%
\AgdaModule{Data.List.Relation.Unary.Any}\AgdaSpace{}%
\AgdaKeyword{using}\AgdaSpace{}%
\AgdaSymbol{(}\AgdaFunction{tail}\AgdaSpace{}%
\AgdaSymbol{;}\AgdaSpace{}%
\AgdaInductiveConstructor{here}\AgdaSpace{}%
\AgdaSymbol{;}\AgdaSpace{}%
\AgdaInductiveConstructor{there}\AgdaSymbol{)}\<%
\\
\>[0]\AgdaKeyword{open}\AgdaSpace{}%
\AgdaKeyword{import}\AgdaSpace{}%
\AgdaModule{Data.List.Relation.Unary.All}\AgdaSpace{}%
\AgdaSymbol{as}\AgdaSpace{}%
\AgdaModule{All}\AgdaSpace{}%
\AgdaKeyword{using}\AgdaSpace{}%
\AgdaSymbol{(}\AgdaDatatype{All}\AgdaSpace{}%
\AgdaSymbol{;}\AgdaSpace{}%
\AgdaFunction{all?}\AgdaSpace{}%
\AgdaSymbol{;}\AgdaSpace{}%
\AgdaFunction{lookup}\AgdaSymbol{)}\<%
\\
\>[0]\AgdaKeyword{open}\AgdaSpace{}%
\AgdaKeyword{import}\AgdaSpace{}%
\AgdaModule{Data.List.Relation.Binary.Disjoint.Propositional}\AgdaSpace{}%
\AgdaKeyword{using}\AgdaSpace{}%
\AgdaSymbol{(}\AgdaFunction{Disjoint}\AgdaSymbol{)}\<%
\\
\>[0]\AgdaKeyword{open}\AgdaSpace{}%
\AgdaKeyword{import}\AgdaSpace{}%
\AgdaModule{Relation.Binary.PropositionalEquality}\AgdaSpace{}%
\AgdaKeyword{using}\AgdaSpace{}%
\AgdaSymbol{(}\AgdaFunction{decSetoid}\AgdaSpace{}%
\AgdaSymbol{;}\AgdaSpace{}%
\AgdaFunction{trans}\AgdaSpace{}%
\AgdaSymbol{;}\AgdaSpace{}%
\AgdaFunction{sym}\AgdaSpace{}%
\AgdaSymbol{;}\AgdaSpace{}%
\AgdaFunction{subst}\AgdaSpace{}%
\AgdaSymbol{;}\AgdaSpace{}%
\AgdaFunction{cong}\AgdaSymbol{)}\<%
\\
\>[0]\AgdaKeyword{open}\AgdaSpace{}%
\AgdaKeyword{import}\AgdaSpace{}%
\AgdaModule{Data.List.Membership.DecSetoid}\AgdaSpace{}%
\AgdaSymbol{(}\AgdaFunction{decSetoid}\AgdaSpace{}%
\AgdaOperator{\AgdaFunction{\AgdaUnderscore{}≟\AgdaUnderscore{}}}\AgdaSymbol{)}\AgdaSpace{}%
\AgdaKeyword{using}\AgdaSpace{}%
\AgdaSymbol{(}\AgdaUnderscore{}∈\AgdaUnderscore{} \AgdaSymbol{;} \AgdaUnderscore{}∈?\AgdaUnderscore{} \AgdaSymbol{;} \AgdaUnderscore{}∉\AgdaUnderscore{}\AgdaSymbol{)}\<%
\\
\>[0]\AgdaKeyword{open}\AgdaSpace{}%
\AgdaKeyword{import}\AgdaSpace{}%
\AgdaModule{Data.List.Membership.Propositional.Properties}\AgdaSpace{}%
\AgdaKeyword{using}\AgdaSpace{}%
\AgdaSymbol{(}\AgdaFunction{∈-++⁺ˡ}\AgdaSpace{}%
\AgdaSymbol{;}\AgdaSpace{}%
\AgdaFunction{∈-++⁺ʳ}\AgdaSymbol{)}\<%
\\
\>[0]\AgdaKeyword{open}\AgdaSpace{}%
\AgdaKeyword{import}\AgdaSpace{}%
\AgdaModule{Relation.Nullary}\AgdaSpace{}%
\AgdaKeyword{using}\AgdaSpace{}%
\AgdaSymbol{(}\AgdaInductiveConstructor{yes}\AgdaSpace{}%
\AgdaSymbol{;}\AgdaSpace{}%
\AgdaInductiveConstructor{no}\AgdaSpace{}%
\AgdaSymbol{;}\AgdaSpace{}%
\AgdaRecord{Dec}\AgdaSymbol{)}\<%
\\
\>[0]\AgdaKeyword{open}\AgdaSpace{}%
\AgdaKeyword{import}\AgdaSpace{}%
\AgdaModule{Relation.Unary}\AgdaSpace{}%
\AgdaKeyword{using}\AgdaSpace{}%
\AgdaSymbol{(}\AgdaFunction{Decidable}\AgdaSymbol{)}\<%
\\
\>[0]\AgdaKeyword{open}\AgdaSpace{}%
\AgdaKeyword{import}\AgdaSpace{}%
\AgdaModule{Relation.Nullary.Negation}\AgdaSpace{}%
\AgdaKeyword{using}\AgdaSpace{}%
\AgdaSymbol{(}\AgdaFunction{contradiction}\AgdaSymbol{)}\<%
\\
\>[0]\AgdaKeyword{open}\AgdaSpace{}%
\AgdaKeyword{import}\AgdaSpace{}%
\AgdaModule{Data.List.Relation.Binary.Subset.Propositional}\AgdaSpace{}%
\AgdaKeyword{using}\AgdaSpace{}%
\AgdaSymbol{(}\AgdaOperator{\AgdaFunction{\AgdaUnderscore{}⊆\AgdaUnderscore{}}}\AgdaSymbol{)}\<%
\\
\>[0]\AgdaKeyword{open}\AgdaSpace{}%
\AgdaKeyword{import}\AgdaSpace{}%
\AgdaModule{Data.List.Relation.Binary.Subset.Propositional.Properties}\AgdaSpace{}%
\AgdaKeyword{using}\AgdaSpace{}%
\AgdaSymbol{(}\AgdaFunction{Any-resp-⊆}\AgdaSpace{}%
\AgdaSymbol{;}\AgdaSpace{}%
\AgdaFunction{⊆-reflexive}\AgdaSymbol{)}\<%
\\
\\[\AgdaEmptyExtraSkip]%
\>[0]\AgdaFunction{maybeAll}\AgdaSpace{}%
\AgdaSymbol{:}\AgdaSpace{}%
\AgdaSymbol{\{}\AgdaBound{sys}\AgdaSpace{}%
\AgdaSymbol{:}\AgdaSpace{}%
\AgdaFunction{FileSystem}\AgdaSymbol{\}}\AgdaSpace{}%
\AgdaSymbol{->}\AgdaSpace{}%
\AgdaSymbol{(}\AgdaBound{ls}\AgdaSpace{}%
\AgdaSymbol{:}\AgdaSpace{}%
\AgdaDatatype{List}\AgdaSpace{}%
\AgdaSymbol{(}\AgdaFunction{FileName}\AgdaSpace{}%
\AgdaOperator{\AgdaFunction{×}}\AgdaSpace{}%
\AgdaDatatype{Maybe}\AgdaSpace{}%
\AgdaFunction{FileContent}\AgdaSymbol{))}\AgdaSpace{}%
\AgdaSymbol{->}\AgdaSpace{}%
\AgdaDatatype{Maybe}\AgdaSpace{}%
\AgdaSymbol{(}\AgdaDatatype{All}\AgdaSpace{}%
\AgdaSymbol{(λ}\AgdaSpace{}%
\AgdaSymbol{(}\AgdaBound{f₁}\AgdaSpace{}%
\AgdaOperator{\AgdaInductiveConstructor{,}}\AgdaSpace{}%
\AgdaBound{v₁}\AgdaSymbol{)}\AgdaSpace{}%
\AgdaSymbol{→}\AgdaSpace{}%
\AgdaBound{sys}\AgdaSpace{}%
\AgdaBound{f₁}\AgdaSpace{}%
\AgdaOperator{\AgdaDatatype{≡}}\AgdaSpace{}%
\AgdaBound{v₁}\AgdaSymbol{)}\AgdaSpace{}%
\AgdaBound{ls}\AgdaSymbol{)}\<%
\\
\>[0]\AgdaFunction{maybeAll}\AgdaSpace{}%
\AgdaSymbol{\{}\AgdaBound{sys}\AgdaSymbol{\}}\AgdaSpace{}%
\AgdaBound{ls}\AgdaSpace{}%
\AgdaSymbol{=}\AgdaSpace{}%
\AgdaFunction{decToMaybe}\AgdaSpace{}%
\AgdaSymbol{(}\AgdaFunction{g₁}\AgdaSpace{}%
\AgdaBound{ls}\AgdaSymbol{)}\<%
\\
\>[0][@{}l@{\AgdaIndent{0}}]%
\>[2]\AgdaKeyword{where}%
\>[225I]\AgdaFunction{g₁}\AgdaSpace{}%
\AgdaSymbol{:}\AgdaSpace{}%
\AgdaSymbol{(}\AgdaBound{ls}\AgdaSpace{}%
\AgdaSymbol{:}\AgdaSpace{}%
\AgdaDatatype{List}\AgdaSpace{}%
\AgdaSymbol{(}\AgdaFunction{FileName}\AgdaSpace{}%
\AgdaOperator{\AgdaFunction{×}}\AgdaSpace{}%
\AgdaDatatype{Maybe}\AgdaSpace{}%
\AgdaFunction{FileContent}\AgdaSymbol{))}\AgdaSpace{}%
\AgdaSymbol{->}\AgdaSpace{}%
\AgdaRecord{Dec}\AgdaSpace{}%
\AgdaSymbol{(}\AgdaDatatype{All}\AgdaSpace{}%
\AgdaSymbol{(λ}\AgdaSpace{}%
\AgdaSymbol{(}\AgdaBound{f₁}\AgdaSpace{}%
\AgdaOperator{\AgdaInductiveConstructor{,}}\AgdaSpace{}%
\AgdaBound{v₁}\AgdaSymbol{)}\AgdaSpace{}%
\AgdaSymbol{→}\AgdaSpace{}%
\AgdaBound{sys}\AgdaSpace{}%
\AgdaBound{f₁}\AgdaSpace{}%
\AgdaOperator{\AgdaDatatype{≡}}\AgdaSpace{}%
\AgdaBound{v₁}\AgdaSymbol{)}\AgdaSpace{}%
\AgdaBound{ls}\AgdaSymbol{)}\<%
\\
\>[.][@{}l@{}]\<[225I]%
\>[8]\AgdaFunction{g₁}\AgdaSpace{}%
\AgdaBound{ls}\AgdaSpace{}%
\AgdaSymbol{=}\AgdaSpace{}%
\AgdaFunction{all?}\AgdaSpace{}%
\AgdaSymbol{(λ}\AgdaSpace{}%
\AgdaSymbol{(}\AgdaBound{f₁}\AgdaSpace{}%
\AgdaOperator{\AgdaInductiveConstructor{,}}\AgdaSpace{}%
\AgdaBound{v₁}\AgdaSymbol{)}\AgdaSpace{}%
\AgdaSymbol{→}\AgdaSpace{}%
\AgdaFunction{≡-dec}\AgdaSpace{}%
\AgdaOperator{\AgdaFunction{\AgdaUnderscore{}≟\AgdaUnderscore{}}}\AgdaSpace{}%
\AgdaSymbol{(}\AgdaBound{sys}\AgdaSpace{}%
\AgdaBound{f₁}\AgdaSymbol{)}\AgdaSpace{}%
\AgdaBound{v₁}\AgdaSymbol{)}\AgdaSpace{}%
\AgdaBound{ls}\<%
\\
\\[\AgdaEmptyExtraSkip]%
\>[0]\AgdaFunction{get}\AgdaSpace{}%
\AgdaSymbol{:}\AgdaSpace{}%
\AgdaSymbol{(}\AgdaBound{x}\AgdaSpace{}%
\AgdaSymbol{:}\AgdaSpace{}%
\AgdaFunction{Cmd}\AgdaSymbol{)}\AgdaSpace{}%
\AgdaSymbol{->}\AgdaSpace{}%
\AgdaSymbol{(}\AgdaBound{ls}\AgdaSpace{}%
\AgdaSymbol{:}\AgdaSpace{}%
\AgdaFunction{Memory}\AgdaSymbol{)}\AgdaSpace{}%
\AgdaSymbol{->}\AgdaSpace{}%
\AgdaBound{x}\AgdaSpace{}%
\AgdaOperator{\AgdaFunction{∈}}\AgdaSpace{}%
\AgdaFunction{map}\AgdaSpace{}%
\AgdaField{proj₁}\AgdaSpace{}%
\AgdaBound{ls}\AgdaSpace{}%
\AgdaSymbol{->}\AgdaSpace{}%
\AgdaDatatype{List}\AgdaSpace{}%
\AgdaSymbol{(}\AgdaFunction{FileName}\AgdaSpace{}%
\AgdaOperator{\AgdaFunction{×}}\AgdaSpace{}%
\AgdaDatatype{Maybe}\AgdaSpace{}%
\AgdaFunction{FileContent}\AgdaSymbol{)}\<%
\\
\>[0]\AgdaFunction{get}\AgdaSpace{}%
\AgdaBound{x}\AgdaSpace{}%
\AgdaSymbol{((}\AgdaBound{x₁}\AgdaSpace{}%
\AgdaOperator{\AgdaInductiveConstructor{,}}\AgdaSpace{}%
\AgdaBound{fs}\AgdaSymbol{)}\AgdaSpace{}%
\AgdaOperator{\AgdaInductiveConstructor{∷}}\AgdaSpace{}%
\AgdaBound{ls}\AgdaSymbol{)}\AgdaSpace{}%
\AgdaBound{x∈}\AgdaSpace{}%
\AgdaKeyword{with}\AgdaSpace{}%
\AgdaBound{x}\AgdaSpace{}%
\AgdaOperator{\AgdaFunction{≟}}\AgdaSpace{}%
\AgdaBound{x₁}\<%
\\
\>[0]\AgdaSymbol{...}\AgdaSpace{}%
\AgdaSymbol{|}\AgdaSpace{}%
\AgdaInductiveConstructor{yes}\AgdaSpace{}%
\AgdaBound{x≡x₁}\AgdaSpace{}%
\AgdaSymbol{=}\AgdaSpace{}%
\AgdaBound{fs}\<%
\\
\>[0]\AgdaSymbol{...}\AgdaSpace{}%
\AgdaSymbol{|}\AgdaSpace{}%
\AgdaInductiveConstructor{no}\AgdaSpace{}%
\AgdaBound{¬x≡x₁}\AgdaSpace{}%
\AgdaSymbol{=}\AgdaSpace{}%
\AgdaFunction{get}\AgdaSpace{}%
\AgdaBound{x}\AgdaSpace{}%
\AgdaBound{ls}\AgdaSpace{}%
\AgdaSymbol{(}\AgdaFunction{tail}\AgdaSpace{}%
\AgdaBound{¬x≡x₁}\AgdaSpace{}%
\AgdaBound{x∈}\AgdaSymbol{)}\<%
\end{code}

\newcommand{\runHuh}{%
\begin{code}%
\>[0]\AgdaFunction{run?}\AgdaSpace{}%
\AgdaSymbol{:}\AgdaSpace{}%
\AgdaFunction{Cmd}\AgdaSpace{}%
\AgdaSymbol{->}\AgdaSpace{}%
\AgdaFunction{State}\AgdaSpace{}%
\AgdaSymbol{->}\AgdaSpace{}%
\AgdaDatatype{Bool}\<%
\\
\>[0]\AgdaFunction{run?}\AgdaSpace{}%
\AgdaBound{x}\AgdaSpace{}%
\AgdaSymbol{(}\AgdaBound{s}\AgdaSpace{}%
\AgdaOperator{\AgdaInductiveConstructor{,}}\AgdaSpace{}%
\AgdaBound{m}\AgdaSymbol{)}\AgdaSpace{}%
\AgdaKeyword{with}\AgdaSpace{}%
\AgdaBound{x}\AgdaSpace{}%
\AgdaOperator{\AgdaFunction{∈?}}\AgdaSpace{}%
\AgdaFunction{map}\AgdaSpace{}%
\AgdaField{proj₁}\AgdaSpace{}%
\AgdaBound{m}\<%
\\
\>[0]\AgdaSymbol{...}\AgdaSpace{}%
\AgdaSymbol{|}\AgdaSpace{}%
\AgdaInductiveConstructor{no}\AgdaSpace{}%
\AgdaBound{x∉}\AgdaSpace{}%
\AgdaSymbol{=}\AgdaSpace{}%
\AgdaInductiveConstructor{Bool.true}\<%
\\
\>[0]\AgdaSymbol{...}\AgdaSpace{}%
\AgdaSymbol{|}\AgdaSpace{}%
\AgdaInductiveConstructor{yes}\AgdaSpace{}%
\AgdaBound{x∈}\AgdaSpace{}%
\AgdaSymbol{=}\AgdaSpace{}%
\AgdaFunction{is-nothing}\AgdaSpace{}%
\AgdaSymbol{(}\AgdaFunction{maybeAll}\AgdaSpace{}%
\AgdaSymbol{\{}\AgdaBound{s}\AgdaSymbol{\}}\AgdaSpace{}%
\AgdaSymbol{(}\AgdaFunction{get}\AgdaSpace{}%
\AgdaBound{x}\AgdaSpace{}%
\AgdaBound{m}\AgdaSpace{}%
\AgdaBound{x∈}\AgdaSymbol{))}\<%
\end{code}}

\newcommand{\doRun}{%
\begin{code}%
\>[0]\AgdaComment{-- store extends the Memory with a new entry}\<%
\\
\>[0]\AgdaFunction{doRun}\AgdaSpace{}%
\AgdaSymbol{:}\AgdaSpace{}%
\AgdaFunction{State}\AgdaSpace{}%
\AgdaSymbol{->}\AgdaSpace{}%
\AgdaFunction{Cmd}\AgdaSpace{}%
\AgdaSymbol{->}\AgdaSpace{}%
\AgdaFunction{State}\<%
\\
\>[0]\AgdaFunction{doRun}\AgdaSpace{}%
\AgdaSymbol{(}\AgdaBound{s}\AgdaSpace{}%
\AgdaOperator{\AgdaInductiveConstructor{,}}\AgdaSpace{}%
\AgdaBound{m}\AgdaSymbol{)}\AgdaSpace{}%
\AgdaBound{x}\AgdaSpace{}%
\AgdaSymbol{=}%
\>[350I]\AgdaKeyword{let}\AgdaSpace{}%
\AgdaBound{s₂}\AgdaSpace{}%
\AgdaSymbol{=}\AgdaSpace{}%
\AgdaFunction{St.run}\AgdaSpace{}%
\AgdaBound{x}\AgdaSpace{}%
\AgdaBound{s}\AgdaSpace{}%
\AgdaKeyword{in}\<%
\\
\>[350I][@{}l@{\AgdaIndent{0}}]%
\>[19]\AgdaSymbol{(}\AgdaBound{s₂}\AgdaSpace{}%
\AgdaOperator{\AgdaInductiveConstructor{,}}\AgdaSpace{}%
\AgdaFunction{store}\AgdaSpace{}%
\AgdaBound{x}\AgdaSpace{}%
\AgdaSymbol{(}\AgdaFunction{cmdReadNames}\AgdaSpace{}%
\AgdaBound{x}\AgdaSpace{}%
\AgdaBound{s}\AgdaSymbol{)}\AgdaSpace{}%
\AgdaBound{s₂}\AgdaSpace{}%
\AgdaBound{m}\AgdaSymbol{)}\<%
\end{code}}

\newcommand{\runF}{%
\begin{code}%
\>[0]\AgdaFunction{runF}%
\>[365I]\AgdaSymbol{:}\AgdaSpace{}%
\AgdaFunction{Cmd}\AgdaSpace{}%
\AgdaSymbol{→}\AgdaSpace{}%
\AgdaSymbol{(}\AgdaFunction{FileSystem}\AgdaSpace{}%
\AgdaOperator{\AgdaFunction{×}}\AgdaSpace{}%
\AgdaFunction{Memory}\AgdaSymbol{)}\<%
\\
\>[.][@{}l@{}]\<[365I]%
\>[5]\AgdaSymbol{→}\AgdaSpace{}%
\AgdaSymbol{(}\AgdaFunction{FileSystem}\AgdaSpace{}%
\AgdaOperator{\AgdaFunction{×}}\AgdaSpace{}%
\AgdaFunction{Memory}\AgdaSymbol{)}\<%
\\
\>[0]\AgdaFunction{runF}\AgdaSpace{}%
\AgdaBound{cmd}\AgdaSpace{}%
\AgdaBound{st}\AgdaSpace{}%
\AgdaSymbol{=}%
\>[377I]\AgdaOperator{\AgdaFunction{if}}\AgdaSpace{}%
\AgdaSymbol{(}\AgdaFunction{run?}\AgdaSpace{}%
\AgdaBound{cmd}\AgdaSpace{}%
\AgdaBound{st}\AgdaSymbol{)}\<%
\\
\>[377I][@{}l@{\AgdaIndent{0}}]%
\>[15]\AgdaOperator{\AgdaFunction{then}}\AgdaSpace{}%
\AgdaFunction{doRun}\AgdaSpace{}%
\AgdaBound{st}\AgdaSpace{}%
\AgdaBound{cmd}\<%
\\
\>[15]\AgdaOperator{\AgdaFunction{else}}\AgdaSpace{}%
\AgdaBound{st}\<%
\end{code}}

\newcommand{\forward}{%
\begin{code}%
\>[0]\AgdaFunction{fabricate}%
\>[385I]\AgdaSymbol{:}\AgdaSpace{}%
\AgdaFunction{Build}\AgdaSpace{}%
\AgdaSymbol{→}\AgdaSpace{}%
\AgdaSymbol{(}\AgdaFunction{FileSystem}\AgdaSpace{}%
\AgdaOperator{\AgdaFunction{×}}\AgdaSpace{}%
\AgdaFunction{Memory}\AgdaSymbol{)}\<%
\\
\>[.][@{}l@{}]\<[385I]%
\>[10]\AgdaSymbol{→}\AgdaSpace{}%
\AgdaSymbol{(}\AgdaFunction{FileSystem}\AgdaSpace{}%
\AgdaOperator{\AgdaFunction{×}}\AgdaSpace{}%
\AgdaFunction{Memory}\AgdaSymbol{)}\<%
\\
\>[0]\AgdaFunction{fabricate}\AgdaSpace{}%
\AgdaInductiveConstructor{[]}\AgdaSpace{}%
\AgdaBound{st}\AgdaSpace{}%
\AgdaSymbol{=}\AgdaSpace{}%
\AgdaBound{st}\<%
\\
\>[0]\AgdaFunction{fabricate}\AgdaSpace{}%
\AgdaSymbol{(}\AgdaBound{x}\AgdaSpace{}%
\AgdaOperator{\AgdaInductiveConstructor{∷}}\AgdaSpace{}%
\AgdaBound{b}\AgdaSymbol{)}\AgdaSpace{}%
\AgdaBound{st}\AgdaSpace{}%
\AgdaSymbol{=}\AgdaSpace{}%
\AgdaFunction{fabricate}\AgdaSpace{}%
\AgdaBound{b}\AgdaSpace{}%
\AgdaSymbol{(}\AgdaFunction{runF}\AgdaSpace{}%
\AgdaBound{x}\AgdaSpace{}%
\AgdaBound{st}\AgdaSymbol{)}\<%
\end{code}}

\begin{code}[hide]%
\>[0]\AgdaComment{\{- Goal: prove if a command is in the memory, and the recorded files and the values of the files in the system are the same as the values in the memory, then
   running the command is equivalent to not running the command. 

 Or similarly/equivalently if a command is in the memory and the recorded files match what is in the system, then the files/values that the oracle says would be the output files, are already in the system. 

What does it mean for the command to be in the memory? The command is in the memory list, so it has
a corresponding list of files and maybe file contents, which in the case of the forward build, are
the files and their values read by the ocmmand when it was recorded as being run.  
-\}}\<%
\end{code}

\begin{code}[hide]%
\>[0]\AgdaSymbol{\{-\#}\AgdaSpace{}%
\AgdaKeyword{OPTIONS}\AgdaSpace{}%
\AgdaPragma{--allow-unsolved-metas}\AgdaSpace{}%
\AgdaSymbol{\#-\}}\<%
\\
\>[0]\AgdaKeyword{open}\AgdaSpace{}%
\AgdaKeyword{import}\AgdaSpace{}%
\AgdaModule{Functional.State}\AgdaSpace{}%
\AgdaKeyword{using}\AgdaSpace{}%
\AgdaSymbol{(}\AgdaFunction{Oracle}\AgdaSpace{}%
\AgdaSymbol{;}\AgdaSpace{}%
\AgdaFunction{FileSystem}\AgdaSpace{}%
\AgdaSymbol{;}\AgdaSpace{}%
\AgdaFunction{Memory}\AgdaSpace{}%
\AgdaSymbol{;}\AgdaSpace{}%
\AgdaFunction{Cmd}\AgdaSpace{}%
\AgdaSymbol{;}\AgdaSpace{}%
\AgdaFunction{extend}\AgdaSpace{}%
\AgdaSymbol{;}\AgdaSpace{}%
\AgdaFunction{save}\AgdaSymbol{)}\<%
\\
\\[\AgdaEmptyExtraSkip]%
\>[0]\AgdaKeyword{module}\AgdaSpace{}%
\AgdaModule{Functional.Forward.Properties}\AgdaSpace{}%
\AgdaSymbol{(}\AgdaBound{oracle}\AgdaSpace{}%
\AgdaSymbol{:}\AgdaSpace{}%
\AgdaFunction{Oracle}\AgdaSymbol{)}\AgdaSpace{}%
\AgdaKeyword{where}\<%
\\
\\[\AgdaEmptyExtraSkip]%
\>[0]\AgdaKeyword{open}\AgdaSpace{}%
\AgdaKeyword{import}\AgdaSpace{}%
\AgdaModule{Functional.State.Helpers}\AgdaSpace{}%
\AgdaSymbol{(}\AgdaBound{oracle}\AgdaSymbol{)}\AgdaSpace{}%
\AgdaSymbol{as}\AgdaSpace{}%
\AgdaModule{St}\AgdaSpace{}%
\AgdaKeyword{hiding}\AgdaSpace{}%
\AgdaSymbol{(}\AgdaFunction{run}\AgdaSymbol{)}\<%
\\
\>[0]\AgdaKeyword{open}\AgdaSpace{}%
\AgdaKeyword{import}\AgdaSpace{}%
\AgdaModule{Functional.Build}\AgdaSpace{}%
\AgdaSymbol{(}\AgdaBound{oracle}\AgdaSymbol{)}\AgdaSpace{}%
\AgdaKeyword{using}\AgdaSpace{}%
\AgdaSymbol{(}DisjointBuild \AgdaSymbol{;} Cons \AgdaSymbol{;} PreCond\AgdaSymbol{)}\<%
\\
\>[0]\AgdaKeyword{open}\AgdaSpace{}%
\AgdaKeyword{import}\AgdaSpace{}%
\AgdaModule{Functional.State.Properties}\AgdaSpace{}%
\AgdaSymbol{(}\AgdaBound{oracle}\AgdaSymbol{)}\AgdaSpace{}%
\AgdaSymbol{as}\AgdaSpace{}%
\AgdaModule{StP}\AgdaSpace{}%
\AgdaKeyword{using}\AgdaSpace{}%
\AgdaSymbol{(}\AgdaFunction{lemma3}\AgdaSpace{}%
\AgdaSymbol{;}\AgdaSpace{}%
\AgdaFunction{lemma4}\AgdaSpace{}%
\AgdaSymbol{;}\AgdaSpace{}%
\AgdaFunction{run-≡}\AgdaSymbol{)}\<%
\\
\>[0]\AgdaKeyword{open}\AgdaSpace{}%
\AgdaKeyword{import}\AgdaSpace{}%
\AgdaModule{Data.Bool}\AgdaSpace{}%
\AgdaKeyword{using}\AgdaSpace{}%
\AgdaSymbol{(}\AgdaInductiveConstructor{false}\AgdaSpace{}%
\AgdaSymbol{;}\AgdaSpace{}%
\AgdaOperator{\AgdaFunction{if\AgdaUnderscore{}then\AgdaUnderscore{}else\AgdaUnderscore{}}}\AgdaSymbol{)}\<%
\\
\>[0]\AgdaKeyword{open}\AgdaSpace{}%
\AgdaKeyword{import}\AgdaSpace{}%
\AgdaModule{Data.Product}\AgdaSpace{}%
\AgdaKeyword{using}\AgdaSpace{}%
\AgdaSymbol{(}\AgdaField{proj₁}\AgdaSpace{}%
\AgdaSymbol{;}\AgdaSpace{}%
\AgdaField{proj₂}\AgdaSpace{}%
\AgdaSymbol{;}\AgdaSpace{}%
\AgdaOperator{\AgdaInductiveConstructor{\AgdaUnderscore{},\AgdaUnderscore{}}}\AgdaSymbol{)}\<%
\\
\>[0]\AgdaKeyword{open}\AgdaSpace{}%
\AgdaKeyword{import}\AgdaSpace{}%
\AgdaModule{Data.String}\AgdaSpace{}%
\AgdaKeyword{using}\AgdaSpace{}%
\AgdaSymbol{(}\AgdaPostulate{String}\AgdaSymbol{)}\<%
\\
\>[0]\AgdaKeyword{open}\AgdaSpace{}%
\AgdaKeyword{import}\AgdaSpace{}%
\AgdaModule{Agda.Builtin.Equality}\<%
\\
\>[0]\AgdaKeyword{open}\AgdaSpace{}%
\AgdaKeyword{import}\AgdaSpace{}%
\AgdaModule{Functional.File}\AgdaSpace{}%
\AgdaKeyword{using}\AgdaSpace{}%
\AgdaSymbol{(}\AgdaFunction{FileName}\AgdaSpace{}%
\AgdaSymbol{;}\AgdaSpace{}%
\AgdaFunction{FileContent}\AgdaSpace{}%
\AgdaSymbol{;}\AgdaSpace{}%
\AgdaFunction{Files}\AgdaSymbol{)}\<%
\\
\>[0]\AgdaKeyword{open}\AgdaSpace{}%
\AgdaKeyword{import}\AgdaSpace{}%
\AgdaModule{Functional.Script.Exec}\AgdaSpace{}%
\AgdaSymbol{(}\AgdaBound{oracle}\AgdaSymbol{)}\AgdaSpace{}%
\AgdaKeyword{using}\AgdaSpace{}%
\AgdaSymbol{(}script\AgdaSymbol{)}\<%
\\
\>[0]\AgdaKeyword{open}\AgdaSpace{}%
\AgdaKeyword{import}\AgdaSpace{}%
\AgdaModule{Functional.Forward.Exec}\AgdaSpace{}%
\AgdaSymbol{(}\AgdaBound{oracle}\AgdaSymbol{)}\AgdaSpace{}%
\AgdaKeyword{using}\AgdaSpace{}%
\AgdaSymbol{(}runF \AgdaSymbol{;} run? \AgdaSymbol{;} maybeAll \AgdaSymbol{;} get \AgdaSymbol{;} fabricate\AgdaSymbol{)}\<%
\\
\>[0]\AgdaKeyword{open}\AgdaSpace{}%
\AgdaKeyword{import}\AgdaSpace{}%
\AgdaModule{Data.Maybe}\AgdaSpace{}%
\AgdaSymbol{as}\AgdaSpace{}%
\AgdaModule{Maybe}\AgdaSpace{}%
\AgdaKeyword{using}\AgdaSpace{}%
\AgdaSymbol{(}\AgdaDatatype{Maybe}\AgdaSpace{}%
\AgdaSymbol{;}\AgdaSpace{}%
\AgdaInductiveConstructor{just}\AgdaSpace{}%
\AgdaSymbol{;}\AgdaSpace{}%
\AgdaInductiveConstructor{nothing}\AgdaSymbol{)}\<%
\\
\>[0]\AgdaKeyword{open}\AgdaSpace{}%
\AgdaKeyword{import}\AgdaSpace{}%
\AgdaModule{Data.Maybe.Relation.Binary.Pointwise}\AgdaSpace{}%
\AgdaKeyword{using}\AgdaSpace{}%
\AgdaSymbol{(}\AgdaFunction{dec}\AgdaSpace{}%
\AgdaSymbol{;}\AgdaSpace{}%
\AgdaDatatype{Pointwise}\AgdaSymbol{)}\<%
\\
\>[0]\AgdaKeyword{open}\AgdaSpace{}%
\AgdaKeyword{import}\AgdaSpace{}%
\AgdaModule{Data.String.Properties}\AgdaSpace{}%
\AgdaKeyword{using}\AgdaSpace{}%
\AgdaSymbol{(}\AgdaOperator{\AgdaFunction{\AgdaUnderscore{}≟\AgdaUnderscore{}}}\AgdaSpace{}%
\AgdaSymbol{;}\AgdaSpace{}%
\AgdaOperator{\AgdaFunction{\AgdaUnderscore{}==\AgdaUnderscore{}}}\AgdaSymbol{)}\<%
\\
\>[0]\AgdaKeyword{open}\AgdaSpace{}%
\AgdaKeyword{import}\AgdaSpace{}%
\AgdaModule{Data.List.Relation.Binary.Equality.DecPropositional}\AgdaSpace{}%
\AgdaOperator{\AgdaFunction{\AgdaUnderscore{}≟\AgdaUnderscore{}}}\AgdaSpace{}%
\AgdaKeyword{using}\AgdaSpace{}%
\AgdaSymbol{(}\AgdaUnderscore{}≡?\AgdaUnderscore{}\AgdaSymbol{)}\<%
\\
\>[0]\AgdaKeyword{open}\AgdaSpace{}%
\AgdaKeyword{import}\AgdaSpace{}%
\AgdaModule{Data.List}\AgdaSpace{}%
\AgdaKeyword{using}\AgdaSpace{}%
\AgdaSymbol{(}\AgdaDatatype{List}\AgdaSpace{}%
\AgdaSymbol{;}\AgdaSpace{}%
\AgdaInductiveConstructor{[]}\AgdaSpace{}%
\AgdaSymbol{;}\AgdaSpace{}%
\AgdaFunction{map}\AgdaSpace{}%
\AgdaSymbol{;}\AgdaSpace{}%
\AgdaFunction{foldr}\AgdaSpace{}%
\AgdaSymbol{;}\AgdaSpace{}%
\AgdaOperator{\AgdaInductiveConstructor{\AgdaUnderscore{}∷\AgdaUnderscore{}}}\AgdaSpace{}%
\AgdaSymbol{;}\AgdaSpace{}%
\AgdaOperator{\AgdaFunction{\AgdaUnderscore{}++\AgdaUnderscore{}}}\AgdaSpace{}%
\AgdaSymbol{;}\AgdaSpace{}%
\AgdaFunction{concatMap}\AgdaSymbol{)}\<%
\\
\>[0]\AgdaKeyword{open}\AgdaSpace{}%
\AgdaKeyword{import}\AgdaSpace{}%
\AgdaModule{Relation.Nullary}\AgdaSpace{}%
\AgdaKeyword{using}\AgdaSpace{}%
\AgdaSymbol{(}\AgdaInductiveConstructor{yes}\AgdaSpace{}%
\AgdaSymbol{;}\AgdaSpace{}%
\AgdaInductiveConstructor{no}\AgdaSymbol{)}\<%
\\
\>[0]\AgdaKeyword{open}\AgdaSpace{}%
\AgdaKeyword{import}\AgdaSpace{}%
\AgdaModule{Relation.Binary.PropositionalEquality}\AgdaSpace{}%
\AgdaKeyword{using}\AgdaSpace{}%
\AgdaSymbol{(}\AgdaFunction{decSetoid}\AgdaSpace{}%
\AgdaSymbol{;}\AgdaSpace{}%
\AgdaFunction{trans}\AgdaSpace{}%
\AgdaSymbol{;}\AgdaSpace{}%
\AgdaFunction{sym}\AgdaSpace{}%
\AgdaSymbol{;}\AgdaSpace{}%
\AgdaFunction{subst}\AgdaSpace{}%
\AgdaSymbol{;}\AgdaSpace{}%
\AgdaFunction{cong}\AgdaSpace{}%
\AgdaSymbol{;}\AgdaSpace{}%
\AgdaFunction{cong₂}\AgdaSymbol{)}\<%
\\
\>[0]\AgdaKeyword{open}\AgdaSpace{}%
\AgdaKeyword{import}\AgdaSpace{}%
\AgdaModule{Data.Product}\AgdaSpace{}%
\AgdaKeyword{using}\AgdaSpace{}%
\AgdaSymbol{(}\AgdaOperator{\AgdaFunction{\AgdaUnderscore{}×\AgdaUnderscore{}}}\AgdaSpace{}%
\AgdaSymbol{;}\AgdaSpace{}%
\AgdaFunction{∃-syntax}\AgdaSymbol{)}\<%
\\
\>[0]\AgdaKeyword{open}\AgdaSpace{}%
\AgdaKeyword{import}\AgdaSpace{}%
\AgdaModule{Data.List.Relation.Unary.All}\AgdaSpace{}%
\AgdaSymbol{as}\AgdaSpace{}%
\AgdaModule{All}\AgdaSpace{}%
\AgdaKeyword{using}\AgdaSpace{}%
\AgdaSymbol{(}\AgdaDatatype{All}\AgdaSpace{}%
\AgdaSymbol{;}\AgdaSpace{}%
\AgdaFunction{all?}\AgdaSpace{}%
\AgdaSymbol{;}\AgdaSpace{}%
\AgdaFunction{lookup}\AgdaSymbol{)}\<%
\\
\>[0]\AgdaKeyword{open}\AgdaSpace{}%
\AgdaKeyword{import}\AgdaSpace{}%
\AgdaModule{Data.List.Membership.DecSetoid}\AgdaSpace{}%
\AgdaSymbol{(}\AgdaFunction{decSetoid}\AgdaSpace{}%
\AgdaOperator{\AgdaFunction{\AgdaUnderscore{}≟\AgdaUnderscore{}}}\AgdaSymbol{)}\AgdaSpace{}%
\AgdaKeyword{using}\AgdaSpace{}%
\AgdaSymbol{(}\AgdaUnderscore{}∈\AgdaUnderscore{} \AgdaSymbol{;} \AgdaUnderscore{}∈?\AgdaUnderscore{} \AgdaSymbol{;} \AgdaUnderscore{}∉\AgdaUnderscore{}\AgdaSymbol{)}\<%
\\
\>[0]\AgdaKeyword{open}\AgdaSpace{}%
\AgdaKeyword{import}\AgdaSpace{}%
\AgdaModule{Data.List.Relation.Unary.Any}\AgdaSpace{}%
\AgdaKeyword{using}\AgdaSpace{}%
\AgdaSymbol{(}\AgdaFunction{tail}\AgdaSpace{}%
\AgdaSymbol{;}\AgdaSpace{}%
\AgdaInductiveConstructor{here}\AgdaSpace{}%
\AgdaSymbol{;}\AgdaSpace{}%
\AgdaInductiveConstructor{there}\AgdaSymbol{)}\<%
\\
\>[0]\AgdaKeyword{open}\AgdaSpace{}%
\AgdaKeyword{import}\AgdaSpace{}%
\AgdaModule{Data.List.Relation.Binary.Disjoint.Propositional}\AgdaSpace{}%
\AgdaKeyword{using}\AgdaSpace{}%
\AgdaSymbol{(}\AgdaFunction{Disjoint}\AgdaSymbol{)}\<%
\\
\>[0]\AgdaKeyword{open}\AgdaSpace{}%
\AgdaKeyword{import}\AgdaSpace{}%
\AgdaModule{Function.Base}\AgdaSpace{}%
\AgdaKeyword{using}\AgdaSpace{}%
\AgdaSymbol{(}\AgdaOperator{\AgdaFunction{\AgdaUnderscore{}∘\AgdaUnderscore{}}}\AgdaSymbol{)}\<%
\\
\>[0]\AgdaKeyword{open}\AgdaSpace{}%
\AgdaKeyword{import}\AgdaSpace{}%
\AgdaModule{Data.List.Membership.Propositional.Properties}\AgdaSpace{}%
\AgdaKeyword{using}\AgdaSpace{}%
\AgdaSymbol{(}\AgdaFunction{∈-++⁺ˡ}\AgdaSpace{}%
\AgdaSymbol{;}\AgdaSpace{}%
\AgdaFunction{∈-++⁺ʳ}\AgdaSpace{}%
\AgdaSymbol{;}\AgdaSpace{}%
\AgdaFunction{∈-++⁻}\AgdaSymbol{)}\<%
\\
\>[0]\AgdaKeyword{open}\AgdaSpace{}%
\AgdaKeyword{import}\AgdaSpace{}%
\AgdaModule{Data.List.Properties}\AgdaSpace{}%
\AgdaKeyword{using}\AgdaSpace{}%
\AgdaSymbol{(}\AgdaFunction{∷-injective}\AgdaSymbol{)}\<%
\\
\>[0]\AgdaKeyword{open}\AgdaSpace{}%
\AgdaKeyword{import}\AgdaSpace{}%
\AgdaModule{Functional.Script.Hazard}\AgdaSpace{}%
\AgdaSymbol{(}\AgdaBound{oracle}\AgdaSymbol{)}\AgdaSpace{}%
\AgdaKeyword{using}\AgdaSpace{}%
\AgdaSymbol{(}HazardFree \AgdaSymbol{;} \AgdaUnderscore{}∷\AgdaUnderscore{} \AgdaSymbol{;} files \AgdaSymbol{;} filesRead\AgdaSymbol{;} Hazard \AgdaSymbol{;} ReadWrite \AgdaSymbol{;} WriteWrite\AgdaSymbol{)}\AgdaSpace{}%
\AgdaKeyword{renaming}\AgdaSpace{}%
\AgdaSymbol{(}save \AgdaSymbol{to} rec\AgdaSymbol{)}\<%
\\
\>[0]\AgdaKeyword{open}\AgdaSpace{}%
\AgdaKeyword{import}\AgdaSpace{}%
\AgdaModule{Functional.Script.Hazard.Properties}\AgdaSpace{}%
\AgdaSymbol{(}\AgdaBound{oracle}\AgdaSymbol{)}\AgdaSpace{}%
\AgdaKeyword{using}\AgdaSpace{}%
\AgdaSymbol{(}hf-still\AgdaSymbol{)}\AgdaSpace{}%
\AgdaKeyword{renaming}\AgdaSpace{}%
\AgdaSymbol{(}g₂ \AgdaSymbol{to} ∉toAll\AgdaSymbol{)}\<%
\\
\>[0]\AgdaKeyword{open}\AgdaSpace{}%
\AgdaKeyword{import}\AgdaSpace{}%
\AgdaModule{Data.List.Relation.Unary.Unique.Propositional}\AgdaSpace{}%
\AgdaKeyword{using}\AgdaSpace{}%
\AgdaSymbol{(}\AgdaDatatype{Unique}\AgdaSymbol{)}\<%
\\
\>[0]\AgdaKeyword{open}\AgdaSpace{}%
\AgdaKeyword{import}\AgdaSpace{}%
\AgdaModule{Data.List.Relation.Unary.AllPairs}\AgdaSpace{}%
\AgdaKeyword{using}\AgdaSpace{}%
\AgdaSymbol{(}\AgdaOperator{\AgdaInductiveConstructor{\AgdaUnderscore{}∷\AgdaUnderscore{}}}\AgdaSymbol{)}\<%
\\
\>[0]\AgdaKeyword{open}\AgdaSpace{}%
\AgdaKeyword{import}\AgdaSpace{}%
\AgdaModule{Functional.Script.Properties}\AgdaSpace{}%
\AgdaSymbol{(}\AgdaBound{oracle}\AgdaSymbol{)}\AgdaSpace{}%
\AgdaKeyword{using}\AgdaSpace{}%
\AgdaSymbol{(}dsj-≡\AgdaSymbol{)}\<%
\\
\>[0]\AgdaKeyword{open}\AgdaSpace{}%
\AgdaKeyword{import}\AgdaSpace{}%
\AgdaModule{Data.List.Relation.Binary.Subset.Propositional}\AgdaSpace{}%
\AgdaKeyword{using}\AgdaSpace{}%
\AgdaSymbol{(}\AgdaOperator{\AgdaFunction{\AgdaUnderscore{}⊆\AgdaUnderscore{}}}\AgdaSymbol{)}\<%
\\
\>[0]\AgdaKeyword{open}\AgdaSpace{}%
\AgdaKeyword{import}\AgdaSpace{}%
\AgdaModule{Data.Sum}\AgdaSpace{}%
\AgdaKeyword{using}\AgdaSpace{}%
\AgdaSymbol{(}\AgdaOperator{\AgdaDatatype{\AgdaUnderscore{}⊎\AgdaUnderscore{}}}\AgdaSpace{}%
\AgdaSymbol{;}\AgdaSpace{}%
\AgdaInductiveConstructor{inj₁}\AgdaSpace{}%
\AgdaSymbol{;}\AgdaSpace{}%
\AgdaInductiveConstructor{inj₂}\AgdaSymbol{)}\<%
\\
\\[\AgdaEmptyExtraSkip]%
\>[0]\AgdaKeyword{open}\AgdaSpace{}%
\AgdaKeyword{import}\AgdaSpace{}%
\AgdaModule{Relation.Nullary.Negation}\AgdaSpace{}%
\AgdaKeyword{using}\AgdaSpace{}%
\AgdaSymbol{(}\AgdaFunction{contradiction}\AgdaSymbol{)}\<%
\\
\\[\AgdaEmptyExtraSkip]%
\\[\AgdaEmptyExtraSkip]%
\>[0]\AgdaComment{-- running the cmd will have no effect on the state}\<%
\\
\>[0]\AgdaFunction{CmdIdempotent}\AgdaSpace{}%
\AgdaSymbol{:}\AgdaSpace{}%
\AgdaFunction{Cmd}\AgdaSpace{}%
\AgdaSymbol{->}\AgdaSpace{}%
\AgdaDatatype{List}\AgdaSpace{}%
\AgdaSymbol{(}\AgdaFunction{FileName}\AgdaSpace{}%
\AgdaOperator{\AgdaFunction{×}}\AgdaSpace{}%
\AgdaDatatype{Maybe}\AgdaSpace{}%
\AgdaFunction{FileContent}\AgdaSymbol{)}\AgdaSpace{}%
\AgdaSymbol{->}\AgdaSpace{}%
\AgdaFunction{FileSystem}\AgdaSpace{}%
\AgdaSymbol{->}\AgdaSpace{}%
\AgdaPrimitiveType{Set}\<%
\\
\>[0]\AgdaFunction{CmdIdempotent}\AgdaSpace{}%
\AgdaBound{x}\AgdaSpace{}%
\AgdaBound{fs}\AgdaSpace{}%
\AgdaBound{sys}\AgdaSpace{}%
\AgdaSymbol{=}\AgdaSpace{}%
\AgdaDatatype{All}\AgdaSpace{}%
\AgdaSymbol{(λ}\AgdaSpace{}%
\AgdaSymbol{(}\AgdaBound{f₁}\AgdaSpace{}%
\AgdaOperator{\AgdaInductiveConstructor{,}}\AgdaSpace{}%
\AgdaBound{v₁}\AgdaSymbol{)}\AgdaSpace{}%
\AgdaSymbol{→}\AgdaSpace{}%
\AgdaBound{sys}\AgdaSpace{}%
\AgdaBound{f₁}\AgdaSpace{}%
\AgdaOperator{\AgdaDatatype{≡}}\AgdaSpace{}%
\AgdaBound{v₁}\AgdaSymbol{)}\AgdaSpace{}%
\AgdaBound{fs}\AgdaSpace{}%
\AgdaSymbol{->}\AgdaSpace{}%
\AgdaSymbol{∀}\AgdaSpace{}%
\AgdaBound{f₁}\AgdaSpace{}%
\AgdaSymbol{→}\AgdaSpace{}%
\AgdaFunction{St.run}\AgdaSpace{}%
\AgdaBound{x}\AgdaSpace{}%
\AgdaBound{sys}\AgdaSpace{}%
\AgdaBound{f₁}\AgdaSpace{}%
\AgdaOperator{\AgdaDatatype{≡}}\AgdaSpace{}%
\AgdaBound{sys}\AgdaSpace{}%
\AgdaBound{f₁}\<%
\\
\\[\AgdaEmptyExtraSkip]%
\>[0]\AgdaKeyword{data}\AgdaSpace{}%
\AgdaDatatype{IdempotentCmd}\AgdaSpace{}%
\AgdaSymbol{:}\AgdaSpace{}%
\AgdaSymbol{(}\AgdaFunction{Cmd}\AgdaSpace{}%
\AgdaSymbol{->}\AgdaSpace{}%
\AgdaFunction{FileSystem}\AgdaSpace{}%
\AgdaSymbol{->}\AgdaSpace{}%
\AgdaDatatype{List}\AgdaSpace{}%
\AgdaFunction{FileName}\AgdaSymbol{)}\AgdaSpace{}%
\AgdaSymbol{->}\AgdaSpace{}%
\AgdaFunction{Cmd}\AgdaSpace{}%
\AgdaSymbol{->}\AgdaSpace{}%
\AgdaDatatype{List}\AgdaSpace{}%
\AgdaSymbol{(}\AgdaFunction{FileName}\AgdaSpace{}%
\AgdaOperator{\AgdaFunction{×}}\AgdaSpace{}%
\AgdaDatatype{Maybe}\AgdaSpace{}%
\AgdaFunction{FileContent}\AgdaSymbol{)}\AgdaSpace{}%
\AgdaSymbol{->}\AgdaSpace{}%
\AgdaFunction{FileSystem}\AgdaSpace{}%
\AgdaSymbol{->}\AgdaSpace{}%
\AgdaPrimitiveType{Set}\AgdaSpace{}%
\AgdaKeyword{where}\<%
\\
\>[0][@{}l@{\AgdaIndent{0}}]%
\>[1]\AgdaInductiveConstructor{IC}\AgdaSpace{}%
\AgdaSymbol{:}\AgdaSpace{}%
\AgdaSymbol{\{}\AgdaBound{f}\AgdaSpace{}%
\AgdaSymbol{:}\AgdaSpace{}%
\AgdaSymbol{(}\AgdaFunction{Cmd}\AgdaSpace{}%
\AgdaSymbol{->}\AgdaSpace{}%
\AgdaFunction{FileSystem}\AgdaSpace{}%
\AgdaSymbol{->}\AgdaSpace{}%
\AgdaDatatype{List}\AgdaSpace{}%
\AgdaFunction{FileName}\AgdaSymbol{)\}}\AgdaSpace{}%
\AgdaSymbol{(}\AgdaBound{x}\AgdaSpace{}%
\AgdaSymbol{:}\AgdaSpace{}%
\AgdaFunction{Cmd}\AgdaSymbol{)}\AgdaSpace{}%
\AgdaSymbol{(}\AgdaBound{fs}\AgdaSpace{}%
\AgdaSymbol{:}\AgdaSpace{}%
\AgdaDatatype{List}\AgdaSpace{}%
\AgdaSymbol{(}\AgdaFunction{FileName}\AgdaSpace{}%
\AgdaOperator{\AgdaFunction{×}}\AgdaSpace{}%
\AgdaDatatype{Maybe}\AgdaSpace{}%
\AgdaFunction{FileContent}\AgdaSymbol{))}\AgdaSpace{}%
\AgdaSymbol{(}\AgdaBound{sys}\AgdaSpace{}%
\AgdaSymbol{:}\AgdaSpace{}%
\AgdaFunction{FileSystem}\AgdaSymbol{)}\AgdaSpace{}%
\AgdaSymbol{->}\AgdaSpace{}%
\AgdaFunction{map}\AgdaSpace{}%
\AgdaField{proj₁}\AgdaSpace{}%
\AgdaBound{fs}\AgdaSpace{}%
\AgdaOperator{\AgdaDatatype{≡}}\AgdaSpace{}%
\AgdaBound{f}\AgdaSpace{}%
\AgdaBound{x}\AgdaSpace{}%
\AgdaBound{sys}\AgdaSpace{}%
\AgdaSymbol{->}\AgdaSpace{}%
\AgdaFunction{CmdIdempotent}\AgdaSpace{}%
\AgdaBound{x}\AgdaSpace{}%
\AgdaBound{fs}\AgdaSpace{}%
\AgdaBound{sys}\AgdaSpace{}%
\AgdaSymbol{->}\AgdaSpace{}%
\AgdaDatatype{IdempotentCmd}\AgdaSpace{}%
\AgdaBound{f}\AgdaSpace{}%
\AgdaBound{x}\AgdaSpace{}%
\AgdaBound{fs}\AgdaSpace{}%
\AgdaBound{sys}\<%
\\
\\[\AgdaEmptyExtraSkip]%
\>[0]\AgdaKeyword{data}\AgdaSpace{}%
\AgdaDatatype{IdempotentState}\AgdaSpace{}%
\AgdaSymbol{:}\AgdaSpace{}%
\AgdaSymbol{(}\AgdaFunction{Cmd}\AgdaSpace{}%
\AgdaSymbol{->}\AgdaSpace{}%
\AgdaFunction{FileSystem}\AgdaSpace{}%
\AgdaSymbol{->}\AgdaSpace{}%
\AgdaDatatype{List}\AgdaSpace{}%
\AgdaFunction{FileName}\AgdaSymbol{)}\AgdaSpace{}%
\AgdaSymbol{->}\AgdaSpace{}%
\AgdaFunction{FileSystem}\AgdaSpace{}%
\AgdaSymbol{->}\AgdaSpace{}%
\AgdaFunction{Memory}\AgdaSpace{}%
\AgdaSymbol{->}\AgdaSpace{}%
\AgdaPrimitiveType{Set}\AgdaSpace{}%
\AgdaKeyword{where}\<%
\\
\>[0][@{}l@{\AgdaIndent{0}}]%
\>[2]\AgdaInductiveConstructor{[]}\AgdaSpace{}%
\AgdaSymbol{:}\AgdaSpace{}%
\AgdaSymbol{\{}\AgdaBound{f}\AgdaSpace{}%
\AgdaSymbol{:}\AgdaSpace{}%
\AgdaSymbol{(}\AgdaFunction{Cmd}\AgdaSpace{}%
\AgdaSymbol{->}\AgdaSpace{}%
\AgdaFunction{FileSystem}\AgdaSpace{}%
\AgdaSymbol{->}\AgdaSpace{}%
\AgdaDatatype{List}\AgdaSpace{}%
\AgdaFunction{FileName}\AgdaSymbol{)\}}\AgdaSpace{}%
\AgdaSymbol{\{}\AgdaBound{s}\AgdaSpace{}%
\AgdaSymbol{:}\AgdaSpace{}%
\AgdaFunction{FileSystem}\AgdaSymbol{\}}\AgdaSpace{}%
\AgdaSymbol{->}\AgdaSpace{}%
\AgdaDatatype{IdempotentState}\AgdaSpace{}%
\AgdaBound{f}\AgdaSpace{}%
\AgdaBound{s}\AgdaSpace{}%
\AgdaInductiveConstructor{[]}\<%
\\
\>[2]\AgdaInductiveConstructor{Cons}\AgdaSpace{}%
\AgdaSymbol{:}\AgdaSpace{}%
\AgdaSymbol{\{}\AgdaBound{f}\AgdaSpace{}%
\AgdaSymbol{:}\AgdaSpace{}%
\AgdaSymbol{(}\AgdaFunction{Cmd}\AgdaSpace{}%
\AgdaSymbol{->}\AgdaSpace{}%
\AgdaFunction{FileSystem}\AgdaSpace{}%
\AgdaSymbol{->}\AgdaSpace{}%
\AgdaDatatype{List}\AgdaSpace{}%
\AgdaFunction{FileName}\AgdaSymbol{)\}}\AgdaSpace{}%
\AgdaSymbol{\{}\AgdaBound{x}\AgdaSpace{}%
\AgdaSymbol{:}\AgdaSpace{}%
\AgdaFunction{Cmd}\AgdaSymbol{\}}\AgdaSpace{}%
\AgdaSymbol{\{}\AgdaBound{fs}\AgdaSpace{}%
\AgdaSymbol{:}\AgdaSpace{}%
\AgdaDatatype{List}\AgdaSpace{}%
\AgdaSymbol{(}\AgdaFunction{FileName}\AgdaSpace{}%
\AgdaOperator{\AgdaFunction{×}}\AgdaSpace{}%
\AgdaDatatype{Maybe}\AgdaSpace{}%
\AgdaFunction{FileContent}\AgdaSymbol{)\}}\AgdaSpace{}%
\AgdaSymbol{\{}\AgdaBound{s}\AgdaSpace{}%
\AgdaSymbol{:}\AgdaSpace{}%
\AgdaFunction{FileSystem}\AgdaSymbol{\}}\AgdaSpace{}%
\AgdaSymbol{\{}\AgdaBound{mm}\AgdaSpace{}%
\AgdaSymbol{:}\AgdaSpace{}%
\AgdaFunction{Memory}\AgdaSymbol{\}}\AgdaSpace{}%
\AgdaSymbol{->}\AgdaSpace{}%
\AgdaDatatype{IdempotentCmd}\AgdaSpace{}%
\AgdaBound{f}\AgdaSpace{}%
\AgdaBound{x}\AgdaSpace{}%
\AgdaBound{fs}\AgdaSpace{}%
\AgdaBound{s}\AgdaSpace{}%
\AgdaSymbol{->}\AgdaSpace{}%
\AgdaDatatype{IdempotentState}\AgdaSpace{}%
\AgdaBound{f}\AgdaSpace{}%
\AgdaBound{s}\AgdaSpace{}%
\AgdaBound{mm}\AgdaSpace{}%
\AgdaSymbol{->}\AgdaSpace{}%
\AgdaDatatype{IdempotentState}\AgdaSpace{}%
\AgdaBound{f}\AgdaSpace{}%
\AgdaBound{s}\AgdaSpace{}%
\AgdaSymbol{((}\AgdaBound{x}\AgdaSpace{}%
\AgdaOperator{\AgdaInductiveConstructor{,}}\AgdaSpace{}%
\AgdaBound{fs}\AgdaSymbol{)}\AgdaSpace{}%
\AgdaOperator{\AgdaInductiveConstructor{∷}}\AgdaSpace{}%
\AgdaBound{mm}\AgdaSymbol{)}\<%
\\
\\[\AgdaEmptyExtraSkip]%
\>[0]\AgdaFunction{lemma1}\AgdaSpace{}%
\AgdaSymbol{:}\AgdaSpace{}%
\AgdaSymbol{(}\AgdaBound{f₁}\AgdaSpace{}%
\AgdaSymbol{:}\AgdaSpace{}%
\AgdaFunction{FileName}\AgdaSymbol{)}\AgdaSpace{}%
\AgdaSymbol{->}\AgdaSpace{}%
\AgdaSymbol{(}\AgdaBound{ls}\AgdaSpace{}%
\AgdaSymbol{:}\AgdaSpace{}%
\AgdaFunction{Files}\AgdaSymbol{)}\AgdaSpace{}%
\AgdaSymbol{->}\AgdaSpace{}%
\AgdaBound{f₁}\AgdaSpace{}%
\AgdaOperator{\AgdaFunction{∈}}\AgdaSpace{}%
\AgdaFunction{map}\AgdaSpace{}%
\AgdaField{proj₁}\AgdaSpace{}%
\AgdaBound{ls}\AgdaSpace{}%
\AgdaSymbol{->}\AgdaSpace{}%
\AgdaFunction{∃[}\AgdaSpace{}%
\AgdaBound{v}\AgdaSpace{}%
\AgdaFunction{]}\AgdaSymbol{(∀}\AgdaSpace{}%
\AgdaBound{s}\AgdaSpace{}%
\AgdaSymbol{→}\AgdaSpace{}%
\AgdaFunction{foldr}\AgdaSpace{}%
\AgdaFunction{extend}\AgdaSpace{}%
\AgdaBound{s}\AgdaSpace{}%
\AgdaBound{ls}\AgdaSpace{}%
\AgdaBound{f₁}\AgdaSpace{}%
\AgdaOperator{\AgdaDatatype{≡}}\AgdaSpace{}%
\AgdaInductiveConstructor{just}\AgdaSpace{}%
\AgdaBound{v}\AgdaSymbol{)}\<%
\\
\>[0]\AgdaFunction{lemma1}\AgdaSpace{}%
\AgdaBound{f₁}\AgdaSpace{}%
\AgdaSymbol{((}\AgdaBound{f}\AgdaSpace{}%
\AgdaOperator{\AgdaInductiveConstructor{,}}\AgdaSpace{}%
\AgdaBound{v}\AgdaSymbol{)}\AgdaSpace{}%
\AgdaOperator{\AgdaInductiveConstructor{∷}}\AgdaSpace{}%
\AgdaBound{ls}\AgdaSymbol{)}\AgdaSpace{}%
\AgdaBound{f₁∈}\AgdaSpace{}%
\AgdaKeyword{with}\AgdaSpace{}%
\AgdaBound{f}\AgdaSpace{}%
\AgdaOperator{\AgdaFunction{≟}}\AgdaSpace{}%
\AgdaBound{f₁}\<%
\\
\>[0]\AgdaSymbol{...}\AgdaSpace{}%
\AgdaSymbol{|}\AgdaSpace{}%
\AgdaInductiveConstructor{yes}\AgdaSpace{}%
\AgdaBound{f≡f₁}\AgdaSpace{}%
\AgdaSymbol{=}\AgdaSpace{}%
\AgdaBound{v}\AgdaSpace{}%
\AgdaOperator{\AgdaInductiveConstructor{,}}\AgdaSpace{}%
\AgdaSymbol{(λ}\AgdaSpace{}%
\AgdaBound{s}\AgdaSpace{}%
\AgdaSymbol{→}\AgdaSpace{}%
\AgdaInductiveConstructor{refl}\AgdaSymbol{)}\<%
\\
\>[0]\AgdaSymbol{...}\AgdaSpace{}%
\AgdaSymbol{|}\AgdaSpace{}%
\AgdaInductiveConstructor{no}\AgdaSpace{}%
\AgdaBound{¬f≡f₁}\AgdaSpace{}%
\AgdaSymbol{=}\AgdaSpace{}%
\AgdaFunction{lemma1}\AgdaSpace{}%
\AgdaBound{f₁}\AgdaSpace{}%
\AgdaBound{ls}\AgdaSpace{}%
\AgdaSymbol{(}\AgdaFunction{tail}\AgdaSpace{}%
\AgdaSymbol{(λ}\AgdaSpace{}%
\AgdaBound{f≡f₁}\AgdaSpace{}%
\AgdaSymbol{→}\AgdaSpace{}%
\AgdaBound{¬f≡f₁}\AgdaSpace{}%
\AgdaSymbol{(}\AgdaFunction{sym}\AgdaSpace{}%
\AgdaBound{f≡f₁}\AgdaSymbol{))}\AgdaSpace{}%
\AgdaBound{f₁∈}\AgdaSymbol{)}\<%
\\
\\[\AgdaEmptyExtraSkip]%
\\[\AgdaEmptyExtraSkip]%
\>[0]\AgdaFunction{cmdIdempotent}\AgdaSpace{}%
\AgdaSymbol{:}\AgdaSpace{}%
\AgdaSymbol{(}\AgdaBound{sys}\AgdaSpace{}%
\AgdaSymbol{:}\AgdaSpace{}%
\AgdaFunction{FileSystem}\AgdaSymbol{)}\AgdaSpace{}%
\AgdaSymbol{(}\AgdaBound{x}\AgdaSpace{}%
\AgdaSymbol{:}\AgdaSpace{}%
\AgdaFunction{Cmd}\AgdaSymbol{)}\AgdaSpace{}%
\AgdaSymbol{->}\AgdaSpace{}%
\AgdaFunction{Disjoint}\AgdaSpace{}%
\AgdaSymbol{(}\AgdaFunction{cmdReadNames}\AgdaSpace{}%
\AgdaBound{x}\AgdaSpace{}%
\AgdaBound{sys}\AgdaSymbol{)}\AgdaSpace{}%
\AgdaSymbol{(}\AgdaFunction{cmdWriteNames}\AgdaSpace{}%
\AgdaBound{x}\AgdaSpace{}%
\AgdaBound{sys}\AgdaSymbol{)}\AgdaSpace{}%
\AgdaSymbol{->}\AgdaSpace{}%
\AgdaSymbol{∀}\AgdaSpace{}%
\AgdaBound{f₁}\AgdaSpace{}%
\AgdaSymbol{→}\AgdaSpace{}%
\AgdaFunction{St.run}\AgdaSpace{}%
\AgdaBound{x}\AgdaSpace{}%
\AgdaSymbol{(}\AgdaFunction{St.run}\AgdaSpace{}%
\AgdaBound{x}\AgdaSpace{}%
\AgdaBound{sys}\AgdaSymbol{)}\AgdaSpace{}%
\AgdaBound{f₁}\AgdaSpace{}%
\AgdaOperator{\AgdaDatatype{≡}}\AgdaSpace{}%
\AgdaFunction{St.run}\AgdaSpace{}%
\AgdaBound{x}\AgdaSpace{}%
\AgdaBound{sys}\AgdaSpace{}%
\AgdaBound{f₁}\<%
\\
\>[0]\AgdaFunction{cmdIdempotent}\AgdaSpace{}%
\AgdaBound{sys}\AgdaSpace{}%
\AgdaBound{x}\AgdaSpace{}%
\AgdaBound{dsj}\AgdaSpace{}%
\AgdaBound{f₁}\AgdaSpace{}%
\AgdaKeyword{with}\AgdaSpace{}%
\AgdaField{proj₂}\AgdaSpace{}%
\AgdaSymbol{(}\AgdaBound{oracle}\AgdaSpace{}%
\AgdaBound{x}\AgdaSymbol{)}\AgdaSpace{}%
\AgdaBound{sys}\AgdaSpace{}%
\AgdaSymbol{(}\AgdaFunction{St.run}\AgdaSpace{}%
\AgdaBound{x}\AgdaSpace{}%
\AgdaBound{sys}\AgdaSymbol{)}\AgdaSpace{}%
\AgdaSymbol{(λ}\AgdaSpace{}%
\AgdaBound{f₂}\AgdaSpace{}%
\AgdaBound{x₁}\AgdaSpace{}%
\AgdaSymbol{→}\AgdaSpace{}%
\AgdaSymbol{(}\AgdaFunction{lemma3}\AgdaSpace{}%
\AgdaSymbol{\{}\AgdaBound{sys}\AgdaSymbol{\}}\AgdaSpace{}%
\AgdaBound{f₂}\AgdaSpace{}%
\AgdaSymbol{(}\AgdaField{proj₂}\AgdaSpace{}%
\AgdaSymbol{(}\AgdaField{proj₁}\AgdaSpace{}%
\AgdaSymbol{(}\AgdaBound{oracle}\AgdaSpace{}%
\AgdaBound{x}\AgdaSymbol{)}\AgdaSpace{}%
\AgdaBound{sys}\AgdaSymbol{))}\AgdaSpace{}%
\AgdaSymbol{λ}\AgdaSpace{}%
\AgdaBound{x₂}\AgdaSpace{}%
\AgdaSymbol{→}\AgdaSpace{}%
\AgdaBound{dsj}\AgdaSpace{}%
\AgdaSymbol{(}\AgdaBound{x₁}\AgdaSpace{}%
\AgdaOperator{\AgdaInductiveConstructor{,}}\AgdaSpace{}%
\AgdaBound{x₂}\AgdaSymbol{)))}\<%
\\
\>[0]\AgdaSymbol{...}\AgdaSpace{}%
\AgdaSymbol{|}\AgdaSpace{}%
\AgdaBound{≡₁}\AgdaSpace{}%
\AgdaKeyword{with}\AgdaSpace{}%
\AgdaBound{f₁}\AgdaSpace{}%
\AgdaOperator{\AgdaFunction{∈?}}\AgdaSpace{}%
\AgdaFunction{cmdWriteNames}\AgdaSpace{}%
\AgdaBound{x}\AgdaSpace{}%
\AgdaBound{sys}\<%
\\
\>[0]\AgdaSymbol{...}\AgdaSpace{}%
\AgdaSymbol{|}\AgdaSpace{}%
\AgdaInductiveConstructor{no}\AgdaSpace{}%
\AgdaBound{f₁∉}\AgdaSpace{}%
\AgdaSymbol{=}\AgdaSpace{}%
\AgdaFunction{sym}\AgdaSpace{}%
\AgdaSymbol{(}\AgdaFunction{lemma3}\AgdaSpace{}%
\AgdaSymbol{\{}\AgdaFunction{foldr}\AgdaSpace{}%
\AgdaFunction{extend}\AgdaSpace{}%
\AgdaBound{sys}\AgdaSpace{}%
\AgdaSymbol{(}\AgdaField{proj₂}\AgdaSpace{}%
\AgdaSymbol{(}\AgdaField{proj₁}\AgdaSpace{}%
\AgdaSymbol{(}\AgdaBound{oracle}\AgdaSpace{}%
\AgdaBound{x}\AgdaSymbol{)}\AgdaSpace{}%
\AgdaBound{sys}\AgdaSymbol{))\}}\AgdaSpace{}%
\AgdaBound{f₁}\AgdaSpace{}%
\AgdaSymbol{(}\AgdaField{proj₂}\AgdaSpace{}%
\AgdaSymbol{(}\AgdaField{proj₁}\AgdaSpace{}%
\AgdaSymbol{(}\AgdaBound{oracle}\AgdaSpace{}%
\AgdaBound{x}\AgdaSymbol{)}\AgdaSpace{}%
\AgdaSymbol{(}\AgdaFunction{foldr}\AgdaSpace{}%
\AgdaFunction{extend}\AgdaSpace{}%
\AgdaBound{sys}\AgdaSpace{}%
\AgdaSymbol{(}\AgdaField{proj₂}\AgdaSpace{}%
\AgdaSymbol{(}\AgdaField{proj₁}\AgdaSpace{}%
\AgdaSymbol{(}\AgdaBound{oracle}\AgdaSpace{}%
\AgdaBound{x}\AgdaSymbol{)}\AgdaSpace{}%
\AgdaBound{sys}\AgdaSymbol{)))))}\AgdaSpace{}%
\AgdaSymbol{(}\AgdaFunction{subst}\AgdaSpace{}%
\AgdaSymbol{(λ}\AgdaSpace{}%
\AgdaBound{x₁}\AgdaSpace{}%
\AgdaSymbol{→}\AgdaSpace{}%
\AgdaBound{f₁}\AgdaSpace{}%
\AgdaOperator{\AgdaFunction{∉}}\AgdaSpace{}%
\AgdaBound{x₁}\AgdaSymbol{)}\AgdaSpace{}%
\AgdaSymbol{(}\AgdaFunction{cong}\AgdaSpace{}%
\AgdaSymbol{((}\AgdaFunction{map}\AgdaSpace{}%
\AgdaField{proj₁}\AgdaSymbol{)}\AgdaSpace{}%
\AgdaOperator{\AgdaFunction{∘}}\AgdaSpace{}%
\AgdaField{proj₂}\AgdaSymbol{)}\AgdaSpace{}%
\AgdaBound{≡₁}\AgdaSymbol{)}\AgdaSpace{}%
\AgdaBound{f₁∉}\AgdaSymbol{))}\<%
\\
\>[0]\AgdaSymbol{...}\AgdaSpace{}%
\AgdaSymbol{|}\AgdaSpace{}%
\AgdaInductiveConstructor{yes}\AgdaSpace{}%
\AgdaBound{f₁∈}\AgdaSpace{}%
\AgdaKeyword{with}\AgdaSpace{}%
\AgdaFunction{lemma1}\AgdaSpace{}%
\AgdaBound{f₁}\AgdaSpace{}%
\AgdaSymbol{(}\AgdaField{proj₂}\AgdaSpace{}%
\AgdaSymbol{(}\AgdaField{proj₁}\AgdaSpace{}%
\AgdaSymbol{(}\AgdaBound{oracle}\AgdaSpace{}%
\AgdaBound{x}\AgdaSymbol{)}\AgdaSpace{}%
\AgdaBound{sys}\AgdaSymbol{))}\AgdaSpace{}%
\AgdaBound{f₁∈}\<%
\\
\>[0]\AgdaSymbol{...}\AgdaSpace{}%
\AgdaSymbol{|}\AgdaSpace{}%
\AgdaBound{v}\AgdaSpace{}%
\AgdaOperator{\AgdaInductiveConstructor{,}}\AgdaSpace{}%
\AgdaBound{∀₁}\AgdaSpace{}%
\AgdaKeyword{with}\AgdaSpace{}%
\AgdaFunction{subst}\AgdaSpace{}%
\AgdaSymbol{(λ}\AgdaSpace{}%
\AgdaBound{x₁}\AgdaSpace{}%
\AgdaSymbol{→}\AgdaSpace{}%
\AgdaFunction{foldr}\AgdaSpace{}%
\AgdaFunction{extend}\AgdaSpace{}%
\AgdaSymbol{(}\AgdaFunction{St.run}\AgdaSpace{}%
\AgdaBound{x}\AgdaSpace{}%
\AgdaBound{sys}\AgdaSymbol{)}\AgdaSpace{}%
\AgdaBound{x₁}\AgdaSpace{}%
\AgdaBound{f₁}\AgdaSpace{}%
\AgdaOperator{\AgdaDatatype{≡}}\AgdaSpace{}%
\AgdaSymbol{\AgdaUnderscore{})}\AgdaSpace{}%
\AgdaSymbol{(}\AgdaFunction{cong}\AgdaSpace{}%
\AgdaField{proj₂}\AgdaSpace{}%
\AgdaBound{≡₁}\AgdaSymbol{)}\AgdaSpace{}%
\AgdaSymbol{(}\AgdaBound{∀₁}\AgdaSpace{}%
\AgdaSymbol{(}\AgdaFunction{St.run}\AgdaSpace{}%
\AgdaBound{x}\AgdaSpace{}%
\AgdaBound{sys}\AgdaSymbol{))}\<%
\\
\>[0]\AgdaSymbol{...}\AgdaSpace{}%
\AgdaSymbol{|}\AgdaSpace{}%
\AgdaBound{≡₂}\AgdaSpace{}%
\AgdaSymbol{=}\AgdaSpace{}%
\AgdaFunction{trans}\AgdaSpace{}%
\AgdaBound{≡₂}\AgdaSpace{}%
\AgdaSymbol{(}\AgdaFunction{sym}\AgdaSpace{}%
\AgdaSymbol{(}\AgdaBound{∀₁}\AgdaSpace{}%
\AgdaBound{sys}\AgdaSymbol{))}\<%
\\
\\[\AgdaEmptyExtraSkip]%
\>[0]\AgdaFunction{cmdReadWrites}\AgdaSpace{}%
\AgdaSymbol{:}\AgdaSpace{}%
\AgdaFunction{Cmd}\AgdaSpace{}%
\AgdaSymbol{->}\AgdaSpace{}%
\AgdaFunction{FileSystem}\AgdaSpace{}%
\AgdaSymbol{->}\AgdaSpace{}%
\AgdaDatatype{List}\AgdaSpace{}%
\AgdaFunction{FileName}\<%
\\
\>[0]\AgdaFunction{cmdReadWrites}\AgdaSpace{}%
\AgdaBound{x}\AgdaSpace{}%
\AgdaBound{sys}\AgdaSpace{}%
\AgdaSymbol{=}\AgdaSpace{}%
\AgdaSymbol{(}\AgdaFunction{cmdReadNames}\AgdaSpace{}%
\AgdaBound{x}\AgdaSpace{}%
\AgdaBound{sys}\AgdaSymbol{)}\AgdaSpace{}%
\AgdaOperator{\AgdaFunction{++}}\AgdaSpace{}%
\AgdaSymbol{(}\AgdaFunction{cmdWriteNames}\AgdaSpace{}%
\AgdaBound{x}\AgdaSpace{}%
\AgdaBound{sys}\AgdaSymbol{)}\<%
\\
\\[\AgdaEmptyExtraSkip]%
\>[0]\AgdaFunction{lemma2}\AgdaSpace{}%
\AgdaSymbol{:}\AgdaSpace{}%
\AgdaSymbol{\{}\AgdaBound{sys}\AgdaSpace{}%
\AgdaSymbol{:}\AgdaSpace{}%
\AgdaFunction{FileSystem}\AgdaSymbol{\}}\AgdaSpace{}%
\AgdaSymbol{\{}\AgdaBound{ls}\AgdaSpace{}%
\AgdaSymbol{:}\AgdaSpace{}%
\AgdaFunction{Files}\AgdaSymbol{\}}\AgdaSpace{}%
\AgdaSymbol{(}\AgdaBound{fs}\AgdaSpace{}%
\AgdaSymbol{:}\AgdaSpace{}%
\AgdaDatatype{List}\AgdaSpace{}%
\AgdaSymbol{(}\AgdaFunction{FileName}\AgdaSpace{}%
\AgdaOperator{\AgdaFunction{×}}\AgdaSpace{}%
\AgdaDatatype{Maybe}\AgdaSpace{}%
\AgdaFunction{FileContent}\AgdaSymbol{))}\AgdaSpace{}%
\AgdaSymbol{->}\AgdaSpace{}%
\AgdaDatatype{All}\AgdaSpace{}%
\AgdaSymbol{(λ}\AgdaSpace{}%
\AgdaSymbol{(}\AgdaBound{f₁}\AgdaSpace{}%
\AgdaOperator{\AgdaInductiveConstructor{,}}\AgdaSpace{}%
\AgdaBound{v₁}\AgdaSymbol{)}\AgdaSpace{}%
\AgdaSymbol{→}\AgdaSpace{}%
\AgdaFunction{foldr}\AgdaSpace{}%
\AgdaFunction{extend}\AgdaSpace{}%
\AgdaBound{sys}\AgdaSpace{}%
\AgdaBound{ls}\AgdaSpace{}%
\AgdaBound{f₁}\AgdaSpace{}%
\AgdaOperator{\AgdaDatatype{≡}}\AgdaSpace{}%
\AgdaBound{v₁}\AgdaSymbol{)}\AgdaSpace{}%
\AgdaBound{fs}\AgdaSpace{}%
\AgdaSymbol{->}\AgdaSpace{}%
\AgdaFunction{Disjoint}\AgdaSpace{}%
\AgdaSymbol{(}\AgdaFunction{map}\AgdaSpace{}%
\AgdaField{proj₁}\AgdaSpace{}%
\AgdaBound{fs}\AgdaSymbol{)}\AgdaSpace{}%
\AgdaSymbol{(}\AgdaFunction{map}\AgdaSpace{}%
\AgdaField{proj₁}\AgdaSpace{}%
\AgdaBound{ls}\AgdaSymbol{)}\AgdaSpace{}%
\AgdaSymbol{->}\AgdaSpace{}%
\AgdaDatatype{All}\AgdaSpace{}%
\AgdaSymbol{(λ}\AgdaSpace{}%
\AgdaSymbol{(}\AgdaBound{f₁}\AgdaSpace{}%
\AgdaOperator{\AgdaInductiveConstructor{,}}\AgdaSpace{}%
\AgdaBound{v₁}\AgdaSymbol{)}\AgdaSpace{}%
\AgdaSymbol{→}\AgdaSpace{}%
\AgdaBound{sys}\AgdaSpace{}%
\AgdaBound{f₁}\AgdaSpace{}%
\AgdaOperator{\AgdaDatatype{≡}}\AgdaSpace{}%
\AgdaBound{v₁}\AgdaSymbol{)}\AgdaSpace{}%
\AgdaBound{fs}\<%
\\
\>[0]\AgdaFunction{lemma2}\AgdaSpace{}%
\AgdaInductiveConstructor{[]}\AgdaSpace{}%
\AgdaBound{all₁}\AgdaSpace{}%
\AgdaBound{dsj}\AgdaSpace{}%
\AgdaSymbol{=}\AgdaSpace{}%
\AgdaInductiveConstructor{All.[]}\<%
\\
\>[0]\AgdaFunction{lemma2}\AgdaSpace{}%
\AgdaSymbol{\{}\AgdaBound{sys}\AgdaSymbol{\}}\AgdaSpace{}%
\AgdaSymbol{\{}\AgdaBound{ls}\AgdaSymbol{\}}\AgdaSpace{}%
\AgdaSymbol{((}\AgdaBound{f₁}\AgdaSpace{}%
\AgdaOperator{\AgdaInductiveConstructor{,}}\AgdaSpace{}%
\AgdaBound{v₁}\AgdaSymbol{)}\AgdaSpace{}%
\AgdaOperator{\AgdaInductiveConstructor{∷}}\AgdaSpace{}%
\AgdaBound{fs}\AgdaSymbol{)}\AgdaSpace{}%
\AgdaSymbol{(}\AgdaBound{px}\AgdaSpace{}%
\AgdaOperator{\AgdaInductiveConstructor{All.∷}}\AgdaSpace{}%
\AgdaBound{all₁}\AgdaSymbol{)}\AgdaSpace{}%
\AgdaBound{dsj}\AgdaSpace{}%
\AgdaSymbol{=}\AgdaSpace{}%
\AgdaSymbol{(}\AgdaFunction{trans}\AgdaSpace{}%
\AgdaSymbol{(}\AgdaFunction{lemma3}\AgdaSpace{}%
\AgdaBound{f₁}\AgdaSpace{}%
\AgdaBound{ls}\AgdaSpace{}%
\AgdaSymbol{λ}\AgdaSpace{}%
\AgdaBound{x}\AgdaSpace{}%
\AgdaSymbol{→}\AgdaSpace{}%
\AgdaBound{dsj}\AgdaSpace{}%
\AgdaSymbol{(}\AgdaInductiveConstructor{here}\AgdaSpace{}%
\AgdaInductiveConstructor{refl}\AgdaSpace{}%
\AgdaOperator{\AgdaInductiveConstructor{,}}\AgdaSpace{}%
\AgdaBound{x}\AgdaSymbol{))}\AgdaSpace{}%
\AgdaBound{px}\AgdaSymbol{)}\AgdaSpace{}%
\AgdaOperator{\AgdaInductiveConstructor{All.∷}}\AgdaSpace{}%
\AgdaSymbol{(}\AgdaFunction{lemma2}\AgdaSpace{}%
\AgdaBound{fs}\AgdaSpace{}%
\AgdaBound{all₁}\AgdaSpace{}%
\AgdaSymbol{λ}\AgdaSpace{}%
\AgdaBound{x₁}\AgdaSpace{}%
\AgdaSymbol{→}\AgdaSpace{}%
\AgdaBound{dsj}\AgdaSpace{}%
\AgdaSymbol{(}\AgdaInductiveConstructor{there}\AgdaSpace{}%
\AgdaSymbol{(}\AgdaField{proj₁}\AgdaSpace{}%
\AgdaBound{x₁}\AgdaSymbol{)}\AgdaSpace{}%
\AgdaOperator{\AgdaInductiveConstructor{,}}\AgdaSpace{}%
\AgdaField{proj₂}\AgdaSpace{}%
\AgdaBound{x₁}\AgdaSymbol{))}\<%
\\
\\[\AgdaEmptyExtraSkip]%
\>[0]\AgdaFunction{∈-resp-≡}\AgdaSpace{}%
\AgdaSymbol{:}\AgdaSpace{}%
\AgdaSymbol{\{}\AgdaBound{ls}\AgdaSpace{}%
\AgdaBound{ls₁}\AgdaSpace{}%
\AgdaSymbol{:}\AgdaSpace{}%
\AgdaDatatype{List}\AgdaSpace{}%
\AgdaPostulate{String}\AgdaSymbol{\}}\AgdaSpace{}%
\AgdaSymbol{\{}\AgdaBound{v}\AgdaSpace{}%
\AgdaSymbol{:}\AgdaSpace{}%
\AgdaPostulate{String}\AgdaSymbol{\}}\AgdaSpace{}%
\AgdaSymbol{->}\AgdaSpace{}%
\AgdaBound{v}\AgdaSpace{}%
\AgdaOperator{\AgdaFunction{∈}}\AgdaSpace{}%
\AgdaBound{ls}\AgdaSpace{}%
\AgdaSymbol{->}\AgdaSpace{}%
\AgdaBound{ls}\AgdaSpace{}%
\AgdaOperator{\AgdaDatatype{≡}}\AgdaSpace{}%
\AgdaBound{ls₁}\AgdaSpace{}%
\AgdaSymbol{->}\AgdaSpace{}%
\AgdaBound{v}\AgdaSpace{}%
\AgdaOperator{\AgdaFunction{∈}}\AgdaSpace{}%
\AgdaBound{ls₁}\<%
\\
\>[0]\AgdaFunction{∈-resp-≡}\AgdaSpace{}%
\AgdaBound{v∈ls}\AgdaSpace{}%
\AgdaBound{ls≡ls₁}\AgdaSpace{}%
\AgdaSymbol{=}\AgdaSpace{}%
\AgdaFunction{subst}\AgdaSpace{}%
\AgdaSymbol{(λ}\AgdaSpace{}%
\AgdaBound{x}\AgdaSpace{}%
\AgdaSymbol{→}\AgdaSpace{}%
\AgdaSymbol{\AgdaUnderscore{}}\AgdaSpace{}%
\AgdaOperator{\AgdaFunction{∈}}\AgdaSpace{}%
\AgdaBound{x}\AgdaSymbol{)}\AgdaSpace{}%
\AgdaBound{ls≡ls₁}\AgdaSpace{}%
\AgdaBound{v∈ls}\<%
\\
\\[\AgdaEmptyExtraSkip]%
\>[0]\AgdaFunction{runPreserves}\AgdaSpace{}%
\AgdaSymbol{:}\AgdaSpace{}%
\AgdaSymbol{\{}\AgdaBound{sys}\AgdaSpace{}%
\AgdaSymbol{:}\AgdaSpace{}%
\AgdaFunction{FileSystem}\AgdaSymbol{\}}\AgdaSpace{}%
\AgdaSymbol{(}\AgdaBound{x}\AgdaSpace{}%
\AgdaBound{x₁}\AgdaSpace{}%
\AgdaSymbol{:}\AgdaSpace{}%
\AgdaFunction{Cmd}\AgdaSymbol{)}\AgdaSpace{}%
\AgdaSymbol{->}\AgdaSpace{}%
\AgdaSymbol{(}\AgdaBound{fs}\AgdaSpace{}%
\AgdaSymbol{:}\AgdaSpace{}%
\AgdaDatatype{List}\AgdaSpace{}%
\AgdaSymbol{(}\AgdaFunction{FileName}\AgdaSpace{}%
\AgdaOperator{\AgdaFunction{×}}\AgdaSpace{}%
\AgdaDatatype{Maybe}\AgdaSpace{}%
\AgdaFunction{FileContent}\AgdaSymbol{))}\AgdaSpace{}%
\AgdaSymbol{->}\AgdaSpace{}%
\AgdaFunction{Disjoint}\AgdaSpace{}%
\AgdaSymbol{(}\AgdaFunction{cmdWriteNames}\AgdaSpace{}%
\AgdaBound{x₁}\AgdaSpace{}%
\AgdaBound{sys}\AgdaSymbol{)}\AgdaSpace{}%
\AgdaSymbol{(}\AgdaFunction{cmdReadWrites}\AgdaSpace{}%
\AgdaBound{x}\AgdaSpace{}%
\AgdaBound{sys}\AgdaSymbol{)}\AgdaSpace{}%
\AgdaSymbol{->}\AgdaSpace{}%
\AgdaDatatype{IdempotentCmd}\AgdaSpace{}%
\AgdaFunction{cmdReadNames}\AgdaSpace{}%
\AgdaBound{x}\AgdaSpace{}%
\AgdaBound{fs}\AgdaSpace{}%
\AgdaBound{sys}\AgdaSpace{}%
\AgdaSymbol{->}\AgdaSpace{}%
\AgdaDatatype{IdempotentCmd}\AgdaSpace{}%
\AgdaFunction{cmdReadNames}\AgdaSpace{}%
\AgdaBound{x}\AgdaSpace{}%
\AgdaBound{fs}\AgdaSpace{}%
\AgdaSymbol{(}\AgdaFunction{St.run}\AgdaSpace{}%
\AgdaBound{x₁}\AgdaSpace{}%
\AgdaBound{sys}\AgdaSymbol{)}\<%
\\
\>[0]\AgdaFunction{runPreserves}\AgdaSpace{}%
\AgdaSymbol{\{}\AgdaBound{sys}\AgdaSymbol{\}}\AgdaSpace{}%
\AgdaBound{x}\AgdaSpace{}%
\AgdaBound{x₁}\AgdaSpace{}%
\AgdaBound{fs}\AgdaSpace{}%
\AgdaBound{dsj}\AgdaSpace{}%
\AgdaSymbol{(}\AgdaInductiveConstructor{IC}\AgdaSpace{}%
\AgdaDottedPattern{\AgdaSymbol{.}}\AgdaDottedPattern{\AgdaBound{x}}\AgdaSpace{}%
\AgdaDottedPattern{\AgdaSymbol{.}}\AgdaDottedPattern{\AgdaBound{fs}}\AgdaSpace{}%
\AgdaDottedPattern{\AgdaSymbol{.}}\AgdaDottedPattern{\AgdaBound{sys}}\AgdaSpace{}%
\AgdaBound{x₂}\AgdaSpace{}%
\AgdaBound{ci}\AgdaSymbol{)}\AgdaSpace{}%
\AgdaSymbol{=}\AgdaSpace{}%
\AgdaInductiveConstructor{IC}\AgdaSpace{}%
\AgdaBound{x}\AgdaSpace{}%
\AgdaBound{fs}\AgdaSpace{}%
\AgdaSymbol{(}\AgdaFunction{St.run}\AgdaSpace{}%
\AgdaBound{x₁}\AgdaSpace{}%
\AgdaBound{sys}\AgdaSymbol{)}\AgdaSpace{}%
\AgdaFunction{≡₁}\AgdaSpace{}%
\AgdaSymbol{λ}\AgdaSpace{}%
\AgdaBound{all₁}\AgdaSpace{}%
\AgdaBound{f₁}\AgdaSpace{}%
\AgdaSymbol{→}\AgdaSpace{}%
\AgdaFunction{≡₂}\AgdaSpace{}%
\AgdaBound{f₁}\AgdaSpace{}%
\AgdaBound{all₁}\<%
\\
\>[0][@{}l@{\AgdaIndent{0}}]%
\>[2]\AgdaKeyword{where}%
\>[842I]\AgdaFunction{≡₀}\AgdaSpace{}%
\AgdaSymbol{:}\AgdaSpace{}%
\AgdaField{proj₁}\AgdaSpace{}%
\AgdaSymbol{(}\AgdaBound{oracle}\AgdaSpace{}%
\AgdaBound{x}\AgdaSymbol{)}\AgdaSpace{}%
\AgdaSymbol{(}\AgdaFunction{St.run}\AgdaSpace{}%
\AgdaBound{x₁}\AgdaSpace{}%
\AgdaBound{sys}\AgdaSymbol{)}\AgdaSpace{}%
\AgdaOperator{\AgdaDatatype{≡}}\AgdaSpace{}%
\AgdaField{proj₁}\AgdaSpace{}%
\AgdaSymbol{(}\AgdaBound{oracle}\AgdaSpace{}%
\AgdaBound{x}\AgdaSymbol{)}\AgdaSpace{}%
\AgdaBound{sys}\<%
\\
\>[.][@{}l@{}]\<[842I]%
\>[8]\AgdaFunction{≡₀}\AgdaSpace{}%
\AgdaSymbol{=}\AgdaSpace{}%
\AgdaFunction{sym}\AgdaSpace{}%
\AgdaSymbol{(}\AgdaField{proj₂}\AgdaSpace{}%
\AgdaSymbol{(}\AgdaBound{oracle}\AgdaSpace{}%
\AgdaBound{x}\AgdaSymbol{)}\AgdaSpace{}%
\AgdaBound{sys}\AgdaSpace{}%
\AgdaSymbol{(}\AgdaFunction{St.run}\AgdaSpace{}%
\AgdaBound{x₁}\AgdaSpace{}%
\AgdaBound{sys}\AgdaSymbol{)}\AgdaSpace{}%
\AgdaSymbol{λ}\AgdaSpace{}%
\AgdaBound{f₁}\AgdaSpace{}%
\AgdaBound{x₃}\AgdaSpace{}%
\AgdaSymbol{→}\AgdaSpace{}%
\AgdaFunction{lemma3}\AgdaSpace{}%
\AgdaBound{f₁}\AgdaSpace{}%
\AgdaSymbol{(}\AgdaField{proj₂}\AgdaSpace{}%
\AgdaSymbol{(}\AgdaField{proj₁}\AgdaSpace{}%
\AgdaSymbol{(}\AgdaBound{oracle}\AgdaSpace{}%
\AgdaBound{x₁}\AgdaSymbol{)}\AgdaSpace{}%
\AgdaBound{sys}\AgdaSymbol{))}\AgdaSpace{}%
\AgdaSymbol{λ}\AgdaSpace{}%
\AgdaBound{x₄}\AgdaSpace{}%
\AgdaSymbol{→}\AgdaSpace{}%
\AgdaBound{dsj}\AgdaSpace{}%
\AgdaSymbol{(}\AgdaBound{x₄}\AgdaSpace{}%
\AgdaOperator{\AgdaInductiveConstructor{,}}\AgdaSpace{}%
\AgdaFunction{∈-++⁺ˡ}\AgdaSpace{}%
\AgdaBound{x₃}\AgdaSymbol{))}\<%
\\
\>[8]\AgdaFunction{≡₁}\AgdaSpace{}%
\AgdaSymbol{:}\AgdaSpace{}%
\AgdaFunction{map}\AgdaSpace{}%
\AgdaField{proj₁}\AgdaSpace{}%
\AgdaBound{fs}\AgdaSpace{}%
\AgdaOperator{\AgdaDatatype{≡}}\AgdaSpace{}%
\AgdaFunction{cmdReadNames}\AgdaSpace{}%
\AgdaBound{x}\AgdaSpace{}%
\AgdaSymbol{(}\AgdaFunction{St.run}\AgdaSpace{}%
\AgdaBound{x₁}\AgdaSpace{}%
\AgdaBound{sys}\AgdaSymbol{)}\<%
\\
\>[8]\AgdaFunction{≡₁}\AgdaSpace{}%
\AgdaSymbol{=}\AgdaSpace{}%
\AgdaFunction{trans}\AgdaSpace{}%
\AgdaBound{x₂}\AgdaSpace{}%
\AgdaSymbol{(}\AgdaFunction{sym}\AgdaSpace{}%
\AgdaSymbol{(}\AgdaFunction{cong}\AgdaSpace{}%
\AgdaSymbol{(}\AgdaFunction{map}\AgdaSpace{}%
\AgdaField{proj₁}\AgdaSpace{}%
\AgdaOperator{\AgdaFunction{∘}}\AgdaSpace{}%
\AgdaField{proj₁}\AgdaSymbol{)}\AgdaSpace{}%
\AgdaFunction{≡₀}\AgdaSymbol{))}\<%
\\
\>[8]\AgdaFunction{≡₂}\AgdaSpace{}%
\AgdaSymbol{:}\AgdaSpace{}%
\AgdaSymbol{(}\AgdaBound{f₁}\AgdaSpace{}%
\AgdaSymbol{:}\AgdaSpace{}%
\AgdaFunction{FileName}\AgdaSymbol{)}\AgdaSpace{}%
\AgdaSymbol{->}\AgdaSpace{}%
\AgdaDatatype{All}\AgdaSpace{}%
\AgdaSymbol{(λ}\AgdaSpace{}%
\AgdaSymbol{(}\AgdaBound{f₁}\AgdaSpace{}%
\AgdaOperator{\AgdaInductiveConstructor{,}}\AgdaSpace{}%
\AgdaBound{v₁}\AgdaSymbol{)}\AgdaSpace{}%
\AgdaSymbol{→}\AgdaSpace{}%
\AgdaFunction{St.run}\AgdaSpace{}%
\AgdaBound{x₁}\AgdaSpace{}%
\AgdaBound{sys}\AgdaSpace{}%
\AgdaBound{f₁}\AgdaSpace{}%
\AgdaOperator{\AgdaDatatype{≡}}\AgdaSpace{}%
\AgdaBound{v₁}\AgdaSymbol{)}\AgdaSpace{}%
\AgdaBound{fs}\AgdaSpace{}%
\AgdaSymbol{->}\AgdaSpace{}%
\AgdaFunction{St.run}\AgdaSpace{}%
\AgdaBound{x}\AgdaSpace{}%
\AgdaSymbol{(}\AgdaFunction{St.run}\AgdaSpace{}%
\AgdaBound{x₁}\AgdaSpace{}%
\AgdaBound{sys}\AgdaSymbol{)}\AgdaSpace{}%
\AgdaBound{f₁}\AgdaSpace{}%
\AgdaOperator{\AgdaDatatype{≡}}\AgdaSpace{}%
\AgdaFunction{St.run}\AgdaSpace{}%
\AgdaBound{x₁}\AgdaSpace{}%
\AgdaBound{sys}\AgdaSpace{}%
\AgdaBound{f₁}\<%
\\
\>[8]\AgdaFunction{≡₂}\AgdaSpace{}%
\AgdaBound{f₁}\AgdaSpace{}%
\AgdaBound{all₁}\AgdaSpace{}%
\AgdaKeyword{with}\AgdaSpace{}%
\AgdaBound{f₁}\AgdaSpace{}%
\AgdaOperator{\AgdaFunction{∈?}}\AgdaSpace{}%
\AgdaFunction{cmdWriteNames}\AgdaSpace{}%
\AgdaBound{x}\AgdaSpace{}%
\AgdaSymbol{(}\AgdaFunction{St.run}\AgdaSpace{}%
\AgdaBound{x₁}\AgdaSpace{}%
\AgdaBound{sys}\AgdaSymbol{)}\<%
\\
\>[8]\AgdaSymbol{...}\AgdaSpace{}%
\AgdaSymbol{|}\AgdaSpace{}%
\AgdaInductiveConstructor{no}\AgdaSpace{}%
\AgdaBound{f₁∉}\AgdaSpace{}%
\AgdaSymbol{=}\AgdaSpace{}%
\AgdaFunction{sym}\AgdaSpace{}%
\AgdaSymbol{(}\AgdaFunction{lemma3}\AgdaSpace{}%
\AgdaBound{f₁}\AgdaSpace{}%
\AgdaSymbol{(}\AgdaField{proj₂}\AgdaSpace{}%
\AgdaSymbol{(}\AgdaField{proj₁}\AgdaSpace{}%
\AgdaSymbol{(}\AgdaBound{oracle}\AgdaSpace{}%
\AgdaBound{x}\AgdaSymbol{)}\AgdaSpace{}%
\AgdaSymbol{(}\AgdaFunction{St.run}\AgdaSpace{}%
\AgdaBound{x₁}\AgdaSpace{}%
\AgdaBound{sys}\AgdaSymbol{)))}\AgdaSpace{}%
\AgdaBound{f₁∉}\AgdaSymbol{)}\<%
\\
\>[8]\AgdaSymbol{...}\AgdaSpace{}%
\AgdaSymbol{|}\AgdaSpace{}%
\AgdaInductiveConstructor{yes}\AgdaSpace{}%
\AgdaBound{f₁∈}\AgdaSpace{}%
\AgdaKeyword{with}\AgdaSpace{}%
\AgdaBound{f₁}\AgdaSpace{}%
\AgdaOperator{\AgdaFunction{∈?}}\AgdaSpace{}%
\AgdaFunction{cmdWriteNames}\AgdaSpace{}%
\AgdaBound{x₁}\AgdaSpace{}%
\AgdaBound{sys}\<%
\\
\>[8]\AgdaSymbol{...}\AgdaSpace{}%
\AgdaSymbol{|}\AgdaSpace{}%
\AgdaInductiveConstructor{yes}\AgdaSpace{}%
\AgdaBound{f₁∈₁}\AgdaSpace{}%
\AgdaSymbol{=}\AgdaSpace{}%
\AgdaFunction{contradiction}\AgdaSpace{}%
\AgdaSymbol{(}\AgdaBound{f₁∈₁}\AgdaSpace{}%
\AgdaOperator{\AgdaInductiveConstructor{,}}\AgdaSpace{}%
\AgdaFunction{∈-++⁺ʳ}\AgdaSpace{}%
\AgdaSymbol{\AgdaUnderscore{}}\AgdaSpace{}%
\AgdaSymbol{(}\AgdaFunction{∈-resp-≡}\AgdaSpace{}%
\AgdaBound{f₁∈}\AgdaSpace{}%
\AgdaSymbol{(}\AgdaFunction{cong}\AgdaSpace{}%
\AgdaSymbol{(}\AgdaFunction{map}\AgdaSpace{}%
\AgdaField{proj₁}\AgdaSpace{}%
\AgdaOperator{\AgdaFunction{∘}}\AgdaSpace{}%
\AgdaField{proj₂}\AgdaSymbol{)}\AgdaSpace{}%
\AgdaFunction{≡₀}\AgdaSymbol{)))}\AgdaSpace{}%
\AgdaBound{dsj}\<%
\\
\>[8]\AgdaSymbol{...}\AgdaSpace{}%
\AgdaSymbol{|}\AgdaSpace{}%
\AgdaInductiveConstructor{no}\AgdaSpace{}%
\AgdaBound{f₁∉}\AgdaSpace{}%
\AgdaSymbol{=}\AgdaSpace{}%
\AgdaFunction{trans}%
\>[990I]\AgdaSymbol{(}\AgdaFunction{trans}\AgdaSpace{}%
\AgdaSymbol{(}\AgdaFunction{lemma4}\AgdaSpace{}%
\AgdaSymbol{\{}\AgdaFunction{St.run}\AgdaSpace{}%
\AgdaBound{x₁}\AgdaSpace{}%
\AgdaBound{sys}\AgdaSymbol{\}}\AgdaSpace{}%
\AgdaSymbol{\{}\AgdaBound{sys}\AgdaSymbol{\}}\AgdaSpace{}%
\AgdaSymbol{(}\AgdaField{proj₂}\AgdaSpace{}%
\AgdaSymbol{(}\AgdaField{proj₁}\AgdaSpace{}%
\AgdaSymbol{(}\AgdaBound{oracle}\AgdaSpace{}%
\AgdaBound{x}\AgdaSymbol{)}\AgdaSpace{}%
\AgdaSymbol{(}\AgdaFunction{St.run}\AgdaSpace{}%
\AgdaBound{x₁}\AgdaSpace{}%
\AgdaBound{sys}\AgdaSymbol{)))}\AgdaSpace{}%
\AgdaBound{f₁}\AgdaSpace{}%
\AgdaBound{f₁∈}\AgdaSymbol{)}\AgdaSpace{}%
\AgdaSymbol{(}\AgdaFunction{cong₂}\AgdaSpace{}%
\AgdaSymbol{(}\AgdaFunction{foldr}\AgdaSpace{}%
\AgdaFunction{extend}\AgdaSpace{}%
\AgdaBound{sys}\AgdaSpace{}%
\AgdaOperator{\AgdaFunction{∘}}\AgdaSpace{}%
\AgdaField{proj₂}\AgdaSymbol{)}\AgdaSpace{}%
\AgdaFunction{≡₀}\AgdaSpace{}%
\AgdaInductiveConstructor{refl}\AgdaSymbol{))}\<%
\\
\>[.][@{}l@{}]\<[990I]%
\>[29]\AgdaSymbol{(}\AgdaFunction{trans}\AgdaSpace{}%
\AgdaSymbol{(}\AgdaBound{ci}\AgdaSpace{}%
\AgdaSymbol{(}\AgdaFunction{lemma2}\AgdaSpace{}%
\AgdaBound{fs}\AgdaSpace{}%
\AgdaBound{all₁}\AgdaSpace{}%
\AgdaSymbol{λ}\AgdaSpace{}%
\AgdaBound{x₃}\AgdaSpace{}%
\AgdaSymbol{→}\AgdaSpace{}%
\AgdaBound{dsj}\AgdaSpace{}%
\AgdaSymbol{(}\AgdaField{proj₂}\AgdaSpace{}%
\AgdaBound{x₃}\AgdaSpace{}%
\AgdaOperator{\AgdaInductiveConstructor{,}}\AgdaSpace{}%
\AgdaFunction{∈-++⁺ˡ}\AgdaSpace{}%
\AgdaSymbol{(}\AgdaFunction{∈-resp-≡}\AgdaSpace{}%
\AgdaSymbol{(}\AgdaField{proj₁}\AgdaSpace{}%
\AgdaBound{x₃}\AgdaSymbol{)}\AgdaSpace{}%
\AgdaBound{x₂}\AgdaSymbol{)))}\AgdaSpace{}%
\AgdaBound{f₁}\AgdaSymbol{)}\AgdaSpace{}%
\AgdaSymbol{(}\AgdaFunction{lemma3}\AgdaSpace{}%
\AgdaBound{f₁}\AgdaSpace{}%
\AgdaSymbol{(}\AgdaField{proj₂}\AgdaSpace{}%
\AgdaSymbol{(}\AgdaField{proj₁}\AgdaSpace{}%
\AgdaSymbol{(}\AgdaBound{oracle}\AgdaSpace{}%
\AgdaBound{x₁}\AgdaSymbol{)}\AgdaSpace{}%
\AgdaBound{sys}\AgdaSymbol{))}\AgdaSpace{}%
\AgdaBound{f₁∉}\AgdaSymbol{))}\<%
\\
\\[\AgdaEmptyExtraSkip]%
\\[\AgdaEmptyExtraSkip]%
\>[0]\AgdaFunction{preserves2}\AgdaSpace{}%
\AgdaSymbol{:}\AgdaSpace{}%
\AgdaSymbol{\{}\AgdaBound{sys}\AgdaSpace{}%
\AgdaSymbol{:}\AgdaSpace{}%
\AgdaFunction{FileSystem}\AgdaSymbol{\}}\AgdaSpace{}%
\AgdaSymbol{\{}\AgdaBound{mm}\AgdaSpace{}%
\AgdaSymbol{:}\AgdaSpace{}%
\AgdaFunction{Memory}\AgdaSymbol{\}}\AgdaSpace{}%
\AgdaSymbol{(}\AgdaBound{x}\AgdaSpace{}%
\AgdaSymbol{:}\AgdaSpace{}%
\AgdaFunction{Cmd}\AgdaSymbol{)}\AgdaSpace{}%
\AgdaSymbol{->}\AgdaSpace{}%
\AgdaDatatype{IdempotentState}\AgdaSpace{}%
\AgdaFunction{cmdReadNames}\AgdaSpace{}%
\AgdaBound{sys}\AgdaSpace{}%
\AgdaBound{mm}\AgdaSpace{}%
\AgdaSymbol{->}\AgdaSpace{}%
\AgdaDatatype{All}\AgdaSpace{}%
\AgdaSymbol{(λ}\AgdaSpace{}%
\AgdaBound{x₁}\AgdaSpace{}%
\AgdaSymbol{→}\AgdaSpace{}%
\AgdaFunction{Disjoint}\AgdaSpace{}%
\AgdaSymbol{(}\AgdaFunction{cmdWriteNames}\AgdaSpace{}%
\AgdaBound{x}\AgdaSpace{}%
\AgdaBound{sys}\AgdaSymbol{)}\AgdaSpace{}%
\AgdaSymbol{(}\AgdaFunction{cmdReadWrites}\AgdaSpace{}%
\AgdaBound{x₁}\AgdaSpace{}%
\AgdaBound{sys}\AgdaSymbol{))}\AgdaSpace{}%
\AgdaSymbol{(}\AgdaFunction{map}\AgdaSpace{}%
\AgdaField{proj₁}\AgdaSpace{}%
\AgdaBound{mm}\AgdaSymbol{)}\AgdaSpace{}%
\AgdaSymbol{->}\AgdaSpace{}%
\AgdaDatatype{IdempotentState}\AgdaSpace{}%
\AgdaFunction{cmdReadNames}\AgdaSpace{}%
\AgdaSymbol{(}\AgdaFunction{St.run}\AgdaSpace{}%
\AgdaBound{x}\AgdaSpace{}%
\AgdaBound{sys}\AgdaSymbol{)}\AgdaSpace{}%
\AgdaBound{mm}\<%
\\
\>[0]\AgdaFunction{preserves2}\AgdaSpace{}%
\AgdaBound{x}\AgdaSpace{}%
\AgdaInductiveConstructor{[]}\AgdaSpace{}%
\AgdaBound{all₁}\AgdaSpace{}%
\AgdaSymbol{=}\AgdaSpace{}%
\AgdaInductiveConstructor{[]}\<%
\\
\>[0]\AgdaFunction{preserves2}\AgdaSpace{}%
\AgdaSymbol{\{}\AgdaBound{sys}\AgdaSymbol{\}}\AgdaSpace{}%
\AgdaSymbol{\{(}\AgdaBound{x₅}\AgdaSpace{}%
\AgdaOperator{\AgdaInductiveConstructor{,}}\AgdaSpace{}%
\AgdaSymbol{\AgdaUnderscore{})}\AgdaSpace{}%
\AgdaOperator{\AgdaInductiveConstructor{∷}}\AgdaSpace{}%
\AgdaBound{mm}\AgdaSymbol{\}}\AgdaSpace{}%
\AgdaBound{x}\AgdaSpace{}%
\AgdaSymbol{(}\AgdaInductiveConstructor{Cons}\AgdaSpace{}%
\AgdaBound{ic}\AgdaSpace{}%
\AgdaBound{is}\AgdaSymbol{)}\AgdaSpace{}%
\AgdaSymbol{(}\AgdaBound{px}\AgdaSpace{}%
\AgdaOperator{\AgdaInductiveConstructor{All.∷}}\AgdaSpace{}%
\AgdaBound{all₁}\AgdaSymbol{)}\AgdaSpace{}%
\AgdaSymbol{=}\AgdaSpace{}%
\AgdaInductiveConstructor{Cons}\AgdaSpace{}%
\AgdaSymbol{(}\AgdaFunction{runPreserves}\AgdaSpace{}%
\AgdaBound{x₅}\AgdaSpace{}%
\AgdaBound{x}\AgdaSpace{}%
\AgdaSymbol{\AgdaUnderscore{}}\AgdaSpace{}%
\AgdaBound{px}\AgdaSpace{}%
\AgdaBound{ic}\AgdaSymbol{)}\AgdaSpace{}%
\AgdaSymbol{(}\AgdaFunction{preserves2}\AgdaSpace{}%
\AgdaBound{x}\AgdaSpace{}%
\AgdaBound{is}\AgdaSpace{}%
\AgdaBound{all₁}\AgdaSymbol{)}\<%
\\
\\[\AgdaEmptyExtraSkip]%
\\[\AgdaEmptyExtraSkip]%
\>[0]\AgdaFunction{preserves}\AgdaSpace{}%
\AgdaSymbol{:}\AgdaSpace{}%
\AgdaSymbol{\{}\AgdaBound{sys}\AgdaSpace{}%
\AgdaSymbol{:}\AgdaSpace{}%
\AgdaFunction{FileSystem}\AgdaSymbol{\}}\AgdaSpace{}%
\AgdaSymbol{\{}\AgdaBound{mm}\AgdaSpace{}%
\AgdaSymbol{:}\AgdaSpace{}%
\AgdaFunction{Memory}\AgdaSymbol{\}}\AgdaSpace{}%
\AgdaSymbol{(}\AgdaBound{x}\AgdaSpace{}%
\AgdaSymbol{:}\AgdaSpace{}%
\AgdaFunction{Cmd}\AgdaSymbol{)}\AgdaSpace{}%
\AgdaSymbol{->}\AgdaSpace{}%
\AgdaFunction{Disjoint}\AgdaSpace{}%
\AgdaSymbol{(}\AgdaFunction{cmdReadNames}\AgdaSpace{}%
\AgdaBound{x}\AgdaSpace{}%
\AgdaBound{sys}\AgdaSymbol{)}\AgdaSpace{}%
\AgdaSymbol{(}\AgdaFunction{cmdWriteNames}\AgdaSpace{}%
\AgdaBound{x}\AgdaSpace{}%
\AgdaBound{sys}\AgdaSymbol{)}\AgdaSpace{}%
\AgdaSymbol{->}\AgdaSpace{}%
\AgdaDatatype{All}\AgdaSpace{}%
\AgdaSymbol{(λ}\AgdaSpace{}%
\AgdaBound{x₁}\AgdaSpace{}%
\AgdaSymbol{→}\AgdaSpace{}%
\AgdaFunction{Disjoint}\AgdaSpace{}%
\AgdaSymbol{(}\AgdaFunction{cmdWriteNames}\AgdaSpace{}%
\AgdaBound{x}\AgdaSpace{}%
\AgdaBound{sys}\AgdaSymbol{)}\AgdaSpace{}%
\AgdaSymbol{(}\AgdaFunction{cmdReadWrites}\AgdaSpace{}%
\AgdaBound{x₁}\AgdaSpace{}%
\AgdaBound{sys}\AgdaSymbol{))}\AgdaSpace{}%
\AgdaSymbol{(}\AgdaFunction{map}\AgdaSpace{}%
\AgdaField{proj₁}\AgdaSpace{}%
\AgdaBound{mm}\AgdaSymbol{)}\AgdaSpace{}%
\AgdaSymbol{->}\AgdaSpace{}%
\AgdaDatatype{IdempotentState}\AgdaSpace{}%
\AgdaFunction{cmdReadNames}\AgdaSpace{}%
\AgdaBound{sys}\AgdaSpace{}%
\AgdaBound{mm}\AgdaSpace{}%
\AgdaSymbol{->}\AgdaSpace{}%
\AgdaDatatype{IdempotentState}\AgdaSpace{}%
\AgdaFunction{cmdReadNames}\AgdaSpace{}%
\AgdaSymbol{(}\AgdaFunction{St.run}\AgdaSpace{}%
\AgdaBound{x}\AgdaSpace{}%
\AgdaBound{sys}\AgdaSymbol{)}\AgdaSpace{}%
\AgdaSymbol{(}\AgdaFunction{save}\AgdaSpace{}%
\AgdaBound{x}\AgdaSpace{}%
\AgdaSymbol{(}\AgdaFunction{cmdReadNames}\AgdaSpace{}%
\AgdaBound{x}\AgdaSpace{}%
\AgdaBound{sys}\AgdaSymbol{)}\AgdaSpace{}%
\AgdaSymbol{(}\AgdaFunction{St.run}\AgdaSpace{}%
\AgdaBound{x}\AgdaSpace{}%
\AgdaBound{sys}\AgdaSymbol{)}\AgdaSpace{}%
\AgdaBound{mm}\AgdaSymbol{)}\<%
\\
\>[0]\AgdaFunction{preserves}\AgdaSpace{}%
\AgdaSymbol{\{}\AgdaBound{sys}\AgdaSymbol{\}}\AgdaSpace{}%
\AgdaBound{x}\AgdaSpace{}%
\AgdaBound{dsj}\AgdaSpace{}%
\AgdaBound{all₁}\AgdaSpace{}%
\AgdaBound{is}\AgdaSpace{}%
\AgdaSymbol{=}\AgdaSpace{}%
\AgdaInductiveConstructor{Cons}\AgdaSpace{}%
\AgdaFunction{ic}\AgdaSpace{}%
\AgdaSymbol{(}\AgdaFunction{preserves2}\AgdaSpace{}%
\AgdaBound{x}\AgdaSpace{}%
\AgdaBound{is}\AgdaSpace{}%
\AgdaBound{all₁}\AgdaSymbol{)}\<%
\\
\>[0][@{}l@{\AgdaIndent{0}}]%
\>[2]\AgdaKeyword{where}%
\>[1170I]\AgdaFunction{g₁}\AgdaSpace{}%
\AgdaSymbol{:}\AgdaSpace{}%
\AgdaSymbol{(}\AgdaBound{xs}\AgdaSpace{}%
\AgdaBound{ys}\AgdaSpace{}%
\AgdaSymbol{:}\AgdaSpace{}%
\AgdaDatatype{List}\AgdaSpace{}%
\AgdaFunction{FileName}\AgdaSymbol{)}\AgdaSpace{}%
\AgdaSymbol{->}\AgdaSpace{}%
\AgdaBound{xs}\AgdaSpace{}%
\AgdaOperator{\AgdaDatatype{≡}}\AgdaSpace{}%
\AgdaBound{ys}\AgdaSpace{}%
\AgdaSymbol{->}\AgdaSpace{}%
\AgdaFunction{map}\AgdaSpace{}%
\AgdaField{proj₁}\AgdaSpace{}%
\AgdaSymbol{(}\AgdaFunction{map}\AgdaSpace{}%
\AgdaSymbol{(λ}\AgdaSpace{}%
\AgdaBound{f}\AgdaSpace{}%
\AgdaSymbol{→}\AgdaSpace{}%
\AgdaBound{f}\AgdaSpace{}%
\AgdaOperator{\AgdaInductiveConstructor{,}}\AgdaSpace{}%
\AgdaFunction{St.run}\AgdaSpace{}%
\AgdaBound{x}\AgdaSpace{}%
\AgdaBound{sys}\AgdaSpace{}%
\AgdaBound{f}\AgdaSymbol{)}\AgdaSpace{}%
\AgdaBound{xs}\AgdaSymbol{)}\AgdaSpace{}%
\AgdaOperator{\AgdaDatatype{≡}}\AgdaSpace{}%
\AgdaBound{ys}\<%
\\
\>[.][@{}l@{}]\<[1170I]%
\>[8]\AgdaFunction{g₁}\AgdaSpace{}%
\AgdaInductiveConstructor{[]}\AgdaSpace{}%
\AgdaBound{ys}\AgdaSpace{}%
\AgdaBound{≡₁}\AgdaSpace{}%
\AgdaSymbol{=}\AgdaSpace{}%
\AgdaBound{≡₁}\<%
\\
\>[8]\AgdaFunction{g₁}\AgdaSpace{}%
\AgdaSymbol{(}\AgdaBound{x}\AgdaSpace{}%
\AgdaOperator{\AgdaInductiveConstructor{∷}}\AgdaSpace{}%
\AgdaBound{xs}\AgdaSymbol{)}\AgdaSpace{}%
\AgdaSymbol{(}\AgdaBound{x₁}\AgdaSpace{}%
\AgdaOperator{\AgdaInductiveConstructor{∷}}\AgdaSpace{}%
\AgdaBound{ys}\AgdaSymbol{)}\AgdaSpace{}%
\AgdaBound{≡₁}\AgdaSpace{}%
\AgdaKeyword{with}\AgdaSpace{}%
\AgdaFunction{∷-injective}\AgdaSpace{}%
\AgdaBound{≡₁}\<%
\\
\>[8]\AgdaSymbol{...}\AgdaSpace{}%
\AgdaSymbol{|}\AgdaSpace{}%
\AgdaBound{x≡x₁}\AgdaSpace{}%
\AgdaOperator{\AgdaInductiveConstructor{,}}\AgdaSpace{}%
\AgdaBound{xs≡ys}\AgdaSpace{}%
\AgdaSymbol{=}\AgdaSpace{}%
\AgdaFunction{cong₂}\AgdaSpace{}%
\AgdaOperator{\AgdaInductiveConstructor{\AgdaUnderscore{}∷\AgdaUnderscore{}}}\AgdaSpace{}%
\AgdaBound{x≡x₁}\AgdaSpace{}%
\AgdaSymbol{(}\AgdaFunction{g₁}\AgdaSpace{}%
\AgdaBound{xs}\AgdaSpace{}%
\AgdaBound{ys}\AgdaSpace{}%
\AgdaBound{xs≡ys}\AgdaSymbol{)}\<%
\\
\>[8]\AgdaFunction{g₂}\AgdaSpace{}%
\AgdaSymbol{:}\AgdaSpace{}%
\AgdaFunction{cmdReadNames}\AgdaSpace{}%
\AgdaBound{x}\AgdaSpace{}%
\AgdaBound{sys}\AgdaSpace{}%
\AgdaOperator{\AgdaDatatype{≡}}\AgdaSpace{}%
\AgdaFunction{cmdReadNames}\AgdaSpace{}%
\AgdaBound{x}\AgdaSpace{}%
\AgdaSymbol{(}\AgdaFunction{St.run}\AgdaSpace{}%
\AgdaBound{x}\AgdaSpace{}%
\AgdaBound{sys}\AgdaSymbol{)}\<%
\\
\>[8]\AgdaFunction{g₂}\AgdaSpace{}%
\AgdaKeyword{with}\AgdaSpace{}%
\AgdaField{proj₂}\AgdaSpace{}%
\AgdaSymbol{(}\AgdaBound{oracle}\AgdaSpace{}%
\AgdaBound{x}\AgdaSymbol{)}\AgdaSpace{}%
\AgdaBound{sys}\AgdaSpace{}%
\AgdaSymbol{(}\AgdaFunction{St.run}\AgdaSpace{}%
\AgdaBound{x}\AgdaSpace{}%
\AgdaBound{sys}\AgdaSymbol{)}\AgdaSpace{}%
\AgdaSymbol{(λ}\AgdaSpace{}%
\AgdaBound{f₁}\AgdaSpace{}%
\AgdaBound{x₁}\AgdaSpace{}%
\AgdaSymbol{→}\AgdaSpace{}%
\AgdaSymbol{(}\AgdaFunction{lemma3}\AgdaSpace{}%
\AgdaBound{f₁}\AgdaSpace{}%
\AgdaSymbol{(}\AgdaField{proj₂}\AgdaSpace{}%
\AgdaSymbol{(}\AgdaField{proj₁}\AgdaSpace{}%
\AgdaSymbol{(}\AgdaBound{oracle}\AgdaSpace{}%
\AgdaBound{x}\AgdaSymbol{)}\AgdaSpace{}%
\AgdaBound{sys}\AgdaSymbol{))}\AgdaSpace{}%
\AgdaSymbol{λ}\AgdaSpace{}%
\AgdaBound{x₂}\AgdaSpace{}%
\AgdaSymbol{→}\AgdaSpace{}%
\AgdaBound{dsj}\AgdaSpace{}%
\AgdaSymbol{(}\AgdaBound{x₁}\AgdaSpace{}%
\AgdaOperator{\AgdaInductiveConstructor{,}}\AgdaSpace{}%
\AgdaBound{x₂}\AgdaSymbol{)))}\<%
\\
\>[8]\AgdaSymbol{...}\AgdaSpace{}%
\AgdaSymbol{|}\AgdaSpace{}%
\AgdaBound{≡₁}\AgdaSpace{}%
\AgdaSymbol{=}\AgdaSpace{}%
\AgdaFunction{cong}\AgdaSpace{}%
\AgdaSymbol{(}\AgdaFunction{map}\AgdaSpace{}%
\AgdaField{proj₁}\AgdaSpace{}%
\AgdaOperator{\AgdaFunction{∘}}\AgdaSpace{}%
\AgdaField{proj₁}\AgdaSymbol{)}\AgdaSpace{}%
\AgdaBound{≡₁}\<%
\\
\>[8]\AgdaFunction{ic}\AgdaSpace{}%
\AgdaSymbol{:}\AgdaSpace{}%
\AgdaDatatype{IdempotentCmd}\AgdaSpace{}%
\AgdaFunction{cmdReadNames}\AgdaSpace{}%
\AgdaBound{x}\AgdaSpace{}%
\AgdaSymbol{(}\AgdaFunction{map}\AgdaSpace{}%
\AgdaSymbol{(λ}\AgdaSpace{}%
\AgdaBound{f}\AgdaSpace{}%
\AgdaSymbol{→}\AgdaSpace{}%
\AgdaBound{f}\AgdaSpace{}%
\AgdaOperator{\AgdaInductiveConstructor{,}}\AgdaSpace{}%
\AgdaFunction{St.run}\AgdaSpace{}%
\AgdaBound{x}\AgdaSpace{}%
\AgdaBound{sys}\AgdaSpace{}%
\AgdaBound{f}\AgdaSymbol{)}\AgdaSpace{}%
\AgdaSymbol{(}\AgdaFunction{cmdReadNames}\AgdaSpace{}%
\AgdaBound{x}\AgdaSpace{}%
\AgdaBound{sys}\AgdaSymbol{))}\AgdaSpace{}%
\AgdaSymbol{(}\AgdaFunction{St.run}\AgdaSpace{}%
\AgdaBound{x}\AgdaSpace{}%
\AgdaBound{sys}\AgdaSymbol{)}\<%
\\
\>[8]\AgdaFunction{ic}\AgdaSpace{}%
\AgdaSymbol{=}\AgdaSpace{}%
\AgdaInductiveConstructor{IC}\AgdaSpace{}%
\AgdaBound{x}%
\>[1292I]\AgdaSymbol{(}\AgdaFunction{map}\AgdaSpace{}%
\AgdaSymbol{(λ}\AgdaSpace{}%
\AgdaBound{f}\AgdaSpace{}%
\AgdaSymbol{→}\AgdaSpace{}%
\AgdaBound{f}\AgdaSpace{}%
\AgdaOperator{\AgdaInductiveConstructor{,}}\AgdaSpace{}%
\AgdaFunction{St.run}\AgdaSpace{}%
\AgdaBound{x}\AgdaSpace{}%
\AgdaBound{sys}\AgdaSpace{}%
\AgdaBound{f}\AgdaSymbol{)}\AgdaSpace{}%
\AgdaSymbol{(}\AgdaFunction{cmdReadNames}\AgdaSpace{}%
\AgdaBound{x}\AgdaSpace{}%
\AgdaBound{sys}\AgdaSymbol{))}\<%
\\
\>[.][@{}l@{}]\<[1292I]%
\>[18]\AgdaSymbol{(}\AgdaFunction{St.run}\AgdaSpace{}%
\AgdaBound{x}\AgdaSpace{}%
\AgdaBound{sys}\AgdaSymbol{)}\<%
\\
\>[18]\AgdaSymbol{(}\AgdaFunction{g₁}\AgdaSpace{}%
\AgdaSymbol{(}\AgdaFunction{cmdReadNames}\AgdaSpace{}%
\AgdaBound{x}\AgdaSpace{}%
\AgdaBound{sys}\AgdaSymbol{)}\AgdaSpace{}%
\AgdaSymbol{(}\AgdaFunction{cmdReadNames}\AgdaSpace{}%
\AgdaBound{x}\AgdaSpace{}%
\AgdaSymbol{(}\AgdaFunction{St.run}\AgdaSpace{}%
\AgdaBound{x}\AgdaSpace{}%
\AgdaBound{sys}\AgdaSymbol{))}\AgdaSpace{}%
\AgdaSymbol{(}\AgdaFunction{g₂}\AgdaSymbol{))}\<%
\\
\>[18]\AgdaSymbol{λ}\AgdaSpace{}%
\AgdaBound{\AgdaUnderscore{}}\AgdaSpace{}%
\AgdaSymbol{→}\AgdaSpace{}%
\AgdaFunction{cmdIdempotent}\AgdaSpace{}%
\AgdaBound{sys}\AgdaSpace{}%
\AgdaBound{x}\AgdaSpace{}%
\AgdaBound{dsj}\<%
\\
\\[\AgdaEmptyExtraSkip]%
\>[0]\AgdaFunction{getCmdIdempotent}\AgdaSpace{}%
\AgdaSymbol{:}\AgdaSpace{}%
\AgdaSymbol{\{}\AgdaBound{f}\AgdaSpace{}%
\AgdaSymbol{:}\AgdaSpace{}%
\AgdaSymbol{(}\AgdaFunction{Cmd}\AgdaSpace{}%
\AgdaSymbol{->}\AgdaSpace{}%
\AgdaFunction{FileSystem}\AgdaSpace{}%
\AgdaSymbol{->}\AgdaSpace{}%
\AgdaDatatype{List}\AgdaSpace{}%
\AgdaFunction{FileName}\AgdaSymbol{)\}}\AgdaSpace{}%
\AgdaSymbol{\{}\AgdaBound{sys}\AgdaSpace{}%
\AgdaSymbol{:}\AgdaSpace{}%
\AgdaFunction{FileSystem}\AgdaSymbol{\}}\AgdaSpace{}%
\AgdaSymbol{(}\AgdaBound{mm}\AgdaSpace{}%
\AgdaSymbol{:}\AgdaSpace{}%
\AgdaFunction{Memory}\AgdaSymbol{)}\AgdaSpace{}%
\AgdaSymbol{(}\AgdaBound{x}\AgdaSpace{}%
\AgdaSymbol{:}\AgdaSpace{}%
\AgdaFunction{Cmd}\AgdaSymbol{)}\AgdaSpace{}%
\AgdaSymbol{->}\AgdaSpace{}%
\AgdaDatatype{IdempotentState}\AgdaSpace{}%
\AgdaBound{f}\AgdaSpace{}%
\AgdaBound{sys}\AgdaSpace{}%
\AgdaBound{mm}\AgdaSpace{}%
\AgdaSymbol{->}\AgdaSpace{}%
\AgdaSymbol{(}\AgdaBound{x∈}\AgdaSpace{}%
\AgdaSymbol{:}\AgdaSpace{}%
\AgdaBound{x}\AgdaSpace{}%
\AgdaOperator{\AgdaFunction{∈}}\AgdaSpace{}%
\AgdaFunction{map}\AgdaSpace{}%
\AgdaField{proj₁}\AgdaSpace{}%
\AgdaBound{mm}\AgdaSymbol{)}\AgdaSpace{}%
\AgdaSymbol{->}\AgdaSpace{}%
\AgdaFunction{CmdIdempotent}\AgdaSpace{}%
\AgdaBound{x}\AgdaSpace{}%
\AgdaSymbol{(}\AgdaFunction{get}\AgdaSpace{}%
\AgdaBound{x}\AgdaSpace{}%
\AgdaBound{mm}\AgdaSpace{}%
\AgdaBound{x∈}\AgdaSymbol{)}\AgdaSpace{}%
\AgdaBound{sys}\<%
\\
\>[0]\AgdaFunction{getCmdIdempotent}\AgdaSpace{}%
\AgdaSymbol{((}\AgdaBound{x₁}\AgdaSpace{}%
\AgdaOperator{\AgdaInductiveConstructor{,}}\AgdaSpace{}%
\AgdaBound{fs}\AgdaSymbol{)}\AgdaSpace{}%
\AgdaOperator{\AgdaInductiveConstructor{∷}}\AgdaSpace{}%
\AgdaBound{mm}\AgdaSymbol{)}\AgdaSpace{}%
\AgdaBound{x}\AgdaSpace{}%
\AgdaSymbol{(}\AgdaInductiveConstructor{Cons}\AgdaSpace{}%
\AgdaSymbol{(}\AgdaInductiveConstructor{IC}\AgdaSpace{}%
\AgdaDottedPattern{\AgdaSymbol{.}}\AgdaDottedPattern{\AgdaBound{x₁}}\AgdaSpace{}%
\AgdaDottedPattern{\AgdaSymbol{.}}\AgdaDottedPattern{\AgdaBound{fs}}\AgdaSpace{}%
\AgdaSymbol{\AgdaUnderscore{}}\AgdaSpace{}%
\AgdaBound{≡₂}\AgdaSpace{}%
\AgdaBound{x₃}\AgdaSymbol{)}\AgdaSpace{}%
\AgdaBound{is}\AgdaSymbol{)}\AgdaSpace{}%
\AgdaBound{x∈}\AgdaSpace{}%
\AgdaKeyword{with}\AgdaSpace{}%
\AgdaBound{x}\AgdaSpace{}%
\AgdaOperator{\AgdaFunction{≟}}\AgdaSpace{}%
\AgdaBound{x₁}\<%
\\
\>[0]\AgdaSymbol{...}\AgdaSpace{}%
\AgdaSymbol{|}\AgdaSpace{}%
\AgdaInductiveConstructor{yes}\AgdaSpace{}%
\AgdaBound{x≡x₁}\AgdaSpace{}%
\AgdaSymbol{=}\AgdaSpace{}%
\AgdaFunction{subst}\AgdaSpace{}%
\AgdaSymbol{(λ}\AgdaSpace{}%
\AgdaBound{x₂}\AgdaSpace{}%
\AgdaSymbol{→}\AgdaSpace{}%
\AgdaFunction{CmdIdempotent}\AgdaSpace{}%
\AgdaBound{x₂}\AgdaSpace{}%
\AgdaSymbol{\AgdaUnderscore{}}\AgdaSpace{}%
\AgdaSymbol{\AgdaUnderscore{})}\AgdaSpace{}%
\AgdaSymbol{(}\AgdaFunction{sym}\AgdaSpace{}%
\AgdaBound{x≡x₁}\AgdaSymbol{)}\AgdaSpace{}%
\AgdaBound{x₃}\<%
\\
\>[0]\AgdaSymbol{...}\AgdaSpace{}%
\AgdaSymbol{|}\AgdaSpace{}%
\AgdaInductiveConstructor{no}\AgdaSpace{}%
\AgdaBound{¬x≡x₁}\AgdaSpace{}%
\AgdaSymbol{=}\AgdaSpace{}%
\AgdaFunction{getCmdIdempotent}\AgdaSpace{}%
\AgdaBound{mm}\AgdaSpace{}%
\AgdaBound{x}\AgdaSpace{}%
\AgdaBound{is}\AgdaSpace{}%
\AgdaSymbol{(}\AgdaFunction{tail}\AgdaSpace{}%
\AgdaBound{¬x≡x₁}\AgdaSpace{}%
\AgdaBound{x∈}\AgdaSymbol{)}\<%
\\
\\[\AgdaEmptyExtraSkip]%
\\[\AgdaEmptyExtraSkip]%
\>[0]\AgdaFunction{run≡}\AgdaSpace{}%
\AgdaSymbol{:}\AgdaSpace{}%
\AgdaSymbol{(}\AgdaBound{sys₁}\AgdaSpace{}%
\AgdaBound{sys₂}\AgdaSpace{}%
\AgdaSymbol{:}\AgdaSpace{}%
\AgdaFunction{FileSystem}\AgdaSymbol{)}\AgdaSpace{}%
\AgdaSymbol{(}\AgdaBound{mm}\AgdaSpace{}%
\AgdaSymbol{:}\AgdaSpace{}%
\AgdaFunction{Memory}\AgdaSymbol{)}\AgdaSpace{}%
\AgdaSymbol{(}\AgdaBound{x}\AgdaSpace{}%
\AgdaSymbol{:}\AgdaSpace{}%
\AgdaFunction{Cmd}\AgdaSymbol{)}\AgdaSpace{}%
\AgdaSymbol{->}\AgdaSpace{}%
\AgdaSymbol{(∀}\AgdaSpace{}%
\AgdaBound{f₁}\AgdaSpace{}%
\AgdaSymbol{→}\AgdaSpace{}%
\AgdaBound{sys₁}\AgdaSpace{}%
\AgdaBound{f₁}\AgdaSpace{}%
\AgdaOperator{\AgdaDatatype{≡}}\AgdaSpace{}%
\AgdaBound{sys₂}\AgdaSpace{}%
\AgdaBound{f₁}\AgdaSymbol{)}\AgdaSpace{}%
\AgdaSymbol{->}\AgdaSpace{}%
\AgdaDatatype{IdempotentState}\AgdaSpace{}%
\AgdaFunction{cmdReadNames}\AgdaSpace{}%
\AgdaBound{sys₂}\AgdaSpace{}%
\AgdaBound{mm}\AgdaSpace{}%
\AgdaSymbol{->}\AgdaSpace{}%
\AgdaSymbol{∀}\AgdaSpace{}%
\AgdaBound{f₁}\AgdaSpace{}%
\AgdaSymbol{→}\AgdaSpace{}%
\AgdaFunction{St.run}\AgdaSpace{}%
\AgdaBound{x}\AgdaSpace{}%
\AgdaBound{sys₁}\AgdaSpace{}%
\AgdaBound{f₁}\AgdaSpace{}%
\AgdaOperator{\AgdaDatatype{≡}}\AgdaSpace{}%
\AgdaField{proj₁}\AgdaSpace{}%
\AgdaSymbol{(}\AgdaFunction{runF}\AgdaSpace{}%
\AgdaBound{x}\AgdaSpace{}%
\AgdaSymbol{(}\AgdaBound{sys₂}\AgdaSpace{}%
\AgdaOperator{\AgdaInductiveConstructor{,}}\AgdaSpace{}%
\AgdaBound{mm}\AgdaSymbol{))}\AgdaSpace{}%
\AgdaBound{f₁}\<%
\\
\>[0]\AgdaFunction{run≡}\AgdaSpace{}%
\AgdaBound{sys₁}\AgdaSpace{}%
\AgdaBound{sys₂}\AgdaSpace{}%
\AgdaBound{mm}\AgdaSpace{}%
\AgdaBound{x}\AgdaSpace{}%
\AgdaBound{∀₁}\AgdaSpace{}%
\AgdaBound{is}\AgdaSpace{}%
\AgdaBound{f₁}\AgdaSpace{}%
\AgdaKeyword{with}\AgdaSpace{}%
\AgdaBound{x}\AgdaSpace{}%
\AgdaOperator{\AgdaFunction{∈?}}\AgdaSpace{}%
\AgdaFunction{map}\AgdaSpace{}%
\AgdaField{proj₁}\AgdaSpace{}%
\AgdaBound{mm}\<%
\\
\>[0]\AgdaSymbol{...}\AgdaSpace{}%
\AgdaSymbol{|}\AgdaSpace{}%
\AgdaInductiveConstructor{no}\AgdaSpace{}%
\AgdaBound{x∉}\AgdaSpace{}%
\AgdaSymbol{=}\AgdaSpace{}%
\AgdaFunction{StP.lemma2}\AgdaSpace{}%
\AgdaSymbol{\{}\AgdaBound{sys₁}\AgdaSymbol{\}}\AgdaSpace{}%
\AgdaSymbol{\{}\AgdaBound{sys₂}\AgdaSymbol{\}}\AgdaSpace{}%
\AgdaSymbol{(}\AgdaField{proj₂}\AgdaSpace{}%
\AgdaSymbol{(}\AgdaBound{oracle}\AgdaSpace{}%
\AgdaBound{x}\AgdaSymbol{)}\AgdaSpace{}%
\AgdaBound{sys₁}\AgdaSpace{}%
\AgdaBound{sys₂}\AgdaSpace{}%
\AgdaSymbol{λ}\AgdaSpace{}%
\AgdaBound{f₂}\AgdaSpace{}%
\AgdaBound{\AgdaUnderscore{}}\AgdaSpace{}%
\AgdaSymbol{→}\AgdaSpace{}%
\AgdaBound{∀₁}\AgdaSpace{}%
\AgdaBound{f₂}\AgdaSymbol{)}\AgdaSpace{}%
\AgdaSymbol{(}\AgdaBound{∀₁}\AgdaSpace{}%
\AgdaBound{f₁}\AgdaSymbol{)}\<%
\\
\>[0]\AgdaSymbol{...}\AgdaSpace{}%
\AgdaSymbol{|}\AgdaSpace{}%
\AgdaInductiveConstructor{yes}\AgdaSpace{}%
\AgdaBound{x∈}\AgdaSpace{}%
\AgdaKeyword{with}\AgdaSpace{}%
\AgdaFunction{maybeAll}\AgdaSpace{}%
\AgdaSymbol{\{}\AgdaBound{sys₂}\AgdaSymbol{\}}\AgdaSpace{}%
\AgdaSymbol{(}\AgdaFunction{get}\AgdaSpace{}%
\AgdaBound{x}\AgdaSpace{}%
\AgdaBound{mm}\AgdaSpace{}%
\AgdaBound{x∈}\AgdaSymbol{)}\<%
\\
\>[0]\AgdaSymbol{...}\AgdaSpace{}%
\AgdaSymbol{|}\AgdaSpace{}%
\AgdaInductiveConstructor{nothing}\AgdaSpace{}%
\AgdaSymbol{=}\AgdaSpace{}%
\AgdaFunction{StP.lemma2}\AgdaSpace{}%
\AgdaSymbol{\{}\AgdaBound{sys₁}\AgdaSymbol{\}}\AgdaSpace{}%
\AgdaSymbol{\{}\AgdaBound{sys₂}\AgdaSymbol{\}}\AgdaSpace{}%
\AgdaSymbol{(}\AgdaField{proj₂}\AgdaSpace{}%
\AgdaSymbol{(}\AgdaBound{oracle}\AgdaSpace{}%
\AgdaBound{x}\AgdaSymbol{)}\AgdaSpace{}%
\AgdaBound{sys₁}\AgdaSpace{}%
\AgdaBound{sys₂}\AgdaSpace{}%
\AgdaSymbol{λ}\AgdaSpace{}%
\AgdaBound{f₂}\AgdaSpace{}%
\AgdaBound{\AgdaUnderscore{}}\AgdaSpace{}%
\AgdaSymbol{→}\AgdaSpace{}%
\AgdaBound{∀₁}\AgdaSpace{}%
\AgdaBound{f₂}\AgdaSymbol{)}\AgdaSpace{}%
\AgdaSymbol{(}\AgdaBound{∀₁}\AgdaSpace{}%
\AgdaBound{f₁}\AgdaSymbol{)}\<%
\\
\>[0]\AgdaSymbol{...}\AgdaSpace{}%
\AgdaSymbol{|}\AgdaSpace{}%
\AgdaInductiveConstructor{just}\AgdaSpace{}%
\AgdaBound{all₁}\AgdaSpace{}%
\AgdaKeyword{with}\AgdaSpace{}%
\AgdaFunction{getCmdIdempotent}\AgdaSpace{}%
\AgdaBound{mm}\AgdaSpace{}%
\AgdaBound{x}\AgdaSpace{}%
\AgdaBound{is}\AgdaSpace{}%
\AgdaBound{x∈}\<%
\\
\>[0]\AgdaSymbol{...}\AgdaSpace{}%
\AgdaSymbol{|}\AgdaSpace{}%
\AgdaBound{ci}\AgdaSpace{}%
\AgdaSymbol{=}\AgdaSpace{}%
\AgdaFunction{trans}%
\>[1522I]\AgdaSymbol{(}\AgdaFunction{StP.lemma2}\AgdaSpace{}%
\AgdaSymbol{\{}\AgdaBound{sys₁}\AgdaSymbol{\}}\AgdaSpace{}%
\AgdaSymbol{\{}\AgdaBound{sys₂}\AgdaSymbol{\}}\AgdaSpace{}%
\AgdaSymbol{(}\AgdaField{proj₂}\AgdaSpace{}%
\AgdaSymbol{(}\AgdaBound{oracle}\AgdaSpace{}%
\AgdaBound{x}\AgdaSymbol{)}\AgdaSpace{}%
\AgdaBound{sys₁}\AgdaSpace{}%
\AgdaBound{sys₂}\AgdaSpace{}%
\AgdaSymbol{λ}\AgdaSpace{}%
\AgdaBound{f₂}\AgdaSpace{}%
\AgdaBound{\AgdaUnderscore{}}\AgdaSpace{}%
\AgdaSymbol{→}\AgdaSpace{}%
\AgdaBound{∀₁}\AgdaSpace{}%
\AgdaBound{f₂}\AgdaSymbol{)}\AgdaSpace{}%
\AgdaSymbol{(}\AgdaBound{∀₁}\AgdaSpace{}%
\AgdaBound{f₁}\AgdaSymbol{))}\<%
\\
\>[1522I][@{}l@{\AgdaIndent{0}}]%
\>[24]\AgdaSymbol{(}\AgdaBound{ci}\AgdaSpace{}%
\AgdaBound{all₁}\AgdaSpace{}%
\AgdaBound{f₁}\AgdaSymbol{)}\<%
\\
\\[\AgdaEmptyExtraSkip]%
\>[0]\AgdaFunction{helper}\AgdaSpace{}%
\AgdaSymbol{:}\AgdaSpace{}%
\AgdaSymbol{∀}\AgdaSpace{}%
\AgdaSymbol{\{}\AgdaBound{s}\AgdaSymbol{\}}\AgdaSpace{}%
\AgdaBound{ls}\AgdaSpace{}%
\AgdaBound{ls₁}\AgdaSpace{}%
\AgdaBound{x}\AgdaSpace{}%
\AgdaSymbol{→}\AgdaSpace{}%
\AgdaFunction{Disjoint}\AgdaSpace{}%
\AgdaSymbol{(}\AgdaFunction{cmdWriteNames}\AgdaSpace{}%
\AgdaBound{x}\AgdaSpace{}%
\AgdaBound{s}\AgdaSymbol{)}\AgdaSpace{}%
\AgdaBound{ls₁}\AgdaSpace{}%
\AgdaSymbol{->}\AgdaSpace{}%
\AgdaSymbol{(}\AgdaFunction{concatMap}\AgdaSpace{}%
\AgdaSymbol{(λ}\AgdaSpace{}%
\AgdaBound{x₁}\AgdaSpace{}%
\AgdaSymbol{→}\AgdaSpace{}%
\AgdaFunction{cmdReadWriteNames}\AgdaSpace{}%
\AgdaBound{x₁}\AgdaSpace{}%
\AgdaBound{s}\AgdaSymbol{)}\AgdaSpace{}%
\AgdaBound{ls}\AgdaSymbol{)}\AgdaSpace{}%
\AgdaOperator{\AgdaFunction{⊆}}\AgdaSpace{}%
\AgdaBound{ls₁}\AgdaSpace{}%
\AgdaSymbol{->}\AgdaSpace{}%
\AgdaDatatype{All}\AgdaSpace{}%
\AgdaSymbol{(λ}\AgdaSpace{}%
\AgdaBound{x₁}\AgdaSpace{}%
\AgdaSymbol{→}\AgdaSpace{}%
\AgdaFunction{Disjoint}\AgdaSpace{}%
\AgdaSymbol{(}\AgdaFunction{cmdWriteNames}\AgdaSpace{}%
\AgdaBound{x}\AgdaSpace{}%
\AgdaBound{s}\AgdaSymbol{)}\AgdaSpace{}%
\AgdaSymbol{(}\AgdaFunction{cmdReadWriteNames}\AgdaSpace{}%
\AgdaBound{x₁}\AgdaSpace{}%
\AgdaBound{s}\AgdaSymbol{))}\AgdaSpace{}%
\AgdaBound{ls}\<%
\\
\>[0]\AgdaFunction{helper}\AgdaSpace{}%
\AgdaInductiveConstructor{[]}\AgdaSpace{}%
\AgdaBound{ls₁}\AgdaSpace{}%
\AgdaBound{x}\AgdaSpace{}%
\AgdaBound{dsj}\AgdaSpace{}%
\AgdaBound{⊆₁}\AgdaSpace{}%
\AgdaSymbol{=}\AgdaSpace{}%
\AgdaInductiveConstructor{All.[]}\<%
\\
\>[0]\AgdaFunction{helper}\AgdaSpace{}%
\AgdaSymbol{\{}\AgdaBound{sys}\AgdaSymbol{\}}\AgdaSpace{}%
\AgdaSymbol{(}\AgdaBound{x₁}\AgdaSpace{}%
\AgdaOperator{\AgdaInductiveConstructor{∷}}\AgdaSpace{}%
\AgdaBound{ls}\AgdaSymbol{)}\AgdaSpace{}%
\AgdaBound{ls₁}\AgdaSpace{}%
\AgdaBound{x}\AgdaSpace{}%
\AgdaBound{dsj}\AgdaSpace{}%
\AgdaBound{⊆₁}\AgdaSpace{}%
\AgdaSymbol{=}\AgdaSpace{}%
\AgdaSymbol{(λ}\AgdaSpace{}%
\AgdaBound{x₂}\AgdaSpace{}%
\AgdaSymbol{→}\AgdaSpace{}%
\AgdaBound{dsj}\AgdaSpace{}%
\AgdaSymbol{((}\AgdaField{proj₁}\AgdaSpace{}%
\AgdaBound{x₂}\AgdaSymbol{)}\AgdaSpace{}%
\AgdaOperator{\AgdaInductiveConstructor{,}}\AgdaSpace{}%
\AgdaBound{⊆₁}\AgdaSpace{}%
\AgdaSymbol{(}\AgdaFunction{∈-++⁺ˡ}\AgdaSpace{}%
\AgdaSymbol{(}\AgdaField{proj₂}\AgdaSpace{}%
\AgdaBound{x₂}\AgdaSymbol{))))}\AgdaSpace{}%
\AgdaOperator{\AgdaInductiveConstructor{All.∷}}\AgdaSpace{}%
\AgdaSymbol{(}\AgdaFunction{helper}\AgdaSpace{}%
\AgdaBound{ls}\AgdaSpace{}%
\AgdaBound{ls₁}\AgdaSpace{}%
\AgdaBound{x}\AgdaSpace{}%
\AgdaBound{dsj}\AgdaSpace{}%
\AgdaSymbol{λ}\AgdaSpace{}%
\AgdaBound{x₂}\AgdaSpace{}%
\AgdaSymbol{→}\AgdaSpace{}%
\AgdaBound{⊆₁}\AgdaSpace{}%
\AgdaSymbol{(}\AgdaFunction{∈-++⁺ʳ}\AgdaSpace{}%
\AgdaSymbol{(}\AgdaFunction{cmdReadWrites}\AgdaSpace{}%
\AgdaBound{x₁}\AgdaSpace{}%
\AgdaBound{sys}\AgdaSymbol{)}\AgdaSpace{}%
\AgdaBound{x₂}\AgdaSymbol{))}\<%
\\
\\[\AgdaEmptyExtraSkip]%
\\[\AgdaEmptyExtraSkip]%
\>[0]\AgdaFunction{∃hazard}\AgdaSpace{}%
\AgdaSymbol{:}\AgdaSpace{}%
\AgdaSymbol{∀}\AgdaSpace{}%
\AgdaSymbol{\{}\AgdaBound{v}\AgdaSymbol{\}\{}\AgdaBound{s}\AgdaSymbol{\}\{}\AgdaBound{x}\AgdaSymbol{\}}\AgdaSpace{}%
\AgdaBound{ls}\AgdaSpace{}%
\AgdaSymbol{\{}\AgdaBound{b}\AgdaSymbol{\}}\AgdaSpace{}%
\AgdaSymbol{→}\AgdaSpace{}%
\AgdaBound{v}\AgdaSpace{}%
\AgdaOperator{\AgdaFunction{∈}}\AgdaSpace{}%
\AgdaFunction{cmdWriteNames}\AgdaSpace{}%
\AgdaBound{x}\AgdaSpace{}%
\AgdaBound{s}\AgdaSpace{}%
\AgdaSymbol{→}\AgdaSpace{}%
\AgdaBound{v}\AgdaSpace{}%
\AgdaOperator{\AgdaFunction{∈}}\AgdaSpace{}%
\AgdaFunction{files}\AgdaSpace{}%
\AgdaBound{ls}\AgdaSpace{}%
\AgdaSymbol{→}\AgdaSpace{}%
\AgdaDatatype{Hazard}\AgdaSpace{}%
\AgdaBound{s}\AgdaSpace{}%
\AgdaBound{x}\AgdaSpace{}%
\AgdaBound{b}\AgdaSpace{}%
\AgdaBound{ls}\<%
\\
\>[0]\AgdaFunction{∃hazard}\AgdaSpace{}%
\AgdaBound{ls}\AgdaSpace{}%
\AgdaBound{v∈ws}\AgdaSpace{}%
\AgdaBound{v∈files}\AgdaSpace{}%
\AgdaKeyword{with}\AgdaSpace{}%
\AgdaFunction{∈-++⁻}\AgdaSpace{}%
\AgdaSymbol{(}\AgdaFunction{filesRead}\AgdaSpace{}%
\AgdaBound{ls}\AgdaSymbol{)}\AgdaSpace{}%
\AgdaBound{v∈files}\<%
\\
\>[0]\AgdaSymbol{...}\AgdaSpace{}%
\AgdaSymbol{|}\AgdaSpace{}%
\AgdaInductiveConstructor{inj₁}\AgdaSpace{}%
\AgdaBound{∈₁}\AgdaSpace{}%
\AgdaSymbol{=}\AgdaSpace{}%
\AgdaInductiveConstructor{ReadWrite}\AgdaSpace{}%
\AgdaBound{v∈ws}\AgdaSpace{}%
\AgdaBound{∈₁}\<%
\\
\>[0]\AgdaSymbol{...}\AgdaSpace{}%
\AgdaSymbol{|}\AgdaSpace{}%
\AgdaInductiveConstructor{inj₂}\AgdaSpace{}%
\AgdaBound{∈₂}\AgdaSpace{}%
\AgdaSymbol{=}\AgdaSpace{}%
\AgdaInductiveConstructor{WriteWrite}\AgdaSpace{}%
\AgdaBound{v∈ws}\AgdaSpace{}%
\AgdaBound{∈₂}\<%
\\
\\[\AgdaEmptyExtraSkip]%
\>[0]\AgdaFunction{helper3}\AgdaSpace{}%
\AgdaSymbol{:}\AgdaSpace{}%
\AgdaSymbol{∀}\AgdaSpace{}%
\AgdaSymbol{\{}\AgdaBound{s₁}\AgdaSymbol{\}}\AgdaSpace{}%
\AgdaSymbol{\{}\AgdaBound{x₁}\AgdaSymbol{\}}\AgdaSpace{}%
\AgdaBound{x}\AgdaSpace{}%
\AgdaSymbol{→}\AgdaSpace{}%
\AgdaFunction{Disjoint}\AgdaSpace{}%
\AgdaSymbol{(}\AgdaFunction{cmdWriteNames}\AgdaSpace{}%
\AgdaBound{x₁}\AgdaSpace{}%
\AgdaBound{s₁}\AgdaSymbol{)}\AgdaSpace{}%
\AgdaSymbol{(}\AgdaFunction{cmdReadWriteNames}\AgdaSpace{}%
\AgdaBound{x}\AgdaSpace{}%
\AgdaBound{s₁}\AgdaSymbol{)}\AgdaSpace{}%
\AgdaSymbol{→}\AgdaSpace{}%
\AgdaFunction{cmdReadWrites}\AgdaSpace{}%
\AgdaBound{x}\AgdaSpace{}%
\AgdaSymbol{(}\AgdaFunction{St.run}\AgdaSpace{}%
\AgdaBound{x₁}\AgdaSpace{}%
\AgdaBound{s₁}\AgdaSymbol{)}\AgdaSpace{}%
\AgdaOperator{\AgdaDatatype{≡}}\AgdaSpace{}%
\AgdaFunction{cmdReadWrites}\AgdaSpace{}%
\AgdaBound{x}\AgdaSpace{}%
\AgdaBound{s₁}\<%
\\
\>[0]\AgdaFunction{helper3}\AgdaSpace{}%
\AgdaSymbol{\{}\AgdaBound{s₁}\AgdaSymbol{\}}\AgdaSpace{}%
\AgdaSymbol{\{}\AgdaBound{x₁}\AgdaSymbol{\}}\AgdaSpace{}%
\AgdaBound{x}\AgdaSpace{}%
\AgdaBound{dsj}\AgdaSpace{}%
\AgdaSymbol{=}\AgdaSpace{}%
\AgdaFunction{cong₂}\AgdaSpace{}%
\AgdaOperator{\AgdaFunction{\AgdaUnderscore{}++\AgdaUnderscore{}}}\AgdaSpace{}%
\AgdaFunction{≡₁}\AgdaSpace{}%
\AgdaFunction{≡₂}\<%
\\
\>[0][@{}l@{\AgdaIndent{0}}]%
\>[2]\AgdaKeyword{where}%
\>[1694I]\AgdaFunction{≡₀}\AgdaSpace{}%
\AgdaSymbol{:}\AgdaSpace{}%
\AgdaField{proj₁}\AgdaSpace{}%
\AgdaSymbol{(}\AgdaBound{oracle}\AgdaSpace{}%
\AgdaBound{x}\AgdaSymbol{)}\AgdaSpace{}%
\AgdaSymbol{(}\AgdaFunction{St.run}\AgdaSpace{}%
\AgdaBound{x₁}\AgdaSpace{}%
\AgdaBound{s₁}\AgdaSymbol{)}\AgdaSpace{}%
\AgdaOperator{\AgdaDatatype{≡}}\AgdaSpace{}%
\AgdaField{proj₁}\AgdaSpace{}%
\AgdaSymbol{(}\AgdaBound{oracle}\AgdaSpace{}%
\AgdaBound{x}\AgdaSymbol{)}\AgdaSpace{}%
\AgdaBound{s₁}\<%
\\
\>[.][@{}l@{}]\<[1694I]%
\>[8]\AgdaFunction{≡₀}\AgdaSpace{}%
\AgdaSymbol{=}\AgdaSpace{}%
\AgdaFunction{sym}\AgdaSpace{}%
\AgdaSymbol{(}\AgdaField{proj₂}\AgdaSpace{}%
\AgdaSymbol{(}\AgdaBound{oracle}\AgdaSpace{}%
\AgdaBound{x}\AgdaSymbol{)}\AgdaSpace{}%
\AgdaBound{s₁}\AgdaSpace{}%
\AgdaSymbol{(}\AgdaFunction{St.run}\AgdaSpace{}%
\AgdaBound{x₁}\AgdaSpace{}%
\AgdaBound{s₁}\AgdaSymbol{)}\AgdaSpace{}%
\AgdaSymbol{λ}\AgdaSpace{}%
\AgdaBound{f₁}\AgdaSpace{}%
\AgdaBound{x₂}\AgdaSpace{}%
\AgdaSymbol{→}\AgdaSpace{}%
\AgdaSymbol{(}\AgdaFunction{lemma3}\AgdaSpace{}%
\AgdaBound{f₁}\AgdaSpace{}%
\AgdaSymbol{(}\AgdaFunction{cmdWrites}\AgdaSpace{}%
\AgdaBound{x₁}\AgdaSpace{}%
\AgdaBound{s₁}\AgdaSymbol{)}\AgdaSpace{}%
\AgdaSymbol{λ}\AgdaSpace{}%
\AgdaBound{x₃}\AgdaSpace{}%
\AgdaSymbol{→}\AgdaSpace{}%
\AgdaBound{dsj}\AgdaSpace{}%
\AgdaSymbol{(}\AgdaBound{x₃}\AgdaSpace{}%
\AgdaOperator{\AgdaInductiveConstructor{,}}\AgdaSpace{}%
\AgdaFunction{∈-++⁺ˡ}\AgdaSpace{}%
\AgdaBound{x₂}\AgdaSymbol{)))}\<%
\\
\>[8]\AgdaFunction{≡₁}\AgdaSpace{}%
\AgdaSymbol{:}\AgdaSpace{}%
\AgdaFunction{cmdReadNames}\AgdaSpace{}%
\AgdaBound{x}\AgdaSpace{}%
\AgdaSymbol{(}\AgdaFunction{St.run}\AgdaSpace{}%
\AgdaBound{x₁}\AgdaSpace{}%
\AgdaBound{s₁}\AgdaSymbol{)}\AgdaSpace{}%
\AgdaOperator{\AgdaDatatype{≡}}\AgdaSpace{}%
\AgdaFunction{cmdReadNames}\AgdaSpace{}%
\AgdaBound{x}\AgdaSpace{}%
\AgdaBound{s₁}\<%
\\
\>[8]\AgdaFunction{≡₁}\AgdaSpace{}%
\AgdaSymbol{=}\AgdaSpace{}%
\AgdaFunction{cong}\AgdaSpace{}%
\AgdaSymbol{(}\AgdaFunction{map}\AgdaSpace{}%
\AgdaField{proj₁}\AgdaSpace{}%
\AgdaOperator{\AgdaFunction{∘}}\AgdaSpace{}%
\AgdaField{proj₁}\AgdaSymbol{)}\AgdaSpace{}%
\AgdaFunction{≡₀}\<%
\\
\>[8]\AgdaFunction{≡₂}\AgdaSpace{}%
\AgdaSymbol{:}\AgdaSpace{}%
\AgdaFunction{cmdWriteNames}\AgdaSpace{}%
\AgdaBound{x}\AgdaSpace{}%
\AgdaSymbol{(}\AgdaFunction{St.run}\AgdaSpace{}%
\AgdaBound{x₁}\AgdaSpace{}%
\AgdaBound{s₁}\AgdaSymbol{)}\AgdaSpace{}%
\AgdaOperator{\AgdaDatatype{≡}}\AgdaSpace{}%
\AgdaFunction{cmdWriteNames}\AgdaSpace{}%
\AgdaBound{x}\AgdaSpace{}%
\AgdaBound{s₁}\<%
\\
\>[8]\AgdaFunction{≡₂}\AgdaSpace{}%
\AgdaSymbol{=}\AgdaSpace{}%
\AgdaFunction{cong}\AgdaSpace{}%
\AgdaSymbol{(}\AgdaFunction{map}\AgdaSpace{}%
\AgdaField{proj₁}\AgdaSpace{}%
\AgdaOperator{\AgdaFunction{∘}}\AgdaSpace{}%
\AgdaField{proj₂}\AgdaSymbol{)}\AgdaSpace{}%
\AgdaFunction{≡₀}\<%
\\
\\[\AgdaEmptyExtraSkip]%
\>[0]\AgdaFunction{helper2}\AgdaSpace{}%
\AgdaSymbol{:}\AgdaSpace{}%
\AgdaSymbol{∀}\AgdaSpace{}%
\AgdaSymbol{\{}\AgdaBound{s₁}\AgdaSymbol{\}}\AgdaSpace{}%
\AgdaSymbol{\{}\AgdaBound{x₂}\AgdaSymbol{\}}\AgdaSpace{}%
\AgdaBound{ls}\AgdaSpace{}%
\AgdaSymbol{→}\AgdaSpace{}%
\AgdaDatatype{All}\AgdaSpace{}%
\AgdaSymbol{(λ}\AgdaSpace{}%
\AgdaBound{x₁}\AgdaSpace{}%
\AgdaSymbol{→}\AgdaSpace{}%
\AgdaFunction{Disjoint}\AgdaSpace{}%
\AgdaSymbol{(}\AgdaFunction{cmdWriteNames}\AgdaSpace{}%
\AgdaBound{x₂}\AgdaSpace{}%
\AgdaBound{s₁}\AgdaSymbol{)}\AgdaSpace{}%
\AgdaSymbol{(}\AgdaFunction{cmdReadWriteNames}\AgdaSpace{}%
\AgdaBound{x₁}\AgdaSpace{}%
\AgdaBound{s₁}\AgdaSymbol{))}\AgdaSpace{}%
\AgdaBound{ls}\AgdaSpace{}%
\AgdaSymbol{→}\AgdaSpace{}%
\AgdaFunction{concatMap}\AgdaSpace{}%
\AgdaSymbol{(λ}\AgdaSpace{}%
\AgdaBound{x₁}\AgdaSpace{}%
\AgdaSymbol{→}\AgdaSpace{}%
\AgdaFunction{cmdReadWrites}\AgdaSpace{}%
\AgdaBound{x₁}\AgdaSpace{}%
\AgdaSymbol{(}\AgdaFunction{St.run}\AgdaSpace{}%
\AgdaBound{x₂}\AgdaSpace{}%
\AgdaBound{s₁}\AgdaSymbol{))}\AgdaSpace{}%
\AgdaBound{ls}\AgdaSpace{}%
\AgdaOperator{\AgdaDatatype{≡}}\AgdaSpace{}%
\AgdaFunction{concatMap}\AgdaSpace{}%
\AgdaSymbol{(λ}\AgdaSpace{}%
\AgdaBound{x₁}\AgdaSpace{}%
\AgdaSymbol{→}\AgdaSpace{}%
\AgdaFunction{cmdReadWrites}\AgdaSpace{}%
\AgdaBound{x₁}\AgdaSpace{}%
\AgdaBound{s₁}\AgdaSymbol{)}\AgdaSpace{}%
\AgdaBound{ls}\<%
\\
\>[0]\AgdaFunction{helper2}\AgdaSpace{}%
\AgdaInductiveConstructor{[]}\AgdaSpace{}%
\AgdaBound{all₁}\AgdaSpace{}%
\AgdaSymbol{=}\AgdaSpace{}%
\AgdaInductiveConstructor{refl}\<%
\\
\>[0]\AgdaFunction{helper2}\AgdaSpace{}%
\AgdaSymbol{(}\AgdaBound{x}\AgdaSpace{}%
\AgdaOperator{\AgdaInductiveConstructor{∷}}\AgdaSpace{}%
\AgdaBound{ls}\AgdaSymbol{)}\AgdaSpace{}%
\AgdaSymbol{(}\AgdaBound{px}\AgdaSpace{}%
\AgdaOperator{\AgdaInductiveConstructor{All.∷}}\AgdaSpace{}%
\AgdaBound{all₁}\AgdaSymbol{)}\AgdaSpace{}%
\AgdaKeyword{with}\AgdaSpace{}%
\AgdaFunction{helper2}\AgdaSpace{}%
\AgdaBound{ls}\AgdaSpace{}%
\AgdaBound{all₁}\<%
\\
\>[0]\AgdaSymbol{...}\AgdaSpace{}%
\AgdaSymbol{|}\AgdaSpace{}%
\AgdaBound{a}\AgdaSpace{}%
\AgdaSymbol{=}\AgdaSpace{}%
\AgdaFunction{cong₂}\AgdaSpace{}%
\AgdaOperator{\AgdaFunction{\AgdaUnderscore{}++\AgdaUnderscore{}}}\AgdaSpace{}%
\AgdaSymbol{(}\AgdaFunction{helper3}\AgdaSpace{}%
\AgdaBound{x}\AgdaSpace{}%
\AgdaBound{px}\AgdaSymbol{)}\AgdaSpace{}%
\AgdaBound{a}\<%
\\
\\[\AgdaEmptyExtraSkip]%
\>[0]\AgdaFunction{correct-inner}\AgdaSpace{}%
\AgdaSymbol{:}\AgdaSpace{}%
\AgdaSymbol{∀}\AgdaSpace{}%
\AgdaSymbol{\{}\AgdaBound{s₁}\AgdaSymbol{\}}\AgdaSpace{}%
\AgdaSymbol{\{}\AgdaBound{s₂}\AgdaSymbol{\}}\AgdaSpace{}%
\AgdaSymbol{\{}\AgdaBound{ls}\AgdaSymbol{\}}\AgdaSpace{}%
\AgdaBound{m}\AgdaSpace{}%
\AgdaBound{b}\AgdaSpace{}%
\AgdaSymbol{\{}\AgdaBound{b₁}\AgdaSymbol{\}}\AgdaSpace{}%
\AgdaSymbol{→}\AgdaSpace{}%
\AgdaSymbol{(∀}\AgdaSpace{}%
\AgdaBound{f₁}\AgdaSpace{}%
\AgdaSymbol{→}\AgdaSpace{}%
\AgdaBound{s₁}\AgdaSpace{}%
\AgdaBound{f₁}\AgdaSpace{}%
\AgdaOperator{\AgdaDatatype{≡}}\AgdaSpace{}%
\AgdaBound{s₂}\AgdaSpace{}%
\AgdaBound{f₁}\AgdaSymbol{)}\AgdaSpace{}%
\AgdaSymbol{→}\AgdaSpace{}%
\AgdaDatatype{DisjointBuild}\AgdaSpace{}%
\AgdaBound{s₁}\AgdaSpace{}%
\AgdaBound{b}\AgdaSpace{}%
\AgdaSymbol{→}\AgdaSpace{}%
\AgdaDatatype{Unique}\AgdaSpace{}%
\AgdaBound{b}\AgdaSpace{}%
\AgdaSymbol{→}\AgdaSpace{}%
\AgdaDatatype{Unique}\AgdaSpace{}%
\AgdaSymbol{(}\AgdaFunction{map}\AgdaSpace{}%
\AgdaField{proj₁}\AgdaSpace{}%
\AgdaBound{ls}\AgdaSymbol{)}\AgdaSpace{}%
\AgdaSymbol{→}\AgdaSpace{}%
\AgdaFunction{Disjoint}\AgdaSpace{}%
\AgdaBound{b}\AgdaSpace{}%
\AgdaSymbol{(}\AgdaFunction{map}\AgdaSpace{}%
\AgdaField{proj₁}\AgdaSpace{}%
\AgdaBound{ls}\AgdaSymbol{)}\AgdaSpace{}%
\AgdaSymbol{→}\AgdaSpace{}%
\AgdaFunction{concatMap}\AgdaSpace{}%
\AgdaSymbol{(λ}\AgdaSpace{}%
\AgdaBound{x₁}\AgdaSpace{}%
\AgdaSymbol{→}\AgdaSpace{}%
\AgdaFunction{cmdReadWrites}\AgdaSpace{}%
\AgdaBound{x₁}\AgdaSpace{}%
\AgdaBound{s₁}\AgdaSymbol{)}\AgdaSpace{}%
\AgdaSymbol{(}\AgdaFunction{map}\AgdaSpace{}%
\AgdaField{proj₁}\AgdaSpace{}%
\AgdaBound{m}\AgdaSymbol{)}\AgdaSpace{}%
\AgdaOperator{\AgdaFunction{⊆}}\AgdaSpace{}%
\AgdaFunction{files}\AgdaSpace{}%
\AgdaBound{ls}\AgdaSpace{}%
\AgdaSymbol{→}\AgdaSpace{}%
\AgdaDatatype{IdempotentState}\AgdaSpace{}%
\AgdaFunction{cmdReadNames}\AgdaSpace{}%
\AgdaBound{s₁}\AgdaSpace{}%
\AgdaBound{m}\AgdaSpace{}%
\AgdaSymbol{→}\AgdaSpace{}%
\AgdaDatatype{HazardFree}\AgdaSpace{}%
\AgdaBound{s₁}\AgdaSpace{}%
\AgdaBound{b}\AgdaSpace{}%
\AgdaBound{b₁}\AgdaSpace{}%
\AgdaBound{ls}\AgdaSpace{}%
\AgdaSymbol{→}\AgdaSpace{}%
\AgdaSymbol{(∀}\AgdaSpace{}%
\AgdaBound{f₁}\AgdaSpace{}%
\AgdaSymbol{→}\AgdaSpace{}%
\AgdaField{proj₁}\AgdaSpace{}%
\AgdaSymbol{(}\AgdaFunction{fabricate}\AgdaSpace{}%
\AgdaBound{b}\AgdaSpace{}%
\AgdaSymbol{(}\AgdaBound{s₁}\AgdaSpace{}%
\AgdaOperator{\AgdaInductiveConstructor{,}}\AgdaSpace{}%
\AgdaBound{m}\AgdaSymbol{))}\AgdaSpace{}%
\AgdaBound{f₁}\AgdaSpace{}%
\AgdaOperator{\AgdaDatatype{≡}}\AgdaSpace{}%
\AgdaFunction{script}\AgdaSpace{}%
\AgdaBound{b}\AgdaSpace{}%
\AgdaBound{s₂}\AgdaSpace{}%
\AgdaBound{f₁}\AgdaSymbol{)}\<%
\\
\>[0]\AgdaFunction{correct-inner}\AgdaSpace{}%
\AgdaBound{m}\AgdaSpace{}%
\AgdaInductiveConstructor{[]}\AgdaSpace{}%
\AgdaBound{∀₁}\AgdaSpace{}%
\AgdaSymbol{\AgdaUnderscore{}}\AgdaSpace{}%
\AgdaSymbol{\AgdaUnderscore{}}\AgdaSpace{}%
\AgdaSymbol{\AgdaUnderscore{}}\AgdaSpace{}%
\AgdaSymbol{\AgdaUnderscore{}}\AgdaSpace{}%
\AgdaSymbol{\AgdaUnderscore{}}\AgdaSpace{}%
\AgdaBound{is}\AgdaSpace{}%
\AgdaBound{hf}\AgdaSpace{}%
\AgdaSymbol{=}\AgdaSpace{}%
\AgdaBound{∀₁}\<%
\\
\>[0]\AgdaFunction{correct-inner}\AgdaSpace{}%
\AgdaSymbol{\{}\AgdaBound{s₁}\AgdaSymbol{\}}\AgdaSpace{}%
\AgdaSymbol{\{}\AgdaBound{s₂}\AgdaSymbol{\}}\AgdaSpace{}%
\AgdaSymbol{\{}\AgdaBound{ls}\AgdaSymbol{\}}\AgdaSpace{}%
\AgdaBound{m}\AgdaSpace{}%
\AgdaSymbol{(}\AgdaBound{x}\AgdaSpace{}%
\AgdaOperator{\AgdaInductiveConstructor{∷}}\AgdaSpace{}%
\AgdaBound{b}\AgdaSymbol{)}\AgdaSpace{}%
\AgdaBound{∀₁}\AgdaSpace{}%
\AgdaSymbol{(}\AgdaInductiveConstructor{Cons}\AgdaSpace{}%
\AgdaBound{x}\AgdaSpace{}%
\AgdaBound{dc}\AgdaSpace{}%
\AgdaBound{b}\AgdaSpace{}%
\AgdaBound{dsb}\AgdaSymbol{)}\AgdaSpace{}%
\AgdaSymbol{(}\AgdaBound{px}\AgdaSpace{}%
\AgdaOperator{\AgdaInductiveConstructor{∷}}\AgdaSpace{}%
\AgdaBound{ub}\AgdaSymbol{)}\AgdaSpace{}%
\AgdaBound{uls}\AgdaSpace{}%
\AgdaBound{dsj}\AgdaSpace{}%
\AgdaBound{⊆₁}\AgdaSpace{}%
\AgdaBound{is}\AgdaSpace{}%
\AgdaSymbol{(}\AgdaBound{¬hz}\AgdaSpace{}%
\AgdaOperator{\AgdaInductiveConstructor{∷}}\AgdaSpace{}%
\AgdaBound{hf}\AgdaSymbol{)}\AgdaSpace{}%
\AgdaBound{f₁}\AgdaSpace{}%
\AgdaKeyword{with}\AgdaSpace{}%
\AgdaBound{x}\AgdaSpace{}%
\AgdaOperator{\AgdaFunction{∈?}}\AgdaSpace{}%
\AgdaFunction{map}\AgdaSpace{}%
\AgdaField{proj₁}\AgdaSpace{}%
\AgdaBound{m}\<%
\\
\>[0]\AgdaSymbol{...}\AgdaSpace{}%
\AgdaSymbol{|}\AgdaSpace{}%
\AgdaInductiveConstructor{no}\AgdaSpace{}%
\AgdaBound{x∉mem}\AgdaSpace{}%
\AgdaSymbol{=}%
\>[1950I]\AgdaFunction{correct-inner}\AgdaSpace{}%
\AgdaSymbol{(}\AgdaFunction{save}\AgdaSpace{}%
\AgdaBound{x}\AgdaSpace{}%
\AgdaSymbol{(}\AgdaFunction{cmdReadNames}\AgdaSpace{}%
\AgdaBound{x}\AgdaSpace{}%
\AgdaBound{s₁}\AgdaSymbol{)}\AgdaSpace{}%
\AgdaSymbol{(}\AgdaFunction{St.run}\AgdaSpace{}%
\AgdaBound{x}\AgdaSpace{}%
\AgdaBound{s₁}\AgdaSymbol{)}\AgdaSpace{}%
\AgdaBound{m}\AgdaSymbol{)}\AgdaSpace{}%
\AgdaBound{b}\AgdaSpace{}%
\AgdaSymbol{(}\AgdaFunction{run-≡}\AgdaSpace{}%
\AgdaBound{x}\AgdaSpace{}%
\AgdaBound{∀₁}\AgdaSymbol{)}\AgdaSpace{}%
\AgdaBound{dsb}\AgdaSpace{}%
\AgdaBound{ub}\<%
\\
\>[.][@{}l@{}]\<[1950I]%
\>[17]\AgdaSymbol{(}\AgdaFunction{∉toAll}\AgdaSpace{}%
\AgdaSymbol{\AgdaUnderscore{}}\AgdaSpace{}%
\AgdaSymbol{(λ}\AgdaSpace{}%
\AgdaBound{x₂}\AgdaSpace{}%
\AgdaSymbol{→}\AgdaSpace{}%
\AgdaBound{dsj}\AgdaSpace{}%
\AgdaSymbol{((}\AgdaInductiveConstructor{here}\AgdaSpace{}%
\AgdaInductiveConstructor{refl}\AgdaSymbol{)}\AgdaSpace{}%
\AgdaOperator{\AgdaInductiveConstructor{,}}\AgdaSpace{}%
\AgdaBound{x₂}\AgdaSymbol{))}\AgdaSpace{}%
\AgdaOperator{\AgdaInductiveConstructor{∷}}\AgdaSpace{}%
\AgdaBound{uls}\AgdaSymbol{)}\<%
\\
\>[17]\AgdaSymbol{(λ}\AgdaSpace{}%
\AgdaBound{x₂}\AgdaSpace{}%
\AgdaSymbol{→}\AgdaSpace{}%
\AgdaBound{dsj}\AgdaSpace{}%
\AgdaSymbol{((}\AgdaInductiveConstructor{there}\AgdaSpace{}%
\AgdaSymbol{(}\AgdaField{proj₁}\AgdaSpace{}%
\AgdaBound{x₂}\AgdaSymbol{))}\AgdaSpace{}%
\AgdaOperator{\AgdaInductiveConstructor{,}}\AgdaSpace{}%
\AgdaFunction{tail}\AgdaSpace{}%
\AgdaSymbol{(λ}\AgdaSpace{}%
\AgdaBound{x₃}\AgdaSpace{}%
\AgdaSymbol{→}\AgdaSpace{}%
\AgdaFunction{lookup}\AgdaSpace{}%
\AgdaBound{px}\AgdaSpace{}%
\AgdaSymbol{(}\AgdaField{proj₁}\AgdaSpace{}%
\AgdaBound{x₂}\AgdaSymbol{)}\AgdaSpace{}%
\AgdaSymbol{(}\AgdaFunction{sym}\AgdaSpace{}%
\AgdaBound{x₃}\AgdaSymbol{))}\AgdaSpace{}%
\AgdaSymbol{(}\AgdaField{proj₂}\AgdaSpace{}%
\AgdaBound{x₂}\AgdaSymbol{)))}\<%
\\
\>[17]\AgdaFunction{⊆₂}\AgdaSpace{}%
\AgdaFunction{is₂}\AgdaSpace{}%
\AgdaBound{hf}\AgdaSpace{}%
\AgdaBound{f₁}\<%
\\
\>[0][@{}l@{\AgdaIndent{0}}]%
\>[2]\AgdaKeyword{where}%
\>[1999I]\AgdaFunction{all₁}\AgdaSpace{}%
\AgdaSymbol{:}\AgdaSpace{}%
\AgdaDatatype{All}\AgdaSpace{}%
\AgdaSymbol{(λ}\AgdaSpace{}%
\AgdaBound{x₁}\AgdaSpace{}%
\AgdaSymbol{→}\AgdaSpace{}%
\AgdaFunction{Disjoint}\AgdaSpace{}%
\AgdaSymbol{(}\AgdaFunction{cmdWriteNames}\AgdaSpace{}%
\AgdaBound{x}\AgdaSpace{}%
\AgdaBound{s₁}\AgdaSymbol{)}\AgdaSpace{}%
\AgdaSymbol{(}\AgdaFunction{cmdReadWriteNames}\AgdaSpace{}%
\AgdaBound{x₁}\AgdaSpace{}%
\AgdaBound{s₁}\AgdaSymbol{))}\AgdaSpace{}%
\AgdaSymbol{(}\AgdaFunction{map}\AgdaSpace{}%
\AgdaField{proj₁}\AgdaSpace{}%
\AgdaBound{m}\AgdaSymbol{)}\<%
\\
\>[.][@{}l@{}]\<[1999I]%
\>[8]\AgdaFunction{all₁}\AgdaSpace{}%
\AgdaSymbol{=}\AgdaSpace{}%
\AgdaFunction{helper}\AgdaSpace{}%
\AgdaSymbol{(}\AgdaFunction{map}\AgdaSpace{}%
\AgdaField{proj₁}\AgdaSpace{}%
\AgdaBound{m}\AgdaSymbol{)}\AgdaSpace{}%
\AgdaSymbol{\AgdaUnderscore{}}\AgdaSpace{}%
\AgdaBound{x}\AgdaSpace{}%
\AgdaSymbol{(λ}\AgdaSpace{}%
\AgdaBound{x₂}\AgdaSpace{}%
\AgdaSymbol{→}\AgdaSpace{}%
\AgdaBound{¬hz}\AgdaSpace{}%
\AgdaSymbol{(}\AgdaFunction{∃hazard}\AgdaSpace{}%
\AgdaSymbol{\AgdaUnderscore{}}\AgdaSpace{}%
\AgdaSymbol{(}\AgdaField{proj₁}\AgdaSpace{}%
\AgdaBound{x₂}\AgdaSymbol{)}\AgdaSpace{}%
\AgdaSymbol{(}\AgdaField{proj₂}\AgdaSpace{}%
\AgdaBound{x₂}\AgdaSymbol{)))}\AgdaSpace{}%
\AgdaBound{⊆₁}\<%
\\
\>[8]\AgdaFunction{is₂}\AgdaSpace{}%
\AgdaSymbol{:}\AgdaSpace{}%
\AgdaDatatype{IdempotentState}\AgdaSpace{}%
\AgdaFunction{cmdReadNames}\AgdaSpace{}%
\AgdaSymbol{(}\AgdaFunction{St.run}\AgdaSpace{}%
\AgdaBound{x}\AgdaSpace{}%
\AgdaBound{s₁}\AgdaSymbol{)}\AgdaSpace{}%
\AgdaSymbol{(}\AgdaFunction{save}\AgdaSpace{}%
\AgdaBound{x}\AgdaSpace{}%
\AgdaSymbol{(}\AgdaFunction{cmdReadNames}\AgdaSpace{}%
\AgdaBound{x}\AgdaSpace{}%
\AgdaBound{s₁}\AgdaSymbol{)}\AgdaSpace{}%
\AgdaSymbol{(}\AgdaFunction{St.run}\AgdaSpace{}%
\AgdaBound{x}\AgdaSpace{}%
\AgdaBound{s₁}\AgdaSymbol{)}\AgdaSpace{}%
\AgdaBound{m}\AgdaSymbol{)}\<%
\\
\>[8]\AgdaFunction{is₂}\AgdaSpace{}%
\AgdaSymbol{=}\AgdaSpace{}%
\AgdaFunction{preserves}\AgdaSpace{}%
\AgdaBound{x}\AgdaSpace{}%
\AgdaBound{dc}\AgdaSpace{}%
\AgdaFunction{all₁}\AgdaSpace{}%
\AgdaBound{is}\<%
\\
\>[8]\AgdaComment{-- ( is}\<%
\\
\>[8]\AgdaFunction{≡₀}\AgdaSpace{}%
\AgdaSymbol{:}\AgdaSpace{}%
\AgdaField{proj₁}\AgdaSpace{}%
\AgdaSymbol{(}\AgdaBound{oracle}\AgdaSpace{}%
\AgdaBound{x}\AgdaSymbol{)}\AgdaSpace{}%
\AgdaSymbol{(}\AgdaFunction{St.run}\AgdaSpace{}%
\AgdaBound{x}\AgdaSpace{}%
\AgdaBound{s₁}\AgdaSymbol{)}\AgdaSpace{}%
\AgdaOperator{\AgdaDatatype{≡}}\AgdaSpace{}%
\AgdaField{proj₁}\AgdaSpace{}%
\AgdaSymbol{(}\AgdaBound{oracle}\AgdaSpace{}%
\AgdaBound{x}\AgdaSymbol{)}\AgdaSpace{}%
\AgdaBound{s₁}\<%
\\
\>[8]\AgdaFunction{≡₀}\AgdaSpace{}%
\AgdaSymbol{=}\AgdaSpace{}%
\AgdaFunction{sym}\AgdaSpace{}%
\AgdaSymbol{(}\AgdaField{proj₂}\AgdaSpace{}%
\AgdaSymbol{(}\AgdaBound{oracle}\AgdaSpace{}%
\AgdaBound{x}\AgdaSymbol{)}\AgdaSpace{}%
\AgdaBound{s₁}\AgdaSpace{}%
\AgdaSymbol{(}\AgdaFunction{St.run}\AgdaSpace{}%
\AgdaBound{x}\AgdaSpace{}%
\AgdaBound{s₁}\AgdaSymbol{)}\AgdaSpace{}%
\AgdaSymbol{λ}\AgdaSpace{}%
\AgdaBound{f₂}\AgdaSpace{}%
\AgdaBound{x₁}\AgdaSpace{}%
\AgdaSymbol{→}\AgdaSpace{}%
\AgdaFunction{lemma3}\AgdaSpace{}%
\AgdaBound{f₂}\AgdaSpace{}%
\AgdaSymbol{(}\AgdaFunction{cmdWrites}\AgdaSpace{}%
\AgdaBound{x}\AgdaSpace{}%
\AgdaBound{s₁}\AgdaSymbol{)}\AgdaSpace{}%
\AgdaSymbol{λ}\AgdaSpace{}%
\AgdaBound{x₂}\AgdaSpace{}%
\AgdaSymbol{→}\AgdaSpace{}%
\AgdaBound{dc}\AgdaSpace{}%
\AgdaSymbol{(}\AgdaBound{x₁}\AgdaSpace{}%
\AgdaOperator{\AgdaInductiveConstructor{,}}\AgdaSpace{}%
\AgdaBound{x₂}\AgdaSymbol{))}\<%
\\
\>[8]\AgdaFunction{≡₁}\AgdaSpace{}%
\AgdaSymbol{:}\AgdaSpace{}%
\AgdaFunction{cmdReadNames}\AgdaSpace{}%
\AgdaBound{x}\AgdaSpace{}%
\AgdaSymbol{(}\AgdaFunction{St.run}\AgdaSpace{}%
\AgdaBound{x}\AgdaSpace{}%
\AgdaBound{s₁}\AgdaSymbol{)}\AgdaSpace{}%
\AgdaOperator{\AgdaDatatype{≡}}\AgdaSpace{}%
\AgdaFunction{cmdReadNames}\AgdaSpace{}%
\AgdaBound{x}\AgdaSpace{}%
\AgdaBound{s₁}\<%
\\
\>[8]\AgdaFunction{≡₁}\AgdaSpace{}%
\AgdaSymbol{=}\AgdaSpace{}%
\AgdaFunction{cong}\AgdaSpace{}%
\AgdaSymbol{(}\AgdaFunction{map}\AgdaSpace{}%
\AgdaField{proj₁}\AgdaSpace{}%
\AgdaOperator{\AgdaFunction{∘}}\AgdaSpace{}%
\AgdaField{proj₁}\AgdaSymbol{)}\AgdaSpace{}%
\AgdaFunction{≡₀}\<%
\\
\>[8]\AgdaFunction{≡₂}\AgdaSpace{}%
\AgdaSymbol{:}\AgdaSpace{}%
\AgdaFunction{cmdWriteNames}\AgdaSpace{}%
\AgdaBound{x}\AgdaSpace{}%
\AgdaSymbol{(}\AgdaFunction{St.run}\AgdaSpace{}%
\AgdaBound{x}\AgdaSpace{}%
\AgdaBound{s₁}\AgdaSymbol{)}\AgdaSpace{}%
\AgdaOperator{\AgdaDatatype{≡}}\AgdaSpace{}%
\AgdaFunction{cmdWriteNames}\AgdaSpace{}%
\AgdaBound{x}\AgdaSpace{}%
\AgdaBound{s₁}\<%
\\
\>[8]\AgdaFunction{≡₂}\AgdaSpace{}%
\AgdaSymbol{=}\AgdaSpace{}%
\AgdaFunction{cong}\AgdaSpace{}%
\AgdaSymbol{(}\AgdaFunction{map}\AgdaSpace{}%
\AgdaField{proj₁}\AgdaSpace{}%
\AgdaOperator{\AgdaFunction{∘}}\AgdaSpace{}%
\AgdaField{proj₂}\AgdaSymbol{)}\AgdaSpace{}%
\AgdaFunction{≡₀}\<%
\\
\>[8]\AgdaFunction{⊆₂}\AgdaSpace{}%
\AgdaSymbol{:}\AgdaSpace{}%
\AgdaFunction{concatMap}\AgdaSpace{}%
\AgdaSymbol{(λ}\AgdaSpace{}%
\AgdaBound{x₁}\AgdaSpace{}%
\AgdaSymbol{→}\AgdaSpace{}%
\AgdaFunction{cmdReadWrites}\AgdaSpace{}%
\AgdaBound{x₁}\AgdaSpace{}%
\AgdaSymbol{(}\AgdaFunction{St.run}\AgdaSpace{}%
\AgdaBound{x}\AgdaSpace{}%
\AgdaBound{s₁}\AgdaSymbol{))}\AgdaSpace{}%
\AgdaSymbol{(}\AgdaBound{x}\AgdaSpace{}%
\AgdaOperator{\AgdaInductiveConstructor{∷}}\AgdaSpace{}%
\AgdaFunction{map}\AgdaSpace{}%
\AgdaField{proj₁}\AgdaSpace{}%
\AgdaBound{m}\AgdaSymbol{)}\AgdaSpace{}%
\AgdaOperator{\AgdaFunction{⊆}}\AgdaSpace{}%
\AgdaFunction{files}\AgdaSpace{}%
\AgdaSymbol{(}\AgdaFunction{rec}\AgdaSpace{}%
\AgdaBound{s₁}\AgdaSpace{}%
\AgdaBound{x}\AgdaSpace{}%
\AgdaBound{ls}\AgdaSymbol{)}\<%
\\
\>[8]\AgdaFunction{⊆₂}\AgdaSpace{}%
\AgdaBound{x₂}\AgdaSpace{}%
\AgdaKeyword{with}\AgdaSpace{}%
\AgdaFunction{∈-++⁻}\AgdaSpace{}%
\AgdaSymbol{(}\AgdaFunction{cmdReadWrites}\AgdaSpace{}%
\AgdaBound{x}\AgdaSpace{}%
\AgdaSymbol{(}\AgdaFunction{St.run}\AgdaSpace{}%
\AgdaBound{x}\AgdaSpace{}%
\AgdaBound{s₁}\AgdaSymbol{))}\AgdaSpace{}%
\AgdaBound{x₂}\<%
\\
\>[8]\AgdaSymbol{...}\AgdaSpace{}%
\AgdaSymbol{|}\AgdaSpace{}%
\AgdaInductiveConstructor{inj₁}\AgdaSpace{}%
\AgdaBound{∈₁}\AgdaSpace{}%
\AgdaKeyword{with}\AgdaSpace{}%
\AgdaFunction{∈-++⁻}\AgdaSpace{}%
\AgdaSymbol{(}\AgdaFunction{cmdReadNames}\AgdaSpace{}%
\AgdaBound{x}\AgdaSpace{}%
\AgdaSymbol{(}\AgdaFunction{St.run}\AgdaSpace{}%
\AgdaBound{x}\AgdaSpace{}%
\AgdaBound{s₁}\AgdaSymbol{))}\AgdaSpace{}%
\AgdaBound{∈₁}\<%
\\
\>[8]\AgdaSymbol{...}\AgdaSpace{}%
\AgdaSymbol{|}\AgdaSpace{}%
\AgdaInductiveConstructor{inj₁}\AgdaSpace{}%
\AgdaBound{∈rs}\AgdaSpace{}%
\AgdaSymbol{=}\AgdaSpace{}%
\AgdaFunction{∈-++⁺ˡ}\AgdaSpace{}%
\AgdaSymbol{(}\AgdaFunction{∈-++⁺ˡ}\AgdaSpace{}%
\AgdaSymbol{(}\AgdaFunction{subst}\AgdaSpace{}%
\AgdaSymbol{(λ}\AgdaSpace{}%
\AgdaBound{x₃}\AgdaSpace{}%
\AgdaSymbol{→}\AgdaSpace{}%
\AgdaSymbol{\AgdaUnderscore{}}\AgdaSpace{}%
\AgdaOperator{\AgdaFunction{∈}}\AgdaSpace{}%
\AgdaBound{x₃}\AgdaSymbol{)}\AgdaSpace{}%
\AgdaFunction{≡₁}\AgdaSpace{}%
\AgdaBound{∈rs}\AgdaSymbol{))}\<%
\\
\>[8]\AgdaSymbol{...}\AgdaSpace{}%
\AgdaSymbol{|}\AgdaSpace{}%
\AgdaInductiveConstructor{inj₂}\AgdaSpace{}%
\AgdaBound{∈ws}\AgdaSpace{}%
\AgdaSymbol{=}\AgdaSpace{}%
\AgdaFunction{∈-++⁺ʳ}\AgdaSpace{}%
\AgdaSymbol{(}\AgdaFunction{filesRead}\AgdaSpace{}%
\AgdaSymbol{(}\AgdaFunction{rec}\AgdaSpace{}%
\AgdaBound{s₁}\AgdaSpace{}%
\AgdaBound{x}\AgdaSpace{}%
\AgdaBound{ls}\AgdaSymbol{))}\AgdaSpace{}%
\AgdaSymbol{(}\AgdaFunction{∈-++⁺ˡ}\AgdaSpace{}%
\AgdaSymbol{(}\AgdaFunction{subst}\AgdaSpace{}%
\AgdaSymbol{(λ}\AgdaSpace{}%
\AgdaBound{x₃}\AgdaSpace{}%
\AgdaSymbol{→}\AgdaSpace{}%
\AgdaSymbol{\AgdaUnderscore{}}\AgdaSpace{}%
\AgdaOperator{\AgdaFunction{∈}}\AgdaSpace{}%
\AgdaBound{x₃}\AgdaSymbol{)}\AgdaSpace{}%
\AgdaFunction{≡₂}\AgdaSpace{}%
\AgdaBound{∈ws}\AgdaSymbol{))}\<%
\\
\>[8]\AgdaFunction{⊆₂}\AgdaSpace{}%
\AgdaBound{x₂}\AgdaSpace{}%
\AgdaSymbol{|}\AgdaSpace{}%
\AgdaInductiveConstructor{inj₂}\AgdaSpace{}%
\AgdaBound{∈₂}\AgdaSpace{}%
\AgdaKeyword{with}\AgdaSpace{}%
\AgdaFunction{∈-++⁻}\AgdaSpace{}%
\AgdaSymbol{(}\AgdaFunction{filesRead}\AgdaSpace{}%
\AgdaBound{ls}\AgdaSymbol{)}\AgdaSpace{}%
\AgdaSymbol{(}\AgdaBound{⊆₁}\AgdaSpace{}%
\AgdaSymbol{(}\AgdaFunction{subst}\AgdaSpace{}%
\AgdaSymbol{(λ}\AgdaSpace{}%
\AgdaBound{x₃}\AgdaSpace{}%
\AgdaSymbol{→}\AgdaSpace{}%
\AgdaSymbol{\AgdaUnderscore{}}\AgdaSpace{}%
\AgdaOperator{\AgdaFunction{∈}}\AgdaSpace{}%
\AgdaBound{x₃}\AgdaSymbol{)}\AgdaSpace{}%
\AgdaSymbol{(}\AgdaFunction{helper2}\AgdaSpace{}%
\AgdaSymbol{(}\AgdaFunction{map}\AgdaSpace{}%
\AgdaField{proj₁}\AgdaSpace{}%
\AgdaBound{m}\AgdaSymbol{)}\AgdaSpace{}%
\AgdaFunction{all₁}\AgdaSymbol{)}\AgdaSpace{}%
\AgdaBound{∈₂}\AgdaSymbol{))}\<%
\\
\>[8]\AgdaSymbol{...}\AgdaSpace{}%
\AgdaSymbol{|}\AgdaSpace{}%
\AgdaInductiveConstructor{inj₁}\AgdaSpace{}%
\AgdaBound{∈₁}\AgdaSpace{}%
\AgdaSymbol{=}\AgdaSpace{}%
\AgdaFunction{∈-++⁺ˡ}\AgdaSpace{}%
\AgdaSymbol{(}\AgdaFunction{∈-++⁺ʳ}\AgdaSpace{}%
\AgdaSymbol{(}\AgdaFunction{cmdReadNames}\AgdaSpace{}%
\AgdaBound{x}\AgdaSpace{}%
\AgdaBound{s₁}\AgdaSymbol{)}\AgdaSpace{}%
\AgdaBound{∈₁}\AgdaSymbol{)}\<%
\\
\>[8]\AgdaSymbol{...}\AgdaSpace{}%
\AgdaSymbol{|}\AgdaSpace{}%
\AgdaInductiveConstructor{inj₂}\AgdaSpace{}%
\AgdaBound{∈₁}\AgdaSpace{}%
\AgdaSymbol{=}\AgdaSpace{}%
\AgdaFunction{∈-++⁺ʳ}\AgdaSpace{}%
\AgdaSymbol{(}\AgdaFunction{filesRead}\AgdaSpace{}%
\AgdaSymbol{(}\AgdaFunction{rec}\AgdaSpace{}%
\AgdaBound{s₁}\AgdaSpace{}%
\AgdaBound{x}\AgdaSpace{}%
\AgdaBound{ls}\AgdaSymbol{))}\AgdaSpace{}%
\AgdaSymbol{(}\AgdaFunction{∈-++⁺ʳ}\AgdaSpace{}%
\AgdaSymbol{(}\AgdaFunction{cmdWriteNames}\AgdaSpace{}%
\AgdaBound{x}\AgdaSpace{}%
\AgdaBound{s₁}\AgdaSymbol{)}\AgdaSpace{}%
\AgdaBound{∈₁}\AgdaSymbol{)}\<%
\\
\>[0]\AgdaSymbol{...}\AgdaSpace{}%
\AgdaSymbol{|}\AgdaSpace{}%
\AgdaInductiveConstructor{yes}\AgdaSpace{}%
\AgdaBound{x∈mem}\AgdaSpace{}%
\AgdaKeyword{with}\AgdaSpace{}%
\AgdaFunction{maybeAll}\AgdaSpace{}%
\AgdaSymbol{\{}\AgdaBound{s₁}\AgdaSymbol{\}}\AgdaSpace{}%
\AgdaSymbol{(}\AgdaFunction{get}\AgdaSpace{}%
\AgdaBound{x}\AgdaSpace{}%
\AgdaBound{m}\AgdaSpace{}%
\AgdaBound{x∈mem}\AgdaSymbol{)}\<%
\\
\>[0]\AgdaComment{-- yes, it is extremely stupid that this is duplicated. i will fix it later}\<%
\\
\>[0]\AgdaSymbol{...}\AgdaSpace{}%
\AgdaSymbol{|}\AgdaSpace{}%
\AgdaInductiveConstructor{nothing}\AgdaSpace{}%
\AgdaSymbol{=}%
\>[2261I]\AgdaFunction{correct-inner}\AgdaSpace{}%
\AgdaSymbol{(}\AgdaFunction{save}\AgdaSpace{}%
\AgdaBound{x}\AgdaSpace{}%
\AgdaSymbol{(}\AgdaFunction{cmdReadNames}\AgdaSpace{}%
\AgdaBound{x}\AgdaSpace{}%
\AgdaBound{s₁}\AgdaSymbol{)}\AgdaSpace{}%
\AgdaSymbol{(}\AgdaFunction{St.run}\AgdaSpace{}%
\AgdaBound{x}\AgdaSpace{}%
\AgdaBound{s₁}\AgdaSymbol{)}\AgdaSpace{}%
\AgdaBound{m}\AgdaSymbol{)}\AgdaSpace{}%
\AgdaBound{b}\AgdaSpace{}%
\AgdaSymbol{(}\AgdaFunction{run-≡}\AgdaSpace{}%
\AgdaBound{x}\AgdaSpace{}%
\AgdaBound{∀₁}\AgdaSymbol{)}\AgdaSpace{}%
\AgdaBound{dsb}\AgdaSpace{}%
\AgdaBound{ub}\<%
\\
\>[2261I][@{}l@{\AgdaIndent{0}}]%
\>[17]\AgdaSymbol{(}\AgdaFunction{∉toAll}\AgdaSpace{}%
\AgdaSymbol{\AgdaUnderscore{}}\AgdaSpace{}%
\AgdaSymbol{(λ}\AgdaSpace{}%
\AgdaBound{x₂}\AgdaSpace{}%
\AgdaSymbol{→}\AgdaSpace{}%
\AgdaBound{dsj}\AgdaSpace{}%
\AgdaSymbol{((}\AgdaInductiveConstructor{here}\AgdaSpace{}%
\AgdaInductiveConstructor{refl}\AgdaSymbol{)}\AgdaSpace{}%
\AgdaOperator{\AgdaInductiveConstructor{,}}\AgdaSpace{}%
\AgdaBound{x₂}\AgdaSymbol{))}\AgdaSpace{}%
\AgdaOperator{\AgdaInductiveConstructor{∷}}\AgdaSpace{}%
\AgdaBound{uls}\AgdaSymbol{)}\<%
\\
\>[17]\AgdaSymbol{(λ}\AgdaSpace{}%
\AgdaBound{x₂}\AgdaSpace{}%
\AgdaSymbol{→}\AgdaSpace{}%
\AgdaBound{dsj}\AgdaSpace{}%
\AgdaSymbol{((}\AgdaInductiveConstructor{there}\AgdaSpace{}%
\AgdaSymbol{(}\AgdaField{proj₁}\AgdaSpace{}%
\AgdaBound{x₂}\AgdaSymbol{))}\AgdaSpace{}%
\AgdaOperator{\AgdaInductiveConstructor{,}}\AgdaSpace{}%
\AgdaFunction{tail}\AgdaSpace{}%
\AgdaSymbol{(λ}\AgdaSpace{}%
\AgdaBound{x₃}\AgdaSpace{}%
\AgdaSymbol{→}\AgdaSpace{}%
\AgdaFunction{lookup}\AgdaSpace{}%
\AgdaBound{px}\AgdaSpace{}%
\AgdaSymbol{(}\AgdaField{proj₁}\AgdaSpace{}%
\AgdaBound{x₂}\AgdaSymbol{)}\AgdaSpace{}%
\AgdaSymbol{(}\AgdaFunction{sym}\AgdaSpace{}%
\AgdaBound{x₃}\AgdaSymbol{))}\AgdaSpace{}%
\AgdaSymbol{(}\AgdaField{proj₂}\AgdaSpace{}%
\AgdaBound{x₂}\AgdaSymbol{)))}\<%
\\
\>[17]\AgdaFunction{⊆₂}\AgdaSpace{}%
\AgdaFunction{is₂}\AgdaSpace{}%
\AgdaBound{hf}\AgdaSpace{}%
\AgdaBound{f₁}\<%
\\
\>[0][@{}l@{\AgdaIndent{0}}]%
\>[2]\AgdaKeyword{where}%
\>[2310I]\AgdaFunction{all₁}\AgdaSpace{}%
\AgdaSymbol{:}\AgdaSpace{}%
\AgdaDatatype{All}\AgdaSpace{}%
\AgdaSymbol{(λ}\AgdaSpace{}%
\AgdaBound{x₁}\AgdaSpace{}%
\AgdaSymbol{→}\AgdaSpace{}%
\AgdaFunction{Disjoint}\AgdaSpace{}%
\AgdaSymbol{(}\AgdaFunction{cmdWriteNames}\AgdaSpace{}%
\AgdaBound{x}\AgdaSpace{}%
\AgdaBound{s₁}\AgdaSymbol{)}\AgdaSpace{}%
\AgdaSymbol{(}\AgdaFunction{cmdReadWriteNames}\AgdaSpace{}%
\AgdaBound{x₁}\AgdaSpace{}%
\AgdaBound{s₁}\AgdaSymbol{))}\AgdaSpace{}%
\AgdaSymbol{(}\AgdaFunction{map}\AgdaSpace{}%
\AgdaField{proj₁}\AgdaSpace{}%
\AgdaBound{m}\AgdaSymbol{)}\<%
\\
\>[.][@{}l@{}]\<[2310I]%
\>[8]\AgdaFunction{all₁}\AgdaSpace{}%
\AgdaSymbol{=}\AgdaSpace{}%
\AgdaFunction{helper}\AgdaSpace{}%
\AgdaSymbol{(}\AgdaFunction{map}\AgdaSpace{}%
\AgdaField{proj₁}\AgdaSpace{}%
\AgdaBound{m}\AgdaSymbol{)}\AgdaSpace{}%
\AgdaSymbol{\AgdaUnderscore{}}\AgdaSpace{}%
\AgdaBound{x}\AgdaSpace{}%
\AgdaSymbol{(λ}\AgdaSpace{}%
\AgdaBound{x₂}\AgdaSpace{}%
\AgdaSymbol{→}\AgdaSpace{}%
\AgdaBound{¬hz}\AgdaSpace{}%
\AgdaSymbol{(}\AgdaFunction{∃hazard}\AgdaSpace{}%
\AgdaSymbol{\AgdaUnderscore{}}\AgdaSpace{}%
\AgdaSymbol{(}\AgdaField{proj₁}\AgdaSpace{}%
\AgdaBound{x₂}\AgdaSymbol{)}\AgdaSpace{}%
\AgdaSymbol{(}\AgdaField{proj₂}\AgdaSpace{}%
\AgdaBound{x₂}\AgdaSymbol{)))}\AgdaSpace{}%
\AgdaBound{⊆₁}\<%
\\
\>[8]\AgdaFunction{is₂}\AgdaSpace{}%
\AgdaSymbol{:}\AgdaSpace{}%
\AgdaDatatype{IdempotentState}\AgdaSpace{}%
\AgdaFunction{cmdReadNames}\AgdaSpace{}%
\AgdaSymbol{(}\AgdaFunction{St.run}\AgdaSpace{}%
\AgdaBound{x}\AgdaSpace{}%
\AgdaBound{s₁}\AgdaSymbol{)}\AgdaSpace{}%
\AgdaSymbol{(}\AgdaFunction{save}\AgdaSpace{}%
\AgdaBound{x}\AgdaSpace{}%
\AgdaSymbol{(}\AgdaFunction{cmdReadNames}\AgdaSpace{}%
\AgdaBound{x}\AgdaSpace{}%
\AgdaBound{s₁}\AgdaSymbol{)}\AgdaSpace{}%
\AgdaSymbol{(}\AgdaFunction{St.run}\AgdaSpace{}%
\AgdaBound{x}\AgdaSpace{}%
\AgdaBound{s₁}\AgdaSymbol{)}\AgdaSpace{}%
\AgdaBound{m}\AgdaSymbol{)}\<%
\\
\>[8]\AgdaFunction{is₂}\AgdaSpace{}%
\AgdaSymbol{=}\AgdaSpace{}%
\AgdaFunction{preserves}\AgdaSpace{}%
\AgdaBound{x}\AgdaSpace{}%
\AgdaBound{dc}\AgdaSpace{}%
\AgdaFunction{all₁}\AgdaSpace{}%
\AgdaBound{is}\<%
\\
\>[8]\AgdaComment{-- ( is}\<%
\\
\>[8]\AgdaFunction{≡₀}\AgdaSpace{}%
\AgdaSymbol{:}\AgdaSpace{}%
\AgdaField{proj₁}\AgdaSpace{}%
\AgdaSymbol{(}\AgdaBound{oracle}\AgdaSpace{}%
\AgdaBound{x}\AgdaSymbol{)}\AgdaSpace{}%
\AgdaSymbol{(}\AgdaFunction{St.run}\AgdaSpace{}%
\AgdaBound{x}\AgdaSpace{}%
\AgdaBound{s₁}\AgdaSymbol{)}\AgdaSpace{}%
\AgdaOperator{\AgdaDatatype{≡}}\AgdaSpace{}%
\AgdaField{proj₁}\AgdaSpace{}%
\AgdaSymbol{(}\AgdaBound{oracle}\AgdaSpace{}%
\AgdaBound{x}\AgdaSymbol{)}\AgdaSpace{}%
\AgdaBound{s₁}\<%
\\
\>[8]\AgdaFunction{≡₀}\AgdaSpace{}%
\AgdaSymbol{=}\AgdaSpace{}%
\AgdaFunction{sym}\AgdaSpace{}%
\AgdaSymbol{(}\AgdaField{proj₂}\AgdaSpace{}%
\AgdaSymbol{(}\AgdaBound{oracle}\AgdaSpace{}%
\AgdaBound{x}\AgdaSymbol{)}\AgdaSpace{}%
\AgdaBound{s₁}\AgdaSpace{}%
\AgdaSymbol{(}\AgdaFunction{St.run}\AgdaSpace{}%
\AgdaBound{x}\AgdaSpace{}%
\AgdaBound{s₁}\AgdaSymbol{)}\AgdaSpace{}%
\AgdaSymbol{λ}\AgdaSpace{}%
\AgdaBound{f₂}\AgdaSpace{}%
\AgdaBound{x₁}\AgdaSpace{}%
\AgdaSymbol{→}\AgdaSpace{}%
\AgdaFunction{lemma3}\AgdaSpace{}%
\AgdaBound{f₂}\AgdaSpace{}%
\AgdaSymbol{(}\AgdaFunction{cmdWrites}\AgdaSpace{}%
\AgdaBound{x}\AgdaSpace{}%
\AgdaBound{s₁}\AgdaSymbol{)}\AgdaSpace{}%
\AgdaSymbol{λ}\AgdaSpace{}%
\AgdaBound{x₂}\AgdaSpace{}%
\AgdaSymbol{→}\AgdaSpace{}%
\AgdaBound{dc}\AgdaSpace{}%
\AgdaSymbol{(}\AgdaBound{x₁}\AgdaSpace{}%
\AgdaOperator{\AgdaInductiveConstructor{,}}\AgdaSpace{}%
\AgdaBound{x₂}\AgdaSymbol{))}\<%
\\
\>[8]\AgdaFunction{≡₁}\AgdaSpace{}%
\AgdaSymbol{:}\AgdaSpace{}%
\AgdaFunction{cmdReadNames}\AgdaSpace{}%
\AgdaBound{x}\AgdaSpace{}%
\AgdaSymbol{(}\AgdaFunction{St.run}\AgdaSpace{}%
\AgdaBound{x}\AgdaSpace{}%
\AgdaBound{s₁}\AgdaSymbol{)}\AgdaSpace{}%
\AgdaOperator{\AgdaDatatype{≡}}\AgdaSpace{}%
\AgdaFunction{cmdReadNames}\AgdaSpace{}%
\AgdaBound{x}\AgdaSpace{}%
\AgdaBound{s₁}\<%
\\
\>[8]\AgdaFunction{≡₁}\AgdaSpace{}%
\AgdaSymbol{=}\AgdaSpace{}%
\AgdaFunction{cong}\AgdaSpace{}%
\AgdaSymbol{(}\AgdaFunction{map}\AgdaSpace{}%
\AgdaField{proj₁}\AgdaSpace{}%
\AgdaOperator{\AgdaFunction{∘}}\AgdaSpace{}%
\AgdaField{proj₁}\AgdaSymbol{)}\AgdaSpace{}%
\AgdaFunction{≡₀}\<%
\\
\>[8]\AgdaFunction{≡₂}\AgdaSpace{}%
\AgdaSymbol{:}\AgdaSpace{}%
\AgdaFunction{cmdWriteNames}\AgdaSpace{}%
\AgdaBound{x}\AgdaSpace{}%
\AgdaSymbol{(}\AgdaFunction{St.run}\AgdaSpace{}%
\AgdaBound{x}\AgdaSpace{}%
\AgdaBound{s₁}\AgdaSymbol{)}\AgdaSpace{}%
\AgdaOperator{\AgdaDatatype{≡}}\AgdaSpace{}%
\AgdaFunction{cmdWriteNames}\AgdaSpace{}%
\AgdaBound{x}\AgdaSpace{}%
\AgdaBound{s₁}\<%
\\
\>[8]\AgdaFunction{≡₂}\AgdaSpace{}%
\AgdaSymbol{=}\AgdaSpace{}%
\AgdaFunction{cong}\AgdaSpace{}%
\AgdaSymbol{(}\AgdaFunction{map}\AgdaSpace{}%
\AgdaField{proj₁}\AgdaSpace{}%
\AgdaOperator{\AgdaFunction{∘}}\AgdaSpace{}%
\AgdaField{proj₂}\AgdaSymbol{)}\AgdaSpace{}%
\AgdaFunction{≡₀}\<%
\\
\>[8]\AgdaFunction{⊆₂}\AgdaSpace{}%
\AgdaSymbol{:}\AgdaSpace{}%
\AgdaFunction{concatMap}\AgdaSpace{}%
\AgdaSymbol{(λ}\AgdaSpace{}%
\AgdaBound{x₁}\AgdaSpace{}%
\AgdaSymbol{→}\AgdaSpace{}%
\AgdaFunction{cmdReadWrites}\AgdaSpace{}%
\AgdaBound{x₁}\AgdaSpace{}%
\AgdaSymbol{(}\AgdaFunction{St.run}\AgdaSpace{}%
\AgdaBound{x}\AgdaSpace{}%
\AgdaBound{s₁}\AgdaSymbol{))}\AgdaSpace{}%
\AgdaSymbol{(}\AgdaBound{x}\AgdaSpace{}%
\AgdaOperator{\AgdaInductiveConstructor{∷}}\AgdaSpace{}%
\AgdaFunction{map}\AgdaSpace{}%
\AgdaField{proj₁}\AgdaSpace{}%
\AgdaBound{m}\AgdaSymbol{)}\AgdaSpace{}%
\AgdaOperator{\AgdaFunction{⊆}}\AgdaSpace{}%
\AgdaFunction{files}\AgdaSpace{}%
\AgdaSymbol{(}\AgdaFunction{rec}\AgdaSpace{}%
\AgdaBound{s₁}\AgdaSpace{}%
\AgdaBound{x}\AgdaSpace{}%
\AgdaBound{ls}\AgdaSymbol{)}\<%
\\
\>[8]\AgdaFunction{⊆₂}\AgdaSpace{}%
\AgdaBound{x₂}\AgdaSpace{}%
\AgdaKeyword{with}\AgdaSpace{}%
\AgdaFunction{∈-++⁻}\AgdaSpace{}%
\AgdaSymbol{(}\AgdaFunction{cmdReadWrites}\AgdaSpace{}%
\AgdaBound{x}\AgdaSpace{}%
\AgdaSymbol{(}\AgdaFunction{St.run}\AgdaSpace{}%
\AgdaBound{x}\AgdaSpace{}%
\AgdaBound{s₁}\AgdaSymbol{))}\AgdaSpace{}%
\AgdaBound{x₂}\<%
\\
\>[8]\AgdaSymbol{...}\AgdaSpace{}%
\AgdaSymbol{|}\AgdaSpace{}%
\AgdaInductiveConstructor{inj₁}\AgdaSpace{}%
\AgdaBound{∈₁}\AgdaSpace{}%
\AgdaKeyword{with}\AgdaSpace{}%
\AgdaFunction{∈-++⁻}\AgdaSpace{}%
\AgdaSymbol{(}\AgdaFunction{cmdReadNames}\AgdaSpace{}%
\AgdaBound{x}\AgdaSpace{}%
\AgdaSymbol{(}\AgdaFunction{St.run}\AgdaSpace{}%
\AgdaBound{x}\AgdaSpace{}%
\AgdaBound{s₁}\AgdaSymbol{))}\AgdaSpace{}%
\AgdaBound{∈₁}\<%
\\
\>[8]\AgdaSymbol{...}\AgdaSpace{}%
\AgdaSymbol{|}\AgdaSpace{}%
\AgdaInductiveConstructor{inj₁}\AgdaSpace{}%
\AgdaBound{∈rs}\AgdaSpace{}%
\AgdaSymbol{=}\AgdaSpace{}%
\AgdaFunction{∈-++⁺ˡ}\AgdaSpace{}%
\AgdaSymbol{(}\AgdaFunction{∈-++⁺ˡ}\AgdaSpace{}%
\AgdaSymbol{(}\AgdaFunction{subst}\AgdaSpace{}%
\AgdaSymbol{(λ}\AgdaSpace{}%
\AgdaBound{x₃}\AgdaSpace{}%
\AgdaSymbol{→}\AgdaSpace{}%
\AgdaSymbol{\AgdaUnderscore{}}\AgdaSpace{}%
\AgdaOperator{\AgdaFunction{∈}}\AgdaSpace{}%
\AgdaBound{x₃}\AgdaSymbol{)}\AgdaSpace{}%
\AgdaFunction{≡₁}\AgdaSpace{}%
\AgdaBound{∈rs}\AgdaSymbol{))}\<%
\\
\>[8]\AgdaSymbol{...}\AgdaSpace{}%
\AgdaSymbol{|}\AgdaSpace{}%
\AgdaInductiveConstructor{inj₂}\AgdaSpace{}%
\AgdaBound{∈ws}\AgdaSpace{}%
\AgdaSymbol{=}\AgdaSpace{}%
\AgdaFunction{∈-++⁺ʳ}\AgdaSpace{}%
\AgdaSymbol{(}\AgdaFunction{filesRead}\AgdaSpace{}%
\AgdaSymbol{(}\AgdaFunction{rec}\AgdaSpace{}%
\AgdaBound{s₁}\AgdaSpace{}%
\AgdaBound{x}\AgdaSpace{}%
\AgdaBound{ls}\AgdaSymbol{))}\AgdaSpace{}%
\AgdaSymbol{(}\AgdaFunction{∈-++⁺ˡ}\AgdaSpace{}%
\AgdaSymbol{(}\AgdaFunction{subst}\AgdaSpace{}%
\AgdaSymbol{(λ}\AgdaSpace{}%
\AgdaBound{x₃}\AgdaSpace{}%
\AgdaSymbol{→}\AgdaSpace{}%
\AgdaSymbol{\AgdaUnderscore{}}\AgdaSpace{}%
\AgdaOperator{\AgdaFunction{∈}}\AgdaSpace{}%
\AgdaBound{x₃}\AgdaSymbol{)}\AgdaSpace{}%
\AgdaFunction{≡₂}\AgdaSpace{}%
\AgdaBound{∈ws}\AgdaSymbol{))}\<%
\\
\>[8]\AgdaFunction{⊆₂}\AgdaSpace{}%
\AgdaBound{x₂}\AgdaSpace{}%
\AgdaSymbol{|}\AgdaSpace{}%
\AgdaInductiveConstructor{inj₂}\AgdaSpace{}%
\AgdaBound{∈₂}\AgdaSpace{}%
\AgdaKeyword{with}\AgdaSpace{}%
\AgdaFunction{∈-++⁻}\AgdaSpace{}%
\AgdaSymbol{(}\AgdaFunction{filesRead}\AgdaSpace{}%
\AgdaBound{ls}\AgdaSymbol{)}\AgdaSpace{}%
\AgdaSymbol{(}\AgdaBound{⊆₁}\AgdaSpace{}%
\AgdaSymbol{(}\AgdaFunction{subst}\AgdaSpace{}%
\AgdaSymbol{(λ}\AgdaSpace{}%
\AgdaBound{x₃}\AgdaSpace{}%
\AgdaSymbol{→}\AgdaSpace{}%
\AgdaSymbol{\AgdaUnderscore{}}\AgdaSpace{}%
\AgdaOperator{\AgdaFunction{∈}}\AgdaSpace{}%
\AgdaBound{x₃}\AgdaSymbol{)}\AgdaSpace{}%
\AgdaSymbol{(}\AgdaFunction{helper2}\AgdaSpace{}%
\AgdaSymbol{(}\AgdaFunction{map}\AgdaSpace{}%
\AgdaField{proj₁}\AgdaSpace{}%
\AgdaBound{m}\AgdaSymbol{)}\AgdaSpace{}%
\AgdaFunction{all₁}\AgdaSymbol{)}\AgdaSpace{}%
\AgdaBound{∈₂}\AgdaSymbol{))}\<%
\\
\>[8]\AgdaSymbol{...}\AgdaSpace{}%
\AgdaSymbol{|}\AgdaSpace{}%
\AgdaInductiveConstructor{inj₁}\AgdaSpace{}%
\AgdaBound{∈₁}\AgdaSpace{}%
\AgdaSymbol{=}\AgdaSpace{}%
\AgdaFunction{∈-++⁺ˡ}\AgdaSpace{}%
\AgdaSymbol{(}\AgdaFunction{∈-++⁺ʳ}\AgdaSpace{}%
\AgdaSymbol{(}\AgdaFunction{cmdReadNames}\AgdaSpace{}%
\AgdaBound{x}\AgdaSpace{}%
\AgdaBound{s₁}\AgdaSymbol{)}\AgdaSpace{}%
\AgdaBound{∈₁}\AgdaSymbol{)}\<%
\\
\>[8]\AgdaSymbol{...}\AgdaSpace{}%
\AgdaSymbol{|}\AgdaSpace{}%
\AgdaInductiveConstructor{inj₂}\AgdaSpace{}%
\AgdaBound{∈₁}\AgdaSpace{}%
\AgdaSymbol{=}\AgdaSpace{}%
\AgdaFunction{∈-++⁺ʳ}\AgdaSpace{}%
\AgdaSymbol{(}\AgdaFunction{filesRead}\AgdaSpace{}%
\AgdaSymbol{(}\AgdaFunction{rec}\AgdaSpace{}%
\AgdaBound{s₁}\AgdaSpace{}%
\AgdaBound{x}\AgdaSpace{}%
\AgdaBound{ls}\AgdaSymbol{))}\AgdaSpace{}%
\AgdaSymbol{(}\AgdaFunction{∈-++⁺ʳ}\AgdaSpace{}%
\AgdaSymbol{(}\AgdaFunction{cmdWriteNames}\AgdaSpace{}%
\AgdaBound{x}\AgdaSpace{}%
\AgdaBound{s₁}\AgdaSymbol{)}\AgdaSpace{}%
\AgdaBound{∈₁}\AgdaSymbol{)}\<%
\\
\>[0]\AgdaSymbol{...}\AgdaSpace{}%
\AgdaSymbol{|}\AgdaSpace{}%
\AgdaInductiveConstructor{just}\AgdaSpace{}%
\AgdaBound{all₁}\AgdaSpace{}%
\AgdaKeyword{with}\AgdaSpace{}%
\AgdaFunction{getCmdIdempotent}\AgdaSpace{}%
\AgdaBound{m}\AgdaSpace{}%
\AgdaBound{x}\AgdaSpace{}%
\AgdaBound{is}\AgdaSpace{}%
\AgdaBound{x∈mem}\<%
\\
\>[0]\AgdaSymbol{...}\AgdaSpace{}%
\AgdaSymbol{|}\AgdaSpace{}%
\AgdaBound{ci}\AgdaSpace{}%
\AgdaSymbol{=}\AgdaSpace{}%
\AgdaFunction{correct-inner}\AgdaSpace{}%
\AgdaBound{m}\AgdaSpace{}%
\AgdaBound{b}\AgdaSpace{}%
\AgdaFunction{∀₂}\AgdaSpace{}%
\AgdaSymbol{(}\AgdaFunction{dsj-≡}\AgdaSpace{}%
\AgdaSymbol{(}\AgdaFunction{St.run}\AgdaSpace{}%
\AgdaBound{x}\AgdaSpace{}%
\AgdaBound{s₁}\AgdaSymbol{)}\AgdaSpace{}%
\AgdaBound{s₁}\AgdaSpace{}%
\AgdaBound{b}\AgdaSpace{}%
\AgdaFunction{∀₃}\AgdaSpace{}%
\AgdaBound{dsb}\AgdaSymbol{)}\AgdaSpace{}%
\AgdaBound{ub}\AgdaSpace{}%
\AgdaBound{uls}\AgdaSpace{}%
\AgdaSymbol{(λ}\AgdaSpace{}%
\AgdaBound{x₂}\AgdaSpace{}%
\AgdaSymbol{→}\AgdaSpace{}%
\AgdaBound{dsj}\AgdaSpace{}%
\AgdaSymbol{(}\AgdaInductiveConstructor{there}\AgdaSpace{}%
\AgdaSymbol{(}\AgdaField{proj₁}\AgdaSpace{}%
\AgdaBound{x₂}\AgdaSymbol{)}\AgdaSpace{}%
\AgdaOperator{\AgdaInductiveConstructor{,}}\AgdaSpace{}%
\AgdaField{proj₂}\AgdaSpace{}%
\AgdaBound{x₂}\AgdaSymbol{))}\AgdaSpace{}%
\AgdaBound{⊆₁}\AgdaSpace{}%
\AgdaBound{is}\AgdaSpace{}%
\AgdaFunction{hf₂}\AgdaSpace{}%
\AgdaBound{f₁}\<%
\\
\>[0][@{}l@{\AgdaIndent{0}}]%
\>[2]\AgdaKeyword{where}%
\>[2599I]\AgdaFunction{∀₃}\AgdaSpace{}%
\AgdaSymbol{:}\AgdaSpace{}%
\AgdaSymbol{∀}\AgdaSpace{}%
\AgdaBound{f₁}\AgdaSpace{}%
\AgdaSymbol{→}\AgdaSpace{}%
\AgdaBound{s₁}\AgdaSpace{}%
\AgdaBound{f₁}\AgdaSpace{}%
\AgdaOperator{\AgdaDatatype{≡}}\AgdaSpace{}%
\AgdaFunction{St.run}\AgdaSpace{}%
\AgdaBound{x}\AgdaSpace{}%
\AgdaBound{s₁}\AgdaSpace{}%
\AgdaBound{f₁}\<%
\\
\>[.][@{}l@{}]\<[2599I]%
\>[8]\AgdaFunction{∀₃}\AgdaSpace{}%
\AgdaBound{f₁}\AgdaSpace{}%
\AgdaSymbol{=}\AgdaSpace{}%
\AgdaSymbol{(}\AgdaFunction{sym}\AgdaSpace{}%
\AgdaSymbol{(}\AgdaBound{ci}\AgdaSpace{}%
\AgdaBound{all₁}\AgdaSpace{}%
\AgdaBound{f₁}\AgdaSymbol{))}\<%
\\
\>[8]\AgdaFunction{∀₂}\AgdaSpace{}%
\AgdaSymbol{:}\AgdaSpace{}%
\AgdaSymbol{∀}\AgdaSpace{}%
\AgdaBound{f₁}\AgdaSpace{}%
\AgdaSymbol{→}\AgdaSpace{}%
\AgdaBound{s₁}\AgdaSpace{}%
\AgdaBound{f₁}\AgdaSpace{}%
\AgdaOperator{\AgdaDatatype{≡}}\AgdaSpace{}%
\AgdaFunction{St.run}\AgdaSpace{}%
\AgdaBound{x}\AgdaSpace{}%
\AgdaBound{s₂}\AgdaSpace{}%
\AgdaBound{f₁}\<%
\\
\>[8]\AgdaFunction{∀₂}\AgdaSpace{}%
\AgdaBound{f₁}\AgdaSpace{}%
\AgdaSymbol{=}\AgdaSpace{}%
\AgdaFunction{trans}\AgdaSpace{}%
\AgdaSymbol{(}\AgdaFunction{∀₃}\AgdaSpace{}%
\AgdaBound{f₁}\AgdaSymbol{)}\AgdaSpace{}%
\AgdaSymbol{(}\AgdaFunction{StP.lemma2}\AgdaSpace{}%
\AgdaSymbol{(}\AgdaField{proj₂}\AgdaSpace{}%
\AgdaSymbol{(}\AgdaBound{oracle}\AgdaSpace{}%
\AgdaBound{x}\AgdaSymbol{)}\AgdaSpace{}%
\AgdaBound{s₁}\AgdaSpace{}%
\AgdaBound{s₂}\AgdaSpace{}%
\AgdaSymbol{λ}\AgdaSpace{}%
\AgdaBound{f₃}\AgdaSpace{}%
\AgdaBound{x₂}\AgdaSpace{}%
\AgdaSymbol{→}\AgdaSpace{}%
\AgdaBound{∀₁}\AgdaSpace{}%
\AgdaBound{f₃}\AgdaSymbol{)}\AgdaSpace{}%
\AgdaSymbol{(}\AgdaBound{∀₁}\AgdaSpace{}%
\AgdaBound{f₁}\AgdaSymbol{))}\<%
\\
\>[8]\AgdaFunction{hf₂}\AgdaSpace{}%
\AgdaSymbol{:}\AgdaSpace{}%
\AgdaDatatype{HazardFree}\AgdaSpace{}%
\AgdaBound{s₁}\AgdaSpace{}%
\AgdaBound{b}\AgdaSpace{}%
\AgdaSymbol{\AgdaUnderscore{}}\AgdaSpace{}%
\AgdaSymbol{\AgdaUnderscore{}}\<%
\\
\>[8]\AgdaFunction{hf₂}\AgdaSpace{}%
\AgdaSymbol{=}\AgdaSpace{}%
\AgdaFunction{hf-still}\AgdaSpace{}%
\AgdaBound{b}\AgdaSpace{}%
\AgdaInductiveConstructor{[]}\AgdaSpace{}%
\AgdaSymbol{((}\AgdaBound{x}\AgdaSpace{}%
\AgdaOperator{\AgdaInductiveConstructor{,}}\AgdaSpace{}%
\AgdaFunction{cmdReadNames}\AgdaSpace{}%
\AgdaBound{x}\AgdaSpace{}%
\AgdaBound{s₁}\AgdaSpace{}%
\AgdaOperator{\AgdaInductiveConstructor{,}}\AgdaSpace{}%
\AgdaFunction{cmdWriteNames}\AgdaSpace{}%
\AgdaBound{x}\AgdaSpace{}%
\AgdaBound{s₁}\AgdaSymbol{)}\AgdaSpace{}%
\AgdaOperator{\AgdaInductiveConstructor{∷}}\AgdaSpace{}%
\AgdaInductiveConstructor{[]}\AgdaSymbol{)}\AgdaSpace{}%
\AgdaSymbol{\AgdaUnderscore{}}\AgdaSpace{}%
\AgdaSymbol{(λ}\AgdaSpace{}%
\AgdaBound{f₂}\AgdaSpace{}%
\AgdaBound{x₂}\AgdaSpace{}%
\AgdaSymbol{→}\AgdaSpace{}%
\AgdaFunction{sym}\AgdaSpace{}%
\AgdaSymbol{(}\AgdaFunction{∀₃}\AgdaSpace{}%
\AgdaBound{f₂}\AgdaSymbol{))}\AgdaSpace{}%
\AgdaBound{ub}\AgdaSpace{}%
\AgdaSymbol{(}\AgdaFunction{∉toAll}\AgdaSpace{}%
\AgdaSymbol{\AgdaUnderscore{}}\AgdaSpace{}%
\AgdaSymbol{(λ}\AgdaSpace{}%
\AgdaBound{x₂}\AgdaSpace{}%
\AgdaSymbol{→}\AgdaSpace{}%
\AgdaBound{dsj}\AgdaSpace{}%
\AgdaSymbol{((}\AgdaInductiveConstructor{here}\AgdaSpace{}%
\AgdaInductiveConstructor{refl}\AgdaSymbol{)}\AgdaSpace{}%
\AgdaOperator{\AgdaInductiveConstructor{,}}\AgdaSpace{}%
\AgdaBound{x₂}\AgdaSymbol{))}\AgdaSpace{}%
\AgdaOperator{\AgdaInductiveConstructor{∷}}\AgdaSpace{}%
\AgdaBound{uls}\AgdaSymbol{)}\AgdaSpace{}%
\AgdaSymbol{(λ}\AgdaSpace{}%
\AgdaBound{x₂}\AgdaSpace{}%
\AgdaSymbol{→}\AgdaSpace{}%
\AgdaBound{dsj}\AgdaSpace{}%
\AgdaSymbol{(}\AgdaInductiveConstructor{there}\AgdaSpace{}%
\AgdaSymbol{(}\AgdaField{proj₁}\AgdaSpace{}%
\AgdaBound{x₂}\AgdaSymbol{)}\AgdaSpace{}%
\AgdaOperator{\AgdaInductiveConstructor{,}}\AgdaSpace{}%
\AgdaFunction{tail}\AgdaSpace{}%
\AgdaSymbol{(λ}\AgdaSpace{}%
\AgdaBound{x₃}\AgdaSpace{}%
\AgdaSymbol{→}\AgdaSpace{}%
\AgdaFunction{lookup}\AgdaSpace{}%
\AgdaBound{px}\AgdaSpace{}%
\AgdaSymbol{(}\AgdaField{proj₁}\AgdaSpace{}%
\AgdaBound{x₂}\AgdaSymbol{)}\AgdaSpace{}%
\AgdaSymbol{(}\AgdaFunction{sym}\AgdaSpace{}%
\AgdaBound{x₃}\AgdaSymbol{))}\AgdaSpace{}%
\AgdaSymbol{(}\AgdaField{proj₂}\AgdaSpace{}%
\AgdaBound{x₂}\AgdaSymbol{)))}\AgdaSpace{}%
\AgdaBound{hf}\<%
\end{code}

\begin{code}[hide]%
\>[0]\AgdaFunction{correct}\AgdaSpace{}%
\AgdaSymbol{:}\AgdaSpace{}%
\AgdaSymbol{∀}\AgdaSpace{}%
\AgdaSymbol{\{}\AgdaBound{s}\AgdaSymbol{\}}\AgdaSpace{}%
\AgdaBound{b}\AgdaSpace{}%
\AgdaBound{ls}\AgdaSpace{}%
\AgdaSymbol{→}\AgdaSpace{}%
\AgdaDatatype{DisjointBuild}\AgdaSpace{}%
\AgdaBound{s}\AgdaSpace{}%
\AgdaBound{b}\AgdaSpace{}%
\AgdaSymbol{→}\AgdaSpace{}%
\AgdaDatatype{Unique}\AgdaSpace{}%
\AgdaBound{b}\AgdaSpace{}%
\AgdaSymbol{→}\AgdaSpace{}%
\AgdaDatatype{Unique}\AgdaSpace{}%
\AgdaSymbol{(}\AgdaFunction{map}\AgdaSpace{}%
\AgdaField{proj₁}\AgdaSpace{}%
\AgdaBound{ls}\AgdaSymbol{)}\AgdaSpace{}%
\AgdaSymbol{→}\AgdaSpace{}%
\AgdaFunction{Disjoint}\AgdaSpace{}%
\AgdaBound{b}\AgdaSpace{}%
\AgdaSymbol{(}\AgdaFunction{map}\AgdaSpace{}%
\AgdaField{proj₁}\AgdaSpace{}%
\AgdaBound{ls}\AgdaSymbol{)}\AgdaSpace{}%
\AgdaSymbol{→}\AgdaSpace{}%
\AgdaDatatype{HazardFree}\AgdaSpace{}%
\AgdaBound{s}\AgdaSpace{}%
\AgdaBound{b}\AgdaSpace{}%
\AgdaBound{b}\AgdaSpace{}%
\AgdaBound{ls}\AgdaSpace{}%
\AgdaSymbol{→}\AgdaSpace{}%
\AgdaSymbol{(∀}\AgdaSpace{}%
\AgdaBound{f₁}\AgdaSpace{}%
\AgdaSymbol{→}\AgdaSpace{}%
\AgdaField{proj₁}\AgdaSpace{}%
\AgdaSymbol{(}\AgdaFunction{fabricate}\AgdaSpace{}%
\AgdaBound{b}\AgdaSpace{}%
\AgdaSymbol{(}\AgdaBound{s}\AgdaSpace{}%
\AgdaOperator{\AgdaInductiveConstructor{,}}\AgdaSpace{}%
\AgdaSymbol{\AgdaUnderscore{}))}\AgdaSpace{}%
\AgdaBound{f₁}\AgdaSpace{}%
\AgdaOperator{\AgdaDatatype{≡}}\AgdaSpace{}%
\AgdaFunction{script}\AgdaSpace{}%
\AgdaBound{b}\AgdaSpace{}%
\AgdaBound{s}\AgdaSpace{}%
\AgdaBound{f₁}\AgdaSymbol{)}\<%
\\
\>[0]\AgdaFunction{correct}\AgdaSpace{}%
\AgdaBound{b}\AgdaSpace{}%
\AgdaBound{ls}\AgdaSpace{}%
\AgdaBound{dsb}\AgdaSpace{}%
\AgdaBound{ub}\AgdaSpace{}%
\AgdaBound{uls}\AgdaSpace{}%
\AgdaBound{dsj}\AgdaSpace{}%
\AgdaBound{hf}\AgdaSpace{}%
\AgdaSymbol{=}\AgdaSpace{}%
\AgdaFunction{correct-inner}\AgdaSpace{}%
\AgdaInductiveConstructor{[]}\AgdaSpace{}%
\AgdaBound{b}\AgdaSpace{}%
\AgdaSymbol{(λ}\AgdaSpace{}%
\AgdaBound{f₁}\AgdaSpace{}%
\AgdaSymbol{→}\AgdaSpace{}%
\AgdaInductiveConstructor{refl}\AgdaSymbol{)}\AgdaSpace{}%
\AgdaBound{dsb}\AgdaSpace{}%
\AgdaBound{ub}\AgdaSpace{}%
\AgdaBound{uls}\AgdaSpace{}%
\AgdaBound{dsj}\AgdaSpace{}%
\AgdaSymbol{(λ}\AgdaSpace{}%
\AgdaSymbol{())}\AgdaSpace{}%
\AgdaInductiveConstructor{[]}\AgdaSpace{}%
\AgdaBound{hf}\<%
\end{code}

\newcommand{\correctF}{%
\begin{code}%
\>[0]\AgdaFunction{correct-fabricate}\AgdaSpace{}%
\AgdaSymbol{:}\AgdaSpace{}%
\AgdaSymbol{∀}\AgdaSpace{}%
\AgdaSymbol{\{}\AgdaBound{s}\AgdaSymbol{\}}\AgdaSpace{}%
\AgdaBound{b}\AgdaSpace{}%
\AgdaSymbol{→}\AgdaSpace{}%
\AgdaFunction{PreCond}\AgdaSpace{}%
\AgdaBound{s}\AgdaSpace{}%
\AgdaBound{b}\AgdaSpace{}%
\AgdaBound{b}\AgdaSpace{}%
\AgdaSymbol{→}\AgdaSpace{}%
\AgdaDatatype{HazardFree}\AgdaSpace{}%
\AgdaBound{s}\AgdaSpace{}%
\AgdaBound{b}\AgdaSpace{}%
\AgdaBound{b}\AgdaSpace{}%
\AgdaInductiveConstructor{[]}\AgdaSpace{}%
\AgdaSymbol{→}\AgdaSpace{}%
\AgdaSymbol{(∀}\AgdaSpace{}%
\AgdaBound{f₁}\AgdaSpace{}%
\AgdaSymbol{→}\AgdaSpace{}%
\AgdaField{proj₁}\AgdaSpace{}%
\AgdaSymbol{(}\AgdaFunction{fabricate}\AgdaSpace{}%
\AgdaBound{b}\AgdaSpace{}%
\AgdaSymbol{(}\AgdaBound{s}\AgdaSpace{}%
\AgdaOperator{\AgdaInductiveConstructor{,}}\AgdaSpace{}%
\AgdaInductiveConstructor{[]}\AgdaSymbol{))}\AgdaSpace{}%
\AgdaBound{f₁}\AgdaSpace{}%
\AgdaOperator{\AgdaDatatype{≡}}\AgdaSpace{}%
\AgdaFunction{script}\AgdaSpace{}%
\AgdaBound{b}\AgdaSpace{}%
\AgdaBound{s}\AgdaSpace{}%
\AgdaBound{f₁}\AgdaSymbol{)}\<%
\end{code}}
\begin{code}[hide]%
\>[0]\AgdaFunction{correct-fabricate}\AgdaSpace{}%
\AgdaBound{b}\AgdaSpace{}%
\AgdaSymbol{(}\AgdaBound{dsb}\AgdaSpace{}%
\AgdaOperator{\AgdaInductiveConstructor{,}}\AgdaSpace{}%
\AgdaBound{ub}\AgdaSpace{}%
\AgdaOperator{\AgdaInductiveConstructor{,}}\AgdaSpace{}%
\AgdaSymbol{\AgdaUnderscore{})}\AgdaSpace{}%
\AgdaBound{hf}\AgdaSpace{}%
\AgdaSymbol{=}\AgdaSpace{}%
\AgdaFunction{correct}\AgdaSpace{}%
\AgdaBound{b}\AgdaSpace{}%
\AgdaInductiveConstructor{[]}\AgdaSpace{}%
\AgdaBound{dsb}\AgdaSpace{}%
\AgdaBound{ub}\AgdaSpace{}%
\AgdaInductiveConstructor{Data.List.Relation.Unary.AllPairs.[]}\AgdaSpace{}%
\AgdaFunction{g₁}\AgdaSpace{}%
\AgdaBound{hf}\<%
\\
\>[0][@{}l@{\AgdaIndent{0}}]%
\>[2]\AgdaKeyword{where}%
\>[2825I]\AgdaFunction{g₁}\AgdaSpace{}%
\AgdaSymbol{:}\AgdaSpace{}%
\AgdaFunction{Disjoint}\AgdaSpace{}%
\AgdaBound{b}\AgdaSpace{}%
\AgdaInductiveConstructor{[]}\<%
\\
\>[.][@{}l@{}]\<[2825I]%
\>[8]\AgdaFunction{g₁}\AgdaSpace{}%
\AgdaSymbol{()}\<%
\end{code}

\setcopyright{rightsretained}
\acmPrice{}
\acmDOI{10.1145/3497775.3503687}
\acmYear{2022}
\copyrightyear{2022}
\acmSubmissionID{poplws22cppmain-p29-p}
\acmISBN{978-1-4503-9182-5/22/01}
\acmConference[CPP '22]{Proceedings of the 11th ACM SIGPLAN International Conference on Certified Programs and Proofs}{January 17--18, 2022}{Philadelphia, PA, USA}
\acmBooktitle{Proceedings of the 11th ACM SIGPLAN International Conference on Certified Programs and Proofs (CPP '22), January 17--18, 2022, Philadelphia, PA, USA}
\begin{document}

\newcommand{\Make}{\textsc{Make}\xspace}
\newcommand{\Rattle}{\textsc{Rattle}\xspace}
\newcommand{\Fabricate}{\textsc{Fabricate}\xspace}
\newcommand{\Bazel}{\textsc{Bazel}\xspace}
\newcommand{\Buck}{\textsc{Buck}\xspace}
\newcommand{\Shake}{\textsc{Shake}\xspace}
\newcommand{\Bigbro}{\textsc{BigBro}\xspace}
\newcommand{\Fac}{\textsc{Fac}\xspace}
\newcommand{\Fsatrace}{\textsc{Fsatrace}\xspace}
\newcommand{\tracedfs}{\textsc{Traced-Fs}\xspace}
\newcommand{\BuildXL}{\textsc{BuildXL}\xspace}
\newcommand{\Nix}{\textsc{Nix}\xspace}
\newcommand{\Memoize}{\textsc{Memoize}\xspace}
\newcommand{\Stroll}{\textsc{Stroll}\xspace}
\newcommand{\Pluto}{\textsc{Pluto}\xspace}
\newcommand{\PIE}{\textsc{PIE}\xspace}
\newcommand{\Agda}{\textsc{Agda}\xspace}
\newcommand{\Haskell}{\textsc{Haskell}\xspace}
\newcommand{\ChezScheme}{\textsc{Chez Scheme}\xspace}
\newcommand{\LaForge}{\textsc{LaForge}\xspace}
\newcommand{\Shell}{\textsc{Shell}\xspace}
\newcommand{\CloudMake}{\textsc{CloudMake}\xspace}
\newcommand{\Script}{\textsc{Script}\xspace}

\title{Forward Build Systems, Formally}

\author{Sarah Spall}
\affiliation{
  \institution{Indiana University}
  \country{USA}
}
\email{sjspall@iu.edu}

\author{Neil Mitchell}
\affiliation{
  \institution{Meta}
  \country{UK}
}
\email{ndmitchell@gmail.com}

\author{Sam Tobin-Hochstadt}
\affiliation{
  \institution{Indiana University}
  \country{USA}
}
\email{samth@indiana.edu}

\begin{abstract}
Build systems are a fundamental part of software construction, but
their correctness has received comparatively little attention,
relative to more prominent parts of the toolchain. In this paper, we
address the correctness of \emph{forward build systems}, which
automatically determine the dependency structure of the build, rather
than having it specified by the programmer.

We first define what it means for a forward build system to be
correct---it must behave identically to simply executing the
programmer-specified commands in order. Of course, realistic build
systems avoid repeated work, stop early when possible, and run
commands in parallel, and we prove that these optimizations, as
embodied in the recent forward build system \textsc{Rattle}, preserve
our definition of correctness. Along the way, we show that other
forward build systems, such as \textsc{Fabricate} and \textsc{Memoize},
are also correct.

We carry out all of our work in \Agda, and describe in detail the
assumptions underlying both \textsc{Rattle} itself and our modeling of it.
\end{abstract}


\maketitle

\section{Introduction}
\label{sec:intro}
Build systems are used by everyone.  They provide the powerful ability to describe how complex projects should be built and to build them in a repeatable and efficient fashion.
Two of a build systems most important features are \emph{incrementality} and \emph{parallelism}.  To provide both \emph{incrementality} and \emph{parallelism} build systems like \Make~\cite{make}
require the user to declare the targets to build, what those targets depend on, and how to build those targets.  Assuming the user specified all the dependencies of the project correctly, the
software project will be built correctly; during re-builds only those targets whose dependencies have changed will be built again; and targets which do not depend on one another can be built in
parallel.  But, getting the dependencies of a software project correct is not so easy, as \citet{detecting_incorrect_build_rules} show. Omitted or incorrect dependencies can lead to consequences ranging from missed opportunities for parallelism to failure to rebuild in the presence of changes (often requiring a ``clean'' step) to outright incorrect results.

An alternative to a build system such as \Make is a \emph{forward build system}, where a user writes a program that says how to build their software project, without declaring targets
or dependencies.  A \emph{forward build system} is, conceptually, just a simple command interpreter; it takes a sequence of commands and executes them, without checking for things such as targets or dependencies, as if the build was a \Shell script.
Unlike a \Shell however, it can provide \emph{incrementality} and even \emph{parallelism}.  Let's take an example:

\begin{verbatim}
gcc -c file.c
gcc -c string.c
gcc -c print.c
gcc file.c string.c print.c -o program
\end{verbatim}

A \Shell script would run each of the commands in this script every time the script was run.  A \emph{forward build system} however, by using \emph{system tracing}, runs each command and
records the files the command read or wrote during its execution.  When the build is re-run, perhaps after a change to one of the files, it can use this information to decide if it should run a command again, much how \Make decides if it should re-build a
target by checking if its dependencies have changed.  Forward build system implementations include \Memoize \cite{memoize}, \Fabricate \cite{fabricate} and \Rattle \cite{rattle}.

Like our notional \Shell script, forward build systems are typically embedded in full-fledged languages to provide control structures, libraries, and other conveniences: \Memoize and \Fabricate build scripts are Python programs, and \Rattle build scripts are Haskell programs.  In this paper we model \emph{forward build systems} in general, and then continue with \Rattle in more detail.  Our reason for focusing on \Rattle is that it contains two unique features---first, it provides support for implicit parallelism via \emph{speculation}, and second it defines a notion of whether the commands in a forward build system are valid or conflict with each other via \emph{hazards}.

Hazards allow stating correctness for Rattle in a way that is not
possible for prior forward build systems. In particular, a build
process that writes to some of the \emph{inputs} to the build cannot
be correctly executed based solely on remembering past commands, since
there's no one answer as to what the past state was. In this
situation as well as others, \Rattle detects the hazard and reports an error to the
user.

\renewcommand{\mkbegdispquote}[2]{\itshape}
We thus aim for the following notion of correctness:
\begin{displayquote}
  A forward build system is correct if, for every build, it either
  produces the same result as running the commands in order, or
  reports an error.
\end{displayquote}

Our central contribution is a simple yet formal account of what it
means for a forward build system to be correct, based on the above
idea. We demonstrate the utility of our approach with application to
models of both simple and sophisticated forward build systems,
including hazard detection, memoization, and parallel speculation.

This paper informally introduces \Rattle, the optimisations is provides, and what correctness means in \S\ref{sec:rattle}. We then move to \Agda, providing:

\begin{itemize}
    \item The key concepts and definitions underpinning all of our models of forward build systems in \S\ref{sec:model}.
    \item A formal model of \Fabricate, as a representative of simple forward build systems, and proof of its correctness in \S\ref{sec:forward}.
    \item A formal model of \Rattle and proof of its correctness in \S\ref{sec:proof}. We prove the correctness of \Rattle first without speculation, and then including speculation.
\end{itemize}

In the process of formalising these proofs we found a small bug in
\Rattle hazard computations, which we fixed in the proof, see \S\ref{sec:bug}. The same
fix can be applied to \Rattle itself, showing the value of proving
these complex concepts formally.

\section{Rattle}
\label{sec:rattle}

\begin{figure}
\begin{verbatim}
-- The Run monad
data Run a = ...
rattle :: Run a -> IO a

-- Running commands
cmd :: CmdArguments args => args
data CmdOption = Cwd FilePath | ...

class CmdArguments args
-- instances to allow any number
-- of String/[String]/CmdOption values
\end{verbatim}
\caption{Part of the Rattle API.}
\label{fig:api}
\end{figure}
\Rattle is a \emph{forward build system} that is implemented in Haskell and whose build scripts are Haskell programs which use the \Rattle API, given in Figure~\ref{fig:api}. A \Rattle build script comprises of control logic in Haskell and calls to \texttt{cmd} which execute external processes. Taking our example from \S\ref{sec:intro}, we can write the following \Rattle program, which uses both the \Rattle API and the Haskell library \texttt{System.FilePath}.

\begin{verbatim}
main = rattle $ do
   let cs = ["file.c" , "string.c"
            , "print.c"]
   forM cs (\c -> cmd "gcc -c" c)
   let to0 x = takeBaseName x <.> "o"
   cmd "gcc -o program" (map to0 cs)
\end{verbatim}

This program is made up of regular Haskell logic, such as \texttt{forM} and \texttt{let}, which is not visible to \Rattle, and the calls invoked by \texttt{cmd}, which are visible to \Rattle. The control logic is executed every time the script is run, so \Rattle's view of the script is:

\begin{verbatim}
gcc -c file.c
gcc -c string.c
gcc -c print.c
gcc -o program file.o string.o print.o
\end{verbatim}

Whenever \Rattle executes a command, it uses \emph{tracing} to record which files were read (inputs) and written (outputs) by the command, and the contents of those files. In future executions, if \Rattle sees a command which previously ran with the current value of the inputs and outputs, it skips execution, assuming that the command would have no effect. If \Rattle sees a command which previously ran with the current inputs, but where \Rattle has access to a copy of the previous outputs, it copies those outputs to where the command would put them, and doesn't run the command. If the command has never been seen before with these inputs, it will be run afresh with tracing.

One weakness of this approach is that the execution is single-threaded, only running one command at a time. In our example, it should be possible to run all the \texttt{gcc -c} commands in parallel, since they work on disjoint source files and destinations. To overcome that weakness, \Rattle uses \emph{speculation}, where it predicts commands that are likely to be required in future but also unlikely to conflict with other commands, and runs them \emph{before} the script requests them. The commands \Rattle considers for speculation are simply those that executed in the previous run. Of more interest is the way \Rattle decides whether a command is likely to conflict, which it does using a concept called hazards.  \Rattle chooses a command to speculate next by picking a command which would not cause a hazard with any command already completed, or a read conflict with any command currently running, according to the tracing data from the commands' previous run.

\subsection{Hazards}
\label{sec:hazards}

A build system has reached a fixed point if on a subsequent rebuild, no work is done, because every command is already up to date. However, that's not true for all shell scripts -- consider the script:

\begin{verbatim}
gcc -c foo.c
echo X >> foo.c
\end{verbatim}

On the first execution the file \texttt{foo.c} is compiled and then, after it has been used as the input to \texttt{gcc}, is modified. The \Rattle paper \cite{rattle} introduced \emph{hazards}, which detect bad behaviour, where the absence of any hazards implies the build has reached a fixed point. In particular, writing to a file after it has already been read from, as in the \texttt{echo X} command above, is a read-before-write hazard. The three kinds of hazards defined by \Rattle are:

\begin{description}
\item [Read-before-write hazard] One command reads a file which a later command writes to.  On a subsequent rebuild the first command will need to run again because the second command wrote to its dependency.
\item [Write-before-write hazard] A command wrote to a file a later command writes to the same file again.  On a subsequent rebuild the first command will need to run again because the second command changed its output file, and that first command rerunning will likely trigger the second command to run again.\footnote{Note that file moves as commands are likely to cause hazards, under these definitions, unless the moved file was in the input, which would make the build fundamentally non-idempotent. Moves can be combined with the command that produced the moved file, however, to produce a command that Rattle accepts.}
\item [Speculative write-before-read hazard] When \Rattle runs commands requested by the build, it marks them as \emph{required}. If a command was run speculatively and hasn't been marked as \emph{required} yet, it is \emph{speculated}.  If a \emph{speculated} command writes to a file later read by a \emph{required} command then \Rattle has run commands in the wrong order, and a speculative write before read hazard has occurred.
\end{description}

\subsection{Assumptions}
\label{sec:assumptions}

In order for a hazard-free build to have reached a fixed point, \Rattle makes certain assumptions which it does not check:

\begin{description}
\item[Determinism of commands] \Rattle assumes all commands are deterministic, although it doesn't enforce this property. When a command is not deterministic it assumes all possible outputs are equivalent.  \Rattle supports using \emph{hash forwarding} for non-deterministic commands, hashing the inputs of a command rather than the outputs. If a build contains non-deterministic commands then it might have a different result each time it runs, which is not the desired behavior of a build system. Note that this is assumed by \emph{all} practical build systems, including \Make---the alternative is to abandon the value of build systems entirely.
\item[Disjoint reads and writes] \Rattle assumes commands do not write to their own inputs (aka all writes first truncate).  If a command writes to its own inputs, \Rattle  cannot record the true value of its inputs since the tracing used only captures their value \emph{after} the command has completed.
\item[Tracing data is correct] \Rattle assumes tracing data is complete and correct, if it is not, then \Rattle might make the wrong decision about when to re-run commands and which commands it is safe to run in parallel.  \Rattle supports a number of tracing backends per platform (using \texttt{LD\_LIBRARY\_PRELOAD}, hooking system calls, etc), which vary in their precision vs performance trade-offs, but we assume an ideal model where they are correct.
  For example, \Rattle does not track directory operations or operations that are not reads or writes; our model does not contain directories or any such file operations. 
  A detailed discussion of Rattle's approach to tracing is given by \citet[\S3.5]{rattle}
\end{description}

\subsection{Correctness}
\label{sec:correctness_informal}

The \Rattle implementation executes a build with speculation, and if that raises a hazard, repeats the build without speculation. Some build scripts are considered flawed, and \emph{always} raise a hazard, even without speculation. We consider the \Rattle approach to be correct if every execution \emph{either} raises a hazard \emph{or} produces a result equivalent to that of the shell script.

In the following sections we model \Rattle in \Agda, then prove that this model, including its approach to speculation, meets this definition of correctness.

\begin{figure}
\begin{flushleft}
\AgdaFunction{FileName} \AgdaSymbol{:} \AgdaFunction{Set}\\
\AgdaFunction{FileName} \AgdaSymbol{=} \AgdaFunction{String}\\
\AgdaFunction{FileContent} \AgdaSymbol{:} \AgdaFunction{Set}\\
\AgdaFunction{FileContent} \AgdaSymbol{=} \AgdaFunction{String}\\
\AgdaFunction{File} \AgdaSymbol{:} \AgdaFunction{Set}\\
\AgdaFunction{File} \AgdaSymbol{=} \AgdaFunction{FileName} \AgdaSymbol{×} \AgdaFunction{FileContent}\\
\AgdaFunction{FileSystem} \AgdaSymbol{:} \AgdaFunction{Set}\\
\AgdaFunction{FileSystem} \AgdaSymbol{=} \AgdaFunction{FileName} \AgdaSymbol{→} \AgdaFunction{Maybe FileContent}\\
\AgdaFunction{MaybeFile} \AgdaSymbol{:} \AgdaFunction{Set}\\
\AgdaFunction{MaybeFile} \AgdaSymbol{=} \AgdaFunction{FileName} \AgdaSymbol{×} \AgdaFunction{Maybe} \AgdaFunction{FileContent}\\
\AgdaFunction{Memory} \AgdaSymbol{:} \AgdaFunction{Set}\\
\AgdaFunction{Memory} \AgdaSymbol{=} \AgdaFunction{List} \AgdaSymbol{(} \AgdaFunction{Cmd} \AgdaSymbol{×} \AgdaFunction{List} \AgdaFunction{MaybeFile} \AgdaSymbol{)}\\
\bigskip
\AgdaFunction{Cmd} \AgdaSymbol{:} \AgdaFunction{Set}\\
\AgdaFunction{Cmd} \AgdaSymbol{=} \AgdaFunction{String}\\
\AgdaFunction{Build} \AgdaSymbol{:} \AgdaFunction{Set}\\
\AgdaFunction{Build} \AgdaSymbol{=} \AgdaFunction{List Cmd}\\
\end{flushleft}
\caption{Table of key types used in our model.}
\label{fig:types}
\end{figure}

\section{Modeling Forward Build Systems}
\label{sec:model}

In this section we describe the framework in which we model both the forward build systems \Fabricate (\S\ref{sec:forward}) and \Rattle (\S\ref{sec:proof}). In particular we aim to prove that the build systems are equivalent to running the underlying shell script, which we define as the function \AgdaFunction{shell}. All these definitions are in \Agda \cite{agda} \footnote{The full source is available at \url{https://github.com/spall/rattle-model/tree/paper_version_final}}. The most important types are given in Figure \ref{fig:types}.

\subsection{Modeling Files}

For any build system, files are an inherently important aspect. Therefore we define:

\begin{enumerate}
  \item \AgdaFunction{FileName}, being a \AgdaFunction{String} which represents a path to a specific file on the file system.
  \item \AgdaFunction{FileContent}, being a \AgdaFunction{String} which represents the contents of a specific file on the file system.
  \item \AgdaFunction{File}, being a pair of \AgdaFunction{FileName} and \AgdaFunction{FileContent} which describes a particular file.
  \item \AgdaFunction{FileSystem}, being a mapping from a \AgdaFunction{FileName} to a \AgdaFunction{Maybe FileContent}, where \AgdaFunction{Nothing} represents that the file does not exist in the \AgdaFunction{FileSystem}.
\end{enumerate}

\subsection{Modeling Commands}

From the view of \Rattle, a build is just a series of commands it is given to execute, so we model a \AgdaFunction{Build} as a list of commands.  We define a command as \AgdaFunction{Cmd}, represented as a \AgdaFunction{String}, although the choice of representation is not too important -- they need equality but little else.  The actions of a command, such as \texttt{gcc -c file.c} depend on the \AgdaFunction{FileSystem} it is run on.  It isn't sufficient for a command to be modeled by something static, the result of the command depends on \texttt{gcc} and \texttt{file.c}, and in particular the header files accessed depend on the contents of \texttt{file.c}.  We model the effect of a \AgdaFunction{Cmd} as a function, \AgdaFunction{CmdFunction}, which describes the actions of the command and reports the files read and written to by the command when run on a \AgdaFunction{FileSystem}.

\cmdF{}

There is an \AgdaFunction{Oracle} for converting a \AgdaFunction{Cmd} to a \AgdaFunction{CmdFunction}.  \AgdaFunction{CmdFunction}s are deterministic, so the result of a \AgdaFunction{CmdFunction} will be equivalent for equivalent \AgdaFunction{FileSystem}s.  But, our intuition tells us a command will have the same result when run on many different \AgdaFunction{FileSystem} values, as long as the files the command depends on are the same.  So, the \AgdaFunction{Oracle} maps a \AgdaFunction{Cmd} to a dependent product of a \AgdaFunction{CmdFunction} and a \AgdaFunction{CmdProof}.  \AgdaFunction{CmdProof} provides us with evidence that for any two \AgdaFunction{FileSystem}s,  $s_1$ and $s_2$, the \AgdaFunction{CmdFunction} $f$ will produce an equivalent result when run on both $s_1$ and $s_2$ if the files read by the command according to $f$ have the same value in both $s_1$ and $s_2$.
Although \Rattle allows adjusting equality comparison using hash forwarding, thus supporting commands that are only weakly deterministic, our model assumes strong determinism (that is, that commands produce identical outputs on identical inputs) for \\ command outputs because weak determinism would provide us with no additional expressiveness. Modeling weak determinism  would require adding a function which maps all equivalent \AgdaFunction{FileContent}s to the same \AgdaFunction{FileContent}, but no other changes to the model.     

\oracle{}

\cmdP{}

Stated alternatively, while the \AgdaFunction{CmdFunction} takes the entire \AgdaFunction{FileSystem}, its result is only dependent on the subset of the \AgdaFunction{FileSystem} it claims to use. This property matches the assumptions of determinism and accurate tracing from \S\ref{sec:assumptions}.

Running a \AgdaFunction{Cmd}, using \AgdaFunction{run}, extends the \AgdaFunction{FileSystem} with the files the \AgdaFunction{CmdFunction} writes to.

\run{}

Our reference behavior is that which happens when a build is executed as if it was a script, i.e. when the commands are executed with no incrementality or other optimizations. To express that concept, we define a function \AgdaFunction{script}, which executes a \AgdaFunction{Build} by calling \AgdaFunction{run} on each command in the build.  This \AgdaFunction{script} function is used to prove that optimisations preserve the reference behavior.

\script{}

\subsection{Preconditions and Simplifications}

Our model has certain preconditions and simplifications based on \Rattle's behavior, which we outline here.

\begin{description}
\item[Builds cannot contain duplicate commands] Some of our lemmas explicitly require that a \AgdaFunction{Build} contain no duplicate commands.  This assumption is sufficient to model \Rattle builds because \Rattle does not run duplicate commands.  If a command occurs more than once in a build it will be skipped on subsequent appearances.
\item[Builds are static lists] Our model makes the simplifying assumption that builds are static lists.  \Rattle provides support for monadic builds, meaning the result of previous commands can influence the future commands run. While convenient for users, monadic commands complicate the proof, and if the proof is considered as a consequence of the initial \AgdaFunction{FileSystem}, provides no additional expressive power.
\item[Commands are run sequentially] \Rattle can support parallel builds, but our model does not explicitly model parallelism.  Instead, builds are a list of commands that run sequentially, each modifying a \AgdaFunction{FileSystem} in sequence. In reality, \Rattle is only able to detect which files were accessed after a command completes, so to be conservative and report all possible hazards, it assumes all reads happened at the beginning of the command and all writes at the end.  We simplify our model to ignore parallelism, but any successful parallel interleaving can be encoded by having multiple distinct commands writing to temporary files, which composed together form the full command (capturing the early read and late write that \Rattle assumes).  If the commands are not atomic, and thus have a different result when run sequentially, then there must be a \emph{hazard} in the overall build, which would be detected in the sequentialization as well as in the actual parallel execution. 
\end{description}

In our \Agda model some of these preconditions are captured as \AgdaFunction{PreCond} \AgdaBound{s} \AgdaBound{br} \AgdaBound{bs}, where \AgdaBound{s} is a \AgdaFunction{FileSystem}, \AgdaBound{bs} represents the \AgdaFunction{Build} provided as the script, and \AgdaBound{br} represents the \AgdaFunction{Build} that was actually run (which may be different from \AgdaBound{bs} if speculation was involved).

\subsection{Hazards}

\begin{figure*}[t]
\cmdRead{}
\cmdWrote{}
\save{}
\hazard{}
\hazardfree{}
\caption{Data structures capturing hazards and the absence of hazards and associated helper functions.}
\label{fig:haz}
\end{figure*}

In \S\ref{sec:correctness_informal} we introduced \Rattle's notion of correctness. We say a build system is correct if executing the build script gives identical results to the shell script and is also idempotent, or raises a hazard.  We also showed there are builds which won't be idempotent because they contain sequences of commands, which modify each others dependencies in a way that can cause those commands to re-run on an immediate subsequent rebuild.  \Rattle considers such sequences of commands to be \emph{hazardous}, and defines two hazards to describe these problematic sequences, \emph{read before write hazards} and \emph{write before write hazards}.

Hazards also enable \Rattle to detect when it has made an error with speculation.  When \Rattle speculates commands it does so assuming the command is part of the build script and that its dependencies have not changed since it was last run.  But, a speculated command might no longer be part of the build, or its dependencies might have changed, potentially leading to a \emph{speculative write before read hazard}, where a \emph{speculated} command wrote to a file a later non-speculated command read from, indicating \Rattle did not run the commands in the order intended by the build author.  For example:

\begin{verbatim}
gcc -c file.c
gcc -c string.c
gcc -c print.c
gcc file.c string.c print.c -o program
\end{verbatim}

Let's say \Rattle ran the above list of commands, speculating \emph{gcc -c file.c}, which is no longer part of the user's build script.  Then, \emph{gcc file.c string.c print.c -o program} would read a version of \emph{file.o} unintended by the build script's author.  Through speculation \Rattle inadvertently ran a command it was not meant to run, causing a later command to potentially read the wrong data.

In this section we present two data types, \AgdaFunction{Hazard} and \AgdaFunction{HazardFree}, which provide evidence when one of the three hazards has occurred, or evidence that a build contains no hazards, respectively.

The \AgdaFunction{Hazard} data type (Figure \ref{fig:haz}), represents a hazard occuring in the build.  \AgdaFunction{Hazard} is indexed on a \AgdaFunction{FileSystem}, \AgdaFunction{Cmd}, \AgdaFunction{Build}, and \AgdaFunction{FileInfo}, and provides evidence of a hazard occurring after the \AgdaFunction{Cmd} has run on the \AgdaFunction{FileSystem}.  The \AgdaFunction{Build} \AgdaFunction{Hazard} is indexed on, is the \emph{script} build, the one meant to be executed, and is specified for the purpose of deciding if a \emph{speculative write before read hazard} has occurred.  \AgdaFunction{FileInfo} is a list, used to record the \AgdaFunction{Cmd}s run so far, and the files they read and wrote to, for the purpose of detecting hazards.

\fileinfo{}

\AgdaFunction{Hazard} has a constructor for each of the three types of hazard.  \AgdaFunction{ReadWrite} and \AgdaFunction{WriteWrite} construct evidence of \emph{read before write} or \emph{write before write} hazards respectively.  A \emph{read before write} or \emph{write before write} hazard has occurred if the command writes to a file a previous command read or wrote to.  \AgdaFunction{ReadWrite} constructs a \AgdaFunction{Hazard} by showing the writes of the \AgdaFunction{Cmd} intersect with the files read by the previous \AgdaFunction{Cmd}s run, as recorded in the \AgdaFunction{FileInfo}.  \AgdaFunction{WriteWrite} constructs a \AgdaFunction{Hazard} by showing the writes of the \AgdaFunction{Cmd} intersect with the files written to by previous \AgdaFunction{Cmd}s, as recorded in the \AgdaFunction{FileInfo}.

Constructing evidence of a \emph{speculative write before read} hazard is more complicated, because whether or not a \emph{speculative write before read} hazard has occurred depends on the order the commands in the build were \emph{meant} to run.  \AgdaFunction{Speculative} constructs evidence of a \emph{speculative write before read} hazard by showing there are two commands, $x_1$ and $x_2$, from those run so far (i.e. recorded in the \AgdaFunction{FileInfo}, or the \AgdaFunction{Cmd} just run, $x$) where the later command, $x_2$ read a file, which the earlier command $x_1$ wrote to, but $x_1$ was not meant to run before $x_2$ , possibly because $x_1$ was never meant to run (\AgdaFunction{x before y ∈ ls} says \AgdaFunction{x} is before \AgdaFunction{y} in the list \AgdaFunction{ls}).
Unlike \AgdaFunction{ReadWrite} and \AgdaFunction{WriteWrite}, which say there exists a \AgdaFunction{Hazard} explicitly involving \AgdaFunction{Cmd} \AgdaFunction{Speculative} says there was a speculative hazard somewhere in the history of the build. When coming up with a representation for \emph{speculative write before read} hazards we realized currently \Rattle does not correctly detect all \emph{speculative write before read} hazards.  Our model assumes a fixed version of \Rattle which correctly detects all speculative hazards. We discuss the specifics of how \Rattle was wrong in \S\ref{sec:bug}.

The converse of the \AgdaFunction{Hazard} data type is the \AgdaFunction{HazardFree} data type (Figure \ref{fig:haz}), which provides evidence a build contains no hazards.  \AgdaFunction{HazardFree} is indexed on a \AgdaFunction{FileSystem}, two \AgdaFunction{Build}s, and a \AgdaFunction{FileInfo}.  The \AgdaFunction{FileSystem} is the one we are running the first \AgdaFunction{Build} in, the second \AgdaFunction{Build} is the \emph{script} build, required to prove there are no \emph{speculative write before read} hazards, and the \AgdaFunction{FileInfo} is the record of the commands run so far.  \AgdaFunction{HazardFree} is an inductive data type with an empty constructor , $[]$, which trivially says that an empty \AgdaFunction{Build} is \AgdaFunction{HazardFree}, and a constructor, $\_::\_$ which says the first command in the build is hazard free, \AgdaFunction{¬ Hazard}, and the rest of the build is \AgdaFunction{HazardFree} after running the first command.

\section{Correctness of \Fabricate}
\label{sec:forward}

\begin{figure*}[t]
  \correctF{}
  \caption{A correctness lemma for \Fabricate.}
  \label{fig:fabcorrect}
 \end{figure*}

A forward build system is fundamentally just a script with support for incrementality.  \Fabricate traces the commands it runs, and records the files they read to decide whether or not they should be run on re-builds.  In this section we present a model of the forward build system \Fabricate \\ \cite{fabricate}, as well as a \emph{correctness} lemma and proof in \Agda. Given the similarity to \Memoize \cite{memoize}, the same model and proofs apply identically.

\subsection{Modeling \Fabricate}

We model \Fabricate by extending \AgdaFunction{script}, described in \S\ref{sec:model} with the use of a \AgdaFunction{Memory} to support incrementality.  \AgdaFunction{Memory} is a list of \AgdaFunction{Cmd}s and a list of \AgdaFunction{File} values (see Figure \ref{fig:types}), which records the commands run and which files they read.
Before we state the definition of \AgdaFunction{fabricate} we define a new function \AgdaFunction{runF} for running commands.  \AgdaFunction{runF} checks if a command should be run, using \AgdaFunction{run?}, which checks if the \AgdaFunction{Cmd} is in the \AgdaFunction{Memory} and the \AgdaFunction{File}s recorded have unchanged values in the \AgdaFunction{FileSystem} and thus would have no effect if run(assuming determinism, see \S\ref{sec:assumptions}).  If \AgdaFunction{run?} says the \AgdaFunction{Cmd} should be run, either because there is no entry for it in the \AgdaFunction{Memory} or the values of the \AgdaFunction{FileName}s stored have changed in the current \AgdaFunction{FileSystem} (\AgdaFunction{get} retrieves the files stored in the \AgdaFunction{Memory} for $x$ and \AgdaFunction{maybeAll} checks if they have changed in the \AgdaFunction{FileSystem}), \AgdaFunction{runF} calls \AgdaFunction{doRun}.  Function{doRun} calls \AgdaFunction{run} defined in \S\ref{sec:model} and extends the Memory with a new entry for the \AgdaFunction{Cmd} run and the files it read; only the files read are recorded just as \Fabricate does.

\runHuh{}

\doRun{}

\runF{}

Finally, below we define \AgdaFunction{fabricate}.  It takes a \AgdaFunction{Build}, \AgdaFunction{FileSystem}, and \AgdaFunction{Memory} and returns a \AgdaFunction{FileSystem} and \AgdaFunction{Memory}, using \AgdaFunction{runF} to run commands.

\forward{}

\subsection{Proving \Fabricate Correct}

In \S\ref{sec:correctness_informal}, we informally stated that a forward build system is correct if for all hazard free builds, executing a build with the forward build system has the same effect as executing it as a script.  So, we state the following correctness theorem for \Fabricate:

\begin{displayquote}
\Fabricate is correct if for all hazard free builds, executing a build with \Fabricate has the same effect as running it as a script.
\end{displayquote}

Figure \ref{fig:fabcorrect} shows the corresponding lemma in \Agda, \AgdaFunction{correct-fabricate}.  It says for a \AgdaFunction{FileSystem}, $s$, and a \AgdaFunction{Build}, $b$, for which we know our pre-conditions are true, and which is \AgdaFunction{HazardFree}, the \AgdaFunction{FileSystem} produced by executing $b$ with \AgdaFunction{fabricate} is equivalent to the \AgdaFunction{FileSystem} produced by executing $b$ with \AgdaFunction{script}.  Two \AgdaFunction{FileSystem}s are equivalent if for all \AgdaFunction{FileName}s, the values are the same in both \AgdaFunction{FileSystem}s.  Essentially, \AgdaFunction{fabricate} and its usage of a \AgdaFunction{Memory} preserves the behavior of \AgdaFunction{script} for \AgdaFunction{HazardFree} builds.

We omit the details of the proof here, but it proves that for each \AgdaFunction{Cmd}, $c$, in the \AgdaFunction{Build}, $b$, the \AgdaFunction{FileSystem} produced by running $c$ with \AgdaFunction{runF} is equivalent to the \AgdaFunction{FileSystem} produced by running $c$ with \AgdaFunction{run}, because if $c$ has been run before then running it is equivalent to not running it because its writes could not have changed in the \AgdaFunction{FileSystem} after $c$ was last run, otherwise there would be a \AgdaFunction{WriteWrite} \AgdaFunction{Hazard}.

Of note, the proof does not require the \AgdaFunction{ReadWrite} hazard free property, although we would require such a property if we were to prove idempotence. The proof also doesn't require \AgdaFunction{Speculative} either as \Fabricate does not perform speculation.

\section{Correctness of Sequential \Rattle}
\label{sec:proof}

In this section we extend our model from \S\ref{sec:model} to \Rattle, and state and prove a correctness lemma for sequential \Rattle.

\subsection{Modeling \Rattle}

We model two variants of \Rattle in order to express the necessary proofs -- \AgdaFunction{rattle-unchecked} which doesn't check for hazards, and \AgdaFunction{rattle} which does. Like \Fabricate, \Rattle offers incrementality through the use of memory, but unlike \Fabricate, \Rattle stores both the files a command read and wrote, rather than just those read. %

We begin by defining a new function for \emph{running} commands, \AgdaFunction{runR}, which is identical to \AgdaFunction{runF}, except it calls \\ \AgdaFunction{doRunR} rather than \AgdaFunction{doRun}.  We omit the body of \AgdaFunction{doRunR} here, because it is the same as \AgdaFunction{doRun} except it records the \AgdaFunction{Cmd}'s writes in the \AgdaFunction{Memory} in addition to the reads.
We then use \AgdaFunction{runR} to define \AgdaFunction{rattle-unchecked} almost identically to \AgdaFunction{fabricate}; it uses \AgdaFunction{runR} rather than \AgdaFunction{run}.

\runR{}

\Rexec{}

When \Rattle executes a build, after each command finishes it checks for the hazards described in \S\ref{sec:hazards}.  \Rattle keeps a record of the files accessed so far in the build, which command accessed the file as well as a timestamp of when the file was accessed, for a read the command's starting timestamp is recorded, and for a write the command's finishing timestamp is recorded.  \Rattle also keeps a record of which commands have been \emph{required} by the build so far, meaning the build script has requested them to run, they were not just run via speculation.  To check for hazards \Rattle compares the files the command which just completed accessed to the files accessed so far.  If the current command wrote to a file after another command read or wrote to that file, then a \emph{read write} hazard or a \emph{write write} hazard has been detected, respectively.  To detect if a \emph{speculative write before read} hazard has occurred, \Rattle checks if a previous command, which was not meant to run before the current command, wrote to a file it read.  \Rattle uses the list of \emph{required} commands to learn the order commands were meant to run in.  If both commands are in the \emph{required} list, and the current command is first, then the commands ran in the wrong order and a \emph{speculative write before read} hazard has occurred.  If only the current command is in the \emph{required} list, then a \emph{speculative write before read} hazard has occurred.  Commands are added to the required list when they are \emph{required} by the build script, regardless of whether or not they were run speculatively.

Because \AgdaFunction{rattle-unchecked} does not check for hazards we also define \AgdaFunction{rattle}, which checks for the hazards described in \S\ref{sec:hazards}, to more closely model \Rattle.  
To facilitate this we define a new function for \emph{running} commands, \AgdaFunction{runWError}.  Just as \AgdaFunction{runR}, \AgdaFunction{runWError} uses \AgdaFunction{run?}, but now it also now performs hazard checking using \AgdaFunction{checkHazard}, before calling \AgdaFunction{doRunR} and adding a new entry to the \AgdaFunction{FileInfo} with \AgdaFunction{rec}.  We omit the implementation of \AgdaFunction{checkHazard} here, but it looks for \AgdaFunction{ReadWrite} and \AgdaFunction{WriteWrite} hazards by checking if the \AgdaFunction{Cmd} just \emph{run} wrote to any files recorded in the \AgdaFunction{FileInfo}.  It checks for \AgdaFunction{Speculative} hazards by seeing if two \AgdaFunction{Cmd}s exist, including the \AgdaFunction{Cmd} just run, where the first \AgdaFunction{Cmd} run wrote to a file the later \AgdaFunction{Cmd} read, but the first \AgdaFunction{Cmd} was not meant to run before the later \AgdaFunction{Cmd}.
Unlike \Rattle, the model has perfect information about which commands were meant to run and the order they were meant to run in.
So, the model could detect all \emph{speculative write before read} hazards as soon as they happen, but to more closely simulate \Rattle, the model waits until the latter command could be detectably \emph{required} by \Rattle to report a \emph{Speculative} hazard.  The model checks if a \AgdaFunction{Cmd} has been \emph{required} by seeing if all of the \AgdaFunction{Cmd}s meant to run before it, are recorded in the \AgdaFunction{FileInfo}.

\checkHazard{}

\runWError{}

Finally,  we define \AgdaFunction{rattle}, which returns either a \AgdaFunction{Hazard} (proving that a hazard was reached), or a \AgdaFunction{FileSystem} and \AgdaFunction{Memory}.  \AgdaFunction{∃Hazard} is an abbreviation for existentials and a \AgdaFunction{Hazard}, because \AgdaFunction{Hazard} is indexed on things we cannot include in \AgdaFunction{rattle}'s type signature.. To support speculation, \AgdaFunction{rattle} takes \emph{two} builds, the build to run, $br$, and the script build supplied by the user, $bs$. In the case of sequential \Rattle, these two are the same.

\exhaz{}

\rattle{}

\begin{figure*}[t]
  \lemmasr{}
  \caption{A lemma stating rattle-unchecked and script produce equivalent FileSystems.}
  \label{lem:equiv}
\end{figure*}

\begin{figure*}[t]
\eqtoscript{}
\correct{}
\soundness{}
\completeness{}
\caption{The correctness lemma for sequential \Rattle, correct-rattle. As well as the soundness and completeness lemmas used to prove it.}
\label{lem:correct}
\label{lem:sound}
\label{lem:complete}
\end{figure*}

\subsection{Correctness of \AgdaFunction{rattle-unchecked}}
\label{subsub:equiv}

Before we prove \AgdaFunction{rattle} is correct, we state and prove \AgdaFunction{rattle-unchecked} is equivalent to \AgdaFunction{script}; that \Rattle which does not check for hazards is equivalent to the \Script.  The lemma, \AgdaFunction{script≡rattle-unchecked}, is stated formally in Figure \ref{lem:equiv}, and says:

\begin{displayquote}
For all builds, $b$, where \AgdaFunction{Cmd}s in $b$ do not write to their reads, executing $b$ with \AgdaFunction{rattle-unchecked} produces a \AgdaFunction{FileSystem} equivalent to the one produced by executing $b$ with \AgdaFunction{script}.
\end{displayquote}

This is analogous to the correctness of \Fabricate in \S\ref{sec:forward}, but notably does \emph{not} include a requirement that the build be hazard free.  This is because \AgdaFunction{rattle-unchecked} records both the files a command read and wrote to, and will re-run a command if the reads or writes of a command have changed since it was last run, rather than just the reads.  Therefore, there is no need to prove the writes of a command haven't changed since it was last run.  Evidence is still required that commands do not write to their own inputs, which is provided by \AgdaFunction{DisjointBuild}, because we need to know the values of the files read by a command are accurately recorded in the Memory, since those values are recorded after a command has \emph{completed}.

\subsection{Correctness of Sequential \Rattle}

We now wish to prove the \emph{correctness} lemma for sequential \Rattle. Following the correctness definition for a forward build system stated in \S\ref{sec:correctness_informal}, we provide the following informal correctness definition for sequential \Rattle:

\begin{displayquote}
Sequential \Rattle is correct if running a build, $b$ with \Rattle, either results in a hazard or a file system equivalent to the one produced by running $b$ as a script.
\end{displayquote}

The formal correctness lemma in \Agda, \AgdaFunction{correct-rattle} is in figure \ref{lem:correct}.

This definition of correctness is faithful to the one in \S\ref{sec:correctness_informal}, because the only hazards in a sequential build, are those caused by problematic sequences of commands in the build script and not due to speculation.  In the remainder of this section we prove soundness and completeness before finally proving the correctness of sequential \Rattle.

\subsubsection{Soundness}

\Rattle is sound if it preserves the semantics of the \Script.  Applied to our \Agda model of \Rattle, \AgdaFunction{rattle} is sound if when it does not find a \AgdaFunction{Hazard}, the resulting \AgdaFunction{FileSystem} is equivalent to the one produced by \AgdaFunction{script}.  In Figure \ref{lem:sound} we formally state a soundness lemma for \AgdaFunction{rattle}:

\begin{displayquote}
\AgdaFunction{rattle} is sound if, when executing a \AgdaFunction{Build}, $br$, if it produces a \AgdaFunction{FileSystem} and not a \AgdaFunction{Hazard}, then \AgdaFunction{script} produces an equivalent \AgdaFunction{FileSystem} when executing $br$.
\end{displayquote}

We omit the details of the proof here, but \AgdaFunction{rattle} is sound because \AgdaFunction{runWError} is sound, meaning the \AgdaFunction{FileSystem} \AgdaFunction{runWError} produces is the same one \AgdaFunction{runR} produces.  The proof of \AgdaFunction{runWError}'s soundness is trivial, so we also omit the details here.

\subsubsection{Completeness}

The final lemma we introduce and prove is the completeness of \AgdaFunction{rattle}.  \Rattle is complete if for all builds with no hazards it does not discover a hazard.  We state a corresponding lemma in our \Agda model for \AgdaFunction{rattle} in Figure \ref{lem:complete}:

\begin{displayquote}
\AgdaFunction{rattle} is complete if for any \AgdaFunction{Build}, $br$, where the standard preconditions are true, and which is \AgdaFunction{HazardFree}, executing $br$ with \AgdaFunction{rattle} produces a \AgdaFunction{FileSystem} and \AgdaFunction{Memory}, and not a \AgdaFunction{Hazard}.
\end{displayquote}

At a high level, the proof of \emph{completeness} shows for each \AgdaFunction{Cmd} in $br$, that \AgdaFunction{runWError} produces a new \AgdaFunction{FileSystem} and \AgdaFunction{Memory} and not a \AgdaFunction{Hazard}, because it would be a contradiction to produce a \AgdaFunction{Hazard}.  Therefore, \AgdaFunction{rattle} will produce a \AgdaFunction{FileSystem} and \AgdaFunction{Memory} rather than a \AgdaFunction{Hazard}.

\subsubsection{Correctness}
The proof of \emph{correctness}, \AgdaFunction{correct-rattle} in figure \ref{lem:correct}, easily follows from \emph{soundness} and \emph{completeness}.  

\begin{figure*}[t]
  \correctS{}
  \caption{A total correctness lemma for \Rattle with speculation, which says the script build has a hazard, or running the re-ordered build has the same effect as running script build. We \emph{cannot} prove this lemma for the current implementation of \Rattle.}
  \label{fig:correct2}
\end{figure*}

\section{Correctness of Speculative \Rattle}

In the previous section we presented a correctness definition for sequential \Rattle and proved sequential \Rattle is correct.  In this section we present a correctness definition for \Rattle with speculation, where \Rattle is executing an ordering of commands other than the one specified by the build author, and prove in its current state that \Rattle is only partly correct.

\Rattle achieves parallelism by \emph{speculating} commands before they are \emph{required} by the build script.  A command is \emph{required} when the build script passes it to the build system with a call to \texttt{cmd}.  If \Rattle already speculated the command it can skip it and immediately continue.  \Rattle uses the commands the build \emph{required} last time it was executed to decide which commands to speculate.  Things can go wrong when running a build with speculation.  First, if the build script has changed since it was last run \Rattle could speculate commands that are no longer part of the build.  Second, the dependencies of the recorded commands could have changed since they were last run causing \Rattle to speculate commands when it shouldn't have.  We limit ourselves to proving what happens in the case where \Rattle only runs commands in the build, but we will briefly discuss how we could extend the lemmas for the case where \Rattle runs unnecessary commands in \S\ref{sec:extra_commands}.

Rather than encode the speculation algorithm used by \Rattle, we instead model two builds, the \emph{script} build, $bs$, which is the one the user wants to execute and the \emph{speculative} build, $br$, which is the build \Rattle actually executes.  The total correctness lemma and the partial correctness lemma's we state below are for all builds, where $br$ is a permutation of $bs$, and thus regardless of the strategy \Rattle chooses for speculating builds, our model and lemmas are correct. These proofs leave \Rattle free to speculate any way it chooses, balancing performance against the likelihood of hazards.

\subsection{Total Correctness (Not Provable)}

We first state a total correctness lemma for \Rattle with speculation, which we cannot prove:

\begin{displayquote}
Speculative \Rattle is correct if either the script build contains a \emph{read before write} or \emph{write before write} hazard, or executing the speculative build with \Rattle is equivalent to executing the script build as a script.
\end{displayquote}

Informally, if the script build has no \emph{read before write} or \emph{write before write} hazards, then whatever speculatively re-ordered build \Rattle decides to run produces a result equivalent to executing the script build.
We state this lemma in \Agda, \AgdaFunction{correct-speculation} in Figure \ref{fig:correct2}, but are not able to prove it.

In practice, \Rattle first runs a build with speculation, and if a hazard occurs, reruns the build without speculation in the hope the hazard was a consequence of speculation rather than inherent in the build. However, it is possible in rare circumstances that speculation might mess up the \emph{inputs} to the build (probably because in previous runs they were outputs). As a result, there might be builds that have hazards caused by speculation, but no real hazards, and cannot be executed equivalent to a script. There are techniques discussed in the \Rattle paper \cite{rattle} to address these (using \texttt{git} for inputs, segregating inputs from outputs), but none are yet implemented in the \Rattle tool. If \Rattle were to address this issue, we believe \Rattle could be proven to be correct by the above definition, but as it is, we seek to prove partial correctness of \Rattle.

\subsection{Partial Correctness}
Our partial correctness theorem weakens the total correctness theorem with an additional clause:

\begin{displayquote}
  Speculative \Rattle is partially correct if either the script build contains a \emph{read before write} or \emph{write before write} hazard, or the speculative build contains a hazard, or executing the speculative build with \Rattle is equivalent to executing the script build as a script.
\end{displayquote}

Specifically, we consider a hazard caused by speculation to be a partially correct execution.  See Figure \ref{fig:semicorrect} for the \emph{partial correctness} lemma and proof in \Agda, \emph{semi-correct}.

\begin{figure*}[t]
\reordered{}
\caption{The reordering lemma for \AgdaFunction{script}, which says for two builds which are permutations of one another, and are \AgdaFunction{HazardFree}, executing both with \AgdaFunction{script} is equivalent.}
\label{fig:reordering}
\end{figure*}

\begin{figure*}[t]
  \correctP{}
  \caption{The partial correctness lemma for speculative \Rattle; which says $br$ has a hazard, or $bs$ has a hazard or running $br$ has the same effect as running $bs$.}
  \label{fig:semicorrect}
\end{figure*}

Before we prove \emph{partial correctness} we require an additional lemma stated below, and in \Agda in figure \ref{fig:reordering}:

\begin{displayquote}
For two builds $br$ and $bs$, which are permutations, where $br$ is \AgdaFunction{HazardFree} with respect to $bs$, executing $br$ with \AgdaFunction{script} produces a \AgdaFunction{FileSystem} equivalent to the one produced by executing $bs$ with \AgdaFunction{script}.
\end{displayquote}

We prove this \emph{reordering} lemma (\AgdaFunction{reordered≡} in Figure \ref{fig:reordering}) by assuming it is true for $br$ and $bs$ with $x$, the last \AgdaFunction{Cmd} in $bs$ removed from both.  And then showing that adding $x$ back to both $br$ and $bs$ still results in equivalent \AgdaFunction{FileSystem}s because $x$ reads and writes to the same files when run as part of $br$ and $bs$, and $x$ does not write to any file another \AgdaFunction{Cmd} in either $br$ or $bs$ read or wrote to, because if it did there would be a \AgdaFunction{Speculative} hazard.

We can now explain our proof of \emph{partial correctness}.  Using the decidability of \AgdaFunction{HazardFree}, we can case on whether or not the speculative build, $br$ is \AgdaFunction{HazardFree} with respect to $bs$.  If it is \AgdaFunction{HazardFree}, we can trivially use \AgdaFunction{soundness}, \AgdaFunction{completeness} and \AgdaFunction{reordered≡} to show that \AgdaFunction{rattle} executing $br$ is equivalent to \AgdaFunction{script} executing $bs$.

\subsection{Correctness with Extra Commands}
\label{sec:extra_commands}

The lemmas in this section state correctness if \Rattle executes a permutation of the script build, but not what happens when \Rattle introduces extra commands.  When \Rattle is executing unnecessary commands we consider the ``output'' of the build to be those files written to by the script build.  We would then say executing a build with extra commands is equivalent to executing the script build, if the files written to by the script build have the same values in both \AgdaFunction{FileSystem}s.  All of our lemmas could be adjusted to support this definition of correctness by stating two \AgdaFunction{FileSystem}s are equivalent for the set of files written to by the script build rather than equivalent for all files.  Future work involves proving these alternative correctness lemmas.

\subsection{Bug Detecting \emph{speculative write before read} Hazards}
\label{sec:bug}

\emph{Speculative write read} hazards occur when a command required by the build script, reads a file, which a \emph{speculated} command has already written to. The implementation of \Rattle currently checks for hazards when a command finishes running, and in order to provide an efficient implementation (hazard checking can be quite expensive) only considers the files that were read or written by that command. Moreover, when a command is required but skipped because it was already run, \Rattle does not recheck for hazards. As a consequence, there is a bug that if the command, $x$, was marked as \emph{speculated} when it went through hazard checking, but later become \emph{required}, \Rattle will potentially fail to detect a \emph{speculative write read} hazard involving $x$.

When attempting to prove and model \Rattle in \Agda it became obvious that if the commands were only considered when they were run, some speculative write read hazards would be missed. In particular, a \AgdaFunction{required?} predicate needed to be tested with the full set of commands, not a prefix, as \Rattle was doing. The fix in the model was to see if the command is ever required, and the fix in \Rattle would be to retest for hazards when a command changes required status.

This flaw in \Rattle was far from obvious, and the tension between efficient implementation and simplicity made it hard to spot. Testing for the correctness properties is hard, especially around speculative hazards, as \Rattle is quite careful as to what it speculates -- making it hard to find a sequence of edits that would cause a problem with a speculative hazard. In contrast, the approach of modeling speculation as producing all permutations makes the bug trivial to find with \Agda (or any proof-based technique).

\section{Related Work}
\label{sec:related}

This paper formalises forward build systems and proves certain properties about these systems. While forward build systems have been around a long time, they are undoubtedly less popular than the more traditional backward build systems. Backward build systems were classified in the Build Systems \`a la Carte paper \cite{build_systems_a_la_carte}, which contains small representative implementations for each type of build system, along with definitions of correctness (that the result is equivalent to rebuilding everything) and minimality (only actions whose dependencies have changed are run). However, the implementations were in Haskell, and the definitions were specified informally. Those definitions were subsequently formalised \cite{coq_a_la_carte}, using Coq to reimplement the build systems. While working towards correctness and minimality (not yet achieved when that paper was published) it was determined that the notion of acyclic tasks (tasks which do not depend on themselves, including indirectly) was an important constraint to prove termination.

In addition to the formalisation of generic build systems, we are aware of two specific backwards build systems that have been had properties proven about them in more depth -- specifically the systems \Pluto and \CloudMake. Starting with \Pluto, the paper introducing \Pluto \cite{erdweg2015sound} included a simplified rebuild algorithm (without cycle support), and then provided manual proofs of soundness (what we call correctness) and optimality (also known as minimality). Compared to our system, soundness property S3 most closely matches our definition, with the property that everything that was built must now be up to date. For their minimality property, the proof relies on a cache so nothing will be built twice (which \Rattle shares) alongside a proof that nothing not required is built. For \Rattle, because speculation might introduce unnecessary work, we do not expect minimality under these definitions. These proofs were reused and extended to apply them to model-driven development \cite{pluto_models}, showing the utility of these proofs. We are aware of an existing project to mechanise these Pluto proofs\footnote{\url{https://wp.doc.ic.ac.uk/vetssannualreport/mechanising-the-theory-of-build-systems/}} and eagerly await the results.

The \CloudMake build system was modeled formally by \citet{formal_cloudmake} using \textsc{Dafny} \cite{dafny}. A \CloudMake build is driven by a script in a mutation-free subset of JavaScript, which is then modeled in \textsc{Dafny}, along with the rebuild algorithm. In particular there is an operation \texttt{exec} that executes an external program, with specific axiomatized properties, which has similarities our \texttt{Cmd} definition. The authors focus is not on proving that the build is correct, but proving that certain optimisations performed by \CloudMake do not change the semantics. Using the framework the authors are able to prove that functions can be evaluated in parallel and that operations are able to be cached, giving the key properties of parallelism and incrementality as an optimisation over the standard build algorithm.

In this paper we have focused on the \Rattle build system, but also included \Memoize and \Fabricate, which can be viewed as a subset of \Rattle without speculation. There are other forward build systems, which contain different features that \Rattle does not. Both \Fac \cite{fac} and \Stroll \cite{stroll} do not require a complete order to be given, meaning there is no corresponding \Shell evaluation order that can be matched. The \LaForge \cite{laforge} system uses tracing to observe execution at a more fine grained level than the user specifies, meaning that commands are sometimes executed in part. A command can be broken into sub-commands, which can be executed in place of the complete command. We believe the our high-level definition of correctness for forward build systems is applicable to these systems, but more work is needed to determine how to model them and prove that they meet this definition. For example, \LaForge's partial execution of commands might result in additional hazards in typical cases. 

\section{Conclusion}
\label{sec:conclusion}

Compilers, operating systems, and web servers need to be correct, and
substantial effort has gone into verifying them. But a humble build
system can sit in between all of these,\footnote{\url{https://www.fireeye.com/blog/threat-research/2020/12/evasive-attacker-leverages-solarwinds-supply-chain-compromises-with-sunburst-backdoor.html}} and thus must be correct as
well.  We modeled \Rattle, a state-of-the-art forward build system,
and proved it correct. In doing so we found a bug which we have a proposed fix for in \Rattle.

While we have used this model to figure out if \Rattle as it stands is
correct, we haven't yet talked about what corrective action \Rattle
could take to recover from hazards. For \emph{speculative} hazards,
\Rattle's current strategy is to rerun without speculation and thus without parallelism, which can
be expensive. The reason for this simplistic strategy is that it's not yet clear what would be both better and correct. Our formal model
will let us figure that out formally in \Agda initially, before transferring the
knowledge to \Rattle itself.

\bibliography{paper}

\end{document}